%% file: dissertation.tex
\documentclass[dissertation,CC-BY-ND]{uathesis}
\usepackage{graphicx}
\usepackage{natbib}	
					
\usepackage{titlesec}

\usepackage{longtable}

\usepackage{bm}

\titleformat{\section}
{\normalfont\normalsize\bfseries}{\thesection}{1em}{}
\titlespacing*{\section}
{0pt}{3.25ex plus 1ex minus .2ex}{1.5ex plus .2ex}

\titleformat{\subsection}
{\normalfont\normalsize\bfseries}{\thesubsection}{1em}{}
\titlespacing*{\subsection}
{0pt}{3.25ex plus 1ex minus .2ex}{1.5ex plus .2ex}

\titleformat{\paragraph}
{\normalfont\normalsize\bfseries}{\theparagraph}{1em}{}
\titlespacing*{\paragraph}
{0pt}{3.25ex plus 1ex minus .2ex}{1.5ex plus .2ex}

\usepackage{amsmath}

\usepackage{amssymb}

\newcommand{\mcO}{\mathcal{O}}
\newcommand{\mcOa}[1]{\mathcal{O}(\alpha_s^{#1})}

\newcommand{\mcL}{\mathcal{L}}
\newcommand{\mcC}{\mathcal{C}}
\newcommand{\mcY}{\mathcal{Y}}

\newcommand{\lalo}{large logarithms}

\newcommand{\tr}{\text{tr}}

\newcommand{\paren}[1]{\left(#1\right)}
\newcommand{\oneov}[1]{\frac{1}{#1}}
\newcommand{\opn}[1]{\operatorname{#1}}
\newcommand{\fpp}[2]{\frac{\partial #1}{\partial #2}}
\newcommand{\sqparen}[1]{\left[#1\right]}
\newcommand{\vpc}{\vspace{1pc}}
\newcommand{\nt}[1]{\noindent \textbf{#1}}

\newcommand{\dg}{\dagger}

\def\Dslash{D\!\!\!\!\slash}

\def\nslash{n\!\!\!\slash}
\def\bnslash{\bar n\!\!\!\slash}

\newcommand{\nn}{\nonumber} 
\newcommand{\bn}{{\bar n}}

\newcommand{\mcdot}{\!\cdot\!}

\newcommand{\be}{\begin{equation}}
\newcommand{\ee}{\end{equation}}

\newcommand{\SCETa}{\mbox{${\rm SCET}_{\rm I}$ }}
\newcommand{\SCETb}{\mbox{${\rm SCET}_{\rm II}$ }}

\newcommand{\vect}[1]{\mathbf{#1}}
\newcommand{\abs}[1]{\left\lvert #1\right\rvert}
\newcommand{\bra}[1]{\left\langle #1\right\rvert}
\newcommand{\ket}[1]{\left\lvert #1\right\rangle}
\newcommand{\braket}[2]{\langle #1\vert #2 \rangle }

\newcommand{\minus}{\!-\!}
\newcommand{\plus}{\!+\!}

\newcommand{\df}{\mathrm{d}}
\newcommand{\Lqcd}{\Lambda_{\text{QCD}}}

\newcommand{\as}{\alpha_s}
\newcommand{\MSbar}{\overline{\text{MS}}}

\newcommand{\cusp}{\mathrm{cusp}}

\newcommand{\Ga}{\Gamma}

\newcommand{\cO}{\mathcal{O}}
\newcommand{\cM}{\mathcal{M}}
\newcommand{\cC}{\mathcal{C}}
\newcommand{\cI}{\mathcal{I}}
\newcommand{\cT}{\mathcal{T}}

\newcommand{\cH}{\mathcal{H}}

\newcommand{\cL}{\mathcal{L}}
\newcommand{\cA}{\mathcal{A}}
\newcommand{\cS}{\mathcal{S}}

\newcommand{\cP}{\mathcal{P}}
\newcommand{\cY}{\mathcal{Y}}

\newcommand{\cV}{\mathcal{V}}

\newcommand{\e}{\epsilon}
\newcommand{\ve}{\varepsilon}

\newcommand{\tL}{\tilde L}
\newcommand{\wt}{\widetilde}

\newcommand{\eq}[1]{Eq.~\eqref{eq:#1}}
\newcommand{\eqs}[2]{Eqs.~\eqref{eq:#1} and \eqref{eq:#2}}
\newcommand{\eqss}[3]{Eqs.~\eqref{eq:#1}, \eqref{eq:#2}, and \eqref{eq:#3}}
\renewcommand{\sec}[1]{Sec.~\ref{sec:#1}}

\newcommand{\appx}[1]{\ref{appx:#1}}
\newcommand{\fig}[1]{Fig.~\ref{fig:#1}}
\newcommand{\figs}[2]{Figs.~\ref{fig:#1} and \ref{fig:#2}}
\newcommand{\figss}[3]{Figs.~\ref{fig:#1}, \ref{fig:#2}, and  \ref{fig:#3}}
\newcommand{\tab}[1]{Table~\ref{tab:#1}}

\newcommand{\qB}{q_B}
\newcommand{\qJ}{q_J}
\newcommand{\qBJ}{q_{B,J}}

\newcommand{\nB}{n_B}
\newcommand{\nJ}{n_J}

\newcommand{\wB}{\omega_B}
\newcommand{\wJ}{\omega_J}

\newcommand{\eec}{$ee$}
\newcommand{\ep}{$ep$}
\newcommand{\pp}{$pp$}
\newcommand{\eeeppp}{\mbox{$ee,\,ep,\,{\rm and}\, pp$ }}

\DeclareMathOperator{\Tr}{Tr}
\DeclareMathOperator{\Li}{Li}
\DeclareMathOperator{\PV}{P.\!V.}
\newcommand{\F}{{}_2F_{1}}
\newcommand{\Fiii}{{}_3F_{2}}

\usepackage{slashed}

\newcommand{\tft}{\tilde{f}^\perp}
\newcommand{\tS}{\tilde{S}}

\newcommand{\LQCD}{\Lambda_{\text{QCD}}}

\newcommand{\req}[1]{Eq.\,\eqref{#1}}

\newcommand{\beq}{\begin{equation}}
\newcommand{\eeq}{\end{equation}}

\newcommand{\sceti}{\mbox{${\rm SCET}_{\rm I}$ }}
\newcommand{\scetii}{\mbox{${\rm SCET}_{\rm II}$ }}

\newcommand{\bea}{\begin{eqnarray}}
\newcommand{\eea}{\end{eqnarray}}

\newcommand{\lqcd}{\Lambda_{\text{QCD}}}

\newcommand{\arccosh}{\text{arccosh}}

\newcommand{\mev}{\text{MeV}}

\newcommand{\pd}{\partial}

\newcommand{\mba}{\mathbb{A}}
\newcommand{\mbv}{\mathbb{V}}
\newcommand{\msl}{\mathcal{L}}

\newcommand{\vpa}{\vec P_a}
\newcommand{\vpad}{\vec P_a^\dg}
\newcommand{\vpb}{\vec P_b}

\newcommand{\hmpi}{\hat M_\pi}
\newcommand{\hmpid}{\hat M_\pi^\dg}
\newcommand{\vnab}{\vec{\nabla}}
\newcommand{\lnlo}{\mathcal{L}_{N.L.O.}}

\newcommand{\olra}[1]{\overleftrightarrow{#1}}

\newcommand{\Aslash}{A\!\!\!\!\!\!\not\,\,}

\usepackage{lineno}

\usepackage{adjustbox}
\usepackage{multirow}

\makeatletter
\def\makeLineNumberLeft{%
  \linenumberfont\llap{\hb@xt@\linenumberwidth{\LineNumber\hss}\hskip\linenumbersep}
  \hskip\columnwidth
  \rlap{\hskip\linenumbersep\hb@xt@\linenumberwidth{\hss\LineNumber}}\hss}
\leftlinenumbers
\makeatother

\usepackage{blindtext}

\newcommand\Tstrut{\rule{0pt}{2.6ex}}       
\newcommand\Bstrut{\rule[-1.2ex]{0pt}{0pt}} 

\def\dbar{{\mathchar'26\mkern-12mu d}^{\mkern4mu 3}}

\newcommand{\ynot}{\left(i\frac{y_0}{\sqrt{2}}\right)^2}

\newcommand{\lpds}{\Lambda_{P.D.S.}}
\newcommand{\mdd}{m_{DD^*}}
\newcommand{\sqrtke}{\sqrt{-\vec{k}^2-i\epsilon}}
\newcommand{\ppie}{\left(\vec{p}_\pi\cdot\vec{\epsilon}\right)}
\newcommand{\ddbarex}{\left(\vec{p}_D\leftrightarrow \vec{p}_{\bar D}\right)}
\newcommand{\re}{\text{Re}}
\newcommand{\oneovponeov}{\left(\frac{1}{\vec{k}^2-\vec{p}_D^2}+\frac{1}{\vec{k}^2-\vec{p}_{\bar D}^2}\right)}

\newcommand{\eps}{\epsilon}
\newcommand{\lam}{\lambda}
\newcommand{\gam}{\gamma}

\newcommand{\sig}{\sigma}
\newcommand{\del}{\delta}
\newcommand{\Del}{\Delta}

\newcommand{\Sig}{\Sigma}

\newcommand{\Gam}{\Gamma}
\newcommand{\Lam}{\Lambda}

\newcommand{\paf}[2]{\paren{\frac{#1}{#2}}}
\newcommand{\ov}[1]{\frac{1}{#1}}

\usepackage{pifont}

\usepackage{multirow}

\newcommand{\lo}{{\rm{L.O.}}}
\newcommand{\nlo}{{\rm{N.L.O.}}}
\newcommand{\nnlo}{{\rm{N.N.L.O.}}}
\newcommand{\full}{{\rm{Full}}}
\newcommand{\dds}{{DD^*}}
\newcommand{\dsd}{{D^*D}}
\newcommand{\pds}{{\rm{P.D.S.}}}

\newcommand{\mbr}{\mathbb{R}}

\usepackage{ mathrsfs }

\completetitle{Effective Field Theories for Quantum Chromo- and Electrodynamics}
\fullname{Ou Zhang}			
\degreename{Doctor of Philosophy}	

\begin{document}

\setboolean{Copyright}{true}
\maketitlepage
{DEPARTMENT OF PHYSICS}	
{2016}							

\approval
{20 July 2016}		
{Sean P. Fleming}	
{Sean P. Fleming}	
{Ubirajara van Kolck}		
{William D. Toussaint}		
{Bjorn Hegelich}		

\statementbyauthor

\incacknowledgements{acknowledgements}

\incdedication{dedication}

\tableofcontents

\listoffigures

\listoftables

\incabstract{abstract}

\include{chapter_0}

\include{chapter_1}

\include{chapter_2}

\include{chapter_3}
\include{chapter_4}

\renewcommand{\baselinestretch}{1}		
\small\normalsize						

\bibliographystyle{uabibnat}
\bibliography{bibliography}

\renewcommand{\baselinestretch}{1.4}		
\small\normalsize						

\appendix
\include{appendix_B}
\include{appendix_C}
\include{appendix_ChPT}
\include{appendix_D}
\include{appendix_E}

\end{document}

%% file: chapter_0.tex
\chapter{Introduction\label{part0}}


\section{Overview} 

In this chapter, I introduce the motivations, principles and methods of Effective Field Theory, which is the foundation of the work in this dissertation. I then present general motivations for my research projects and the organization of this dissertation.

\section{Introduction of Effective Field Theory}

We develop effective field theories on the premise that interactions at low energies, or long distances, are insensitive to the details of the interactions at high energies, or short distances.  When we describe an observable/experiment involving only low energies, we can focus on constructing the Lagrangian with only the degrees of freedom relevant to the dynamics accessed by the energy scale of the experiment. 

For example, in the hydrogen atom, the typical momentum scale is the inverse of the Bohr radius $1/a_0=m_e\alpha$, and the typical energy scale is $m_e\alpha^2$, where $\alpha(m_e)=\frac{e^2}{4\pi}=\frac{1}{137}$ is called the hyperfine coupling constant in this system.  At these momentum and energy scales,
 we do not need to know the mass of the boson mediating nuclear interactions, the pion.  In fact, we  calculate the hydrogen energy levels by implicitly ignoring the physics above a certain momentum scale $\Lambda$, which is larger than $m_e$, with an error of order $m_e\alpha/\Lambda$.  As the energy scale of the experiment/probe increases, or the desired accuracy increases, we may add the $\alpha m_e/\Lambda$ corrections to the hydrogen atom system.

This expansion in small ratios of physical scales has a rigorous mathematical basis in the analytic structure of correlation functions.  The complete set of correlation functions encode all the content of a quantum field theory but in general depend on the momenta of external states in a complicated way.  
Their intricate structure (in the space of incoming and outgoing momenta) arises from poles and branch cuts corresponding to possible intermediate states allowed by kinematics.  They can be somewhat simplified when the external momenta are far away from the kinematic region capable of producing some intermediate states, and in such regions, far away from the corresponding nonanalyticities, we expect that correlation functions are well-approximated by the first few terms of a power series expansion in the momenta. For instance, consider neutron decay $n\to p+e^-+\bar\nu_e$ at tree level, which in the Standard Model is the result of a $d$ quark decay $d\to u+e^-+\bar\nu_e$ through a $W$ boson.   The mass difference between the proton and the neutron is about 1.3 MeV, an energy scale ($10^{-5}$) one millionth of the mass of $W$ boson $m_W$, and it is natural that the $W$ propagator can be expanded according to
\begin{equation}\label{Wpropagatorexpansion}
\frac{1}{p^2-m_W^2}=-\frac{1}{m_W^2}-\frac{p^2}{m_W^4}+\mathcal{O}\left(\frac{p^4}{m_W^6}\right)\,.
\end{equation}
Here $p^2$ is much too small to create a real $W$ boson, and the $W$ remains virtual, being far away from its mass shell and far away from pole of the $W$ propagator.

Expanding the propagator in a power series simplifies the calculation far away from the pole corresponding to the on-mass-shell $W$ ($p^2=m_W^2$).  However, in general there are other non-analyticities in the complex $p^2$-plane that arise from multiple-particle states and other nonperturbative physics.  Taking these into account requires a more systematic approach.  Instead of correlation functions, we turn our attention to the Lagrangian  and expand it in powers of the (small) external momenta divided by a (large) momentum scale with terms of the series composed of local operators involving only the low-momentum degrees of freedom. This expanded Lagrangian is called an {\bf effective field theory} and in practice we truncate according to desired accuracy for the process of interest.

Systematic elaboration of this philosophy makes the effective field theory method a very important and powerful tool for exploring unknown physics. In general there are two approaches to develop an effective field theory: the top-down approach and the bottom-up approach.  This dissertation exhibits two examples of the top-down approach, which is constructed starting from an underlying theory, also sometimes called the ``full theory.''  The underlying theory  is well-understood at some high energy scale but too complicated or even impossible to calculate to fixed perturbative order at a low energy scale.  Top-down effective theories provide a simpler framework capturing the most essential physics in a manner that can be corrected to arbitrary precision, especially for physical processes involving several hierarchically-separated energy scales. The bottom-up approach is a completely different view of constructing effective field theory because we may not know whether an underlying theory at high energy scale exists.  
This approach is based on a `folk theorem' of \cite{weinberg1979phenomenological}:  `If one writes down the most general possible Lagrangian, including all terms consistent with assumed symmetry principles, and then calculates matrix elements with this Lagrangian to any given order of perturbation theory, the result will simply be the most general possible S-matrix consistent with analyticity, perturbative unitarity, cluster decomposition, and the assumed symmetry principles.'  This allows constructing a theory and predicting dynamics knowing only the observed low-energy degrees of freedom and symmetries. In this thesis, I also present one example of this type of EFT.

Both approaches are governed by one set of principles concerning the interactions appearing in the effective field theory (\cite{georgilectures, kaplanlectures, manoharlectures}).
These general principles are:
\begin{enumerate}
\item The nature of the underlying theory is encoded as interaction couplings and symmetries in the low energy effective field theory
\item The effective field theory has an infinite number of operators which are the terms composing a series expansion of the underlying theory. For a given precision, determined by power-counting, one only needs a finite subset of these operators, therefore truncating this series. At this fixed precision, this finite subset of operators closes under renormalization.
\item The effective field theory exhibits the same low-energy (infrared) physics as the underlying theory but generally different high-energy (ultraviolet) properties.
\end{enumerate}



\section{General Method for EFT Construction}\label{sect:I.2} 
These principles suggest a general procedure for the construction of effective field theories.

First, one must separate physical scales in the processes of interest. In this step, one usually sets a typical momentum scale $p$ and a much larger scale $\Lambda$ as the scale that the corresponding effective field theory breaks down, such that one can expand the series of operators in powers of $p/\Lambda$. However a physical process often involves many scales below the breaking scale $\Lambda$, so that one must keep track of all these scales and find expansion parameters to construct the operators connecting with observables. The systematic organization of expansion parameters is called {\bf power counting}.

After setting up the large breaking scale $\Lambda$, one can focus on determining relevant physical degrees of freedom for momenta below the scale $\Lambda$, which is mentioned previously as the low energy region. To be rigorous, when writing a top-down effective field theory, we start with the action and degrees of freedom of the underlying theory, whose scattering amplitudes and observables are obtained from a generating functional,
\begin{align}
\mathcal{Z}=\int\mathcal{D}\Psi\:\exp\left(i\int d^Dx \cL_{\rm High}[\Psi]\right),
\end{align}
the `High' subscript indicating the lagrangian contains the dynamics of the (set of) $\Psi$ fields in the underlying theory at/above the high-energy scale $\Lambda$.
To obtain an effective field theory action, we project the heavy fields or field components scaling as $\Lambda$ relative to other fields in the underlying theory, decouple them from the light components, and integrate them out by performing the path integral.  We obtain a new generating functional for the light (effective) fields $\psi$
\begin{align}
\mathcal{Z} = \int\mathcal{D}\psi\:\exp\left(i\int d^Dx \cL_{\rm Low}[\psi]\right),
\end{align}
with the effective lagrangian given by
\begin{align}
\int d^Dx \cL_{\rm Low}[\psi] = -i\ln\int\mathcal{D}\Psi_h\:\exp\left(i\int d^Dx \cL_{\rm High}[\Psi]\right),
\end{align}
where $\Psi_h$ are the heavy fields or field components in the set of $\Psi$ fields of the underlying theory.
 Indeed, in three examples encountered below (soft collinear effective theory, heavy quark effective theory, and laser effective theory), the two Lagrangians are related by integrating out two components of the Dirac spinors appearing in $\mathcal{L}_{\rm High}$. If the underlying theory is unknown or when it is too difficult to integrate out heavy particles from the full theory, we take the bottom-up approach to obtain relevant effective operators, namely, identifying all symmetries preserved below the large scale $\Lambda$ and construct the interactions accordingly.

In both approaches, the result is a series expansion in a set of fixed, small power-counting parameters $\{\lambda_i\}$ that are composed of the ratios of different scales,
\begin{align}
\mathcal{L}_{\text{Low}}&=\mathcal{L}_{\text{Low}}^{(0)}(\{\lambda_i\},\{c_i\})+\mathcal{L}_{\text{Low}}^{(1)}(\{\lambda_i\},\{c_i\})+\ldots\nn\\
&=\sum_{n=0}^\infty \mathcal{L}_{\text{Low}}^{(n)}(\{\lambda_i\},\{c_i\})  \,, \label{eq:ChapterI.2-general-expansion}
\end{align}
where $\mathcal{L}_{\text{Low}}^{(n)}$s are the low-energy effective field theory Lagrangians, and $\{c_i\}$ are the coefficients of the interaction terms containing the physics from the high energy scale $\Lambda$.  After expanding, the fields appearing in $\mathcal{L}_{\rm High}$ and $\mathcal{L}_{\rm Low}$ need not be the same. 

Because the effective field theory is built below the scale $\Lambda$, dynamics in the underlying theory above the scale $\Lambda$, both leading and subleading orders, must be encoded into the low-energy effective field theory. This procedure is called {\bf matching}, and the physics is encoded into the low energy theory through the set of coefficients $\{c_i\}$ in \eqref{eq:ChapterI.2-general-expansion}, also known as {\bf Wilson coefficients}.  However, seeing that the series in \eqref{eq:ChapterI.2-general-expansion} continues to infinite order, adding progressively more high-scale physics at each order, one may wonder how to consistently truncate the series keeping both the calculation simple and the dynamics accurate.  It is a very important and nontrivial outcome that the truncation is achieved by power-counting these operators up to the desired precision with the help of dimensional analysis and the renormalization group.

After matching, one needs to connect the effective Lagrangian to the observables at the required low energy scale by solving the renormalization group equation. This step re-organizes the perturbation series of underlying-theory matrix elements, to connect to physical observables by extracting and summing non-converging or even non-perturbative terms in the original series, and creating a new perturbative expansion series around the set of parameters $\{\lambda_i\}$.  We often refer to this step as ``resummation.''

\section{EFTs on Quantum Chromo- and Electro-Dynamics}

In this dissertation, we will use effective field theories of both types discussed in Section \ref{sect:I.2}: bottom-up and top-down.

A well-known example of bottom-up type is Chiral Perturbation Theory (ChPT), which is an EFT for hadrons interacting at momenta $p\lesssim$ 1\,GeV.  Hadrons, the experimentally relevant degrees of freedom at this low-energy scale, are known to be composed of quarks and gluons thanks to high-energy scattering experiments that probe their internal structure.  However, the momentum region below 1\,GeV is highly non-perturbative with respect to quark and gluon interactions, which are described in Quantum Chromodynamics (QCD).  Consequently, 
the matching from QCD to ChPT is very difficult to carry out. Instead, we write down all possible interactions consistent with the parity, charge conjugation and approximate chiral symmetries of QCD to study the hadron dynamics (\cite{Leutwyler1994}). This shows how a hierarchy of effective theories may be necessary, each theory incorporating symmetries and certain properties of the underlying theory, but possibly based on different expansions at each step in the hierarchy.
Another example of a bottom-up EFT is the Standard Model of particle physics, though in this case the underlying theory is unknown.

In this dissertation, I present a non-relativistic effective field theory based on heavy hadron chiral perturbation theory (HH$\chi$PT).  Similar to ChPT, HH$\chi$PT describes low-momentum ($p\lesssim$ 1\,GeV) interactions between hadrons, of which each contains one heavy quark. Using this effective field theory, I describe the scattering between heavy $D$ mesons (composed of one charm quark and one light quark) and light mesons (composed of two light quarks).  I also use this theory to investigate the exotic hadronic state $X(3872)$ as a hadronic molecule.  In this context, we will also encounter the need for multiple resummations.  

Top-down EFTs in both the non-relativstic and ultra-relativistic momentum regions are well-developed starting from QCD and QED as the underlying theory. For non-relativistic particles, these are NRQCD and NRQED to describe the heavy quark-antiquark and electron-positron bound states respectively. For highly relativistic particles, there is soft collinear effective theory (SCET) for the energetic hadronic jets created during collisions in particle accelerators, and an electron-laser effective theory for the radiation by energetic electrons accelerated by high-intensity laser fields.  Here, we encounter large separations of scales, leading to large logarithms, because we wish to calculate corrections due to radiation at energies much smaller than the collision energy or the final energy of the accelerated electron. 

In this dissertation,  I use soft-collinear effective theory  to describe the parton-level QCD interactions in collider experiments in the energy region between 1\,GeV and electroweak scale 90\,GeV.  I focus on soft radiation and elastic limits of the scattering processes where the momentum of the radiation becomes much smaller than the center-of-mass energy. Finally, I present a new effective field theory that I developed for QED processes by ultra-relativistic electrons traveling in a strong classical laser background.


%% file: chapter_1.tex
\chapter{Soft Collinear Effective Theory \label{partI}}

\section{Overview} 

Experiments probing the structure and interactions of hadrons typically involve physics from several different momentum scales, especially due to the confinement and strong-coupling properties of the underlying theory of quantum chromodynamics (QCD) that describes dynamics of the quarks and gluons inside hadrons.  As a result, effective field theories have assumed a critical role in deriving predictions from QCD.  As we introduced in the previous chapter, EFTs have been established using power expansions in forward scattering limit, in heavy quark mass and more generally utilizing hierachically-separated kinematic scales. For QCD predictions in collider experiments, the EFT of choice is the Soft Collinear Effective Field Theory (SCET).

In this chapter, I first review challenges that arise in using perturbative QCD to predict observables associated with high-energy scattering processes in colliders.  I then present the motivations for constructing soft-collinear effective theory from QCD for scattering processes involving highly-relativistic particles, and then connect SCET with observables using factorization theorems that manifest the separation of physical scales. 
After this, I present my work under two versions of SCET associated with different power-counting schemes, the first focusing on inclusive and the second on exclusive and semi-inclusive processes.

\section{General Physics of Perturbative QCD} \label{app:QCDbasics}  

Quantum chromodynamics is a gauge theory describing the interactions between particles with color charge: quarks and gluons.  The gauge group of QCD is $SU(3)$, which has a non-Abelian Lie algebra.\footnote{Textbooks reviewing these topics include \cite{field1989applications,Sterman:1991ay,peskin1995introduction,Weinberg:1995mt,nakamura2010review}.}  As a result, the degrees of freedom associated with the gauge field, the gluons, also carry color charge and interact with other gluons.  The QCD Lagrangian reads
\begin{equation}\label{QCDlagrangian}
\mathcal{L}=\sum_q\bar\psi_{q}(i\gamma^\mu D_{\mu}-m_q)\psi_{q}-\frac{1}{4}F_{\mu\nu}^C F^{C,\mu\nu}
\end{equation}
The quark fields $\psi_{q}$ are gauge-covariantly coupled to gluons via the covariant derivative 
\begin{equation}\label{covariantD}
D_{\mu}=\partial_\mu-ig_s t^C A_\mu^C
\end{equation} 
in which $g_s$ is the QCD coupling and $A_\mu^C$ are the gluon fields.  
The combination \req{covariantD} ensures the lagrangian \req{QCDlagrangian} is invariant under local gauge transformations.  $q$ is the quark flavor index, and $C=1,\ldots,8$ is the color index for gluons, which live in the 8-dimensional adjoint representation of $SU(3)$. 
The $\gamma^\mu$ are Dirac matrices, and $t^C$ are the 8 generators of the Lie algebra $\mathfrak{su}(3)$ in the 3-dimensional fundamental representation. Intuitively, in the gluon-quark interaction, the gluon `rotates' the color of quark in the fundamental representation of $SU(3)$.  
$F_{\mu\nu}^C$ is the field strength tensor of the gluon fields, defined through 
\begin{align}
[D_\mu,D_\nu]&=-ig_st^C F_{\mu\nu}^C \\ \label{gluontensor}
F_{\mu\nu}^A&=\partial_\mu A_\nu^A-\partial_\nu A_\mu^A-g_s f_{ABC}A_\mu^B A_\nu^C
\end{align}
where $f_{ABC}$ are structure constants of the Lie algebra $\mathfrak{su}(3)$ defined by
\begin{equation}
[t^A,t^B]=if_{ABC}t^C
\end{equation}
From \req{gluontensor}, it is easy to see that interaction vertices in Feynman rules of QCD include a 3-gluon vertex with coefficient $g_s$, a 4-gluon vertex with coefficient $g_s^2$, a consequence of the non-Abelian algebra.  See \cite{ellis2003qcd} for an enumeration of the Feynman rules.

Although quarks and gluons are the fundamental dynamical degrees of freedom of QCD, they are not observed as free particles.  Only baryons or mesons that live in the trivial representation of the gauge group are observed (called color-singlet or color neutral). This phenomenon is called confinement. Lattice gauge theory and perturbative QCD (pQCD), expanding with respect to the coupling $g_s$, are the only first-principles methods for making predictions in QCD.  However, as we will see, relating predictions from lattice or pQCD to experimental observables often requires introducing an effective theory.

\subsection{Parameters of the QCD Lagrangian: Running coupling and quark mass} 

Observables in perturbative QCD are expressed in terms of the renormalized coupling constant $\alpha_s(\mu_R^2)$ that depends on a renormalization scale $\mu_R$.  The scale is arbitrary and physical observables do not depend on the choice of $\mu_R$.  On the other hand, for $Q$ a typical momentum transfer in a process, $\alpha_s(Q^2)$ is representative of the effective magnitude of the strong interaction in that process. The dependence of the coupling on momentum scale obeys a renormalization group equation (RGE):
\begin{equation}\label{eq:QCDRGE}
\mu_R^2\frac{d\alpha_s}{d\mu_R^2}=\beta(\alpha_s)=-(b_0\alpha_s^2+b_1\alpha_s^3+b_2\alpha_s^4+\ldots)
\end{equation}
where 
\begin{equation}
b_0=\frac{11C_A-4n_fT_R}{12\pi}\,, 
\quad
b_1=\frac{17C_A^2-n_fT_R(10C_A+6C_F)}{24\pi^2} 
\end{equation}
are the 1-loop and 2-loop $\beta$ function coefficients respectively.  The color structures in the corresponding loop diagrams give rise to the group theoretic constants
\begin{equation}\label{eq:QCDcolorfactors}
C_F = \frac{N_c^2-1}{2N_c},\qquad
C_A = N_c,\qquad
T_R = \frac{1}{2}
\end{equation}
with $N_c=3$ the number of colors.  The number of flavors $n_f$ depends on the momentum scale, but never exceeds 6, meaning that $b_0,b_1$ are positive.
$b_2, b_3$ and higher orders depend on the renormalization scheme, of which the most popular in QCD is the modified minimal subtraction (MS) scheme (\cite{bardeen1978deep}).
This dependence of the coupling strength on momentum scale is referred to as {\bf running coupling}. 

The minus sign in \eqref{eq:QCDRGE} combined with the positivity of $b_0,b_1$ indicates {\bf asymptotic freedom} (\cite{gross1973ultraviolet}, \cite{politzer1973reliable}): the coupling strength $\alpha_s$ weakens at large momentum transfer i.e. hard processes.  For example, $\alpha_s\sim 0.1$ for $Q\sim 100\:\text{GeV}-1\:\text{TeV}$ while $\alpha_s$ becomes fairly large around and below 1 GeV.

Due to the confinement property of strong interactions, all $u,d,s,c,b$ quarks form meson or baryons on a timescale $\sim 1/\LQCD$, a process known as hadronization (the $t$ quark decays on timescale shorter than $1/\LQCD$). This adds significant complication to the concept of quark mass. One may define quark mass perturbatively by the location of the pole in the propagator; though close to the physical depiction of mass, this prescription is afflicted by considerable ambiguities (see for example \cite{beneke1999renormalons}). Using the $\overline{MS}$ mass $\bar m_q(\mu_R^2)$, a function of renormalization scale $\mu_R$, is a good alternative, but means quark masses, like the coupling, exhibit dependence on $\mu_R$ that must be removed in observables.

On the other hand, when calculating QCD scattering processes it is conventional to treat as massless the quarks whose masses are much lower than the momentum transfer in the process.

\subsection{Structure of QCD predictions} 

In most theoretical calculations for collider experiments, the main ingredient is a factorization theorem.  Most factorizations are quite intuitive: based on the principles of EFT, one considers dynamics on separate momentum scales as independent of each other. For instance, for lepton pair production from a proton-proton ($pp$) collision, the Drell-Yan process, the factorization theorem reads
\begin{equation}
\sigma=\sum_{i,j}\hat\sigma_{ij}\otimes f_{i/P}\otimes f_{j/P}
\end{equation}
where $\hat\sigma_{ij}$ is the partonic cross section for producing two leptons from two initial partons $i$ and $j$.  $f_{i/P}, f_{j/P}$ are parton distribution functions (PDFs), representing the probability of finding partons $i$ and $j$ in the proton. The PDFs and the partonic cross section are separately functions of the partonic momentum fraction and $\otimes$ denotes convolution over these momentum fraction variables. Factorization exists for other processes (e.g. jet production, involving hadrons in the final state) but are more complicated than for the Drell-Yan process.

I now review the basic collider experiments involving strong interaction dynamics, with a rough factorization scheme, as a guideline for the rest of the chapter in which more specialized aspects of these processes are studied in detail using effective field theory.

\subsubsection{Fully inclusive cross sections: $e^+e^-$, $ep$ and $pp$ collisions} 
In positron-electron ($e^+e^-$) collisions, observables that sum over the details of the final state are said to be fully inclusive, and are the easiest in pQCD, because there is also no dependence on hadronic details of the initial state (having no hadrons in the initial state).  An example is the total cross section of electron positron annihilation with hadrons in final state at center of mass energy $Q$.  Normalized to the electroweak prediction,
\begin{equation}
\frac{\sigma(e^+e^-\to \text{hadrons}, Q)}{\sigma(e^+e^-\to \mu^+\mu^-, Q)}\equiv R(Q)=R_{EW}(Q)(1+\delta_{QCD}(Q)),
\end{equation}
where $R_{EW}(Q)$ is the prediction from electroweak theory for the ratio and $\delta_{QCD}(Q)$ the correction from QCD. We confine our discussion to momenta much smaller than the $Z$-boson mass $Q\ll m_Z$ for simplicity, so that the process is governed by photon exchange, and 
\begin{equation}\label{eq:9.8}
\delta_{QCD}(Q)=\sum_{n=1}^\infty c_n\cdot\left(\frac{\alpha_s(Q^2)}{\pi}\right)^n+\mathcal{O}\left(\frac{\Lambda^4}{Q^4}\right)
\end{equation}
where $\{c_n\}$ are the coefficients of the power series expansion in $\alpha_s$ and $\Lambda/Q$ is a ratio of momentum scales characterizing power-suppressed contributions.  The coupling constant depends on the momentum scale, which at first we consider to be the scale of the hard collision $Q^2$.

The numbers of terms in the power series grows quickly from $n$th order to $(n+1)$th order: the coefficients $\{c_n\}$ typically grow factorially and consequently computations in pQCD converge slower than anticipated from the size of $\alpha_s$.  The power corrections $\mathcal{O}(\Lambda^4/Q^4)$ in \req{eq:9.8} incorporate non-perturbative contributions, for example here the neglected $Z$ exchange. Such corrections are present in all high-energy QCD computations, though the power of $\Lambda/Q$ is dependent on the specific observable.  One typically sets up an operator product expansion (OPE) for many processes and observables to count operators corresponding to power-suppressed contributions.  Although this sounds like a bit of the effective field theory principle, only a systematic counting of contributions, from physics at all relevant energy scales, would constitute an effective theory approach.

In \req{eq:9.8} we have evaluated $\alpha_s$ at the scale $Q$ but the result can also be written at some other renormalization scale $\mu_R$ by writing
\begin{equation}\label{eq:9.10}
\delta_{QCD}(Q)=\sum_{n=1}^\infty \bar c_n\left(\frac{\mu_R^2}{Q^2}\right)\cdot\left(\frac{\alpha_s(\mu_R^2)}{\pi}\right)^n+\mathcal{O}\left(\frac{\Lambda^4}{Q^4}\right)
\end{equation}
where $\bar c_1(\mu_R^2/Q^2)\equiv c_1, \bar c_2(\mu_R^2/Q^2)=c_2+\pi b_0c_1\ln(\mu_R^2/Q^2)$ and so on. In the infinite expansion in powers of $\alpha_s$, the $\mu_R$ dependence eventually cancels out and the physical observable is again independent of the unphysical scale $\mu_R$.  Truncating at some finite order $n=N$ leaves residual dependence on $\mu_R$.  The error is of same order as neglected terms, $\sim\mathcal{O}(\alpha_s^{N+1})$, and it is conventional to use this $\mu_R$-scale dependence as a measure of error due to neglected terms. To avoid slow convergence due to large logarithms $\ln^{n-1}(\mu_R^2/Q^2)$ in $\bar c_n$ coefficients when $\mu_R\ll$ or $\gg Q$, it is convenient to choose a central value for $\mu_R\sim Q$. One may then compare predictions for $2\mu_R$ and $\mu_R/2$ to this central value to estimate the uncertainties.


\noindent {\bf Deep Inelastic Scattering (DIS)} demonstrates crucial properties of QCD cross sections in processes involving hadrons in the initial state. In DIS, an electron with four momentum $k$ collides with a proton with momentum $p$ via a highly off-shell photon with momentum $q$ much larger than the proton mass $m_p$.  We define a positive photon virtuality $Q^2\equiv -q^2>0$ with $m_Z\gg Q\gg m_p$ and write a general decomposition
\begin{equation} \label{eq:9.11}
\frac{d^2\sigma}{dx dQ^2}=\frac{4\pi \alpha}{2xQ^4}\left[(1+(1-y)^2)F_2(x,Q^2)-y^2F_L(x,Q^2)\right]
\end{equation}
where $F_2(x,Q^2)$, $F_L(x,Q^2)$ are the structure functions, defined by the interaction of the proton with transversely and longitudinally polarized photons, respectively.  The kinematic variables $x=Q^2/(2p\cdot q)$, $y=(q\cdot p)/(k\cdot p)$, and $\alpha$ is the fine structure constant of electromagnetism. 

The structure functions cannot be obtained from perturbative computations in QCD. Instead, to 0th order ($\alpha_s^0$), we write the structure functions as parton distribution functions (PDFs) $f_{q/p}(x)$ which are non-perturbative in nature,
\begin{equation}\label{eq:9.12}
F_2(x,Q^2)=x\sum_q e_q^2f_{q/p}(x), \qquad F_L(x,Q^2)=0
\end{equation}
The PDF $f_{q/p}(x)$ is interpreted as the number density of type $q$ quark with fraction $x$ of the light-like momentum of the proton.  This definition is based on the `quark-parton model' with incoherent, elastic scattering between the electron and point-like constituents of the proton called partons (\cite{bjorken1969inelastic}).  We are only beginning to extract phenomenological information about them from lattice QCD (c.f \cite{lin2015flavor}, \cite{alexandrou2015strangeness}), and for current, practical purposes, PDFs are derived from data (c.f. \cite{fortewatt2013progress}).  Although inspired by a model, we shall see how matrix elements corresponding to the PDF arise naturally in soft-collinear effective theory.  

Adding corrections higher order in $\alpha_s$, the structure function is
\begin{equation}\label{eq:9.13}
F_2(x,Q^2)=x\sum_{n=0}^\infty \frac{\alpha_s^n(\mu_R^2)}{(2\pi)^n}\sum_{i=q,g}\int_x^1\frac{dz}{z}C_{2,i}^{(n)}(z,Q^2,\mu_R^2,\mu_F^2)f_{i/p}\left(\frac{x}{z},\mu_F^2\right)+\mathcal{O}\left(\frac{\Lambda^2}{Q^2}\right)
\end{equation}
with the zeroth order coefficients $C_{2,q}^{(0)}=e_q^2\delta(1-z), C_{2,g}^{(0)}=0$ corresponding to \req{eq:9.12}.
Similar to \req{eq:9.10}, it is a series in $\alpha_s(\mu_R^2)$ with coefficients $C_{2,i}^{(n)}$ obtained by perturbative computation.  However, before interacting with the photon, the parton may emit a gluon.  As a consequence, the $C_{2,i}^{(n)}$ are also functions of $z$, the ratio of the parton momentum before and after gluon emission, and $z$ must be integrated over to account for all possible ratios. 

As we shall justify below looking at QCD amplitudes, most emissions that change parton momentum are collinear, parallel to the parton.  Regarding the specifically collinear splittings as a change in the structure of proton, one may formally separate the transverse dynamics.  Computing the perturbative corrections in \req{eq:9.13} using dimensional regularization and the $\overline{MS}$ scheme involves an arbitrary choice of scale $\mu_F$ that can be interpreted as a separation scale: transverse momenta greater than $\mu_F$ are incorporated in $C_{2,q}^{(n)}(z,Q^2,\mu_R^2,\mu_F^2)$ and those less than $\mu_F$ are incorporated into PDFs $f_{i/p}(x,\mu_F^2)$. Although this ``collinear factorization'' is commonly viewed as acceptable for sufficiently inclusive observables in processes with a large collisions scale $Q\gg m_p$, one must be cautious about whether such factorization is exhaustive, especially for hadron-collider processes as we shall discuss below (see also \cite{collinssopersterman2004factorization} which examines the proofs of factorization theorems in detail).  

The $\mu_F$-dependence of the PDFs satisfies the Dokshitzer-Gribov-Lipatov-Altarelli-Parisi (DGLAP) equations (\cite{gribov1972deep, Lipatov:1974qm, altarelli1977asymptotic, Dokshitzer:1977sg}) which to leading order (LO) are
\begin{equation} \label{eq:9.14}
\mu_F^2\frac{\partial f_{i/p}(x,\mu_F^2)}{\partial \mu_F^2}=\sum_j \frac{\alpha_s(\mu_F^2)}{2\pi}\int_x^1 \frac{dz}{z} P_{i\rightarrow j}^{(1)}(z)f_{j/p}\left(\frac{x}{z},\mu_F^2\right),
\end{equation}
where $P_{i\rightarrow j}^{(1)}(z)$ is the $\cO(\alpha_s)$ splitting function for a parton of type $i$ to produce a parton of type $j$.  For instance, $P_{q\rightarrow g}^{(1)}(z)=T_R(z^2+(1-z)^2)$.  Choice of $\mu_F$ is arbitrary; just as for renormalization scale, the $\mu_F$-dependence of coefficient functions and of PDFs cancel each other in the full perturbative series with infinitely many terms. Truncation at $N$ terms in the series again results in a residual uncertainty of $\mathcal{O}(\alpha_s^{N+1})$ related to uncertainty in the choice of $\mu_F$. Similar to $\mu_R$, one estimates uncertainty of predictions by changing $\mu_F$. 

``Collinear factorization,'' widely used in pQCD, does not satisfy the stricter definition applied to the factorization theorems in SCET and studied in this dissertation.  As we develop SCET below, we shall see how the separation of transverse and collinear momenta is undertaken at the lagrangian level, allowing for systematic power counting and running of the separation scale.

\noindent {\bf Hadron-hadron collisions} are described by extending the above method to two initial-state hadrons.  The paradigmatic process in this study is the Drell-Yan process, in which a two hadrons $h_1,h_2$ collide to create a lepton pair $\ell\bar\ell$ plus other (unmeasured) particles collectively denoted $X$.  The total inclusive cross section for the Drell-Yan process is
\begin{align}
\sigma(h_1h_2\to \ell\bar\ell+X)&=\sum_{n=0}^\infty \alpha_s^n(\mu_R^2)\sum_{i,j}\int dx_1dx_2 f_{i/h_1}\left(x_1,\mu_F^2\right) f_{j/h_2}\left(x_2,\mu_F^2\right)\nn\\
&\quad \times \hat\sigma_{ij\to \ell\bar\ell+X}^{(n)}\left(x_1x_2s,\mu_R^2,\mu_F^2\right)+\mathcal{O}\left(\frac{\Lambda^2}{M_{\ell\bar\ell}^4}\right)\label{eq:9.15}
\end{align}
where $s=(p_1+p_2)^2$ is squared center-of-mass energy.  In the narrow $W$ boson width approximation, the first $n=0$  term in  the hard partonic cross section $\hat\sigma_{ij\to W+X}^{(0)}(x_1x_ss,\mu_R^2,\mu_F^2)$ is proportional to $\delta(x_1x_2s-M_{\ell\bar\ell}^2)$ and nonvanishing only for choices of indices $i,j$ permitting direct lepton pair creation, e.g. $i=u, j=\bar u$. Higher orders $n\ge 1$ receive contributions from other partonic channels such as $gq$, relaxing the restriction $x_1x_2s=M_{\ell\bar\ell}^2$.

Similar to \req{eq:9.13}, \req{eq:9.15} separates PDFs from hard cross sections.  The PDFs are universal, encoding the non-perturbative distribution of quarks inside the hadron, typically associated with the scale $\LQCD$.  Provided one applies the same factorization scheme in DIS and hadron-hadron collisions, the PDFs derived from DIS processes can be applied without modification to $pp$, $p\bar p$ collisions (\cite{collins1985factorization, collinssopersterman2004factorization}), with the anti-quark distribution in an anti-proton identical to the quark distribution in a proton by invoking CP symmetry.

\subsubsection{Non-fully inclusive cross sections} 
Taking into account details of the final state, we must address the fact that pQCD degrees of freedom are partons whereas QCD final states always consist of color-neutral hadrons.  High energy partons in the final state split into numerous additional partons (called showering) which in turn hadronize. 
Despite the complication, the above fully-inclusive cross sections are not affected, because showering and hadronization are approximately unitary processes: their impact on the overall probability of hard scattering is power-suppressed by the ratio of time scales $\LQCD/Q$.

Less inclusive measurements may be influenced by the final-state dynamics.  For the same reason that naive ``collinear factorization'' can be applied, showering and hadronization do not significantly alter overall energy flow.  Fixed-order perturbation theory may suffice for observables insensitive to anything but the main direction of energy flow (jet rates, event shapes etc.), and using a small number of partons, one can make predictions agreeing well with measurements of the same observable implemented for hadrons.  Later in this chapter, we shall see how SCET improves predictions of event shapes by re-organzing the perturbation series and summing large logarithms in $\LQCD/Q$ that arise from these real-radiation corrections in showering dynamics.

\subsubsection{Soft and collinear limits} 

An important ingredient in the pQCD approach to PDFs as well as the development of SCET is the behaviour of QCD matrix elements in the soft and collinear limits. Let us examine the process $e^+e^-\to n$ partons with momenta $p_1,\ldots,p_n$: the tree-level squared-matrix element is $|M_n(p_1,\ldots,p_n)|^2$ and the phase-space is denoted by $d\Phi_n$. Assume the $n$th particle is a gluon that becomes collinear to another parton $i$ with momentum approaching zero (i.e. becomes soft).  In this limit, the matrix element can be written as
\begin{equation}\label{eq:9.16}
\lim_{\theta_{in}\to 0,E_n\to 0}d\Phi_n|M_n(p_1,\ldots,p_n)|^2=d\Phi_{n-1}|M_{n-1}(p_1,\ldots,p_{n-1})|^2\frac{\alpha_s C_i}{\pi}\frac{d\theta_{in}^2}{\theta_{in}^2}\frac{dE_n}{E_n}
\end{equation}
with $C_i=C_F$ for $i$ a quark and $C_i=C_A$ for $i$ a gluon. The divergences in the inter-parton angle $\theta_{in}\to 0$ and gluon energy $E_n\to 0$ are not integrable. This is also reflected in the divergence structure of the loop diagram. These divergences are important not only because they dominate the typical structure of events (e.g. emissions at a small angle to the hard partons or with low energy), but also because they control the range of observables computable with pQCD.

Although there is a clear separation of scales between the soft and collinear limits lending itself to an effective theory treatment, direct computation within the framework of perturbative QCD is possible.  This is fixed-order perturbation theory, and it is useful to review its techniques and challenges.

\subsubsection{Fixed-order predictions in pQCD}\label{pQCD} 

Perturbative QCD can be used to calculate both inclusive and non-inclusive observables, provided they are defined independent of unmeasurable quantities such as the number of partons in the final state.  To this end, we consider the cross section for events $\sigma_{\mathcal{O}}$ weighted by observable $\mathcal{O}$ that depends on the parton or hadron four-momenta $p_1,\ldots,p_n$ in the final state.  Setting $\mathcal{O}_n\equiv 1$ for all $n$, we recover the total cross section.  To go beyond the total cross section and describe events with greater nuance, we use in $\mathcal{O}_n$ event-shape variables, such as thrust $\tau$, defined below.  Setting $\mathcal{O}_n\equiv \tau(p_1,\ldots,p_n)$, we obtain the average thrust $\langle \tau\rangle=\sigma_{\mathcal{O}}/\sigma_{tot}$.  We can extract the cross section differential in an event-shape variable by setting (for thrust as an example) $\mathcal{O}_n\equiv \delta\big(\tau-\tau(p_1,\ldots,p_n)\big)$, which yields $\sigma_{\mathcal{O}}=d\sigma/d\tau$.

In addition to corrections from QCD radiation, there are power corrections going as $\sim(\LQCD/Q)^n$ to the underlying hard process.  For example, an electroweak process below $m_W$ receives corrections in powers of $p^2/m_W^2$ as indicated by \req{Wpropagatorexpansion}. In what follows we suppress such non-perturbative power correction terms. For most observables we study this is proportional to a single power of $\Lambda/Q$.

For $e^+e^-$ annihilation, the leading order QCD prediction for the cross section weighted by an observable that is vanishing for less than $n$ final state particles reads
\begin{equation}\label{eq:9.17}
\sigma_{\mathcal{O},LO}=\alpha_s^{n-2}(\mu_R^2)\int d\Phi_n|M_n(p_1,\ldots,p_n)|^2 \mathcal{O}_n(p_1,\ldots,p_n)\,,
\end{equation}
where $|M_n|^2$ with related symmetry factors is summed over all subprocesses. In this case, we put all factors of $\alpha_s$ in front, while for processes other than electron-positron annihilation we usually put them inside integral as the center-of-mass energy in the leading-order process is usually not fixed and different renormalization scales $\mu_R$ are chosen for different events according to their kinematics.

For the same observable that is only nonvanishing with $\ge n$ final state particles, the NLO prediction is obtained by adding two pieces to the LO formula \eqref{eq:9.17}:  the $2\to (n+1)$ squared matrix element $|M_{n+1}|^2$ at tree-level, and the mixing between $2\to n$ tree-level and $2\to n$ 1-loop amplitudes, $2\text{Re}(M_nM_{n,\text{1-loop}}^*)$.  In sum,
\begin{align}
\sigma_{\mathcal{O}}^{NLO}&=\sigma_{\mathcal{O}}^{LO}+\alpha_s^{n-1}(\mu_R^2)\int d\Phi_{n+1}|M_{n+1}^2(p_1,\ldots,p_{n+1})|\mathcal{O}_{n+1}(p_1,\ldots,p_{n+1})\nn\\
&\,\,\, +\alpha_s^{n-1}(\mu_R^2)\int d\Phi_n 2\text{Re}\left[ M_n(p_1,\ldots,p_n) M_{n,\text{1-loop}}^*(p_1,\ldots,p_n)\right]\mathcal{O}_n(p_1,\ldots,p_n)\,.
\label{eq:9.18}
\end{align}
Loop amplitudes diverge in $d=4$. Tree-level amplitudes are finite but diverge after being integrated, because of divergences in the phase space measure \eqref{eq:9.16}.
Both divergences originate in the soft and collinear limits, and they cancel each other after integration provided the observable $\mathcal{O}$ satisfies the conditions 
\begin{align}
\mathcal{O}_{n+1}(p_1,\ldots,p_s,\ldots,p_n)&\to \mathcal{O}_n(p_1,\ldots,p_n), \text{ if }p_s\to 0,\nn\\
\mathcal{O}_{n+1}(p_1,\ldots,p_a,p_b,\ldots,p_n)&\to \mathcal{O}_n(p_1,\ldots,p_a+p_b,\ldots,p_n),\text{ if }p_a||p_b,\label{eq:9.19}
\end{align}
which are called infrared and collinear safety.

An event shape distribution is an important example of an infrared-safe quantity that we will utilize in this work. On the other hand, examples of unsafe quantities are: observables that require no radiation in some region of phase space (violated by soft emissions), momentum distributions of the hardest QCD particles (affected by collinear splitting) and particle multiplicity (changed by both soft and collinear emission). Considering that the LO calculation suffices if and only if higher-order terms are shown to be sufficientlly suppressed, non-cancellation of infrared and collinear divergences in unsafe observables jeopardises not only the NLO calculations themselves but also the LO calculations. In addition, infrared and collinear unsafety commonly also suggest large non-perturbative effects. 

\subsection{Resummation} 
Emission in the final state can be strongly constrained by applying stringent conditions to the infrared safe observables, such as requiring the thrust in $e^+e^-$ events is smaller than a certain value $\tau\ll 1$.  Examples we will study extensively in this dissertation are DIS and Drell-Yan processes near the kinematic endpoint, where very little energy is free to contribute to real emission.

Such restrictions cut out a large part of the integration domain over the soft and collinear divergences of \eqref{eq:9.16}. Consequently each power of $\alpha_s$ is multiplied by a large coefficient $\sim L^2$ (where $L$ represents a logarithm, for instance $L=\ln \tau$), because the cancellation is only partial between real emission corrections, which are subject to the restriction, and loop (virtual) corrections which have no restrictions. 
The result is a series with terms $\sim (\alpha_sL^2)^n$, where often $\alpha_sL^2\gg 1$.  Such a power series will not converge very well or may not converge at all. To enhance convergence one performs a procedure called `resummation': by investigating the known properties of virtual corrections to all orders and matrix elements for multiple soft and collinear emissions, one identifies dominant logarithmically enhanced terms to all orders in $\alpha_s$.

When two powers of logarithms accompany each power of $\alpha_s$, we say the series has double logarithmic enhancements or {\bf Sudakov logarithms}. In this case there are two classification schemes for resummation accuracy. The series can be written
\begin{equation}\label{eq:9.20}
\sigma(L)\simeq \sigma_{tot}\sum_{n=0}^\infty \sum_{k=0}^{2n}R_{nk}\alpha_s^n(\mu_R)^2 L^k\,,\,\,\,\, L\gg 1,
\end{equation}
where $\sigma(L)$ is cross section including the phase-space restriction and $\sigma_{tot}$ is the total unconstrained cross section. Leading log (LL) resummation incorporates all terms with $k=2n$, next-to-leading-log (NLL) in addition accounts for terms with $k=2n-1$ etc. For many observables, $\sigma(L)$ or its Fourier/Melin transform  exponentiates, in which case we can write
\begin{equation}\label{eq:9.21}
\sigma(L)\sim \sigma_{tot}\exp\left[\sum_{n=1}^\infty \sum_{k=0}^{n+1}G_{nk}\alpha_s^n(\mu_R^2)L^k\right]\, ,\,\,\, L\gg 1,
\end{equation}
where now the sum on $k$ runs to $n+1$ rather than $2n$ as in \eq{9.20}. 
Compared to the original series, the power of this resummation lies in the fact that $G_{12}$ alone reproduces the full LL sub-series in \req{eq:9.20}. In the context of \eqref{eq:9.21}, LL means one accounts for all terms with $k=n+1$ and NLL accounts for all terms with $k=n$ etc.

Although there are further resummation methods in pQCD to address logarithmically-enhanced contributions, we will now show how they are efficiently handled in soft-collinear effective theory.

\section{Soft-Collinear Effective Theory} 

SCET is an example of a highly-relativistic particle EFT, derived from QCD for the description of interactions between light-like (collinear) and soft particles as corrections to a hard interaction, for example as in jet production at the LHC (\cite{bauer2001effective}, \cite{bauer2009factorization}, \cite{bauer2004enhanced}, \cite{bauer2004shape}).
In this part we introduce the construction, notation and Lagrangian of SCET, and demonstrate the procedure of field redefinition which can be used to factorize soft and collinear radiation.  These topics are reviewed in \cite{1410.1892v2,Agashe:2014kda,Petrov:2016azi}.

\subsection{Power counting}
Consider a highly-relativistic quark moving in the $\vec n$ direction with virtuality $\ll Q$ where $Q\gg m $ is its energy.  Its 4-momentum $p_n^\mu$ is parameterized in light-cone coordinates as
\begin{equation}\label{eq:lightconep}
p_n^\mu\equiv (p^+,p^-,p_\perp)=n\cdot p\frac{\bar n^\mu}{2}+\bar n\cdot p\frac{n^\mu}{2}+p_\perp^\mu
\end{equation}
where the light-like 4-vectors $n^\mu=(1,\vec n), \bar n^\mu=(1,-\vec n)$ are defined by the unit 3-vector $\vec n$.  The 4-vectors satisfy $n^2=\bar n^2=0$, $n\cdot \bar n=2$, showing that $\bar n\cdot p=p^-$, $n\cdot p=p^+$.  This implies $n\cdot p_n^\perp=\bar n\cdot p_n^\perp=0$.
 $(p_n^-,p_n^+,p_n^\perp)=(\bar n\cdot p_n,n\cdot p_n,p_n^\perp)$, recall \req{eq:lightconep}.  In SCET, it is conventional to denote the expansion parameter as $\lambda^2=p^2/Q^2$, where $p^2(\gg m^2)$ is the invariant mass-squared scale of the fluctuations considered in the theory.  With this notation, the 4-momentum scales as  $(p_n^-,p_n^+,p_n^\perp)\sim Q(1,\lambda^2,\lambda)$.\footnote{Other scalings either correspond to $\bar n$ direction particles $p_{\bar n}^\mu\sim Q(\lambda^2,1,\lambda)$, or are highly off-shell (e.g. $q^\mu\sim Q(1,\lambda,\lambda)$ has $q^2\sim \lambda\gg \lambda^2$), or belong to a different invariant mass hyperbola $p^2\sim\lambda^4$ as for soft.} When the collinear particle interacts with soft particles with $(p_s^-,p_s^+,p_s^\perp)\sim Q(\lambda^2,\lambda^2,\lambda^2)$ or energetic particles travelling in the same direction $n^\mu$, the virtuality of this energetic particle is unchanged parametrically. SCET is established to reproduce the low-energy ($\ll Q$) dynamics originating from interactions between collinear and soft degrees of freedom.

In the above power counting we have $p_s^\perp\ll p_n^\perp$ since the transverse momenta of soft degrees of freedom scales as $p_s^\perp\sim Q\lambda^2$ while transverse momenta of collinear fields scales as $p_n^\perp\sim Q\lambda$. This theory with this scaling
\begin{equation}
p_n\sim Q(1,\lambda^2,\lambda), \,\, p_s\sim Q(\lambda^2,\lambda^2,\lambda^2) 
\qquad(\SCETa)
\end{equation}
is called $\SCETa$. On the other hand when external kinematics require $p_n^\perp\sim p_s^\perp$, the corresponding scaling are
\begin{equation}\label{SCETbscaling}
p_n\sim Q(1,\lambda^2,\lambda), \,\, p_s\sim Q(\lambda,\lambda,\lambda),
\qquad(\SCETb)
\end{equation}
and the theory is referred to as $\SCETb$. $\SCETb$ is called for in exclusive hadronic decays (e.g. $\bar B\to D\pi$) in which both soft and collinear virtuality is set by $\Lambda_{QCD}$ as well as in transverse momentum contributions at colliders.

\subsection{Leading-order Lagrangian}\label{sec:SCETlagrangian}
SCET is a top-down effective theory, derived from QCD as the underlying theory.  We begin with a single collinear sector, separating the full QCD field into components $q_n(x)=\psi_n(x)+\Xi_n(x)$ with
\begin{equation}
\psi_n(x)=\frac{\nslash\bnslash}{4}q_n(x),\quad \Xi_n(x)=\frac{\bnslash\nslash}{4}q_n(x).
\end{equation}
Solving the equation of motion for the field $\Xi_n$ shows it suppressed by $(i\bar n\cdot D)^{-1}\slashed{D}_\perp\sim\lambda $ relative to the $\psi_n$, and implying that its fluctuations would be very off-shell involving virtuality $(i\bn\cdot D)^2\sim Q^2$.  Therefore one removes $\Xi_n$ using its equation of motion, yielding
\begin{equation}\label{eq:26eq17.12}
\mcL_n=\bar\psi_n(x)\left[in\cdot D+i\slashed{D}^\perp\frac{1}{i\bar n\cdot D}i\slashed{D}^\perp\right]\frac{\bnslash}{2}\psi_n(x).
\end{equation}
Next we decompose collinear momentum into a `label' and a `residual' momentum $p^\mu=P^\mu+k^\mu$ with $n\cdot P=0$ and $k^\mu \sim (\lambda^2,\lambda^2,\lambda^2)Q$. Then we redefine the collinear field by $\psi_n(x)=e^{iP\cdot x}\xi_n(x)$ and separate the gluon field into collinear and soft sectors 
\begin{equation}\label{SCETgluondecomp}
A_\mu=A_\mu^{n}+A_\mu^{\bar n}+A_\mu^{s}+\ldots
\end{equation}
scaling as $A_\mu^{n}\sim (1,\lambda^2,\lambda)Q$, $A_\mu^{\bar n}\sim (\lambda^2,1,\lambda)Q$ and $A_\mu^{s}\sim (\lambda^2,\lambda^2,\lambda^2)Q$ respectively.  The additional terms `...' in this decomposition are determined by gauge invariance. Label momentum in SCET is not conserved, and we define a label operator $\mathcal{P}^\mu$ acting as $\mathcal{P}^\mu\xi_n(x)=P^\mu \xi_n(x)$ (\cite{bauer2001invariant}) and a covariant label operator $i\mathcal{D}_n^\mu=\mathcal{P}^\mu+gA_{n}^\mu(x)$.  Spatial derivatives now only pick out the residual momentum from $\xi_n(x)$.   Soft gluons do not appear in $i\mathcal{D}_n^\mu$ at leading order in the power counting. Therefore the SCET Lagrangian can be written as (\cite{bauer2001invariant, Bauer:2001yt, Chay:2002vy, Beneke:2002ph, Bauer:2002nz})
\begin{equation}\label{eq:26eq17.13}
\mcL_n=\bar \xi_n(x)\left[in\cdot D_n+gn\cdot A_s+i\slashed{\mathcal{D}}_n^\perp\frac{1}{i\bar n\cdot\mathcal{D}_n}i\slashed{\mathcal{D}}_n^\perp\right]\frac{\bnslash}{2}\xi_n(x)+\ldots
\end{equation}
where $in\cdot D$ is decomposed into a soft ($gn\cdot A_s$) and a collinearly-covariant piece ($in\cdot D_n=in\cdot\partial+gn\cdot A_{n}$).  The soft piece $gn\cdot A_s$ induces only interaction between soft gluons and collinear quark at leading power in $\lambda$. The \ldots represent interactions between soft and collinear degrees of freedom that are higher-order in $\lambda$.

Distinct collinear sectors can be defined for directions satisfying $n_1\cdot n_2>\lambda^2$ and the Lagrangian involving different collinear sectors is the sum $\mcL=\sum_n\mcL_n$ with the same soft-gluon fields coupling to all $n$ directions. Expressing the SCET Lagrangian in position space is another way to see the separation between large and small momentum components (\cite{Beneke:2002ph}) where no label operators appear and dependence on short-distance effects enters through non-localities.  Short-distance physics enters into SCET in two cases:
\begin{enumerate}
\item different collinear sectors are coupled by some external current, for instance the hard collision producing separated jets at electron-positron or hadron colliders;
\item a heavy particle decaying in its rest frame produces collinear particles (e.g. $B$ decays).
\end{enumerate}
Here short-distance fluctuations are incorporated in the Wilson coefficients of external source operators.

\subsection{Symmetries of SCET}
By exploiting the symmetries of SCET, one can write down all possible operators at each order of power counting (\cite{Petrov:2016azi}).

\subsubsection{Reparametrization Invariance}\label{sec:SCETRPI}

Reparameterization invariance (RPI) refers to invariance under the artificial separation of collinear momentum into `label' and `residual' parts.  This symmetry is a remnant of Lorentz invariance of the full theory.  RPIs in SCET take three different forms, written in terms of the change in the collinear direction vector $n^\mu$,
\begin{align}
\rm{Type~I}:\quad & n^\mu\to n^\mu+\Delta_\perp^\mu ,\quad \bn^\mu~\rm{ unchanged,}\label{eq:91eq7.37}\\
\rm{Type~II}:\quad & \bn^\mu \to\bn^\mu +\eps_\perp^\mu, \quad n^\mu ~\rm{ unchanged,}\label{eq:91eq7.38}\\
\rm{Type~III}:\quad & n^\mu\to e^\alpha n^\mu, \qquad \bn^\mu \to e^{-\alpha}\bn^\mu\label{eq:91eq7.39}
\end{align}
where $\Delta_\perp\sim \lam$ and $\alpha,\eps_\perp\sim\lam^0$. To keep the conditions $n^2=\bn^2=0$ and $\bn\cdot n=2$ the parameters of an RPI transformation must be perpendicular vectors. Note also that an arbitrary 4-vector
\be
V^\mu=\frac{\bn^\mu}{2}(n\cdot V)+\frac{n^\mu}{2}(\bn\cdot V)+V_\perp^\mu
\ee
is unchanged under all three types of transformations.

It is easy to see that invariance under Type III (\eq{91eq7.39}) above requires any $n^\mu$ in an operator be accompanied either by an $\bn^\mu$ or an $n^\mu$ in the denominator. This rules out operators like $\bar\xi(i\bn\cdot D)\xi$ with a scaling $\lambda^{-2}$ in the action, because the $\bn$ must be accompanied by $\nslash$ which annihilates $\xi$: $\nslash \xi=0$ as we have seen above.

The SCET fields transform as
\begin{align}
n\cdot D&\to n\cdot D+\Delta_\perp\cdot D_\perp\label{eq:91eq7.40}\\
D_\perp^\mu& \to D_\perp^\mu-\frac{\Delta_\perp^\mu}{2}\bn\cdot D-\frac{\bn^\mu}{2}\Delta_\perp\cdot D_\perp\label{eq:91eq7.41}\\
\xi&\to \paren{1+\frac{\slashed{\Delta}_\perp\slashed{\bn}}{4}}\xi_n\label{eq:91eq7.42}
\end{align}
and $\bn\cdot D$ and $W$ are invariant under Type I (\eq{91eq7.37}) above. On the other hand the transformation rules under Type II (\eq{91eq7.38}) are as follows:
\begin{align}
\bn\cdot D&\to \bn\cdot D+\eps_\perp\cdot D_\perp\label{eq:91eq7.43}\\
D_\perp^\mu&\to D_\perp^\mu-\frac{\eps_\perp^\mu}{2}n\cdot D-\frac{n^\mu}{2}\eps_\perp\cdot D_\perp\label{eq:91eq7.44}\\
\xi_n&\to \paren{1+\frac{\slashed{\eps}_\perp}{2}\oneov{i\bn\cdot D}i\slashed{D}_\perp}\xi_n\label{eq:91eq7.45}\\
W_n&\to \paren{1-\frac{1}{i\bn\cdot D}i\eps_\perp\cdot D_\perp}W_n\label{eq:91eq7.46}
\end{align}
and $n\cdot D$ is invariant.

\subsubsection{Gauge Invariance}

Local gauge invariance is a crucial feature of SCET. The gauge invariance of QCD breaks into individual invariances for collinear and soft degrees of freedom, which enable us to determine the forms of currents and operators in the effective theory.  The effective Lagrangian for each sector is invariant only under residual gauge transformations because the effective theory field operators only describe modes with specific momentum scaling. The residual gauge symmetries satisfy either the soft scaling $(\bar n\cdot\partial,n\cdot\partial,\partial^\perp)U_s(x)\sim Q(\lambda^2,\lambda^2,\lambda^2)U_s(x)$ or the collinear scaling $(\bar n\cdot\partial,n\cdot\partial,\partial^\perp)U_{n}(x)\sim Q(1,\lambda^2,\lambda)U_{n}(x)$.

The idea is that, because there is no coupling between collinear and soft degrees of freedom at leading order in the Lagrangian, the gauge transformations can be decomposed into a collinear part acting only on $A_n^\mu$ and a soft part acting only on $A_{s}^\mu$. We can derive the rules of transformation for all the fields involved by looking at the full gauge symmetries of all the operators, given in \tab{91tab7.1}

As we can see from \tab{91tab7.1}, each field transforms differently under soft or collinear gauge transformations.  The requirement of invariance under both transformations restricts the form of operators. 
In particular, the decomposition of gauge symmetry by sector implies that distinct collinear sectors must be separately gauge invariant since collinear fields in distinct directions transform under the distinct gauge symmetries.  It is convenient to introduce the collinear Wilson line (\cite{bauer2001invariant})
\begin{equation}
W_n(x)=P\exp\left[-ig\int_{-\infty}^0ds\bar n\cdot A_n(s\bar n+x)\right]
\end{equation}
with the transformation law $W_n\to U_{n}W_n$ under collinear gauge symmetry. Since $\psi_n$ obeys the same transformation law, the combination $\chi_n=W_n^\dagger\psi_n$ is gauge invariant. Correspondingly, a collinear gauge-invariant gluon field is defined as (\cite{Bauer:2002nz,Hill:2002vw})
\begin{equation}
B_n^\mu=g^{-1}W_n^\dagger iD_n^\mu W_n\,.
\end{equation}
One uses these gauge-invariant combinations to build collinear operators in SCET.

\begin{table}
\centering
\begin{tabular}{lcccc}
\hline
& Field & Scaling & $U_{n}$& $U_{s}$ \\
 & $\xi_n$ & $\lambda$ & $U_{n}\xi_n$ & $U_{s}\xi_n$ \\
collinear & $A_{n}^\mu$ & $(\lambda^2,1,\lambda)$ & $U_{n}A_{n}^\mu U_{n}^\dag+\frac{i}{g}U_{n}[iD_n^\mu,U_{n}^\dag]$ & $U_{s}A_n^\mu U_{us}^\dag$ \\
 & $W_n$ & 1 & $U_{n}W_n$ & $U_{s}W_nU_{s}^\dag$ \\
\hline
& $q$ & $\lambda^3$ & $q$ & $U_{s}q$ \\
soft & $A_{s}^\mu$ & $\lam^2$ & $A_{s}^\mu$ & $U_{s}(A_{s}^\mu+\frac{i}{g}\pd^\mu)U_{s}^\dag$ \\
& $Y$ & 1 & $Y$ & $U_{s}Y$\\
\hline
\end{tabular}
\caption{Rules for collinear and soft gauge transformations and scaling dimensions of SCET fields}
\label{tab:91tab7.1}
\end{table}

\subsection{Derivation of Factorization Theorems}
SCET may be applied to understand factorization theorems, namely how cross sections involving high momentum particles moving in different directions factorize into simpler pieces that can be computed in pQCD or measured from experiment. Although factorization theorems have been studied even before the birth of SCET (\cite{Collins:1989gx}), effective field theories such as SCET provide a conceptually clearer picture for certain factorization theorems as most reductions happen already at Lagrangian level. In this part we discuss the derivation of some of these factorization theorems, valid to leading order in power counting.

As seen above, collinear particles traveling in different directions do not couple to each other in the SCET Lagrangian and soft gluons couple to them only through $\bar \xi_n gn\cdot A_s\frac{\bnslash}{2}\xi_n$ (see \req{eq:26eq17.13}). One can remove this term by performing a field redefinition (\cite{Bauer:2001yt})
\begin{equation}\label{eq:26eq17.17}
\psi_n(x)=Y_n(x)\psi_n^{(0)}(x), \quad A_n^a(x)=Y_n^{ab}(x)A_n^{b(0)}(x),
\end{equation}
where $Y_n$ is matrix in fundamental representation while $Y_n^{ab}$ is in adjoint representation. This simplification clarifies the proofs of factorization theorems greatly in SCET. 

Consider a typical QCD scattering process involving two incoming collinear particles ($a,b$ directions) and $N$ outgoing particles ($n_1,...,n_N$ directions).  In matching the corresponding QCD  operator $\hat O(x)$ to SCET,  the redefinition \req{eq:26eq17.17} separates all the soft gluon interactions from the $N+2$ collinear operators, resulting in
\begin{equation}\label{eq:26eq17.18}
\hat O(x)=C_{\hat O}(\mu)\mcC_{n_a,\mu}^{(0)}(x)\mcC_{n_b,\mu}^{(0)}(x)\mcC_{n_1,\mu}^{(0)}(x)\ldots \mcC_{n_N,\mu}^{(0)}(x)[\mathcal{Y}_{n_a}\mathcal{Y}_{n_b}\mathcal{Y}_{n_1}\ldots\mathcal{Y}_{n_N}]_\mu(x).
\end{equation}
where $\mcC^{(0)}(x)$ are collinear-gauge-invariant collinear operators, $C_O$ is the hard-matching coefficient incorporating short-distance effects at scale $Q$, and $\mathcal{Y}$s are soft Wilson lines that can either be in color triplet or octet representation.  The subscript $\mu$ on the operator indicates dependence on the renormalization scale $\mu$, which carries over to the squared matrix element.  One squares the matrix element and integrates over phase space of all final-state particles to obtain the cross section. Since collinear particles in distinct directions do not interact, the forward matrix element factorizes
\begin{align}
\bra{in}O(x)O^\dagger(0)\ket{in}=&\:|C_{\hat O}(\mu)|^2\bra{in_a}\mcC_{n_a}(x)\mcC_{n_a}^\dagger(0)\ket{in_a}_\mu\bra{in_b}\mcC_{n_b}(x)\mcC_{n_b}^\dagger(0)\ket{in_b}_\mu \nn\\
&\times\bra{0}\mcC_{n_1}(x)\mcC_{n_1}^\dagger(0)\ket{0}_\mu\times\ldots\bra{0}\mcC_{n_N}(x)\mcC_{n_N}^\dagger(0)\ket{0}_\mu\nn\\
&\times\bra{0}[\mcY_{n_a}\ldots\mcY_{n_N}](x)[\mcY_{n_a}\ldots\mcY_{n_N}]^\dagger(0)\ket{0}_\mu
\,,\label{eq:26eq17.19}
\end{align}
and each of the simpler factors can now be computed separately. There are several types of vacuum matrix elements appearing:
\begin{enumerate}
\item vacuum matrix element of outgoing collinear particles are specified by jet functions $J_i(\mu)$ ($i=1...N$) which can be computed perturbatively when the relevant momentum scale (e.g. jet mass) is large enough;
\item vacuum matrix elements of incoming collinear particles are non-perturbative quantities $B_{p/N}(\mu)$ called {\bf beam functions} for parton of type $p$ in hadron of type $N$ (see \cite{Stewart:2009yx}). In general these beam functions can be expressed perturbatively in terms of parton distribution functions (PDFs);
\item vacuum matrix elements of soft Wilson lines are defined into soft functions $S_{ab\ldots N}(\mu)$.
\end{enumerate}
The dependence on $x$ in \req{eq:26eq17.19} implies that in momentum space the corresponding momenta are convolved. Care must be taken when deriving this momentum space expression as appropriate treatment is needed in the phase space integration involving the large and residual component of each momentum. 

To sum up, the differential cross section with $N$ jets can be expressed
\begin{equation}\label{eq:26eq17.20}
d\sigma\simeq \sum_{ab}H_{ab}(\mu)[B_{a/P}(\mu)B_{b/P}(\mu)]\otimes[J_1(\mu)\ldots J_N(\mu)]\otimes S_{ab\ldots N}(\mu).
\end{equation}
where $H_{ab}(\mu)=|C_{\hat O}(\mu)|^2$ is the hard function, the norm squared of the matching coefficient $C_{\hat O}$.  More generally, allowing $n_B=0,1,2$ hadronic beams in the initial state for $e^+e^-$ collisions, DIS, and Drell-Yan (DY) processes respectively, factorization theorems take the generic form, for $N$ outgoing jets,
\begin{equation}\label{generalizedfactorization}
\sigma = Tr(HS ) \otimes J_1 \times \ldots J_N \otimes \Pi_{i=1}^{n_B} B_i ,
\end{equation}
where $H, S$ are the hard and soft functions, that are in general matrices in color space.
 The most difficult ingredient in traditional proofs of factorization theorems involve the delicate analysis of so-called Glauber gluons (\cite{collins1985factorization}), which have not yet been thoroughly studied for the SCET proofs.

\subsection{Resummation of Large Logarithms}
One application of SCET is the all $\alpha_s$-orders summmation of \lalo\, of ratios of different scales in a problem that arise in perturbative calculations.  Sudakov logarithms commonly arise in processes whose initial and final states have energy far exceeding their masses. As described above, these logs ruin convergence of fixed-order perturbation calculation if the scales in the problem are well-separated, such that there are ratios of scales $r\ll 1$. It is therefore necessary to rearrange the expansion for fixed $\alpha_s \ln r=\mcO(1)$; after summing logarithmically-enhanced contributions $\alpha_s^n (\ln r)^m$ with $m\le n+1$, the result should be a series of the form
\begin{equation}\label{logimprovedseries}
\ln \sigma\sim L\,g_0(\alpha_sL)+g_1(\alpha_s L)+\alpha_s \,g_2(\alpha_sL)+\ldots
\end{equation}
where $g_n$ are functions of $\alpha_sL$ (fixed) that need to be determined.

The resummation is enabled by the fact that cross sections in SCET factorize into pieces, each containing only a single physical scale.  Logarithms in each piece can only depend on the ratio of this scale to the renormalization scale $\mu$. Therefore one can set a renormalization scale $\mu'$ in each separate factor for which \lalo\, do not appear. Since the factorization theorem requires a common $\mu$, each factor is evolved from the common scale $\mu$ to the factor's ``natural'' scale $\mu'$ using the renormalization group equation (RGE). In contrast to other effective theories, SCET anomalous dimensions in the RGE for jet, hard, beam and soft functions in a factorization formula (e.g. \req{eq:26eq17.20}) contain only a single logarithm. For instance the anomalous dimension of the hard function $\gamma_H$ takes the form
\begin{equation}\label{eq:26eq17.21}
\gamma_H(\mu)=c_H\Gamma_{cusp}(\alpha_s)\ln\frac{Q^2}{\mu^2}+\gamma(\alpha_s)
\end{equation}
with $c_H$ a coefficient depending on specific process of interest and $\Gamma_{cusp}$ the so-called cusp anomalous dimension (\cite{korchemsky1987renormalization, korchemskaya1992light}). The remaining piece(s) of the anomalous dimension are called non-cusp and vary from process to process. A crucial feature of Sudakov problems is that the anomalous dimension involves a logarithm, seen here, that arises since the perturbative series involves double logarithms of scale ratios. 

The functions $g_n(\alpha_s\ln r)$ in \req{logimprovedseries} are written in ratios of the running coupling as a result of solving the RGE and thus one resums all \lalo\, of scale ratios in the factorized cross section. To solve the RGE and compute the first two terms $Lg_0(\alpha_sL)+g_1(\alpha_sL)$ (i.e. the next-to-leading logarithmic (NNL) approximation)  we need the two-loop cusp anomalous dimension and beta function, the one-loop non-cusp part of the anomalous dimension and the tree-level matching conditions for all component functions at their respective scales. To derive the next term $\alpha_sg_2(\alpha_sL)$ i.e. NNLL order, higher-order loop expansions are needed.

\subsection{Factorization and Resummation in \texorpdfstring{$\SCETb$}{SCETII}}
Recall the momentum scaling in \SCETb, \req{SCETbscaling}. The virtuality of both soft and collinear particles are small: $p_n^2\sim p_s^2\sim Q^2\lambda^2$ but they differ in rapidity. Cross sections for events where the transverse momentum of particles are restricted by external kinematics constitute an important family of observables with \SCETb momentum scaling.  An example is the kinematic endpoint in DIS or Drell-Yan processes, because the tranverse momentum of the collinear initial state is $\LQCD\sim\lambda Q$ while the soft radiation must also be on the $p_s^2\sim\LQCD^2\sim\lambda^2Q^2$ hyperbola. Because the virtualities of collinear and soft degrees of freedom are far below the hard scale $Q$, logarithms emerge that are controlled by RGE of the effective theory. However since rapidities of collinear and soft modes are differ parametrically, more \lalo\, emerge that must be factorized in the cross section and resummed.

To sum these logs, we will use the `rapidity renormalization group equation' derived by introducing a new scale parameter $\nu$ separating phase space in the $(p_-,p_+)$-plane into collinear and soft regions along a fixed-invariant mass hyperbola (see \cite{Chiu:2012ir}). Because different contributions have the same virtuality, there is no running coupling in the rapidity-RG equation contrary to ordinary RGE.

$\SCETb$ is also useful for investigating factorization of various exclusive $B$ decays including $\bar B\to \pi\ell\nu, \bar B\to K^*\gamma, \bar B\to\pi\pi$ where both virtuality of energetic final-state particle and the scale of soft light particles in $B$ meson initial are of order $\sim \Lambda_{QCD}$.

\section{\sceti and soft radiation in inclusive QCD processes}\label{sec:sceti}

The major motivation to improve the precision of QCD calculations using in particular $\SCETa$ is to aid discovery of new physics, which is sought in the hadronic collision processes deep-inelastic scattering (DIS), proton-proton scattering (Drell-Yan, DY) and electron-positron ($e^+e^-$) scattering.  QCD collisions are the primary source of background and in deep-inelastic and proton-proton processes will also contribute radiative corrections.  $\SCETa$ facilitates the proof of factorization theorems and the computation of infrared and collinear-safe observables to any desired order.  

The main class of observables I focus on is event-shape variables $\tau$ that, when $\tau\ll 1$, isolate events with collinear particles in two separable (outgoing or incoming) directions.  The corresponding differential cross-sections $d\sigma/d\tau$ can be factorized into hard, beam, jet and soft functions that separate contributions respectively from the heavy-particle production, collinear initial state, collinear final state and lower-momentum soft particles interacting with both and initial and final states.  The soft function for $e^+e^-$ scattering has been computed in the effective theory up to $\mathcal{O}(\as^2)$.  Together with the hard function and collinear jet function, this piece provides enough information to predict event shapes for $e^+e^-\to$ two jets to an unprecedented next-to-next-to-next-to-leading-logarithmic accuracy.  However, the event-shape cross sections for DIS and DY processes, which are currently more directly relevant for new physics searches, have not achieved this level accuracy, in part because the corresponding soft functions were not known to the same $\cO(\as^2)$.  The hard and jet functions appear to be the same, but the soft functions could in principle be different.
 
In this section, I will prove to two-loop and three-loop order that the soft radiation is universal across DIS, DY and $e^+e^-$ processes.  To setup the proof, I first briefly review deep-inelastic scattering, proton-proton scattering and electron-positron scattering.  I will also introduce jet and event shape observables for these processes and use $\SCETa$ to prove the corresponding factorization theorems.  My proof improves the accuracy of the QCD contribution to these scattering processes to next-to-next-to-next-to-leading-logarithmic order.  This proof has been published in \cite{Kang:2015moa}.

\subsection{Hadronic Final-States Observables for QCD Experiments}
\label{sec:I.4.2}
A key task in connecting QCD to experiment is to construct a dictionary between observables at hadronic and partonic levels, and this requires considering only infrared- and collinear-safe observables.  
Although total cross sections are the simplest examples, we often want to investigate the structure or composition of the final state and therefore must define differential cross sections in terms of infrared- and collinear-safe observables.  Final states from hard QCD scatterings are usually dominated by collimated particle bunches called {\bf jets}. Roughly speaking, a jet can be viewed as a high energy parton that went through subsequent soft and collinear showering and hadronization. By measuring jet shapes and identifying subjets, one can break down jet substructure, which is strongly affected by the details of QCD radiation and shower development inside the jet.  In this way, jets provide a testing ground for high-energy QCD predictions and also enable investigation of the hard partonic structure in the decay processes of massive quarks.

I will now introduce an important class of such observables and the corresponding factorization theorems for each of the processes $e^+e^-$, DIS, and proton-proton collisions.  These factorization theorems define the soft function, describing the low-energy radiation in each case.

{\bf Event-shape variables} describe the topology of the energy flow in an event using the four momenta of the final state particles.
An example is {\bf thrust} in $e^+e^-$ annihilation (\cite{brandt1964principal}, \cite{farhi1977quantum}),
\begin{equation}\label{eq:thrust}
T = \frac{1}{Q} \min_{\vect{\hat t}}\sum_{i\in X} \abs{\vect{\hat t} \cdot \vect{p}_i} \,,\quad \tau \equiv 1 - T\,,
\end{equation}
where $\vec p_i$ are momenta of particles or jets in final state, and the axis $\hat t$ where the maximum is achieved is called the ``thrust axis''. For $\bar q q$ pair production in the Born limit, $\tau\to 1$ for a perfect back-to-back pair and $\tau\to 1/2$ for a perfectly spherical many-particle configuration.  Because gluon emission alters the topology of energy flow, event-shape variables are sensitive to QCD radiation and hence to the strong coupling. Other examples of event-shape variables have been measured at LEP and HERA (see \cite{ellis2003qcd}, \cite{dissertori2003quantum}, \cite{dasgupta2004event}, \cite{biebel2001experimental}, \cite{kluth2006tests}) and at electron-positron colliders (\cite{basham1978energy}).  Recently a new event-shape variable, $N$-jettiness, has been introduced to quantify how much final state hadrons align along $N$ jet axes or beam directions  (\cite{stewart2010n}).  In the limit of exactly $N$ infinitely narrow jets the $N$-jettiness vanishes.

Event shapes have a wide range of applications which include studying analytic models of hadronization, identifying potential decay of new particles from QCD events, tuning parameters of Monte Carlo simulations and measuring the strong coupling.



\subsection{Factorization using SCET for $e^+e^-$, $ep$, $pp$ collisions}
\label{sec:I.4.3}

Event shapes can be used to isolate events with collinear particles in two separate (outgoing or incoming) directions.\footnote{This section is based on unpublished work in collaboration with D. Kang and C. Lee.}  Due to the differing content of the initial and final states, the observables, their operator definitions, and their factorization theorems differ between $e^+e^-$, $ep$ and $pp$ collisions.  For $e^+e^−$ collisions, one considers events such that the thrust \req{eq:thrust} $\tau\ll 1$ (\cite{farhi1977quantum}).  There are corresponding kinematic limits expressed in terms of 1-jettiness (\cite{kang2012n,Kang:2013nha}) or DIS thrust (\cite{antonelli2000resummation}) for $ep$ collisions, and 0-jettiness or beam thrust for DY (\cite{Stewart:2009yx,Stewart:2010tn,Stewart:2010pd}).

Predictions of thrust in pQCD show large logarithms $\alpha_s^n\ln^k\tau$ in the two-jet endpoint region. Other event shape variables have similar behavior. Hence in this region one must systematically sum the large logarithms that appear at each fixed order in $\alpha_s$ in order to get convergent and physical results (\cite{catani1993resummation, catani1991thrust}). Techniques for resumming the logarithms are constructed from factorization and renormalization group equations, either in pQCD (\cite{contopanagos1997sudakov}) or in effective field theory, especially SCET (\cite{bauer2000summing, bauer2001effective, bauer2001invariant, Bauer:2001yt, Bauer:2002nz}). The two methods are essentially equivalent, although specific applications to a given order of accuracy may yield different results (see \cite{almeida2014comparing}).

\subsubsection{Electron-position event shapes}
\label{sec:I.4.3.1}

A prediction for the thrust distribution in the two-jet limit can be derived used the factorization method (\cite{berger2003event, bauer2008factorization, almeida2014comparing})
\begin{equation}\label{eq:softeq5}
\frac{1}{\sigma_0}\frac{d\sigma}{d\tau}=H(Q^2,\mu)\int dt_n dt_{\bar n}dk_s\,\delta\!\left(\tau-\frac{t_n+t_{\bar n}}{Q^2-\frac{k_s}{Q}}\right)J_n(t_n,\mu)J_{\bar n}(t_{\bar n},\mu)S_2^{ee}(k_s,\mu),
\end{equation}
where $H(Q^2,\mu)=|C(Q^2,\mu)|^2$ is the squared Wilson coefficient obtained by matching the QCD current $\bar q\Gamma^\mu q$ onto the two-jet operator $\mathcal{O}_2=\bar\chi_{\bar n}\Gamma^\mu\chi_n$ with $\chi_{n,\bar n}$ collinear jet fields in SCET (see \cite{bauer2003enhanced, bauer2004enhanced}). Here the collinear directions are $n^\mu,\bar n^\mu=(1,\pm \vec n_\tau)$ with $\vec n_\tau$ the thrust axis. Definitions  of the jet functions $J_n$ and $J_{\bar n}$, which depend on the invariant masses $t_n, t_{\bar n}$, are given in  \cite{bauer2008factorization} and \cite{abbate2011thrust}. See \cite{lunghi2003factorization, bauer2004shape} for $\mathcal{O}(\alpha_s)$ computation and \cite{Becher:2006qw} for $\mathcal{O}(\alpha_s^2)$ computation of these jet functions. The soft function in \req{eq:softeq5} can be written as
\begin{equation}\label{eq:softeq6}
S_2^{ee}(k,\mu)=\int d\ell_1 d\ell_2\:\delta(k-\ell_1-\ell_2)S_2^{ee}(\ell_1,\ell_2,\mu),
\end{equation}
which is the projection of the hemisphere soft function $S_2^{ee}(\ell_1,\ell_2,\mu)$ with two variables $\ell_{1,2}$ being the small light-cone components of the soft radiation in the two hemispheres defined by collinear axes $n,\bar n$ respectively. Note the same notation $S_2$ is used to denote both the hemisphere soft function and its projection in \req{eq:softeq6} which will not be ambiguous since they differ in number of arguments. The hemisphere soft function is given by
\begin{align}
S_2^{ee}(\ell_1,\ell_2,\mu)=&\frac{1}{N_C}\tr\sum_{i\in X_s}\left|\langle X_s|T[T_n^{+\dagger}(0)Y_{\bar n}^+(0)]|0\rangle \right|^2 \delta\!\left(\ell_1-\!\sum_{i\in X_s}\theta(\bar n\!\cdot\! k_i-n\!\cdot\! k_i)\,n\!\cdot\! k_i\right)\nn\\
&\times\delta\!\left(\ell_2-\sum_{i\in X_s}\theta(n\cdot k_i-\bar n\cdot k_i)\bar n\cdot k_i\right),\label{eq:softeq7}
\end{align}
i.e. the matrix element of Wilson lines arising from the field redefinition of collinear fields that decouples soft and collinear interactions at leading power in SCET Lagrangian.  The trace is over color indices, with $N_C$ the number of colors and $T$ the time-ordering operator. 
This form of the hemisphere soft function is derived from general consideration of the event shape in Appendix \ref{app:hemispheresoft}. 
The Wilson lines $Y_n$, $Y_{\bar n}$ in $e^+e^-$ soft functions are in the $n, \bar n$ directions corresponding to radiation from outgoing jets,
\begin{align}
Y_n^{+\dagger}(x)&=P\exp\left[i g\int_0^\infty ds n\cdot A_s(ns+x)\right],\nn\\
Y_{\bar n}^+(x)&=\bar P\exp\left[ -ig\int_0^\infty ds\bar n\cdot A_s(\bar ns+x)\right]\,.\label{eq:softeq8}
\end{align}
Alternatively we can write the hemisphere soft function as
\begin{equation}\label{eq:softeq9}
S_2^{ee}(\ell_1,\ell_2,\mu)=\frac{1}{N_C} \Tr  \bra{0} \overline Y_\bn^{+T} (0) Y_n^+ (0)  \delta(\ell_1 - \Theta_R n\cdot \partial)\delta(\ell_2 - \Theta_L\bn\cdot \partial)Y_n^{+\dag}(0) \overline Y_\bn^{+*}(0)\ket{0}
\end{equation}
with
\begin{equation}\label{eq:softeq10}
\Theta_R=\theta(\bar n\cdot\partial-n\cdot\partial), \qquad \Theta_L=\theta(n\cdot\partial-\bar n\cdot\partial)
\end{equation}
restricting to one of the two hemispheres. Here $R$, $L$ stands for right and left hemispheres with $n_\tau$ pointing to right. One can also write the measurement $\delta$-functions in \req{eq:softeq9} using energy flow operators or energy-momentum tensor (\cite{Korchemsky:1999kt, Belitsky:2001ij,bauer2008factorization}). Using the transformation of Wilson lines under time reversal
\begin{align}
T(Y_\bn^+)^T&=\bar Y_\bn^{+\dagger}\nn\\
\bar T(Y_\bn^{+\dagger})^T&=\bar Y_\bn^+\,,
\end{align}
we can eliminiate the time ordering and write
\begin{equation}\label{eq:softeq11}
\bar Y_\bn^+(x)=\bar P\exp\left[-ig\int_0^\infty ds\bar n\cdot \bar A_s(\bar ns+x)\right],
\end{equation}
with $\bar A_s=A_s^A\bar T^A$ where $\bar T^A$ are generators of the anti-fundamental representation of $SU(N_C)$ (\cite{bauer2004enhanced,Monni:2011gb}).

\subsubsection{1-jettiness in DIS}
\label{sec:I.4.3.2}
We also use event shapes to study electron-proton scattering. Consider $e+p\to X+e$ with initial and final state electron momenta $k, k'$ respectively, proton momentum $P$ and denote by $p_X$ momentum of hadron final state $X$. Define
\begin{equation}\label{eq:softeq12}
q=k-k', \,\,\, Q^2=-q^2, \,\,\, x=\frac{Q^2}{2P\cdot q}.
\end{equation}
One can make extra measurements to detect jet-like structures in the hadronic final state $X$ similar to thrust in electron-position collisions.  Ways to define thrust in DIS have been discussed in \cite{antonelli2000resummation, dasgupta2004event} including one which is specifically suited for the Breit frame (in which momentum transfer is purely spacelike: $q^\mu=(0,0,0,Q)$). In the Breit frame one defines
\begin{equation}\label{eq:softeq13}
\tau_Q=1-\frac{2}{Q}\sum_{i\in \mathcal{H}_C}p_z^i
\end{equation}
with $\mathcal{H}_C$ the hemisphere centered around $+\hat z$ axis back-to-back with the proton beam in Breit frame. $\tau_Q$ defined above is a special case in a family of event shaps in DIS called {\bf 1-jettiness}. More generally the $N$-jettiness quantifies the amount of collimation of final-state hadrons into $N$ distinct lightlike directions, in addition to radiation along incoming hadron beam (\cite{stewart2010n}).  Therefore 1-jettiness in DIS measures the collimation of final state hadrons along one jet direction $q_J$ and incoming proton direction $q_B$ (\cite{Kang:2013nha})
\begin{equation}\label{eq:softeq14}
\tau_1=\frac{2}{Q^2}\sum_{i\in X}\min\{q_B\cdot p_i,q_J\cdot p_i\}
\end{equation}
with the minimum taken over hadrons in the final state that are separated into two regions by the beam directions $q_B$ and $q_J$. There are many choices for these two directions, and in \cite{Kang:2013nha} three choices are discussed
\begin{align}
\tau_1^a: ~~&q_B=xP,\,\, q_J=P_J\nn\\
\tau_1^b: ~~&q_B=xP,\,\, q_J=q+xP\nn\\
\tau_1^c: ~~&q_B=xP,\,\, q_J=k\label{eq:softeq15}
\end{align}
with $k,P,q,x$ defined in \req{eq:softeq12} and $P_J$ the jet axis. The first choice corresponds to $q_J$ being along the physical jet axis while second and third choices correspond to separating particles into groups that are hemispheres in the Breit and center-of-mass frames respectively. The second choice $\tau_1^b$ is equivalent to $\tau_Q$ given in \req{eq:softeq13}. By using the light-like directions $n_J, n_B$ given by $q_J, q_B$ in the first choice in \req{eq:softeq15} but different normalizations, one can consider different beam sizes and jet regions and therefore another definition of 1-jettiness (\cite{kang2012n, Kang:2013lga}).

The factorization theorem for the the first choice $\tau_1^a$ is given by (\cite{Kang:2013nha}),
\begin{align}
\frac{d\sigma}{dx\,dQ^2\,d\tau_1^a} &= \frac{d\sigma_0}{dx\,dQ^2} \int dt_J dt_B dk_s\, \delta\Bigl(\tau_1^a - \frac{t_J+t_B}{Q^2} - \frac{k_S}{Q}\Bigr)\nn  \\
& \times \sum_{i= q,\bar q} H_i^a (y,Q^2,\mu) B_i(t_B,x,\mu) J_i(t_J,\mu) S_2^{ep}(k_s,\mu), \label{eq:softeq16}
\end{align}
where the sum is over all quark and antiquark flavors and $d\sigma_0/dxdQ^2$ is the Born cross section and $H_i^a$ is a hard function given in \cite{Kang:2013nha}, $J_i$ the (anti-)quark jet function and $S_2^{ep}$ given by
\begin{equation}\label{eq:softeq17}
S_2^{ep}(k,\mu) = \int dk_J \,dk_B \, \delta(k - k_J - k_B)S_2(k_J,k_B,\mu)\,,
\end{equation}
is again the projection of the hemisphere soft function for DIS. Similar to above we can write
\begin{align}
\label{eq:softeq18}
S_2^{ep}(k_J,k_B,\mu) =& \frac{1}{N_C}\Tr \sum_{X_s} \abs{\bra{X_s} Y_{n}^{+\dagger} Y_{\bn}^- (0)\ket{0}}^2 \delta\!\left( k_J \minus \!\sum_{i\in X_s} \theta(\bn\!\cdot\! k_i \minus n\!\cdot\! k_i)\, n\!\cdot\! k_i\right) \nn\\
&\times  \delta\Bigl( k_B \,\minus \sum_{i\in X_s} \theta(n\cdot k_i \minus \bn\cdot k_i) \bn\cdot k_i\Bigr)\,. 
\end{align}
where $n,\bar n$ are along the original light-like directions of $q_{J,B}=\omega_{J,B}n_{J,B}/2$,
\begin{equation}
\label{eq:softeq19}
n = \frac{n_J}{R_J} \,,\, \bn = \frac{n_B}{R_B} \, ; \quad 
R_J^2 = \frac{\omega_B n_J\cdot n_B}{2\omega_J} \,,\, R_B^2 = \frac{\omega_J n_J\cdot n_B}{2\omega_B} \,.
\end{equation}
Although $n, \bar n$ are not necessarily in opposite directions, the rescaling of $n_J, n_B$ makes sure that $n\cdot\bar n=2$. Mathematically \req{eq:softeq18} is equivalent to a soft function with back-to-back $n,\bar n$ since the step function $\theta(x)$ depends on $n,\bn$ in exactly the same way as in \req{eq:softeq18} and its dependence is only through the dot product $n\cdot\bn=2$. Note there is no need to have explicit time-ordering operator in \req{eq:softeq18} as both Wilson lines are already time-ordered and commute.

In general $q_J, q_B$ are not back-to-back and hence do not separate the final state into exact hemispheres. Thus it may seem surprising that hemisphere soft function should appear in \req{eq:softeq16}. Nonetheless the generic soft function $S(k_J,k_B,q_J,q_B,\mu)$ for 1-jettiness in DIS is related to back-to-back hemisphere soft function \req{eq:softeq18} via rescaling \req{eq:softeq19} (\cite{Kang:2013nha}).

The only difference between the hemisphere soft function in \req{eq:softeq18} and that of electron-positron collision in \req{eq:softeq17} is the $\bar n$-Wilson line,
\begin{equation}\label{eq:softeq20}
Y_\bn^-(x)=P\exp\left[ig\int_{-\infty}^0\bar n\cdot A_s(\bar ns+x)\right]\,.
\end{equation}
The other Wilson line $Y_n^{+\dagger}$ is the same as in $e^+e^-$ and given in \req{eq:softeq8}. The two Wilson lines are essentially different.  To see whether the observable soft functions differ, one needs to compare their explicit perturbation expansions to test this order by order.

\subsubsection{Beam Thrust in pp Collisions}
\label{sec:I.4.3.3}
Beam thrust (\cite{Stewart:2009yx, Stewart:2010pd}) or 0-jettiness (\cite{stewart2010n}) in $pp$ collisions measures the collimation of the hadronic final-state along beam direction. This can be used to veto jets in the central region for DY processes $pp\to\ell^+\ell^-X$ which are crucial to searches for Higgs and BSM particle decay into a specific number of jets. This quantity also facilitates the study of initial state radiation.

The beam thrust with respect to light-like vectors $n_{a,b}$ along incident the proton directions is given by (\cite{stewart2010n}) 
\begin{equation}\label{eq:softeq21}
q_{a,b}^\mu=\frac12 x_{a,b}E_{cm}n_{a,b}^\mu
\end{equation}
with $n_{a,b}^\mu=n,\bar n=(1,\pm\hat z)$ in the center-of-mass frame and $x_{a,b}$ is determined from $Y$, the dilepton rapidity and the invariant mass $q^2$, transverse momentum $\vec q_T$ via
\begin{equation}
x_{a,b}E_{cm}=e^{\pm Y}\sqrt{q^2+\vec q_T^{\,2}}\,.
\end{equation}
The 0-jettiness with respect to $q_{a,b}$ is defined by
\begin{equation}\label{eq:softeq22}
\tau_0\equiv\frac{2}{Q^2}\sum_{k\in X}\min\{q_a\cdot p_k,q_b\cdot p_k\}=\frac{1}{Q}\sum_{k\in X}\left|\vec p_k^{\,T}\right|\min\{e^{Y-\eta_k},e^{-Y+\eta_k}\}.
\end{equation}
We will cite the factorization theorem in terms of the beam thrust $\tau_B$, which is related to the 0-jettiness by a rescaling,
\begin{equation}
\tau_B=\tau_0\sqrt{1+\frac{\vec q_T^{\,2}}{q^2}}\,.
\end{equation}
The factorization theorem for beam thrust $\tau_B$ is (\cite{Stewart:2009yx, stewart2010n, Stewart:2010pd})
\begin{align}
\frac{d\sigma}{dQ\,dY\,d\tau_B} &= \frac{d\sigma_0}{dQ^2\,dY} \int\!\! dt_a \,dt_b\,dk_a\,dk_b \delta\Bigl(\tau_B - \frac{t_a + t_b}{Q^2} - \frac{k_s}{Q}\Bigr) \nn \\
& \times \sum_{ij} H_{ij}(Q^2,\mu) B_i(t_a,x_a,\mu)B_j(t_b,x_b,\mu) S_2^{pp}(k_s,\mu)\,,
\label{eq:softeq23}
\end{align}
where $S_2^{pp}(k,\mu)$ is again the projection of the hemisphere soft function,
\begin{equation}\label{eq:softeq24}
S_2^{pp}(k,\mu) = \int dk\,\delta(k - k_a - k_b) S_2^{pp}(k_a,k_b,\mu)\,,
\end{equation}
with 
\begin{align}
S_2^{pp} (k_a,k_b,\mu) &= \frac{1}{N_C} \Tr\sum_{X_s} \abs{ \bra{X_s} T[ Y_{n}^{-\dag} Y_{\bn}^-]\ket{0}}^2\,\delta\Bigl( k_a \minus \!\sum_{i\in X_s} \theta(q_b\mcdot k_i \minus q_a\mcdot k_i)\,n\mcdot k_i\Bigr) \nn \\
&\times  \delta\Bigl( k_b \minus \sum_{i\in X_s} \theta(q_a\mcdot k_i \minus q_b\mcdot k_i)\,\bn\mcdot k_i\Bigr),\label{eq:softeq25}
\end{align}
with the Wilson lines 
\begin{align}
Y_{n}^-(x) &= P\exp\Bigl[ig\int_{-\infty}^0 ds\,n\mcdot A_s(n_a s + x)\Bigr]\nn \\
Y_{\bn}^{-\dagger}(x) &= \overline P\exp\Bigl[ - ig \int_{-\infty}^0 ds\,\bn\cdot A_s(n_b s+x)\Bigr]\,.\label{eq:softeq26}
\end{align}
The soft function in \req{eq:softeq25} does not separate the final state into exact hemispheres, but by performing a boost along the beam direction into the partonic center-of-mass frame
\begin{equation}\label{eq:softeq27}
q_{a,b}\cdot k_i \to e^{\mp Y} q_{a,b}\cdot k_i = \sqrt{q^2 + \vec{q}_T^{\,2}}\: n_{a,b}\cdot k_i\,.
\end{equation}
(under which the Wilson lines are unchanged), we have
\begin{equation}\label{eq:softeq28}
S_2^{ep}(e^Yk_a,e^{-Y}k_b;Y)=S_2^{pp}(k_a,k_b,0)\,.
\end{equation}
Following \cite{Stewart:2009yx}, the third argument denotes the rapidity of the boundary separating the two regions, and $Y=0$ corresponds to the back-to-back hemisphere soft function for incoming lines.

In this part we have examined how soft functions in factorization theorems for thrust in $e^+e^-$ collisions, 1-jettiness in DIS and beam thrust/0-jettiness in $pp$ collisions can all be derived from hemisphere soft functions where soft radiation is separated into two exact back-to-back hemispheres. The definitions of the respective soft functions \req{eq:softeq7}, \req{eq:softeq18}, and \req{eq:softeq25} only differ in the directions of the Wilson lines.  


\subsection{Soft Radiation in Scattering Processes}
\label{sec:I.4.4}

In this section I recount my proofs of the equality of the soft functions for the three processes $e^+e^-\to\text{dijets}$, DIS 1-jettiness, and $pp$ beam thrust at  $\mcOa{2}$ (\cite{Kang:2015moa}) and $\mcOa{3}$. 
The effect of switching the direction of a soft Wilson line from incoming to outgoing or vice versa flips the sign of the $i\e$ in the eikonal propagators formed by emission or absorption of soft gluons, {e.g.} \req{eq:softeq35}. This can potentially affect the value of soft gluon diagrams. Nevertheless, we will show that the value remains the same up to $\mcOa{2}$ and $\mcOa{3}$.

\subsubsection{Structure of Perturbative Computation}
\label{sec:I.4.4.1}
First we set up some of the notation we will use in our proof. 
The perturbative computation of the soft functions in \eqss{softeq7}{softeq18}{softeq25} can be performed by either of two methods.  One may compute cut diagrams with four Wilson lines with an appropriate measurement delta function, which fixes the soft momenta in the left and right hemisphere to be convolved with the other factors in \eqs{softeq5}{softeq6}  (see e.g. \cite{Hornig:2009vb}).  Or, one may directly compute amplitudes for emission of $n=0,1,2,\dots$ particles up to the appropriate order in $\as$ and performing the phase space integrals implicit in the sum over particles in \eqss{softeq7}{softeq18}{softeq25}. We will take the latter approach here. 

The result of computing \eqss{softeq7}{softeq18}{softeq25} up to $\cO(\as^N)$ in perturbation theory takes the generic form,
\be
\label{eq:softeq29}
\begin{split}
S_2(\ell_1,\ell_2) &= \frac{1}{N_C}\Tr \sum_{n=0}^N \int d\Pi_n \cM(\ell_1,\ell_2;\{k_n\})
\sum_{i,j}\cA_i(\{k_n\})\cA_j(\{k_n\})^\dag\,,
\end{split}
\ee
where $\cA_i(\{k_n\})$ is an amplitude to emit $n$ particles in the final state with momenta $k_1,\cdots,\, k_n$ . The sum over amplitudes $i,j$ includes sums over final-state spins, polarizations, and colors and goes over those pairs of amplitudes which can in fact be paired together and whose total order in $\as$ is $\leq N$. The phase space integration measure is given by
\be
\label{eq:softeq30}
d\Pi_n = \prod_{i=1}^n \frac{d^D k_i}{(2\pi)^D} 2\pi\delta(k_i^2)\theta(k_i^0)\,,
\ee
and the measurement function $\cM$ for the hemisphere soft function is given by
\be
\label{eq:softeq31}
\cM(\ell_1,\ell_2;\{k_n\}) = 
\delta\Bigl(\ell_1 - \sum_{i=1}^n k^+_i \, \theta(k_i^- - k_i^+)\Bigr)\,  \delta\Bigl(\ell_2 - \sum_{i=1}^n  k_i^- \, \theta(k^+_i - k^-_i)\Bigr)\,,
\ee
where $k^\pm$ are defined as
\be \label{eq:softeq32}
k^+=n\mcdot k \quad\text{and}\quad k^-= \bn \mcdot k.
\ee

\subsubsection{One-loop soft function}
\label{sec:I.4.4.2}

\begin{figure}
\centering
\includegraphics[width=.52\textwidth]{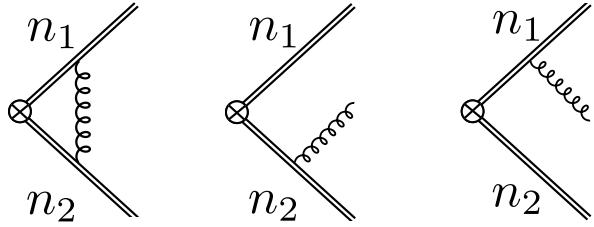}
\caption{virtual and real diagrams for one-loop soft function.}
\label{fig:softfig1}
\end{figure}

The one-loop result for the soft function $S_2$ can be computed from the diagrams illustrated in \fig{softfig1}. There is a tree-level, 0-gluon amplitude, not drawn, which simply takes the value
\be \label{eq:softeq33}
\cA_0^{(0)} = 1.
\ee
The virtual 0-gluon amplitude, $\cA_0^{(1)}$ represented by the first diagram on the left in \fig{softfig1}, is  scaleless and zero in dimensional regularization (DR), only playing the role of converting IR to UV divergences (e.g. \cite{Manohar:2006nz,Hornig:2009kv}). 

The first nontrivial amplitudes are the real 1-gluon amplitudes in \fig{softfig1}. For emission from two outgoing lines as in $e^+e^-$ \eq{softeq7}, the amplitudes for emitting a gluon of momentum $k$ can be written
\be
\label{eq:softeq34}
\cA_{1n} = -g\mu^\epsilon \frac{n\cdot \varepsilon(k)}{n\cdot k + i\epsilon}  \,, \quad \cA_{1\bn} = g\mu^\epsilon \frac{\bn\cdot \varepsilon(k)}{\bn\cdot k + i\epsilon} \,,
\ee
where $\varepsilon(k)= \varepsilon^A(k) T^A$ is the polarization vector for a final-state gluon of momentum $k$.
Switching an outgoing line to an incoming line changes $+i\epsilon$ to $-i\epsilon$.
The change occurs in the amplitude $\cA_{1\bn}$ for DIS and  in both $\cA_{1 n}$ and $\cA_{1\bn}$ for $pp$.
These signs are determined by the regulation of the integration limit at $\pm \infty$ in the path of the Wilson line, as seen considering
\be
\label{eq:softeq35}
\begin{split}
-ig \int_0^\infty ds\, e^{(ik\cdot \bn -\epsilon) s} &= -ig\frac{-1}{i\bn\cdot k - \epsilon} = \frac{g}{\bn\cdot k + i\epsilon}\,, \\
  ig \int_{-\infty}^0 ds\, e^{(ik\cdot \bn + \epsilon) s} &= ig\frac{1}{i\bn\cdot k + \epsilon} = \frac{g}{\bn\cdot k - i\epsilon}\,.
\end{split}
\ee
The measurement function for one real gluon is a special case of \req{eq:softeq31}, given by 
\be \label{eq:softeq36}
\cM(\ell_1,\ell_2;k) = \theta(k^- \minus k^+) \,\delta(\ell_2)\delta(\ell_1 \minus k^+) +\theta(k^+ \minus k^-) \,\delta(\ell_1) \delta(\ell_2 \minus k^-)\,,
\ee
where $k^\pm$ is defined in \eq{softeq32}. 

The sum over squared amplitudes in \eq{softeq29} up to $\cO(\as)$ is very easily evaluated and gives the well-known result in the $\MSbar$ scheme,
\be
\label{eq:softeq37}
S_2^{(1)}(\ell_1,\ell_2) = \frac{\as C_F}{\pi} \frac{(\mu^2 e^{\gamma_E})^{\epsilon}}{\Gamma(1-\epsilon)} \frac{1}{\epsilon} \bigl[ \ell_1^{-1-2\epsilon} \delta(\ell_2) + \ell_2^{-1-2\epsilon} \delta(\ell_1)\bigr]\,,
\ee
which does not depend on the signs of the $i\e$'s in \eqs{softeq34}{softeq35}.

\subsubsection{Two-loop Soft Function}
\label{sec:I.4.4.3}
At $\cO(\as^2)$, an explicit computation has been given only for the $e^+e^-$ soft function \eq{softeq7} (\cite{Kelley:2011ng,Monni:2011gb,Hornig:2011iu}). 
The $\cO(\as^2)$ contributions to the soft function are given by the appropriate terms contained in \eq{softeq29}. The products of amplitudes contributing at this order are:
\begin{enumerate}
\item 2-loop virtual times tree-level (0-gluon), $\cA_0^{(2)}\,\cA_0^{(0)\dag}$,  
\item 1-loop virtual times its conjugate, $\cA_0^{(1)}\cA_0^{(1)\dag}$, 
\item 2-real-gluon emission times conjugated 2-real-gluon emissions, $\cA_{2g}^{\text{tree}}\cA_{2g}^{\text{tree}\dag}$, 
\item $q\bar{q}$ and ghost-ghost pairs from a gluon splitting times their conjugates, $\cA_{q\bar q}\cA_{q\bar q}^\dag$ and $\cA_{\text{ghost}}\cA_{\text{ghost}}^\dag$, 
\item real-gluon vacuum polarization times a 1-real-gluon emission, $\cA_n^{\text{vac}}\cA_{1\bn}^\dag$ and $\cA_\bn^{\text{vac}}\cA_{1n}^\dag$, 
\item 1-real-gluon emission with another gluon loop times a 1-real-gluon emission, $ \cA_{1}^{(1)}\, \cA_{1n,1\bn}^\dag$, 
\item 1-real-gluon emission with a 3-gluon vertex times a 1-real-gluon emission, $\cA_1^\cT \cA_{1n,1\bn}^\dag$, 
\end{enumerate}
and their complex conjugates.  The diagrams will be displayed below.  All types have been computed in \cite{Kelley:2011ng}, and we will not repeat the results for individual classes of diagrams. To prove equivalence of the $ee,ep,pp$ soft functions, we will actually only need to look at diagrams in category 7 in detail, and we defer this to \sec{I.4.4}. The complete result of summing all $\cO(\as^2)$ contributions 1--7 is summarized in Appendix \appx{soft3gluon}. 

The proof does not depend on the momenta $k_i$ of the final states in \eq{softeq29}, nor on the measurement function $\cM(\ell_1,\ell_2;\{k_n\})$, but only on properties of the amplitudes $\cA_i$ themselves. Therefore, our proof applies to various classes of observables including event shapes, $P_T$ of massive particles, and exclusive observables defined by jet algorithms even near kinematic thresholds.

\paragraph{Types of Terms at \texorpdfstring{$\mcOa{2}$}{O(alpha\^2)}}
\label{sec:I.4.4.3.1}

In proving the equivalence, we group the seven categories of amplitudes above according to the number of particles in the final state:
\begin{itemize}
\item{categories 1--2:  purely virtual amplitudes in \figs{softfig1}{softfig2}}
\item{categories 3--4: 2-gluon/quark emission amplitudes in \figs{softfig3}{softfig4}}
\item{categories 5--7: 1-gluon emission amplitudes in \figss{softfig5}{softfig6}{softfig7}} 
\end{itemize}
The virtual diagrams in categories 1 and 2 are scaleless and zero in DR.  The 2-gluon/quark emission amplitudes in categories 3 and 4  are  manifestly real and independent of $i\e$'s in the propagators.  Therefore, the 0- and 2-particle amplitudes (categories 1-4) are trivially equivalent for \eeeppp.  We need only examine in detail the amplitudes belonging to the third group.  For completeness, we will discuss all 3 groups step-by-step below. 

To reduce the analysis further, it is most convenient to give results for the $\cO(\as^2)$ soft function in terms of the integrated or cumulative soft function,
\be \label{eq:softeq38}
S_c(\ell_1,\ell_2,\mu) = \int_0^{\ell_1}\int_0^{\ell_2} d\ell_1' d\ell_2' \,S_2(\ell_1',\ell_2',\mu)\,.
\ee
Then the contributions to $S_c$ from categories 3--7 with one or two real gluons in the final state can be written
\be
\label{eq:softeq39}
S_c^{(2)}(\ell_1,\ell_2,\mu) = \frac{\as(\mu)^2}{4\pi^2}\Bigl[ R_c^{(2)}(\ell_1,\ell_2,\mu) + S_{\text{NG}}^{(2)}(\ell_1,\ell_2)  + c_S^{(2)} \Bigr] \,,
\ee
where $R_c$ contains $\mu$-dependent logs associated with the anomalous dimension \eq{softeqA.5}, $S_{\text{NG}}$ contains the ``non-global'' terms arising from two soft gluons entering opposite hemispheres and depends non-trivially on both $\ell_1,\ell_2$ simultaneously, and the last term $c_S^{(2)}$ is a constant.
In this form, we can deduce which pieces of \eq{softeq39} must be equal for the \eec, \ep, and \pp\ hemisphere soft functions. 

The logarithmic terms in $R_c^{(2)}$ in \eq{softeqA.19} are the same for all three soft functions, \eec, \ep, and \pp, since they have the same anomalous dimension. This follows from the factorization theorems \eqss{softeq5}{softeq16}{softeq23} in which the soft functions appear.  RG-invariance (that is, $\mu$-independence) of the cross sections requires that the anomalous dimensions
\be \label{eq:softeq40}
\gamma_H = \frac{d\ln H}{d\ln \mu} \,, \ \tilde\gamma_J = \frac{d\ln \tilde J}{d\ln \mu} \,,\  \tilde\gamma_B = \frac{d\ln \tilde B}{d\ln \mu}\,,\tilde\gamma_S = \frac{d\ln \tilde S_2}{d\ln \mu}\,,
\ee
for the position-space jet $\tilde J$, beam $\tilde B$ and soft $\tilde S_2$ functions, satisfy a consistency condition
\be \label{eq:softeq41}
\gamma_H + 2\tilde\gamma_{J,B} + \tilde\gamma_S = 0\,.
\ee
Since the hard functions for \eeeppp all have the same anomalous dimension, and the jet/beam functions likewise all have the same anomalous dimension, $\tilde\gamma_J = \tilde \gamma_B$, the three soft functions must all have the same anomalous dimensions.  Thus the $\mu$-dependent part of the soft function, $R_c^{(2)}$, is the same for all \eeeppp\!.

The non-global terms in $S_{\text{NG}}^{(2)}$ in \eq{softeqA.20} are also the same, since they are entirely determined by the graphs with two real gluons, as discussed in \cite{Hornig:2011iu}. 
Since these amplitudes are manifestly real, the signs of the $i\e$s in the eikonal propagators do not matter, and they are the same for \eeeppp\!.

The only terms that could potentially differ for the three soft functions are the constant terms in $c_S^{(2)}$ in \eq{softeqA.24}. By examining the pole structure of the Feynman diagrams that can contribute, we find in fact that they are also the same.

\paragraph{Amplitudes Contributing to \texorpdfstring{$\mcOa{2}$}{O(alpha\^2)} Soft Functions}
\label{sec:I.4.4.3.2}
\nt{Virtual diagrams}

\vpc

\begin{figure}[thb]
\centering
\includegraphics[width=.45\textwidth]{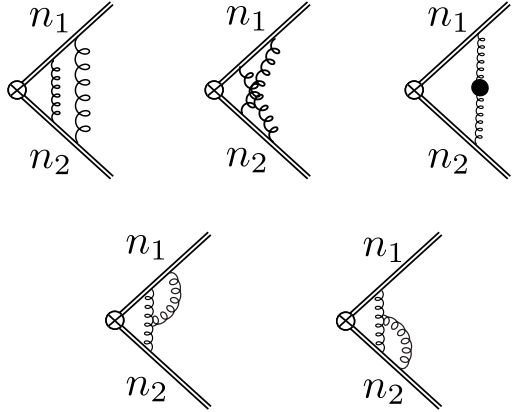}
\caption{2-loop virtual diagrams. The solid bubble represents gluon, fermion, or ghost loop.}
\label{fig:softfig2}
\end{figure}

The diagrams in \fig{softfig2} are purely virtual, with no real particles in the final state.  The loop integrals in these diagrams are scaleless and thus zero in DR.  This  is unaffected by the signs of the $i\epsilon$'s in the eikonal propagators, and therefore holds for \eeeppp soft functions.

\vpc

\nt{2-gluon/quark emission diagrams}

\vpc

\begin{figure}[tbh]
\centering
\includegraphics[width=.45\textwidth]{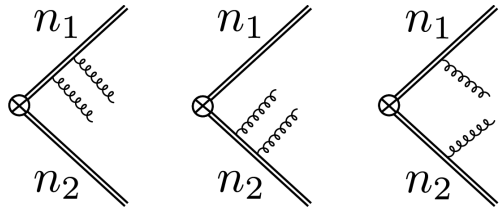}
\caption{Diagrams for 2 real gluon emission} 
\label{fig:softfig3}
\end{figure}

\begin{figure}[tbh]
\centering
\includegraphics[width=.55\textwidth]{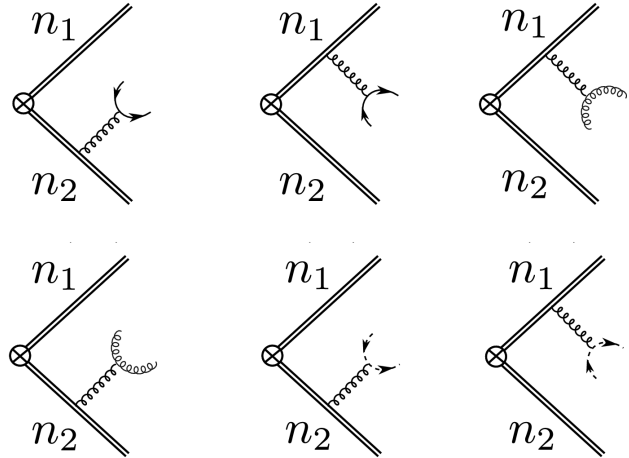}
\caption{Diagrams for 2 particle emission from gluon splitting. Solid and dashed lines are quark and ghost lines respectively.}
\label{fig:softfig4}
\end{figure}

The diagrams in \figs{softfig3}{softfig4} all have two real gluons, quark/antiquark or ghosts in the final state.  The factors in the formula \eq{softeq29} that could differ amongst \eeeppp soft functions are the eikonal propagators, which all take the form  $\sim 1/(p^+ \pm i\e)$ or $1/(p^- \pm i\e)$, where $p = k_1,k_2$ or $k_1+k_2$.
For these diagrams, each of the emitted particles is put on shell via the phase space delta functions in the phase space measures \eq{softeq30}, $\delta(k_i^2) \theta(k_i^0)$ where $i=1,2$. This ensures that $k_{1,2}^\pm \geq 0$, and the integrals over $k_{1,2}$ in \eq{softeq30} do not cross the poles in the eikonal propagators. Therefore the $i\e$s can be dropped in these terms.  

The same property holds for the gluon propagator $1/[(k_1+k_2)^2 + i\e]$ in each of the diagrams in \fig{softfig4} and is the same for each of the \eeeppp soft functions. The phase space measures functions again ensure that $(k_1+k_2)^2\geq 0$, and the $i\e$ can be dropped here as well,
As the $i\e$s do not affect the integration, the diagrams in \figs{softfig3}{softfig4} contribute the same, manifestly real results for the two real particle contributions to the $\cO(\as^2)$ \eeeppp soft functions.
\vpc

\nt{1-gluon emission diagrams}

\vpc

The remaining diagrams to check are those with one real gluon in the final state, in \figss{softfig5}{softfig6}{softfig7}.

\begin{figure}[tbh]
\centering
\includegraphics[width=.33\textwidth]{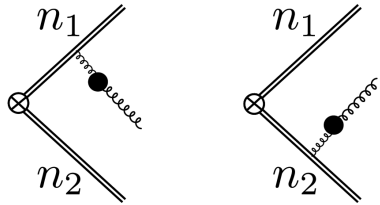}
\caption{Diagrams for 1 gluon emission: vacuum polarization loop}
\label{fig:softfig5}
\end{figure}

{\bf Vacuum polarization diagrams.}  The vacuum polarization diagrams in \fig{softfig5} have the same eikonal propagators as the single real gluon graphs at $\cO(\as)$ in \fig{softfig1}.  The $i\e$s in these propagators can be dropped because the phase space factors $\delta(k^2)\theta(k^0)$ in \eq{softeq30} place the momentum $k$ flowing through the propagator on shell and ensure the components in the eikonal propagators $k^\pm\geq 0$. The uncut gluon propagator and any propagators in the bubbles in \fig{softfig5} remain the same for \eeeppp soft functions. Therefore the diagrams in \fig{softfig5} make the same contribution to all three soft functions.


\begin{figure}[tbh]
\centering
\includegraphics[width=.55\textwidth]{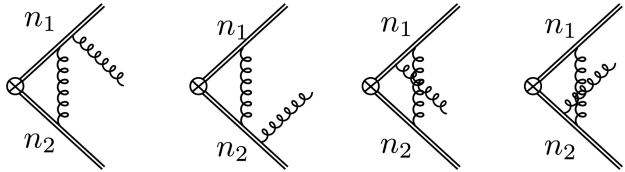}
\caption{diagrams for 1 gluon emission: independent emission and virtual loop}
\label{fig:softfig6}
\end{figure}

{\bf Independent emission diagrams.} The independent emission diagrams in \fig{softfig6} with one real gluon in the final state have left over a virtual loop which is scaleless and thus zero in DR independent of the sign of the $i\e$ in the propagators. They do not contribute to any of the \eec,\ \ep\ and \pp\ soft functions.


\begin{figure}
\centering
\includegraphics[width=.5\textwidth]{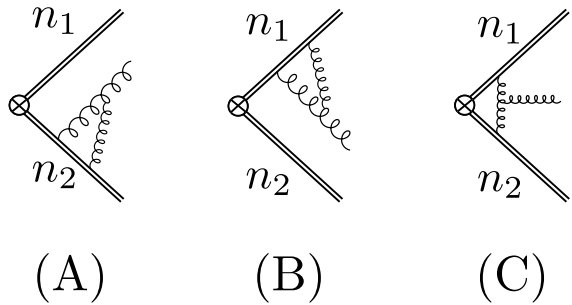}
\caption{diagrams for 1 gluon emission with 3-gluon vertex}
\label{fig:softfig7}
\end{figure}

{\bf 3-gluon vertex diagrams.}  The diagrams (A)$\sim$(C) in \fig{softfig7} involve a loop with eikonal propagator and we will investigate this integral carefully to show equality across the \eeeppp processes.

In diagrams (\ref{fig:softfig7}A) and (\ref{fig:softfig7}B), both virtual gluons are attached to the same eikonal line. The signs of the $i\epsilon$s in these eikonal propagators change when flipping from incoming to outgoing lines. \cite{Kelley:2011ng} observe that the loop integrals associated with these diagrams are scaleless and thus zero in DR for the $e^+e^-$ soft function, when both lines are outgoing. We now provide a few details of the argument for this and show that it is independent of the direction of the Wilson lines.

Let us consider first the diagram (\ref{fig:softfig7}B) for $e^+e^-$, where both gluons are attached to the $Y_n^{+\dag}$ Wilson line in \eq{softeq7}. The associated amplitude is
\be
\label{eq:softeq42}
\cA_{1(B)}^{\cT}(k) = \frac{i}{2} g^3\mu^{3\epsilon} C_A 
\frac{\varepsilon^+(k)}{k^+ +i\epsilon} 
\int \frac{d^D q}{(2\pi)^D} \frac{(2q-k)^+}{q^++i\epsilon}  \frac{1}{q^2 + i\epsilon} \frac{1}{(q-k)^2+i\epsilon}\,,
\ee
where $\ell^+=n\mcdot \ell$ and $\ell^- = \bn \mcdot \ell$.
The two eikonal propagators containing $k^+$ and $q^+$ change in the sign of $i\epsilon$ for an incoming line. We can, however, evaluate the $q$ integral by first performing a contour integral over $q^-$, the light-cone component not associated with the eikonal propagators whose $i\epsilon$ changes sign. The poles are  $q^-$ are at
\be
\label{eq:softeq43}
q^- = \frac{\vect{q}_\perp^2 - i\e }{q^+}\,, 
\frac{\vect{q}_\perp^2 - 2\vect{k}_\perp\mcdot \vect{q}_\perp - k^- q^+ - k^2  - i\epsilon}{(q-k)^+}\,.
\ee
Now, we know from the phase space integration \eq{softeq30} that $k^2 = 0$ and that $k^+ \geq 0$. These poles both lie in the lower half plane if $q^+ >k^+$. They both lie in the upper half plane for $q^+ < 0$. The contour integral over $q^-$ is nonzero only if $0<q^+<k^+$ where the two poles are in opposite planes. Closing the $q^-$ contour in the lower half-plane in this region, we obtain
\begin{align}
\label{eq:softeq44}
\cA_{1(B)}^{\cT}(k) &= \frac{g^3\mu^{3\epsilon} C_A }{4(2\pi)^{D-1}}
\frac{\varepsilon^+(k)}{k^++i\epsilon} 
\int_0^{k^+} \frac{dq^+}{q^+}\frac{(k-2q)^+}{k^+} \\ 
&\times \int\!{d^d q_\perp} 
\Bigl[ \vect{q}_\perp^2 - 2\frac{q^+}{k^+} \vect{q}_\perp\mcdot \vect{k}_\perp 
        + \Bigl(\frac{q^+}{k^+}\Bigr)^2 \vect{k}_\perp^2 - i\epsilon
       \Bigr]^{-1}\,,\nn
\end{align}
where $d=D-2=2-2\e$. To obtain the very last term we used the on-shell condition $k^- = \vect{k}_\perp^2 /k^+$ in the last propagator of \eq{softeq42}. The $q_\perp$ integrand in \eq{softeq44} is thus a perfect square, and in DR, we obtain
\be
\label{eq:softeq45}
\begin{split}
\cA_{1(B)}^{\cT}(k) &\propto \int\!{d^d q_\perp} \Bigl[\Bigl(\vect{q}_\perp - \frac{q^+}{k^+}  \vect{k}_\perp\Bigr)^2 - i\epsilon\Bigr]^{-1} \\
&\to \int \! d^d q_\perp' \frac{1}{{q_\perp'}^2 - i\e } = 0.
\end{split}
\ee
In the last step we shifted the $q_\perp$ integration variable and obtained a scaleless integral. Note that this result is independent of the signs of the $i\e$s in the eikonal propagators in \eq{softeq42} that could change between \eeeppp\!, and also of the actual measurement performed on the final state gluon $k$  in \eq{softeq29}.  Obviously the same conclusion holds for diagram (\ref{fig:softfig7}A). We have thus verified that these diagrams vanish not only for the $ee$ soft function but also for the $ep,pp$ soft functions.

Now we turn our attention to diagram (\ref{fig:softfig7}C), the only case where equivalence among \eeeppp diagrams is nontrivial in DR. The result for the amplitude in diagram (\ref{fig:softfig7}C) for $ee$ can be written
\be
\label{eq:softeq46}
\begin{split}
\cA_{1(C)}^{\cT ee}(k) &= \frac{ig^3\mu^{3\e} C_A}{2(2\pi)^D}
 \!\!\int \!\! \frac{d^D q}{q^2 \plus i\e} 
 \frac{1}{(k \minus q)^2 +i\e} \frac{1 }{ (k\minus q)^+ +i\e} \frac{1}{q^- +i\e}\\
&\qquad \times 
\biggl\{ \varepsilon^-(k)  (2k- q)^+ -  \varepsilon^+(k) (k+q)^- 
     - 2{\boldmath{\mbox{$\varepsilon$}}}_\perp\mcdot (\vect{k}_\perp
     \minus 2 \vect{q}_\perp)
      \biggr\}\,,
\end{split}
\ee
In the $\cO(\as^2)$ soft function, this amplitude is multiplied by one of the one-gluon tree-level amplitudes in \fig{softfig1}, which are proportional to $\varepsilon^+$ or $\varepsilon^-$.  Therefore, in the sum over gluon polarizations in \eq{softeq29}, the term with ${\boldmath{\mbox{$\varepsilon$}}}_\perp$ in \eq{softeq46} vanishes, and we drop it from here on.

The remaining terms in \eq{softeq46} can be split into a scaleless, and thus zero, part and a nonzero part. The scaleless part comes from the term in the numerator containing $(k-q)^+$ in the first term and $q^-$ in the second term, as each cancels one of the eikonal propagators on the first line. Together with a change of variables $q\to k-q$ in one of the terms, the scaleless part can be written
\be
\label{eq:softeq47}
\begin{split}
\cA_{1(C)}^{\cT\text{scaleless}}(k) &= \frac{ig^3\mu^{3\e} C_A}{2(2\pi)^D} \!\!\int \!\! \frac{d^D q}{q^2 \plus i\e} \frac{1}{(k \minus q)^2 +i\e}
\biggl\{ \frac{\varepsilon^-(k)}{q^- + i\e} - \frac{\varepsilon^+(k)}{ q^+ + i\e}\biggr\}
\\&=0\,.
\end{split}
\ee
This we see has the same analytic structure as the integral in \eq{softeq42} for diagram (\ref{fig:softfig7}B), and thus is scaleless and zero as in \eq{softeq45}. This conclusion holds for each of the \eeeppp soft functions. 

The nonzero part of \eq{softeq46} can be written
\be
\label{eq:softeq48}
\cA_{1(C)}^{\cT ee}(k) = \frac{i}{2}g^3\mu^{3\e} C_A  \Bigl[ \varepsilon^-(k)  k^+ -  \varepsilon^+(k) k^- \Bigr]\, \cI^{\cT ee}_C(k) \,,
\ee
where we have defined the integral
\be
\label{eq:softeq49}
\cI^{\cT ee}_C \equiv  \int \!\! \frac{d^D q}{(2\pi)^D} 
\frac{1}{q^2 \plus i\e}\frac{1}{(k \minus q)^2 +i\e} \frac{1 }{(k \minus q)^++i\e} \frac{1}{q^- +i\e}
\ee
for the $e^+e^-$ soft function.
The corresponding integrals for $ep$ and $pp$ are 
\begin{align}
\label{eq:softeq50}
\cI^{\cT ep}_C &\equiv  \int \!\! \frac{d^D q}{(2\pi)^D} 
\frac{1}{q^2 \plus i\e}\frac{1}{(k \minus q)^2 +i\e} \frac{1 }{(k \minus q)^+ +i\e} \frac{1}{q^- -i\e}\,,
\\
\label{eq:softeq51}
\cI^{\cT pp}_C &\equiv  \int \!\! \frac{d^D q}{(2\pi)^D} 
\frac{1}{q^2 \plus i\e}\frac{1}{(k \minus q)^2 +i\e} \frac{1 }{(k \minus q)^+ -i\e} \frac{1}{q^- -i\e}\,.
\end{align}
Compared to $e^+e^-$, the eikonal propagator $q^--i\e$ in $ep$ and
both $q^--i\e$ and $(k-q)^+-i\e$ in $pp$ have opposite sign $i\e$ terms.
Inserting \eqs{softeq50}{softeq51} into \eq{softeq48} we obtain corresponding amplitudes $\cA_{1(C)}^{\cT ep},\cA_{1(C)}^{\cT pp}$.

We must show the integrals \eqss{softeq49}{softeq50}{softeq51} are equal.
We will first find that the difference $\cI^{\cT ee}_C-\cI^{\cT ep}_C$ is zero, and hence the amplitudes for $ee$ and $ep$ are equal. The difference $\cI^{\cT ep}_C-\cI^{\cT pp}_C$ is nonzero and purely real. Nevertheless, at the level of squared amplitudes in \eq{softeq29} their difference vanishes when adding the hermitian conjugate, that is
\begin{align}
\label{eq:softeq52}
\left( \cA_{1(C)}^{\cT ep} - \cA_{1(C)}^{\cT pp} \right) \cA^\dag_{1n, 1\bn} +h.c.&=0\,,
\end{align}
where $\cA^\dag_{1n, 1\bn}$ is the 1-gluon amplitude given in \eq{softeq34}.  

First, we compute the difference $\cI^{\cT ee}_C-\cI^{\cT ep}_C$.  With the change of variable $q\to k-q$ in \eqs{softeq49}{softeq50}, everything remains the same except for the eikonal propagators $q^-\to (k-q)^-$ and $(k-q)^+\to q^+$. Only the sign of the $i\e$ in the $(k-q)^-$ eikonal propagator differs. In taking the difference, then, we can use the identity
\be\label{eq:softeq53}
\frac{1}{(k-q)^- + i\e} - \frac{1}{(k-q)^- - i\e} = -2\pi i\, \delta(k^--q^-)\,.
\ee
Also using the on-shell condition $k^2=0$ for the outgoing gluon, we obtain
\be
\label{eq:softeq54}
\cI_C^{\cT ee}\! - \cI_C^{\cT ep} 
= \frac{i}{2}  \int \frac{d^d q_\perp}{(2\pi)^{D-1}}
\frac{1}{(\vect{q}_\perp-\vect{k}_\perp)^2 - i\e} 
\int_{-\infty}^\infty dq^+\, \frac{1}{q^+ + i\e}\,\frac{1}{q^+k^- -\vect{q}_\perp^2 +i\e}\,.
\ee
Now we can do the remaining light-cone integral over $q^+$ by contours. Both poles in $q^+$ are in the lower-half plane:
\be\label{eq:softeq55}
q^+ = - i\e\,,\quad \frac{ \vect{q}_\perp^2 - i\e}{k^-}\,,
\ee
since $k^-\geq 0$ by the on shell condition in \eq{softeq30}. Closing the contour in the lower half plane shows the integral is zero. This establishes that
\be
\label{eq:softeq56}
\cA_{1(C)}^{\cT ee} = \cA_{1(C)}^{\cT ep} \,.
\ee

For the second equality in \eq{softeq52}, the argument is slightly more involved.  The integrals in \eqs{softeq50}{softeq51} differs in eikonal propagator $(k-q)^+\pm i\e$.  In their differences we use the identity
\be\label{eq:softeq57}
\frac{1}{(k-q)^+ + i\e} - \frac{1}{(k-q)^+ - i\e} = -2\pi i\delta(k^+-q^+)\,,
\ee
and obtain the difference
\be
\label{eq:softeq58}
\cI_C^{\cT ep}\! - \cI_C^{\cT pp} 
= \frac{i}{2}  \int \frac{d^d q_\perp}{(2\pi)^{D-1}}
\frac{1}{(\vect{q}_\perp-\vect{k}_\perp)^2 - i\e} 
\int_{-\infty}^\infty dq^-\, \frac{1}{q^- - i\e}\,\frac{1}{q^-k^+ -\vect{q}_\perp^2 +i\e}\,.
\ee
This time the poles in the $q^-$ contour integral are on opposite sides of the real axis,
\be\label{eq:softeq59}
q^- = + i\e\,,\quad \frac{\vect{q}_\perp^2 - i\e}{k^+}\,.
\ee
The integral \eq{softeq58} is finite and non-zero:
\be
\label{eq:softeq60}
\begin{split}
\cI_C^{\cT ep} - \cI_C^{\cT pp} 
&= \frac{1}{2(2\pi)^{D-2}} \int \frac{d^d q_\perp}{\vect{q}_\perp^2 - i\e} \frac{1}{(\vect{q}_\perp - \vect{k}_\perp)^2 -i\e}
\\&=\frac{(4\pi)^{\e}}{8\pi}\frac{\Gamma(-\e)^2\Gamma(1+\e)}{\Gamma(-2\e)}\,\frac{1}{\vect{k}_\perp^{2+2\e}}\,,
\end{split}
\ee
closing the $q^-$ contour in \eq{softeq58} in the upper half plane. The remaining integral over $q_\perp$ does not cross any poles in the remaining propagators, meaning \eq{softeq60} is real. The prefactor in front of the amplitudes in \eq{softeq48}, however, is purely imaginary (which comes from the color factor identity $f^{ABC}T^A T^B = \frac{i}{2}C_A T^C$ where $T^C$ is absorbed into $\ve^\pm(k)= \ve_C^\pm(k)\, T^C$ and will turn into real number in the polarization sum). Therefore when we add the relevant products of amplitudes to their complex conjugates, we obtain for the contributions containing the amplitudes in \fig{softfig7},
\be
\label{eq:softeq61}
\cS_{1(C)}^{\cT ep} - \cS_{1(C)}^{\cT pp} = 0\,,
\ee
where
\be\label{eq:softeq62}
\cS_{1(C)}^{\cT} \equiv \cA_{1(C)}^\cT (\cA_{ 1n}+\cA_{ 1\bn})^\dag + (\cA_{ 1n}+\cA_{1\bn} )\, \cA_{1(C)}^{\cT \dag}\,,
\ee
where $\cA_{ 1n, 1\bn}$ are purely real and given in \eq{softeq34}.  

Together \eqs{softeq56}{softeq61} establish the final result that the total perturbative soft functions for \eeeppp are equal up to $\cO(\as^2)$:
\be
\label{eq:softeq64}
S_2^{(2)ee} = S_2^{(2)ep} = S_2^{(2)pp}.
\ee
We have examined all the diagrams that could contribute to the $\cO(\as^2)$ soft functions in \eq{softeq64} and found that their contributions to the soft function are equal for \eeeppp\!. 

In Appendix \appx{soft3gluon} we provide an explicit computation of the contribution of diagram (\ref{fig:softfig7}C) to the $\cO(\as^2)$ soft function. The result was given in \cite{Kelley:2011ng} for $e^+e^-$ without sufficient details to verify easily whether the result is the same for $ep$ and $pp$. We provide our own derivation of the results for $e^+e^-,ep$ and $pp$, also confirming as observed above that the amplitudes all have a nonzero imaginary part which is the same for $e^+e^-,ep$ but opposite sign for $pp$:
\be
\label{eq:softeq65}
\begin{split}
\cI_C^{\cT} (k) &= \frac{i}{16\pi^2} (4\pi)^\epsilon \Gamma(1+\e) (\vect{k}_\perp^2)^{-1-\e}  \\
& \times \biggl[ \frac{2}{\e^2} - \pi^2 - 4\zeta_3\e + \frac{\pi^4}{60}\e^2 \pm i\pi \Bigl( \frac{2}{\e} - \frac{\pi^2}{3}\e - 4\zeta_3 \e^2\Bigr)\biggr] \,,
\end{split}
\ee
where the $+i\pi$ sign is for $ee,ep$ and $-i\pi$ is for $pp$.
The $\pm i\pi$ terms cancel in the sum over complex conjugate diagrams in the soft function, giving the result \eq{softeq64}.


\subsubsection{Gluon-soft Functions}
\label{sec:I.4.4.4}

So far, our discussion has focussed on quark soft functions, the building blocks of which are Wilson lines with gluon fields in the fundamental representation, $A_\mu(x)=A^a_{\mu}(x)\, T^a$.  Wilson lines for gluon soft functions are defined in terms of the adjoint representation, with the gluon field written as
\be\label{eq:softeq66}
A_\mu(x)=A^a_{\mu}(x)\,  \cT^a\,, \quad (\cT^a)_{bc}=-i\, f^{abc}\,.
\ee
For example, the counterpart of the out-going Wilson line $Y_n^\dag(x)$ in \eq{softeq67} is
\be
\cY_n^\dag(x) = P\exp\biggl[ig \int_0^\infty ds\,n\cdot A^a_s(ns+x) \,\cT^a \biggr]
\,.\label{eq:softeq67}
\ee
Since the gluon soft functions differ from quark soft functions only in color factors, most of the proofs in \sec{I.4.4} apply to the gluon soft functions.  The exception is \eq{softeq61} where we used that the color factor $i\,C_A/2$ multiplying \eq{softeq48} is purely imaginary.  This factor could differ for the gluon-soft functions.
 
The color factors in the amplitude $\cA_{1(C)}^{\cT}$ for quark and gluon Wilson lines are
\be\label{eq:softeq68}
\left.\cA_{1(C)}^{\cT}\right|_\text{color} =
\ \begin{cases}
f^{ABC} T^A  T^B = \frac{i}{2}C_A \,T^C, & \mbox{for quark} \\
f^{ABC} \cT^A  \cT^B = \frac{i}{2}C_A \,\cT^C,& \mbox{for gluon.}  
 \end{cases}
\ee
These factors differs only in the color matrices $T^C$ and $\cT^C$.  In \eq{softeq48}, $T^C$ was absorbed into the polarization vector $\varepsilon^\pm(k) = \varepsilon^\pm_C(k)\, T^C$.  Other than the substitution $\varepsilon^\pm = \varepsilon^\pm_C\, \cT^C$, the amplitude for a gluon Wilson line has the same form.  Because the color factor remains pure imaginary, the argument used to obtain \eq{softeq61} is valid also for gluon Wilson lines, and the equality at $\cO(\as^2)$ in \eq{softeq64} also holds for the gluon soft functions.


\subsubsection{Multi-jet Soft Functions at \texorpdfstring{$\mcOa{2}$}{O(alpha\^2)}}
\label{sec:I.4.4.5}

\begin{figure}
\centering
\includegraphics[width=.56\textwidth]{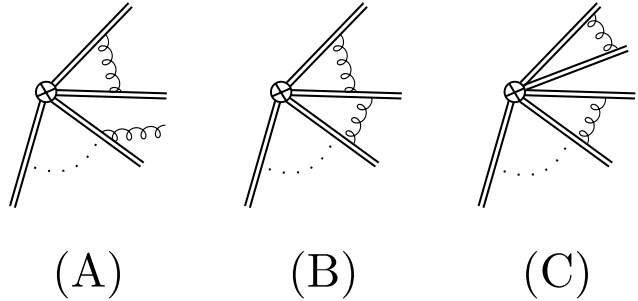}
\caption{
Examples of $\cO(\as^2)$ diagrams with 3 Wilson lines (A and B) and 4 Wilson lines (C) involved.
Dots represent Wilson lines not emitting soft gluons.}
\label{fig:softfig8}
\end{figure}

We can extend our discussion on the equality of \eeeppp soft functions to multi-jets functions defined by more than 2 Wilson lines.  A generic form of the multi-jet soft function can be written as
\be 
S_\text{multi-jet} =
 \bra{0}  \bar{T} [Y_\bn^\dag{\bf \hat{Y}}^\dag  Y_n(0)]\, \hat{\cM}\,  T [Y_n^\dag {\bf \hat{Y}}  Y_\bn(0)] \ket{0},
\label{eq:softeq69}
\ee
where ${\bf \hat{Y}}$ is a product of Wilson lines for outgoing $q, \bar q$, or $g$ and $\hat{\cM}$ is a measurement operator that measures momenta of final state particles\footnote{Here, we use the measurement operator for simplicity of expression. Instead, as in \eq{softeq7} a measurement function in terms of final momenta can be used by inserting the sum of final states.}.

The Wilson lines $Y_\bn$ and $Y_n^\dag$ in \eq{softeq69} are different for three collisions \eeeppp as in \eqss{softeq8}{softeq20}{softeq26}, while  the product ${\bf \hat{Y}}$ remains the same for different collisions.  Therefore, differences in multi-jet soft functions should come from  differences in Wilson lines $Y_\bn$ and $Y_n^\dag$.  The differences are trivially zero for tree diagrams and for any loop diagrams not involving eikonal propagators from $Y_\bn$ and $Y_n^\dag$. The non-trivial diagrams to investigate are the loop diagrams involving eikonal propagators on one or two of $Y_\bn$ and $Y_n^\dag$, which we call relevant diagrams or relevant loops.

At $\cO(\as)$ diagrams for multi-jets are essentially same as those in \fig{softfig1} except that the $n$ and $\bar n$ Wilson lines are replaced by any of the Wilson lines in \eq{softeq69}.  These relevant diagrams are purely virtual and zero in DR.  The 1-gluon real diagrams are equal by the same arguments as above. 

Similarly, diagrams at $\cO(\as^2)$ include all amplitudes in \sec{I.4.4.1} with $n$ and $\bar n$  lines replaced by $n_1$ and $n_2$ lines, which are any of 2 lines in \eq{softeq69} and also include amplitudes with 3 and 4 Wilson lines as shown in \fig{softfig8}(B) and (C). Diagrams with 3 and 4 Wilson lines involve only virtual loops and are zero in DR.  Among the diagrams in \sec{I.4.4.1} the relevant diagrams are \figss{softfig2}{softfig6}{softfig7}.  The diagrams in \figs{softfig2}{softfig6} involve purely virtual loop and are zero in DR as described in \sec{I.4.4.3.2}. The amplitude \ref{fig:softfig7}(B) with an $n_1$ Wilson line is
\begin{align}
\cA_{1(B)}^{\cT}(k;\delta_1) &= \frac{i}{2} g^3\mu^{3\e} C_A  \frac{n_1\mcdot \ve^C(k)\,t^C}{n_1\mcdot k +\delta_1} 
\int \frac{d^D q}{(2\pi)^D} \frac{n_1\mcdot (2q-k)}{n_1\mcdot q +\delta_1}  \frac{1}{q^2 + i\e} \frac{1}{(q-k)^2+i\e}
\nn \\ \label{eq:softeq70}
&=0\,,
\end{align} 
where $t^C =T^C$ for $q/\bar q$ or $t^C=\cT^C$ for $g$ Wilson lines.  $\delta_1=\pm i\e$ when the $n_1$ line is an outgoing/incoming $q,\bar q,g$ line.

The integral in \eq{softeq70} is exactly same as the integral in \eq{softeq42} with $q^+,\, k^+$ replaced by $n_1\mcdot q,\, n_1\mcdot k$ and $i\e$ replaced by $\delta_1$. Hence, after the $n_2 \mcdot q $ contour integral, we obtain the scaleless $q_\perp$ integral as in \eq{softeq45}, which is zero in DR. This result is independent of $\delta_1=\pm i\e$ and valid for the \eeeppp soft functions. The amplitude \ref{fig:softfig7}(A) is also zero in DR for the three collisions.

The amplitude \ref{fig:softfig7}(C) with Wilson lines $n_1$ and $n_2$ is given by
\be
\label{eq:softeq71}
\begin{split}
\cA_{1(C)}^{\cT}(k;\delta_1,\delta_2) &= -\Delta\frac{g^3\mu^{3\e}  }{(2\pi)^D}  f^{ABC} t^A_{a_1 a_2} t^B_{a_3 a_4}\\
 &\times\int \!\! \frac{d^D q}{q^2 \plus i\e} 
 \frac{1}{(k \minus q)^2 +i\e} \frac{1}{n_1\mcdot (k-q)+\delta_1}\frac{1}{n_2\mcdot q+\delta_2}\\
& \times 
\biggl[n_1\mcdot (2k- q)\,  n_2 - n_2\mcdot (k+q)\,  n_1 +n_1\mcdot n_2 \,({k}_\perp - 2 {q}_\perp)
      \biggr]\mcdot \ve^C(k) \,,
\end{split}
\ee
where, independent of the process, $\Delta=+1$ when the $n_{1,2}$ Wilson lines are outgoing $q$ or $g$ or both are outgoing $\bar q$ and $\Delta=-1$ otherwise.  The color matrix $t^A_{ab} =T^A_{ab}$ or $\cT^A_{ab}$ for $q/\bar q$ or $g$ Wilson lines, respectively.  Unlike \eq{softeq46}, the color factor in \eq{softeq71} is not simplified because the multi-jet soft function \eq{softeq69} is usually a matrix in color space that can only be simplified once the color structure of the hard coefficient is specified.  In the eikonal propagators, $\delta_{1,2}=\pm i\e$  for an outgoing/incoming line so that the pair $(\delta_1,\delta_2)$ is one of four combinations: 
\be\label{eq:softeq72}
(\delta_1,\delta_2)\in \{(+i\e,\, +i\e),\, (+i\e,\, -i\e),\, (-i\e,\, -i\e),\, (-i\e,\, +i\e)\}.
\ee
The first 3 combinations have the same sign $i\e$ terms as in \eqss{softeq49}{softeq50}{softeq51}.
The last, $(-i\e,\, +i\e)$, is the same as \eq{softeq50} upon changing variables $q\to k-q$ and directions $n_1\leftrightarrow n_2$.  Therefore, the differences of $q$ integrals between the $(+i\e,\, +i\e)$ and $(+i\e,\, -i\e)$ cases and between the $(+i\e,\, +i\e)$ and $(-i\e,\, +i\e)$ cases vanish for the same reason as in \eq{softeq54}:
\be\label{eq:softeq73} 
\cA_{1(C)}^{\cT}(k;+i\e,+i\e)=\cA_{1(C)}^{\cT}(k;+i\e,-i\e)=\cA_{1(C)}^{\cT}(k;-i\e,+i\e).
\ee
In the $e^+e^-$ soft function, the $n_{1,2}$ lines are always outgoing and $(\delta_1,\delta_2)_{ee}=(+i\e, +i\e)$.  In the $ep$ soft function, $n_2$ can be either incoming or outgoing while $n_1$ is outgoing, implying $(\delta_1,\delta_2)_{ep}=(+i\e, \pm i\e)$.  In the $pp$ soft function, $n_{1,2}$ can be incoming or outgoing and all 4 combinations in \eq{softeq72} are possible.  Therefore, the difference between amplitudes for $e^+e^-$ and $ep$ is always zero by \eq{softeq73}, which proves equality of the $e^+e^-$ and $ep$ soft functions for multi-jet final states at $\cO(\as^2)$:
\be\label{softeq74}
S_{\text{multi-jet}}^{(2)\, ee} = S_{\text{multi-jet}}^{(2)\, ep}\,.
\ee

To prove the last equality $S_{\text{multi-jet}}^{(2)\, pp} = S_{\text{multi-jet}}^{(2)\, ep}$, we consider the difference between amplitudes for the $ (+i\e,\, -i\e)$ and $(-i\e,\, -i\e)$ cases in \eq{softeq72} is non-zero as in \eq{softeq58},
\begin{align}\label{eq:softeq75} 
\cA_{1(C)}^{\cT}(k;+i\e,-i\e)-\cA_{1(C)}^{\cT}(k;-i\e,-i\e)
&=-\Delta\, g^3\mu^{3\e}\,f^{ABC} t^A_{a_1 a_2} t^B_{a_3 a_4}\frac{\Gamma(-\e)^2\Gamma(1+\e)}{\Gamma(-2\e)}
\nn \\ & \times
\frac{[n_1\mcdot k\, n_2 - n_2\mcdot k\,  n_1]\cdot \ve^C(k)}{n_1\mcdot n_2 \,(4\pi)^{1-\e} }
\frac{1}{ \vect{k}_\perp^{2+2\e}} \,,
\end{align}
and the difference between $ep$ and $pp$ amplitudes is nonzero.  Above, we used that the color factor of the amplitude product $\cA_{1(C)}^{\cT}\, \cA_{1n, 1\bn}^\dag$ is pure imaginary so that the non-vanishing term in \eq{softeq58} cancels when adding the complex conjugate.  

The color matrices can be reduced to numbers in all of the soft functions involving $q\,\bar{q}\,g$ Wilson lines whether there are 1, 2 or 3 jets, because the hard coefficient has a simple color structure: $(C_H)^a_{\alpha \beta}=T^a_{\alpha \beta}$,
where $\alpha\,, \beta$, and $a$ are color indices of the three partons $q^\alpha\,\bar{q}^\beta\,g^a$, respectively. 
Then, the color factor of amplitude product $\cA_{1(C)}^{\cT}\, \cA_{1n_3}^\dag$ multiplied by $(C_H)^a_{\alpha\beta}$ can be written
as $ (C_H)^a_{\alpha\beta}\, f^{ABC} t^A_{a_1 a_2} t^B_{a_3 a_4}  t^C_{a_5 a_6} $,
where unmatched color indicies $\alpha,\beta, a_i$ are properly contracted when the color charges $t^{A,B,C}$ of 3 vertices on the $q\bar q g$ Wilson lines.
The 3 gluons from 3-gluon vertex in diagram \ref{fig:softfig7}(C) can be attached on $q\bar q g$ Wilson lines in
 7 different ways: $q\bar{q}g$,
$qgg$, $\bar{q}gg$, $qqg$, $\bar{q}\bar{q}g$, $q\bar{q}\bar{q}$, and $\bar{q}qq$.
(We exclude cases like $qqq, \bar{q}\bar{q}\bar{q}, ggg$, which are associated with diagrams \ref{fig:softfig7}(A) and \ref{fig:softfig7}(B) and the amplitudes of which are zero in DR.)
The color factors of the 7 configurations as 
\begin{align*}
 &q\bar{q}g:&  f^{ABC} T^A C_H^e T^B \cT^C_{ae} &= 0\, C_H^a
\,,\\
& qgg:         & f^{ABC}  T^A C_H^e \cT^B_{fe} \cT^C_{af} &= +i\,\left(\tfrac{C_A}{2}\right)^2 C_H^a
\,,\\
& \bar{q}gg:&  f^{ABC}  C_H^e T^A \cT^B_{fe} \cT^C_{af} &=  -i\,\left(\tfrac{C_A}{2}\right)^2 C_H^a
\,,\\ 
& qqg:         &  f^{ABC}  T^A T^B C_H^e \cT^C_{ae} &= -i\,\left(\tfrac{C_A}{2}\right)^2 C_H^a
\,,\\ 
&\bar{q}\bar{q} g: & f^{ABC}   C_H^e T^B T^A \cT^C_{ae} &= -i\,\left(\tfrac{C_A}{2}\right)^2 C_H^a
\,,\\ 
& qq\bar{q}:   & f^{ABC}  T^A T^B C_H^a T^C &= -i\,\left(\tfrac{C_A}{2}\right)\left(\tfrac{C_A}{2}-C_F\right)  C_H^a
\,,\\ 
&\bar{q}\bar{q}q:  &  f^{ABC}  T^A C_H^a T^B T^C &= -i\,\left(\tfrac{C_A}{2}\right)\left(\tfrac{C_A}{2}-C_F\right)  C_H^a
\,,\end{align*}
where $\cT^b_{ac}=i\,f^{abc}$ is the adjoint representation.
Therefore, the color factors are pure imaginary and the $q\bar{q}g$ soft functions for \eeeppp are all equal up to $\cO(\as^2)$.
\be\label{softeq76}
S^{(2)\, ee}_{q\bar q g}= S^{(2)\, ep}_{q\bar q g}= S^{(2)\, pp}_{q\bar q g}\,.
\ee
One can play similar game for other multi-jet soft functions once the color structure of their hard coefficients are known.
If their color factors reduce to imaginary numbers, the equality of $ep$ and $pp$ soft functions is proved.
Even if the color factors are not imaginary, it is straightforward to calculate differences between $ep$ and $pp$ soft functions by using \eq{softeq75}.  


\subsubsection{Equality of Soft Functions at \texorpdfstring{$\mathcal{O}(\alpha_s^3)$}{O(alpha\^3)}}
\label{sec:I.4.4.6}
In this section we show equality of the soft functions for the three processes $e^+e^-, ep$ and $pp$ at $\mathcal{O}(\alpha_s^3)$.\footnote{This section is based on unpublished work in collaboration with D. Kang.}  The amplitudes at $\mathcal{O}(\alpha_s^3)$ include 3-emission, 2-emission with 1-loop, 1-emission with 2-loop, and 3-loop virtual amplitudes.

The 3-emission amplitudes have the same eikonal propagators for $e^+e^-$, $ep$, and $pp$ because the $i\epsilon$ in the propagators can be dropped. The 3-loop virtual loop is zero in DR. The nontrivial pieces to check are the 1- and 2-emission amplitudes.

\paragraph{1-emission 2-loop contribution}
\label{sec:I.4.4.6.1}

\begin{figure}
\centering
\includegraphics[width=.8\textwidth]{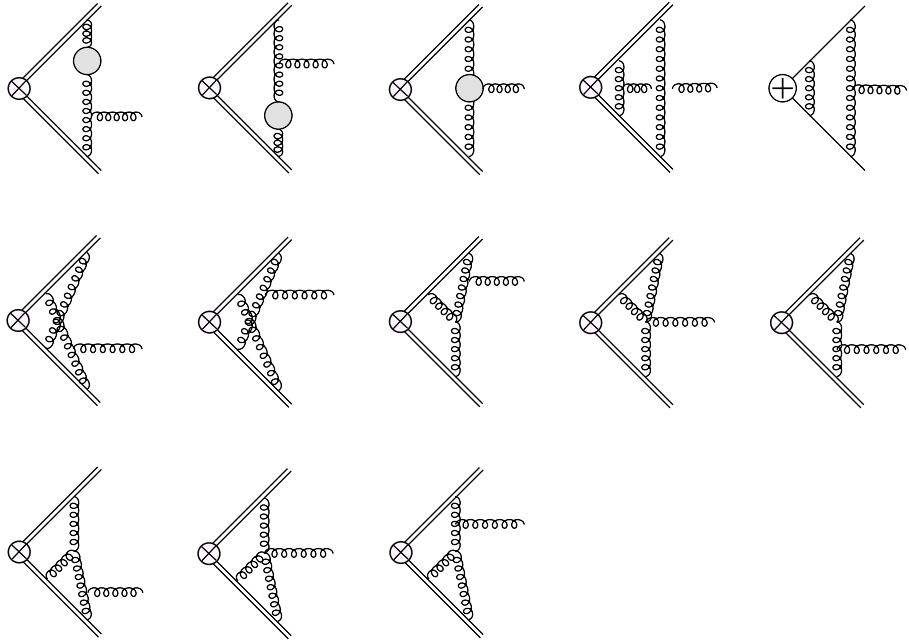}
\caption{Two-loop non-vanishing diagrams for single soft gluon emission (\cite{Li:2013lsa}).}
\label{fig:25fig5}
\end{figure}

The result for single-gluon emission with 2 loops is (\cite{Li:2013lsa})
\begin{equation}\label{eq:25eq8}
\varepsilon_\mu(k)\mathcal{A}^\mu_{g-2loop}(k)=-S_{12}^{(2)}(k)=\left(\frac{\alpha_s}{4\pi}\right)^2e^{2i\pi \sigma_{12}\epsilon}\left(\frac{\mu^2}{k^+k^-}\right)^{2\epsilon}f(k^\pm,\mu),
\end{equation}
where $\sigma_{12}=-1$ for $pp$ and $\sigma_{12}=+1$ otherwise ($e^+e^-$ and $ep$).  $f(k^\pm,\mu)$ is a real-valued function depending also on the DR parameter $\epsilon$  and $N_c, N_f$. Therefore the amplitudes for the diagrams in \fig{25fig5} are the same for $e^+e^-$ and $ep$ 
\begin{equation}\label{eq:25eq9}
\varepsilon(k)\cdot\mathcal{A}_{g-2loop}^{ee}(k)=\varepsilon(k)\cdot\mathcal{A}_{g-2loop}^{ep}(k)\,.
\end{equation}
The difference between $ep$ and $pp$ amplitudes is purely imaginary, and after adding the complex conjugate, vanishes
\begin{equation}\label{eq:25eq10}
(\mathcal{A}_{g-2loop}^{ep}-\mathcal{A}_{g-2loop}^{pp})\mathcal{A}_{1R}^{n,\bar n\dagger}+h.c.=0\,.
\end{equation}
The square of the 1-loop 1-emission amplitude $|\mathcal{A}_{\mathcal{T}}|^2$ also contributes at $\mcOa{3}$. The difference between $e^+e^-$ and $ep$ is zero, because the imaginary part of these amplitudes are the same with opposite sign: $\text{Im}\mathcal{A}_{\mathcal{T}}^{ep}=-\text{Im}\mathcal{A}_{\mathcal{T}}^{pp}$. Therefore, $|\mathcal{A}_{\mathcal{T}}^{ee}|^2=|\mathcal{A}_{\mathcal{T}}^{ep}|^2=|\mathcal{A}_{\mathcal{T}}^{pp}|^2$.

\begin{figure}
\centering
\includegraphics[width=.53\textwidth]{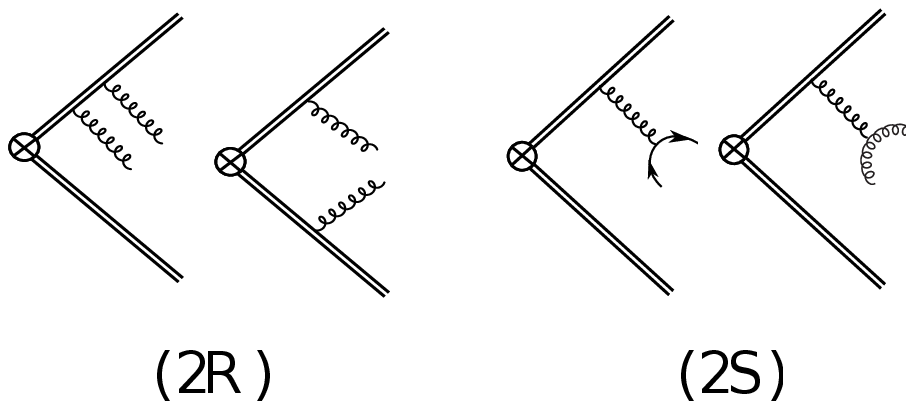}
\caption{2-emission tree-level diagrams interfering with the 1-loop diagrams in \fig{25fig6} at $\cO(\as^3)$.}
\label{fig:25fig2}
\end{figure}

\begin{figure}
\centering
\includegraphics[width=\textwidth]{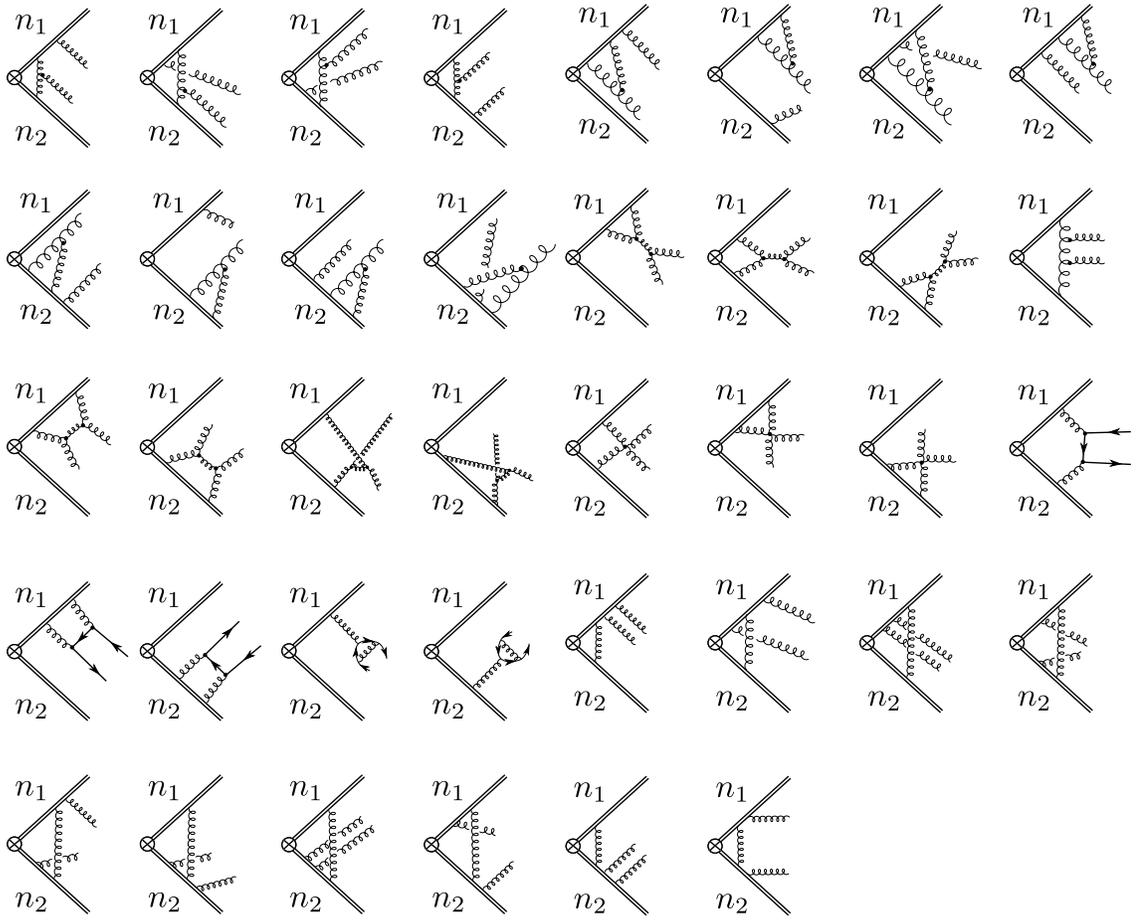}
\caption{1-loop 2-emission diagrams}
\label{fig:25fig6}
\end{figure}

\paragraph{2-emission 1-loop contribution}
\label{sec:I.4.4.6.2}
The 2-emission with 1-loop results were obtained in \cite{Li:2014bfa} for Higgs boson and Drell-Yan at threshold resummation, corresponding to the $pp$ soft function with a measurement of the final-state energy. However, we also need the results for $e^+e^-$ and $ep$ to compare. In this part, we study the differences between $e^+e^-, ep$ and $pp$ in the amplitudes for the 2-emission with 1-loop diagrams. There are 5 types of nontrivial diagrams classified by number and type of vertices:
\begin{enumerate}
\item single 3-gluon vertex
\item two 3-gluon vertices
\item single 4-gluon vertex
\item 2 emissions plus an independent loop with no 3- and 4- gluon vertex
\item $q\bar q$ emission with 1- and 2-current vertices
\end{enumerate}
Fig.\,\ref{fig:25fig6} shows all non-vanishing diagrams listed above in Feynman gauge and in DR scheme.

As was true at $\mcOa{2}$, we find that the difference between the $e^+e^-$ and $ep$ amplitudes is zero for all the diagrams in Fig.\ref{fig:25fig6},
\begin{equation}\label{eq:25eq11}
\mathcal{A}_i^{ee}(k_1,k_2)-\mathcal{A}_i^{ep}(k_1,k_2)=0
\end{equation}
where index $i$ run over the individual diagrams in Fig.\ref{fig:25fig6} and $k_1$ and $k_2$ are momenta of the final states. The differences between $ep$ and $pp$ diagrams are nonzero. For convenience we rewrite the amplitudes in terms of color and momentum dependent parts as 
\begin{equation}\label{eq:25eq12}
\mathcal{A}_i(k_1,k_2)=\begin{cases}
\varepsilon_\mu^\alpha(k_1)\varepsilon_\nu^\beta(k_2)C_i^{\alpha\beta}A_i^{\mu\nu}(k_1,k_2) & \text{ for }gg\text{ emission}\\
\bar u(k_1)C_iA_i(k_1,k_2)u(k_2) & \text{ for }q\bar q\text{ emission}.
\end{cases}
\end{equation}
$C_i^{\alpha\beta}$ and $C_i$ are color factors, and $A^{\mu\nu}(k_1,k_2)$ and $A_i(k_1,k_2)$ contain the dependence on the final momenta $k_{1,2}$. $\alpha,\beta$ are color indices of the final gluons. The color index for the quark and antiquark in the matrix $C$ and on the spinors $u,\bar u$ and spinor index are implicit, following common convention. In $ep$ and $pp$, the color factors are the same and only the $A$s differ. Full expressions for $C$ and the differences $A^{ep}-A^{pp}$ are given in the appendix.

The amplitudes in Fig.\ref{fig:25fig6} are multiplied by the tree-level amplitudes $\mathcal{A}_{i}^{tree}$ in Fig.\,\ref{fig:25fig2}, which also can be written in the same form of \req{eq:25eq12} in terms of a cr factor $C_j^{tree}$ and momentum-dependent function $A_j^{tree}$, with $j$ labelling the diagram in Fig.\,\ref{fig:25fig6}. The differences between products of amplitudes become
 \be\label{eq:25eq13}
 \Tr[\cA_j^{tree\, \dagger} ( \cA_i^{ep}-\cA_i^{pp})] =
 \begin{cases}
 \Tr [C^{tree,ba}_j C^{ab}_i] \, A^{tree}_{j\,\mu\nu}  A^{\mu \nu}_i \\
 \Tr [C^{tree}_j C_i] \, \tr[A^{tree}_{j} \slashed{k}_1  A_i \slashed{k}_2]\,. 
 \end{cases}
 \ee
The momentum-dependent parts  $A^{tree}_{j\,\mu\nu}  A^{\mu \nu}_i $ and $\tr[A^{tree}_{j} \slashed{k}_1 A_i \slashed{k}_2]$ are real-valued as shown by their explicit expressions in the appendix and the fact that the trace of Dirac matrices is real-valued.
 
For $gg$ emission, the color factor $C_j^{tree, ab}$ contains the structure $T^a T^b$, $T^b T^a$ and for $q \bar q$ emission the color part $C_j^{tree\, kl}$ is $T^a_{kl}\, T^a$  where $k,l$ are quark color indices.  All possible color factors can be enumerated:
 \begin{align}
\frac{1}{N_c} \Tr [C_i^{ab}\, T^aT^b] \in &
 \Big\{- \tfrac i2 C_A C_F^2 \,, \tfrac i4 C_A^2 C_F \,, -\tfrac i2 C_A^2 C_F\,, \tfrac i4 C_F\,, \nn\\
 &\quad \tfrac i4 C_F/C_A^2\,,  \tfrac i4 C_F(1+C_A^2)/C_A^2\,, 0
 \Big\}\,,\\
\frac{1}{N_c} \Tr [C_i^{ab}\, T^bT^a] \in &
 \Big\{\tfrac i2 C_A C_F^2 \,, \pm\tfrac i4 C_A^2 C_F \,, -\tfrac i2 C_A^2 C_F\,, \pm\tfrac i4 C_F\,, \nn\\
 &\quad \tfrac i4 C_F/C_A^2\,,-\tfrac i2 C_F/C_A^2
 \Big\}\,,\\ 
\frac{1}{N_c} \Tr [C_i^{kl}\, T^a_{kl} T^a] \in & \Big\{  -\tfrac i4 C_A C_F\,, \tfrac i2 C_F/C_A   \Big\}\,.
 \label{eq:25eq16}
 \end{align}
for all diagrams $i\in {A,B,C}$ in \fig{softfig8}. All color factors are purely imaginary, hence \eq{25eq13} is purely imaginary.  Adding \eq{25eq13} to its hermitian conjugate, we see the difference between $ep$ and $pp$ vanishes
 \be
 \Tr[\cA_j^{tree\, \dagger} ( \cA_i^{ep}-\cA_i^{pp})]+h.c.=0\,,
 \label{eq:25eq17}
 \ee
 for all $i$ and $j$.

Therefore, for all diagrams including tree-level 3-emission, 1-loop 2-emission, 2-loop 1-emission and 1-loop 1-emission squared, the differences between \eeeppp all vanish at the level of squared amplitudes, and we conclude the $\cO(\as^3)$ hemisphere soft functions are same the for \eeeppp
\be
S^{(3)}_{ee}(k_1,k_2,k_3)=S^{(3)}_{ep}(k_1,k_2,k_3)=S^{(3)}_{pp}(k_1,k_2,k_3)
\label{eq:25eq18}
\ee

In Appendix \ref{appx:B.2}, we prove the soft function associated with single-jet-axis event shape variables is universal to all orders in $\alpha_s$.  It is natural to expect the hemisphere soft function, generalizing this event shape only slightly by an additional jet axis, should also be universal to all orders in $\alpha_s$.  To complete such a proof, it suffices to show that the hemisphere soft function is time-reversal invariant (corresponding to the incoming versus outgoing hadron jets in $e^+e^-$, $ep$ and $pp$).  If so, we can further conclude that it is only sensitive to the total energy in the soft radiation sector and the requirement of color neutrality across the jets. 

This concludes the study $\SCETa$ in this dissertation, and we move now to the conditions addressed by $\SCETb$.

\section{\texorpdfstring{$\SCETb$}{SCETII} and Semi-inclusive Processes}
\label{sec:I.5}

In collisions near the elastic limit, known as semi-inclusive processes, there is a long-existing issue of a large logarithm arising from the large separation of scales between the initial state center-of-mass energy and the final state radiation.  
In electron-proton scatttering, the elastic limit is easily understood as the limit in which the transferred momentum in the center-of-mass frame only changes the momentum of the proton, rather than breaks it into pieces.  Studying the DIS process near this limit improves our understanding of the hadronization process and confinement phenomenon, because it connects parton-level inelastic scattering (away from the elastic limit) to the region where the proton recoil involves only emission of hadrons.

Using the second version of soft-collinear effective field theory, $\SCETb$\!, I show how to resolve this issue in deep-inelastic scattering and proton-proton scattering near the elastic limits, both theoretically and phenomenologically. I discuss two applications of resumming these semi-inclusive rapidity logarithmic corrections: the parton fragmentation function near the elastic limit and the long-distance initial state radiation in vector boson fusion.  To start, I provide the basics to address these issues: an introduction to $\SCETb$ and rapidity regulators with a side application, resumming rapidity logarithmic corrections in the transverse momentum-dependent parton distribution function.

\subsection{Rapidity Divergences and Logarithmic Corrections in \texorpdfstring{$\SCETb$}{SCETII}}
\label{sec:I.5.2} 

In this section, I first discuss the origin of the specific type of large logarithm associated with semi-inclusive processes, namely rapidity logarithms.  Second, I will give the general prescription for resumming such logarithms.  Third, I will use the $\eta$-regulator method to discuss the transverse momentum distribution (TMD) in Drell-Yan processes and demonstrate how rapidity logarithms are resummed.  The introduction to rapidity divergences and the $\eta$-regulator are based on \cite{Chiu:2012ir}.

\subsubsection{The Origin of Rapidity Divergences}
\label{sec:I.5.2.1}
Effective field theories, such as \SCETa and HH$\chi$PT, together with the standard RGE aid the treatment of  large logarithms that arise from fixed scales such as masses.  However less inclusive observables could contain large logarithms involving kinematic factors, e.g. large hierachies emerging when investigating corners of phase space. 
One example in Drell-Yan processes is jet broadening (\cite{catani1992jet,Dokshitzer:1998kz}).  

The traditional EFT approach, in which the RGE sums \lalo\, that have been re-expressed as arising from UV divergences, is not suited for these large kinematic logs.  In particular, the standard RGE cannot be applied when the EFT contains several fields on the same invariant mass hyperbola, as in $\SCETb$ (\cite{Bauer:2002aj}) in which the light-like momenta scale according to \req{SCETbscaling}: soft momenta $\sim(\lambda,\lambda,\lambda)$ and collinear $\sim(1,\lambda^2,\lambda)$. With such scaling one encounters a divergence from large rapidity instead of UV or IR divergences (\cite{Manohar:2006nz}).  Such rapidity divergences are discussed under different guises in the context of transverse momentum distribution functions and Sudakov form factors (see \cite{Collins:1992tv,collins1981back,Collins:2008ht}).

Rapidity divergences emerge from the momentum region with large $k_+/k_-$ (or $k_-/k_+$) but fixed invariant mass $k^2$. Since these divergences do not appear in full theory they are not associated to IR divergences, and because they arise at both the upper and lower bounds of the momentum integral, they are also not UV in nature. This type of divergence originates in the eikonal propagators that are the leading order in the multipole expansion in the effective theory (\cite{Grinstein:1997gv}). In addition, although rapidity divergences appear in factorized IR sectors, the sum of EFT sectors is free from rapidity divergences  (collinear or soft).  This implies that rapidity divergences are an effect of factorization that are indispensable to summing large logarithms and saving perturbation theory. 

Only some QCD observables give rise to rapidity divergences in SCET.  Rapidity divergences appear in observables receiving important contributions from degrees of freedom carrying similar invariant mass but very different rapidities.  With $p_T\ll Q$, we replace $p_T$ with $\mu_L$ above and large rapidity logarithms appear that need to be resummed. However, rapidity divergences can also arise in exclusive processes, and I will address those associated with the end point singularity, pointed out by \cite{Manohar:2006nz}.  

\subsubsection{Zero-Bin Subtractions and Rapidity Divergences: \texorpdfstring{$\SCETa$ vs $\SCETb$}{SCETI vs SCETII}}

As is customary for EFT, scale separation in SCET occurs at Lagrangian level, which systematizes the power counting by introducing operators with definite scaling in powers of $\mu_L/Q$ with $\mu_L$ some relevant low energy scale.  
To achieve clean separation of scales, we introduced a dynamical label formalism (\cite{Luke:1999kz}), applied to SCET in \cite{bauer2001effective} in which the full QCD field is written as
\begin{equation}\label{eq:28eq3.2}
\psi(x)=\sum_{n\cdot p,p_\perp}e^{-i(n\cdot p)(\bar n\cdot x)+ip_\perp\cdot x_\perp}\xi_{n\cdot p,p_\perp}(x)\,.
\end{equation}
This ensures derivatives scale as $\lambda^2$ when acting on $\xi$. Instead of an integral, in \req{eq:28eq3.2} we have a sum  over bins that tesselate the space of large momenta and whose size scales with the residual momentum $\sim\lambda^2$.
Lagrangian interactions can alter the large momentum components of fields, e.g. in splitting of collinear gluons, meaning there are loops in Feynman diagrams in which labels must be summed over.  In the loop integral, we encounter regions of this summation where labels become parametrically small and modes start to overlap each other.  This overlap obscures the physics underlying the EFT computations.  The overlap is explicitly removed by studying diagrams of a particular mode, expanding the mode-sum around the overlap region, and subtracting their contribution to the complete integral. In $\SCETa$, this `zero-bin' subtraction is equivalent to dividing by the matrix element of Wilson lines in pQCD factorization formulas (\cite{Collins:1989bt}), which can generally be identified with the reciprocal of the soft function in SCET factorization theorems (\cite{Lee:2006nr, Idilbi:2007yi, Idilbi:2007ff}). 

For $\SCETa$ all modes have parametrically different virtualities (except collinear modes in different directions), and all divergences are regulated using dimensional regularization or off-shellness. Zero-bin subtractions are applied relatively easily to remove any double counting, and consequently $\SCETa$ is not plagued by rapidity divergences.  

For transverse momentum distributions with $p_T\ll Q$ or similar observables, one must take into account real radiation with momentum scaling $(p_\perp,p_\perp,p_\perp)$. Considering $p_\perp$ the infrared scale, collinear momentum scales as $(Q,p_\perp^2/Q,p_\perp)$ and both soft and collinear modes have the same invariant mass, which is $\SCETb$ scaling \req{SCETbscaling}.  Here too, label and residual momentum regions can overlap in loop integrals, necessitating zero-bin subtractions.  However, the soft and collinear modes are distinguished only by a boost, i.e. relative rapidity.   
To factorize physical observables, one must break the boost invariance of the operators to differentiate between different sectors. Consequently factorization in $\SCETb$ will lead to additional divergences in sectors that are not regulated by dimensional regularization or off-shellness (\cite{Beneke:2003pa}) and will not cancel within the sector as in $\SCETa$.  This non-cancellation changes the renormalization group evolution, but also enables resummation of rapidity logarithms.  To approach such logarithms, zero-bin subtraction is a possible but not quite convenient regulator.


\subsubsection{Rapidity Divergences in \texorpdfstring{$\SCETb$}{SCETII}}
\label{sec:I.5.2.3}

For a concrete example, consider the one-loop correction to emission of a massive gauge boson, in the frame where the boson momentum is purely space-like and the transverse momenta of the incoming and outgoing fermion can be set to zero.  This vertex correction is written as a Sudakov form factor, which is factorized in SCET as follows (\cite{Bauer:2010cc})
\begin{align}\label{eq:29eq4.1}
\bar u(p_n)\gamma_\mu^\perp u(p_\bn) F(Q^2,M^2)&\approx \bra{p_n}\bar\xi_n W_nS_n^\dag\gamma_\mu^\perp C(n\cdot \cP, \bn \cdot \cP)S_\bn W_\bn^\dag \xi_\bn \ket{p_\bn}\nn\\
&=H(Q^2,\mu)J_n(M;\mu,\nu/Q)\gamma_\mu^\perp J_\bn(M;\mu,\nu/Q)S(M;\mu,\nu/M)
\end{align}
where $\approx$ means equality at leading power in $\lambda$.  The functions $J_n,J_\bn$ and $S$ are $\SCETb$ matrix elements
\begin{align}
S(M;\mu,\nu/M)&=\bra{0}S^\dag S\ket{0}\nn\\
J_n(M;\mu,\nu/Q)&=\bra{p_n}\bar\xi_n W_n\ket{0}\nn\\
J_\bn(M;\mu,\nu/Q)&=\bra{0}\bar W_\bn^\dag\xi_\bn\ket{p_\bn}\label{eq:29eq4.2}
\end{align}
where $S$ and $W$ are Wilson lines of soft and collinear degrees of freedom.

\begin{figure}
\centering
\includegraphics[width=.4\textwidth]{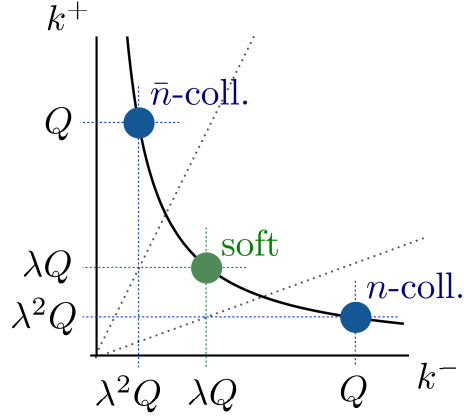}
\caption{The mass-shell hyperbola separating different sectors (\cite{Manohar:2006nz,Chiu:2012ir}). Arbitrariness in separation between soft and collinear modes leads to rapidity divergences. The two distinct rapidity UV divergences arising from soft sector must cancel one rapidity IR divergence from each collinear sector.
}
\label{fig:29fig1}
\end{figure}

Dispensing with the Dirac structure, the loop integral in the fully theory is
\begin{equation}\label{eq:29eq4.3}
I_f=\int [d^nk]\frac{1}{k^2-M^2}\frac{1}{(k^2-(n\cdot k)(\bar n\cdot p_1)+i\epsilon)}\frac{1}{(k^2-(\bar n\cdot k)(n\cdot p_2)-i\epsilon)}
\end{equation}
which is finite in both UV and IR regions. The EFT separates it into three contributions: two collinear integrals $I_{n,\bn}$ both in the form
\begin{equation}\label{eq:29eq4.5}
I_n=\int[d^nk]\frac{1}{(k^2-M^2)}\frac{1}{(k^2-(n\cdot k)(\bar n\cdot p_1)-i\epsilon)}\frac{1}{(-\bar n\cdot k+i\epsilon)}\,,
\end{equation}
and a soft integral resulting from the limit $k^\mu\to (M,M,M)$ in light-cone coordinates,
\begin{equation}\label{eq:29eq4.4}
I_S=\int [d^nk]\frac{1}{(k^2-M^2)}\frac{1}{(-n\cdot k+i\epsilon)}\frac{1}{(-\bar n\cdot k+i\epsilon)}\,.
\end{equation}
As the full theory diagram is infrared finite, so must be the sum of EFT diagrams. Integrating the soft diagram over $k_\perp$,
\begin{equation}\label{eq:29eq4.6}
I_s\sim \int[d^2k]((n\cdot k)(\bar n\cdot k)-M^2)^{-2\epsilon}\frac{1}{(-n\cdot k+i\epsilon)}\frac{1}{(-\bar n\cdot k+i\epsilon)}
\end{equation}
shows that $(n\cdot k)(\bar n\cdot k)\sim M^2$ is the relevant region of phase space.  This hyperbola is shown in \fig{29fig1}. The integral is scaleless away from the hyperbola, and the divergence arises when $(n\cdot k)/(\bar n\cdot k)\to \infty$ or $\to 0$. These limits correspond to rapidity divergences that occur when the soft integral overlaps with the two collinear rapidity regions and they are not regulated by dimensional regularization, which only regulates the divergence from the invariant mass hyperbola going to infinity. The $n$-collinear integral contains only one divergence arising from the limit $(n\cdot k)/(\bar n\cdot k)\to \infty$. Similarly the $\bn$-collinear diagram diverges only in the limit with $(n\cdot k)/(\bn\cdot k)\to 0$.  Each divergence is associated to a rapidity-separation between soft and collinear sectors.

There are now a variety of methods to regulate rapidity divergences.  We will explain the $\eta$-regulator (\cite{Chiu:2011qc}) and $\Delta$ regulator (\cite{Chiu:2009mg}), which we will use in our calculations later in this thesis.

\paragraph{\texorpdfstring{$\eta$}{eta}-Regulator Scheme for Rapidity Divergences}
\label{sec:I.5.2.3.1}
Since effective field theories are designed to resum large logarithms it is beneficial to regulate the theory in a way that manifests the renormalization group evolution.  This is achieved efficiently by a rapidity regulator analogous to dimensional regularization suggested by \cite{Chiu:2011qc}.  The rapidity regulator is applied by rewriting the mometum space Wilson line as
\begin{align}
W_n&=\sum_{\text{perms}}\exp\left[-\frac{gw^2}{\bn\cdot \cP}\frac{|\bn\cdot\cP_g|^{-\eta}}{\nu^{-\eta}}\bn\cdot A_n\right]\label{eq:30eq4.8}\\
S_n&=\sum_{\text{perms}}\exp\left[-\frac{gw}{n\cdot\cP}\frac{|2\cP_{g3}|^{-\eta/2}}{\nu^{-\eta/2}}n\cdot A_s\right]\label{eq:30eq4.9}
\end{align}
The new parameter $\nu$ acts as a rapidity cut-off. $\cP^\mu$ is the momentum operator and since $|2\cP_3|\to|\bn\cdot\cP|$ in the collinear limit the longitudinal momenta are in principle regulated. The power of $\eta$ differs between soft and collinear Wilson lines, with the correct power determined by requiring cancellation of rapidity divergences to all orders.  The parameter $w$ acts like a coupling constant to facilitate derivation of the RGE and is taken to $1$ at the end.  

The effective theory with the above regulators exhibits divergences in both $\eta\to 0$ and $\epsilon\to 0$ (from  dimensional regularization) limit. The order of limits is essential. The proper order, considering the origin of rapidity divergences, is first $\eta\to 0$ then $\epsilon\to 0$ such that $\eta/\epsilon^n\to 0$ for all $n>0$. This ordering ensures the rapidity cut-off remains on a fixed invariant mass hyperbola when the limit is taken. 


\paragraph{\texorpdfstring{$\Delta$}{Delta}-Regulator Scheme for Rapidity Divergences}
\label{sec:I.5.2.3.2}
The $\Delta$-regulator is defined by rewriting propagator of particle $i$ as
\begin{equation}\label{eq:31eq8}
\oneov{(p_i+k)^2-m_i^2}\to\oneov{(p_i+k)^2-m_i^2-\Delta_i}
\end{equation}
This is equivalent to a shift in the particle mass and hence can be carried out at the level of Lagrangian. The on-shell condition will still be $p_i^2=m_i^2$. The corresponding collinear Wilson lines are given by
\begin{align}
W_n&=\sum_{\text{perm}}\exp\left[-\frac{g}{\bn\cdot \cP-\delta_1}\bn\cdot A_n\right]\nn\\
W_\bn^\dag&=\sum_{\text{perm}}\exp\left[-\frac{g}{n\cdot\cP-\delta_2}n\cdot A_\bn\right]\,,\label{eq:31eq122}
\end{align}
while the soft Wilson lines are given for DIS by
\begin{align}
\tilde Y_\bn^\dag&=\sum_{\text{perm}}\exp\left[-\frac{g}{n\cdot \cP_s-\delta_2+i0}\bn\cdot A_s\right]\nn\\
Y_n&=\sum_{\text{perm}}\exp\left[-\frac{g}{\bn\cdot \cP_s-\delta_1-i0}n\cdot A_s\right]\label{eq:31eq123}
\end{align}
and for Drell-Yan process by
\begin{align}
\tilde Y_n^\dag&=\sum_{\text{perm}}\exp\left[-\frac{g}{n\cdot \cP_s-\delta_2-i0}\bn\cdot A_s\right]\nn\\
Y_n&=\sum_{\text{perm}}\exp\left[-\frac{g}{\bn\cdot\cP_s-\delta_1-i0}n\cdot A_s\right]\,,\label{eq:31eq124}
\end{align}
where $\delta_1=\Delta_1/p^+$, $\delta_2=\Delta_2/p^-$ with $p^\pm$ the $n$, $\bn$ direction component of collinear momentum.

In SCET, the denominator of a collinear propagator is replaced according to \eq{31eq8}. If particle $j$ couples with a collinear gluon in direction $n_i$ and goes off-shell, we have
\begin{equation}\label{eq:31eq9}
\oneov{(p_i+k)^2-m_j^2-\Delta_j}\to \oneov{\oneov{2}(\bar n_i\cdot k)(\bn_j\cdot p_j)(n_i\cdot n_j)-\Delta_j}\,,
\end{equation}
with momentum $k$ being $n$-collinear. Corresponding to this shifted denominator in the collinear propagator, the denominator of the eikonal gluon line must be shifted
\begin{align}
\frac{\epsilon\cdot n_j}{k\cdot n_j}\to\frac{\epsilon\cdot\bn_i}{k\cdot \bn_i-\delta_{j,n_i}}
\,, \qquad 
\delta_{j,n_i}\equiv \frac{2\Delta_j}{(n_i\cdot n_j)(\bn_j\cdot p_j)}\,.\label{eq:31eq10}
\end{align}
After the zero-bin subtraction, the particle index ($j$-dependence) in the shifts $\delta_{j,n_i}$ disappears, and one can combine $n_i$-collinear gluon emission from different particles into a single Wilson line in $\bn_i$ direction.  Thus combining the $\Delta$-regulator with the zero-bin subtraction completes the separation of soft and collinear modes in $\SCETb$ integrals.

\subsubsection{Rapidity Renormalization Group}
\label{sec:I.5.2.4}

The $\eta$-regulator has  the advantange of a simple form of RGE.  We continue with the example of the Sudakov form factor in \eq{29eq4.1}.  The RGE follow from the set of equations true for any observables:
\begin{equation}\label{eq:32eq4.15}
\frac{d}{d\ln[\mu]}(J_n,S)^{\text{bare}}=\frac{d}{d\ln[\nu]}(J_n,S)^{\text{bare}}=0\,.
\end{equation}
In addition, the scales $\mu$ and $\nu$ are independent of each other, implying that
\begin{equation}\label{eq:32eq4.16}
\left[\frac{d}{d\ln[\mu]},\frac{d}{d\ln[\nu]}\right]=0\,.
\end{equation}
The anomalous dimensions are given by
\begin{align}
\gamma_\mu^{n,S}&=-Z_{n,S}^{-1}\left(\fpp{}{\ln[\mu]}+\beta\fpp{}{g}\right)Z_{n,S}\,,\label{eq:32eq4.17}\\
\gamma_\nu^{n,S}&=-Z_{n,S}^{-1}\fpp{}{\ln\nu}Z_{n,S}\,,\label{eq:32eq4.18}
\end{align}
 and from \eq{32eq4.16} it follows that
\begin{equation}\label{eq:32eq4.19}
\paren{\fpp{}{\ln[\mu]}+\beta\fpp{}{g}}\gamma_\nu=\frac{d}{d\ln[\nu]}\gamma_\mu=\mathbb{Z}\Gamma_{\text{cusp}}\,,
\end{equation}
a restriction holding for all observables of interest. The last step is due to the fact that $\mu$-anomalous dimension is consistent with the hard anomalous dimension, and the latter is linear in the logarithm with coefficient $\Gamma_{\text{cusp}}$. The value of the integer coefficient $\mathbb{Z}$ is determined by whether one considers amplitude or squared-amplitude and for Sudakov form factor is 1 or 2. Due to the fact that the hard function is independent of $\nu$, we have the universal relation between the collinear and soft anomalous dimensions
\begin{equation}\label{eq:32eq4.20}
-2\mathbb{Z}_c=\mathbb{Z}_S
\end{equation}
which will be revisited in later chapters.

\begin{figure}
\centering
\includegraphics[width=.2\textwidth]{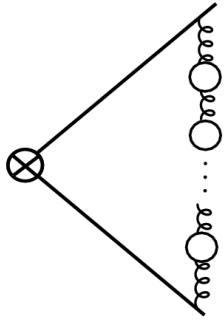}
\caption{Contributions to anomalous dimension $\gamma_\nu$ from renormalization of coupling. Note this is not present in the fixed order one-loop result. \label{fig:29fig2}}
\end{figure}

Renormalization group equations
\begin{align}
\mu\frac{d}{d\mu}(J_n,S)&=\gamma_\mu^{n,S}(J_n,S)\,,\nn\\
\nu\frac{d}{d\nu}(J_n,S)&=\gamma_\nu^{n,S}(J_n,S)\,.\label{eq:32eq4.29}
\end{align}
can be now applied to resum large logarithms arising from both large invariant mass ratios and rapidity ratios. The evolution in $\mu-\nu$ parameter space is independent of path thanks to \eq{32eq4.19}. Nonetheless one must be careful when solving the $\nu$ renormalization group flow since $\gamma_\nu$ contains a term $\alpha_s^n(\mu)(\ln(\mu/M))^m$ with $m\le n$. For example \fig{29fig2} shows that one loop results will be accompanied by a product of logarithms $\sum_n(\beta_0\alpha_s\ln(\mu/M))^n$ that can be large and needs to be resummed when e.g. $\mu\gg M$. One can show this by solving \eq{32eq4.19} to a required order in perturbation theory:
\begin{align}
\gamma_\nu&=\int^{\ln\mu} d\ln(\mu')\frac{d}{d\ln(\nu)}\gamma_\mu(\mu')+\text{const.}\nn\\
&\propto\int^{\ln\mu}d\ln(\mu')\Gamma_{\text{cusp}}(\mu')+\text{const.}\,,\label{eq:32eq4.30}
\end{align}
where the integration constant is determined by the fixed-order anomalous dimension and correponds to its non-cusp part.  (In our case, the non-cusp piece vanishes at 1-loop order.)  \eq{32eq4.30} determines entirely the logarithmic structure of $\gamma_\nu$ to all orders in perturbation theory (expanded with respect to $\alpha_s(\mu)$) and therefore provides validation for higher order computation. $\gamma_\nu$ in its integrated form resums the diagrams renormalizing the coupling. In the Abelian case these diagrams are given by bubble chains in \fig{29fig2}. 

\begin{figure}
\centering
\includegraphics[width=.375\textwidth]{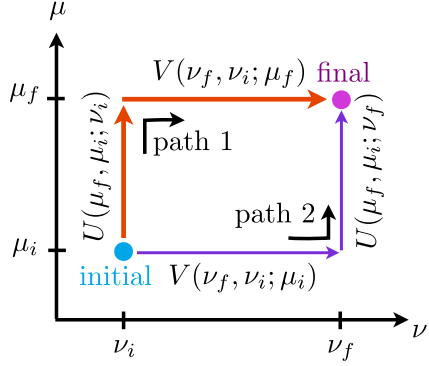}
\caption{The two choices of evolution paths in $\mu-\nu$ space which give equivalent results due to commutativity of the two evolutions (\cite{Chiu:2012ir}).}
\label{fig:29fig3}
\end{figure}

In our example of the Sudakov form factor, there are large logarithms in $\gamma_\nu$ besides rapidity logarithms that need to be resummed when $\mu_f\gg M$.  The RG evolution is depicted in \fig{29fig3}  where $U, V$ are evolution factors in $\mu,\nu$ respectively and $\mu_i,\nu_i$ are scales for initial conditions, e.g. $U(\mu_f,\mu_i;\nu_a)$ means evolving from $\mu_i$ to $\mu_f$ along a path with fixed $\nu=\nu_a$.  When evolution is traced through path 1 shown in \fig{29fig3} with $\mu_i\sim\nu_i\sim M\ll\mu_f\sim \nu_f$, we evolve $\mu$ first and $\nu$ last and can use fixed order form of $\gamma_\nu$.  Along path 2 in \fig{29fig3}, we evolve $\nu$ first and $\mu$ last and the integrated form \eq{32eq4.30} is necessary.  Since $\mu$- and $\nu$-evolutions commute we have
\begin{equation}\label{eq:32eq4.31}
V(\nu_f,\nu_i;\mu_f)U(\mu_f,\mu_i;\nu_i)=U(\mu_f,\mu_i;\nu_i)V(\nu_f,\nu_i;\mu_f)
\end{equation}
Note that the resummed form of $\gamma_\nu$ must be used when computing $V(\nu_f,\nu_i;\mu_f)$ to ensure the above relation.

It is worth noting that the low energy parameter $M$ shows up in the anomalous dimension which should not normally be the case. Nonetheless in the context of rapidity divergences $M$ is simply the invariant mass hyperbola on which the rapidity RGE runs, not really a low energy parameter.

\begin{figure}[t]
\centering
\includegraphics[width=.5\textwidth]{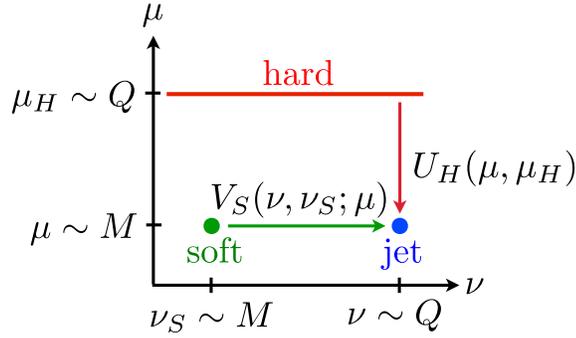}
\caption{The easiest strategy of evolution to resum all large logarithms arising in the Sudakov Form Factor \cite{Chiu:2012ir}}
\label{fig:29fig4}
\end{figure}

The first step in summing large logarithms is to identify the natural scales for hard functions $(\mu_H)$, soft functions $(\mu_S,\nu_S)$ and jet functions $(\mu_J,\nu_J)$ 
\begin{equation}\label{eq:32eq4.32}
\mu_H\sim Q,~~\mu_S\sim\nu_S\sim\mu_J\sim M,~~\nu_J\sim Q\,.
\end{equation}
Next we evolve $\mu,\nu$ to certain fixed scale and evaluate the fixed order functions at their natural scales to eliminate large logarithms,
\begin{align}
S(\mu,\nu)&=V_S(\nu,\nu_S;\mu)(U_S(\mu,\mu_S;\nu_S)S(\mu_S,\nu_S))\nn\\
J_n(\mu,\nu)&=V_J(\nu,\nu_J;\mu)(U_J(\mu,\mu_J;\nu_J)J_n(\mu_J,\nu_J))\nn\\
H(\mu)&=H(\mu_H)U(\mu,\mu_H)\,, \label{eq:32eq4.33}
\end{align}
where $U_{J,S}, V_{n,S}$ are $\mu$, $\nu$ evolution factors for jet and soft functions having run $\mu$ first and $\nu$ last. To resum all large logarithms due to running coupling it is necessary to use $\gamma_\nu$ in the integrand form in \eq{32eq4.30}. We have
\begin{align}
U_S(\mu,\mu_S;\nu_S)&=\exp\sqparen{-\frac{8\pi C_F}{\beta_0^2}\paren{\oneov{\alpha(\mu)}-\oneov{\alpha(\mu_S)}-\oneov{\alpha(\nu_S)}\ln\frac{\alpha(\mu)}{\alpha(\mu_S)}}}\label{eq:32eq4.34}\\
V_S(\nu,\nu_S;\mu)&=\exp\sqparen{\frac{2C_F}{\beta_0}\ln\paren{\frac{\alpha(\mu)}{\alpha(M)}}\ln\paren{\frac{\nu^2}{\nu_S^2}}}\label{eq:32eq4.35}\\
U_S(\mu,\mu_J;\nu_J)&=\exp\sqparen{-\frac{2C_F}{\beta_0}\paren{\frac34+\oneov{2}\ln\paren{\frac{\nu_J^2}{Q^2}}}\ln\frac{\alpha(\mu)}{\alpha(\mu_J)}}\label{eq:32eq4.36}\\
V_J(\nu,\nu_J;\mu)&=\exp\sqparen{-\frac{C_F}{\beta_0}\ln\paren{\frac{\alpha(\mu)}{\alpha(M)}}\ln\paren{\frac{\nu^2}{\nu_J^2}}}\label{eq:32eq4.37}\\
U_H(\mu,\mu_H)&=\exp\sqparen{-\frac{8\pi C_F}{\beta_0^2}\paren{\oneov{\alpha(\mu_H)}-\oneov{\alpha(\mu)}-\oneov{\alpha(Q)}\ln\frac{\alpha(\mu)}{\alpha(\mu_H)}}}\label{eq:32eq4.38}\\
S(\mu_S,\nu_S)&=1+\frac{\alpha(\mu_S)C_F}{\pi}\sqparen{\ln^2\paren{\frac{\mu_S}{M}}-2\ln\paren{\frac{\mu_S}{M}}\ln\paren{\frac{\nu_S}{M}}-\frac{\pi^2}{24}}\label{eq:32eq4.39}\\
J_n(\mu_J,\nu_J)&=1+\frac{\alpha(\mu_J)C_F}{\pi}\sqparen{\ln\paren{\frac{\mu_J}{M}}\ln\paren{\frac{\nu_J}{n\cdot p_1}}+\frac{3}{4}\ln\paren{\frac{\mu_J}{M}}-\frac{\pi^2}{12}+\frac12}\,.\label{eq:32eq4.40}
\end{align}
At the order we are working we can utilize relations \eq{32eq4.34} through \eq{32eq4.37} to directly show commutativity \eq{32eq4.31}. For an arbitrary choice of scales $\mu$ and $\nu$, \eq{32eq4.33} through \eq{32eq4.39} perform the resummation. It is convenient to choose $\mu=\mu_J=\mu_S\sim M$ and $\nu=\nu_J\sim Q$ where it is only necessary to evolve hard function in $\mu$ and soft function in $\nu$ to the natural scale of jet function as is illustrated in \fig{29fig4}. Here it is not necessary to use integrated form \eq{32eq4.30} and the fixed order form of $\gamma_\nu$ will suffice.

\begin{figure}[t]
\centering
\includegraphics[width=.4\textwidth]{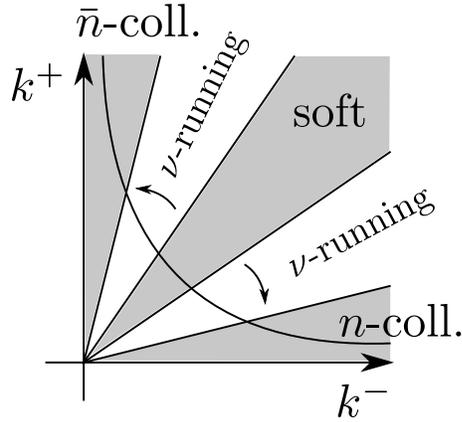}
\caption{Flow along the hyerbola of invariant mass corresponding to $\nu$-running.}
\label{fig:29fig5}
\end{figure}

\fig{29fig5} shows how rapidity renormalization group (RRG) flow in soft and collinear regions is reflected by a change in the scale $\nu$. The collinear function is at the scale $Q$ and soft function at the scale $n\cdot k\sim \bn\cdot k\sim M$.  To sum the logarithms, one pushes the cutoff of the soft function along the invariant mass hyperbola to a point in the collinear sectors where the scale $\nu$ minimizes the logarithms.

The RRG allows proving that the resummed form factor is independent of $\mu$. In our example, one can merge the evolution factors and write the completely resummed form factor as
\begin{align}
F(Q^2,M^2)&=E(\mu,\mu_H,\mu_J,\mu_S;\nu_J,\nu_S)H(Q^2,\mu_H^2)J_n(\mu_J,\nu_J;M;Q)\nn\\
&\times J_\bn(\mu_J,\nu_J;M;Q)S(\mu_S,\nu_S;M) \label{eq:32eq4.41}
\end{align}
The dependence of jet and soft functions on the scale $M$ and $Q$ is explicit here. The this defines the total evolution factor $E$, whose dependence on $\mu$ is always sub-leading and will cancel in exact result. To show that $\mu$-dependence cancels to any given working order we keep $\mu$ an arbitrary scale. The choice of parameters
\begin{align}\label{eq:33eq4.43}
\mu_H&=Q, \quad \mu_J=\mu_S=M\sim \mu, \quad
\nu_J=Q, \quad \nu_S=M
\end{align}
minimizes logarithms in all sectors. At 1-loop, we then have
\begin{equation}\label{eq:33eq4.44}
E=\exp\left[-\frac{8\pi C_F}{\beta_0^2}\paren{\oneov{\alpha(Q)}-\oneov{\alpha(\mu)}-\oneov{\alpha(Q)}\ln\frac{\alpha(\mu)}{\alpha(Q)}}+2\frac{C_F}{\beta_0}\ln\frac{\alpha(\mu)}{\alpha(M)}\ln\frac{Q^2}{M^2}
\right]
\end{equation}
where the first term comes from $\mu$-renormalization and the second term comes from rapidity renormalization.  Without the second term, logarithms of $Q/M$ would have remained uncancelled in the form factor, whereas now the exponent of the form factor is independent of $\mu$ to our working order in logarithms. 

Cancellation of the hard double logarithm is enabled by summing the large logarithm that arose in the low-scale matrix element due to the two-scale nature of the $\SCETb$ problem.  To see this, consider a generic soft-collinear factorization of a physical process with large double logarithms,
\begin{equation}\label{eq:33eq4.45}
\sigma_{\text{resum}}=\exp\sqparen{\Gamma[\alpha]L^2-2\Gamma[\alpha]L\bar L+\ldots}f(\tilde L)
\end{equation}
where $L=\ln(Q/\mu)$ and $\tilde L=\ln(M/\mu)$ with $Q$ the hard scale, $M$ the infrared scale and we assume $\mu\sim M$ and hence $L\gg \tilde L$. Here $f$, the low scale matrix element of soft and collinear sectors, contains no large logaithms. The running in $\alpha_s$ is suppressed, as it does not alter the essence of this discussion. Varying \eq{33eq4.45} with respect to $\mu$, we have
\begin{equation}\label{eq:33eq4.48}
\delta_\mu\sigma_{\text{resum}}=\sigma_{\text{resum}}\delta_\mu\paren{\Gamma[\alpha]L^2-2\Gamma[\alpha]L\tilde L}=0+\ldots
\end{equation}
Since $\delta L=\delta\tilde L$ the exponent is in the correct form to cancel leading variation. Generally, an RRG summation of the form $2\Gamma_{n-1}\alpha^n L\tilde L$ cancels terms of the form $\Gamma_{n-1}\alpha^nL^2$ in the resummed exponent. Additional sub-leading variations that scale as $\alpha^n\tilde L$ or $\alpha^n$ are cancelled within the low scale matrix elements as they do not contain large logarithms.


\subsubsection{Application: Drell-Yan at Small \texorpdfstring{$p_T$}{pT} with \texorpdfstring{$\eta$}{eta} Regulator}
\label{sec:I.5.2.6}

In this section I study transverse momentum dependence (TMD) in Drell-Yan processes and calculate the quark tranverse-momentum-dependent parton distribution function (TMDPDF).\footnote{The work in this section is in collaboration with D. Kang and C. Lee.}  The TMDPDF is essential to measure physics observables with non-trivial transverse momentum dependence in hadron colliders, especially for the proper interpretation of signals of beyond Standard Model physics and Higgs boson searches.  My calculation of the TMDPDF in DY processes is original and unpublished.

For observables sensitive to the transverse momentum of the final state, large rapidity logarithms in the ratio $p_\perp/Q$ arise when the total transverse momentum of the lepton pair is small because the lepton pair recoils off QCD radiation with equal and opposite transverse momentum.  On the other hand to remain in the perturbative region, we assume $p_\perp\gg\LQCD$.  In this case, $p_\perp$ is the infrared scale and $p_\perp/Q\sim\lambda$, the $\SCETb$ momentum scaling.  

I first derive a factorization theorem for DY processses with unintegrated proton transverse momentum.  In this derivation, I consider the cross section differentiated with respect to the Bjorken-$x$ of the two protons and the transverse component of the momentum transferred between them.  Then I calculate the two collinear factors and one soft function to $\cO(\as)$, using the $\eta$-regulator introduced in the previous section.  After this, I explicitly demonstrate the resummation of rapidity logarithms via the rapidity renormalization group (RRG) equation.  I also express the TMD PDF in impact parameter $b$-space where the RG evolution is simpler.  Finally I give the cross section with the resummed large logarithms.

\paragraph{Factorization for $p_\perp$-dependent Drell-Yan cross section}

I denote $e_q$ as the quark electric charge, $p$ and $\bar p$ as the two incoming hadron momenta with $s=(p+\bar p)^2$. The hard photon momentum $q$ scales as $Q(1,1,\lambda)$ with $\lambda=q_\perp/Q$.  The Bjorken-$x$s for the two protons are
\begin{align}
x_1 = \frac{q^2}{2q\cdot p},\qquad x_2 = \frac{q^2}{2q\cdot \bar p}.
\end{align}
We can write the factorized cross-section of Drell-Yan at small $q_\perp$
\begin{align}
d\sigma&=\frac{4\pi\alpha^2}{3N_cq^2}\frac{dx_1dx_2d^2\vec q_\perp}{2(2\pi)^4}H(Q^2/\mu^2)\sum_q e_q^2\int \!d^2\vec k_{n\perp}d^2\vec k_{\bar n\perp}d^2\vec k_{s\perp}\,\nn\\
&\times \delta^{(2)}(\vec q_\perp\!-\vec k_{n\perp}\!-\vec k_{\bar n\perp}\!-\vec k_{s\perp})f_n(x_1;\vec k_{n\perp})f_{\bar n}(x_2;\vec k_{\bar n\perp})\phi(\vec k_{s\perp})\,,\label{eq:34eq2.1}
\end{align}
where $H(Q^2/\mu^2)=|C(Q^2/\mu^2)|^2$ is the hard function matching onto QCD at $Q^2$, and the two quark TMDPDFs and the soft function are
\begin{align}\label{eq:TMDPDFfn}
f_n(x_1,\vec k_{n\perp})&=\frac12\int\frac{dr^- d^2\vec r_\perp}{(2\pi)^3}e^{-i(\frac12 r^-x_1p^+-\vec r_\perp\cdot\vec k_{n\perp})}f_n(0^+,r^-,\vec r_\perp)\\
\label{eq:TMDPDFfbarn}
f_{\bar n}(x_2,\vec k_{\bar n\perp})&=\frac12\int\frac{dr^+d^2\vec r_\perp}{(2\pi)^3}e^{-i(\frac12 r^+x_2\bar p^--\vec r_\perp\cdot\vec k_{\bar n\perp})}f_{\bar n}(r^+,0^-,\vec r_\perp)\\
\label{eq:TMDPDFsoft}
\phi(\vec k_{s\perp})&=\int\frac{d^2\vec r_\perp}{(2\pi)^2}e^{i\vec r_\perp\cdot\vec k_{s\perp}}\phi(0^+,0^-,\vec r_\perp)\,,
\end{align}
with
\begin{align}
\bar f_n(0^+,r^-,\vec r_\perp)&=\left\langle p\bigg\vert [\bar \xi_n W_n](0^+,r^-,\vec r_\perp)\frac{\bnslash}{2}[W_n^\dg\xi_n](0)\bigg\vert p\right\rangle \label{eq:34eq2.2a}\\
\bar f_{\bar n}(r^+,0^-,\vec r_\perp)&=\left\langle \bar p\bigg\vert [\bar \xi_{\bar n} W_{\bar n}](0)\frac{\nslash}{2}[W_{\bar n}^\dg\xi_{\bar n}](r^+,0^-,\vec r_\perp)\bigg\vert \bar p\right\rangle
\label{eq:34eq2.2b} \\
\phi(0^+,0^-,\vec r_\perp)&=\left\langle 0\bigg\vert \tr[S_n^\dg S_{\bar n}](0^+,0^-,\vec r_\perp)[S_{\bar n}^\dg S_n](0)\bigg\vert 0\right\rangle.\label{eq:34eq2.3}
\end{align}
In the following subsections, we will calculate the $\cO(\alpha_s)$ corrections to each piece, convert to $b$-space, combine them and present the resummed result for the differential cross section \eq{34eq2.1}.

\paragraph{Calculation of Quark TMD PDF: $q\to q$ process}
\label{sec:I.5.2.6.1}

\begin{figure}[h!]
\centering
\includegraphics[width=.8\textwidth]{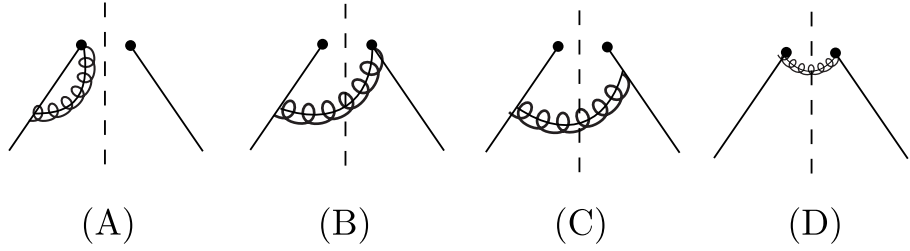}
\caption{$\mathcal{O}(\alpha_s)$ collinear function correction.  Mirror image diagrams not shown.}
\label{fig:34fig1}
\end{figure}

In this section I calculate the collinear functions in \eq{34eq2.1} to $\mathcal{O}(\alpha_s)$. We use the $\eta$-regulator to regulate the rapidity divergences brought in by matching \eqs{34eq2.2a}{34eq2.2b} from $\text{SCET}_{\text{I}}$ to $\text{SCET}_{\text{II}}$. The $\eta$-regulator-implemented Wilson lines in momentum space are given in \eq{30eq4.8} and \eq{30eq4.9}.
The $n$-collinear function $\mathcal{O}(\alpha_s)$ correction diagrams are shown in \fig{34fig1}
where diagrams \ref{fig:34fig1}(A) and (B) also have mirror images.  The diagram \ref{fig:34fig1}(D) vanishes because it is proportional to $\bar n\cdot\bar n=0$.

The sum of the virtual contribution from Fig.\,\ref{fig:34fig1}(A) and its mirror image is
\begin{align}
\hat f_{n\perp}^{(1A)}&=\delta(1-x_1)\delta^{(2)}(\ell_\perp)(-2ig^2C_F)\mu^{2\epsilon}\nu^\eta\int\frac{d^dk}{(2\pi)^d}\frac{n\cdot(p+k)}{[(p^++k^+)k^--k_\perp^2]+i0}\frac{|k^+|^{-\eta}}{k^++i0}\nn\\
&\times\oneov{k^+k^--k_\perp^2+i0}+h.c.\,,\label{eq:34eq2.6}
\end{align}
and it is scaleless.

The sum of the real contribution from Fig.\,\ref{fig:34fig1}(B) and its mirror image is
\begin{align}
\hat f_{n\perp}^{(1B)}&=(-4\pi g^2C_F)p^+\mu^{2\epsilon}\nu^\eta\int\frac{d^dk}{(2\pi)^d}\frac{p^+-k^+}{-k^-(p^+-k^+)-k_\perp^2+i0}\frac{|k^+|^{-\eta}}{k^++i0}\nn\\
&\times\delta(k^+-(1-x_1)p^+)\delta^{(2)}(\vec k_\perp+\vec k_{n\perp})\delta(k^+k^--k_\perp^2)+h.c.\nn\\
&=\frac{4\alpha_sc_F}{(2\pi)^{2-2\epsilon}}p^+\frac{\Gamma(\frac12-\epsilon)}{\sqrt{\pi}\Gamma(1-\epsilon)}\frac{\mu^{2\epsilon}\nu^\eta}{|\vec k_{\perp n}^2|^{1+\epsilon}}\oneov{((1-x_1)p^+)^{1+\eta}}\,.\label{eq:34eq2.7}
\end{align}
The real contribution from Fig.\,\ref{fig:34fig1}(C) is
\begin{align}
\hat f_{n\perp}^{(1C)}&=(2\pi g^2C_F)p^+\mu^{2\epsilon}\nu^{\eta}\int\frac{d^dk}{(2\pi)^d}\delta(k^2)\delta((1-x_1)p^+-k^+)\delta^{(2)}(\vec k_\perp+\vec k_{n\perp})\nn\\
&\times\frac{-p^+k^-+k^+k^-}{[(p-k)^2+i0][(p-k)^2-i0]}\nn\\
&=\frac{2\alpha_sc_F}{(2\pi)^{2-2\epsilon}}\frac{\Gamma(\frac12-\epsilon)}{\sqrt{\pi}\Gamma(1-\epsilon)}(1-\epsilon)(1-x)\oneov{|\vec k_{\perp n}^2|^{1+\epsilon}}\,.\label{eq:34eq2.8}
\end{align}
The zero-bin contributions for each of these diagrams are zero.  The zero bin of Fig.\,\ref{fig:34fig1}(A) is obtained by considering the limit $p^+\gg k^+$,
\begin{align}
\hat f_{n\perp}^{(1A)\phi}&=\delta(1-x_1)\delta^{(2)}(\ell_\perp)(-2ig^2C_F)\mu^{2\epsilon}\nu^\eta\nn\\
&\times\int\frac{d^dk}{(2\pi)^d}\frac{1}{k^-+i0}\frac{|k^+|^{-\eta}}{k^++i0}\frac{1}{k^+k^--k_\perp^2+i0}+\text{h.c.}\nn\\
&=0\,.\label{eq:34eq2.9}
\end{align}
The zero bins of Fig.\,\ref{fig:34fig1}(B) and its mirror image are obtained under the same condition $p^+\gg k^+$
\begin{align}
\hat f_{n\perp}^{(1B)\phi}&=-4\pi g^2C_Fp^+\mu^{2\epsilon}\nu^\eta\!\int\!\frac{d^dk}{(2\pi)^d}\frac{1}{k^-\!-i0}\frac{|k^+|^{-\eta}}{k^+\!+i0}\,\delta^{(2)}(\vec k_\perp\!+\vec k_{n\perp})\delta(k^+k^-\!-k_\perp^2)+\text{h.c.}\nn\\
&=0\,.\label{eq:34eq2.10}
\end{align}
The zero bin of Fig.\,\ref{fig:34fig1}(C) is
\begin{align}
\hat f_{n\perp}^{(1C)\phi}&=(2\pi g^2C_F)p^+\mu^{2\epsilon}\nu^\eta\int\frac{d^dk}{(2\pi)^d}\delta(k^2)\delta^{(2)}(\vec k_\perp+\vec k_{n\perp})\frac{p^+k^-}{(-p^+k^-+i0)(p^+k^-+i0)}\nn\\
&=0\,.\label{eq:34eq2.11}
\end{align}

\paragraph{Calculation of Quark TMD PDF: Calculation of $q\to g$ process}

\begin{figure}[b]
\centering
\includegraphics[width=.45\textwidth]{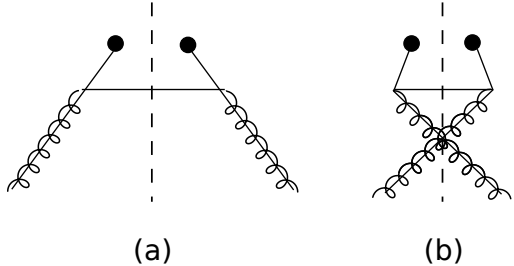}
\caption{$\mathcal{O}(\alpha_s)$ quark mixing in gluon TMDPDF}
\label{fig:34fig2}
\end{figure}

At $\cO(\alpha_s)$ the quark and gluon PDFs mix, and I must include the $q\to g$ process.  The gluon TMDPDF is defined as
\begin{align}
\hat f_{\perp g\to q}^{\mu\nu}(x^+,\vec x_\perp)&=\int\frac{dz}{4\pi}e^{\frac{i}{2}z(x^+p^-)}\int\frac{d^2\vec p_\perp}{(2\pi)^2}e^{i\vec x_\perp\cdot\vec p_\perp}f^{\mu\nu}_{\perp g/p}(z,\vec p_\perp)\nn\\
f_{\perp g\to q}^{\mu\nu}(p_z,\vec p_\perp)&=(\vec n\cdot p)\left\langle
p\bigg\vert [B_{n\perp}]^{A\mu}(0)
\delta(p_z-\overline{\mathcal{P}})\delta^{(2)}(\vec p_\perp-\vec{\mathcal{P}}_\perp)B_{n\perp}^{A\nu}(0)\bigg\vert\right\rangle\label{eq:34eq2.12}
\end{align}
Up to $\mathcal{O}(\alpha_s)$, the quark mixing contribution to the gluon TMDPDF is shown in Fig.\,\ref{fig:34fig2}.  

\noindent The amplitude of Fig.\ref{fig:34fig2}(a) is
\begin{align}
f_{\perp g\to q}^{\mu\nu}&=q^2\frac{\mu^{2\epsilon\nu^\eta}}{(d-2)}\frac{T_F}{p^+}\int\frac{d^dq}{(2\pi)^d}(2\pi)\delta(q^2)\theta(q^0)\delta(1-x_1-q^+/p_n^+)\delta^{(2)}(\vec p_\perp+\vec q_\perp)\nn\\
&\times 4q^+(-\vec q_\perp^{\:2})\cdot\left[+\frac{1}{q^+}\frac{1}{q^+-p^+}+(d-2)\left(\frac{1}{q^+}-\frac{1}{q^+-p^+}\right)^2\right]\nn\\
&=\frac{\alpha_sT_F}{2\pi^2}\bigg\{-\delta(\vec p_\perp^{\:2})\left[\theta(1-x_1)(x_1^2+(1-x_1)^2)\left(\frac{1}{\epsilon}+1\right)-1\right]\nn\\
&+\frac{1}{\mu^2}\theta(1-x_1)x_1^2+(1-x_1)^2)\left(\frac{\mu^2}{\vec p_\perp^{\:2}}\right)_+\bigg\}\,. \label{eq:34eq2.13}
\end{align}
The zero bin is obtained from the limit $p^+\gg q^+$ and is zero.  The amplitude of Fig.\,\ref{fig:34fig2}(b) is
\begin{align}
f_{\perp g\to q}^{\mu\nu}&=g^2\frac{\mu^{2\epsilon}\nu^\eta}{(d-2)}\frac{T_F}{(-p^+)}\int\frac{d^dq}{(2\pi)^d}\delta(q^2)\theta(q^0)\delta(1-x_1+q^+/p_n^+)\delta^{(2)}(-\vec p_\perp+\vec q_\perp)\nn\\
&\times 4q^+(-\vec q_\perp^{\:2})\left[\frac{1}{q^+}\frac{1}{q^++p^+}+(d-2)\left(\frac{1}{q^+}-\frac{1}{q^++p^+}\right)^2\right]\,,
\end{align}
which does not contribute for $x_1>0$.

\paragraph{Calculation of Quark TMD PDF: Result in $b$ space}

In this section, I transform the $\cO(\as)$ collinear function result into impact paramber $b$-space for simplification.
In $b$ space, $\mathcal{O}(\alpha_s)$ order collinear function is
\begin{align}
f^{(1n)}(\vec b)&=\frac{\alpha_sC_F}{(2\pi)^2}\bigg\{-\frac{2}{\eta}\Gamma(-2\epsilon)\paren{\frac{b\mu e^{\gamma_E}}{2}}^{2\epsilon}e^{-\gamma_E \epsilon}
\nn\\&
+\frac{2}{\epsilon}\left[\ln\frac{\nu}{p^+}-\paren{\oneov{1-x_1}}_+-\frac12(1-x_1)\right]\nn\\
&+\paren{\ln\frac{b\mu}{2}}\left[2\ln\frac{\nu}{p^+}-2\paren{\oneov{1-x_1}}_+-(1-x_1)\right]+1\bigg\}\,.\label{eq:34eq2.14}
\end{align}
The counter term is
\begin{align}
Z_c(\vec b)&=1-\frac{\alpha_sC_F}{(2\pi)^2}\bigg\{-\frac{2}{\eta}\Gamma(-2\epsilon)\paren{\frac{\mu}{\mu_b}}^{\!2\epsilon}\! e^{-\gamma_E \epsilon}
\nn\\&
+\oneov{\epsilon}\left[-2\ln\frac{\nu}{p^+}+2\paren{\oneov{1-x_1}}_+ +(1-x_1)\right]\bigg\}\,,\label{eq:34eq2.15}
\end{align}
in which I have defined $\mu_b=2/(b e^{\gamma_E})$ for notational convenience.  The renormalized collinear function is
\begin{equation}\label{eq:34eq2.16}
f^{(1n)}(\vec b)=\frac{\alpha_sC_F}{(2\pi)^2}\left\{
\ln\frac{b\mu}{2}\left[2\ln\frac{\nu}{p^+}-2\paren{\oneov{1-x}}_+-(1-x)\right]+1
\right\}\,,
\end{equation}
and the anomalous dimensions are
\begin{align}
\mu_c^\mu&=-\frac{\alpha_sc_F}{\pi}\left[-2\ln\frac{\nu}{p^+}+2\paren{\oneov{1-x}}_++(1-x)\right]\,,
\label{eq:34eq2.17}\\
\mu_c^\nu&=\frac{\alpha_sc_F}{\pi}\ln\frac{\mu^2}{\mu_b^2}\,.\label{eq:34eq2.18}
\end{align}
The renormalized collinear function $f_{1n}^{(\tilde R)}$ satisfies the RG equations
\begin{align}
\mu\frac{d}{d\mu}f_{1n}^{(\tilde R)}(\mu,\nu,\vec b)&=\gamma(\mu)f_{1n}^{(\tilde R)}(\mu,\nu,\vec b)\label{eq:34eq2.19}\\
\nu\frac{d}{d\nu}f_{1n}^{(\tilde R)}(\mu,\nu,\vec b)&=\gamma(\nu)f_{1n}^{(\tilde R)}(\mu,\nu,\vec b)\,,\label{eq:34eq2.20}
\end{align}
with the solutions 
\begin{align}
f_{1n}^{(\tilde R)}(\mu,\nu,\vec b)&=f_{1n}^{(\vec R)}(\vec b,\mu_0,\nu)\paren{\frac{\mu}{\mu_0}}^{-\frac{\alpha_sC_F}{\pi}[-2\ln\frac{\nu}{p^+}+2\paren{\oneov{1-x_1}}_+ +(1-x_1)]}\,,\label{eq:34eq2.21}\\
f_{1n}^{(\tilde R)}(\mu,\nu,\vec b)&=f_{1n}^{(\vec R)}(\vec b,\mu,\nu_0)\paren{\frac{\mu^2}{\mu_b^2}}^{-\paren{\ln\frac{\nu}{\nu_0}}\frac{\alpha_sC_F}{\pi}}\,.\label{eq:34eq2.22}
\end{align}

Next I match the TMDPDF onto the ordinary PDF $I_{q\to q}$ in $b$ space with $q_\perp$ integrated over. Up to the $\mathcal{O}(\alpha_s)$, the bare, $j$th order (superscript $B(j)$) TMDPDF can be written 
\begin{align}
f_{q\to q}^{B(0)}&=I_{q\to q}^{B(0)}\otimes f_q^{B(0)}(z)=I_{q\to q}^{B(0)}(z,\vec b)=\frac{1}{(2\pi)^2}\delta(1-z)\label{eq:34eq2.23}\\
f_{q\to q}^{B(1)}&=I_{q\to q}^{B(0)}\otimes zf_q^{B(1)}(z)+I_{q\to q}^{B(1)}\otimes zf_q^{B(0)}(z)\,,\label{eq:34eq2.24}
\end{align}
where $\otimes$ represents convolution with respect to $z$ and
\begin{align}
I_{q\to q}^{B(1)}(z,\vec b)&=f_q^{B(1)}-\frac{1}{2\pi}\left(\frac{\alpha_sC_F}{2\pi^2}\right)\frac{P_{qq}(z)}{\epsilon}\nn\\
&=f_q^{\eta\text{-div}}+\tilde f_q^{\epsilon\text{-div}}+f_q^{\text{fin}}\,.\label{eq:34eq2.25}
\end{align}
$f_q^{\eta\text{-div}}$ is the $\eta$-divergent part, $f_q^{\text{fin}}$ is the finite part, and $\tilde f_q^{\epsilon\text{-div}}$ is the $\epsilon$-divergent part of the TMDPDF.  I obtain
\begin{align}
\tilde f_q^{\epsilon\text{-div}}&=\frac{\alpha_sC_F}{(2\pi)^2}\frac{2}{\epsilon}\left[P_{qq}(z)+2\delta(1-z)\ln\frac{\nu}{p^+}-2\left(\frac{1}{1-z}\right)_++(1-z)\right]\nn\\
&=\frac{\alpha_sC_F}{(2\pi)^2}\frac{2}{\epsilon}\left[\frac32+2\ln\frac{\nu}{p^+}\right]\delta(1-z)\,,
\end{align}
where $P_{q\to q}(z)=\left(\oneov{1-z}\right)_++(1+z^2)+\frac32\delta(1-z)$.  The renormalized $j$th order (superscript $R(j)$) ordinary PDF is related to the bare PDF by
\begin{align}
I_{q\to q}^{B(0)}(z,\vec b)&=Z^{(0)}\otimes I^{R(0)}\nn\\
I_{q\to q}^{B(1)}(z,\vec b)&=Z^{(1)}\otimes I^{R(0)}(z,\vec b)+Z^{(0)}\otimes I^{R(1)}(z,\vec b)\,,
\end{align}
where the counter-terms are
\begin{align}
Z^{(0)}&=\oneov{2\pi}\nn\\
Z^{(1)}&=\frac{\alpha_sC_F}{(2\pi)^2}\bigg[-\frac{2}{\eta}\Gamma(-2\epsilon)e^{-\gamma_E\epsilon}\left(\frac{\mu}{\mu_b}\right)^{2\epsilon}
+\oneov{\epsilon}\left(\frac23+2\ln\frac{\nu}{p^-}\right)\bigg]\,.
\end{align}

\paragraph{Soft function}
\label{sec:I.5.2.6.2}

\begin{figure}[t!]
\centering
\includegraphics[width=.65\textwidth]{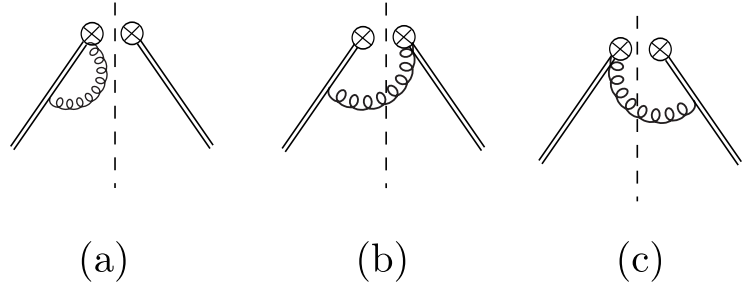}
\caption{$\mathcal{O}(\alpha_s)$ correction to soft function.}\label{TMDPDFsoftdiagrams}
\end{figure}

We reproduce the soft function and anomalous dimensions obtained in \cite{Chiu:2012ir}.  The soft function in the quark TMDPDF factorization formula is 
\begin{equation}\label{eq:34eq3.1}
S(0^+,0^-,\vec y_\perp)=\left\langle 0\bigg\vert \tr[S_n^\dg,S_{\bar n}](0^+,0^-,\vec y_\perp)[S_{\bar n}^\dg, S_n](0)\bigg\vert 0\right\rangle \,.
\end{equation}
Up to $\mathcal{O}(\alpha_s)$ order, the Feynman diagrams of the soft function are shown in Fig.\,\ref{TMDPDFsoftdiagrams}. The virtual diagram Fig.\,\ref{TMDPDFsoftdiagrams} is scaleless and hence zero in our regulation scheme.  The sum of the results for real soft function diagrams Fig.\,\ref{TMDPDFsoftdiagrams}(b) and Fig.\,\ref{TMDPDFsoftdiagrams}(c) is
\begin{align}
S_1^{(b+c)}&=(-igt_a)^2(-i)(-2\pi i)2\mu^{2\epsilon}\nu^\eta\int\frac{d^dk}{(2\pi)^d}\delta(k^2)\nn\\
&\times\delta^{(2)}(\vec k_\perp+\vec k_{n\perp})\frac{i}{k\cdot \bar n+i0}\frac{-i}{k\cdot n-i0}|k^+ -k^-|^\eta+\text{h.c.}\nn\\
&=\frac{4\alpha_sC_F}{(2\pi)^{3-2\epsilon}}\frac{\nu^{2\epsilon}\nu^\eta e^{\epsilon\gamma_E}i}{(\vec p_\perp^{\:2})^{1+\epsilon+\eta/2}}\frac{\Gamma(1+\epsilon+\eta/2)}{\Gamma(1+\eta/2)}\frac{2^\eta\Gamma(1/2-\eta/2)}{\sqrt{\pi}}\Gamma(\eta/2)\,.\label{eq:34eq3.2}
\end{align}
In $b$ space, the result above becomes 
\begin{align}
\tilde S^{(1)}(\vec b)&=\frac{2\alpha_sC_F}{\pi(2\pi)^2}\bigg[\oneov{\eta}\Gamma(-\eta)\paren{\frac{\mu}{\mu_b}}^{2\epsilon}e^{-\gamma_E \epsilon}+\oneov{2\epsilon^2}-\oneov{2\epsilon}\ln\frac{\nu^2}{\mu^2}\nn\\
&-\ln\frac{\nu b}{2}\ln\frac{\mu b}{2}+\ln^2\frac{\mu b}{2}+\frac{\pi^2}{24}\bigg]\,.\label{eq:34eq3.3}
\end{align}
The corresponding renormalization counter term is
\begin{equation}\label{eq:34eq3.4}
Z_s^{(1)}(\vec b)=1-\frac{\alpha_sC_F}{(2\pi)^4\pi}\left[\frac{2}{\eta}\Gamma(-\epsilon)e^{-\gamma_E \epsilon}\paren{\frac{\mu}{\mu_b}}^{2\epsilon}+\oneov{\epsilon^2}+\oneov{\epsilon}\ln\frac{\mu^2}{\nu^2}\right]\,,
\end{equation}
and the renormalized soft function in $b$ space is
\begin{equation}\label{eq:34eq3.5}
\tilde{S}^{R(1)}(\vec b)=\frac{2\alpha_sC_F}{\pi(2\pi)^2}\left[-\ln\frac{\nu b}{2}\ln\frac{\mu b}{2}+\ln^2\frac{\mu b}{2}+\frac{\pi^2}{24}\right]\,.
\end{equation}
The anomalous dimensions in $b$ space are
\begin{align}
\gamma_\mu^s&=\frac{2\alpha_sC_F}{\pi}\ln\frac{\mu^2}{\nu^2}\label{eq:34eq3.6}\\
\gamma_\nu^s&=-\frac{2\alpha_sC_F}{\pi}\ln\frac{\mu^2}{\mu_b^2}\,,\label{eq:34eq3.7}
\end{align}
which determine the RG equations in $\nu$ and $\mu$ respectively,
\begin{align*}
\nu\frac{d\tilde S^{(R)}}{d\nu}&=\gamma_s^\nu\tilde S^{(R)}\,,\\
\mu\frac{d\tilde S^{(R)}}{d\mu}&=\gamma_s^\mu\tilde S^{(R)}\,.
\end{align*}
The solutions to the above RG equations are
\begin{align}
\tilde S(\mu,\nu,\vec b)&=\exp\left[-\ln\paren{\frac{\nu}{\nu_0}}\frac{2\alpha_sC_F}{\pi}\ln\frac{\mu^2}{\mu_b^2}\right]\tilde S(\nu_0,\mu,\vec b)\,,\\
\tilde S(\mu,\nu,\vec b)&=\exp\left[\frac{2\alpha_sC_F}{\pi}\left[\ln^2\paren{\frac{\mu}{\nu}}-\ln^2\paren{\frac{\mu_0}{\nu}}\right]\right]\tilde S(\mu_0,\nu,\vec b)\,,
\end{align}
recalling the definition $\mu_b=2/(b e^{\gamma_E})$.

\paragraph{RG evolution}
\label{sec:I.5.2.6.3}

We can now combine these results and complete the double running in $\mu$ and $\nu$.  The total cross section in $b$ space is written as a product of hard, soft, and 2 PDF functions in Eqs.\,\eqref{eq:34eq2.1},\:\eqref{eq:TMDPDFfn},\:\eqref{eq:TMDPDFfbarn} and \eqref{eq:TMDPDFsoft},
\begin{equation} \label{eq:34eq4.1}
\sigma(\mu,\nu) =H(Q^2,\mu) \tS(\mu,\nu) \tft (\mu,\nu) \tft (\mu,\nu)\,.
\end{equation}
Invariance of \eq{34eq4.1} under change of the scales $\mu$ and $\nu$ leads to consistency conditions between the anomalous dimensions for these functions,
\begin{align} 
\label{eq:34eq4.2}
\frac{\df \sigma (\mu,\nu)}{\df \ln\mu} =0 &\to \gamma_H (Q^2,\mu)+ \gamma_S(\mu,\nu)+ 2\gamma_f(Q,\nu) =0\,,
\\ 
\label{eq:34eq4.3}
\frac{\df \sigma (\mu,\nu)}{\df \ln\nu} = 0 &\to  \gamma_{R\, S}(b,\mu)+ 2\gamma_{R\, f} (b,\mu)=0\,,
\,,\end{align}
where the anomalous dimensions are defined as
\begin{align} \label{eq:34eq4.4}
\gamma_F \equiv \frac{\df}{\df\mu} \ln F(\mu,\nu)\,,\\
\label{eq:34eq4.5}
 \gamma_{R\,F} \equiv \frac{\df}{\df\nu} \ln F(\mu,\nu)\,,
\end{align}
with $F(\mu,\nu)$ standing for the functions {$H$, $\tS$, $\tft$} for $\mu$-anomalous dimension  $\gamma_F$ and {$\tS$, $\tft$} for $\nu$-anomalous dimension $\gamma_{R\,F}$.
 
The RGE and anomalous dimension for the hard function, $\gamma_{H}(Q^2,\mu)$, are obtained from the square of hard Wilson coefficient for the two-quark operator (\cite{Manohar:2003vb, bauer2004enhanced})
\begin{align} \label{eq:34eq4.11}
\mu \frac{\df}{\df\mu} H(Q^2, \mu) &= \gamma_H(Q^2, \mu)\, H(Q^2, \mu)
\,,\\
\label{eq:34eq4.12}
\gamma_H(Q^2, \mu) &= \Gamma_H(\alpha_s) \ln\frac{Q^2}{\mu^2} + \gamma_H(\alpha_s)\,,
\end{align}
where $\Gamma_H=2\Gamma_\cusp$, and $\Gamma_\cusp$ and $\gamma_H$ are respectively the cusp and non-cusp anomalous dimensions, known up to 3 loops. 

The consistency in \eq{34eq4.2} implies that both $\gamma_{S},\gamma_f$ can be written in similar fashion to $\gamma_{H}$ and the cusp and non-cusp terms separately cancel when summed,
\begin{align} 
\label{eq:34eq4.6}
 \Gamma_H (\as) \ln\frac{Q^2}{\mu^2} + \Gamma_S(\as) \ln \frac{\nu}{\mu}+ 2\Gamma_f(\as) \ln \frac{\nu}{Q} =0
\\ 
\label{eq:34eq4.7}
 \gamma_H (\as)+ \gamma_S(\as)+ 2\gamma_f(\as) =0\,.
\end{align}
Cancellation of the cusp pieces requires that $ \Gamma_S  =-2\Gamma_f =-2 \Gamma_H \equiv -4\Gamma_\cusp$.

Applying the fact that the scales $\mu,\nu$ are independent of each other to the soft function and PDF, we obtain relations between $\gamma_S$ and $\gamma_{R\,S}$ and between $\gamma_f$ and $\gamma_{R\, f}$,
\begin{align} 
\label{eq:34eq4.8}
&\Bigg[ \frac{\df}{\df \ln\mu}\,,   \frac{\df }{\df \ln\nu} \Bigg] \tS=0 \to \frac{\df }{\df \ln\mu} \gamma_{R\,S}=\frac{\df }{\df \ln\nu} \gamma_{S} 
=\Gamma_{S}=-4\,\Gamma_\cusp\,,
\\ \label{eq:34eq4.9}
&\Bigg[ \frac{\df }{\df \ln\mu}\,,   \frac{\df }{\df \ln\nu} \Bigg] \tft=0 \to \frac{\df }{\df \ln\mu} \gamma_{R\,f}=\frac{\df }{\df \ln\nu} \gamma_{f}
=\Gamma_{f}=2\,\Gamma_\cusp\,.
\end{align}
The last equalities result from  \eq{34eq4.6}.
Integrating \eqs{34eq4.8}{34eq4.9} with respect to $\mu$, we obtain the $\mu$-dependence of $\gamma_{R\, S,f}$ to all order in $\as$
\begin{equation} 
\label{eq:34eq4.10}
\gamma_{R\, S}(\mu,b) =-2 \gamma_{R\,f}(\mu,b)  =-4 \int^\mu_{\mu_b} \df (\ln \mu') \Gamma_\cusp\big[\as(\mu') \big] +\gamma_{R\,S} \big[\as(\mu_b)\big]\,.
\end{equation}

The solution of the RGE for the hard function, \eq{34eq4.11}, evolved from $\mu_0$ to $\mu$ is
\begin{align} \label{eq:34eq4.13}
H(Q^2, \mu) &= H(Q^2, \mu_0)\, U_H(Q^2, \mu_0, \mu)
\,, \\
\label{eq:34eq4.14}
\ln U_H(Q^2, \mu_0, \mu)
&= {-4K_{\Gamma}(\mu_0,\mu) + K_{\gamma_H}(\mu_0,\mu)}+ {4\eta_{\Gamma}(\mu_0, \mu)} \ln \Bigl(\frac{Q }{\mu_0}\Bigr)
\,,\end{align}
where the functions $K_{\Gamma}(\mu_0, \mu)$, $\eta_{\Gamma}(\mu_0, \mu)$ and $K_\gamma$ are given below in the appendix, \eqs{34eqA.1}{34eqA.3}


The quark PDF in $b$ space was calculated in \sec{I.5.2.6.1}. 
The $\mu$- and $\nu$-RG equations and their anomalous dimensions are written as
\begin{align} \label{eq:34eq4.15}
\mu \frac{\df}{\df\mu} \tft_q(\mu,\nu) &= \gamma_f(Q,\nu)\, \tft_q (\mu,\nu)
\,,\\ \label{eq:34eq4.16}
\nu \frac{\df}{\df\nu} \tft_q(\mu,\nu) &= \gamma_{R\, f}(b,\mu)\, \tft_q(\mu,\nu)
\,,\\ \label{eq:34eq4.17}
\gamma_f(Q,\nu)&=
\Gamma_f(\alpha_s) \ln\frac{\nu}{Q} + \gamma_f(\alpha_s)\,
\,,\\  \label{eq:34eq4.18}
\gamma_{R\, f}(b,\mu)&=
2\,\eta_\Gamma (\mu_b,\mu)+ \gamma_{R\, f}(\alpha_s(\mu_b))\,,
\end{align}
where $\Gamma_f=2\Gamma_\cusp$, $\gamma_{R\, f}=-1/2 \, \gamma_{R\, S}$ expected because of the consistency condition derived from the factorized cross section. 
We express the $\mu$- and $\nu$-dependence in the PDF while we suppress its variables $z$ and $p_\perp$ for simplicity.

The solutions of \eq{34eq4.15} and \eq{34eq4.16} evolved from $(\mu_0,\nu_0)$ to $(\mu,\nu)$ are 
\begin{align} 
\label{eq:34eq4.19}
\tft_q(\mu,\nu) &= U_f(\mu_0, \mu; \nu)\,  V_f(\nu_0,\nu;\mu_0)\,\tft_q(\mu_0,\nu_0) 
\,,\\ 
\label{eq:34eq4.20}
\ln U_f(\mu_0,\mu;\nu)&=
 2\eta_\Gamma(\mu_0,\mu) \ln\frac{\nu}{Q} + K_{\gamma_f}(\mu_0,\mu) 
\,,\\ 
\label{eq:34eq4.21}
\ln V_f(\nu_0,\nu;\mu_0)&=\gamma_{R\, f}\big[\alpha_s(\mu_0) \big]=
\Big[ 2\,\eta_\Gamma (\mu_b,\mu) + \gamma_{R\, f}\big[\alpha_s(\mu_b) \big] \Big] \ln\frac{\nu}{\nu_0}
\,,
\end{align}
where we choose to evolve first along the $\nu$ direction then the $\mu$ direction. 


The gluon soft function at $\cO(\as)$ has been calculated in \cite{Chiu:2012ir} and by replacing the color factor $C_A$ by $C_F$ we obtain soft function for the quark Wilson line at $\cO(\as)$ given in \sec{I.5.2.6.2}. 
Its RGE and anomalous dimension are given by
\begin{align} \label{eq:34eq4.22}
\mu \frac{\df}{\df\mu} \tS (\mu,\nu) &= \gamma_S(\mu,\nu)\, \tS  (\mu,\nu)
\,,\\ \label{eq:34eq4.23}
\nu \frac{\df}{\df\nu} \tS (\mu,\nu) &= \gamma_{R\, S}(b,\mu)\, \tS (\mu,\nu)
\,,\\ \label{eq:34eq4.24}
\gamma_S(\mu,\nu)&=
\Gamma_s(\as)\ln\frac{\nu}{\mu} + \gamma_S (\alpha_s)\,
\,,\\  \label{eq:34eq4.25}
\gamma_{R\, S}(b,\mu)&=
-4\, \eta_\Gamma (\mu_b,\mu) + \gamma_{R\, S}\big[\as(\mu_b )\big]\,,
\end{align}
where $\Gamma_{S}=-4\Gamma_\cusp$ and $\mu_b=2/(b e^{\gamma_E})$.
The solutions of \eqs{34eq4.22}{34eq4.23} that evolve the soft function from $(\mu_0,\nu_0)$ to $(\mu,\nu)$ are given by
\begin{align} 
\label{eq:34eq4.26}
\tS(\mu,\nu) &= U_S(\mu_0, \mu; \nu)\,  V_S(\nu_0,\nu;\mu_0)\,\tS(\mu_0,\nu_0) 
\,,\\ 
\label{eq:34eq4.27}
\ln U_S(\mu_0,\mu;\nu)&=
 4 K_\Gamma(\mu_0,\mu)  - 4\eta_\Gamma(\mu_0,\mu) \ln\frac{\nu}{\mu_0} + K_{\gamma_S}(\mu_0,\mu) 
\,,\\ 
\label{eq:34eq4.28}
\ln V_S(\nu_0,\nu;\mu_0)&=
\Big[ -4\, \eta_\Gamma (\mu_b,\mu_0)+ \gamma_{R\, S}\big[\alpha_s(\mu_b) \big] \Big] \ln\frac{\nu}{\nu_0}
\,.
\end{align}

\paragraph{Resummed cross section}

The reduced cross section \eq{34eq4.1} in $b$ space is a product of hard, soft, and 2 PDFs in \eqss{34eq4.13}{34eq4.19}{34eq4.26} with pieces evolved from their natural scales $\mu_H$, $(\mu_S,\nu_S)$ and $(\mu_f,\nu_f)$ at which no large logarithms exist, to the common scales $(\mu,\nu)$ at which the cross section is evaluated. The natural scales for  $\mu_S$ and $\mu_f$ are $p_\perp\sim 1/b$ and we set $\mu_f=\mu_S$.  On the other hand, the natural sizes of the rapidity scales $\nu_{S,f}$ differ $\nu_S\sim 1/b$ and $\nu_f \sim Q$, producing the large rapidity logarithm.  The resummed cross section is now
\begin{align} 
\label{eq:34eq4.29}
\sigma(b,z_1,z_2;\mu_i,\nu_i;\mu,\nu) &=U_{\text{tot}}(\mu_i,\nu_i;\mu,\nu) \,  H(Q^2,\mu_H)\tS(b;\mu_f,\nu_S)
\nn\\&\times
\tft(b,z_1;\mu_f,\nu_f) \tft(b,z_2;\mu_f,\nu_f)
\,,\\ \nn
\label{eq:34eq4.30}
\ln U_{\text{tot}}(\mu_i,\nu_i,\mu,\nu) &\equiv \ln \Big [U_H(Q^2, \mu_H, \mu) U_S(\mu_f, \mu; \nu)V_S(\nu_S,\nu;\mu_f)\nn\\
&\times U_f^2(\mu_f, \mu; \nu) V_f^2(\nu_f,\nu;\mu_f)\Big] \\ 
&= -4 K_\Gamma(\mu_H,\mu) +4 K_\Gamma(\mu_f, \mu) \nn \\
&+4 \eta_\Gamma(\mu_H,\mu)\ln\frac{Q}{\mu_H} -2 \eta_\Gamma(\mu_f,\mu) \ln\frac{Q}{\mu_f}
\\ &=
-4 K_\Gamma(\mu_f,\mu_H) -4 \eta_\Gamma(\mu_f,\mu_H) \ln\frac{Q}{\mu_f} -K_{\gamma_H}(\mu_f,\mu_H)\nn\\
&+\Big[ -4\, \eta_\Gamma (\mu_b,\mu_f)+ \gamma_{R\, S}\big[\alpha_s(\mu_b) \big] \Big] \ln \frac{\nu_f}{\nu_S}
\,,
\end{align}
where $\mu_b=2/(b e^{\gamma_E})$.  The rapidity renormalization group has resummed the large rapidity logarithm $\ln(\nu_f/\nu_S)\sim\ln(Qb)$ arising from the large rapidity gap between the soft radiation from which the final state recoils and the large collinear momenta in the hard collision.

This completes our simple example of summing rapidity logarithms, in a context similar to that studied by \cite{Chiu:2012ir}.  I will next discuss our new application of rapidity resummation to the elastic limit of hadron scattering.

\subsection{Semi-inclusive Deep Inelastic Scattering}
\label{sec:I.5.3.1}
Scatterings near the elastic limit contain large rapidity logarithms.  In the DIS process, a high energy electron with momentum $k$ strikes a proton with momentum $p$ going to a final state $X(p_X)$, and the elastic limit corresponds to the struck parton carrying momentum fraction $z\to 1$.  In this limit, the total momentum(-squared) of the soft radiation approaches the invariant mass of the initial proton, i.e. $p^2\sim\LQCD^2$.  Consequently, rapidity logarithms arise in the ratio $Q/\LQCD$, since the initial state and final state radiation lie on the same invariant mass hyperbola.

In this section, I apply the prescriptions introduced in the previous section to regulate rapidity divergences associated with the rapidity logarithms and propose a new rapidity scale-free definition of the quark PDF in the endpoint region by properly separating the soft nonperturbative effects from the hard-collinear jet final state.  This work has been published in \cite{Fleming:2012kb,Fleming:2016nhs}.

To begin we denote the square of the momentum transfer is $q^2=(k-k')^2$ where $k'$ is the final state electron momentum. We define $Q^2\equiv -q^2$, and $x=\frac{Q^2}{2p\cdot q}$. With this notation we follow \cite{Manohar:2003vb} and write the differential cross section as
\begin{equation}\label{eq:35eq1}
d\sigma=\frac{d^3\vec k}{2|\vec k'|(2\pi)^3}\frac{\pi e^4}{SQ^4}L_{\mu\nu}(k,k')W^{\mu\nu}(p,q)\,,
\end{equation}
where $s=(p+k)^2$ is the squared invariant mass in the collision, and the lepton tensor is:
\begin{equation}\label{eq:35eq2}
L_{\mu\nu}=2(k_\mu k'_\nu+k_\nu k'_\mu-k\cdot k'g_{\mu\nu})\,.
\end{equation}
$W_{\mu\nu}$ is the DIS hadronic tensor, which at large $x$ will be subject of our analysis.

\subsubsection{Factorization theorem}
\label{sec:I.5.3.1.1}

In this part we first determine the kinematics and power-counting specific to the endpoint. Then we match QCD to $\SCETa$. Next, at an intermediate scale of order the invariant mass of the final state, we match $\SCETa$ onto $\SCETb$. Using rapidity regulator introduced in \cite{Chiu:2011qc,Chiu:2012ir}, we explicitly calculate both the collinear and the soft functions to one-loop in the $\SCETb$. Recombining these factors, we show that non-perturbative infrared effects appearing in the collinear function (as dependence on nonzero gluon mass-squared) are absorbed by the soft function.

\nt{Kinematics.} There are a number of approaches (\cite{Manohar:2003vb,Chay:2004zn,Chay:2013zya}) to separating and scaling momentum components in the $x\sim 1$ regime. The light-cone unit vectors $n^\mu=(1,0,0,1)$ and $\bn^\mu=(1,0,0,-1)$ allow decomposition of the proton momenta $p^\mu=\frac{n^\mu}{2}\bn\cdot p+\frac{\bn^\mu}{2}n\cdot p+p_\perp^\mu$, where $p^+=n\cdot p, p^-=\bn\cdot p$. In the target rest frame, $p=(p^+,p^-,p_\perp)=(M_p,M_p,0)$, and $Q^2=-q^2=-q^+q^-$. The direction of the incoming electron fixes the $z$-axis, and in the target rest frame, $q^-\gg q^+$. In this limit, Bjorken $x$ simplifies,
\begin{equation}\label{eq:35eq3}
x=\frac{Q^2}{2p\cdot q}=-\frac{q^+q^-}{p^+q^-+p^-q^+}\simeq -\frac{q^+}{p^+}\,.
\end{equation}
We can express all momenta in terms of $x$, $M_p$ and $Q$ in the target rest frame, and then boost them along $z$-axis into the Breit frame:
\begin{eqnarray}
q&=&\left(-x M_p,\frac{Q^2}{xM_p}, 0 \right) \xrightarrow{\text{boost}} \left(-Q,Q,0\right)\nonumber \\
p&=&\left(M_p, M_p, 0\right) \xrightarrow{\text{boost}} \left(\frac{Q}{x}, \frac{xM_p^2}{Q}, 0\right)\nonumber \\
p_X&=&p+q=\left(M_p(1-x), q^-, 0\right)\nonumber
\xrightarrow{\text{boost}} \left(\frac{Q(1-x)}{x}, Q,  0\right)\,.
\end{eqnarray}
where $p_X$ is the (total) final state momentum.
In the large-$x$ limit, the large component of the incoming proton is $p^+=\frac{Q}{x}=Q+l^+$, in which $l^+=Q\frac{1-x}{x}$ is a rapidity scale lying between the collinear momentum scale $Q$ and soft momentum scale $\Lambda_{\text{QCD}}$. Correspondingly, we have naturally separated momenta:
\begin{itemize}
\item hard modes with $q\sim \left(-Q, Q, 0 \right)$ and invariant mass $q^2\sim Q^2$ at the hard collision scale;
\item $n$-collinear modes with $p\sim \left(Q, \frac{\Lambda_{QCD}^2}{Q}, \LQCD\right)$ and invariant mass $M_p^2\sim \Lambda_{\text{QCD}}^2$ at the soft scale;
\item final-jet hard collinear modes with $p_X\sim \left(Q\paren{\frac{1-x}{x}}, Q, \Lambda_{\text{QCD}}\right)\sim \left(l^+, Q, \Lambda_{\text{QCD}}\right)$ and invariant mass $p_X^2\sim Ql^+ \gg \Lambda_{\text{QCD}}^2$ at the hard-collinear scale.
\end{itemize}
We first integrate out the hard collision degrees of freedom in QCD at the scale $Q^2$ by matching onto $\text{SCET}_{\text{I}}$ with offshellness $Ql^+$. We then integrate out hard-collinear degrees of freedom at $Ql^+$ by matching onto  $\text{SCET}_{\text{II}}$ with offshellness $\LQCD^2$.  In this case, the final state momentum $p_X^+$ is of order $Q\paren{\frac{1-x}{x}}\sim l^+\ll Q$, so we must resum the large rapidity logarithms from $Q$ to $Q\paren{\frac{1-x}{x}}$.  In this way the matching procedure incorporates the semi-inclusive character of end-point processes.  If on the other hand we had $l^+\sim \frac{\Lambda_{\text{QCD}}^2}{Q}$, then the collision would be exclusive, and with a fixed and distinct final state momentum $p_X^2\sim Q\Lambda_{\text{QCD}}\gg \LQCD^2$, we would be unable to carry the DIS factorization smoothly from moderate $x$ to large $x$.

\nt{Factorization.} In Eq.\eqref{eq:35eq1}, the DIS hadronic tensor is the matrix element of the time-ordered product of  two QCD currents $J^\mu(x)=\bar \psi(x)\gamma^\mu \psi(x)$ between external in- and out- proton states,
\begin{eqnarray}\label{eq:35eq4}
W^{\mu\nu}(p,q)&=&\frac12 \sum_\sigma \int d^4x e^{iq\cdot x} 
\langle h(p,\sigma) \vert J^\mu(x) J^\nu(0) \vert h(p,\sigma) \rangle,
\end{eqnarray}
where $\sigma$ is the spin of the proton.  Matching QCD onto SCET is carried out at the scale $\mu_q\sim Q$, so the SCET current is
\begin{eqnarray}\label{35eq5}
J^\mu(x)\to\sum_{w_1,w_2}C(w_1,w_2;\mu,\mu_q)
\left(e^{-\frac i2 w_1n\cdot x}e^{\frac i2 w_2\bar n\cdot x}\bar \chi_{\bar n,w_2}\gamma_\perp^\mu \chi_{n,w_1}+\text{h.c.}\right)\,,~
\end{eqnarray}
where $\bar\chi_{\bar n,w_2},\chi_{n,w_1}$ are SCET fields.  Correspondingly, the hadronic tensor in $\text{SCET}_{\text{I}}$ is
\begin{eqnarray}
W_{\text{eff}}^{\mu\nu}&=&\sum_{w_1,w_2,w_1',w_2'}C^*(w_1,w_2;\mu_q,\mu)C(w_1',w_2';\mu_q,\mu) \int\frac{d^4x}{4\pi}e^{-\frac i2(Q-w_1)n\cdot x}e^{\frac i2 (Q-w_2)\bar n\cdot x}\nonumber \\
&&\times \oneov{2} \sum_\sigma \langle h_n(p,\sigma)\vert \bar T[\bar \chi_{n,w_1}\gamma_\perp^\mu \chi_{\bar n,w_2}(x)] T[\bar \chi_{\bar n,w_2'}\gamma_\perp^\nu \chi_{n,w_1'}(0)]\vert h_n(p,\sigma)\rangle \nn \\
&=& \frac{- g_\perp^{\mu\nu}}{2} N_c 
 \sum_{\omega'_1, \omega'_2} C^*(Q ,Q;\mu_q, \mu ) C(\omega'_1,\omega'_2;\mu_q, \mu)
 \int \frac{d^4 x}{4 \pi}\frac{1}{2}\nn\\
&&\times \sum_\sigma  \langle{h_{n}(p,\sigma)}|  \bar{\chi}_{n, Q}(x)\frac{\bnslash}{2} \chi_{n, \omega'_1}(0)|{h_{n}(p,\sigma)}\rangle \langle{0} |\frac{\nslash}{2}\chi_{\bn, Q}(x) \bar{\chi}_{\bn, \omega'_2}(0) |0\rangle \nn\\
&&\times   \frac{1}{N_c} \langle{0} |\text{Tr}\bigg(
\bar{T}\bigg[ Y^\dagger_{n}(x) \tilde{Y}_{\bn}(x)\bigg]
T\bigg[ \tilde{Y}^\dagger_{\bn}(0) Y_{n}(0) \bigg]
\bigg)|0\rangle \,. \label{eq:35eq6}
\end{eqnarray}
where $T$ and $\bar T$ denote time ordering and anti-time ordering operations of the soft gluon fields ${Y}_{\bn}$ and $Y_{n}$ respectively. The two collinear sectors and one usoft sector are decoupled by the BPS phase redefinition in \cite{Bauer:2001yt}.

In order to match Eq.\,\eqref{eq:35eq6} onto $\text{SCET}_{\text{II}}$, it is convenient to introduce a jet function as in \cite{Fleming:2007xt},
\begin{eqnarray}
&&\langle{0}|\frac{\bnslash}{2}\chi_{\bn, \omega_2}(x) \bar{\chi}_{\bn, \omega'_2}(0)|0 \rangle
\equiv  Q \delta(\bn\cdot x) \delta^{(2)}(x_{\perp}) \int dr \, e^{-\frac{i}{2}r n \cdot x} J_{\bn}(r;\mu) \,,\label{eq:35eq7}
\end{eqnarray}
which characterizes the final state with $p_X^2\sim Ql^+$. The final state is integrated out at the scale $\mu_c\sim \sqrt{Ql^+}$ and $J_{\bar n}(r;\mu)$ becomes a matching coefficient in $\text{SCET}_{\text{II}}$. 

We define a soft function in $\text{SCET}_{\text{I}}$ as in \cite{bauer2004enhanced}
\begin{align}
\frac{1}{N_c} \langle{0} |\text{Tr}&\bigg(
\bar{T}\bigg[ Y^\dagger_{n}(n\cdot x) \tilde{Y}_{\bn}(n\cdot x) \bigg]
\text{T}\bigg[\tilde{Y}_{\bn}^\dg(0) Y_{n}(0) \bigg]
\bigg)|0\rangle
\equiv  \int d\ell \, e^{-\frac{i}{2}\ell n\cdot x} S^{(DIS)}(\ell;\mu) \,,\label{eq:35eq8}
\end{align}
which describes usoft gluon emission throughout the interaction, from initial to final state.  The Wilson lines are defined as
\begin{align}
Y_n(x)&=P\exp\paren{ig\int_{-\infty}^xds\,\,\, n\cdot A_s(s_n)}\nn\\
\tilde Y_{\bar n}^\dg (x)&=P\exp\paren{ig\int_x^\infty ds\,\,\, \bar n\cdot A_s(s_{\bar n})}. \label{eq:35eq9}
\end{align}
The usoft gluons in $\text{SCET}_{\text{I}}$ with offshellness $p_{us}^2\sim \Lambda_{\text{QCD}}^2$ become soft gluons of $\text{SCET}_{\text{II}}$, so  Eq.\eqref{eq:35eq8} retains its form in matching $\text{SCET}_{\text{I}}$ to $\text{SCET}_{\text{II}}$.

Using label momentum conservation, which is just momentum conservation at fixed (large) $Q$, we simplify the collinear matrix element in the $n$-collinear direction:
\begin{align}
 \langle{h_{n}(p,\sigma)} &|\bar{\chi}_{n,  Q}(x)\frac{\bnslash}{2} \chi_{n, \omega'_1}(0)|{h_{n}(p,\sigma)}\rangle 
=\delta_{ Q,\omega'_1} \,\langle{h_{n}(p,\sigma)}| \bar{\chi}_{n}(x)\frac{\bnslash}{2} \delta_{\bar{\cal P}, 2  Q}  \chi_{n}(0) |{h_{n}(p,\sigma)}\rangle \,.\label{eq:35eq10}
\end{align}
We then define an $n$-direction collinear sector as the n-collinear function and match it onto $\text{SCET}_{\text{II}}$.  We insert an explicit Kronecker delta to ensure the large momentum of the proton $\tilde p\cdot\bar n$ is $Q$ at large $x$,
\begin{align}
{\cal C}_{n}( Q-k;\mu)  
&= \int \frac{d\, n\mcdot x }{4 \pi}\, e^{\frac{i}{2}kn\cdot x} \frac{1}{2} \sum_\sigma  
\delta_{\bn\cdot\tilde p,  Q}\,
 \langle{h_{n}(p,\sigma)}| \bar{\chi}_{n}(n\mcdot x)\frac{\bnslash}{2} \delta_{{\cal \bar P}, 2  Q} \chi_{n}(0) |{h_{n}(p,\sigma)}\rangle \nn \\
&=\frac{1}{2} \sum_\sigma  
\delta_{\bn\cdot\tilde p,  Q}\, \langle{h_{n}(p,\sigma)}| \bar{\chi}_{n}(0)\frac{\bnslash}{2} \delta_{\bar{\cal P}, 2  Q} 
\delta(i\bn\cdot\partial -k) \chi_{n}(0) |{h_{n}(p,\sigma)}\rangle\,.\label{eq:35eq11}
\end{align}
where ${\cal \bar P}= {\bar n} \cdot(\cal P+\cal P^+)$ and $k\sim \Lqcd$ is the residual momentum lying in the $\text{SCET}_{\text{II}}$ soft region.  Label momentum conservation then forces $w_1'=Q$, meaning that the large momenta of the incoming and outgoing protons are both equal to $Q$. 

In $\text{SCET}_{\text{II}}$, soft and collinear fields have the same off-shellness $p^2\sim \Lambda_{\text{QCD}}^2$. Arbitrary separation between these soft and collinear modes leads to rapidity divergences \cite{Chiu:2011qc,Chiu:2012ir}, which we regulate using the rapidity regulator. Since the matching procedure shows that the final state jet function is decoupled from the initial state $n$-collinear function, we can express the $n$-collinear function as $C_n(Q-k;\mu)\to C_n(Q-k;\mu,\nu)$
and the soft function as $S(l,\mu)\to S(l;\mu,\nu)$.  Combining Eq.\eqref{eq:35eq6}, Eq.\eqref{eq:35eq7}, Eq.\eqref{eq:35eq10} and Eq.\eqref{eq:35eq11}, we arrive at the $\text{SCET}_{\text{II}}$ factorized DIS hadronic tensor,
\begin{eqnarray}
W^{\mu \nu}_{\rm eff} = -g_\perp^{\mu\nu}H( Q ;\mu_q, \mu_c)   
 \int d \ell \, J_{\bn}(\ell;\mu_c,\mu) \phi^{ns}_q(Q\paren{\frac{1-x}{x}}+\ell;\mu) \label{eq:35eq12}
\end{eqnarray}
with
\begin{equation}\label{eq:35eq13}
\phi_q^{ns}\paren{\ell;\mu}=\delta_{\tilde n\cdot\tilde p,Q}\mathcal{Z}_n(\mu,\nu)S^{(DIS)}(\ell;\mu,\nu)
\end{equation}
and
\begin{equation}\label{eq:35eq14}
\mathcal{Z}_n(\mu,\nu)=C_n(Q-k;\mu,\nu)\delta(k)\delta_{\bar n\cdot\tilde p,Q}\,.
\end{equation}

\subsubsection{Collinear Function to \texorpdfstring{$\cO(\alpha_s)$}{O(alpha)} for DIS with \texorpdfstring{$\eta$}{eta}-scheme}
\label{sec:I.5.3.1.2a}

\begin{figure}
\centering
\includegraphics[width=.25\textwidth]{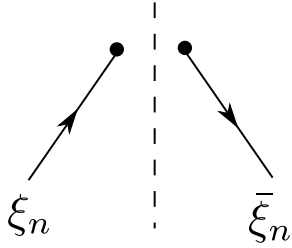}
\caption{$\mathcal{O}(\alpha_s^0)$ Feynman diagram for the $n$-collinear function.}
\label{fig:35fig1}
\end{figure}

The $n$-collinear function in \req{eq:35eq11} has the tree-level Feynman diagram shown in Fig.\,\ref{fig:35fig1}
We consider the explicit calculation of this amplitude using external parton states, and find the $\mathcal{O}(\alpha_s^0)$ result is
\begin{equation}\label{eq:35eq16}
C_n^{(0)}(Q-k)=\delta_{\bar n\cdot \tilde p,Q}\delta(\tilde n\cdot p_r-k)m_0\,,
\end{equation}
where $\bar n\cdot \tilde p$ is the $\mathcal{O}(1)$ quark label momentum at the hard scale $Q$, $p_r$ is the quark residual momentum at the soft scale $\Lambda_{\text{QCD}}$ and
\begin{equation}\label{eq:35eq17}
m_0=\frac12\sum_\sigma\bar\xi_n^\sigma\frac{\bar n \!\!\!\!\!\not\,\,\,\,}{2}\xi_n^\sigma\,,
\end{equation}
where $\xi_n^\sigma$ is the SCET quark spinor in the $n$-direction with spin $\sigma$.

\begin{figure}
\centering
\includegraphics[width=.8\textwidth]{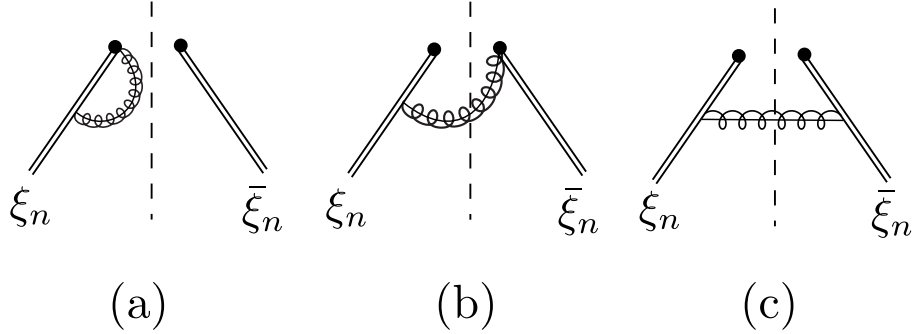}
\caption{$\mathcal{O}(\alpha_s)$ Feynman diagram for the $n$ collinear function (a) is the virtual contribution; (b) and (c) are the real contribution.}
\label{fig:35fig2}
\end{figure}

The $\mathcal{O}(\alpha_s)$ level $n$-collinear function Feynman diagrams are shown in Fig.\,\ref{fig:35fig2}.
Fig.\,\ref{fig:35fig2}(a) shows the virtual contribution, while Figs. \ref{fig:35fig2}(b) and (c) show the real contribution.  We omit the mirror images of the Fig.\,\ref{fig:35fig2}(a) and (b).
With the rapidity regulated collinear Wilson lines, we obtain the naive amplitude corresponding to Fig.~\ref{fig:35fig2}(a),
\begin{align}
i {\tilde m_a^n}&=(im_0)(2g^2C_F)\delta_{\bar n\cdot p,Q}\delta(l^-)\mu^{2\epsilon}\nu^\eta
\int\frac{d^Dq}{(2\pi)^D}\frac{|\bar n\cdot q|^{-\eta}}{\bar n\cdot q}\frac{\bar n\cdot(p-q)}{(p-q)^2+i\epsilon}\oneov{q^2-m_g^2+i\epsilon}   \,,\label{eq:35eq18}
\end{align}
in $D=4-2\epsilon$ dimensions. The Kronecker delta sets the large component of the external quark momentum to $Q$.  The integral in \req{eq:35eq18} overlaps with a region of soft momenta that must be subtracted to avoid double counting, the so-called zero-bin \cite{Manohar:2006nz,Idilbi:2007yi}.  Taking the limit $\bar n\cdot q\ll \bar n\cdot p$ in the collinear gluon loop gives the overlap region, and the zero-bin subtraction for this amplitude is
\begin{align}
im_a^{n\phi}&=im_0(2g^2C_F)\delta_{\bar n\cdot p,Q}\delta(l^-)\mu^{2\epsilon}\nu^\eta
\int\!\frac{d^Dq}{(2\pi)^D}\frac{|\bar n\!\cdot\! q|^{-\eta}}{\bar n\!\cdot\! q}\frac{\bar n\cdot p}{(\bar n\!\cdot\! p)(n\!\cdot\! q)+i\epsilon}\oneov{q^2-m_g^2+i\epsilon}    \,.\label{eq:35eq19}
\end{align}
Eq.~(\ref{eq:35eq19}) is scaleless and thus vanishes. The naive amplitudes of Fig.~\ref{fig:35fig2}(b) and (c) are
\begin{align}
i{\tilde m_b^n}&=(-im_0)(2g^2C_F)\delta_{\bar n\cdot p+\bar n\cdot \tilde q,Q}\delta_{\bar n\cdot p,Q}\mu^{2\epsilon}\nu^\eta\nn\\
&\times\int\!\frac{d^Dq}{(2\pi)^D}(-2\pi i)\delta(q^2)\frac{|\bar n\cdot q|^{-\eta}}{\bar n\cdot q}\frac{\bar n\cdot(p-q)}{(p\!-\!q)^2+i\epsilon}\delta(\bar n\cdot q_r-l^-) \label{eq:35eq20}\\
i{\tilde m_c^n}&=(im_0)(2g^2C_F)\delta_{\bar n\cdot p+\bar n\cdot \tilde q,Q}\delta_{\bar n\cdot p,Q}\mu^{2\epsilon}(D-2)\nn\\
&\times\int\frac{d^Dq}{(2\pi)^D}(-2\pi i)\delta(q^2)\frac{(\bar n\cdot q)(n\cdot q)}{((p-q)^2+i\epsilon)^2}\delta(\bar n\cdot q_r-l^-) \,,\label{eq:35eq21}
\end{align}
where $\tilde q$ is the large component of the collinear gluon momentum which obeys label momentum conservation, and $q=\tilde q+q_r$ with $q_r$ the soft residual momentum. In the $n$-collinear function, the $n$-collinear quarks only couple with $n$-collinear gluons, which means $n\cdot \tilde q=0$ and $n\cdot q=n\cdot q_r$.  The two Kronecker deltas in front of the integrals in both Eq.~(\ref{eq:35eq20}) and Eq.~(\ref{eq:35eq21}) force $\bar n\cdot\tilde q=0$, which implies that gluons emitted from initial to final state only have soft momentum. As a result, Eq.\,(\ref{eq:35eq20}) and Eq.\,(\ref{eq:35eq21}) can be reduced to
\begin{align}
i{\tilde m_b^n}&=(-im_0)(2g^2C_F)\delta_{\bar n\cdot \tilde q,0}\int\frac{d^Dq_r}{(2\pi)^D}(-2\pi i)\delta(q_r^2)
\frac{|\bar n\cdot q_r|^{-\eta}}{\bar n\cdot q_r}\frac{\bar n\cdot(p-q_r)}{(p-q_r)^2+i\epsilon}\delta(\bar n\cdot q_r-l^-)\label{eq:35eq22}\\
i{\tilde m_c^n}&=(im_0)(2g^2C_F)\delta_{\bar n\cdot\tilde q,0}\int\frac{d^Dq_r}{(2\pi)^D}(-2\pi i)
\delta(q_r^2)\frac{(\bar n\cdot q_r)(n\cdot q_r)}{((p-q)^2+i\epsilon)^2}\delta(\bar n\cdot q_r-l^-)\,,\label{eq:35eq23}
\end{align}
which is equal to the zero-bin subtraction. Therefore, after subtracting Eq.~(\ref{eq:35eq22}) and Eq.~(\ref{eq:35eq23}) from Eq.~(\ref{eq:35eq20}) and Eq.~(\ref{eq:35eq21}) respectively, the results vanish.

After computing the virtual collinear function in Eq.~(\ref{eq:35eq18}) and Eq.~(\ref{eq:35eq19}) and adding their mirror amplitudes, we have 
to $\mathcal{O}(\alpha_s)$ the collinear function
\begin{align}
{C}_{n}( Q-k)^{(1)}=\:&\:
{C}_n^{(0)}(Q-k) \, \frac{\alpha_s C_F}{\pi}w^2 
\bigg\{ \frac{e^{\epsilon \gamma_E} \Gamma(\epsilon)}{\eta}\bigg(\frac{\mu^2}{m^2_g}\bigg)^\epsilon +\frac{1}{\epsilon}\bigg[ 1 + \ln\frac{\nu}{\bn\cdot p} \bigg]
\nn\\&
+\ln\frac{\mu^2}{m_g^2}\ln\frac{\nu}{\bar n\cdot p}+\ln\frac{\mu^2}{m_g^2}+1-\frac{\pi^2}{6}\bigg\}\,,\label{eq:35eq24}
\end{align}
which depends on the rapidity regulator. A natural choice of  $\nu\sim \bar n\cdot p=Q$ minimizes the rapidity logarithm. 
The collinear matrix element is obtained by multiplying the above result by the quark wave function renormalization
\begin{equation}
C_{\psi} = 1 - \frac{\alpha_{s}C_{F}}{4 \pi} \bigg( \frac{1}{\epsilon} + \ln\frac{\mu^{2}}{m_g^2}+1\bigg)\,,
\end{equation}
which gives
\begin{align}
{C}_n^{(1)}(Q-k)=\:&\:
{C}_n^{(0)}(Q-k) \, \frac{\alpha_s C_F}{\pi}w^2 
\bigg\{ \frac{e^{\epsilon \gamma_E} \Gamma(\epsilon)}{\eta}\bigg(\frac{\mu^2}{m^2_g}\bigg)^\epsilon +\frac{1}{\epsilon}\bigg[ \frac{3}{4} + \ln\frac{\nu}{\bn\cdot p} \bigg]
\nn\\&
+\ln\frac{\mu^2}{m_g^2}\ln\frac{\nu}{\bar n\cdot p}+\frac{3}{4}\ln\frac{\mu^2}{m_g^2}+\frac{3}{4}-\frac{\pi^2}{6}\bigg\}\,.\label{eq:24temp}
\end{align}

\subsubsection{DIS Collinear Function with \texorpdfstring{$\Delta$}{Delta}-scheme}
\label{sec:I.5.3.1.2b}
The form of substitution 
\begin{equation}
\oneov{(p_i+k)^2-m_i^2}\to\oneov{(p_i+k)^2-m_i^2-\Delta_i}
\end{equation}
makes the $\Delta$ regulator behave like a mass shift for the particle $i$. Correspondingly, the collinear Wilson lines are
\begin{align}
W_n&=\sum_{perm}\exp\left[-\frac{g}{\bar n\cdot \mathcal{P}-\delta_1}\bar n\cdot A_n\right]\nn \\
W_{\bar{n}}^\dg&=\sum_{perm}\exp\left[-\frac{g}{n\cdot\mathcal{P}-\delta_2}n\cdot A_{\bar n}\right]\,,\label{eq:35eq122}
\end{align}
while the soft Wilson lines for DIS are
\begin{align}
\tilde{Y}_{\bar{n}}^\dg&=\sum_{perm}\exp\left[-\frac{g}{n\cdot \mathcal{P}_s-\delta_2+i0}\bar{n}\cdot A_s\right]\nn\\
Y_n&=\sum_{perm}\exp\left[-\frac{g}{\bar n\cdot\mathcal{P}_s-\delta_1-i0}n\cdot A_s\right]\,,\label{eq:35eq123}
\end{align}
and for DY are
\begin{align}
\tilde{Y}_{\bar{n}}^\dg&=\sum_{perm}\exp\left[-\frac{g}{n\cdot \mathcal{P}_s-\delta_2-i0}\bar{n}\cdot A_s\right]\nn\\
Y_n&=\sum_{perm}\exp\left[-\frac{g}{\bar n\cdot\mathcal{P}_s-\delta_1-i0}n\cdot A_s\right]\,,\label{eq:35eq124}
\end{align}
where $\delta_1=\Delta_1/p^+$ and $\delta_2=\Delta_2/p^-$, with $p^+$ or $p^-$ the collinear momentum in the $n$ or $\bar n$ direction.
We have the DIS hadronic tensor in $\text{SCET}_{\text{II}}$ with the Delta regulator
\begin{align}
(W_{\mu\nu})_{\Delta-DIS}^{\rm eff}&=-g_\perp^{\mu\nu}H(Q,\mu)\int _\phi d\ell J_{\bar n}(r;\mu)S(\ell;\mu;\delta_2,m_g^2)C_n(Q-r-\ell;\mu;\delta_2,m_g^2)\,. \label{eq:35eq125}
\end{align}

For DIS, the naive virtual $n$-collinear function shown in Fig.\ref{fig:35fig2}(a) is
\begin{align}
\tilde{C}_n^v=&(2ig^2C_F)\delta(k^-)\mu^{2\epsilon}\int \frac{d^dq}{(2\pi)^d}\frac{1}{-q^-+\delta_1+i\epsilon}
\nn\\&\times
\frac{p^-+q^-}{(p^-+q^-)q^+-q_\perp^2-\Delta_2+i\epsilon}\frac{1}{q^-q^+-q_\perp^2-m_q^2+i\epsilon}\nn\\
=&\left(-\frac{\alpha_s C_F}{2\pi}\right)\delta(k^-)\Bigg(\frac{1}{\epsilon}\left(-\ln\frac{\delta_1}{p^-}-1\right)-\ln\frac{\mu^2}{m_g^2}\left(\ln\frac{\delta_1}{p^-}+1\right)\nn\\
&-\left[\ln\left(1-\frac{\Delta_2}{m_g^2}\right)\ln\frac{\Delta_2}{m_g^2}+1-\frac{\Delta_2/m_g^2}{\frac{\Delta_2}{m_g^2}-1}\ln\frac{\Delta_2}{m_g^2}+Li_2\left(\frac{\Delta_2}{m_g^2}\right)-\frac{\pi^2}{6}\right]\Bigg)\,.\label{eq:35eq127}
\end{align}
We see that $\Delta_2$ is the infrared regulator for the quark propagator, effectively a quark mass in the loop integral.  The zero-bin amplitude for this virtual function is 
\begin{align}
C_{n}^{v\phi}&=(-2ig^2C_F)\delta(k^-)\mu^{2\epsilon}\int \frac{d^dq}{(2\pi)^d}\frac{1}{q^--\delta_1+i\epsilon}\frac{1}{q^+-\delta_2+i\epsilon}\frac{1}{q^2-m_g^2+i\epsilon}\nn\\
&=\left(\frac{-\alpha_sC_F}{2\pi}\right)\delta(k^-)\Bigg(\frac{1}{\epsilon^2}+\frac{1}{\epsilon}\ln\frac{\mu^2}{\delta_1\delta_2}
\nn \\
&~~~+\ln\left(\frac{\mu^2}{m_g^2}\right)\ln\frac{\mu^2}{\delta_1\delta_2}-\frac{1}{2}\ln^2\frac{\mu^2}{m_g^2}
-Li_2\left(1-\frac{\delta_1\delta_2}{m_g^2}\right)+\frac{\pi^2}{12}\Bigg)\,.\label{eq:35eq128}
\end{align}
For the real collinear function, the naive real collinear amplitudes only get contributions from the soft momentum region, which are their exact zero-bin subtraction amplitudes. Thus, after the zero-bin subtractions, the real collinear function amplitudes shown in Fig.\,\ref{fig:35fig2}(b) and (c) vanish,
\begin{equation}\label{eq:35eq129}
\tilde C_n^r=C_n^{r\phi}\Rightarrow C_n^r=\tilde C_n^r-C_n^{r\phi}=0\,.
\end{equation}
After multiplying the calculated amplitudes in Eq.~\eqref{eq:35eq127} and Eq.~\eqref{eq:35eq128} by 2 for their mirror images, we have the collinear function with quark wavefunction renormalization in semi-inclusive DIS with the Delta regulator
\begin{align}
C_n^v=&2(\tilde C_n^v-\tilde C_n^{v\phi})\nn\\
=&2\left(\frac{-\alpha_s C_F}{2\pi}\right)\delta(k^-)
\Bigg(-\frac{1}{\epsilon^2}-\frac{1}{\epsilon}\left(\ln\frac{\mu^2}{\Delta_2}+\frac{3}{4}\right)+\frac{\pi^2}{12}-\frac{3}{4}+\frac12\ln^2\frac{\mu^2}{m_g^2}
\nn\\ &-\ln\frac{\mu^2}{m_g^2}\left(\ln\frac{\mu^2}{\Delta_2}+\frac{3}{4}\right)
+Li_2\left(1-\frac{\delta_1\delta_2}{m_g^2}\right)-Li_2\left(\frac{\Delta_2}{m_g^2}\right)
\nn\\ & +\ln\frac{\Delta_2}{m_g^2}\left(\frac{\Delta_2/m_g^2}{\frac{\Delta_2}{m_g^2}-1}-\ln\left(1-\frac{\Delta_2}{m_g^2}\right)\right)\Bigg)\,.
\label{eq:35eq130}
\end{align}
The infrared part of final result of the $n$-collinear function is independent of $\delta_1$, which is the infrared regulator of the $n$-direction Wilson Line.  In contrast, using the rapidity $\eta$-regulator exhibited rapidity divergences in the $n$-collinear function in Eq.~\eqref{eq:35eq24} brought in by the  $n$-direction Wilson line.


\subsubsection{Soft Function to \texorpdfstring{$\cO(\alpha_s)$}{O(alpha)} for DIS with \texorpdfstring{$\eta$}{eta}-scheme}
\label{sec:I.5.3.1.3a}

\begin{figure}
\centering
\includegraphics[width=.5\textwidth]{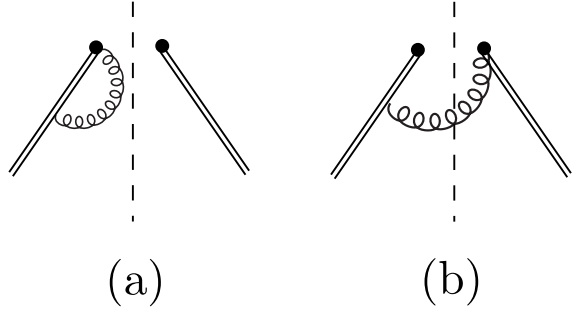}
\caption{$\mathcal{O}(\alpha_s)$ soft function Feynman diagrams: (a) is the virtual contribution; (b) is the real soft function Feynman diagrams: (a) .}
\label{fig:35fig3}
\end{figure}

The soft function, given in \req{eq:35eq8}, at tree level is
\begin{equation}\label{eq:35eq25}
S(l)^{(0)}=\delta(l).
\end{equation}
To $\mathcal{O}(\alpha_s)$, with the $\eta$-regulated soft Wilson line 
, we can explicitly isolate the rapidity poles of the soft function. The Feynman diagrams for the one-loop soft functions are shown in Fig.\,\ref{fig:35fig3}, where Fig.\,\ref{fig:35fig3}(a) is the virtual piece and Fig.\,\ref{fig:35fig3}(b) is the real piece.
The double lines represent the eikonal lines. Here we also omit the mirror images of Fig.\,\ref{fig:35fig3}(a) and (b).

The naive virtual soft function amplitude determined from Fig.\,\ref{fig:35fig3}(a) is 
\begin{align}
\tilde S_v&=(2ig^2C_F)\delta(l)\mu^{2\epsilon}\nu^\eta w^2 
\int d^dk\frac{|2k_3|^{-\eta}}{k^2-m_g^2+i\epsilon}\oneov{k^-+i\epsilon}\oneov{k^++i\epsilon}\nn \\
&=\delta(l)\frac{\alpha_sC_F}{\pi}w^2\bigl[-\frac{e^{\epsilon\gamma_E}\Gamma(\epsilon)}{\eta}\paren{\frac{\mu}{m_g}}^{\!2\epsilon}\!+\oneov{2\epsilon^2}
+\oneov{\epsilon}\ln\frac{\mu}{\nu}+\ln^2\frac{\mu}{m_g}-\ln\frac{\mu^2}{m_g^2}\ln\frac{\nu}{m_g}-\frac{\pi^2}{24}\bigl]. \label{eq:35eq26}
\end{align}
The zero bin subtraction for the naive virtual piece is the overlap with the $n$ and $\bn$ collinear directions:
\begin{align}
S_{v\phi}^{\bar n}&(k^-\gg k^+)=(2ig^2C_F)\delta(l)\mu^{2\epsilon}\nu^\eta\int\frac{d^Dk}{(2\pi)^D}
\frac{|k^-|^{-\eta}}{(k^++i\epsilon)(k^-+i\epsilon)(k^2-m_g^2+i\epsilon)},\label{eq:35eq27}\\
S_{v\phi}^{n}&(k^+\gg k^-)=(2ig^2C_F)\delta(l)\mu^{2\epsilon}\nu^\eta\int\frac{d^Dk}{(2\pi)^D}
\frac{|k^+|^{-\eta}}{(k^++i\epsilon)(k^-+i\epsilon)(k^2-m_g^2+i\epsilon)}.\label{eq:35eq28}
\end{align}
These integrals are scaleless in rapidity regularization and vanish. This must be the case because adding the rapidity regulator to the soft Wilson lines 
restricts the soft function integral to lie only in the soft momentum region.  In other words, in the virtual contributions, the rapidity regulator properly separates soft and collinear modes in $\text{SCET}_{\text{II}}$.   Thus the total virtual soft function is
\begin{equation}
S_v=2\tilde S_v\,.
\end{equation}

The naive real contribution from the diagram in Fig.\,\ref{fig:35fig3}(b) is
\begin{align}
\label{eq:35eq30}
\tilde S_\textrm{r} &= + 4\pi C_F g^2_s \mu^{2 \epsilon} w^2 \nu^\eta \int \frac{d^Dk}{(2\pi)^{D-1}}\delta(k^2-m^2_g) \delta(\ell-k^+)|2 k_3|^{-\eta} \frac{1}{k^+}\frac{1}{k^-}  
\\&= -\frac{\alpha_s C_F}{\pi}\bigg(e^{ \gamma_E} \frac{\mu^2}{m^2_g}\bigg)^\epsilon w^2 \nu^\eta \frac{\theta(\ell)}{\ell^{1+\eta}} \Gamma(\epsilon)\,.\nn
\end{align}
The zero bin subtraction for the real soft function is not zero at this order, because any overlap with the collinear regions in the soft function is not suppressed by the rapidity regulator in the initial state Wilson lines.  Mathematically, we see this by the presence of the scale brought into the integral by the measurement function.  The overlap of the integral in Eq.~(\ref{eq:35eq30}) with the $n$-collinear region is given by taking the limit $k^+\gg k^-$ with $k^+k^- \sim k^2_\perp$
\begin{align}
\label{eq:35eq31}
S_{n\phi} ^{r}&=  - 4\pi C_F g^2_s \mu^{2 \epsilon} w^2 \nu^\eta \int \frac{d^Dk}{(2\pi)^{D-1}}\delta(k^2-m^2_g) \delta(\ell-k^+)|k^+|^{-\eta} \frac{1}{k^+}\frac{1}{k^-}  \\
& = +\frac{\alpha_s C_F}{\pi}\bigg(e^{ \gamma_E} \frac{\mu^2}{m^2_g}\bigg)^\epsilon w^2 \nu^\eta \frac{\theta(\ell)}{\ell^{1+\eta}} \Gamma(\epsilon)\,,\nn
\end{align}
which is the same as the result in Eq.~(\ref{eq:35eq30}). The $\bar n$-collinear subtraction is given by taking the limit $k^-\gg k^+$ with $k^+k^- \sim k^2_\perp$ in the first line of Eq.~(\ref{eq:35eq30}):
\begin{align}
\label{eq:35eq32}
S_{\bar n\phi} ^{r}&=  - 4\pi C_F g^2_s \mu^{2 \epsilon} w^2 \nu^\eta \int \frac{d^Dk}{(2\pi)^{D-1}}\delta(k^2-m^2_g) \delta(\ell-k^+)|k^-|^{-\eta} \frac{1}{k^+}\frac{1}{k^-}  \\
&= -\frac{\alpha_s C_F}{\pi}\bigg(e^{ \gamma_E} \frac{\mu^2}{m^2_g}\bigg)^\epsilon w^2 \bigg(\frac{\nu}{m^2_g}\bigg)^\eta \frac{\theta(\ell)}{\ell^{1-\eta}} 
\frac{\Gamma(\eta+\epsilon)}{\Gamma(1+\eta)}\,.\nn
\end{align}
Comparing \req{eq:35eq30}, \req{eq:35eq31} and \req{eq:35eq32}, we see that the unsubtracted soft function $\tilde S_r$ is dominated by overlap with the $n$-collinear region as \req{eq:35eq31} represents the $n$-collinear modes running into the soft function.  This is due to the measurement being on soft radiation only in the $n$-collinear direction.  Radiation in the $\bar n$-collinear direction has been integrated out in the matching onto $\text{SCET}_{\text{II}}$ and subtracting \req{eq:35eq32} from \req{eq:35eq30} removes the momentum in the soft function that overlaps with the $\bar n$-collinear momentum region.  Thus the zero bin subtracted real contribution, given by the diagrams in Fig.\,\ref{fig:35fig3}(b), is
\begin{align}
S_\textrm{r} &= 2(\tilde S_\textrm{r}-S_{n\phi} ^{r} -S_{\bar n\phi} ^{r})= - 2S_{\bar n\phi} ^{r} \nn \\
&= 2 \frac{\alpha_s C_F}{\pi} w^2 \bigg\{\bigg[\frac{1}{2} \frac{e^{\epsilon \gamma_E} \Gamma(\epsilon)}{\eta}\bigg(\frac{\mu}{m_g}\bigg)^{2\epsilon}-\frac{1}{2\epsilon^2} +\frac{1}{2 \epsilon} \ln\frac{\nu Q}{\mu^2}-\ln^2\frac{\mu}{m_g} 
\nn\\ &
+ \ln\frac{\mu}{m_g}\ln\frac{\nu Q}{m^2_g}+\frac{\pi^2}{24}\bigg]\delta(z)\frac{1}{Q}+\bigg[\frac{1}{2\epsilon}+\ln\frac{\mu}{m_g}\bigg] \frac{1}{z_+}\bigg\} \,, \label{eq:35eq33}
\end{align}
where the plus-function of the dimensionful variable $\ell$ is given in terms of the definition of a dimensionless variable $z = \ell/\kappa$
\begin{align}
\frac{1}{(\ell)_+} = \frac{1}{\kappa (z)_+} + \ln \kappa \, \delta(\kappa \,z) \,,
\end{align}
with
\begin{align}
\frac{1}{(z)_+} \equiv \lim_{\beta \to 0} \bigg[ \frac{\theta(z-\beta)}{z}+\ln \beta \, \delta(z)\bigg] \,.
\end{align}

Adding the virtual and real contributions gives the one loop expression for the soft function
\begin{eqnarray}
S(z)^{(1)} &= \frac{\alpha_s C_F}{\pi Q}w^2\bigg\{ -\frac{e^{\epsilon \gamma_E} \Gamma(\epsilon)}{\eta}\bigg(\frac{\mu}{m_g}\bigg)^{2\epsilon} \delta(z)\nn\\
&+\bigg( \frac{1}{\epsilon}+ \ln\frac{\mu^2}{m^2_g}\bigg)
\bigg(\!-\ln\frac{\nu}{Q}\delta(z)+\left(\frac{1}{z}\right)_+\bigg) \bigg\}\,. \label{eq:35eq36}
\end{eqnarray}
Logarithms in the soft function are minimized by setting $\mu\sim m_g$ and $\nu\sim\ell\sim Qz\sim Q\paren{\frac{1-x}{x}}$.  Note that $Q \paren{\frac{1-x}{x}}$ is an end-point region energy scale, which is however different from what one naturally chooses for the collinear function. Clearly, resumming these logarithms in $\nu$ is needed. 


\subsubsection{Soft Function with \texorpdfstring{$\Delta$}{Delta}-scheme}
\label{sec:I.5.3.1.3b}

The naive virtual soft function for DIS shown in Fig.~\ref{fig:35fig3}(a) is the same as the zero-bin of the virtual collinear function, since the momentum contributing to that integral comes from the same soft region
\begin{align}
\tilde S_v&=\left(-\frac{\alpha_sC_F}{2\pi}\right)\delta(l)\bigl\{\frac{1}{\epsilon^2}+\frac{1}{\epsilon}\ln\frac{\mu^2}{\delta_1\delta_2}+\ln\frac{\mu^2}{m_g^2}\ln\frac{\mu^2}{\delta_1\delta_2}-\frac{1}{2}\ln^2\frac{\mu^2}{m_g^2}\nn\\
&-Li_2\left(1-\frac{\delta_1\delta_2}{m_g^2}\right)+\frac{\pi^2}{12}\bigl\}.\label{eq:35eq131}
\end{align}
The naive real soft function shown in Fig.~\ref{fig:35fig3}(b) is
\begin{align}
\tilde S_r=&(4\pi g^2C_F)\mu^{2\epsilon}\int \frac{d^{4-2\epsilon}k}{(2\pi)^{4-2\epsilon}}\delta(k^2-m_g^2)\delta(l-k^-)\theta(k^0)\frac{1}{k^+-\delta_2}\frac{1}{k^--\delta_1}\nn\\
=&\frac{\alpha_sC_F}{2\pi Q}
\Bigg(-\frac{1}{\epsilon}\delta(z)\ln\frac{-\delta_1}{Q}+\frac{1}{z_+}\left(\frac{1}{\epsilon}+\ln\frac{-\mu^2}{\delta_2Q}\right)-\delta(z)\ln\!\left(-\frac{\delta_1}{Q}\right)\ln\frac{-\mu^2}{\delta_2Q}\Bigg)\,,\label{eq:35eq132}
\end{align}
where $zQ =l$, and $z$ is dimensionless.  We omit the term proportional to $\frac{\ln(1-z)}{(z)_+}$, which contributes a constant in the endpoint limit $z\to 0$.  The Delta regulator restricts the integrals leading to Eq.~\eqref{eq:35eq131} and Eq.~\eqref{eq:35eq132} to the soft momentum region, so we do not need to subtract the collinear overlap. This differs from the prescription with the $\eta$-regulator, which serves as a smooth step function in the loop integral and may leave residual overlap with the collinear function that must be eliminated by subtracting. Multiplying Eq.~\eqref{eq:35eq131} and Eq.~\eqref{eq:35eq132} by 2 for their mirror images, we get the soft function with the Delta regulator
\begin{align}
S=&2(\tilde S_v+\tilde S_r)\nn\\
=&-2\frac{\alpha_sC_F}{2\pi Q}\Bigg(\frac{1}{\epsilon^2}\delta(z)+\frac{1}{\epsilon}\bigl[\delta(z)\ln\frac{\mu^2}{\delta_1\delta_2}+\delta(z)\ln(-\frac{\delta_1}{Q})-\left(\frac{1}{z}\right)_{\!+}\bigl]-\left(\frac{1}{z}\right)_{\!+}\!\ln\frac{-\mu^2}{\delta_2Q}\nn\\
&+\ln(-\frac{\delta_1}{Q})\ln\frac{-\mu^2}{\Delta_2}\delta(z)-\frac{\pi^2}{12}\delta(z)
+\ln\frac{\mu^2}{m_g^2}\ln\frac{\mu Q}{\delta_1 m_g}\delta(z)-Li_2\left(1-\frac{\delta_1\delta_2}{m_g^2}\right)\delta(z)\Bigg)\,. \label{eq:35eq133}
\end{align}
Introducing $\kappa$ to make the arguments of the logarithms dimensionless as in Eq.~\eqref{eq:35eq133} and choosing $-\delta_2Q= m_g^2$, we can recombine logarithms to show that the infrared divergence in the soft function is independent of $\delta_1$.  We can make this choice to relate the regulators, because  in the soft function one of the three infrared Delta regulators, $\delta_1,\delta_2$ and $m_g^2$ is redundant, and the system is under-constrained.  Again this is very different from what we obtain by using the $\eta$-regulator in Eq.~\eqref{eq:35eq36}, where we separate rapidity divergences from infrared divergences and get a result containing both rapidity and IR divergences, each with an appropriate regulator, $\eta$ and $m_g^2$.

\subsubsection{Renormalization Group Running for DIS with \texorpdfstring{$\eta$}{eta}-scheme}
\label{sec:I.5.3.1.4a}

To subtract the divergences in $\epsilon$ and $\eta$ in Eq.~(\ref{eq:35eq24}) and Eq.~(\ref{eq:35eq36}), we introduce counter-terms,
\begin{align}
{C}_{n}( Q-k)^{R} &=Z_{n}^{-1} {C}_{n}( Q-k)^{B} \nn\\
S(z)^{R} &= \int dz' Z_s(z-z')^{-1}S(z')^{B} \,, \nn
\end{align}
with superscripts $R$ and $B$ indicating renormalized and bare. To determine $Z_{n}$ we need the one-loop wave function renormalization 
\beq
Z_\psi = 1 - \frac{\alpha_s C_F}{4 \pi \, \epsilon} \,,
\eeq
and find the resulting one-loop collinear counter-term is
\beq
\label{eq:35eq38}
Z_{n} = 1 + \frac{ \alpha_s C_F}{\pi}w^2\bigg[\frac{e^{\epsilon \gamma_E} \Gamma(\epsilon) }{ \eta}\bigg(\frac{\mu}{m_g}\bigg)^{2 \epsilon}+
\frac{1}{\epsilon}\bigg( \frac{3}{4} + \ln \frac{\nu}{\bn\cdot p}\bigg)\bigg]\,.
\eeq
The one-loop soft counter-term is 
\beq
\label{eq:35eq39}
Z_s(z) = 
\delta(z) + \frac{\alpha_s C_F}{\pi}w^2\bigg\{- \frac{e^{\epsilon \gamma_E} \Gamma(\epsilon) }{ \eta}\bigg(\frac{\mu}{m_g}\bigg)^{2 \epsilon}\delta(z)
+\frac{1}{\epsilon} \bigg[\frac{1}{(z)_+} -\ln\frac{\nu}{Q}\delta(z)\bigg]\bigg\}\,.
\eeq

The counter-terms must obey the consistency condition put forth by \cite{Fleming:2007xt},
\beq\label{eq:35eq40}
Z_H Z_{J_{\bn}}(z) = Z_{n}^{-1} Z_s^{-1}(z)\,,
\eeq
where $Z_{J_{\bn}}(\ell)$ is the jet-function counter-term and $Z_H$ is the square of the counter-term for the SCET DIS current, which has been given at one loop by \cite{bauer2004enhanced} in $4-\epsilon$ dimensions.   Converting the result of \cite{bauer2004enhanced} to $4 - 2 \epsilon$ dimensions and squaring gives
\beq
Z_H = 1 - \frac{\alpha_s C_F}{2\pi} \bigg( \frac{2}{\epsilon^2}+ \frac{3}{\epsilon}+ \frac{2}{\epsilon}\ln \frac{\mu^2}{Q^2}\bigg)\,,
\label{eq:35eq41}
\eeq
where $Q^2 = \bn \cdot p\, n\cdot p_X$.  
The one-loop result for $Z_{J_{\bn}}(z)$ is given by \cite{Manohar:2003vb}
\beq
Z_{J_{\bn}}(z) = \delta(z) +  \frac{\alpha_s C_F}{4\pi}\bigg[ \bigg(\frac{4}{\epsilon^2}+\frac{3}{\epsilon}-\frac{1}{\epsilon}\ln\frac{(n\cdot p)Q}{\mu^2}\bigg)\delta(z) -\frac{4}{\epsilon}\left(\frac{1}{z}\right)_+\bigg] \,.
\eeq
Putting the factors together,
\beq\label{eq:35eq43}
Z_H Z_{J_{\bn}}(z)=   
\delta(z) +  \frac{\alpha_s C_F}{4\pi}\bigg\{ \bigg[-\frac{3}{\epsilon}+\frac{4}{\epsilon}\ln \left(\frac{\bn\cdot p}{Q}\right)\bigg]\delta(z) -\frac{4}{\epsilon}\frac{1}{z_+}\bigg\} \,.
\eeq
which is exactly equal to the product of inverses $Z_{n}^{-1} Z_s^{-1}(z)$, taken from Eq.~(\ref{eq:35eq38}) and Eq.~(\ref{eq:35eq39}).

From the one-loop results, we extract the $\mu$ anomalous dimensions for the collinear and soft function respectively,
\begin{eqnarray}\label{eq:35eq44}
\gamma^\mu_{n} (\mu,\nu)&=& \frac{2 \alpha_s(\mu) C_F}{\pi}\bigg(\frac{3}{4}+\ln\frac{\nu}{\bn\cdot  p}\bigg),\\
\gamma^\mu_s(\ell;\mu,\nu) &=& \frac{2 \alpha_s(\mu) C_F}{ \pi}\bigg[\frac{1}{z_+} -\ln \frac{\nu}{Q} \delta(z)\bigg] \,.\nn
\end{eqnarray}
Note that
\begin{equation}\label{eq:35eq45}
\gamma^\mu=\gamma_n^\mu+\gamma_s^\mu=\frac{2\alpha_s C_F}{\pi}\left(\left(\frac34-\ln\left(\frac{\bar n\cdot p}{Q}\right)\right)\delta(z)+\frac{1}{z_+}\right)   \,,
\end{equation}
which agrees with the known result, and the $\nu$-dependence cancels as expected. However, we can now trace the origin of the large logarithm in the ratio of $\bar n\cdot p\sim Q$ to $Q\left(\frac{1-x}{x}\right)$ to the rapidity region.
The $\nu$ anomalous dimensions for the collinear and soft functions are
\begin{align}
\gamma_n^\nu(\mu,\nu)&=\frac{\alpha_s(\mu)C_F}{\pi}\ln\frac{\mu^2}{m_g^2}   \,,\nn\\
\gamma_s^\nu(\mu,\nu)&=-\frac{\alpha_s(\mu)C_F}{\pi}\ln\frac{\mu^2}{m_g^2}   \,.\label{eq:35eq46}
\end{align}
Adding them together, we have
$\gamma^\nu=\gamma_n^\nu+\gamma_s^\nu=0$, as is dictated by the consistency condition.  The presence of $m_g$ in $\gamma_n^\nu$ and $\gamma_s^\nu$ however suggests that the renormalization group running in $\nu$ is dependent on an infrared scale, and therefore is non-perturbative.

The  $\mu$ and $\nu$ running are independent and can be carried out in either order. 
For the collinear function, the $\mu$-running is given to one loop by
\begin{align}\label{eq:35eq47}
{C}_{n}(Q-k;\mu,\nu_c) &= U(\mu,\mu_0,\nu_c){C}_{n}(Q-k;\mu_0,\nu_c), \\
U(\mu,\mu_0,\nu_c)&=e^{\frac{3}{4}\omega(\mu,\mu_0)} \bigg[\frac{\nu_c}{\bn\cdot p}\bigg]^{\omega(\mu,\mu_0)}
\,, \nn
\end{align}
where $\nu_c$ is the collinear rapidity scale and
\beq
\omega(\mu,\mu_0) = \frac{4 C_F}{\beta_0}\ln\bigg[\frac{\alpha_s(\mu)}{\alpha_s(\mu_0)}\bigg]\,.
\eeq
For the soft function, the one-loop $\mu$-running is
\begin{align}\label{eq:35eq49}
S(\ell;\mu,\nu_s) &= \int d r \, U(\ell-r;\mu,\mu_0,\nu_s) S(\ell;\mu_0,\nu_s),\\
 U(\ell-r;\mu,\mu_0,\nu_s)&=\frac{ \big( e^{2 \gamma_E}\nu_s\big)^{-\omega(\mu,\mu_0)}}{\Gamma(\omega(\mu,\mu_0))}\frac{1}{[(\ell-r)^{1- \omega(\mu,\mu_0)}]_+}\,. \nn
\end{align}
Combining the running factors we find
\begin{equation}
U(\mu,\mu_0,\nu_c)U(\ell-r;\mu,\mu_0,\nu_s) =  \bigg[ \frac{e^{-2 \gamma_E}\nu_c}{\bn\cdot p \, \nu_{s}}\bigg]^{\omega(\mu,\mu_0)}
\frac{e^{\frac{3}{4}\omega(\mu,\mu_0)} }{\Gamma(\omega(\mu,\mu_0))}\left(\frac{1}{(\ell-r)^{1- \omega(\mu,\mu_0)}}\right)_+\,.
\end{equation}

To get a  feel for which logarithms are being summed we will transform the combined running factors into Mellin moment space:
\begin{equation}
\label{mellinrunfac}
U(\mu,\mu_0,\nu_c)U(N;\mu,\mu_0,\nu_s) =  \bigg[ \frac{e^{- \gamma_E}\nu_c}{\bar{N} \, \nu_{s}}\bigg]^{\omega(\mu,\mu_0)}e^{\frac{3}{4}\omega(\mu,\mu_0)} \,.
\end{equation}
The first term on the right-hand  side in square brackets can be expressed as
\begin{align}
 \bigg[ \frac{e^{- \gamma_E}\nu_c}{\bar{N} \, \nu_{s}}\bigg]^{\omega(\mu,\mu_0)} &= \textrm{Exp}\bigg[ \omega(\mu,\mu_0) \ln\bigg( \frac{e^{- \gamma_E}\nu_c}{\bar{N} \, \nu_{s}}\bigg)\bigg]\nn\\
&= \textrm{Exp}\bigg[ \frac{4 C_{F}}{\beta_{0}}\ln\bigg( \frac{\bar{N} \, \nu_{s}}{e^{- \gamma_E}\nu_c}\bigg) \sum^{\infty}_{n=1} \frac{1}{n}\bigg(\frac{\beta_{0} \alpha_{s}(\mu)}{2 \pi} \ln\frac{\mu_{0}}{\mu}\bigg)^{n}\bigg],
\label{logseries}
\end{align}
which, in the exponent,  gives a series in $\alpha^{n}_{s}(\mu)\ln^{n}(\mu_{0}/\mu)$ times a single power of $\ln(\bar{N}\nu_{s}/\nu_{c})$. If we make the choice $\nu_{c}=\nu_{s}$ we reproduce
the standard result of a single logarithmic series multiplied by a single logarithm of $N$. However, if we make the choice for $\nu_{c}$ and $\nu_{s}$ given above then we merely have a single logarithmic series multiplied by an ${\mathcal O}(1)$ quantity. We argue that this is the natural choice from an EFT perspective.

Having widely separated rapidity scales then forces us to consider the rRGE.  Although the $\nu$ running is non-perturbative it is still enlightening to see what the resummation looks like, and we push ahead and determine the soft $\nu$ running factor
\begin{align}
\label{softnurun}
S(\ell;\mu_s,\nu)&= V(\mu_s, \nu,\nu_0) S(\ell;\mu_s,\nu_0),   \\
V(\mu_s, \nu,\nu_0)&= \bigg[\frac{\nu}{\nu_0}\bigg]^{-\frac{\alpha_{s}(\mu)C_{F}}{\pi}\ln\frac{\mu^{2}}{m^{2}_{g}}} \,.\nn
\end{align}
Note that if we choose $\nu = \nu_{c}$ and $\nu_{0}=\nu_{s}$ in the above equations with $\mu_{s}=\mu$ then 
\begin{equation}
V(\mu, \nu_{s},\nu_c) = \bigg[\frac{\nu_{s}}{\nu_c}\bigg]^{\frac{\alpha_{s}(\mu)C_{F}}{\pi}\ln\frac{\mu^{2}}{m^{2}_{g}}} = \bigg[\frac{\nu_{s}}{\nu_c}\bigg]^{\frac{\alpha_{s}(\mu)C_{F}}{\pi}\ln\frac{\mu^{2}}{\mu^{2}_{0}}}\bigg[\frac{\nu_{s}}{\nu_c}\bigg]^{\frac{\alpha_{s}(\mu)C_{F}}{\pi}\ln\frac{\mu_{0}^{2}}{m^{2}_{g}}}\,.
\end{equation}
The first term in square brackets on the far right-hand side cancels the $\nu_{s}/\nu_{c}$ dependence in the first term ($n=0$) of the series in the exponent of Eq.~(\ref{logseries}). The second term in the square brackets is clearly infrared sensitive and we will absorb it into the definition of
the PDF. Finally we express the rapidity running factor as
\begin{equation}
\label{nonpertrrge}
V(\mu_{0}, \nu_{s},\nu_c)= \textrm{Exp}\bigg[\frac{\alpha_{s}(\mu_{0})C_{F}}{\pi}\ln \frac{\mu^{2}}{m^{2}_{g}}\ln \frac{\nu_{s}}{\nu_c}\bigg]\,,
\end{equation}
which, using $\nu_{c}/\nu_{s} = \bar{N}$, makes it clear that what is being summed (in Mellin moment space) by the rRGE is the product $\alpha_{s}(\mu_{0}) \ln(\mu_{0}/m_{g})\ln N$. The large logarithm of $N$ multiplies an infrared logarithm, which explains why no one has tried to sum these terms before. 

Of course, this begs the question of why we should even bother to separate collinear from soft in the PDF. One answer is that we have a consistent EFT formalism that never produces terms that violate power counting. There is, however, more. Currently fits of the PDF produce a very steeply falling function of momentum fraction as the endpoint is approached, with no understanding of why; our result offers an explanation. To see this we define our PDF for large-$x$ in DIS as a modified form of the function $\phi^{ns}_{q}$ in Eq.~(\ref{eq:35eq13}):
\begin{equation}
f_q^{ns}(z;\mu)_{\textrm{endpoint}}= \delta_{\tilde n\cdot\tilde p,Q}\mathcal{Z}_n(\mu,\nu_{c})S^{(DIS)}(\ell;\mu,\nu_{s})V(\mu_{0}, \nu_{s},\nu_c)
\,. \label{eq:016}
\end{equation}
This is the same as the operator definition we give in our previous paper, but we have made the presence of the $V(\mu_{0}, \nu_{s},\nu_c)$ factor explicit. Away from the endpoint $\nu_{c}$ and $\nu_{s}$ must flow together so the PDF in the endpoint matches smoothly onto the usual definition of the PDF.  In Mellin moment space the rapidity factor 
\begin{equation}
V(\mu_{0}, \nu_{s},\nu_c) = \bar{N}^{-\frac{\alpha_{s}(\mu)C_{F}}{\pi}\ln\frac{\mu_{0}^{2}}{m^{2}_{g}}}\,.
\end{equation}
If we transform back into momentum fraction space we find
\begin{equation}
V(\mu_{0}, \nu_{s},\nu_c)  =\frac{1}{\Gamma\big(\frac{\alpha_{s}(\mu)C_{F}}{\pi}\ln\frac{\mu_{0}^{2}}{m^{2}_{g}}\big)} (1-z)^{\frac{\alpha_{s}(\mu)C_{F}}{\pi}\ln\frac{\mu_{0}^{2}}{m^{2}_{g}}-1}\,,\nn
\end{equation}
where the exponent on the $(1-z)$ is nonperturbative and could be large. This offers a possible explanation for why the PDF has a steep fall-off in the endpoint.

\subsubsection{Renormalization Group Running for DIS with \texorpdfstring{$\Delta$}{Delta}-scheme}
\label{sec:I.5.3.1.4b}

The counter-terms that renormalize the soft and collinear functions in Eq.~\eqref{eq:35eq133} and Eq.~\eqref{eq:35eq130} are
\begin{align}
Z_n&=1+\frac{\alpha_sC_F}{\pi}\left[\frac{1}{\epsilon^2}+\frac{1}{\epsilon}\left(\frac{3}{4}+\ln\frac{\mu^2}{\Delta_2}\right)\right],\label{eq:35eq134}\\
Z_s&=\delta(z)-\frac{\alpha_sC_F}{\pi}\left[\frac{1}{\epsilon^2}\delta(z)+\frac{1}{\epsilon}\left[-\frac{1}{(z)_+}+\ln\frac{\mu^2}{\delta_1\delta_2}\delta(z)+\delta(z)\ln(-\frac{\delta_1}{Q})\right]\right] \,.\label{eq:35eq135}
\end{align}
The result \req{eq:35eq133} is consistent with perturbative QCD in the endpoint limit, as we show later in this section; however, it differs from Eq.\:(A.5) in \cite{Chay:2013zya} which is also performed in the Delta-regulator scheme. The last term of Eq.~(A.5) of \cite{Chay:2013zya} is not shown in the body of the paper, as it  should not be included in the combined result to be consistent with QCD.

To check our results with the DIS consistency condition Eq.\,\eqref{eq:35eq40}, we must first calculate the counter-term of the jet function with the Delta regulator.  The calculation is carried out in Appendix \ref{appx:B.3}.  The result is 
\begin{align}\label{eq:35eq136}
Z_J=\delta(z)+\frac{\alpha_s C_F}{2\pi}\left(\frac{1}{\epsilon^2}\delta(z)+\oneov{\epsilon}\left(\delta(z)\left(\frac{3}{4}+\ln\frac{\mu^2}{\Delta_1}+\ln\frac{-\Delta_1}{n\cdot p}\right)-\frac{1}{(z)_+}\right)\right)\,.
\end{align}
Combining this with Eqs.~\eqref{eq:35eq134} and \eqref{eq:35eq135}, we verify the consistency condition Eq.~\eqref{eq:35eq40}.  The anomalous dimensions are
\begin{align}
\gamma^\mu_n=&\frac{2\alpha_s C_F}{\pi}\oneov{\epsilon}\left(\frac{3}{4}+\ln\frac{\mu^2}{\Delta_2}\right), \label{eq:35eq137} \\
\gamma^\mu_s=&-\frac{2\alpha_s C_F}{\pi}\left(\frac{1}{\epsilon^2}\delta(z)+\frac{1}{\epsilon}\left(-\frac{1}{(z)_+}+\delta(z)\ln\frac{\mu^2}{-\Delta_2}\right)\right)\,.   \label{eq:35eq138}
\end{align}
Analogous to Eq.\eqref{eq:35eq44} and Eq.\eqref{eq:35eq46}, we can see that: (1) Because we only treat the rapidity divergences in the semi-inclusive region as one type of infrared divergence, we cannot separate and resum it using the dimensional regularization scale $\mu$.  (2)  Similar to $\eta$-regulator, the sum of the anomalous dimensions $\gamma^\mu=\gamma^{\mu}_n+\gamma^\mu_s$ from Eqs.\,\eqref{eq:35eq136} and \eqref{eq:35eq138} is independent of the additional scale $\Delta_2$.  However, the presence of $\Delta_2$ means the running of both the collinear and soft functions  is non-perturbative.  Since the Delta regulator and $\eta$ regulator both exhibit nonperturbative running, our calculations suggest that the dependence on the infrared physics is independent of the regulator.  As a consequence, combining the collinear and soft functions into the new definition of the PDF 
is justified as a regulator-independent choice. 

With the counter-terms given in Eq.\,\eqref{eq:35eq134} and Eq.\,\eqref{eq:35eq135}, we choose $-\delta_2 Q=m_g^2$, subtract them along with the wave-function renormalization given in Eq.~\eqref{eq:35eqA12} from the collinear function in Eq.~\eqref{eq:35eq130} and soft function in \req{eq:35eq133}, and let $\delta_1\to 0$ to obtain the renormalized collinear and soft functions
\begin{align}
{C}_n^R&=\left(-\frac{\alpha_s C_F}{\pi}\right)\delta(k^-)\left[\frac{\pi^2}{12}-\frac34-\frac12\ln^2\frac{\mu^2}{m_g^2}-\frac34\ln\frac{\mu^2}{m_g^2}\right],\label{eq:35eq139}\\
{S}^R&=\left(-\frac{\alpha_s C_F}{\pi}\right)\left[-\left(\frac{1}{z}\right)_+\ln\frac{\mu^2}{m_g^2}-\frac{\pi^2}{4}\delta(z)+\frac12\ln^2\frac{\mu^2}{m_g^2}\delta(z)\right]\,.\label{eq:35eq140}
\end{align}

Note that the $\Delta$-regulator does not provide a renormalization group-style mechanism to sum the rapidity logarithms.  In this scheme, one would have to sum the series of loop diagrams shown in \fig{29fig2}, as we will perform in the next chapter in the context of HH$\chi$PT.

\subsubsection{Comparing to Perturbative QCD Result}
\label{sec:I.5.3.1.5a}

In this section, we compare the one-loop expression of the hadronic tenor in SCET to that in QCD.  This provides a powerful check that nothing has been missed in the SCET calculation.
Extracting the scalar part of the SCET effective hadronic tensor from Eq.\eqref{eq:35eq12}, we have
\begin{equation}
W_{\text{eff}}^{\mu\nu}=-\frac{g_\perp^{\mu\nu}}{2}W_{\text{eff}},
\end{equation}
where
\begin{equation}\label{eq:35eq53}
W_{\text{eff}}=2QH(Q;\mu_q,\mu_c)\int_x^1 \frac{dw}{w}J_{\bar n}(Qw;\mu_c;\mu)C_n((Q-k);\mu_c;\mu,\nu)S^{\text{DIS}}(Q(1-w);\mu,\nu)\,.
\end{equation}
The renormalized hard function $H^R(Q;\mu_q;\mu_c)$ and jet function $J^R_{\bar n}(Qz;\mu_c;\mu)$ are given in the literature (\cite{Manohar:2003vb, Chay:2013zya, bauer2004enhanced, Becher:2006mr, Hornig:2009kv, Bauer:2010vu}):
\begin{align}
H_{\text{DIS}}^R(Q,\mu)&=1+\frac{\alpha_sC_F}{2\pi}\left(-\ln^2\frac{\mu^2}{Q^2}-3\ln\frac{\mu^2}{Q^2}-8+\frac{\pi^2}{6}
\right),\label{eq:35eq54}\\
J_{\bar n}^R(Q(1-x),\mu)&=\delta(1-x)+\frac{\alpha_sC_F}{2\pi}\bigg\{
\delta(1-x)\left(\frac32 \ln\frac{\mu^2}{Q^2}+\ln^2\frac{\mu^2}{Q^2}+\frac72-\frac{\pi^2}{2}\right)\nn\\
&-\left(\frac{2}{1-x}\right)_+\left(\ln\frac{\mu^2}{Q^2}+\frac34\right)+2\left(\frac{\ln(1-x)}{1-x}\right)_+
\bigg\}   \,.\label{eq:35eq55}
\end{align}
From \req{eq:35eq24} and \req{eq:35eq38}, we obtain the renormalized collinear function,
\begin{align}
C^R(Q-k;\mu,\nu)&=m_0\delta_{\bar n,\tilde p,Q}\delta(k)\bigg[1+\frac{\alpha_sC_F}{\pi}\bigg(\ln\frac{\mu^2}{m_g^2}\ln\frac{\nu}{\bar n\cdot p} 
+\frac34\ln\frac{\mu^2}{m_g^2}+\frac34-\frac{\pi^2}{6}
\bigg)
\bigg]  \,.\label{eq:35eq56}
\end{align}
From \req{eq:35eq36} and \req{eq:35eq39}, we obtain the renormalized soft function,
\begin{equation}\label{eq:35eq57}
S^R(Q(1-x);\mu,\nu)=\frac1Q\delta(1-x)+\frac{\alpha_sC_F}{\pi Q}\left\{\ln\frac{\mu^2}{m_g^2}\left[
\left(\frac{1}{1-x}\right)_+-\ln\frac{\nu}{Q}\delta(1-x)
\right]
\right\}.
\end{equation}
Inserting Eqs.\,\eqref{eq:35eq54}, \eqref{eq:35eq55}, \eqref{eq:35eq56}, and \eqref{eq:35eq57} into \eqref{eq:35eq53}, we arrive at the one-loop expression for the hadronic structure function calculated in SCET which is valid in the endpoint region:
\begin{align}
W_{\text{eff}}&=2m_0\delta_{\bar n,\tilde p,Q}\bigg\{\delta(1-x)
+\frac{\alpha_sC_F}{\pi}\bigg[\left(
-\frac34\ln\frac{m_g^2}{Q^2}-\frac32-\frac{\pi^2}{3}
\right)\delta(1-x)\nn\\
&-\left(\frac{1}{1-x}\right)_+\left(\ln\frac{m_g^2}{Q^2}+\frac34\right)+\left(\frac{\ln(1-x)}{1-x}\right)_+
\bigg]+\ln\frac{\mu^2}{m_g^2}\ln\frac{\nu_c}{\nu_s}\bigg\}
\,,\label{eq:35eq58}
\end{align}
in which we set $\nu_s=\nu_c$.  The quark contribution to the hadronic structure function in  perturbative QCD is given in \cite{field1989applications},
\begin{align}
\mathcal{F}_2(x)&=\int_x^1\frac{dy}{y}(G_{p-q}^{(0)}(y)+G_{p-\bar q}^{(0)}(y))\bigg\{
\delta(1-z) 
+\frac{\alpha_s}{2\pi}P_{q\to gq}(z)\ln\frac{Q^2}{m_g^2}+\alpha_sf_2^{q \text{ DIS}}(z)\bigg\}\,,\label{eq:35eq59}
\end{align}
where
\begin{align}
P_{q\to qg}(z)&=\frac43\left(\frac{1+z^2}{1-z}\right)_+  =\frac{4}{3}\left(\frac{1+z^2}{(1-z)_+}+\frac{3}{2}\delta(1-z)\right),\nn\\
\alpha_sf_2^{q\text{ DIS}}(z)&=\frac{2\alpha_s}{3\pi}\bigg[(1+z^2)\left(\frac{\ln(1-z)}{1-z}\right)_++\frac{1+z^2}{1-z}(-2\ln z)-\frac32\frac{1}{(1-z)_+}\nn\\
&+4z+1-\left(\frac{2\pi^2}{3}+\frac94\right)\delta(1-z)\bigg],\label{eq:35eq60}
\end{align}
$z=x/y$, and $G_{p\to q}^{(0)}+G_{p\to \bar q}^{(0)}=2m_0\delta_{\bar n,\tilde p,Q}$.
As $x\to 1$, we have
\begin{align}
\mathcal{F}_2\xrightarrow{z\to 1}&\left(2m_0\delta_{\bar n\cdot \bar p,Q}\right)\bigg\{ \delta(1-x)+\frac{\alpha_sC_F}{\pi}\bigg(-\frac{1}{(1-x)_+}\left(\ln\frac{m_g^2}{Q^2}+\frac{3}{4}\right)\nn\\
&+\left(\frac{\ln(1-x)}{1-x}\right)_+-\frac34\frac{1}{(1-x)_+}+9-\left(\frac{\pi^2}{3}+\frac94+\frac34\ln\frac{m_g^2}{Q^2}\right)\delta(1-x)\bigg)\bigg\}.\label{eq:35eq61}
\end{align}
Comparing \req{eq:35eq61} to \req{eq:35eq58}, we find the low energy behavior agrees. In particular, by comparing the
jet function and soft function separately in SCET, we can trace the origin of $m_g^2$ dependence in the quark splitting term $P_{q\to qg}\sim \ln\frac{Q^2}{m_g^2}$ to the large scale difference between the collinear gluons and the soft gluons entering the final state jet. The difference between Eq.\,\eqref{eq:35eq61} and Eq.\,\eqref{eq:35eq58} is the constant coefficient of $\delta(1-x)$ and the constant term which are regularization scheme dependent.  Since the SCET calculation uses a different regularization scheme from \cite{field1989applications} this discrepancy is expected.


We insert Eq.~\eqref{eq:35eq141}, Eq.~\eqref{eq:35eq140}, renormalized final-jet function Eq.~\eqref{eq:35eqA15} and the hard function Eq.~\eqref{eq:35eq54} into the hadronic tensor Eq.~\eqref{eq:35eq125} and replace $z$ with $(1-x)$ to obtain
\begin{align}
(W_{\mu\nu})_{\Delta-\text{DIS}}^{\text{eff}}&=2m_0\delta_{\bar n\cdot\tilde p,Q}\bigg\{\delta(1-x)+\frac{\alpha_s C_F}{\pi}\bigg[\left(-\frac34\ln\frac{m_g^2}{Q^2}\right)\delta(1-x) \nn\\
&-\left(\frac{1}{1-x}\right)_+\bigg(\ln\frac{m_g^2}{Q^2}
+\frac34\bigg)+\left(\frac{\ln(1-x)}{1-x}\right)_++\left(\frac{15}{8}-\frac{\pi^2}{12}\right)\delta(1-x)\bigg]\bigg\}\,.
\label{eq:35eq141}
\end{align}
Again, we reproduced the perturbative QCD result except for the constant coefficient of $\delta(1-x)$ term, which depends on the regularization scheme we choose.


\subsubsection{Definition of the Parton Distribution Function}
\label{sec:I.5.3.1.6}


We consider the definition of the parton distribution function. The PDF defined in 
\be
\label{eq:44eq25} 
\phi^{ns}_q(Q(1-x)+\ell;\mu) =  {\cal Z}_{n}(Q;\mu,\nu)S(\ell; \mu, \nu)\,,
\ee
is worrisome because the soft function is sensitive to both the initial and final state (due to the soft Wilson lines running to infinity). This would imply that the PDF is not universal to other process with the same initial state but different final state. 
To keep the PDF universal, we want to require that the PDF only depend on properties of the initial state. 
In this section we show that the soft function in 
\begin{align}
\label{eq:44eq17} 
\frac{1}{N_c} \langle{0} |\textrm{Tr}\bigg(
\bar{\textrm{T}}\bigg[ Y^\dagger_{n}(n\cdot x) \tilde{Y}_{\bn}(n\cdot x) \bigg]
\textrm{T}\bigg[\tilde{Y}^\dagger_{\bn}(0) Y_{n}(0) \bigg]
\bigg)|0\rangle
 \equiv  \int d\ell \, e^{-\frac{i}{2}\ell n\cdot x} S(\ell;\mu) \
\end{align}
can be manipulated into a form which is only sensitive to initial state radiation making our definition of the PDF universal. 

We introduce Wilson lines linking the far past to the far future (\cite{Arnesen:2005nk})
\bea
\tilde Y^{\infty\dag}_\bn=\bar{P}\exp\left(-ig\int_{-\infty}^{\infty}ds \bn\cdot A_s(\bn s)\right) \\
\tilde Y^{\infty}_\bn= P\exp\left(ig\int_{-\infty}^{\infty}ds \bn\cdot A_s(\bn s)\right)
\eea
and insert the identity $\tilde Y^{\infty\dag}_\bn \tilde Y^{\infty}_\bn\equiv 1$ between the time-ordered and anti-time-ordered Wilson lines in the soft function \req{eq:44eq17}.  In Appendix \ref{appx:B.2}, we show that
\begin{align}
\frac{1}{N_c}\langle 0|&\textrm{Tr}\left(\bar{\textrm{T}}\left[Y^\dag_n(n\cdot x)\tilde Y^\dag_\bn(n\cdot x)\right]\tilde Y^{\infty\dag}_\bn\tilde Y^{\infty}_\bn\textrm{T}\left[\tilde Y_\bn(0)Y_n(0)\right]\right)|0\rangle 
\nn \\ 
&=  \frac{1}{N_c}\langle 0|\textrm{Tr}\left(\bar{\textrm{T}}\left[Y^\dag_n(n\cdot x)Y_\bn(n\cdot x)\right]\textrm{T}\left[Y^\dag_\bn(0)Y_n(0)\right]\right)|0\rangle 
\equiv \int d\ell e^{\frac{i}{2}ln\cdot x}S(\ell,\mu), \label{eq:44eq57}
\end{align}
which gives an $S(\ell,\mu)$ that is sensitive only to initial state information, since all four Wilson lines extend from minus infinity to the interaction point. Now the expression for the PDF defined in Eq.~(\ref{eq:44eq25}) has the form
\bea\label{eq:44eq58}
\phi_q^{ns}(z;\mu) &=&
\frac{1}{2} \sum_\sigma  
 \langle{h_{n}(p,\sigma)}| \bar{\chi}_{n}(0)\frac{\bnslash}{2}   \chi_{n}(0) |{h_{n}(p,\sigma)}\rangle\\
 && \times \int \frac{d n\cdot x}{4 \pi} e^{\frac{i}{2}Q z n\cdot x}
 \frac{1}{N_c} \langle{0} |\textrm{Tr}\bigg(
\bar{\textrm{T}}\bigg[ Y^\dagger_{n}(n\cdot x) {Y}_{\bn}(n\cdot x) \bigg]
\textrm{T}\bigg[{Y}^\dagger_{\bn}(0) Y_{n}(0) \bigg]
\bigg)|0\rangle\,, \nn
\eea
which makes the it manifest that the PDF only depends on the initial state. 


Jefferson Lab is undertaking a new set of experiments aimed at determining the PDF in the moderate- and large-$x$ regions.   By expanding our result in powers of $\alpha_s$, we can match smoothly onto the leading-log fixed-order PDF at moderate $x$ and predict the PDF at $x\to 1$ with rapidity logarithms resummed.

\subsection{Drell-Yan Processes Near Endpoint}
\label{sec:I.5.3.2}

We now apply a similar analysis to the Drell-Yan processes.\footnote{This section is based on work in collaboration with S. Fleming (\cite{Fleming:2016nhs}).}  Of particular theoretical interest is the so-called threshold region, where the invariant mass of the lepton pair approaches the center-of-mass energy of the collision. In this regime large Sudakov logarithms must be resummed~(\cite{Sterman:1986aj,Catani:1989ne, Magnea:1990qg,Korchemsky:1993uz}). Similar, but on less rigorous footing is the need for partonic resummation. In this case one is not in the true endpoint region, but rather in the region where the invariant mass of the colliding partons is just above the threshold for the production of the final state. \cite{Appell:1988ie,Catani:1998tm} argue that the sharp fall-off of parton luminosity at large $x$ enhances the partonic threshold region, and thus requires resummation. A quantitative study of this question was carried out in the context of soft collinear effective theory (SCET) by \cite{Becher:2007ty}, who conclude among other things that ``the dynamical enhancement of the threshold contributions remains effective down to moderate values $\tau \approx 0.2$...'', where $\tau =1$ represents the true endpoint.

In the threshold region the large Sudakov logarithms which need to be resummed have a simple form in Mellin moment space, where leading terms appear in perturbation theory as double logarithms $\alpha_{s}^{n}\ln^{2n}(N)$, where $N$ is the Mellin moment.  The threshold region corresponds to the limit of large $N$, so clearly the presence of these types of terms poses problems for a naive perturbative expansion and calls for resummation. Part of this resummation occurs when the parton distribution function (PDF) is evolved using the DGLAP equation (\cite{gribov1972deep,altarelli1977asymptotic,Dokshitzer:1977sg}), which in the threshold region becomes particularly simple. In Mellin moment space the anomalous dimension for the non-singlet quark-to-quark PDF has the form (\cite{Moch:2004pa})
\begin{equation}
\label{mellmomanom}
\gamma_{ns}^{(n)} = -\bigg(\frac{\alpha_{s}(\mu)}{4 \pi}\bigg)^{n+1}\bigg[ A_{n}\log(\bar{N}) - B_{n}\bigg] + {\mathcal O}\bigg(\frac{\ln(N)}{N},\frac{1}{N}\bigg)\,,
\end{equation}
where $\bar{N} = N e^{\gamma_{E}}$, $\gamma_{E}$ being the Euler-Mascheroni constant. At order $n=0$, for example, $A_{0} = 16/3\approx 5.3$ and $B_{0}= 4$.  What is peculiar about this result is that while $A_{n}$ and $B_{n}$ are numbers of the same order, there is the large logarithm of $N$ enhancing the $A_{n}$ term.  From an EFT point of view the large logarithm is problematic because if we have a correct power counting of threshold enhancements we should never encounter such power enhanced terms. 

In \cite{Fleming:2012kb}, we used SCET to show that the PDF in the DIS threshold region can be expressed as the product of a collinear factor and a soft function. Since both the collinear and soft degrees of freedom in the endpoint have an invariant mass of order the hadronic scale such a separation necessitates the introduction of a rapidity regulator to keep the two modes separate. We use the rapidity regulator of \cite{Chiu:2011qc,Chiu:2012ir}, which introduces a rapidity scale $\nu$ to keep different modes on the same mass curve distinct. This tool allows us to reorganize the perturbative expansion of the anomalous dimension for the non-singlet quark-to-quark PDF in the threshold region. We find the leading order anomalous dimension in Mellin moment space to be
\begin{equation}
\gamma_{ns}^{(0)} = -\bigg(\frac{\alpha_{s}(\mu)}{4 \pi}\bigg)^{n+1}\bigg[ A_{0}\ln \bigg(\frac{\nu_{c}\nu_{s}}{Q^{2}/\bar{N}}\bigg) - B_{0}\bigg] \,,
\end{equation}
where $\nu_{c}\approx Q$ is the collinear rapidity scale, and $\nu_{s} \approx Q/\bar{N}$ is the soft rapidity scale. The rapidity scales are set by minimizing logarithms in the collinear and soft anomalous dimensions, and result in a $\gamma_{ns}^{(0)} $ free of a logarithmic enhancement.  Now both terms in the anomalous dimension are of ``natural'' size, ${\mathcal O}(1)$.

This shows that large rapidity logarithms arise in the kinematic limit that the total momentum of the soft radiation has invariant mass similar to that of the initial protons, namely $~\LQCD^2$.  In view of the rapidity-scale-free definition of the PDF introduced in the previous section for DIS, we may extract new information on the elastic limit of hadron-hadron collisions by comparing the DIS-defined PDF to the PDF we obtain here in the Drell-Yan analysis.  Departures of the DY PDF from a direct product of DIS PDFs represent low-energy interferences between the two protons.  Such interference is only expected near the endpoint $x\to 1$; at moderate $x$ the colliding partons are well-separated in time during the collision so that they are uncorrelated. 

We analyze the large-$x$ Drell-Yan process in the same fashion as the previous section: first we integrate out the large scale $\sim Q$ by matching QCD onto $\text{SCET}_{\text{I}}$, then we factorize, and finally we integrate out the scale $Q(1-x)$ by matching onto $\text{SCET}_{\text{II}}$.  We compute each piece in the factorization formula to the first perturbative order and resum large logarithms to NLL order.  Finally, we discuss the PDF for two protons colliding at large $x$.

\subsubsection{Factorization theorem}
\label{sec:I.5.3.2.1}

\nt{Kinematics.}
While we worked through the kinematics of DIS process in both the target rest frame and Breit frame, we consider the Drell-Yan process only in the Breit frame. The proton in the $\bar n$-direction  carries momentum $p^\mu=\frac{\bar n^\mu}{2}n\cdot p+\frac{n^\mu}{2}\bar n\cdot p+\bar p_\perp^\mu$, and proton in the $n$-direction  carries momentum $\bar p^\mu=\frac{n^\mu}{2}\bar n\cdot \bar p+\frac{\bar n^\mu}{2}n\cdot \bar p+\bar p_\perp^\mu$. The invariant mass-squared of the proton-proton collision is $s=(p+\bar p)^2\simeq (n\cdot p)(\bar n\cdot \bar p)$, since $n\cdot p$ and $\bar n\cdot \bar p$ are the large components of $p$ and $\bar p$ respectively. The squared momentum transfer between the two protons is $Q^2=q^2$, so for $p$ we define
\begin{equation}\label{eq:16'}
x=\frac{Q^2}{2p\cdot q}=\frac{Q^2}{(n\cdot p)(\bar n\cdot q)}\simeq \frac{n\cdot q}{n\cdot p}\,,
\end{equation}
while for $\bar p$  we define
\begin{equation}\label{eq:16''}
\bar x=\frac{Q^2}{2\bar p\cdot q}=\frac{Q^2}{(\bar n\cdot \bar p)(n\cdot q)}\simeq \frac{\bar n\cdot q}{\bar n\cdot \bar p}\,.
\end{equation}
Here $\tau=Q^2/s=x\cdot \bar x$ is the fraction of the energy-squared taken by the colliding partons from the protons. The endpoint corresponds to $\tau\to 1$. As in DIS, we define $\frac{Q}{x}=Q+l^+$, $\frac{Q}{\bar x}=Q+\bar l^-$ with lightcone momenta $l^+$ and $\bar l^-$.  The separated scales are
\begin{itemize}
\item hard modes with $q=(Q,Q,0)$ at the hard scale;
\item $n$-collinear modes with $p_c=\left(\frac{Q}{x},\frac{\LQCD^2}{Q},\LQCD\right)\sim (Q+l^+,l^-,\LQCD)$ with invariant mass $p^2\sim\Lambda_{\text{QCD}}^2$;
\item $\bar n$-collinear modes with $\bar p_c=\left(\frac{\LQCD^2}{Q},\frac{Q}{\bar x},\LQCD\right)\sim (\bar l^+,Q+\bar l^-,\LQCD)$ with invariant mass $\bar p^2\sim \Lambda_{\text{QCD}}^2$;
\item soft modes with $p_s\sim(\LQCD,\LQCD,\LQCD)$ at the soft scale.
\end{itemize}
As $x,\bar x\to 1$, the off-shellness of the initial states $Q\frac{(1-x)}{x}\sim l^+$ and $Q\frac{(1-\bar x)}{\bar x}\sim \bar l^-$ go to $\Lambda_{\text{QCD}}$, bringing in new rapidity singularities arising from the fact that both soft and collinear modes have invariant mass squared of  order $\Lambda^2_{\text{QCD}}$.  These singularities are regulated with the covariant $\eta$ regulator, which allows us to resum the rapidity logarithms by running from $Q$ to $Q\frac{(1-x)}{x}\sim l^+\sim Q\frac{(1-\bar x)}{\bar x}\sim \bar l^-$.

\nt{Factorization.}
A number of papers, \cite{Catani:1996yz,Bauer:2002nz,Idilbi:2005ky,Idilbi:2006dg,Becher:2006nr,Becher:2007ty,Chay:2012jr}, have discussed factorization of Drell-Yan using SCET.  Here we follow \cite{Bauer:2002nz}, starting with the unpolarized DY cross section:
\begin{equation}\label{eq:17}
d\sigma=\frac{32\pi^2\alpha^2}{sQ^4}L_{\mu\nu}W^{\mu\nu}\frac{d^3k_1}{(2\pi)^3(2k_1^0)}\frac{d^3k_2}{(2\pi)^3(2k_2^0)}\,,
\end{equation}
where $L_{\mu\nu}$ is the lepton tensor, and $W^{\mu\nu}$ is the DY hadronic tensor.  \req{eq:17} gives
\begin{align}\label{DYdiffcrosssection}
\frac{d\sigma}{dQ^2}=-\frac{2\alpha}{3Q^2s}\int\frac{d^4q}{(2\pi)^3}\delta(q^2-Q^2)\theta(q_0)W(\tau,Q^2),
\end{align}
where $Q^2=\tau s$ is the lepton pair's center of mass energy squared.  Summing over final states, we obtain
\begin{equation}\label{eq:19}
W(\tau,Q^2)=-\frac{1}{4}\sum_{\text{spin}}\int d^4y e^{-iq\cdot y}\langle p\bar p|J^{\mu\dag}(y)J_\mu(0)|p\bar p\rangle\,,
\end{equation}
where $J^\mu(y)$ is the QCD current.  Near the endpoint region, the magnitude of the 3-momentum transferred is
\begin{align}
|\vec q|\leq \frac{Q}{2}(1-\tau),
\end{align}
where $Q=\sqrt{Q^2}$.  As a result, the zero component is
\begin{align}
q_0=Q+\mathcal{O}(1-\tau)\gg|\vec q|.
\end{align}
Therefore the $\delta$-function in \req{DYdiffcrosssection} is expanded
\begin{align}
\delta(q^2-Q^2)=\frac{1}{2Q}\delta(q_0-Q)+\mathcal{O}(1-\tau)^2.
\end{align}
Carrying out the $q_0$ integration, the hadronic structure function becomes
\begin{align}
W(\tau,Q^2)= -\frac{1}{8Q}\sum_{\rm spins}\int\frac{d^3q}{(2\pi)^3}\int d^4y e^{-iQy_0+i\vec q\cdot \vec y}\langle p\bar p|J^{\mu\dag}(y)J_\mu(0)|p\bar p\rangle.
\end{align}

We match $W$ onto $\scetii$, and get
\begin{align}
W^{\text{eff}}& =-\oneov{4}\sum_{\sigma,\sigma'}\int\frac{d^3q}{(2\pi)^3}\int d^4y\oneov{2Q}\sum_{\bar w,w}
C^*(Q,Q';\mu_q,\mu)C(\bar w,w,;\mu_q,\mu) \delta_{\bar n\cdot p_n,Q}\delta_{n\cdot\bar p_{\bar n},Q} \nn \\
& \times
\langle h(p_n,\sigma)\bar h(\bar p_{\bar n},\sigma')|\bar T[\bar \chi_{\bar n,\bar w'}Y_{\bar n}^\dagger \bar Y_n \gamma_\mu^\perp\chi_{n,w'}(y)]\:
\nn \\ &\times 
T[\bar \chi_{n,w}\bar Y_{\bar n}^\dagger Y_n \gamma^\mu_\perp\chi_{\bar n,\bar w}(0)]|h(p_n,\sigma)\bar h(\bar p_{\bar n},\sigma')\rangle \,. \label{eq:20}
\end{align}
Here, we have defined the $\bar n$-direction incoming proton to be carrying momentum $\bar p^\mu=\oneov{2}(n\cdot p_{\bar n}+n\cdot \bar p_r)\bar n^\mu$ with the large component of $\bar p^\mu$ scaling as $n\cdot p_{\bar n}\simeq Q/\bar x \simeq Q$ and the residual momentum $\bar p_r^\mu$ containing the small momentum $\bar p_r^\mu\simeq \bar\ell^-\simeq Q\frac{1-\bar x}{\bar x}$.  Similarly, the $n$-direction incoming proton momentum is $p^\mu=\oneov{2}(\bar n\cdot p_n +\bar n\cdot p_r)n^\mu$ with the large component of $p^\mu$ scaling as $\bar n\cdot p_{n}\simeq Q/x\simeq Q$ and the residual momentum $\bar p_r^\mu$ containing the small momenta $p^\mu_r\simeq \ell^+\simeq Q\frac{1-x}{x}$.
We introduce Kronecker deltas to fix the large components of $p$ and $\bar p$ to be equal to $Q$ and integrate over the residual components of the coordinates in position space.  
The Wilson lines $Y_{\bar n}$ and $\bar Y_n$ are associated with soft radiation from two incoming states,
\begin{align}
Y_n(y)&=P\exp\bigg[ig\int_{-\infty}^y ds \:n\cdot A_{us}(sn)\bigg],\nn \\
\bar Y_{\bar n}^\dagger(y)&=P\exp\bigg[-ig\int_{-\infty}^y ds\:\bar n\cdot A_{us}(s\bar n)\bigg]\,. \label{eq:23}
\end{align}
The hadronic structure function can be factored into the three sectors:
\begin{align}
W^{\text{eff}}&=-\oneov{4}\sum_\sigma\int\frac{d^3q}{(2\pi)^3}\int \frac{d^4y}{2Q}e^{i\vec q\cdot \vec y}\sum_{\bar w,w}  C^*(Q,Q;\mu_q,\mu)C(w,\bar w;\mu_q,\mu)\nn\\
&\times \langle h(p_n,\sigma)\bar h(\bar p_{\bar n},\sigma)|
\bar T[(\bar \chi_{\bar n,Q}^\alpha)^i
\paren{Y_{\bar n}^\dagger (\gamma_{\mu}^\perp)_{\alpha\beta} \bar Y_n}_{ij}(\chi_{n,Q}^\beta)^j(y)]
\delta_{\bar n\cdot p_n,Q}\nn\\
&\times \delta_{n\cdot\bar p_{\bar n},Q}T[(\bar \chi_{n,w}^\rho)^l\paren{\bar Y_{\bar n}^\dagger(\gamma^\mu_{\perp})_{\rho\lambda} Y_n}_{lm}(\chi_{\bar n,w'}^\lambda)^m(0)]|h(p_n,\sigma)\bar h(\bar p_{\bar n},\sigma)\rangle \label{eq:21}\\
&=-\oneov{4}\sum_\sigma\int \frac{d^4y}{2Q} \delta^3(\vec y) \sum_{w,\bar w}C^*(Q,Q;\mu_q,\mu)C(w,\bar w;\mu_q,\mu)
 \delta_{\bar n\cdot p_n,Q}\delta_{n\cdot \bar p_{\bar n},Q}\nn\\
&\times \langle h(p_n,\sigma)|\bar T[(\bar\chi_{\bar n,Q}^\alpha)^i(y)
(\chi_{\bar n,\bar w}^\lambda)^m(0)]|h(p_n,\sigma)\rangle \nn\\
&\times \langle \bar h(\bar p_{\bar n},\sigma)|T[(\bar \chi_{n,w}^\rho)^l(0)(\chi_{n,Q}^\beta)^j(y)]|\bar h(\bar p_{\bar n},\sigma)\rangle \nn\\
&\times \langle 0 |\bar T[(Y_n^\dagger \bar Y_n)_{ij}(y)]T[(\bar Y_{\bar n}^\dagger Y_n)_{lm}(0)]|0\rangle(\gamma_\mu^\perp)_{\alpha\beta}(\gamma^\mu_\perp)_{\rho\lambda}\,.\label{eq:22}
\end{align}
Integrating over $\vec y$, contracting the color indices and averaging the color of the initial states, we have
\begin{align}
W^{\text{eff}}&=
\frac{1}{4}\int \frac{dy_0}{2Q}\oneov{2}\sum_{w,\bar w}C^*(Q,Q;\mu_q,\mu)C(w,\bar w;\mu_q,\mu)
\nn \\&\times
\frac{1}{N_c}\sum_\sigma\langle h(p,\sigma)|\bar \chi_{\bar n,Q}(y_0)\frac{\nslash}{2}\chi_{\bar n,\bar w}(0)|h(p,\sigma)\rangle \delta_{\bar n\cdot p_n,Q}
\nn \\ &\times
\oneov{N_c}\sum_{\sigma'}\langle \bar h(\bar p_{\bar n},\sigma')|\bar \chi_{n,w}{(0)}\frac{\slashed{\bar n}}{2}\chi_{n,Q}(y_0)|\bar h(\bar p_{\bar n},\sigma')\rangle \delta_{\bar p_{\bar n}\cdot n,Q}
\nn \\ &\times
\langle 0|\bar T[(Y_{\bar n}^\dagger \bar Y_n)](y_0)T[(\bar Y_{\bar n}^\dagger Y_n)](0)|0\rangle \,.\label{eq:25}
\end{align}
Due to label momentum conservation, $w=Q=\bar w$, and we rewrite the large component of the matter field as $\bar \chi_{n,w}=\bar \chi_n\delta_{w,Q}$.  We insert the identities
\begin{align}\label{fieldshiftid}
\bar\chi_{\bar n,Q}(y_0)&=e^{i\hat\partial_0y_0}\bar\chi_{\bar n,Q}(0)e^{-i\hat\partial_0y_0}, \\
\chi_{n,Q}(y_0)&=e^{i\hat\partial_0y_0}\chi_{n,Q}(0)e^{-i\hat\partial_0y_0} ,
\end{align}
to shift $\bar\chi_{\bar n}$ and $\chi_n$ to the same spacetime point.  The operator $\hat\partial_0$ is a residual momentum operator that acts on the external states to yield
\begin{align}
\hat\partial_0|\bar h(\bar p_{\bar n},\sigma')\rangle &= \frac{Q}{2}\frac{1-\bar x}{\bar x}|\bar h(\bar p_{\bar n},\sigma')\rangle ,\\
\langle h(p_n,\sigma)|\hat\partial_0&=\langle h(p_n,\sigma)|\frac{Q}{2}\frac{1-x}{x}\,.
\end{align}
 Thus the hadronic structure function is reduced to
 \begin{align}
 W^{\rm eff}&=
|C(Q;\mu_q;\mu)|^2\delta_{\bar n\cdot p_n,Q}\delta_{n\cdot \bar p_{\bar n},Q}
\frac{1}{N_c}\int \frac{dy_0}{2Q}e^{-\frac{i}{2}Q\frac{1-x}{x}y_0}e^{-\frac{i}{2}Q\frac{1-\bar x}{\bar x}y_0}
	\nn \\&\times   
\frac{1}{N_c}\langle 0|\bar T[Y_{\bar n}^\dag \bar Y_n](y_0) T[\bar Y_n^\dag Y_{\bar n}](0)|0\rangle 
		 \notag \\  &\times 
\frac{1}{2}\sum_\sigma\langle h(p_n,\sigma)|\bar\chi_{\bar n}e^{-i\hat\partial_0y_0}\frac{\slashed{n}}{2}\chi_{\bar n}|h(p_n,\sigma)\rangle
         \frac{1}{2}\sum_{\sigma'}\langle\bar h(\bar p_{\bar n},\sigma')|\bar\chi_{n}e^{i\hat\partial_0y_0}\frac{\slashed{\bar n}}{2}\chi_{n}|\bar h(\bar p_{\bar n},\sigma')\rangle\,.
 \end{align}
As in DIS, we define a hard coefficient $H(Q;\mu)=|C(Q;\mu_q,\mu)|^2$, and two collinear functions 
\begin{align}
\frac{1}{2}\sum_\sigma\delta_{\bar n\cdot p_n,Q}\langle h(p_n,\sigma)|\bar\chi_{\bar n}e^{-i\hat\partial_0y_0}\frac{\slashed{n}}{2}\chi_{\bar n}|h(p_n,\sigma)\rangle
&\equiv \int dr e^{-ir y_0}C_{\bar n}(Q+r;\mu),\\
 \frac{1}{2}\sum_{\sigma'}\delta_{n\cdot \bar p_{\bar n},Q}\langle\bar h(\bar p_{\bar n},\sigma')|\bar\chi_{n}e^{i\hat\partial_0y_0}\frac{\slashed{\bar n}}{2}\chi_{n}|\bar h(\bar p_{\bar n},\sigma')\rangle
&\equiv \int d\bar r e^{i\bar r y_0}C_n(Q+\bar r;\mu)   \,.
\end{align}
The SCET hadronic structure function can then be expressed as
\begin{align}\label{Weffstep2}
W^{\rm eff}&=\frac{H(Q,\mu)}{2QN_c}\int dy_0\,e^{-\frac{i}{2}Q(1-\tau)y_0}\int dr d\bar r \,e^{-iry}e^{-i\bar r y}C_{\bar n}(Q+r;\mu)C_n(Q+\bar r;\mu) 
\nonumber \\ & \times 
\frac{1}{N_c}\langle 0|\bar T[Y_{\bar n}^\dag \bar Y_n](y_0) T[\bar Y_n^\dag Y_{\bar n}](0)|0\rangle ,
\end{align}
where $\mu$ is the arbitrary energy scale brought in by matching QCD onto SCET, and its dependence in the hard coefficient $H(Q;\mu)$ introduced by this matching process, is canceled by the dependence in the product of the two collinear functions and one soft function. The collinear functions become collinear factors because momentum conservation forbids collinear radiation into the final state. This then requires an additional rapidity scale $\nu$ to separate soft from collinear modes.  Including the rapidity scale dependence,
\begin{align}
C_{\bar n}(Q+r;\mu) & \to C_{\bar n}(Q+r;\mu,\nu)=\mathcal{Z}_{\bar n}(\mu,\nu)\delta(r)\delta_{\bar n\cdot p_n,Q}, \label{eq:79temp}\\
C_n(Q+\bar r;\mu) & \to C_{n}(Q+\bar r;\mu,\nu)=\mathcal{Z}_n(\mu,\nu)\delta(\bar r)\delta_{n\cdot \bar p_{\bar n},Q}  \,. \label{eq:80temp}
\end{align}
As in DIS, these functions are proportional to $\delta$ functions in $r, \bar r$ because there is no real gluon emission into the final state from either proton.

We redefine the soft Wilson lines analogously to the collinear fields in \req{fieldshiftid}, so that
\begin{align}
\langle 0|\bar T[Y_{\bar n}^\dag \bar Y_n](y) T[\bar Y_n^\dag Y_{\bar n}](0)|0\rangle =\langle 0|\bar T[Y_{\bar n}^\dag \bar Y_n](0)e^{i\hat\partial_0y_0} T[\bar Y_n^\dag Y_{\bar n}](0)|0\rangle \,.
\end{align}
Integrating over $r,\bar r$ in \req{Weffstep2} we obtain
\begin{align}
W^{\rm eff}&=H(Q;\mu)\frac{1}{2QN_c}
\mathcal{Z}_{\bar n}(\mu,\nu)\delta_{\bar n\cdot p_n,Q}
\mathcal{Z}_n(\mu,\nu)\delta_{n\cdot \bar p_{\bar n},Q}
\nonumber \\ &\times
\int dy_0\frac{1}{N_c}\langle 0|\bar T[Y_{\bar n}^\dag \bar Y_n](0)e^{i(\hat\partial_0-\frac{Q}{2}(1-\tau))y_0}T[\bar Y_n^\dag Y_{\bar n}](0)|0\rangle .
\end{align}
We define the DY soft function in momentum space to be:
\begin{align}
S^{\rm (DY)}(1-\tau;\mu,\nu)=\frac{1}{N_c}\langle 0|\mathrm{tr}\bar T[Y_{\bar n}^\dag \bar Y_{n}](0)\delta\left(2\hat\partial_0-Q(1-\tau)\right)T[\bar Y_n^\dag Y_{\bar n}](0)|0\rangle   \,. \label{eq:DYsoftfunction}
\end{align}
The hadronic structure function becomes
\begin{align}
W^{\rm eff}=\frac{2\pi}{QN_c}H(Q;\mu)
\mathcal{Z}_{\bar n}(\mu,\nu)\delta_{\bar n\cdot p_n,Q}\mathcal{Z}_n(\mu,\nu)\delta_{n\cdot \bar p_{\bar n},Q}
S^{\rm (DY)}(1-\tau;\mu,\nu),
\end{align}
and the differential cross section is
\begin{align}
\left(\frac{d\sigma}{dQ^2}\right)_{\text{eff}}=\frac{2\alpha^2}{3Q^2s}\frac{2\pi}{N_c}H(Q;\mu)
\mathcal{Z}_{\bar n}(\mu,\nu)\delta_{\bar n\cdot p_n,Q}\mathcal{Z}_n(\mu,\nu)\delta_{n\cdot \bar p_{\bar n},Q}
\frac{1}{Q}S^{\rm (DY)}(1-\tau;\mu,\nu).\label{eq:effcrosssection}
\end{align}

The soft function and the collinear functions run to the common rapidity scale $\nu$ in the endpoint region, suggesting the soft radiation contains information from both incoming protons.  Since the $n$ direction and $\bar n$ direction collinear functions are each connected to this soft function at low momenta by the rapidity scale $\nu$, they are coupled to each other through the soft radiation. Therefore, in the endpoint region, it does not suffice to identify the PDF of each proton with just the $n$- and $\bar n$-collinear functions.

We introduce a luminosity function that defines the $n$-collinear, $\bar n$-collinear and soft functions all together:
\begin{eqnarray}
\label{DYphi}
\mathbf{L}^{n{\bar n}s}(1-\tau;\mu) =  \delta_{\bar n\cdot p_n,Q} {\cal Z}_{n}(\mu,\nu)\, \delta_{n\cdot \bar p_{\bar n},Q}\mathcal{Z}_{\bar n}(\mu,\nu) S^{\rm (DY)}(1-\tau;\mu,\nu)\,.
\end{eqnarray}
On the right hand side, the $\nu$ dependence of the $n$-collinear, $\bar n$-collinear and soft functions cancels between the three factors.  In order to relate the Drell-Yan luminosity function in \req{DYphi} to the definition of the PDF in DIS, we can express $\mathbf{L}^{n{\bar n}s}$ as:
\begin{align}\label{DYluminosity}
\mathbf{L}^{n\bar ns}(1-\tau;\mu) 
&=\int dx d\bar x\:f^{\bar ns}_{q}(\frac{1\!-\!x}{x};\mu)f^{ns}_{q'}(\frac{1\!-\!\bar x}{\bar x};\mu)I^{\rm (DY)}_{\tau\to1}(1\!-\!\tau-\frac{1\!-\!x}{x}-\frac{1\!-\!\bar x}{\bar x};\mu),
\end{align}
where the two PDFs are defined  in Eq.~(\ref{eq:016}), and $I^{\rm (DY)}_{\tau\to1}(1-\tau;\mu)$
is an interference factor, independent of $\nu$, which represents the effect of the two protons interfering with each other at the DY endpoint.

With this interference function, the  $\scetii$ hadronic structure function is
\begin{align}
W^{\rm eff} &= \frac{2\pi}{QN_c}H( Q ;\mu)   \nn \\
& \times  \int \!dxd\bar x  \:f^{\bar ns}_{q}(\frac{1-x}{x};\mu)\:f^{ns}_{q'}(\frac{1-\bar x}{\bar x};\mu)\:I^{\rm (DY)}_{\tau\to1}(1-\tau-\frac{1-x}{x}-\frac{1-\bar x}{\bar x};\mu)\,.
\label{DYhadtenSCET2}
 \end{align}

\subsubsection{Collinear and Soft Functions to \texorpdfstring{$\cO(\alpha_s)$}{O(alpha)} for DY with \texorpdfstring{$\eta$}{eta}-scheme}
\label{sec:I.5.3.2.2a}

It is easy to show that the collinear functions in DIS and DY are equal. As in DIS, Fig.\,\ref{fig:35fig1} shows the $\mathcal{O}(\alpha_s^0)$ Feynman diagram for the collinear function. The $n$-direction collinear function tree level structure calculated from that diagram is
\begin{equation}
C_n^{\rm DY}(Q+\bar r)^{(0)}=C_n^{DIS(0)}=\delta_{\bar n\cdot p_n,Q}\delta(\bar r)m_n^{(0)} \,,\label{eq:35eq91}
\end{equation}
where $m_n$ is 
\begin{equation}
m_n^{(0)}=\oneov{2}\sum_\sigma\bar\xi_n^\sigma \frac{\bnslash}{2}\xi_n^\sigma\,. \label{eq:35eq92}
\end{equation}
The $\bar n$-direction collinear function at leading order is
\begin{equation}
C_{\bar n}^{\rm DY}(Q+\bar r)^{(0)}=\delta_{n\cdot \bar p_{\bar n},Q}\delta(r)m_{\bar n}^{(0)}, \label{eq:35eq93}
\end{equation}
where
\begin{equation}
m_{\bar n}^{(0)}=\oneov{2}\sum_\sigma\bar\xi_{\bar n}^\sigma\frac{\nslash}{2}\xi_{\bar n}^\sigma\,. \label{eq:35eq94}
\end{equation}
The $\mathcal{O}(\alpha_s)$ $n$-collinear function Feynman diagrams are shown in Fig.~\ref{fig:35fig2}. As discussed in the DIS section, Fig.~\ref{fig:35fig2}(a) is the one-loop virtual correction to the collinear function, while Fig.~\ref{fig:35fig2}(b) and Fig.~\ref{fig:35fig2}(c) are real corrections. We add the diagrams of Fig.~\ref{fig:35fig2}(a) and Fig.~\ref{fig:35fig2}(b) with the mirror diagrams and multiply it by the quark wave function renormalization to obtain
\begin{align}
C_n^{\rm DY}(Q+\bar r;\mu,\nu)^{(1)}&=C_n^{DIS(1)}\nn\\
&=
C_n^{(0)}(Q\!+\!\bar r;\mu,\nu) \, \frac{\alpha_s C_F}{\pi}w^2 
\bigg\{ \frac{e^{\epsilon \gamma_E} \Gamma(\epsilon)}{\eta}\bigg(\frac{\mu^2}{m^2_g}\bigg)^{\!\epsilon}\! +\frac{1}{\epsilon}\bigg[ \frac{3}{4} + \ln\frac{\nu}{\bn\cdot p} \bigg] \nn\\
&+\ln\frac{\mu^2}{m_g^2}\ln\frac{\nu}{\bar n\cdot p}
+\frac{3}{4}\ln\frac{\mu^2}{m_g^2}
+\frac{3}{4}-\frac{\pi^2}{6}\bigg\}\,.
\label{eq:35eq95}
\end{align}
For the $\mathcal{O}(\alpha_s)$ $\bar n$-collinear function, we repeat the whole procedure and get
\begin{align}
C_\bn^{\rm DY}(Q+r;\mu,\nu)^{(1)}
&=
C_n^{(0)}(Q\!+\!r;\mu,\nu) \, \frac{\alpha_s C_F}{\pi}w^2 
\bigg\{ \frac{e^{\epsilon \gamma_E} \Gamma(\epsilon)}{\eta}\bigg(\frac{\mu^2}{m^2_g}\bigg)^{\!\epsilon}\! +\frac{1}{\epsilon}\bigg[ \frac{3}{4} + \ln\frac{\nu}{n\cdot \bar p} \bigg] \nn\\
&+\ln\frac{\mu^2}{m_g^2}\ln\frac{\nu}{n\cdot \bar p}
+\frac{3}{4}\ln\frac{\mu^2}{m_g^2}+\frac{3}{4}-\frac{\pi^2}{6}\bigg\}\,.
\label{eq:35eq96}
\end{align}

Next we turn our attention to the soft function. 
The tree level result is trivial:
\begin{equation}
S(1-\tau)^{(0)}=\frac{\delta(1-\tau)}{Q} \,. \label{eq:35eq97}
\end{equation}
The $\mathcal{O}(\alpha_s)$ soft function Feynman diagrams are shown in Fig.\,\ref{fig:35fig3} (mirror diagrams are not shown).  The soft Wilson lines in \eq{DYsoftfunction} are defined in \eqs{35eq123}{35eq124}.  Comparing these to the soft Wilson lines in DIS in Eq.\,(\ref{eq:35eq9}), we find the $\bar n$-direction gluons are changed from outgoing to incoming. \cite{Kang:2015moa} however show that up to $\mathcal{O}(\alpha_s^2)$, the dijet hemisphere soft function in DIS and DY are equal, so the virtual DY soft function at $\mathcal{O}(\alpha_s)$ is the same as in DIS,
\begin{align}
S_v^{\rm DY} &=\delta(1-\tau)\frac{2\alpha_sC_F}{\pi}w^2\bigg[-\frac{e^{\epsilon\gamma_E}\Gamma(\epsilon)}{\eta}\!\paren{\frac{\mu}{m_g}}^{2\epsilon}\!\!+\oneov{2\epsilon^2}
\nn \\&
+\oneov{\epsilon}\ln\frac{\mu}{\nu}+\ln^2\frac{\mu}{m_g}-\ln\frac{\mu^2}{m_g^2}\ln\frac{\nu}{m_g}-\frac{\pi^2}{24}\bigg] \,.
\label{eq:35eq98}
\end{align}

The naive contribution to the $\mathcal{O}(\alpha_s)$ real DY soft function shown in \fig{35fig3}(b) is
\begin{align}
\tilde S_r^{\rm DY}=&-4C_Fg^2\mu^{2\epsilon}\nu^\eta \int\frac{d^Dk}{(2\pi)^{D-1}}\delta(k^2-m_g^2)\delta(\ell_0-\frac{k^++k^-}{2})\frac{|2k^3|^{-\eta}}{k^+k^-} \label{eq:35eq99}\\
=&-\frac{\alpha_sC_F}{2\pi Q}\left\{2\ln\frac{m_g^2}{Q^2}\paren{\oneov{1-\tau}}_+-4\paren{\frac{\ln (1-\tau)}{1-\tau}}_+ -\paren{\frac{1}{2}\ln^2\frac{Q^2}{m_g^2}}\delta(1-\tau)\right\}\,,
\label{eq:35eq100}
\end{align}
where $\ell_0 = Q (1-\tau)$.The measurement $\delta$-function at the endpoint region of Drell-Yan process requires the soft momentum $\ell$ to be the symmetric sum of $n$ and $\bar n$ gluon momenta, $\ell_0=k^++k^-$.  Consequently, there are neither rapidity divergences nor ultra-violet divergences.  In Appendix \ref{appx:B}, we show that the kinematic constraints in DY imply that no collinear modes overlap with the soft momentum region.  Therefore, there is no zero-bin subtraction for the real soft function, and 
\begin{equation}
S_r^{\text{DY}}=2\tilde S_{r}^{\text{DY}}.
\end{equation}
The $\mathcal{O}(\alpha_s)$ expression of the soft function is given by adding virtual and soft pieces with their mirror amplitudes:
\begin{align}
S(1-\tau;\mu,\nu)^{(1)}&=\frac{\alpha_sC_F}{\pi Q}w^2\Bigg[
\Bigg( -\frac{2\Gamma(\epsilon)}{\eta}\paren{\frac{\mu}{m_g}}^{2\epsilon}+\frac{1}{\epsilon^2}+\frac{2}{\epsilon}\ln\frac{\mu}{\nu} \nonumber \\
&+2\ln^2\frac{\mu}{m_g}-2\ln\frac{\mu^2}{m_g^2}\ln\frac{\nu}{m_g}
-\frac{\pi^2}{12}-2\ln 2+\frac{1}{2}\ln^2\frac{m_g^2}{Q^2}
\Bigg)\delta(1-\tau)\nonumber \\
&+4\paren{\frac{\ln (1-\tau)}{1-\tau}}_+ -2\paren{\ln \frac{m_g^2}{Q^2}}\paren{\oneov{1-\tau}}_+ \Bigg]\,.
\label{eq:35eq102}
\end{align}
Comparing this result with the $\mathcal{O}(\alpha_s)$ $n$ and $\bn$ collinear functions given in \req{eq:35eq95} and \req{eq:35eq96}, we see that the $\nu$-dependence cancels in the cross section at $\mathcal{O}(\alpha_s)$.  Forming the ratio of the DY soft function to the product of the $n$ and $\bn$ DIS soft functions proves that the interference factor \req{DYhadtenSCET2} is independent of $\nu$.


\subsubsection{DY Collinear and Soft Functions with \texorpdfstring{$\Delta$}{Delta}-scheme}
\label{sec:I.5.3.2.2b}


The virtual and real collinear functions of DY are the same as in DIS, with the $\bar n$-collinear function regulated by  $\Delta_1$, 
\begin{align}
C_n^{\rm DY}&=2\paren{-\frac{\alpha_s C_F}{2\pi}}\delta(k^-)\Bigg(-\oneov{\epsilon^2}-\oneov{\epsilon}\paren{\ln\frac{\mu^2}{\Delta_2}+\frac{3}{4}}+\frac{\pi^2}{12}-\frac{3}{4}\nn\\
&
-\ln\frac{\mu^2}{m_g^2}\paren{\ln\frac{\mu^2}{\Delta_2}+\frac{3}{4}}+\frac12\ln^2\frac{\mu^2}{m_g^2}+Li_2\paren{1-\frac{\delta_1\delta_2}{m_g^2}}-Li_2\paren{\frac{\Delta_2}{m_g^2}}
\nn\\&
+\ln\frac{\Delta_2}{m_g^2}\paren{\frac{\Delta_2/m_g^2}{\frac{\Delta_2}{m_g^2}-1}-\ln\paren{1-\Delta_2/m_g}}\Bigg)\,,\label{eq:35eq142}
\end{align}
and the $n$-collinear function regulated by  $\Delta_2$,
\begin{align}
C_{\bar n}^{\rm DY}&=2\paren{-\frac{\alpha_s C_F}{2\pi}}\delta(k^+)\Bigg(-\oneov{\epsilon^2}-\oneov{\epsilon}\paren{\ln\frac{\mu^2}{\Delta_1}+\frac{3}{4}}+\frac{\pi^2}{12}-\frac{3}{4}\nn\\
&-\ln\frac{\mu^2}{m_g^2}\paren{\ln\frac{\mu^2}{\Delta_1}+\frac{3}{4}}+\frac12\ln^2\frac{\mu^2}{m_g^2}
+Li_2\paren{1-\frac{\delta_1\delta_2}{m_g^2}}-Li_2\paren{\frac{\Delta_1}{m_g^2}}
\nn\\&
+\ln\frac{\Delta_1}{m_g^2}\paren{\frac{\Delta_1/m_g^2}{\frac{\Delta_1}{m_g^2}-1}-\ln\paren{1-\frac{\Delta_1}{m_g^2}}}\Bigg)\,.\label{eq:35eq143}
\end{align}
The virtual soft function for DY is also the same as in DIS
\begin{align}
S_v^{\rm DY}&=\,2\!\paren{\!-\frac{\alpha_sC_F}{2\pi Q}}\delta(1-\tau)\Bigg(\oneov{\epsilon^2}+\oneov{\epsilon}\ln\frac{\mu^2}{\delta_1\delta_2}\nn\\
&+\ln\frac{\mu^2}{m_g^2}\ln\frac{\mu^2}{\delta_1\delta_2}-\oneov{2}\ln^2\!\!\paren{\frac{\mu^2}{m_g^2}}+\frac{\pi^2}{12}-Li_2\!\paren{1-\frac{\delta_1\delta_2}{m_g^2}}\!\Bigg)\,.\label{eq:35eq144}
\end{align}
The real piece of the DY soft function is,
\begin{align}
S_r^{\rm DY}&=-2(2\pi g^2C_F)\frac{\mu^{2\epsilon}}{(2\pi)^{4-2\epsilon}}\int dk^+ dk^-\int d\Omega_{1-\epsilon}\frac{d(k_\perp^2)}{2}(k_\perp^2)^{-\epsilon}\nn\\
&\times\frac{\delta(k^+k^--k_\perp^2-m_g^2)\delta(\ell_0-\frac{k^++k^-}{2})}{(k^+-\delta_1)(k^--\delta_2)}
\nonumber \\
&=2\frac{\alpha_sC_F}{2\pi Q}\Bigg(\left(\frac12\ln^2\frac{Q^2}{m_g^2}\right)\delta(1-\tau)-2\left(\frac{1}{1-\tau}\right)_+\ln\left(\frac{m_g^2}{Q^2}\right)+4\left(\frac{\ln (1-\tau)}{1-\tau}\right)_+\Bigg)\,.
\end{align}
The soft function for DY is
\begin{align}
S^{\rm DY}&=S_v^{\rm DY}+S_r^{\rm DY}\nn\\
&=-\frac{\alpha_s C_F}{\pi Q}\Bigg\{ \Bigg[ \frac{1}{\epsilon^2}+\frac{1}{\epsilon}\ln\frac{\mu^2}{\delta_1\delta_2}-\frac12\ln^2\frac{Q^2}{m_g^2}+\frac{\pi^2}{12}-\frac12\ln^2\frac{\mu^2}{m_g^2}+\ln\frac{\mu^2}{m_g^2}\ln\frac{\mu^2}{\delta_1\delta_2}
\nn\\
&-Li_2\!\left(1-\frac{\delta_1\delta_2}{m_g^2}\right)\Bigg]\delta(1-\tau)+2\ln\frac{m_g^2}{Q^2}\left(\frac{1}{1-\tau}\right)_+-4\left(\frac{\ln (1-\tau)}{1-\tau}\right)_+\Bigg\}\,.\label{eq:35eq146}
\end{align}
Therefore, the counter-terms for the DY collinear and soft functions are
\begin{align}
Z_n&=1+\frac{\alpha_s C_F}{\pi}\left[\oneov{\epsilon^2}+\oneov{\epsilon}\paren{\frac34+\ln\frac{\mu^2}{\Delta_2}}\right],\nn\\
Z_{\bar n}&=1+\frac{\alpha_s C_F}{\pi}\left[\oneov{\epsilon^2}+\oneov{\epsilon}\paren{\frac34+\ln\frac{\mu^2}{\Delta_1}}\right],\nn\\
Z_s&=1-\frac{\alpha_s C_F}{\pi}\left[\oneov{\epsilon^2}+\oneov{\epsilon}\ln\frac{\mu^2}{\delta_1\delta_2}\right]\,,\label{eq:35eq147}
\end{align}
which are regulator dependent and satisfy the consistency condition. The anomalous dimensions for the DY collinear and soft functions are
\begin{align}
\gamma_n^\mu&=\frac{2\alpha_sC_F}{\pi}\paren{\frac34+\ln\frac{\mu^2}{\Delta_2}},\nn\\
\gamma_{\bar n}^\mu&=\frac{2\alpha_s C_F}{\pi}\paren{\frac34+\ln\frac{\mu^2}{\Delta_1}},\nn\\
\gamma_s^\mu&=-\frac{2\alpha_s C_F}{\pi}\ln\frac{\mu^2}{\delta_1\delta_2}\,.\label{eq:35eq148}
\end{align}
The Delta regulators cancel in the sum of the anomalous dimensions, and a large logarithm in $(n\cdot p)(\bar n\cdot\bar p)\sim -Q^2$ remains.
Similar to the DIS case, each piece of the collinear and soft functions is dependent on the infrared physics regardless of the regularization scheme.

We can also compute the interference factor defined in Eq.~\eqref{DYhadtenSCET2} with the soft functions in DIS Eq.~\eqref{eq:35eq135} and DY Eq.~\eqref{eq:35eq146} as,
\begin{align}
I^{\rm DY}&=\frac{\alpha_s C_F}{\pi Q}\Bigg\{\left(\frac{1}{\epsilon^2}+\frac{1}{\epsilon}\ln\frac{\mu^2}{Q^2}\right)\delta(1-\tau)-\frac{2}{\epsilon}\left(\frac{1}{1-\tau}\right)_+\nn\\
&+2\ln\frac{Q^2}{\mu^2}\left(\frac{1}{1-\tau}\right)_++4\left(\frac{\ln (1-\tau)}{1-\tau}\right)_++ \frac{1}{2}\ln^2\frac{\mu^2}{Q^2}\delta(1-\tau)\Bigg\}\,.\label{eq:35eq153}
\end{align}
In relating the DY and DIS soft functions, we exploit the redundancy of our IR regulators and set $\delta_1\delta_2=m_g^2$ in the virtual contribution to the DY soft function.  Except for the constant coefficient of $\delta(z)$, we have the exact same interference factor as Eq.~\eqref{eq:35eq112} obtained using rapidity regulator.

We obtain the above result by setting $\delta_1,\delta_2$ to zero, which has the exact form of the real contribution to the soft function in the $\eta$-regulator scheme Eq.\eqref{eq:35eq99}.  This is reasonable because the $\delta_i$ do not regulate any divergences in the integral and the infrared divergence is regulated by $m_g^2$.  Since there is only one infrared divergence, the regulators $\delta_1,\delta_2$ are redundant, similar to the DIS case.


\subsubsection{Anomalous Dimensions for Collinear and Soft Functions}
\label{sec:I.5.3.2.3a}

The divergences in UV and rapidity in the collinear and soft functions Eq.~(\ref{eq:35eq95}), Eq.~(\ref{eq:35eq96}) and Eq.~(\ref{eq:35eq102}) can be subtracted by counter-terms in textbook fashion. We define the relations between the renormalized and the bare functions as
\begin{align}
C_n(Q+\bar r)^R&=Z_n^{-1}C_n(Q+\bar r)^B, \nn \\
C_{\bar n}(Q+r)^R&=Z_{\bar n}^{-1}C_{\bar n}(Q+r)^B,\nn \\
S(1-\tau)^R&=-\int d\tau'\: Z_s(\tau'-\tau)^{-1}\:S(\tau')^B \,.\label{eq:35eq103}
\end{align}
As in DIS, the wave function renormalization factor at $\mathcal{O}(\alpha_s)$ is
\begin{equation}\label{eq:35eq104}
Z_\psi=1-\frac{\alpha_sC_F}{4\pi\epsilon}\,.
\end{equation}
Thus, Eqs.\,\eqref{eq:35eq95},~\eqref{eq:35eq96} and \eqref{eq:35eq102} yield for the $\mathcal{O}(\alpha_s)$ collinear and soft renormalization factors 
\begin{align}
Z_n&=1+\frac{\alpha_sC_F}{\pi}w^2\left[\frac{\Gamma(\epsilon)}{\eta}\paren{\frac{\mu}{m_g}}^{2\epsilon}+\oneov{\epsilon}\paren{\frac{3}{4}+\ln\frac{\nu}{\bar n\cdot p}}\right],\label{eq:35eq105}\\
Z_{\bar n}&=1+\frac{\alpha_sC_F}{\pi}w^2\left[\frac{\Gamma(\epsilon)}{\eta}\paren{\frac{\mu}{m_g}}^{2\epsilon}+\oneov{\epsilon}\paren{\frac{3}{4}+\ln\frac{\nu}{n\cdot p}}\right],\label{eq:35eq106}\\
Z_s&=\delta(\ell_0)+\frac{\alpha_sC_F}{\pi}w^2\left[-\frac{2\Gamma(\epsilon)}{\eta}\paren{\frac{\mu}{m_g}}^{2\epsilon}+\oneov{\epsilon^2}+\frac{2}{\epsilon}\ln\frac{\nu}{\mu}\right]\delta(\ell_0)   \,.\label{eq:35eq107}
\end{align}
These obey the consistency condition for Drell-Yan
\begin{equation}\label{eq:35eq108}
Z_H=Z_{\bar n}^{-1}Z_n^{-1}Z_s^{-1}(1-\tau),
\end{equation}
where $Z_H$ is given in Eq.\,\eqref{eq:35eq41}.
The logarithms in the collinear function are minimized by setting $\nu\sim Q$, while in the soft function $\nu\sim\mu\sim\LQCD$.  Therefore we must resume these logarithms both in $\mu$ and $\nu$.
From Eq.~(\ref{eq:35eq105}) to Eq.~(\ref{eq:35eq107}) we also can extract the $\mathcal{O}(\alpha_s)$ anomalous dimensions. The $\mu$ anomalous dimensions are
\begin{align}
\gamma_n^\mu(\mu,\nu)&=\frac{2\alpha_sC_F}{\pi}\paren{\frac{3}{4}+\ln\frac{\nu}{\bar n\cdot p}},\nn\\
\gamma_{\bar n}^\mu(\mu,\nu)&=\frac{2\alpha_sC_F}{\pi}\paren{\frac{3}{4}+\ln\frac{\nu}{n\cdot \bar p}},\nn\\
\gamma_s^\mu(\mu,\nu)&=\frac{4\alpha_sC_F}{\pi}\ln\frac{\mu}{\nu}\,\delta(1-\tau)\,.
\label{eq:35eq109}
\end{align}
As in DIS, the sum of the collinear and soft $\mu$ anomalous dimensions is independent of the rapidity scale $\nu$, as expected.  However, the sum contains a large logarithm of $(\bar n\cdot p) (n\cdot \bar p)\sim -Q^2$. The $\nu$ anomalous dimensions are 
\begin{align} 
\gamma_n^\nu(\mu,\nu)&=\frac{\alpha_sC_F}{\pi}\ln\frac{\mu^2}{m_g^2},\nn \\
\gamma_{\bar n}^\nu(\mu,\nu)&=\frac{\alpha_sC_F}{\pi}\ln\frac{\mu^2}{m_g^2},\nn\\
\gamma_s^\nu(\mu,\nu)&=-\frac{2\alpha_sC_F}{\pi}\ln\frac{\mu^2}{m_g^2}\,\delta(1-\tau)\,.
\label{eq:35eq110}
\end{align}
Unsurprisingly, the sum of above $\nu$ anomalous dimensions is zero. The presence of $m_g$ suggests the same IR sensitivity as occurred in DIS. As we will see in next section, this IR dependence in anomalous dimensions also shows up in the Delta regulator scheme for the divergences in the endpoint region.

From the $\mu$ anomalous dimensions in Eq.\eqref{eq:35eq109}, we can see the soft function runs to the scale $\nu$ common also to the collinear functions as we have already seen in DIS case.  This is problematic because it means the two collinear functions, which are traditionally identified with the proton PDFs, are not independent from each other.  At moderate $x$, these scales would not run to the same point and the two collinear functions can be separated.  Thus at large $x$ the two collinear functions cannot be separated and we do not have a unique way to define independent (and so universal) PDFs for the colliding protons. Preserving the conventional description of the colliding protons in terms of $n$-collinear, $\bar n$-collinear and soft functions, we arrive at the luminosity function in Eq.~\eqref{DYluminosity}, and at large $x$ the two collinear pieces and one soft piece are related by the common rapidity scale $\nu$.  The luminosity function  Eq.~\eqref{DYluminosity} in position space has the form
\begin{align}
\mathbf{L}_{q\bar q}^{n\bar ns}(\mathcal{Z}_n,\mathcal{Z}_{\bar n};\mu)&=
  \frac{1}{2}\sum_\sigma\langle h_n(p,\sigma)|\bar\chi_n(0)\frac{\slashed{\bar n}}{2}\chi_n(0)|h_n(p,\sigma)\rangle\nn\\
 &\times\frac{1}{2}\sum_{\bar\sigma}\langle \bar h_{\bar n}(\bar p,\bar\sigma)|\bar\chi_{\bar n}(0)\frac{\slashed{n}}{2}\chi_{\bar n}(0)|\bar h_{\bar n}(\bar p,\bar\sigma)\rangle \nn\nonumber\\
	 &\times \int\frac{d(n\cdot x)}{4\pi}\int\frac{d(\bar n\cdot\bar x)}{4\pi}e^{\frac{i}{2}Q\bar z\bar n\cdot\bar x}e^{\frac{i}{2}Qzn\cdot x}S(n\cdot x+\bar n\cdot\bar x;\mu)   \,.
	  \label{eq:35eq111}
\end{align}

Now we connect the running and the resummation results in DY with those in DIS by solving the renormalization equation of the interference factor $I^{\text{DY}}$ we defined in Eq.\eqref{DYhadtenSCET2}. At moderate $x$, this interference factor is unity. At large $x$, to $\mathcal{O}(\alpha_s)$, the interference factor is
\begin{align}
I^{\text{DY}}&=\delta(1-\tau)\frac{\alpha_s C_F}{\pi}\bigg[ \oneov{\epsilon^2}+\oneov{\epsilon}\ln\frac{\mu^2}{Q^2}+2\ln^2\frac{\mu}{Q}-\frac{\pi^2}{12}\bigg]
\notag \\ &
+\frac{\alpha_s C_F}{\pi}\bigg[4\paren{\frac{\ln (1-\tau)}{1-\tau}}_+ -2\ln\frac{\mu^2}{Q^2}\paren{\oneov{1-\tau}}_+ -\frac{2}{\epsilon}\paren{\oneov{1-\tau}}_+\bigg], \label{eq:35eq112}
\end{align}
which is independent of rapidity scale $\nu$. The counter-term is
\begin{equation}\label{eq:35eq113}
Z_I^{\text{DY}}=\delta(1-\tau)+\frac{\alpha_s C_F}{\pi}\bigg\{\left[\oneov{\epsilon^2}+\oneov{\epsilon}\ln\frac{\mu^2}{Q^2}\right]\delta(1-\tau)-\frac{2}{\epsilon}\paren{\oneov{1-\tau}}_+\bigg\},
\end{equation}
and the $\mu$ anomalous dimension is
\begin{equation}\label{eq:35eq114}
\gamma_I^{\text{DY}}(1-\tau,\mu)=\frac{4\alpha_s C_F}{\pi }\left[\ln\frac{\mu}{Q}\delta(1-\tau)-\paren{\oneov{1-\tau}}_+\right],
\end{equation}
through which we can resum the logarithms brought in by the interference effect between the two protons.  Note the appearance of the cusp in $\gamma^{\rm DY}_I$, which resums double logarithm Sudakov logarithms.  This is in contradistinction to DIS where no cusp appears.  To $\mathcal{O}(\alpha_s)$, the renormalized interference factor is
\begin{align}\label{IDYrenorm}
I^{(DY)}(1-\tau;\mu)&=2Q\delta(1-\tau)+2Q\frac{\alpha_s C_F}{\pi}\bigg(\left[\frac{1}{2}\ln^2\frac{\mu^2}{Q^2}-\frac{\pi^2}{12}\right]\delta(1-\tau) \nn \\
&+4\left(\frac{\ln(1-\tau)}{1-\tau}\right)_+ +2\ln\frac{Q^2}{\mu^2}\left(\frac{1}{1-\tau}\right)_+ \bigg).
\end{align}

This interference factor, only present in the $\tau\to 1$ limit, contains information of the long-distance behavior of hadron-hadron scattering.  In contrast, away from the endpoint, at moderate $\tau$ the protons are closer together and the collision is well-represented by the perturbative scattering of partons.  We shall see in the next chapter that long-distance hadron scattering can also be considered to involve exchange a pion -- a set of degrees of freedom completely different from the parton-level theory we have worked with here.


\subsubsection{Comparing to Perturbative QCD Results}
\label{sec:I.5.3.2.4}

The hard function $H^{\text{DY}}$ from \cite{Becher:2007ty} is
\begin{equation}\label{eq:100b}
H^{\text{DY}}(Q,\mu)=1+\frac{\alpha_sC_F}{\pi}\left(-\frac12\ln^2\frac{\mu^2}{Q^2}-\frac32\ln\frac{\mu^2}{Q^2}-4+\frac{7\pi^2}{12}\right)  \,.
\end{equation}
Taking $N_c=3$, and inserting the Drell-Yan collinear and soft functions with Eq.\,\eqref{eq:100b} into \eq{effcrosssection}, we find at $\mathcal{O}(\alpha_s)$ SCET Drell-Yan cross section is,
\begin{align}
\left(\frac{d\sigma}{dQ^2}\right)_{\!\text{eff}}&=m_0^2\delta_{n\cdot\bar p_{\bar n},Q}\delta_{\bar n\cdot p_n,Q}\left(\frac{4\pi \alpha^2}{9Q^4}\right)\frac{\alpha_sC_F}{\pi}\bigg\{\left[\frac32\ln\frac{Q^2}{m_g^2}-\frac{5}{2}-\frac{\pi^2}{6}\right]\delta(1-\tau) \nn \\
& +4\left(\frac{\ln(1-\tau)}{(1-\tau)}\right)_++2\ln\frac{Q^2}{m_g^2}\left(\frac{1}{1-\tau}\right)_+\bigg\} .\label{eq:100c}
\end{align}
To $\mathcal{O}(\alpha_s)$ in QCD, the quark contribution to the DY cross section is (\cite{field1989applications})
\begin{align}
\frac{d\sigma}{dQ^2}&=m_0^2\delta_{n\cdot\bar p_{\bar n},Q}\delta_{\bar n\cdot p_n,Q}\frac{4\pi}{9}\frac{\alpha^2}{Q^4}\int_\tau^1\frac{dx_a}{x_a}\int_{\tau/x_a}^1\frac{dx_b}{x_b}\bigg\{G_{p\to q}^{(0)}(x_a)G_{p\to q}^{(0)}(x_b)\nn\\
&\times\bigg(\frac{\sigma_{\text{tot}}^{\text{DY}}}{\sigma_0}\delta(1-z)+\frac{\alpha_s}{\pi}P_{q\to qg}(z)\ln\frac{Q^2}{m_g^2}+2\alpha_s f^{q\text{ DY}}(z)\bigg)\bigg\}\,,\label{eq:100d}
\end{align}
where $z=\tau/(x_ax_b)$, $G_{p\to q}^{(0)}(x_a), G_{p\to q}^{(0)}(x_b)$ are zero-order PDFs, and
\begin{align}
P_{q\to qg}(\tau)&=C_F\left(\frac{1+z^2}{(1-z)_+}+\frac{3}{2}\delta(1-z)\right),\nn\\
\alpha_s f^{q\text{ DY}}(\tau)&=\frac{\alpha_sC_F}{\pi}\bigg\{(1+z^2)\left(\frac{\ln(1-z)}{1-z}\right)_+-\left(\frac{1+z^2}{1-z}\right)\ln z\nn\\
&-(1-z)-\frac{\pi^2}{3}\delta(1-z)
\bigg\}\,,\nn\\
\frac{\sigma_{\text{tot}}}{\sigma_0}&=1+\left(\frac{8\pi}{9}-\frac{7}{3\pi}\right)\alpha_s+\ldots
\label{eq:100e}
\end{align}
In the endpoint, $z\to 1$ the perturbative QCD Drell-Yan cross section at $\mathcal{O}(\alpha_s)$ becomes
\begin{align}
\frac{d\sigma}{dQ^2}&=m_0^2\delta_{n\cdot\bar p_{\bar n},Q}\delta_{\bar n\cdot p_n,Q}\left(\frac{4\pi\alpha^2}{9Q^4}\right)\frac{\alpha_sC_F}{\pi}
\nn \\ &\times
\bigg\{\left[\frac32 \ln\frac{Q^2}{m_g^2}-\frac{7}{4}\right]\delta(1-\tau)
+4\left(\frac{\ln(1-\tau)}{1-\tau}\right)_+ + 2 \ln\frac{Q^2}{m_g^2} \left(\frac{1}{1-\tau}\right)_++4\bigg\} \,.
\label{eq:100g}
\end{align}
Comparing Eq.\eqref{eq:100g} and Eq.\eqref{eq:100c}, we arrive at the same conclusion as for DIS, that the $\text{SCET}_{\text{II}}$ hadronic structure function reproduces all the low energy physics of the perturbative QCD results in the endpoint region up to constant coefficients of  $\delta(1-\tau)$, which is regularization scheme dependent.  As in DIS, this discrepancy is expected since the SCET and QCD calculations use different regularization schemes.

We can compare regularization schemes by repeating this comparison with our Delta-regulator results.  In the Delta-regulation scheme, the renormalized $n$- and $\bar n$-collinear functions are
\begin{align}
{C}_n^{\text{DY}-R}&=\left(-\frac{\alpha_sC_F}{\pi}\right)\delta(k^-)\bigg[\frac{\pi^2}{12}-\frac34+\frac12\ln^2\frac{\mu^2}{m_g^2}-\ln\frac{\mu^2}{m_g^2}\ln\frac{\mu^2}{\Delta_2}+Li_2\left(1-\frac{\delta_1\delta_2}{m_g^2}\right)
\nn\\&
-Li_2\left(\frac{\Delta_2}{m_g^2}\right)+\ln\frac{\Delta_2}{m_g^2}\left(\frac{\Delta_2/m_g^2}{\frac{\Delta_2}{m_g^2}-1}-\ln\left(1-\frac{\Delta_2}{m_g^2}\right)\right)\bigg],\label{eq:35eq149}\\
{C}_{\bar n}^{\text{DY}-R}&=\left(-\frac{\alpha_s C_F}{\pi}\right)\delta(k^+)\bigg[\frac{\pi^2}{12}-\frac34+\frac12\ln^2\frac{\mu^2}{m_g^2}-\ln\frac{\mu^2}{m_g^2}\ln\frac{\mu^2}{\Delta_1}+Li_2\left(1-\frac{\delta_1\delta_2}{m_g^2}\right)\nn\\
&-Li_2\left(\frac{\Delta_2}{m_g^2}\right)+\ln\frac{\Delta_1}{m_g^2}\left(\frac{\Delta_1/m_g^2}{\frac{\Delta_1}{m_g^2}-1}-\ln\left(1-\frac{\Delta_1}{m_g^2}\right)\right)\bigg]\,.\label{eq:35eq150}
\end{align}
The renormalized soft function is
\begin{align}
{S}^{\text{DY}-R}&=-\frac{\alpha_sC_F}{\pi Q}\bigg(\bigg[\ln\frac{\mu^2}{m_g^2}\ln\frac{\mu^2}{\delta_1\delta_2}-\frac12\ln^2\left(\frac{\mu^2}{m_g^2}\right)+\frac{\pi^2}{12}-Li_2\left(1-\frac{\delta_1\delta_2}{m_g^2}\right)\bigg]\delta(1-\tau) \nn\\
&-\frac12\ln^2\frac{Q^2}{m_g^2}\delta(1-\tau)-\ln\frac{m_g^2}{Q^2}\left(\frac{1}{1-\tau}\right)_+ +2\left(\frac{\ln (1-\tau)}{1-\tau}\right)_+\bigg)\,.\label{eq:35eq151}
\end{align}
Inserting Eq.~\eqref{eq:35eq149}, Eq.~\eqref{eq:35eq150} and Eq.~\eqref{eq:35eq151} with the hard function \eq{100b} into the DY hadronic tensor \req{DYhadtenSCET2}, we obtain
\begin{align}
(W^{\mu\nu})_{\text{DY}-\Delta}^{\text{eff}}&=\frac{\alpha_sC_F}{\pi}\bigg[\left(\frac{2}{1-\tau}\right)_+\ln\frac{Q^2}{m_g^2}+4\left(\frac{\ln(1-\tau)}{1-\tau}\right)_+\nn\\
&+\left(\frac32\ln\frac{Q^2}{m_g^2}+\frac32-\frac{\pi^2}{4}\right)\delta(1-\tau)\bigg]\,,\label{eq:35eq152}
\end{align}
where we replace $z$ with $(1-\tau)$. We can clearly see that Eq.\eqref{eq:35eq152} reproduces the perturbative QCD result up to the constant coefficient of $\delta(1-\tau)$ which is due to the regularization scheme.

\subsubsection{Summary of Drell-Yan Endpoint Resummation}

We have studied the deep inelastic scattering and Drell-Yan processes in the endpoint $x\to 1$ ($\tau\to1$) region using both the $\eta$ rapidity regulator and the $\Delta$-regulator.  In this region, both DIS and DY exhibit a large Sudakov logarithm, arising as the collinear and soft degrees of freedom approach the same invariant mass scale, which becomes much smaller than the collision center-of-mass scale.  Using soft collinear effective theory and the covariant rapidity regulator to separate collinear and soft degrees of freedom, we see this large logarithm as a logarithm of the ratio of collinear and soft rapidity scales.  We showed how the logarithm of rapidity scales corresponds to the well-known threshold logarithm by transforming the result to Mellin space where it is seen as a divergence going as $\ln N$ for $N\gg 1$.  We also confirmed our previous results for DIS by comparing the same calculations in the $\Delta$ regulator scheme and verified agreement with the perturbative QCD result in the limit $x\to 1$.  However, it is notable that the $\Delta$ regulator does not provide a convenient mechanism to resum the logarithmic enhancements, which have been argued to be operative even well away from the true endpoint.

Although separating the parton distribution function in the endpoint region into collinear and soft factors brings in dependence on an infrared scale, the rapidity factorization is rigorous, as proven by its successfully reproducing the standard results.  Indeed, the factorization cures the problematic large logarithm, which would otherwise spoil the convergence of the effective theory expansion in the threshold region.  From this point of view, rapidity factorization (and summation) is necessary, even if the running must at some point be re-absorbed into the function chosen to model the PDF at the hadronic scale.  We remark that our definition of the PDF smoothly goes over to the traditional definition away from the endpoint, and we will fit the experimentally-determined PDF to our factorized form in a future publication.  The tangible gain from our analysis is that the running in rapidity we identify may help explain the steep fall off in the PDFs near the endpoint.

We demonstrated that this rapidity factorization works more generally by performing the same analysis on DY processes.  We resummed the single large rapidity logarithm and compared the resulting factorized collinear functions to the definition of the endpoint-region PDF we obtained in DIS.  Morevoer, we verified the results by calculating again in $\Delta$-regulator scheme and by comparing to the perturbative QCD result.  The success of the resummation establishes that rapidity factorization of the PDF is valid also in DY processes, and the parton luminosity function can be related to the PDFs measured in DIS.  

An interesting outcome of separating the DY collinear functions into soft and collinear factors is that the soft radiation necessarily couples to both incoming $n$ and $\bn$ protons.  Consequently there is only a single soft function and the $n$ and $\bn$ parton distribution functions can only be exhibited as separate factors by defining an interference factor.  The hadronic structure function in $\scetii$ has the form
\begin{align}
W^{\rm eff} =\frac{2\pi}{QN_c}H( Q ;\mu)   
 \int dx d\bar x  \,f_q^{\bar ns}(\frac{1-x}{x};\mu)f_{q'}^{ns}(\frac{1-\bar x}{\bar x};\mu)I^{\rm (DY)}_{\tau\to1}(1-\tau;\mu)\,,~~
 \end{align}
in which each $\phi(q;m)$ is a PDF defined to be identical to the PDF determined from DIS in the endpoint region, and $I^{\rm (DY)}_{\tau\to1}(1-\tau;\mu)$ is the interference factor, whose renormalized form is given in \req{IDYrenorm}.  Calculating its running proves that $I^{\rm (DY)}_{\tau\to 1}$ is a nontrivial function and is independent of the rapidity scale.   The running of the interference factor sums Sudakov logarithms associated with the threshold region, but does not bring in any infrared scale dependence.  Understanding it more thoroughly thus appears a promising route to understanding the transition to the elastic limit of hadron-hadron scattering.


\subsection{Application of \texorpdfstring{$x\to 1$}{x to 1} Drell-Yan Process: Vector Boson Fusion on LHC}
\label{sec:I.5.3.2.5}

The general method we have just introduced for the endpoint of the Drell-Yan process can also be applied to other processes in hadron-hadron colliders such as vector boson fusion (VBF).\footnote{This section is based on unpublished work in collaboration with S. Fleming, W. K. Lai, and A. Leibovich.}  At the LHC, VBF into the Higgs boson is an important production mode to search for low-mass Higgs in combination with its subsequent decay into $W+W$ ($W^*$) with both $W$ bosons decaying leptonically into a lepton and a neutrino. After a quark in each initial state proton emits the virtual $W$ or $Z$, the two protons become jets at small angles to the proton beams. These jets are used as `tag jets'. The indicident $W$s or $Z$s fuse into a Higgs boson that subsequently decays (\cite{Han:1992hr}).  In sum,
\begin{equation}\label{eq:56eq1}
\begin{array}{l}
u+d\to d+H+u\\
H\to W+W(W^*)\to \ell_1+\nu_1+\ell_2+\bar \nu_2\,.
\end{array}
\end{equation}
where $u$ and $d$ denote the up and down quark respectively and $\ell_1,\ell_2,\nu_1,\nu_2$ denote the leptons.
Because the neutrino 3-momenta are generally not measured, the final state cannot be fully reconstructed. In the experimentally-relevant case that the Higgs mass is less than 180 GeV, the $W+W(W^*)$ decay products of the Higgs have small momenta in the Higgs center-of-mass frame.  By assuming the momenta are zero, the final state neutrinos can be reconstructed and useful kinematic cuts can be derived from the reconstruction.

Because the Higgs mass is much larger than the beam invariant mass, $m_H\gg \LQCD$, the endpoint of this process is accessed when the soft QCD radiation has total invariant mass $\sim\LQCD^2$ (\cite{Fleming:2016nhs}).

\nt{Kinematics.}
The center-of-mass energy of the colliding protons is denoted $s=(p_1+p_2)^2$, as usual, and we define the positive squared-momentum transfer in each vector boson $Q_i^2=-q_i^2$ for $i=1,2$.  The partonic momentum fractions are $w_i=\xi_i p_i$.  Finally defining Bjorken $x$ for each vector boson  $x_i=\frac{Q_i^2}{2p_i\cdot q_i}$ allows expressing the endpoint condition as the limit
\begin{align}
m_{X_i}^2&=(p_i+q_i)^2=\frac{Q_i^2}{x_i}(1-x_i)\to 0\quad\text{ for }x_i\to 1.\label{eq:57eq2}
\end{align}
$x_i\to 1$ forces the $i$th jet to have a small mass ($m_{j_i}\sim \lqcd$). In this limit, $\xi_i\to 1$ as well. Therefore emissions from the $i$th quark are either soft or collinear to the final $i$th quark line, and the invariant mass of the $i$th beam remnant is soft.  
In other words, for both $x_1,x_2\to 1$ we have two energetic light jets, a Higgs plus everything else soft expressed by
\begin{equation}\label{eq:57eq3}
E_H+E_{j_1}+E_{j_2}\to \sqrt{s}.
\end{equation}

\begin{figure}
\centering
\includegraphics[width=.75\textwidth]{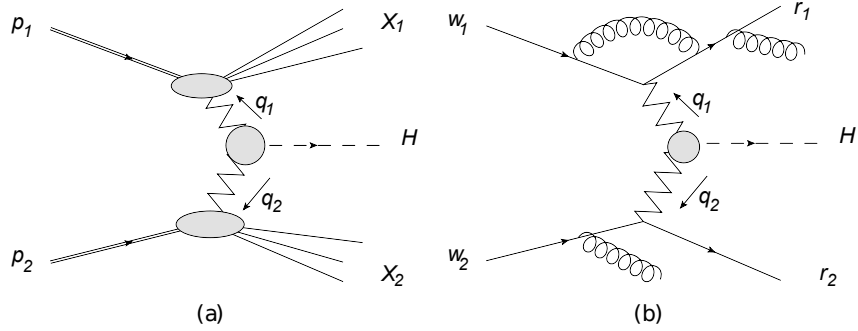}
\caption{VBF diagrams: (a) hadron level, (b) parton level.  \label{fig:57fig1}}
\end{figure}

\nt{Scale separation in SCET.} Consider one of the two DIS vertices embedded in the VBF diagram (Fig.\,\ref{fig:57fig1}). Let $P$ be the proton momentum, $p$ the initial parton momentum. For $x\to 1$, $m_X^2=(P+q)^2=Q^2\frac{(1-x)}{x}+m_p^2\approx Q^2(1-x)\sim Q^2\lambda^2$ with $\lambda\sim \sqrt{1-x}$. Tentatively we put $Q\lambda^2\sim \lqcd$. So
\begin{align}
(p+q)^2&=p^2+2p\cdot q+q^2=2p\cdot q-Q^2+\cO(\lqcd^2)\nn\\
&=2\xi P\cdot q-Q^2+\cO(\lqcd Q)=Q^2\paren{\frac{\xi}{x}-1}+\cO(\lambda^2Q^2).\label{eq:57eq4}
\end{align}
Therefore let $p'$ be collinear to $p+q$, which is a lightcone direction $n'$ distinct from the lightcone directions determined by the initial proton, $n$ or $\bar n$. $p$ is collinear to $P$, we have
\begin{align}
p&=\xi\bn\cdot P\frac{n}{2}+n\cdot p\frac{\bn}{2}+p_\perp\sim Q\paren{1,\frac{\lqcd^2}{Q^2},\frac{\lqcd}{Q}}\nn\\
p'&=\bn'\cdot p'\frac{n'}{2}+n'\cdot p'\frac{\bn'}{2}+p_\perp'\sim Q(1,\lambda^2,\lambda)'.\label{eq:57eq5}
\end{align}
Also, $q\sim Q(1,1,1)\sim Q(1,1,1)'$. Any radiation $k$ must obey
\begin{equation}\label{eq:57eq6}
k=p+q-p'\sim Q(1,\lambda^2,\lambda)'\text{ or }Q(\lambda^2,\lambda^2,\lambda^2)'.
\end{equation}
For the case $x_1,x_2\to 1$, the ultrasoft radiation connecting the two DIS vertices can be factored out.

\begin{figure}[!h]
\centering
\includegraphics[width=.45\textwidth]{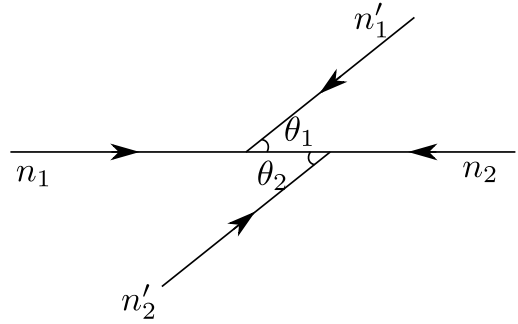}
\caption{Proton-proton collision kinematics.  \label{fig:57afig1}}
\end{figure}

\nt{Factorization in SCET.}
We denote $n_1$ and $n_2$ as two initial jet directions, $n_1'$ and $n_2'$ as two final jet directions, $\theta_1$ and $\theta_2$ are the angles between $n_1, n_1'$ and $n_2, n_2'$ respectively. In the center of mass frame of two initial protons, as shown in Fig.\,\ref{fig:57afig1}, we have
\begin{align}
n_1^\mu&=(1,0,0,1)=\bar n_2^\mu,\\
n_2^\mu&=(1,0,0,-1)=\bar n_1^\mu,\\
(n_1')^\mu&=(1,0,-\sin\theta_1,-\cos\theta_1),\\
(n_2')^\mu&=(1,0,\sin\theta_2,\cos\theta_2)\label{eq:57aeqfourdirections}
\end{align}
We decompose the transferred momentum as
\begin{align}
q_1^\mu&=p_1^{\prime\mu}-p_1^\mu=(\bar n_1\cdot p_1'-\bar n_1\cdot p_1,n_1\cdot p_1'-n_1\cdot p_1,p_{1\perp}'-p_{1\perp})\\
q_2^\mu&=p_2^{\prime\mu}-p_2^\mu=(\bar n_2\cdot p_2'-\bar n_2\cdot p_2,n_2\cdot p_2'-n_2\cdot p_2,p_{2\perp}'-p_{2\perp})
\end{align}

In QCD, the hadronic tensor arising from squaring the matrix element is
\begin{align}
W^{\mu\nu\alpha\beta}=&\int d^4x \int d^4 y\int d^4 z e^{-iq_1\cdot x}e^{-iq_2\cdot y}e^{+iq_1\cdot z}\nn\\
&\times\left\langle p_1p_2\Bigg\vert \bar T[J_2^{\dagger \beta}(x)J_1^{\dagger \alpha}(y)]T[J_1^\mu(z) J_2^\nu(0)]\Bigg\vert p_1p_2\right\rangle. \label{eq:57aeqQCDWtensor}
\end{align}
First we match QCD to $\text{SCET}_{\text{I}}$ at the scale $\mu_q\sim Q\sim 14\:\text{TeV}$ for the current LHC collisions.
The SCET current is,
\begin{align}
J^\mu(x)&\to\sum_{w,\bar w}C(w,\bar w;\mu,\mu_q)(e^{-\frac{i}{2}w_n\cdot x}e^{\frac{i}{2}\bar w_n'\cdot x}\bar\chi_{n',\bar\omega}\gamma_\perp^\mu\chi_{n,w}+h.c.),\nn\\
J^{\mu\dagger}(x)&\to \sum_{w,\bar w}C^*(w,\bar w;\mu,\mu_q)(e^{\frac{i}{2}w_n\cdot x}e^{-\frac{i}{2}\bar w_n'\cdot x}\bar\chi_{n,w}\gamma_\perp^\mu\chi_{n'\cdot \bar w}+h.c.),\label{eq:57aeqSCETIcurrent}
\end{align}
where $\chi_{n,w},\chi_{n',w}, \bar \chi_{n',\bar w}$ and $\bar\chi_{n,w}$ are SCET fields, and $C(w,\bar w;\mu,\mu_q)$, $C^*(w,\bar w;\mu,\mu_q)$ are Wilson coefficients encoding all QCD information from $Q$ to proton mass.

\begin{figure}[!h]
\centering
\includegraphics[width=.45\textwidth]{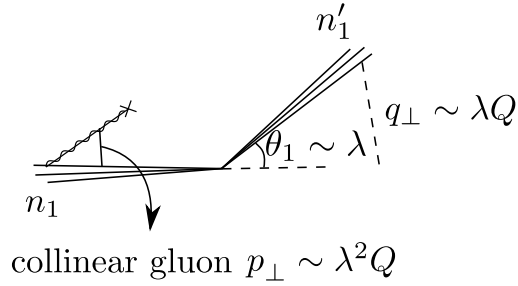}
\caption{Collinear gluons from the initial jet $n_1$ decoupled with final jet $n_1'$}
\label{fig:57afig2}
\end{figure}

Inserting the SCET currents in the hadronic tensor, we obtain
\begin{align}
W^{\mu\nu\alpha\beta}
&=\sum_{\substack{w_1,\bar w_1,w_2,\bar w_2,\\\underline{w}_1,\underline{\bar w},\underline{w}_2,\underline{\bar w}_2} } 
\int d^4x \int d^4y \int d^4z (\delta_{Q,w_2}\delta_{Q,w_1}\delta_{Q,\underline{w}_1})(\delta_{Q,p_1\cdot\bar n_1}\delta_{Q,p_2\cdot\bar n_2})\nn\\
&\times\delta_{Q,p_1'\cdot\bar n_1'}\,\delta_{Q,p_2'\cdot\bar n_2'}\,\delta_{\bar w_1,\underline{\bar w}_1}\,\delta_{\bar w_2,\underline{\bar w}_2}e^{\frac{i}{2}x\cdot n_2'\cdot\cos\theta_2Q}e^{\frac{i}{2}Q\cos\theta_1n_1'\cdot y}e^{-\frac{i}{2}n_2\cdot zQ}
\nn\\
&
\times e^{iq_1^\perp\cdot x_\perp}e^{iq_2^\perp \cdot y_\perp}e^{-iq_1^\perp \cdot z_\perp}\Bigg\langle p_1p_2\Bigg\vert\bar T\Bigg\{ C_2^*(Q,\bar w_2;\mu,\mu_q)e^{-\frac{i}{2}\bar w_2n_2'\cdot x}\bar\chi_{n_2Q}\gamma_\perp^\mu\chi_{n_2'\bar w_2}\nn\\
&
\times C_1^*(Q,\bar w_1;\mu,\mu_q)e^{-\frac{i}{2}\bar w_1\cdot n_1'\cdot y}\bar\chi_{n,w_1}\gamma_\perp^\nu\chi_{n_1'\bar w_1}\Bigg\}T\Bigg\{ C_1(Q,\underline{\bar w}_1;\mu,\mu_q)e^{\frac{i}{2}\underline{\bar w}_1n_1'\cdot z}
\nn\\ &
\times \bar\chi_{n_1'\underline{\bar w}_1}\gamma_\perp^\alpha\chi_{n_1 Q} \,C_2(\underline{w}_2,\underline{\bar w}_2;\mu,\mu_q)\bar\chi_{n_2'\underline{\bar w}_2}\gamma_\perp^\beta\chi_{n_2w_2}\Bigg\}\Bigg\vert p_1p_2\Bigg\rangle , \label{eq:57aeqWtensorSCETIfull}
\end{align}
where we first integrate over the large components of $n_2\cdot x, n_1\cdot y$ and $n_1\cdot z$, to obtain the label-conserving $\delta$-functions $(\delta_{Q,p_1\cdot\bar n_1}\delta_{Q,p_2\cdot\bar n_2})$ and $(\delta_{Q,p_1'\cdot\bar n_1'}\delta_{Q,p_2'\cdot\bar n_2'})$ for initial and final states and $(\delta_{\bar w_1,\underline{\bar w}_1}\delta_{\bar w_2,\underline{\bar w}_2})$ for momentum conservation. Summing over $w_2,\underline{\bar w}_2,w_1,\underline{w}_1,\underline{\bar w}_1$, we have
\begin{align*}
W^{\mu\nu\alpha\beta}&=\sum_{\bar w_1,\bar w_1, \underline{w_2}}\int d^4x \int d^4y \int d^4 z \delta_{Q,p_1\cdot\bar n_1}\,\delta_{Q,p_2\cdot\bar n_2}\,\delta_{Q,p_1'\cdot \bar n_1'}\,\delta_{Q,p_2'\cdot \bar n_2'}\\
&\times \Bigg\langle p_1p_2\Bigg\vert \bar T\Bigg\{ C_2^*(Q,\bar w_2;u,u_q)e^{-\frac{i}{2}x\cdot n_2'(\bar w_2-Q\cos\theta_2)}\bar\chi_{n_2,Q}\gamma_\perp^\mu\chi_{n_2',\bar w_2}\\
& \times C_1^*(Q,\bar w_1;\mu,\mu_q)e^{-\frac{i}{2}n_1'\cdot y(\bar w_1-Q\cos\theta_1)}\bar\chi_{n_1,Q_1}\gamma_\perp^\nu\chi_{n_1',\bar w_1}\Bigg\}\\
& \times T\Bigg\{ C_1(Q,\bar w_1;\mu,\mu_q)e^{\frac{i}{2}n_1'\cdot z(\bar w_1-Q\cos\theta_1)}\bar\chi_{n_1',\bar w_1}\gamma_\perp^\alpha\chi_{n_1,Q}\\
& \times C_2(\underline{w_2},\bar w_2;\mu,\mu_q)\bar\chi_{n_2',\bar w_2}\gamma_\perp^\beta \chi_{n_2,Q}\Bigg\}\Bigg\vert p_1p_2\Bigg\rangle e^{iq_1^\perp\cdot x_\perp}e^{iq_2^\perp\cdot y_\perp}e^{-iq_1^\perp\cdot z_\perp}.
\end{align*}
So far, we have integrated out the large proton collision energy scale $Q\sim $7\:TeV. Next we integrate out the vector boson emission scale $\tilde Q=\lambda Q\sim 100$\,GeV by integrating over the large components of $x_\perp,y_\perp,z_\perp$, 
\begin{align*}
W^{\mu\nu\alpha\beta}&=\sum_{\bar w_1,\bar w_2,\underline{w}_2}\int\! d^4x\int\! d^4y\int\! d^4z\,\delta_{Q,p_1\cdot \bar n_1}\,\delta_{Q,p_2\cdot \bar n_2}\,\delta_{Q,p_1'\cdot \bar n_1'}\,\delta_{Q,p_2'\cdot\bar n_{2'}},\delta_{q_1^\perp,\tilde Q}\,\delta_{q_2^\perp \tilde Q} \nn \\
&\times \Bigg\langle p_1p_2\Bigg\vert \bar T\Bigg\{C_2^*(Q,\bar w_2;\mu,\mu_q)e^{-\frac{i}{2}x\cdot n_1(\bar w_2/\cos\theta_2-Q)}\bar\chi_{n_2,Q}\gamma_\perp^\mu \chi_{n_2',\bar w_2}\\
&\times C_1^*(Q,\bar w_1;\mu,\mu_q)e^{-\frac{i}{2}y\cdot n_2(\bar w_1/\cos\theta_1-Q)}\bar \chi_{n_1,Q}\gamma_\perp^\nu\chi_{n_1'\bar w_1}\Bigg\}\\
&\times T\Bigg\{C_1(Q,\bar w_1;\mu,\mu_q)e^{\frac{i}{2}z\cdot n_2(\bar w_1/\cos\theta_1-Q)}\bar\chi_{n_1'\bar w_1}\gamma_\perp^\alpha\chi_{n_1,Q}\\
&\times C_2(Q,\bar w_2;\mu,\mu_q)\bar\chi_{n_2'\bar w_2}\gamma_\perp^\beta\chi_{n_2,Q}\Bigg\}\Bigg\vert p_1p_2\Bigg\rangle
\end{align*}
Integrating over the large components of $x\cdot n_1,y\cdot n_2,z\cdot n_2$, we have
\begin{align*}
W^{\mu\nu\alpha\beta}&=\sum_{\bar w_1,\bar w_2}\int\! d^4x \int\! d^4y\int\! d^4z\paren{\delta_{Q,p_1\cdot\bar n_1}\delta_{Q,p_2\cdot\bar n_2}}\nn\\
&\times \paren{\delta_{Q,p_1'\cdot\bar n_1'}\delta_{Q,p_2'\cdot \bar n_2'}\delta_{q_1^\perp\tilde Q}\delta_{q_2^\perp\tilde Q}}\paren{\delta_{Q\cos\theta_2\bar w_2}\delta_{Q\cos\theta_1\bar w_1}}\\
&\times\Bigg\langle p_1p_2\Bigg\vert \bar T\Bigg\{ C_2^*(Q,Q;\mu,\mu_q)\bar\chi_{n_2,Q}\gamma_\perp^\mu\chi_{n_2',Q}C_1^*(Q,Q;\mu,\mu_q)\bar\chi_{n_1,Q}\gamma_\perp^\nu\chi_{n_1',Q}\Bigg\}\\
&\times T\Bigg\{C_1(Q,Q;\mu,\mu_q)\bar\chi_{n_1'Q}\gamma_\perp^\alpha \chi_{n_1,Q}
C_2(Q,Q;\mu,\mu_q)\bar\chi_{n_2'Q}\gamma_\perp^\beta\chi_{n_2,Q}\Bigg\}\Bigg\vert p_1p_2\Bigg\rangle
\end{align*}
$\cos\theta_1,\cos\theta_2$ have entered Kronecker $\delta$s for label momentum conservation in the above result. We now power count $\cos\theta_1$, and $\cos\theta_2$ in terms of $\lambda$, and prove that we can safely drop $\cos\theta_1$ and $\cos\theta_2$ from the $\delta$ functions.  At this point, we have two scenarios,
\begin{description}
\item[Case 1:] $\theta_{1,2}\sim \lambda$, $\cos\theta_{1,2}=1-\theta_{1,2}^2/2=1-\cO(\lambda^2)$ so that we have
\begin{align*}
n_1'\cdot x&=n_1\cdot x(1-\lambda^2)+x_\perp\lambda\simeq n_1\cdot x+\lambda x_\perp+\ldots\\
n_2'\cdot x&=\bar n_1\cdot x(1-\lambda^2)+x_\perp\lambda\simeq \bar n_1\cdot x+\lambda x_\perp.
\end{align*}
\item[Case 2:] $\theta_{1,2}\sim \pi/2-\lambda$, and using elementary trigonometric identities $\cos\theta_{1,2}=\sin\lambda\simeq \lambda$ and $\sin\theta_{1,2}=\cos\lambda=1-\lambda^2/2$ so that we have,
\begin{align*}
n_1'\cdot x&=n_1\cdot x\lambda+x_\perp(1-\lambda^2)\simeq x_\perp+n_1\cdot x\lambda\\
n_2'\cdot x&\simeq x_\perp-\bar n_1\cdot x\lambda.
\end{align*}
\end{description}
However since $\theta_{1,2}=\arctan\frac{\tilde Q}{Q}\sim\oneov{70}\sim\lambda$, Case 2 is not consistent with the kinematics of the endpoint.  Dropping $\cos\theta_1,\cos\theta_2$ in the $\delta$s, we write out the separated matrix elements
\begin{align*}
W^{\mu\nu\alpha\beta}&= 
\int d^4x\int d^4y\int d^4z\paren{\delta_{Q,p_1\cdot\bar n_1}\delta_{Q,p_2\cdot \bar n_2}}\paren{\delta_{Q,p_1'\cdot\bar n_1}\delta_{Q,p_2'\cdot\bar n_2'}}\paren{\delta_{q_1^\perp,Q}\delta_{q_2^\perp,Q}}\nn\\
&\times\Bigg\vert C_2(Q;\mu,\mu_q)\Bigg\vert^2\Bigg\vert C_1(Q;\mu,\mu_q)\Bigg\vert^2\\
&\times\Bigg\langle p_1p_2\Bigg\vert \bar T\Bigg\{\left[\paren{\bar\chi_{n_2,Q}^\sigma}^i\paren{Y_{n_2}^\dagger\gamma_{\sigma\lambda\perp}^\mu \bar Y_{n_2'}}_{ij}\paren{\chi_{n_2'Q}^\lambda}^j(x)\right]\nn\\
&\times\left[\paren{\bar\chi_{n_1Q}^\rho}^l\paren{\bar Y_{n_1}^\dagger \gamma_{\rho\delta\perp}^\nu Y_{n_1'}}_{lm}\paren{\chi_{n_1'Q}^\delta}^m\right]\Bigg\}\\
&\times T\Bigg\{\left[\paren{\bar \chi_{n_1'Q}^\epsilon}^k\paren{Y_{n_1'}^\dagger\gamma_{\epsilon\xi\perp}^\alpha\bar Y_{n_1}}_{kn}\paren{\chi_{n_1Q}^\xi}^n(z)\right]\nn\\
&\times\left[\paren{\bar\chi_{n_2'Q}^\xi}^h\paren{\bar Y_{n_2'}^\dagger\gamma_{\xi\eta\perp}^\beta Y_{n_2}}_{hf}\paren{\chi_{n_2Q}^\eta}^f(y)\right]\Bigg\}\Bigg\vert p_1p_2\Bigg\rangle,
\end{align*}
where $k,n,h,f,i,j,l,m$ are color indices and $\sigma,\mu,\nu,\lambda,\rho,\delta, \epsilon,\xi,\eta$ are Lorentz indices.
Now we can safely separate the two initial protons into two beam functions, because we focus on the large-$x$ regime of the colliding protons, and there is no Glauber gluon exchange between them,
\begin{align*}
W^{\mu\nu\alpha\beta}&=\int d^4x\int d^4y \int d^4z \delta_{Q,p_1\cdot\bar n_1}\,\delta_{Q,p_2\cdot\bar n_2}\,\delta_{Q,p_1'\cdot\bar n_1}\delta_{Q,p_2'\cdot\bar n_2'}\nn\\
&\times\paren{\delta_{q_1^\perp Q}\delta_{q_2^\perp Q}}|C_2(Q;\mu,\mu_q)|^2\\
&\times\Bigg\langle p_1\Bigg\vert \bar T\left[ \paren{\bar\chi_{n,Q}^\rho}^l\paren{\bar Y_{n_1}^\dagger \gamma_{\rho\delta\perp}^\nu Y_{n_1'}}_{lm}\paren{\chi_{n_1'Q}^\delta}^m\right]\nn\\
&\times T\left[\paren{\bar \chi_{n_1'Q}^\epsilon}^k\paren{Y_{n_1}^\dagger\gamma_{\epsilon\xi\perp}^\alpha\bar Y_{n_1}}_{kn}\paren{\chi_{n_1Q}^\xi}^n(z)\right]\Bigg\vert p_1\Bigg\rangle\\
&\times\Bigg\langle p_2\Bigg\vert \bar T\left[ \paren{\bar\chi_{n_2Q}^\sigma}^i\paren{Y_{n_2}^\dagger \gamma_{\sigma\lambda\perp}^\mu\bar Y_{n_2'}}_{ij}\paren{\chi_{n_2'Q}^\lambda}^j(x)\right]\nn\\
&T\left[\paren{\chi_{n_2'Q}^\xi}^h\paren{\bar Y_{n_2'}^\dagger\gamma_{\xi\eta\perp}^\beta Y_{n_2}}_{hf}\paren{\chi_{n_2Q}^\eta}^f(y)\right]\Bigg\vert p_2\Bigg\rangle.
\end{align*}
Next we must prove that $n_1$ and $n_1'$, $n_2$ and $n_2'$ are not in the same collinear sectors so that we can factor the $n_1'$ and $n_2'$ final jets out of the beam functions for $p_1$ and $p_2$.  This requires  proving that there are no collinear gluons exchanged between $n_1$ and $n_1'$, $n_2$ and $n_2'$ respectively. Take $n_1$ and $n_1'$; the argument proceeds identically for $n_2$ and $n_2'$.  As shown in \fig{57afig2}, the collinear gluons in the initial beam $n_1$, obey the $\text{SCET}_{\text{I}}$ scaling $(p^+,p^-,p_\perp)\sim(Q,\lambda^4 Q,\lambda^2 Q)$, while the transverse momentum of the final jet $n_2'$ in the initial proton center of mass frame is $q_\perp\sim \tilde Q\sim \lambda Q$. Thus gluons with transverse momentum $p_\perp\sim \lambda^2 Q$ cannot couple with the final jet, which is diffracted by an angle $\theta\sim \lambda$. 

Having separated the $n_i$ and $n_i'$ directions, we can write the proton-proton hadronic tensor as
(including the previously suppressed sum over proton spins)
\begin{align*}
W^{\mu\nu\alpha\beta}&=
\frac14 g_\perp^{\mu\beta}g_\perp^{\alpha\nu}\int d^4x\int d^4y\int d^4z\paren{\delta_{Qp_1\cdot \bar n_1}\delta_{Qp_2\cdot \bar n_2}}\paren{\delta_{Qp_1'\cdot\bar n_1}\delta_{Qp_2'\cdot\bar n_2}}
\nn\\&\times
\paren{\delta_{q_1^\perp\tilde Q}\delta_{q_2^\perp\tilde Q}}|C_1|^2|C_2|^2
\\&\times 
\frac12\sum_\sigma\oneov{N_c}\Bigg\langle p_2\Bigg\vert \bar\chi_{n_2Q}(x)\frac{\bar\nslash_2}{2}\chi_{n_2Q}(0)\Bigg\vert p_2\Bigg\rangle \oneov{2}\sum_\sigma\oneov{N_c}\Bigg\langle p_1\Bigg\vert \bar\chi_{n_1Q}(y)\frac{\bar\nslash_1}{2}\chi_{n_1Q}(z)\Bigg\vert p_1\Bigg\rangle
\\&\times 
\oneov{N_c^2}\Bigg\langle 0\Bigg\vert \chi_{n_2'Q}(x)\frac{\bar \nslash_2'}{2}\bar\chi_{n_2'Q}(0)\chi_{n_1'Q}(y)\frac{\bar \nslash_1'}{2}\bar\chi_{n_1'Q}(z)\Bigg\vert 0\Bigg\rangle
\\&\times 
\Bigg\langle 0\Bigg\vert \bar T\left[\paren{Y_{n_2}^\dagger \bar Y_{n_2'}}(x)\paren{\bar Y_{n_1}^\dagger Y_{n_1'}}(y)\right] T\left[\paren{ Y_{n_1'}^\dagger \bar Y_{n_1}}(z)\paren{\bar Y_{n_2'}^\dagger Y_{n_2}}(0)\right]\Bigg\vert 0\Bigg\rangle.
\end{align*}
We rewrite the matter field as $\bar\chi_{n_1,w}=\bar\chi_n\delta_{Q,w}$ and shift the fields to the origin
\begin{align*}
\bar\chi_n(y)&=e^{\frac{i}{2}y^+p_-^r}e^{\frac{i}{2}y^-p_+^r}e^{-iy_\perp p_\perp}\bar\chi_n(0)e^{-\frac{i}{2}y^+p_-^r}e^{-\frac{i}{2}y^-p_+^r}e^{iy_\perp p_\perp},\\
\chi_n(y)&=e^{\frac{i}{2}y^+p_-^r}e^{\frac{i}{2}y^-p_+^r}e^{-iy_\perp p_\perp}\chi_n(0)e^{-\frac{i}{2}y^+p_-^r}e^{-\frac{i}{2}y^-p_+^r}e^{iy_\perp p_\perp}.
\end{align*}
Because we can only measure the final jets momentum, we project $n_1$ and $n_2$ directions onto $n_1'$ and $n_2'$, expressing $n_1$ and $n_2$ using $\cos\theta_i$ and $\sin\theta_i$.
We define the jet functions
\begin{align*}
\oneov{N_c}\Bigg\langle 0 \Bigg\vert \chi_{n_2'Q}(x)\frac{\bar \nslash_2'}{2}\chi_{n_2'Q}\Bigg\vert 0\Bigg\rangle &\equiv Q\delta(n_2'\cdot x)\delta^{(2)}(n_{2\perp}'\cdot x)\int dr_2'e^{-\frac{i}{2}r_2'\bar n_2'\cdot x}J_{\bar n_2'}(r_2';\mu)\\
\oneov{N_c}\Bigg\langle 0\Bigg\vert \chi_{n_1'Q}(y)\frac{\bar \nslash_1'}{2}\chi_{n_1'Q}(z)\Bigg\vert 0\Bigg\rangle &\equiv Q\delta(n_1'\cdot z)\delta(n_1'\cdot y)\delta^{(2)}(n_{1\perp}'\cdot z)\delta^{(2)}(n_{1\perp}'\cdot y)\\
&\times \int dr_1' e^{-\frac{i}{2}r_1'(\bar n_1'\cdot y-\bar n_1'\cdot z)}J_{\bar n_1'}(r_1;\mu)
\end{align*}
so that,
\begin{align}
W^{\mu\nu\alpha\beta}&=\frac14 g_\perp^{\mu\beta}g_\perp^{\nu\alpha}\int d\bar n_2'\cdot x \int \bar n_1'\cdot y\int \bar n_1'\cdot z\paren{\delta_{Qp_1\cdot\bar n_1}\delta_{Qp_2\cdot \bar n_2}}\paren{\delta_{Qp_1'\cdot\bar n_1'}\delta_{Qp_2'\cdot \bar n_2'}}\nn\\
&\times \paren{\delta_{q_\perp'\tilde Q}\delta_{q_\perp'\tilde Q}}|C_1|^2|C_2|^2\nn\\
&\times\sum_\sigma\oneov{2}\oneov{N_c}\Bigg\langle p_2\Bigg\vert \chi_{n_2Q}\exp\Bigg\{-\frac{i}{2}(\bar n_2'\cdot x)(n_2\cdot p_2)\cos\theta_2\Bigg\}\frac{\bar \nslash_2}{2}\chi_{n_2Q}\Bigg\vert p_2\Bigg\rangle\nn\\
&\times\sum_\sigma\oneov{2}\oneov{N_c}\Bigg\langle p_1\Bigg\vert \bar\chi_{n_1Q}\exp\Bigg\{-\frac{i}{2}\left[(\bar n_1'\cdot y-\bar n_1'\cdot z)\right](n_1\cdot p_1)\cos\theta_1\Bigg\}\frac{\bar \nslash_1}{2}\chi_{n_1Q}\Bigg\vert p_1\Bigg\rangle\nn\\
&\times Q\int dr_2'\exp\Bigg\{-\frac{i}{2}r_2'\bar n_2'\cdot x\Bigg\}J_{\bar n_2'}(r_2';\mu)\nn\\
&\times Q\int dr_1'\exp\Bigg\{-\frac{i}{2}r_1'(\bar n_1'\cdot y-\bar n_1'\cdot z)\Bigg\} J_{\bar n_1'}(r_1';\mu)\nn\\
&\times\Bigg\langle 0\Bigg\vert \bar T\left[Y_{n_2}^\dagger \bar Y_{n_2'}\right]e^{-\frac{i}{2}(\bar n_2'\cdot x)[n_2\cdot p_2\cos\theta_2+n_2'\cdot p_2')}T\left[\bar Y_{n_2'}^\dagger Y_{n_2}\right]\nn\\
&\times\bar T\left[\bar Y_{n_1}^\dagger Y_{n_1'}\right]e^{-\frac{i}{2}(\bar n_1'\cdot y)[n_1\cdot p_1\cos\theta_1+n_1'\cdot p_1']-\bar n_1'\cdot z}T\left[ Y_{n_1'}^\dagger \bar Y_{n_1}\right]\Bigg\vert 0\Bigg\rangle.\label{eq:57aeqhadronic-tensor-with-jet-function}
\end{align}
We change integration variables to 
$\bar n_1'\cdot u=\frac12 (\bar n_1'\cdot y+\bar n_1'\cdot z)$ and 
$\bar n_1'\cdot v=\frac12 (\bar n_1'\cdot y-\bar n_1'\cdot z)$
and define the collinear factors containing the information of initial protons
\begin{align*}
C_{\bar n_2}(k_2;\mu)&=\int\frac{d\bar n_2'\cdot x}{4\pi}e^{+\frac{i}{2}\bar n_2'\cdot xk_2}\Bigg\langle p_2\Bigg\vert \bar\chi_{n_2Q}(\bar n_2'\cdot x)\frac{\bar\nslash_2}{2}\chi_{n_2Q}\Bigg\vert p_2\Bigg\rangle \delta(Q,p_2\cdot\bar n_2)\\
&\times \delta_{\tilde Qq_{2\perp}}\delta(Q,p_2'\cdot\bar n_2')\\
&=\delta_{Qp_2\cdot \bar n_2}\delta_{Qp_2'\cdot \bar n_2'}\delta_{q_\perp^2\tilde Q}\Bigg\langle p_2\Bigg\vert \bar\chi_{n_2Q}(0)\frac{\bar \nslash_2}{2}\delta\paren{\cos\theta_2 in_2\cdot\partial_2+in_2'\cdot\partial_2-k_2}\chi_{n_2Q}\Bigg\vert p_2\Bigg\rangle\\
C_{\bar n_1}(k_1;\mu)&=\int\frac{d\bar n_1'\cdot v}{4\pi}e^{\frac{i}{2}\bar n_1'\cdot v k_1}\Bigg\langle p_1\Bigg\vert \bar \chi_{n_1Q}(\bar n_1'\cdot v)\frac{\bar \nslash_1}{2}\chi_{n_1 Q}\Bigg\vert p_1\Bigg\rangle \delta_{Qp_2\cdot n_2}\delta_{\tilde Q q_{2\perp}}\delta_{Qp_2'\cdot\bar n_2'}\\
&=\delta_{Qp_2\cdot\bar n_2}\delta_{Qp_2'\cdot \bar n_2'}\delta_{q_\perp^2 Q}\Bigg\langle p_1\Bigg\vert \bar\chi_{n_1Q}(0)\frac{\bar\nslash_1}{2}\delta\paren{\cos\theta_1 in_1\cdot\partial_1+in_1'\cdot\partial_1'-k_1}\chi_{n_1Q}\Bigg\vert p_1\Bigg\rangle.
\end{align*}
Integrating over $\bar n_2'\cdot x$ and $\bar n_1\cdot v$ in \eq{57aeqhadronic-tensor-with-jet-function}, the soft piece can be written
\begin{align*}
\Bigg\langle 0\Bigg\vert \ldots \Bigg\vert 0\Bigg\rangle
&=\Bigg\langle 0 \Bigg\vert \bar T\left[Y_{n_2}^\dagger \bar Y_{n_2'}\right] \delta\paren{in_2\cdot\partial_2\cos\theta_2+in_2'\cdot \partial_2'-\ell_2}T\left[\bar Y_{n_2'}^\dagger Y_{n_2}\right]\bar T\left[\bar Y_{n_1}^\dagger Y_{n_1'}\right]\\
&\times \delta\paren{in_1\cdot \partial_1\cos\theta_1+in_1'\cdot\partial_1'-\ell_1}T\left[ Y_{n_1'}^\dagger \bar Y_{n_1}\right]\Bigg\vert 0\Bigg\rangle\\
&=\int dle^{-\frac{i}{2}\ell_2\bar n_2\cdot x'}e^{-\frac{i}{2}\ell_1\bar n_1\cdot v}S(\ell;\mu).
\end{align*}
into a soft function $S(\ell;\mu)$ on the last line.

\begin{figure}
\centering
\includegraphics[width=.75\textwidth]{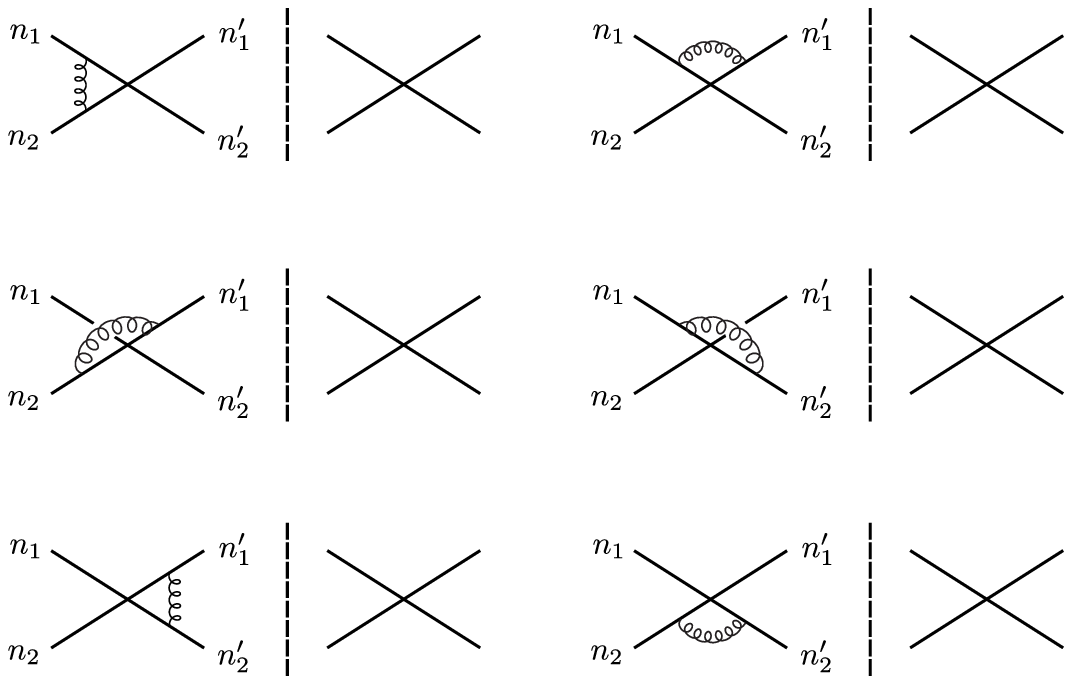}
\caption{Soft function for VBF logs Feynman diagrams (virtual)}
\label{fig:fig8a}
\end{figure}

The two collinear functions containing rapidity divergences are identical to those we calculated in the study of DIS, see \eq{35eq11} in Section \ref{sec:I.5.3.1.1}. The soft function, however, is different. The $\cO(\as)$ Feynman diagrams for the soft function are displayed in \figs{fig8a}{fig8b}

\begin{figure}
\centering
\includegraphics[width=.75\textwidth]{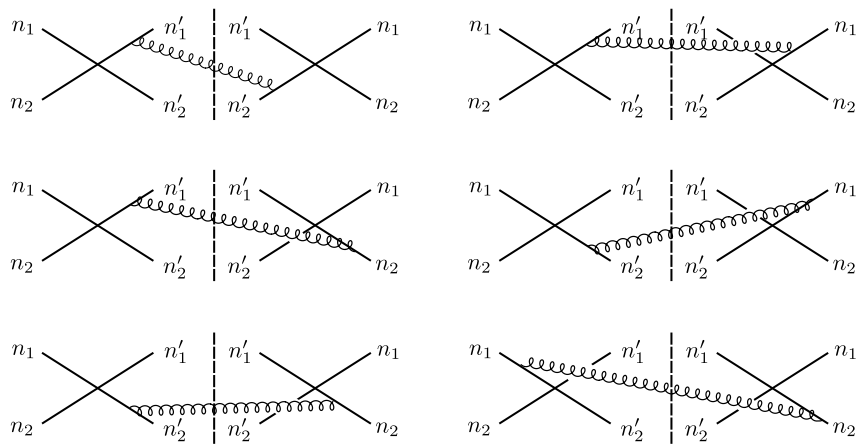}
\caption{Soft function for VBF logs Feynman diagrams (real)}
\label{fig:fig8b}
\end{figure}

\section{Summary of SCET results}

In this chapter, we have seen the difficulties arising in QCD perturbation theory in the form of logarithms of large ratios of physical scales.  We introduced the tools of soft-collinear effective theory to efficiently sum  several forms of these logarithms in inclusive and semi-inclusive processes.  Using the first version of the theory, $\SCETa$, on inclusive observables, we proved that the wide-angle, low-energy radiation accompanying jets is universal across electron-positron, electron-proton and proton-proton collisions.  This result improves the prediction of the hemisphere event-shape variables, used for precision extractions of $\alpha_s$ and controlling backgrounds for beyond standard model searches, to next-to-next-to-next-to-leading-logarithmic order.  Using the second version of the theory, $\SCETb$, we developed the resummation of large rapidity logarithms arising in the elastic limit of electron-proton and proton-proton scattering.  Our results provide a basis to sum large logarithms in near-threshold production mechanisms, such as vector boson fusion at the LHC, an example we began work on in the last section of this chapter.  Our endpoint resummation also offers a new avenue to study the transition from the semi-inclusive process (where partons may be considered the relevant degrees of freedom) to the exactly elastic limit, which is exclusive process (where pions may be the relevant degrees of freedom).

The systematic framework of SCET for summing large logarithms gradually reorganizes QCD perturbation theory to such a point that we may envision a precision description of fragmentation processes and thus of the transition from partonic to hadronic degrees of freedom.

%% file: chapter_2.tex
\chapter{Heavy Hadron Chiral Perturbation Theory and Hadronic Molecules}
\label{sec:II}

\section{Overview} 
In this chapter, I first provide the general background and motivation to study hadron-hadron interactions and hadronic molecules, as well as the motivation of heavy hadron chiral perturbation theory (HHChPT) and a non-relativistic effective field theory XEFT, derived from HHChPT to describe low energy interactions of $D$ mesons and a possible bound molecular state of $D$ mesons.  I provide the first complete derivation of isospin and $SU(2)$ spin-flavor symmetry breaking operators in the HHChPT lagrangian up to next-to-leading order in the expansion.  I explain the XEFT, generalize it to handle the possibilities that interactions between $D$-mesons yield a virtual bound state or resonance, and finally apply the XEFT to describe the $B$-meson decay in the $DD\pi$ channel, in which the possible $D$-meson molecular state was discovered.  These studies of heavy hadrons interacting with pions provide a unique view of the nuclear force.

\section{QCD, Hadrons and the Nuclear Force}
\label{sec:II.2.1}

Colliding hadrons at energy $\ll 1$ GeV do not access the quark and gluon degrees of freedom described by QCD.  Instead, low-momentum hadrons interact via the {\bf nuclear force}, which is a long-distance residual interaction between composite color-singlet states, analogous to the van der Waals force between neutral atoms and molecules.  
Here the momentum scale $\LQCD\simeq 1$ GeV emerges from the asymptotic freedom property as the scale where the coupling constant of QCD $\as$ approaches 1, and QCD becomes nonperturbative.  This strong coupling is believed to contribute to the observational fact that at long-distance quarks and gluons are confined inside hadrons, as either baryons (composed of 3 valence quarks) or mesons (composed of a valence quark-anti-quark pair).  

Extending the chemistry analogy, nuclei can be regarded as hadronic molecules of protons and neutrons, the simplest of which is the deuteron consisting of one proton and one neutron. We are applying the concept of a molecule exactly as used in chemistry, where atoms form a bound state by the residual electromagnetic force between electrically neutral particles.  Hadronic molecules are bound via the residual force resulting from incomplete cancellation of strong forces between colored particles. In contrast to the electromagnetic case where the binding energy can be calculated from first principles, the nuclear force is more intricate and far from being well-understood.  

Some properties of hadrons and their interactions in the low-momentum regime $\ll 1$ GeV can be computed by brute force on a discretized Euclidean space-time lattice, i.e. in lattice QCD, but the computations are costly and impractical for everyday nuclear structure physics.  Rather, following the principles of effective field theory and using symmetries inherited from the underlying theory QCD, we construct a theory of hadrons interacting with momenta $p\ll \LQCD$ and use it to study scattering of heavy hadrons.  The possibility of bound states or resonances is determined by the position of poles in the complex energy plane of the S-matrix.  By applying effective theory methods, we expect that insights from the specific heavy hadron systems we study here are more widely applicable to understand the nuclear force.

Hadrons are commonly classified by their valence quarks and their quantum numbers explained by a scheme known as the {\bf quark model}.  The quark model groups hadrons by representation of the approximate flavor $SU(3)$ symmetry, otherwise known as the `Eightfold Way'.  It received experimental support in the 1960s by successfully categorizing a great number of light hadrons that were being discovered at the time.  At present, the quark model remains a valid, widely-used, effective classification scheme for hadrons and has been incorporated as a part of the Standard Model of particle physics.  However, some experimentally-observed hadrons are not predicted by the quark model and are known as exotic hadrons.

\nt{Exotic Hadrons.}  
Exotic hadrons are described in one of two theoretical models: either color-singlets of $>3$ constituent quarks and gluons, or hadron-level molecules, i.e. subclusters of quarks and gluons such as meson molecules where subclusters are two color-singlet pairs of quark-antiquark. Hadronic molecules that are not nuclei can be viewed as exotic hadrons, for example {\it baryonium} consisting of a baryon and its antibaryon, or meson molecules which have been intensively sought. One way to explore the possibility of exotic hadrons is to search for poles in the S matrix with the corresponding exotic quantum numbers forbidden for ordinary hadrons.  

More practically, one can examine data to focus on specific channels, as recent experiments have suggested the existence of several such exotic hadrons.  
The following are some candidates of exotic hadrons:
\begin{description}
\item[X(3872)] Detected by Belle collaboration 2003, hypothetically a diquark or meson molecule
\item[Y(3940)] Charmonium spectrum predicted by theory fails to fit this particle
\item[Y(4140)] Detected by Fermilab in 2009
\item[Y(4260)] Detected on BaBar collaboration, hypothetically having a $\bar qqg$ composition
\item[Zc(3900)] Detected by Belle and BESIII
\item[Z(4430)$^-$] Detected by Belle collaboration in 2007 (\cite{Choi:2007wga}) and later confirmed by LHCb with 13.9$\sigma$ significance (\cite{Aaij:2014jqa}), interpreted as a tetraquark state
\item[$P_c^+(4380)$ and $P_c^+(4450)$] Observed in 2015 by LHCb (\cite{Aaij:2015tga}) and interpreted as two pentaquark states 
\end{description}
See also the summary table at \cite{WikiExoticHadrons}.
As opposed to baryonic molecules of which all nuclei in nature serve as examples, the existence of mesonic molecules has not been firmly established. The spin-0 mesons $a_0(980), f_0(980)$ have been suggested to be $K\bar K$ meson molecules (\cite{weinstein1983qq, weinstein1990k}), and charm meson molecules consisting of two charm mesons were proposed shortly following the discovery of the $c$ quark (\cite{voloshin1976hydronic,de1977molecular}).
Some of the best candidates to be mesonic molecules are the X(3872) and the Z(4430).  The X(3872), discovered in 2003 by Belle collaboration (\cite{Choi:2003ue}) and later confirmed by several other collaborations, has a mass of $3871.69\pm 0.17\,\text{MeV}$ (\cite{Olive:2016xmw}) and does not fit into quark model due to its exotic quantum numbers. Several theoretical models exist (\cite{Swanson:2006st}), including both mesonic molecules and tetraquark states. The quantum numbers were determined by LHCb to be $J^{PC}=1^{++}$ (\cite{Aaij:2013zoa}).

\subsection{Effective Field Theory Approach to Non-perturbative QCD}
\label{sec:II.2.2}
An important first step in establishing an effective theory is to identify relevant separations of scales.  We will use two.  First, the mass gap between the pion and the chiral-symmetry breaking scale leads to chiral perturbation theory, which is briefly reviewed in Appendix \ref{appx:ChPT}.  Second, the splitting between the non-perturbative scale of QCD and the masses of heavy quarks leads to heavy quark effective theory.


\subsubsection{Heavy Quark Effective Theory}
\label{sec:II.2.2.2}
For heavy quarks with masses much larger than the nonperturbative scale, taking the heavy quark mass $m_Q\to \infty$ is a good approximation under which the theory exhibits spin-flavor heavy quark symmetry and can be used to predict properties of hadrons with a single heavy quark.

Considering a meson containing a heavy and a light anti-quark, $Q\bar q$, with $m_Q\gg \lqcd, m_q\ll \lqcd$, the typical momentum transfer $\Delta p$ between the $Q$ and $\bar q$ is of order the scale of the nonperturbative gluon dynamics, $\Delta p\sim\lqcd$. Consequently, the heavy quark velocity is almost unchanged, since $\Delta v\sim\Delta p/m_Q\sim\lqcd/m_Q\ll 1$.

This motivates considering the heavy quark four-velocity $v^\mu$ as a label that does not change over time in the limit $m_Q\to \infty$. The dynamics of the meson reduce to the interaction of light degrees of freedom with a static external color source that transforms as a color triplet. In the limit $m_Q\to \infty$, $m_Q$ drops out of the Lagrangian and the dynamics is insensitive to the value of $m_Q$.  Therefore all heavy quarks interact in similar ways in mesons, and the dynamics is invariant under a change of the heavy flavor, leading to {\it heavy quark flavor symmetry}. On the other hand corrections of order $1/m_Q$ incorporate effects associated with finite heavy quark mass and do depend on the numerical values of the heavy quark masses. Consequently heavy quark flavor symmetry breaking effects are proportional to the difference, i.e. $1/m_{Q_i}-1/m_{Q_j}$ with $i,j$ labeling the two different flavors. 

The static heavy quark interacts with gluons (there are no quark-quark interactions in Lagrangian) only by its chromoelectric charge which is independent of spin, and therefore dynamics is unchanged under changes of the heavy quark spin, leading to {\it heavy quark spin symmetry}.  At $\cO(1/m_Q)$, the chromomagnetic moment gives the part of the interaction that is spin-dependent. In contrast to heavy quark flavor symmetry, heavy quark spin symmetry breaking is proportional to $1/m_Q$, not the heavy quark mass difference, as it is broken even in the case that all heavy quarks have identical mass. In the $m_Q\to\infty$ limit one can combine the $U(N_h)$ flavor symmetry and $SU(2)$ spin symmetry to embed into a $U(2N_h)$ spin-flavor symmetry under which the $2N_h$ states of $N_h$ flavors of heavy quarks with either spin up or spin down transform in the fundamental representation.

In the limit $m_Q\to \infty$, the QCD Lagrangian does not directly give heavy quark flavor-spin symmetry, and it is necessary to establish an EFT where this symmetry is manifest. This EFT, describing interaction of hadrons containing a single heavy quark, is known as heavy quark effective theory (HQET) and correctly describes physics of momentum transfers much less than $m_Q$. Instead of containing positive powers $m_Q$ term as QCD, the HQET Lagrangian is an expansion in $p_{typ}/m_Q$ for $p_{typ}\sim\lqcd$ the typical momentum transfer  and will only contain terms $\sim (p/m_Q)^n$, $n\geq 0$.

To setup the expansion, we write the heavy quark four-momentum as $p=m_Qv+k$ with $k$ the residual momentum measuring the off-shellness of the quark due to interactions (e.g. $k\sim\lqcd$ for a heavy quark inside a hadron). In the heavy quark limit $m_Q\to \infty$, the quark propagator is 
\begin{equation}\label{eq:69eq2.41}
i\frac{\slashed{p}+m_Q}{p^2-m_Q^2+i\epsilon}=i\frac{m_Q\slashed{v}+m_Q+\slashed{k}}{2m_Qv\cdot k+k^2+i\epsilon}\to i\frac{1+\slashed{v}}{2v\cdot k+i\epsilon}\,,
\end{equation}
which now contains a projection operator depending on heavy quark velocity: $(1+\slashed{v})/2$. In the rest frame of the heavy quark, $v^\mu\to (1,\vec 0)$ and the projection operator becomes
\begin{equation}\label{HQparticleprojector}
\frac{1+\slashed{v}}{2}\to\frac{1+\gamma^0}{2}\,,
\end{equation}
projecting onto the particle components of the four-component Dirac spinor.

The effective Lagrangian is established in terms of velocity-dependent fields $Q_v(x)$ which are related to the original quark field $Q$ at tree level by
\begin{equation}\label{eq:69eq2.43}
Q(x)=e^{-im_Qv\cdot x}[Q_v(x)+\mathfrak{Q}_v(x)]\,,
\end{equation}
with 
\begin{align}
Q_v(x)&=e^{im_Qv\cdot x}\frac{1+\slashed{v}}{2}Q(x)\,,\nn\\
\mathfrak{Q}_v(x)&=e^{im_Qv\cdot x}\frac{1-\slashed{v}}{2}Q(x)\,,\label{eq:69eq2.44}
\end{align}
where the phase factor $e^{im_Qv\cdot x}$ comes from the heavy quark momentum decomposition.  Plugging this decomposition into \eq{69eq2.43} shows that $Q_v$ is of leading order, whereas $\mathfrak{Q}_v$ is suppressed by powers of $m_Q^{-1}$.  The leading order $(p/m_Q)^0$ Lagrangian is just the $\bar Q_v(i\slashed{D})Q_v$ piece, which after using the projection operators again $(1+\slashed{v})/2$ simplifies to
\begin{equation}\label{eq:69eq2.45}
\cL=\bar Q_v(iv\cdot D)Q_v\,.
\end{equation}
This procedure of projecting out heavy degrees of freedom and using their equation of motion to eliminate them from the LO Lagrangian is equivalent to integrating them out of the generating functional in its path integral form.  We show this equivalence explicitly in Appendix \appx{D.1}.

There is no simple relation between the QCD field $Q$ and the EFT field $Q_v$ beyond tree level.  The EFT ensures that on-shell Green's functions agree with those in QCD to a given order in $p_{typ}/m_Q$ and $\as(m_Q)$.  
The Lagrangian \eq{69eq2.45} yields the $Q_v$ propagator
\begin{equation}\label{eq:69eq2.46}
\paren{\frac{1+\slashed{v}}{2}}\frac{i}{v\cdot k+i\epsilon}\,,
\end{equation}
in agreement with the previously derived result of Feynman rules in the $m_Q\to \infty$ limit.  
The gluon vertex is given by $-igT^A\gamma^\mu$ in the full theory and $-igT^Av^\mu$ in the EFT (due to $v\cdot D$ term in \eq{69eq2.45}).  To see that the EFT vertex is equivalent to the full theory vertex to $\cO(p/m_Q)$, recall that the vertex is accompanied by propagators on either side which are both proportional to $(1+\slashed{v})/2$.  Therefore we have the replacement
\begin{equation}\label{eq:69eq2.48}
\gamma^\mu\to \frac{1+\slashed{v}}{2}\gamma^\mu\frac{1+\slashed{v}}{2}=v^\mu\frac{1+\slashed{v}}{2}\to v^\mu\,.
\end{equation}
The EFT Lagrangian at leading order in $\cO(p/m_Q)$ in case of more than one heavy flavor (the total number of heavy flavors denoted by $N_h$) is
\begin{equation}\label{eq:69eq2.49}
\cL_{HQET}=\sum_{i=1}^{N_h}\bar Q_v^{(i)}(iv\cdot D)Q_v^{(i)}\,,
\end{equation}
where $v$ is the common velocity of all heavy quarks. It is clear from \eq{69eq2.49} that the EFT Lagrangian exhibits $U(2N_h)$ heavy quark spin-flavor symmetry with $2N_h$ quark fields transforming in the fundamental representation. Due to the restriction \eq{69eq2.44}, there are only $2N_h$ independent components of quark fields $Q_v^{(i)}$.

\section{\texorpdfstring{$D\pi$}{D-pi} Scattering Near \texorpdfstring{$D^*$}{D-star} Threshold Using X-Effective Field Theory}
\label{sec:II.3}

In this section, I study $D\pi$ scattering near $D^*$ threshold, which is analogous to nucleon-pion ($N\pi$) scattering and offers a unique view of the long-distance nuclear force between hadrons.\footnote{This section is based on unpublished work in collaboration with U. van Kolck and S. Fleming.} I first provide the background required to understand this process, Heavy Hadron Chiral Perturbation Theory (HHChPT) with terms to $\cO(p/\Lambda_\chi)$ where $\Lambda_\chi\sim 1$\:GeV and including both isospin conserving and breaking terms. Later, I reduce HHChPT to the non-relativistic region, and then further to $D\pi$ scattering near the $D^*$ threshold. In this region, the effective field theory is also known as XEFT. Finally I present the required resummation in this process and pole hunting in the framework of the scattering theory.

\subsection{Motivation and Kinematics}
\label{sec:II.3.2}

$\{D^0,D^+\}$ are mesons containing charm quarks in an isospin doublet with anti-particles $\{\overline{D^0},D^-\}$. $\{D^{0*},D^{+*}\}$ have the same quark components but larger masses due to the excited spin states compared to $\{D^0,D^{+}\}$ respectively. Because the mass splitting between $D^*$ and $D$ states is much less than their masses, the heavy quark symmetry is well-justified, providing the basis for Heavy Hadron Chiral Perturbative Theory (HHChPT), which is widely used for the interaction between pseudo-Goldstone bosons and heavy mesons 
(\cite{Burdman:1992gh,Wise:1992hn,Yan:1992gz,grinstein1992chiral,Goity:1992tp,Cheng:1992xi, Cheng:1993gc,Cheng:1993kp,Amundson:1992yp,Cho:1992nt,Boyd:1994pa,Cheng:1993ah,Jenkins:1992hx, Stewart:1998ke,Burdman:1993es}).

Elastic $D\pi$ scattering near the $D^*$ threshold provides a new system in which to study hadron scattering kinematically near to a bound state.  It thus serves as an important laboratory, independent of $N\pi$ scattering, to investigate the bound-state and resonance structures arising from pion exchange.

In the elastic scattering of $D^0$ and $\pi^0$ near the $D^{*0}$ threshold, the masses and mass differences of the involved particles are: 
\begin{align*}
m_{D^0}=1864.6\,&\mev, \,\,\, m_{D^{0*}}=2006.7\,\mev, \,\,\, m_{\pi^0}=134.98\,\mev, \\
&\Delta_{D^{0*}-D^0}=m_{D^{*0}}-m_{D^0}=142.1\,\mev.
\end{align*}
For the reaction $D^0\pi^0\to D^0\pi^0$, in the center of mass frame of $D^0$ and $\pi^0$, by energy conservation we have
\begin{equation} \label{eq:72aeq1}
(E_{\pi^0}+E_{D^0})^2=\left(\sqrt{p_{\pi^0}^2+m_{\pi^0}^2}+\sqrt{p_{D^0}^2+m_{D^0}^2}\right)^2\,,
\end{equation}
and for the elastic scattering of $D^0$ and $\pi^0$ near $D^{*0}$ threshold, we first set the sum of the $D^0$ and $\pi^0$ energies equal to the $D^{0*}$ pole mass to obtain
\begin{equation} \label{eq:72aeq2}
(E_{\pi^0}+E_{D^0})^2=m_{D^{*0}}^2\,.
\end{equation}
Combining the two equations above, we find $|\vec p_{\pi^0}|=|\vec p_{D^0}|=42.82$\,MeV, $E_{D^0}=1865.09$\,MeV, $E_\pi=141.61$\,MeV. Near the $D^{*0}$ threshold, the on-shell $D^{*0}$ can have a maximum kinetic energy
\begin{equation}\label{deltaDstardefn}
\delta=m_{D^{*0}}-m_{D^0}-m_{\pi^0}\simeq 7\,\mathrm{MeV}.
\end{equation} 
Since the momenta of the $D^0$ and $\pi^0$ are much smaller than their masses, and the kinetic energy of the $D^{*0}$ is much smaller than its mass,  we can treat $D^0, \pi^0$ and $D^{*0}$ non-relativistically as in XEFT (\cite{Fleming:2007rp}).

\begin{figure}[th]
  \centering
    \includegraphics[width=.75\textwidth]{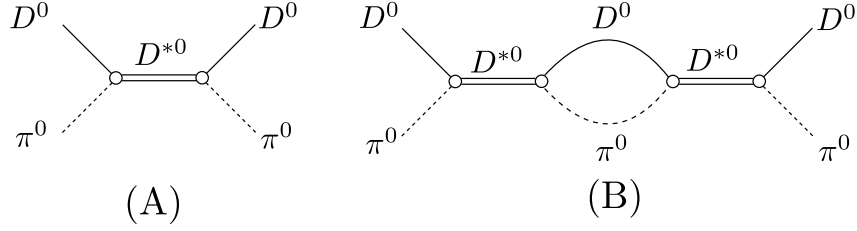}
  \caption{(A) Tree-level diagram for $D^0\pi^0$ to $D^0\pi^0$ scattering; (B) one-loop correction}
  \label{fig:72afig1}
\end{figure}

We are going to use XEFT to study the $D^0\pi^0$ elastic scattering near $D^{0*}$ threshold. If the $D^0\pi^0$ loop correction to the $D^{0*}$ propagator in \fig{72afig1}(B) is the same order as the tree level $D^{0*}$ propagator in \fig{72afig1}(A), we must resum to all orders the $D^0\pi^0$ loop correction to the $D^{0*}$ propagator.  To determine the scattering conditions under which resummation is necessary, we solve for the $D^{0*}$ energy with the ratio of the loop-corrected to the tree-level propagator equal to unity,
\begin{equation}\label{eq:72aeq3}
\frac{\mathcal{M}_B}{\mathcal{M}_A}=\frac{g^2}{f_\pi^2}\frac{(2m_\pi\delta)^{3/2}}{4\pi E_{D^*}}\sim 1\,.
\end{equation}
Here $g\simeq0.6$ is the $D$-meson axial transition coupling, and $E_{D^*}$ is the $D^{0*}$ kinetic energy.  We shall derive the tree and loop amplitudes below in Sec.\:\ref{sec:II.3.5}; the square-root dependence on the mass splitting \req{deltaDstardefn} and pion mass arises from the nonrelativistic treatment of the $D$ and $\pi$.
Solving for $E_{D^*}$ from \eq{72aeq3}, we find that the loop correction is the same order as tree-level when $E_{D^*}\sim 0.2\,\mev$.  Therefore, for $D^0\pi^0$ scattering within $0.2\,\mev$ of the $D^{*0}$ threshold, resummation of the $D^{0*}$ propagator is required.

This effective field theory technique of $D^0\pi^0$ resummation at $D^{*0}$ threshold is useful in the extrapolation of the $D$-meson mass spectrum on lattice.  Several lattice calculations (\cite{Kalinowski:2012re, Mohler:2011ke}) obtain the D-meson mass spectrum with a pion mass much larger than the physical one ($m_\pi^{(phys)}\sim 140$\,MeV).  In \cite{Kalinowski:2012re}, $m_\pi\approx 285, 325$ and $457\,\mev$, and in \cite{Mohler:2011ke}, $m_\pi\approx 141, 316, 418$ and $707\,\mev$.
However, the mass splitting $\Delta$ of the $D$ and $D^*$ states on the lattice is smaller than the pion mass, $\Delta\approx 150, 150$ and $170\,\mev$ in \cite{Kalinowski:2012re}, and $\Delta\approx 131, 153, 148$ and $156\,\mev$ in \cite{Mohler:2011ke}.  Therefore, extrapolating to the physical pion mass in order to determine the physical D-meson masses and their splittings, one must encounter an energy neighborhood around the $D^*$ threshold such that $D\pi$ scattering resummation is necessary. 
In the extrapolation this coincidence occurs around $m_\pi\approx \Delta\approx 145\,\mev$.

We also implement the XEFT with the charged $D,\pi,D^*$ states to study to the isospin violation effects near $D^*$ threshold. We consider the elastic scattering of $\{D^0,D^\pm\}$ with $\{\pi^0,\pi^\pm\}$, and the additional masses and mass differences in these processes are: 
\begin{align*}
m_{D^+}=1869.58,\mev, \quad m_{D^{+*}}=2010.0\, \mev, \quad m_{\pi^+}=139.57 \,\mev\,,\qquad\\
m_{D^{+*}}\!-m_{D^+}=140.7 \,\mev, ~~ m_{D^{0*}}\!-m_{D^+}=137.4\,\mev, \quad m_{D^{+*}}\!-m_{D^0}=145.4\,\mev.
\end{align*}

\begin{figure}[!h]
\centering
\includegraphics[width=.75\textwidth]{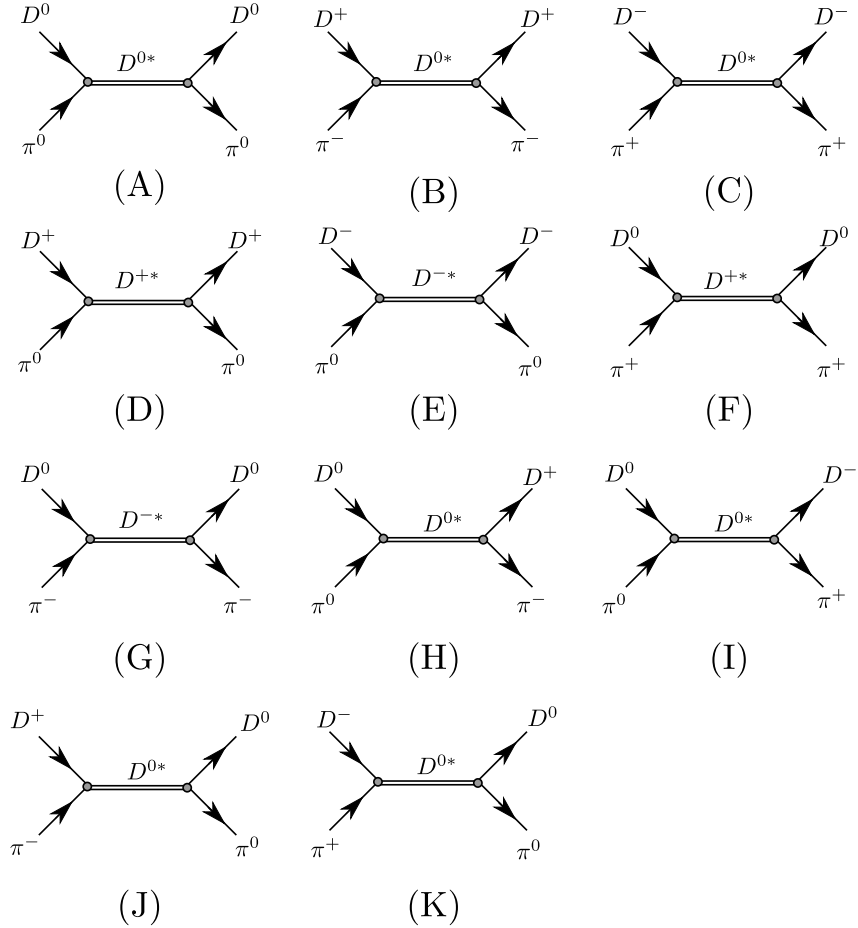}
\caption{Possible $D\pi$ to $D\pi$ scattering diagrams.}
\label{fig:72afig2}
\end{figure}

The tree level Feynman diagrams for $D\pi\to D\pi$ scattering are shown in \fig{72afig2}. Among these scattering processes, the $D^{0*}$ is \fig{72afig2}(B)\,and\,(C) are above $D^+\pi^0$ or $D^-\pi^0$ threshold, while rest of the $D^*$ are below $D\pi$ threshold, meaning the $D^*$ in these processes can be on-shell. 
We start with isospin conserving limit $m_{D^+}=m_{D^0}$ and $m_{D^{*+}}=m_{D^*}$, and treat $D^+$, $D^{*+}$ and their anti-particles as two more sets of degrees of freedom in the $D$-meson isospin space. Then we add isospin breaking terms into the XEFT Lagrangian. The physical splitting between the charged and neutral $D$-mesons is around 4\,MeV, larger than the energy range 0.2\,MeV determined by \eq{72aeq3} within which the $D^*$ resummation becomes necessary. In order to study the impact of the isospin breaking effects on the $D$-meson system, we first assume the isospin breaking terms are small enough to be treated perturbatively, and then gradually approach the physical splitting. This procedure may also facilitate extrapolation of the lattice $D$-meson isospin splitting effects to the physical value, because isospin breaking effects are also introduced perturbatively on the lattice (\cite{deDivitiis:2011eh,Tantalo:2013maa}).

\subsection{Heavy Hadron SU(2) Chiral Perturbation Theory}
\label{sec:II.3.3}

In this part, we describe the standard heavy hadron chiral perturbation Lagrangian, including for the first time the first order of isospin and spin-flavor symmetry breaking parameters for the $SU(2)$ chiral symmetry.  We then reduce it to the effective field Lagrangian for $D\pi\to D\pi$ scattering near the $D^*$ threshold, i.e. to the XEFT. 

In the standard heavy hadron chiral perturbation Lagrangian, the typical momentum $q\sim |\vec p_\pi|\sim|\vec p_D|\sim m_\pi$, and the heavy scale in the theory is either $M_D$ the $D$ meson mass or $4\pi f_\pi\sim \Lambda_\chi$, the chiral symmetry breaking scale $\sim 1$\,GeV.  Adding spin-flavor symmetry breaking terms brings in operators in which $\lqcd$ also appears as a natural heavy scale.  Since the three scales $M_D\sim\Lambda_\chi\sim\lqcd$ differ numerically by factors of order 1, we choose $M_D$ as the heavy scale for all operators, in terms of which reparameterization invariance constraints are directly applied.  Residual differences are then absorbed into the dimensionless coefficients and we define the dimensionless power-counting parameter $\lambda=q/M_D$.

Isospin and spin-flavor symmetry breaking terms first appear at $\cO(\lambda)$ and we derive the complete $\cO(\lambda)$ Lagrangian including these terms.  $e=|e|$ is the unit charge of heavy and light mesons, and is only considered in the isospin-breaking case. We denote $P_a$ as the isospin doublet pseudo-scalar $(D^0,D^+)$ while $P_a^*$ as their excited vector states $(D^{0*},D^{+*})$, where the label $a$ distinguishes the light quark flavor in $D$ mesons. Combining $P_a$ and $P_a^*$ we define the heavy field
\begin{equation}\label{eq:72beq1}
H_a=\frac{1+\slashed{v}}{2}[P_a^{*\mu}\gamma_\mu-P_a\gamma_5]\,,
\end{equation}
which under the unbroken chiral symmetry subgroup $SU(2)_V$ transforms as 
\begin{equation}\label{eq:72beq2}
H_a\to H_bV_{ba}^\dagger\,, \qquad V\in SU(2)_V\,.
\end{equation}
The symmetry group associated with the heavy quark spin is the spin group Spin$(n)$, where $n=\frac{1}{2}$ for the heavy quark being a fermion.  An element of Spin($\frac{1}{2}$) acts on the heavy meson field as
\begin{equation}\label{eq:72beq3}
H_a\to D_QH_a\,,\qquad D_Q\in \mathrm{Spin}(\textstyle{\frac{1}{2}})\,.
\end{equation}
$P_a$ and $P_a^*$ are the heavy meson spin symmetry doublets.

We refer to the heavy meson fields as matter fields, as we will shortly see that interactions with the pseudo-Goldstone bosons can be written into a covariant derivative similar to a gauge field.  In general, it would be useful to write terms that are not consequences of spontaneous chiral symmetry breaking using a form of the heavy meson fields respecting $SU(2)_L\times SU(2)_R$ chiral symmetry. However, by construction of the matter-Goldstone interactions, one always can reduce a general $SU(2)_L\times SU(2)_R$  invariant matter field to a representation $H_a$ obeying Eq.(\ref{eq:72beq2}) by redefining $H_a$ with some combination of
\begin{equation}
\xi_{ab}=\exp(iM_{ab}/f_\pi)=\sqrt{\Sigma_{ab}}, 
\end{equation}
which is chosen to transform as 
\begin{align}
\xi\to V\xi V^\dag,\qquad V\in SU(2)_V,
\end{align}
and where $\Sigma(x)$ is the Goldstone boson field conventionally used in the chiral perturbation lagrangian.
Although this chosen transformation for $\xi$ is not the most general representation of the full $SU(2)_L\times SU(2)_R$ symmetry group, it results in the same physical predictions  (\cite{Manohar:2000dt}). See Appendix \ref{appx:ChPT} for more detail. 

\subsubsection{Lowest order Lagrangian for heavy hadron chiral perturbation theory without isospin breaking}
\label{sec:II.3.3.1}

To construct operators that only involve $V,V^\dagger$ in their chiral symmetry transformation rules, we combine $\xi$ with derivatives as
\begin{align}
\mbv_{ab}^\mu&=\frac i2\paren{\xi^\dagger\pd^\mu\xi+\xi\pd^\mu\xi^\dg}_{ab}\,,\nn \\
\mba_{ab}^\mu&=\frac i2\paren{\xi^\dg\pd^\mu\xi-\xi\pd^\mu\xi^\dg}_{ab}\,,\label{eq:72beq8}
\end{align}
which transform under $SU(2)_L\times SU(2)_R$ as 
\begin{equation}\label{eq:72beq9}
\mbv_\mu\to V\mbv_\mu V^\dg+iV\pd_\mu V^\dg,\,\,\,\, \mba_\mu\to V\mba_\mu V^\dg\,,
\end{equation}
so that $L$ and $R$ do not appear.  
As the matter fields are only associated with $V$, we can use the vector field $\mbv_\mu$ which transforms like a $V$-gauge field to define the chiral covariant derivative,
\begin{equation*}
D_{ab}^\mu=\delta_{ab}\pd_\mu+i\mbv_{ab}^\mu\,.
\end{equation*}
The axial vector field $\mba_\mu$ is coupled with heavy meson fields in a Yukawa form since it transforms in the adjoint representation.  Further, to establish a non-relativistic effective theory similar to HQET, we rescale the matter fields by $e^{-iM_{D}v\cdot x}$ with $v_\mu$ the 4-velocity  and integrate out the heavy components of the field.  Thus, the lowest order heavy meson-pion interaction terms are
\begin{align}
\mathcal{L}_{L.O.}^{D\pi}&=-i\tr[\bar H_av_\mu D_{ab}^\mu H_b] 
+g_\pi\tr[\bar H_aH_b\gamma_\mu\gamma_5\mba_{ab}^\mu] \,.
\label{eq:72beq9p}
\end{align}
$g_\pi$ is the axial heavy meson coupling and $\bar H_a$ is the conjugate field of $H_a$:
\begin{align}
\bar H_a&=\gamma_0 H_a^\dagger\gamma_0=\gamma_0[P_a^{*\mu\dg}\gamma_0\gamma_\mu\gamma_0-P_a^\dg\gamma_5]\frac{1+v^\mu \gamma_0\gamma_\mu\gamma_0}{2}\gamma_0\nn\\
&=[\bar P_a^{*\mu}\gamma_\mu+\bar P_a\gamma_5]\frac{1+\slashed{v}}{2}\,.\label{eq:72beq10}
\end{align}
Because $P_a^*$ is a vector field, $v\cdot P_a^*=0$, and in the rest frame $v_\mu=(1,\vec 0)$ implies that $P_a^{*0}=0$. Thus, in the rest frame we obtain
\begin{align}
H_a&=-\begin{pmatrix}
0 & \vec P_a^*\cdot\vec \sigma+P_a \\
0 & 0
\end{pmatrix}\equiv -\begin{pmatrix}
0 & h_a \\ 0 & 0
\end{pmatrix}\,,\nn\\
\bar H_a&=\begin{pmatrix}0 & 0 \\ \vec{P}_a^{*\dagger}\cdot\vec\sigma+P_a^\dagger & 0
\end{pmatrix}\equiv\begin{pmatrix}
0 & 0 \\ h_a^\dagger & 0
\end{pmatrix}\,,\label{eq:72beq11}
\end{align}
where $\vec\sigma$ are Pauli matrices, and Eq.(\ref{eq:72beq9p}) reduces to 
\begin{equation}\label{eq:72beq12}
\mathcal{L}_{L.O.}^{D\pi}=\tr[h_a^\dagger iD_{ba}^0h_b]-g_\pi\tr[h_a^\dagger h_b\vec\sigma\cdot\vec{\mba}_{ba}]\,,
\end{equation}
where
\begin{align}
iD_{ba}^0&=i\pd^0\delta_{ba}-\mbv_{ba}^0\,,\nn \\
\vec\sigma\cdot\vec{\mba}_{ba}&=-\oneov{f_\pi}\vec\sigma\cdot\vec{\nabla}M_{ba}+... \nn\\
M_{ab}&=\pi_{ab}^i\sigma^i=
\begin{pmatrix}
\pi^0/\sqrt{2} & \pi^+ \\ \pi^- & \pi^0/\sqrt{2}
\end{pmatrix}.
\label{eq:72beq13}
\end{align}

To derive the kinetic and interaction terms for anti-meson matter fields, we start from the full relativistic theory and apply the projection operator $\frac{1-\slashed{v}}{2}$ to the meson field, instead of $\frac{1+\slashed{v}}{2}$.  We denote $P_a^{(anti)*\mu}$ as vector anti-meson states $(\overline{D^{0*}},D^{-*})$ and $P^{(anti)}_a$ as pseudo-scalar anti-meson states $(\overline{D^0},D^-)$, and obtain the heavy anti-meson field $H_a^{(anti)}$ from
\begin{align}\label{antiHdefn}
H_a^{(anti)}&\equiv[P_a^{(anti)*\mu}\gamma_\mu-P_a^{(anti)}\gamma_5]\frac{1-\slashed{v}}{2}\,.
\end{align}
This projection retains the anti-particle components of the relativistic field and leads to construction of the effective Lagrangian based on the same symmetries, integrating out heavy components that now correspond to particle components.  
In Appendix \ref{appx:antiheavymeson}, we show this procedure is consistent with applying the charge conjugation operation $\mathcal{C}$ to the heavy meson field Eq.(\ref{eq:72beq11}).   

As we systematically write the operators breaking heavy meson spin symmetry, we find from heavy quark effective theory (HQET), at $\frac{\Lambda_{QCD}}{m_Q}$ order, the first order of heavy quark spin symmetry violation is brought in by the (anomalous) chromo-magnetic moment operator,
\begin{equation}\label{eq:72beq26}
\mathcal{L}_{(\Lambda_{QCD}/m_Q)}^{HQET}\sim\frac{a g_s}{4m_Q}\bar h_\nu^{(Q)}\sigma_{\mu\nu}T_A h_\nu^{(Q)}F_A^{\mu\nu}\,,
\end{equation}
where $a=(g_Q-2)/2$ and $\Lambda_{QCD}$ is the typical momentum of a gluon in a meson containing one heavy quark. $m_Q$ is the heavy quark mass, and $F_A^{\mu\nu}$ is the gluon field tensor with color index $A$ given in \req{gluontensor}.  $\sigma^{\mu\nu}$ is the $4\times 4$ Lorentz tensor
\begin{equation}\label{eq:72beq28}
\sigma^{\mu\nu}=\frac{i}{2}[\gamma^\mu,\gamma^\nu]\,.
\end{equation}
The constant can be computed explicitly by matching Eq.(\ref{eq:72beq26}) onto the QCD Lagrangian. The color magnetic moment operator transforms as a vector under the heavy quark spin group Spin$(\frac12)$. We construct the corresponding heavy hadron operator by applying the same symmetry. Defining the light quark spin operator $S_l^\mu=(0,\frac12\vec\sigma)$, and heavy quark spin operator $S_Q^\mu=(0,\frac12\vec\sigma)_Q$, we find the mass splitting between vector mesons and pseudo-scalar mesons is, to leading order,
\begin{equation}\label{eq:72beq29}
\mathcal{L}_{L.O.}^{D^*-D}=-\Delta\tr(\bar H_a S_{H}^\alpha H_a S_{l\alpha})=-\frac{\Delta}{8}\tr(\bar H_a\sigma^{\mu\nu}H_a\sigma_{\mu\nu})\,,
\end{equation}
where $\Delta=m_{D^*}-M_D\simeq m_\pi$, making Eq.(\ref{eq:72beq29}) same order as Eq.(\ref{eq:72beq12}). The spin coupling operator,
\begin{equation}\label{eq:72beq30}
S_{H}^\alpha S_{l\alpha}=-\vec S_{H}\cdot\vec S_l=-\frac12\paren{\vec S_{\text{total}}^2-\vec S_{H}^2-\vec S_l^2}\,,
\end{equation}
gives $-3/4$ for the pseudoscalar meson $P_a$ and $1/4$ for the vector meson $P_a^*$. In the rest frame of the heavy mesons, Eq.(\ref{eq:72beq29}) reduces to
\begin{equation}\label{eq:72beq31}
\mathcal{L}_{L.O.}^{D^*-D}=\frac{\Delta}{4}\tr[h_a^\dg \sigma^i h_a\sigma^i]\,.
\end{equation}

\subsubsection{Lowest order Lagrangian for heavy hadron chiral perturbation theory with isospin breaking}
\label{sec:II.3.3.2}

In last section, we did not distinguish the neutral states of $D$ and $\pi$ mesons from the charged states. In this section, we take into account the differences between $u$ and $d$ quarks.
There are several sources of isospin breaking effects. One is from the mass difference of the $u$ and $d$ quark, and it is sensible to describe these effects as proportional to $m_d-m_u\sim 4\,\mev$. In the standard HHChPT power counting, this is much smaller than the pion typical momentum, so one may expect these terms are subleading. 

A second source of isospin breaking is the charge difference of $u$ and $d$ quarks, which results in a significant difference in electromagnetic effects. These terms must show up at both leading order and next-to-leading order, since the uncharged particles are distinguished from the charged ones at both orders.

Defining the charge matrix for the light quarks,
\begin{equation}\label{eq:72beq20} 
Q_{ab}=\begin{pmatrix}
2/3 & 0 \\ 0 & -1/3 \end{pmatrix}\,,
\end{equation}
and the charge matrix for the charm quark
\begin{equation}\label{eq:72beq21} 
Q'=\begin{pmatrix}2/3 & 0 \\ 0 & 2/3\end{pmatrix}\,,
\end{equation}
we write the covariant derivative for the Goldstone bosons 
\begin{equation}\label{eq:72beq22} 
D_\mu\xi_{ab}=\pd_\mu\xi_{ab}+ieB_\mu[Q,\xi]_{ab},
\end{equation}
and the covariant derivative for the matter field
\begin{equation}\label{eq:72beq23} 
D_\mu H_a=\pd_\mu H_a+ieB_\mu(Q'_{bb}H_a-H_bQ_{ba})+i\mbv_{\mu ab}H_b\,,
\end{equation}
where $B_\mu$ is the photon field. If $Q$ were to transform under $SU(2)_L\times SU(2)_R$ as $Q\to LQR^\dg$, the Lagrangian would respect the full chiral symmetry. 

Replacing the derivatives in Eq.(\ref{eq:72beq8}) and Eq.(\ref{eq:72beq12}), we have the lowest order Lagrangian with electromagnetic effects. The leading order photon kinetic Lagrangian is
\begin{equation}\label{eq:72beq24} 
\mathcal{L}_{L.O.}^\gamma=-\oneov{4}F^{\mu\nu}F_{\mu\nu},\,\,\, F_{\mu\nu}=\pd_\mu B_\nu-\pd_\nu B_\mu.
\end{equation}
Because a new parameter $e$ or $\alpha_{em}=e^2/4\pi$ is introduced into our effective field theory, to avoid double power counting, we must power count $e$ in terms of $\lambda=q/M_D$. The baseline for doing so is to consider the pion isospin breaking term 
\begin{equation}\label{eq:72beq80}
\mathcal{L}_{N.L.O.}^{\pi^+\pi^-}=e^2f_\pi^2\,\tr[(Q^\dagger \Sigma+\Sigma Q)^2]
\end{equation}
as a next-to-leading order mass correction compared to its leading-order mass term, $m_\pi^2$.  
Comparing the mass contribution $e^2f_\pi^2$ from \eq{72beq80}, we find
\begin{equation}\label{eq:72beq81-mass}
\frac{e^2f_\pi^2}{m_\pi^2}=\frac{(4\pi f_\pi)^2}{m_\pi^2}\frac{\alpha}{4\pi}\sim\frac{M_D^2}{m_\pi^2}\frac{\alpha}{4\pi}\sim\lambda\sim\frac{q}{M_D}\,.
\end{equation}
In HHChPT, we consider $m_\pi\sim q$ so far, implying $\frac{\alpha_{em}}{4\pi}\sim \lambda^3$.  As a result, we power count the unit charge $e$ as
\begin{equation}\label{eq:72beq81}
e\simeq \sqrt{4\pi\alpha_{em}}\sim\sqrt{(4\pi)^2 \lambda^3}=\sqrt{\frac{(4\pi)^2 p_\pi^3}{(4\pi f_\pi)^3}}\sim \sqrt{\lambda}\frac{p_\pi}{f_\pi}\,,
\end{equation}
with $\sqrt{\lambda}$ considered a $\mathcal{O}(1)$ quantity. This is consistent with Eqs.\,\eqref{eq:72beq22} and \eqref{eq:72beq23} where the photon coupling is a leading order effect keeping the derivatives gauge covariant. 

\subsubsection{Next-to-leading order Heavy Hadron Chiral Perturbative Lagrangian without isospin breaking}
\label{sec:II.3.3.3}

From this section on, we include next-to-leading order corrections to  the HHChPT Lagrangian in the $\lambda=q/M_D$ expansion. Operators arise from the following sources:
\begin{description}
\item[I.] quark mass-induced chiral symmetry breaking terms, which are the heavy hadron mass terms proportional to light quark masses,
\item[II.] velocity reparameterization invariance (V.R.I.)-induced terms, including the heavy hadron kinetic term and heavy-light hadron interaction terms,
\item[III.] heavy quark spin symmetry breaking terms which allow the $D$--$D^*$ transition but are excluded in \eq{72beq1} and \eq{72beq2},
\item[IV.] additional Goldstone boson terms, mainly for the heavy-light interaction terms, which preserves chiral symmetry,
\item[V.] virtual photon-induced terms, due to the inner electromagnetic structure of heavy  hadrons, and
\item[VI.] real photon-induced terms, also due to the inner electromagnetic structure of heavy hadrons.
\end{description}

We start with category I for the investigation of next-to-leading order terms.  These terms are the heavy and light hadron mass terms breaking chiral symmetry and heavy quark spin symmetry, and the heavy hadron mass terms proportional to charm quark charge that preserve all symmetries.  We power count the light quark mass as $m_q\sim \frac{m_\pi^2}{B_0}\sim \lambda m_\pi$, by taking the constant $B_0$ (estimated to be $\sim 3\,$GeV above) at the same order as the large scale $M_D$ in HHChPT. Thus we can write the terms explicitly violating chiral symmetry but preserving heavy quark spin symmetry and isospin symmetry as
\begin{equation}\label{eq:72beq32}
\msl_{N.L.O.}^{M_1}=\sigma_1\tr(\bar H_a H_a M_{\xi bb}^{(+)})\,,
\end{equation}
in which
\begin{equation}\label{eq:72beq33}
M_\xi^{(+)}=(\xi M_q^\dg \xi+\xi^\dg M_q\xi^\dg)\,.
\end{equation}
The subscripts $a$ and $b$ in \eqref{eq:72beq32} do not automatically sum to 1. This term is analogous to the $\sigma$-term in pion-nucleon scattering and provides the overall shift to the heavy meson masses arising from the explicit chiral symmetry breaking by the light quark masses. Although it can be absorbed into the heavy hadron mass term $\tr[\bar H_aH_a]$ by redefining $H_a$ by a phase, it distinguishes itself by containing also a $\pi\pi HH$ four point interaction. This type of interaction provides a dynamical picture how heavy hadron masses vary with light quark masses.

Following the same logic in deriving Eq.\,\eqref{eq:72beq29}, we can write the heavy quark spin violation term
\begin{equation}\label{eq:72beq34}
\msl_{N.L.O.}^{M_2}=-\frac18\Delta^{(\sigma_1)}\tr(\bar H_a\sigma^{\mu\nu} H_a\sigma_{\mu\nu}M_{\xi bb}^{(+)})\,.
\end{equation}
On the other hand, parity invariance discussed in Appendix \ref{app:CPTonHHChPT} excludes terms of the form
\begin{equation}\label{eq:72beq39}
\msl\sim\tr\left[\bar H_aH_a(\xi M_q\xi-\xi^\dg M_q\xi^\dg)\right].
\end{equation}
In the heavy hadron rest frame, we have
\begin{align}
\msl_{N.L.O.}^{M_1}&=-\sigma_1\,\tr(h_a^\dg h_aM_{\xi bb}^{(+)}), \label{eq:72beq35}\\
\msl_{N.L.O.}^{M_2}&=\frac{\Delta^{(\sigma_1)}}{4}\tr(h_a^\dg \sigma^i h_a \sigma_i M_{\xi bb}^{(+)})\,.\label{eq:72beq36}
\end{align}

We also can write the heavy-light meson interaction term that violates chiral symmetry but preserves heavy quark spin symmetry,
\begin{equation}\label{eq:72beq40} 
\msl_{N.N.L.O.}^{D\pi(1)}=\frac{gK_1}{M_D}\tr[\bar H_a H_b \mba_{ba}^\mu \gamma_\mu \gamma_5 M_{\xi cc}^{(+)}]\,,
\end{equation}
where $g$ is the $D$-meson axial coupling and $K_1$ is a dimensionless constant.  This operator corrects the $D$ to $D^*$ transition via the pion term to one more $\lambda$ order, since this term is order $\sim\frac{m_q}{M_D}\sim\frac{\lambda m_\pi}{M_D}\sim\lambda^2$.

Next we turn our attention to Category II.  Because we work in the frame where the heavy hadrons are almost at rest, we must ensure the theory retains Lorentz invariance.  The frame is chosen by separating the particle momentum $p=Mv+k$ and defining the velocity label such that $v\ll 1$ and $k\ll M$.  However, the separation is arbitrary, and observables should not depend on this parameterization of the momentum.  There is therefore a velocity reparameterization invariance (VRI), reviewed in Appendix \ref{appx:VRI} and similar to the reparameterization invariance in SCET explained in Sec.\,\ref{sec:SCETRPI}.  Respecting also the symmetries connecting the quark- and meson-level theories, velocity reparameterization invariance requires the following operators in HHChPT at $\mathcal{O}(q/M_D)$:
\begin{align}
\mathcal{L}^{VRI_1}_{N.L.O.}&=-\frac{1}{2M_D}\tr[\bar H_a(iD)^2_{ba}H_b]\,,\label{eq:72beq61}\\
\mathcal{L}^{VRI_2}_{N.L.O.}&=\frac{g}{M_D}\tr[\bar H_c(i\overleftarrow{D}_{ac}^\mu v\cdot\mathbb{A}_{ba}-iv\cdot\mba_{ac}\overrightarrow{D}_{ba}^\mu)H_b\gamma_\mu\gamma_5]\,.\label{eq:72beq62}
\end{align}
Eqs.\,\eqref{eq:72beq61} and \eqref{eq:72beq62} correspond to the first term and second term of \eq{72beq12} respectively. All other terms, such as
\begin{align}
\mathcal{L}_{N.L.O}^{VRI_3}&\sim-\frac{2}{M_D}\tr[\bar H_a(iv\cdot D)^2_{ba}H_b]\,,\label{eq:72beq61'}\\
\mathcal{L}_{N.L.O}^{VRI_4}&\sim-\frac{g}{M_D}\tr[\bar H_c(iv\cdot \overleftarrow{D}_{ac}\mba_{ba}^\mu\gamma_\mu-\mba_{bc}^\mu\gamma_\mu iv\cdot D_{bd})H_b\gamma_5]\,,\label{eq:72beq63}
\end{align}
are eliminated by the leading order equation of motion.  In Sec.\,\ref{app:CPTonHHChPT}, we show that these operators are invariant under C, P, and T transformations and hermitian conjugation separately, which also proves that \req{eq:72beq62} is the effective Lagrangian for heavy anti-mesons to this order with the replacement of the heavy meson fields $H$ by the heavy anti-meson fields $H^{(anti)}$.

Next we move to the category III of the next-to-leading order operators. When constructing spin operators in the leading order $D^*$ and $D$ mass splitting section, we did not include breaking of spin symmetry by the low energy pion interaction, which arises from the difference between the axial-vector couplings to $D^*D\pi$ and $D^*D^*\pi$ in the spin factor of the operator $\tr[\bar H_a(S_Q \mba_{ba})H_b]$.  In the rest frame of the heavy state, this operator appears as a spin-orbit coupling between the heavy hadron and the moving pion, but can be equally considered as a coupling of the heavy current to the pion spin polarization. 
We write this next-to-leading order heavy meson spin symmetry breaking term as
\begin{equation}\label{eq:72beq73}
\mathcal{L}_{N.L.O.}^{\rm spin}=\frac{2g_2}{M_D}\tr[\bar H_a(S_Q \mba_{ab})H_b]=\frac{g_2}{M_D}\tr[\bar H_a\mba_{ab}^\mu\gamma_\mu\gamma_5 H_b]\,,
\end{equation}
where $g_2$ is a parameter with dimension (mass)$^1$, and we could extract a factor $\lqcd\sim M_D$ from $g_2$ to write $g_2=g_2'M_D$ with $g_2'$ dimensionless.  This term shows that the next-to-leading order correction to the axial $DD^*$ transition results from heavy quark spin symmetry breaking. 


For category IV, we construct next-to-leading order operators involving higher-order couplings to  Goldstone bosons and respecting all symmetries.
First, we consider adding one more derivative to the Goldstone bosons, and write
\begin{align}
\mathcal{L}_{N.L.O.}^{\delta_2,\delta_3}&=\frac{\delta_2}{M_D}\tr[\bar H_aH_b\gamma_\mu \gamma_5 iv\cdot D_{bc}\mba_{ca}^\mu]
+\frac{\delta_3}{M_D}\tr[\bar H_aH_b\gamma_\mu \gamma_5 iD_{bc}^\mu v\cdot\mba_{ca}]\,,\label{eq:72beq77}
\end{align}
using that the natural heavy scale here is $4\pi f_\pi=\Lambda_\chi\sim M_D$.  Integrating \req{eq:72beq77} by parts, we can use the leading order heavy meson equation of motion to eliminate the $\delta_2$ term. The $\delta_3$ term has the same form as \req{eq:72beq62} with different coefficient.  VRI fixes the coefficient in \req{eq:72beq62} and therefore we must set $\delta_3=0$. As a result, besides the terms in \req{eq:72beq62}, the next-to-leading order terms can only contain two Goldstone boson axial operators. The simplest forms of these terms are
\begin{align} 
\mathcal{L}_{N.L.O.}^{\delta_4\delta_5}&=\frac{\delta_4}{M_D}\tr[\bar H_aH_b\mba_{bc}^\mu\mba_{\mu ca}]
+\frac{\delta_5}{M_D}\tr[\bar H_aH_bv\cdot\mba_{bc}v\cdot\mba_{ca}]\,,\label{eq:72beq78}
\end{align}
which preserve the heavy quark spin symmetry.

The heavy quark spin symmetry breaking operators permitted by the C, P and T symmetries are
\begin{align}
\mathcal{L}_{N.L.O.}^{\delta_6,\delta_7}&=\frac{\delta_6}{M_D}\tr[\bar H_a[\sigma^{\mu\nu},\sigma_{\nu\sigma}]\mba_{\mu ba}\mba_{cb}^\sigma H_c]
+\frac{\delta_7}{M_D}\tr[\bar H_a[\sigma^{\mu\nu},\sigma_{\sigma\lambda}]v_\nu v^\lambda \mba_{\mu ba}\mba_{cb}^\sigma H_c]\,.\label{eq:72beq79}
\end{align}
The $\delta_6$ term contains the $D^*D$ transition, which has no analog in the proton-pion scattering effective field theory.

Next we discuss the Category V next-to-leading order terms, which preserve  chiral $SU(2)_V$ symmetry and isospin symmetry. These terms arise from electromagnetic effects of energetic photons interacting with the charged heavy quarks inside the heavy meson, and therefore are suppressed by $\alpha_{em}=\frac{e^2}{4\pi}$.  These terms are 
\begin{align}
\mathcal{L}_{N.N.L.O.}^{\text{virtual }\gamma}&=\beta_3 f_\pi\alpha_{em}q_c^2\tr[\bar H_a H_a ]
-\frac18 \Delta_{\beta_3} f_\pi \alpha_{em}q_c^2\tr[\bar H_a\sigma^{\mu\nu} H_b \sigma_{\mu\nu}]\,,\label{eq:72beq83}
\end{align}
where $q_c=+2/3$ is the charm quark charge, and $\beta_3$ and $\Delta_{\beta_3}$ are dimensionless coefficients.  In making the operator the right dimension, the factor $f_\pi\sim m_\pi$ is chosen corresponding to the momentum scale of the virtual photon.  The first term preserves heavy quark spin symmetry, while the second term breaks it.  However, as indicated by the subscript these terms are next-to-next-leading-order in the power counting in \req{eq:72beq81}.  To see this, we estimate the first term, which is the virtual photon-induced heavy meson mass correction,
\begin{align}
m_{em}&\sim \beta_3 f_\pi \alpha_{em}=\beta_3 f_\pi\frac{e^2}{4\pi}\sim \beta_3\frac{\lambda}{4\pi} \frac{p_\pi^2}{f_\pi^2}\sim\beta_3\lambda^2p_\pi\,,\label{eq:72beq82}
\end{align}
using that $p_\pi/4\pi f_\pi\sim\lambda$.  Being of order $\lambda^2=(q/M_D)^2$, we can safely ignore these terms at NLO.

Finally, we discuss the real photon-induced Category VI corrections. We attach the external real photon to the charm quark in the D meson and construct an operator based on the symmetry of the magnetic moment operator for the heavy quark,
\begin{equation}\label{eq:HQmagneticmoment}
\mathcal{L}_{EM}^{HQ}\sim\frac{e\beta_1}{m_Q}\bar\psi_v\sigma^{\alpha\beta}F_{\alpha\beta}\psi_v\,,
\end{equation}
where $\psi_v$ is a velocity-labelled heavy quark field.  The analogous isospin-invariant operator at hadron-level is
\begin{equation}\label{eq:72beq86}
\mathcal{L}_{N.L.O}^{D\gamma_1}=-\frac{e\beta_1}{4M_D}q_c\tr[\bar H_a\sigma^{\mu\nu}H_a F_{\mu\nu}]\,,
\end{equation}
where $\beta_1$ is a dimensionless coefficient and suppressed by the heavy meson scale $M_D$.  We power count the photon momentum $P_\gamma\sim m_\pi$, and charge $e\sim \sqrt{\lambda}\frac{p_\pi}{f_\pi}$, which is also numerically justified, so that the contribution of \req{eq:72beq86} is $\sim\lambda^{\frac{3}{2}}(p_\gamma/f_\pi)$.

\subsubsection{Next-to-leading order Heavy Hadron Chiral Perturbation Theory with isospin breaking}
\label{sec:II.3.3.4}

Next-to-leading order isospin breaking operators in the effective field theory arise from the following sources:
\begin{description}
\item[I.] electromagnetic covariant derivatives in next-to-leading order operators,
\item[II.] light-quark-mass-difference-induced light flavor symmetry breaking terms,
\item[III.] light-quark-charge-difference-induced light flavor symmetry breaking terms, and
\item[IV.] photons interacting with charged heavy mesons and Goldstone bosons.
\end{description}

In category I, the covariant derivatives in the next-to-leading order operators are already described in Eqs.\,\eqref{eq:72beq64}, \eqref{eq:72beq65}, \eqref{eq:72beq78} and \eqref{eq:72beq79}.

In category II, the isospin-breaking terms are written as
\begin{align}
\mathcal{L}_{N.L.O.}^{M_3,M_4}&=\lambda_1\tr[\bar H_a H_bM_{\xi ba}^{(+)}]   
-\frac18\Delta_{\lambda_1}\tr[\bar H_a\sigma^{\mu\nu}H_b\sigma_{\mu\nu}M_{\xi ba}^{(+)}]\,,
\end{align}
where only the second term breaks heavy quark spin symmetry.

In category III, similar to the previous section, the virtual photon-induced heavy meson isospin breaking operators are suppressed by $\alpha_{em}$ so that they are the next-to-next-to-leading order.  They take the forms
\begin{align}
\mathcal{L}_{N.N.L.O.}^{\text{virtual }\gamma_2}&=a_1\alpha_{em}f_\pi\tr[\bar H_a q_Q H_aQ_{bb}^{\xi (+)}]
-\frac{\Delta_{a_1}}{8}\alpha_{em}f_\pi\tr[\bar H_aq_Q\sigma^{\mu\nu}H_b\sigma^{\mu\nu}Q_{ba}^{\xi (+)}]\nn\\
&+a_2\alpha_{em}f_\pi\tr[\bar H_aH_aQ_{bb}^{\xi (+)} Q_{cc}^{\xi (+)}]
-\frac{\Delta_{a_2}}{8}\alpha_{em}f_\pi\tr[\bar H_aQ_{bb}^{\xi(+)}\sigma^{\mu\nu}\sigma^{\mu\nu}H_c\sigma_{\mu\nu}Q_{dd}^{\xi (+)}]\,,\label{eq:72beq85}
\end{align}
where $Q^{\xi(+)}=\frac12(\xi^\dg Q\xi+\xi Q\xi^\dg)$, and $a_1,a_2,\Delta_{a_1},\Delta_{a_2}$ are dimensionless coefficients.  Thus, only pion isospin breaking terms in \req{eq:72beq80} fit in this category.

In category IV, we construct the next-to-leading order operator for the photon-heavy meson coupling,
\begin{equation}\label{eq:72beq87}
\mathcal{L}_{N.L.O.}^{D\gamma_2}=\frac{e\beta}{4M_D}\tr[\bar H_a H_b\sigma^{\mu\nu}F_{\mu\nu}Q_{ba}^{\xi(+)}]\,,
\end{equation}
where $\beta$ is a dimensionless coefficient. Using the power counting of $e$ \req{eq:72beq81}, and power-counting the photon momentum as $P_\gamma\sim m_\pi$, we estimate this term as 
\begin{equation}\label{eq:72beq88}
\mathcal{L}_{N.L.O.}^{D\gamma_2}\sim e\frac{P_\gamma}{M_D}\sim\lambda^{\frac{3}{2}}\frac{P_\gamma}{f_\pi}\,,
\end{equation}
which justifies considering \req{eq:72beq87} as a next-to-leading order operator. 

All other operators in this category are next-to-next-to leading order. For example, if we apply velocity reparameterization invariance to \req{eq:72beq87}, we have
\begin{align}
\frac{e\beta}{4 f_\pi}&\tr[\bar H_aH_b\sigma^{\mu\nu}F_{\mu\nu}Q_{ba}^{\xi(+)}]\nn\\
&\to \frac{e\beta}{4f_\pi}\tr[\bar H_aH_b\sigma^{\mu\nu}F_{\mu\nu}Q_{ba}^{\xi(+)}]+\frac{e\beta}{4f_\pi}\frac{1}{M}\tr[\bar H_aH_b\sigma^{\mu\nu}F_{\mu\sigma}2iD_\nu v^\sigma Q_{ba}^{\xi(+)}]\,,\label{eq:72beq91}
\end{align}
where the second term contains one more derivative and contributes one more power of $\lambda$ according to this expansion. The same power counting occurs as one more Goldstone boson axial field is added into this operator. Thus we can safely conclude \req{eq:72beq87} is the only allowed term containing isospin symmetry breaking at this order.

\subsubsection{Summary of the Lagrangian for Heavy Hadron Chiral Perturbative Theory}
\label{sec:II.3.3.5}

The leading order Lagrangian without isospin breaking is
\begin{align}
\mathcal{L}_{L.O.}&=\mathcal{L}_{L.O.}^{D\pi}+\mathcal{L}_{L.O.}^{D^*-D}+\mathcal{L}_{L.O.}^{\pi}\label{eq:72beq94}\\
&=-i\tr[\bar H_av_\mu D_{ab}^\mu H_a]+g_\pi \tr[\bar H_a H_b\gamma_\mu \gamma_5\mba_{ab}^\mu]-\frac{\Delta}{8}\tr[\bar H_a\sigma^{\mu\nu}H_a\sigma_{\mu\nu}]\nn\\
&+\frac{f_\pi^2}{8}\tr[\pd^\mu\Sigma \pd_\mu\Sigma]+\frac{f_\pi^2B_0}{4}\tr[m_q^\dg \Sigma+m_q\Sigma^\dg].\label{eq:72beq95}
\end{align}
The leading order Lagrangian with isospin breaking is
\begin{equation}\label{eq:LOLag-isobreaking}
\mathcal{L}_{L.O.}^{\text{isospin}}=\mathcal{L}_{L.O.}+\mathcal{L}_{L.O.}^\gamma,
\end{equation}
with all derivatives in \req{eq:72beq95} replaced by \req{eq:72beq22} and \req{eq:72beq23}. $\mathcal{L}_{L.O.}^\gamma$ is the photon kinematic term.

The next-to-leading order Lagrangian without isospin breaking is
\begin{align}
\mathcal{L}_{N.L.O.}&=\mathcal{L}_{N.L.O.}^{M_1}+\mathcal{L}_{N.L.O.}^{M_2}+\mathcal{L}_{N.L.O.}^{V.R.I.1}+\mathcal{L}_{N.L.O.}^{V.R.I.2}+\mathcal{L}_{N.L.O.}^{\text{spin}}\nn\\
&+\mathcal{L}_{N.L.O.}^{\delta_4,\delta_5}+\mathcal{L}_{N.L.O.}^{\delta_6,\delta_7}+\mathcal{L}_{N.L.O.}^{D\gamma_2}\label{eq:72beq96}\\
&=\sigma_1\tr[\bar H_a H_bM_{\xi bb}^{(+)}]-\oneov{8}\Delta^{(\sigma_1)}\tr[\bar H_a\sigma^{\mu\nu}H_a\sigma_{\mu\nu}M_{\xi bb}^{(+)}]\nn\\
&-\oneov{2M_D}\tr[\bar H_a(iD)^2_{ba}H_b]+\frac{g}{M_D}\tr[\bar H_c\paren{i\overleftarrow{D}_{ac}^\mu v\cdot\mba_{ba}-iv\cdot\mba_{ac}\overrightarrow{D}_{ba}^\mu}H_b\gamma_\mu\gamma_5]\nn\\
&+g_2'\tr[\bar H_a\mba_{ab}^\mu\gamma_\mu\gamma_5 H_b]+\frac{\delta_4}{M_D}\tr[\bar H_aH_b\mba_{bc}^\mu\mba_{\mu ca}]+\frac{\delta_5}{M_D}\tr[\bar H_aH_b v\cdot\mba_{bc}v\cdot\mba_{ca}]\nn\\
&+\frac{\delta_6}{M_D}\tr[\bar H_a[\sigma^{\mu\nu},\sigma_{\nu\sigma}]\mba_{\mu ba}\mba_{cb}^\sigma H_c]+\frac{\delta_7}{M_D}\tr[\bar H_a[\sigma^{\mu\nu},\sigma_{\sigma\lambda}]v_\nu v^\lambda\mba_{\mu ba}\mba_{cb}^\sigma H_c]\nn\\
&-\frac{e\beta_1}{4M_D}q_c\tr[\bar H_a\sigma^{\mu\nu}H_a F_{\mu\nu}].\label{eq:72beq97}
\end{align}
The next-to-leading order isospin-breaking Lagrangian is
\begin{align}
\mathcal{L}_{N.L.O.}^{\text{isospin}}&=\mathcal{L}_{N.L.O.}+\mathcal{L}_{N.L.O.}^{M_3,M_4}+\mathcal{L}_{N.L.O.}^{D\gamma_2}+\mathcal{L}_{N.L.O.}^{\pi^+\pi^-}\label{eq:72beq98}\\
&=\mathcal{L}_{N.L.O.}    
+\lambda_1\tr[\bar H_aH_bM_{\xi ba}^{(+)}]-\oneov{8}\Delta_{\lambda_1}\tr[\bar H_a\sigma^{\mu\nu} H_b\sigma_{\mu\nu}M_{\xi ba}^{(+)}]\nn\\
&+\frac{e\beta}{4M_D}\tr[\bar H_a H_b\sigma^{\mu\nu}F_{\mu\nu}Q_{ba}^{\xi(+)}] +e^2f_\pi^2\tr[(Q^\dg\Sigma+\Sigma Q)^2],
\end{align}
where we must replace all the covariant derivatives with the ones in \eqref{eq:72beq22} and \eqref{eq:72beq23}

\subsection{X Effective Field Theory}
\label{sec:II.3.4}
In this part, we match HHChPT operators to XEFT operators using the XEFT power counting, where the large scale is $m_\pi$. We first look at the isospin-conserving case.

\begin{figure}[!h]
\centering
\includegraphics[width=.75\textwidth]{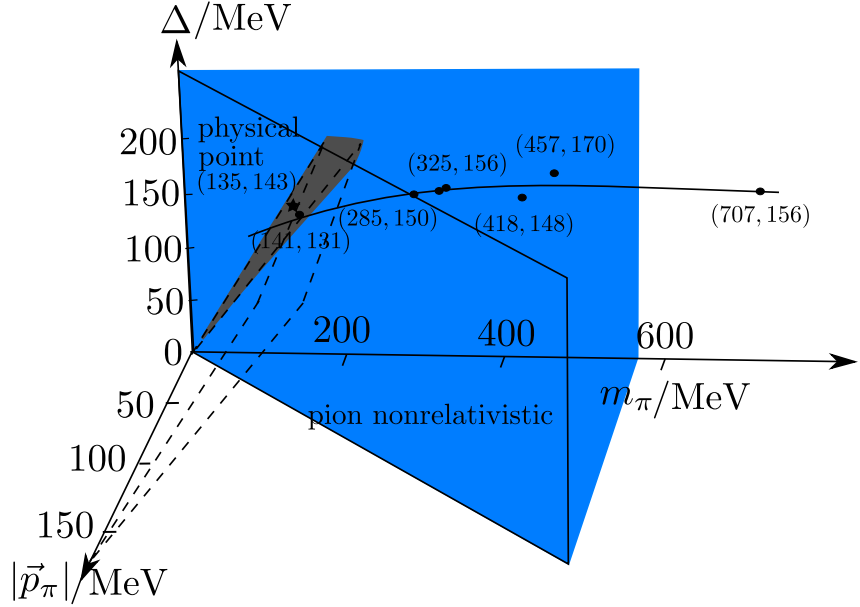}
\caption{XEFT and fine-tuning XEFT energy region.  \label{fig:XEFTregion}}
\end{figure}

The blue-shaded area in \fig{XEFTregion} is the energy area XEFT describes. It envelops the area where pions are non-relativistic $\epsilon=\frac{|\vec p_\pi|}{m_\pi}\lesssim\frac13$.  The fine-tuned XEFT region is where the XEFT area overlaps the splitting $\delta$ between $D^*$ and $D$ such that $\frac{|\Delta-m_\pi|}{m_\pi}=\epsilon^2\lesssim\frac19$. The data points on the $\Delta-m_\pi$ plane are data from lattice calculations. As we can see, XEFT describes the physics of $D\pi$ scattering very close to the $D^*$ threshold.  In our study, we vary the pion mass from 130\,MeV to 150\,MeV and $\Delta$ from 150\,MeV to 139\,MeV.

Another scale that enters the XEFT is the $D$ meson kinetic energy
\begin{equation*}
E_D=\frac{|\vec P_D|^2}{2M_D}\sim\epsilon^2\frac{m_\pi^2}{M_D}=\epsilon^2\lambda m_\pi\,,
\end{equation*}
where $\lambda$ is the HHChPT expansion parameter. In HHChPT $\lambda=\frac{|\vec q|}{M_D}$ with $|\vec q|\sim m_\pi$. In lattice calculations in which the pion mass varies between 130\,MeV and 150\,MeV, the D meson mass remains almost constant.

\subsubsection{XEFT Lagrangian Isospin Symmetric Case}
\label{sec:II.3.4.1.1}

We match the HHChPT Lagrangian without isospin breaking to the XEFT term by term. The leading order heavy meson-pion interaction terms in HHChPT are
\begin{equation}\label{eq:72ceq4}
\mathcal{L}_{L.O.}^{D\pi}=-i\tr[\bar H_a v_\mu D_{ab}^\mu H_b]+g_\pi\tr[\bar H_a H_b\gamma_\mu \gamma_5\mba_{ba}^\mu]\,.
\end{equation}
In the rest frame of the heavy mesons, we have
\begin{align}
\mathcal{L}_{L.O.}^{D\pi}&=[(\overrightarrow{P}_a^{*\dg}\cdot \vec\sigma+P_a^\dg)iD_{ba}^0(\overrightarrow{P}_b^\cdot\vec\sigma+P_b)]\nn\\
&-g_\pi[(\overrightarrow{P}_a^{*\dg}\cdot\vec\sigma+P_a^\dg)(\overrightarrow{P}_b^*\cdot\vec\sigma+P_b)\vec\sigma\cdot\overrightarrow{\mba_{ba}}]\,,\label{eq:72ceq5}
\end{align}
where $D_{ba}^0$ and $\vec\sigma\cdot\vec\mba_{ba}$ are given in \eq{72beq13}.  Using the identities for arbitrary $SU(2)$ vectors $\vec a$ and $\vec b$,
\begin{align}
(\vec a&\cdot\vec \sigma)(\vec b\cdot\vec \sigma)=(\vec a\cdot\vec b)I+(\vec a\times\vec b)\cdot\vec \sigma, \label{eq:72ceq7}\\
\tr&[\vec\sigma\cdot\vec a]=0    \label{eq:72ceq8}\,,
\end{align}
we reduce \req{eq:72ceq5}
\begin{align}
\mathcal{L}_{L.O.}^{D\pi}&=2\vec P_a^\dg\,iD_{ba}^0 \vec P_b+2P_a^\dg (iD_{ba}^0)P_b\nn\\
&+\frac{2g_\pi}{f_\pi}(\vec P_a^\dg\times\vec P_b)\cdot\vec\nabla M_{ba}+\frac{2g_\pi}{f_\pi}(\vec P_a^\dg\cdot P_b\vec\nabla M_{ba})\nn\\
&+\frac{2g_\pi}{f_\pi}(P_a^\dg\vec P_b\cdot\vec\nabla M_{ba}),\label{eq:72ceq9}
\end{align}
where we suppress (*) on the vector particles and the indices $a$ and $b$ are $SU(2)$ flavor indices.  Here in the first term, and below, the scalar product between the vector particles is implied when no other possible contractions are indicated.

In the same way, we reduce the rest of the leading order HHChPT Lagrangian in $SU(2)$ spin space,
\begin{align}
\mathcal{L}_{L.O.}^{D^*D}&=\frac{\Delta}{4}\tr[h_a^\dagger \sigma^i h_a\sigma^i]\nn\\
&=\frac{\Delta}{4}[6P_a^\dagger P_a-2\vec P_a^\dagger\cdot \vec P_a]\,,\label{eq:72ceq10}
\end{align}
\begin{align}
\mathcal{L}_{L.O.}^\pi&=\frac{f_\pi^2}{8}\tr[\partial^\mu\Sigma\partial_\mu\Sigma]+\frac{f_\pi^2 B_0}{4}\tr[m_g^\dagger\Sigma+m_g\Sigma^\dg]\nn\\
&=\pd_\mu\vec \pi\cdot\pd^\mu\vec \pi-\oneov{f_\pi}(\pi^i\pd_\mu \pi^j\pd^\mu\pi^k-\pd_\mu \pi^i(\pd^\mu\pi^j)\pi^k)\epsilon_{ijk}\nn\\
&+\frac{f_\pi^2B_0}{2}(m_u+m_d)-B_0(m_u+m_d)\vec\pi^2-iB_0(m_u-m_d)\epsilon^{ij3}\pi^i\pi^j  
+\ldots\,,\label{eq:72ceq11}
\end{align}
where
\begin{align}
\pi^1=\frac{1}{\sqrt{2}}(\pi^++\pi^-), \quad
\pi^2=-\frac{i}{\sqrt{2}}(\pi^+-\pi^-), \quad
\pi^3=\pi^0\label{eq:72ceq12}\,.
\end{align}
Similarly, for the next-to-leading order mass splitting terms, we have
\begin{align}
\mathcal{L}_{L.O.}^{M_1}&=\sigma_1\tr[h_a^\dg h_aM_{\xi bb}^{(+)}]\nn\\
&=-2\sigma_1(\vec P_a^\dg\cdot\vec P_a+P_a^\dg P_a)\bigl[(m_u+m_d)   
-\frac{4}{f_\pi^2}M_qM_{bb}^2+\mathcal{O}\paren{\frac{1}{f_\pi^4}}\bigl]\label{eq:72ceq13} \\
&\equiv \mathcal{L}_{L.O.}^{M_1(\opn{old})}, \nn \\
\mathcal{L}_{N.L.O.}^{M_2}&=\frac{\Delta^{(\sigma_1)}}{4}\tr[h_a^\dg\sigma_i h_a\sigma_i M_{\xi bb}^{(+)}]\nn\\
&=-\frac{\Delta^{(\sigma_1)}}{2}(\vec P_a^\dg\cdot \vec P_a-3P_a^\dg P_a)\Bigg[m_u+m_d   
-\frac{4}{f_\pi^2}M_qM_{bb}^2+\mathcal{O}\paren{\oneov{f_\pi^4}}\Bigg]\,.\label{eq:72ceq14} \\
&\equiv \mathcal{L}_{N.L.O.}^{M_2(\opn{old})}. \nn 
\end{align}
The next-to-leading order velocity reparameterization invariant terms become
\begin{align}
\mathcal{L}_{N.L.O.}^{VRI_1}&=-\oneov{2M_D}\tr[\bar H_a(iD)_{ab}^2H_b]\nn\\
&=-\oneov{2M_D}\tr[\bar H_a((i\pd)^2\delta_{ab}-i\pd_\mu \mathbb{V}_{ab}^\mu-\mathbb{V}_{ab}^\mu i\pd_\mu)H_b]+\mathcal{O}\paren{\oneov{f_\pi^4}}\nn\\
&=-\oneov{2M_D}2\Bigg\{\vec P_a^\dg(i\pd)^2\vec P_a+P_a^\dg(i\pd)^2P_a\nn\\
&+(i\pd_\mu\vec P_a^\dg)\paren{\frac{i}{f_\pi^2}M_\pi \overleftrightarrow{\pd^\mu}M_\pi}_{ab}\vec P_b
+(i\pd_\mu P_a^\dg)\paren{\frac{i}{f_\pi^2}M_\pi\overleftrightarrow{\pd^\mu}M_\pi}_{ab}P_b\nn\\
&+\mathcal{O}\paren{\oneov{f_\pi^4}}\Bigl\}\,,\label{eq:72ceq15}
\end{align}
where $\overleftrightarrow{\pd^\mu}=\frac12(\overrightarrow{\pd^\mu}-\overleftarrow{\pd^\mu})$ and
\begin{align}
\mathcal{L}_{N.L.O.}^{V.R.I_2}&=\frac{g}{M_D}\tr[\bar H_c(i\overleftarrow{D}_{ac}^\mu v\cdot \mathbb{A}_{ba}-iv\cdot\mathbb{A}_{ac}\vec D_{ba}^\mu)H_b\gamma_\mu\gamma_5]\nn\\
&=-\frac{g}{M_D}\Bigg\{(-i)\paren{\frac{-\pd^0}{f_\pi}}M_{ba}[(i\vec \nabla\times\vec P_a^\dg)\cdot\vec P_b+\vec P_a^\dg\cdot(i\vec\nabla\times\vec P_b)]\nn\\
&+\paren{\frac{-\pd^0}{f_\pi}M_{ba}}[(i\vec\nabla\cdot\vec P_a^\dg)P_b+P_a^\dg(i\vec\nabla\cdot \vec P_b)]\nn\\
&+\paren{\frac{-\pd^0}{f_\pi}M_{ba}}[(i\vec \nabla P_a^\dg)\cdot\vec P_b+\vec P_a^\dg\cdot(i\vec \nabla P_b)]\nn\\
&+(-i\vec P_c^\dg\times\vec P_b+\vec P_c^\dg P_b+P_c^\dg\vec P_b)\cdot\paren{-\frac{\pd^0}{f_\pi}M_{ba}}\paren{\frac{i}{f_\pi^2}M_\pi\overleftrightarrow{\nabla}M_\pi}_{ac}\nn\\
&-(-i\vec P_c^\dg\times\vec P_b+\vec P_c^\dg P_b+P_c^\dg\vec P_b)\cdot\paren{\frac{i}{f_\pi^2}M_\pi\overleftrightarrow{\nabla}M_\pi}_{ba}\paren{-\frac{\pd^0}{f_\pi}M_{ac}}+\mathcal{O}\paren{\oneov{f_\pi^5}}  \nn \\
&\equiv\mathcal{L}_{N.L.O.}^{VRI_2(\opn{old})}. \label{eq:72ceq16}
\end{align}
The spin symmetry breaking term is
\begin{align}
\mathcal{L}_{N.L.O.}^{\rm spin}&=\frac{g_2}{M_D}\tr[\bar H_a\mathbb{A}_{ab}^\mu\gamma_\mu\gamma_5 H_b]\nn\\
&=-\frac{g_2}{M_D}\tr[(\vec P_a^\dg\cdot\vec\sigma+P_a^\dg)\vec\sigma\cdot\vec{\mathbb{A}}_{ab}(\vec P_b\cdot \sigma+P_b)]\nn\\
&=\frac{2g_2}{M_D}(-i\vec P_a^\dg\times\vec P_b+\vec P_a^\dg P_b+P_a^\dg\vec P_b)\cdot\!\paren{\frac{\vec \nabla M_\pi}{f_\pi}}_{ba}\,,\label{eq:72ceq17}
\end{align}
which renormalizes the leading order $D\pi$ interaction term.

The two Goldstone boson axial operators terms are
\begin{align}
\mathcal{L}_{N.L.O.}^{\delta_4,\delta_5}&=\frac{\delta_4}{M_D}\tr[\bar H_a H_b\mba_{bc}^\mu\mba_{ca,\mu}]+\frac{\delta_5}{M_D}\tr[\bar H_a H_b v\cdot\mba_{bc} v\cdot\mba_{ca}]\nn\\
&=-2\frac{\delta_4+\delta_5}{M_D}(\vec P_a^\dg\!\cdot\!\vec P_b+P_a^\dg P_b)\frac{(\pd^0 M)_{ba}^2}{f_\pi^2}
-2\frac{\delta_4}{M_D}(\vec P_a^\dg\!\cdot\!\vec P_b+P_a^\dg P_b)\frac{(\vec \nabla M)_{ba}^2}{f_\pi^2} \nn \\
&\equiv\mathcal{L}_{N.L.O.}^{\delta_4,\delta_5},  \label{eq:72ceq18}
\end{align}
and
\begin{align}
\mathcal{L}_{N.L.O.}^{\delta_6,\delta_7}&=\frac{4(\delta_6+\delta_7)}{M_D}\Bigg\{\vec P_a^\dg\cdot\!\paren{\frac{\vec\nabla M_{ba}}{f_\pi}}\vec P_c\cdot\!\paren{\frac{\vec\nabla M_{cb}}{f_\pi}}
-\vec P_a^\dg\cdot\!\paren{\frac{\vec\nabla M_{cb}}{f_\pi}}\vec P_c\cdot\!\paren{\frac{\vec\nabla M_{ba}}{f_\pi}}\Bigg\}\nn\\
&+\frac{4(\delta_6-\delta_7)}{M_D}i(\vec P_a^\dg P_c+P_a^\dg\vec P_c)\cdot\!\paren{\frac{\vec\nabla M_{ba}\times\vec\nabla M_{cb}}{f_\pi^2}} \nn \\
&\equiv\mathcal{L}_{N.L.O.}^{\delta_6,\delta_7(\opn{old})}. \label{eq:72ceq19}
\end{align}
The photon-$D$ meson interaction term is
\begin{align}
\mathcal{L}_{N.L.O.}^{D\gamma_1}&=-\frac{e\beta_1}{4M_D}\tr[\bar H_a H_b\sigma^{\mu\nu}F_{\mu\nu}]\nn\\
&=+\frac{e\beta_1}{M_D}[2\vec P_{aj}^\dg\vec P_{ak}F_{jk}+i\epsilon_{jkl}(\vec P_{aj}^\dg P_a+P_a^\dg\vec P_{aj})F_{kl}]\,,\label{eq:72ceq20}
\end{align}
in which the vector components of the $D$ multiplet are contracted component-wise with the spatial components of the field tensor.

Now we shift the mass difference between $D^*$ and $D$ by redefining the vector field $\vec P_a$ and the scalar field $P_a$ as
\begin{align}
\vec P_a\to \frac{1}{\sqrt{2}}e^{i\frac{3\Delta}{4}t}\vec P_a\,, \qquad 
P_a\to \frac{1}{\sqrt{2}}e^{i\frac{3}{4}\Delta t}P_a\,.\label{eq:72ceq21}
\end{align}
The following operators remain unchanged except for a factor of $1/2$,
\begin{align}
\msl_{L.O.}^{D^*D}\to\frac12\msl_{L.O.}^{D^*D}, \quad \msl_{N.L.O.}^{M_1}\to\frac12\msl_{N.L.O.}^{M_1},\quad \msl_{N.L.O.}^{M_2}\to\frac12\msl_{N.L.O.}^{M_2}, \nn\\
\msl_{N.L.O.}^{V.R.I_2}\to\frac12\msl_{N.L.O.}^{V.R.I_2},\quad \msl_{N.L.O.}^{D\gamma_1}\to\frac12\msl_{N.L.O.}^{D\gamma_1},\,\,\,\msl_{N.L.O.}^{\text{spin}}\to\frac12\msl_{N.L.O.}^{\text{spin}}, \nn\\
\msl_{N.L.O.}^{\delta_4,\delta_5}\to\frac12\msl_{N.L.O.}^{\delta_4,\delta_5},\quad \msl_{N.L.O.}^{\delta_6,\delta_7}\to\frac12\msl_{N.L.O.}^{\delta_6\delta_7}\,.\label{eq:72ceq22}
\end{align}
The leading order $D^*D$ kinetic term changes as
\begin{align}
\msl_{L.O.}^{D\pi}
&\to-\frac34\Delta(\vec P_a^\dg\cdot\vec P_a+P_a^\dg P_a)+(\vec P_a^\dg(i\pd^0)\vec P_a+P_a^\dg(i\pd^0)P_a),\nn\\
&+\frac{2g_\pi}{f_\pi}(\vec P_a^\dg\times\vec P_b)\cdot\vec\nabla M_{ba}+\frac{g_\pi}{f_\pi}(\vec P_a^\dg\cdot P_b\vec \nabla M_{ba})+\frac{g_\pi}{f_\pi}(P_a^\dg\vec P_a\cdot\vec\nabla M_{ba})\nn\\
&=\frac12\msl_{L.O.}^{D\pi(\opn{old})}-\frac{3\Delta}{4}(\vec P_a^\dg\cdot\vec P_a+P_a^\dg P_a)\,,\label{eq:72ceq23}
\end{align}
so that the leading-order Lagrangian becomes
\begin{equation}\label{eq:72ceq24}
\msl_{L.O.}=\msl_{L.O.}^\pi+\frac12\msl_{L.O.}^{D\pi(\opn{old})}+\frac12\msl_{L.O.}^{D^*D}+\paren{-\frac34}\Delta(\vec P_a^\dg\vec P_a+P_a^\dg P_a)\,,
\end{equation}
where the definition of $\mathcal{L}_{L.O.}^{D\pi(\opn{old})}$ can be found in \eq{72ceq9}.  The double-derivative velocity-reparameterization-invariant term becomes,
\begin{align}
\msl_{N.L.O.}^{V.R.I_1}&\to -\frac{1}{2M_D}\bigg\{\paren{\frac{3\Delta}{4}}^2(\vec P_a^\dg\cdot \vec P_a+P_a^\dg P_a)+\vec P_a^\dg(i\pd)^2\vec P_a+P_a^\dg(i\pd)^2P_a\nn\\
&+\frac{3\Delta}{4}(\vec P_a^\dg\vec P_b+P_a^\dg P_b)(M_\pi\overleftrightarrow{\pd^0}M_\pi)_{ba}-\frac{3\Delta}{2}(\vec P_a^\dg(i\pd^0)\vec P_a+P_a^\dg(i\pd^0)P_a)\nn\\
&-[(-i\pd_\mu \vec P_a^\dg)\vec P_b+(-i\pd_\mu P_a^\dg)P_b](M_\pi\overleftrightarrow{\pd^\mu}M_\pi)_{ba}
\bigg\}\,.\label{eq:72ceq25}
\end{align}

Next we redefine pions to non-relativistic fields by factoring out the large phase
\begin{equation}\label{eq:72ceq26}
\vec \pi\to\frac{1}{\sqrt{2m_\pi}}e^{-im_\pi t}\vec{\hat \pi}+\frac{1}{\sqrt{2m_\pi}}e^{im_\pi t}\vec{\hat{\pi}}^\dg.
\end{equation}
As a result, we have
\begin{align}
M_\pi&\to\begin{pmatrix}
\frac{1}{\sqrt{2}}\frac{1}{\sqrt{2m_\pi}}(e^{-im_\pi t}\hat \pi_0+e^{im_\pi t}\hat \pi_0^\dg) & \frac{1}{\sqrt{2m_\pi}}(e^{-im_\pi t}\hat \pi^++e^{im_\pi t}(\hat \pi^-)^\dg)\\
\frac{1}{\sqrt{2m_\pi}}(e^{-im_\pi t}\hat \pi^-+e^{im_\pi t}(\hat \pi^+)^\dg) & -\frac{1}{\sqrt{2}\sqrt{2m_\pi}}(e^{-im_\pi t}\hat \pi^0+e^{im_\pi t}(\hat \pi^0)^\dg)
\end{pmatrix}\nn\\
&=\frac{1}{\sqrt{2m_\pi}}(e^{-im_\pi t}\hat M_\pi+e^{im_\pi t}\hat M_\pi^\dg),\label{eq:72ceq27}
\end{align}
where
\begin{equation}\label{eq:72ceq28}
M_\pi=\begin{pmatrix}
\oneov{\sqrt{2}}\hat\pi_0 & \hat\pi^+ \\
\hat\pi^- & -\oneov{\sqrt{2}}\hat\pi_0
\end{pmatrix}, \quad
\hat M_\pi^\dg=\begin{pmatrix}
\frac{1}{\sqrt{2}}\hat \pi_0^\dg & (\hat \pi^-)^\dg\\
(\hat \pi^+)^\dg & -\frac{1}{\sqrt{2}}\hat \pi^{0\dg}
\end{pmatrix}.
\end{equation}
The pion axial vector becomes
\begin{align}
\mathbb{A}_{ab}^\mu&\to -\frac{1}{f_\pi}\frac{1}{\sqrt{2m_\pi}}\bigg(e^{-im_\pi t}\pd_\mu \hat M_\pi+e^{im_\mu t}\pd_\mu\hat M_\pi^\dg
-im_\pi e^{-im_\pi t}\hat M_\pi\delta^\mu_0+im_\pi e^{im_\pi t}\hat M_\pi^\dg\delta^\mu_0\bigg),\label{eq:72ceq29}
\end{align}
where $\delta^\mu_0$ is a Kronecker delta function, and the pion vector 
\begin{align}
\mbv_{ab}^\mu&\to\frac{i}{f_\pi^2}(M_\pi\overleftrightarrow{\pd_\mu}M_\pi)\nn\\
&=\frac{i}{f_\pi^2}\oneov{2m_\pi}[im_\pi\hat M_\pi\hat M_\pi^\dg\delta^\mu_0-im_\pi\hat M_\pi^\dg \hat M_\pi\delta^\mu_0+\hat M_\pi^\dg\overleftrightarrow{\pd_\mu}\hat M_\pi+\hat M_\pi\overleftrightarrow{\pd_\mu}\hat M_\pi^\dg].
\label{eq:72ceq29'}
\end{align}
The pion kinetic term becomes
\begin{align}
\msl_{L.O.}^\pi&\to\frac{1}{2m_\pi}\bigg\{
m_\pi^2\vec{\hat \pi}\vec{\hat \pi}^\dg +\vec{\hat \pi}(-im_\pi)(\pd_0\vec{\hat \pi}^\dg)+(\pd_0\vec{\hat\pi})(im_\pi\vec{\hat\pi}^\dg)\nn\\
&+\pd_\mu\vec{\hat\pi}\pd^\mu\vec{\hat\pi}^\dg+m_\pi^2\vec{\hat\pi}\vec{\hat\pi}^\dg+im_\pi\vec{\hat\pi}^\dg\pd_0\vec{\hat\pi}+(\pd_0\vec{\hat\pi}^\dg)(-im_\pi\vec{\hat\pi})\nn\\
&+\pd_\mu\vec{\hat\pi}^\dg\pd^\mu\vec{\hat\pi}\nn\\
&+e^{2im_\pi t}(im_\pi\delta_0^\mu+\pd^\mu)^\dg\vec{\hat\pi}^\dg(im_\pi\delta_\mu^0+\pd_\mu)\vec{\hat\pi}^\dg\nn\\
&+e^{-2im_\pi t}(-im_\pi\delta_0^\mu+\pd^\mu)\vec{\hat\pi}(-im_\pi\delta_\mu^0+\pd_\mu)\vec{\hat\pi}^\dg\bigg\}\nn\\
&-B_0\frac{(m_u+m_d)}{2m_\pi}(\vec{\hat\pi}\vec{\hat\pi}^\dg+\vec{\hat\pi}\vec{\hat\pi})^\dg\nn\\
&-B_0\frac{(m_u+m_d)}{2m_\pi}(e^{-2im_\pi t}(\vec{\hat \pi})^2+e^{2im_\pi t}(\vec{\hat \pi}^\dg)^2).
\label{eq:72ceq30}
\end{align}
The leading-order terms involving $D$ mesons become
\begin{align}
\frac12\msl_{L.O.}^{D\pi(\opn{old})}&=\vec P_a^\dg(i\pd^0)\vec P_a+P_a^\dg(i\pd^0)P_a\nn\\
&+\frac{g_\pi}{f_\pi}(\vec P_a^\dg\times\vec P_b)\cdot\vec\nabla M_{ba}+\frac{g_\pi}{f_\pi}(\vec P_a^\dg\cdot P_b\vec \nabla M_{ba})\nn\\
&+\frac{g_\pi}{f_\pi}(P_a^\dg\vec P_b\cdot\vec\nabla M_{ba})\label{eq:72ceq31}\\
&\to\vec P_a^\dg(i\pd^0)\vec P_a+P_a^\dg(i\pd^0)P_a\nn\\
&+\frac{g_\pi}{f_\pi}(\vec P_a^\dg\times\vec P_b)^\mu\frac{1}{\sqrt{2m_\pi}}(e^{-im_\pi t}\hat M_\pi+e^{im_\pi t}\hat M_\pi^\dg)_{ba}\delta^0_{\mu}\nn\\
&+\frac{g_\pi}{f_\pi}(\vec P_a^\dg\cdot P_a)\vec\nabla\frac{1}{\sqrt{2m_\pi}}(e^{-im_\pi t}\hat M_\pi+e^{im_\pi t}\hat M_\pi^\dg)_{ba}\nn\\
&+\frac{g_\pi}{f_\pi}(P_a^\dg\vec P_b\cdot\vec\nabla)\frac{1}{\sqrt{2m_\pi}}(e^{-im_\pi t}\hat M_\pi+e^{im_\pi t}\hat M_\pi^\dg)_{ba}\,,\label{eq:72ceq32}
\end{align}
where $\msl_{L.O.}^{D\pi\opn{(old)}}$ is defined in \eq{72ceq9}. For the next-to-leading order mass terms, we have
\begin{align}
\frac12\msl_{N.L.O.}^{M_1(\opn{old})}&\to (-1)(\vec P_a^\dg\cdot\vec P_a+P_a^\dg P_a)[(m_u+m_d)\nn\\
&-\frac{4}{f_\pi^2}M_q\frac{1}{2m_\pi}(\hat M_\pi\hat M_\pi^\dg+\hat M_\pi^\dg\hat M_\pi+e^{-2im_\pi t}\hat M_\pi^2+e^{2im_\pi t}(\hat M_\pi^\dg)^2)_{bb}+\mathcal{O}\paren{\frac{1}{f_\pi^2}}],\label{eq:72ceq33}\\
\frac12\msl_{N.L.O.}^{M_2(\opn{old})}&\to\frac{\Delta^{(\sigma_1)}}{4}[(-\vec P_a^\dg\cdot\vec P_a+3P_a^\dg P_a)]\cdot[(m_u+m_d)\nn\\
&-\frac{4}{f_\pi^2}M_q\frac{1}{2m_\pi}(\hat M_\pi\hat M_\pi^\dg+\hat M_\pi^\dg \hat M_\pi+e^{-2im_\pi t}\hat M_\pi^2+e^{2im_\pi t}(\hat M_\pi^2))_{bb}+\mathcal{O}\paren{\oneov{f_\pi^4}}],\label{eq:72ceq34}
\end{align}
where $\msl_{N.L.O.}^{M_1\opn{(old)}}$ and $\msl_{N.L.O.}^{M_2\opn{(old)}}$ are defined in \req{eq:72ceq13} and \req{eq:72ceq14}. The transformed V.R.I. term in \req{eq:72ceq25} becomes
\begin{align}
\msl_{N.L.O.}^{V.R.I_1}&\paren{\{\vec P_a,P_a\}\to e^{3\frac{i\Delta t}{4}}\{\vec P_a,P_a\}}\nn\\
&\to -\oneov{2M_D}\bigg\{\paren{\frac{3\Delta}{4}}^2(\vec P_a^\dg\cdot\vec P_a+P_a^\dg P_a)-\frac{3\Delta}{2}(\vec P_a^\dg(i\pd^0)\vec P_a+P_a^\dg(i\pd^0)P_a)\nn\\
&+\vec P_a^\dg(i\pd)^2\vec P_a+P_a^\dg(i\pd)^2 P_a\nn\\
&+\frac{i}{f_\pi^2}\frac{3\Delta}{4}(\vec P_a^\dg \vec P_b+P_a^\dg P_b)\cdot\frac{1}{2m_\pi}(im_\pi \hat M_\pi\hat M_\pi^\dg-im_\pi\hat M_\pi^\dg \hat M_\pi\nn\\
&+\hat M_\pi^\dg\overleftrightarrow{\pd_0}\hat M_\pi+\hat M_\pi\overleftrightarrow{\pd_0}\hat M_\pi^\dg)_{ba}\nn\\
&-\frac{i}{f_\pi^2}(-i\pd_\mu \vec P_a^\dg)\vec P_b+(-i\pd_\mu P_a^\dg)P_b]\frac{1}{2m_\pi}(im_\pi\hat M_\pi\hat M_\pi^\dg\delta_0^\mu-i\delta_0^\mu m_\pi\hat M_\pi^\dg\hat M_\pi\nn\\
&+\hat M_\pi^\dg\overleftrightarrow{\pd_\mu}\hat M_\pi+\hat M_\pi\overleftrightarrow{\pd_\mu}\hat M_\pi^\dg)_{ba}
\bigg\},\label{eq:72ceq35}
\end{align}
in which we drop the large phase $e^{\pm im_\pi t}$ terms. In \eq{72ceq16}, the other V.R.I. term becomes
\begin{align}
\frac12\msl_{N.L.O.}^{V.R.I_2\opn{(old)}}&\to-\frac{g_\pi}{2M_D}\bigg\{\paren{\frac{i}{f_\pi}}(-im_\pi e^{-im_\pi t}\hat M_\pi +e^{-im_\pi t}\pd_0\hat M_\pi+im_\pi e^{im_\pi t}\hat M_\pi^\dg\nn\\
&+e^{im_\pi t}\pd_0\hat M_\pi^\dg)\frac{1}{\sqrt{2m_\pi}}[(i\vec\nabla\times\vec P_a^\dg)\cdot\vec P_b+\vec P_a^\dg\cdot(i\vec\nabla\times P_b)]\nn\\
&+\paren{-\oneov{f_\pi}}(-im_\pi e^{-im_\pi t}\hat M_\pi+e^{-im_\pi t}\pd_0\hat M_\pi\nn\\
&+im_\pi e^{im_\pi t}\hat M_\pi^\dg+e^{im_\pi t}\pd_0\hat M_\pi^\dg)_{ba}\frac{1}{\sqrt{2m_\pi}}\nn\\
&\cdot[(i\vec \nabla\cdot\vec P_a^\dg)P_a+P_a^\dg(i\vec\nabla\cdot P_b)+(i\vec\nabla P_a^\dg)\cdot \vec P_b+\vec P_a\cdot (i\vec\nabla P_b)]\nn\\
&+\frac{1}{\sqrt{2m_\pi}}(-i\vec P_c^\dg\times \vec P_b+\vec P_c^\dg P_b+P_c^\dg\vec P_b)\cdot\paren{-\oneov{f_\pi}}\nn\\
&(-im_\pi e^{-im_\pi t}\hat M_\pi+e^{-im_\pi t}\pd_0\hat M_\pi+im_\pi e^{im_\pi t}\hat M_\pi^\dg+e^{im_\pi t}\pd_0\hat M_\pi^\dg)_{ba}\nn\\
&\cdot\oneov{2m_\pi}\bigl(\hat M_\pi\overleftrightarrow{\nabla}\hat M_\pi^\dg+\hat M_\pi^\dg\overleftrightarrow{\nabla}\hat M_\pi\nn\\
&+e^{-2im_\pi t}\hat M_\pi\overleftrightarrow{\nabla}\hat M_\pi+e^{2im_\pi t}\hat M_\pi^\dg\overleftrightarrow{\nabla}\hat M_\pi^\dg\bigl)_{ac}\nn\\
&-\oneov{\sqrt{2m_\pi}}(-i\vec P_c^\dg\times\vec P_b+\vec P_c^\dg P_b+P_c^\dg\vec P_b)\cdot\paren{\frac{i}{f_\pi^2}}\frac{i}{2m_\pi}\nn\\
&\cdot\paren{\hat M_\pi\overleftrightarrow{\nabla}\hat M_\pi^\dg+\hat M_\pi^\dg\overleftrightarrow{\nabla}\hat M_\pi+e^{-2im_\pi t}\hat M_\pi\overleftrightarrow{\nabla}\hat M_\pi+e^{2im_\pi t}\hat M_\pi^\dg\overleftrightarrow{\nabla}\hat M_\pi^\dg}_{ba}\nn\\
&\paren{-\oneov{f_\pi}}(-im_\pi e^{-im_\pi t}\hat M_\pi+e^{-im_\pi t}\pd_0\hat M_\pi+im_\pi e^{im_\pi t}\hat M_\pi^\dg+e^{im_\pi t}\pd_0\hat M_\pi^\dg)_{ac}\bigg\}.\label{eq:72ceq36}
\end{align}
The spin symmetry breaking term $\frac12\msl_{N.L.O.}^{\opn{spin}}$ has the same form as the leading order $D\pi$ interaction term. The two Goldstone boson axial operator terms in \eqs{72ceq18}{72ceq19} are
\begin{align}
\frac12\msl_{N.L.O.}^{\delta_4,\delta_5(\opn{old})}&\to(-1)\paren{\frac{\delta_4+\delta_5}{\Lambda}}(\vec P_a^\dg\cdot\vec P_b+P_a^\dg P_b)\paren{\oneov{f_\pi^2}}\nn\\
&\cdot[m_\pi^2(\hat M_\pi\hat M_\pi^\dg+\hat M_\pi^\dg \hat M_\pi)-im_\pi\hat M_\pi\pd_0\hat M_\pi^\dg-im_\pi(\pd_0\hat M_\pi^\dg)\hat M_\pi\nn\\
&+im_\pi\hat M_\pi^\dg\pd_0\hat M_\pi+im_\pi(\pd_0\hat M_\pi)\hat M_\pi^\dg+\pd_0\hat M_\pi\pd_0\hat M_\pi^\dg+\pd_0\hat M_\pi^\dg\pd_0\hat M_\pi]_{ba}\frac{1}{2m_\pi}\nn\\
&+(-1)\frac{\delta_4}{\Lambda}(\vec P_a^\dg\cdot \vec P_b+P_a^\dg P_b)\paren{\oneov{f_\pi^2}}\frac{1}{2m_\pi}\nn\\
&\cdot[\vec\nabla\hat M_\pi\vec\nabla\hat M_\pi^\dg+\vec\nabla\hat M_\pi^\dg\vec\nabla\hat M_\pi]_{ba},\label{eq:72ceq37}
\end{align}
again dropping the large phase $e^{\pm im_\pi t}$ terms.  Next,
\begin{align}
\frac12\msl_{N.L.O.}^{\delta_6,\delta_7\opn{(old)}}&\to\frac{2(\delta_6+\delta_7)}{\Lambda}\bigg\{\vec P_a^\dg\cdot\paren{-\oneov{f_\pi}}[e^{-im_\pi t}\vec\nabla\hat M_\pi+e^{im_\pi t}\vec\nabla \hat M_\pi^\dg]_{ba}\oneov{2m_\pi}\nn\\
&\paren{-\oneov{f_\pi}}\vec P_c\cdot(e^{-im_\pi t}\vec\nabla\hat M_\pi+e^{im_\pi t}\vec\nabla\hat M_\pi^\dg)_{cb}\nn\\
&-\vec P_a^\dg\cdot\paren{-\oneov{f_\pi}}(e^{-im_\pi t}\vec\nabla\hat M_\pi+e^{im_\pi t}\vec\nabla\hat M_\pi^\dg)_{ab}\oneov{2m_\pi}\nn\\
&\paren{-\oneov{f_\pi}}\vec P_c\cdot(e^{-im_\pi t}\vec\nabla\hat M_\pi+e^{im_\pi t}\vec\nabla\hat M_\pi^\dg)_{ba}\bigg\}\nn\\
&+\frac{2(\delta_6-\delta_7)}{\Lambda}\paren{\oneov{f_\pi^2}}\oneov{2m_\pi}\bigg\{i(\vec P_a^\dg P_c+P_a^\dg\vec P_c)\cdot\nn\\
&\cdot\left([e^{-im_\pi t}\vec\nabla\hat M_\pi+e^{im_\pi t}\vec\nabla \hat M_\pi^\dg]_{ba}\times[e^{-im_\pi t}\vec\nabla\hat M_\pi+e^{im_\pi t}\vec\nabla\hat M_\pi^\dg]_{cb}\right)\bigg\}.\label{eq:72ceq38}
\end{align}
The photon-$D$ meson interaction term $\frac12\msl_{N.L.O.}^{D\gamma_1}$ remains unchanged. 

Next we redefine the spin-excited states $\vec P_a\to e^{-im_\pi t}\vec P_a$, dropping the large phase $e^{\pm im_\pi t}$ in the Lagrangian.  This changes the leading order terms by
\begin{align}
\msl_{L.O.}&=\msl_{L.O.}^\pi+\msl_{L.O.}^{D\pi}+\msl_{L.O.}^\gamma,\label{eq:72ceq39}\\
\msl_{L.O.}^\pi&=\oneov{2m_\pi}\bigg\{2m_\pi^2\vec{\hat \pi}\vec{\hat \pi}^\dg+\pd_\mu\vec{\hat \pi}\pd^\mu\vec{\hat\pi}^\dg\nn\\
&+\pd_\mu\vec{\hat\pi}^\dg\pd^\mu\vec{\hat\pi}+2\vec{\hat\pi}(-im_\pi)(\pd_0\vec{\hat \pi}^\dg)+2(\pd_0\vec{\hat\pi})(im_\pi\vec{\hat\pi}^\dg)\bigg\}\nn\\
&-\frac{B_0(m_u+m_d)}{m_\pi}\vec{\hat \pi}\vec{\hat\pi}^\dg,\label{eq:72ceq40}\\
\msl_{L.O.}^{D\pi}&=\vec P_a^\dg(i\pd^0)\vec P_a+P_a^\dg(i\pd^0)P_a+\frac{g_\pi}{f_\pi}\oneov{\sqrt{2m_\pi}}(\vec P_a^\dg P_b)\cdot\vec\nabla\hat M_{ba}\nn\\
&+\vec P_a^\dg(+m_\pi)\vec P_a+\frac{g_\pi}{f_\pi}\oneov{\sqrt{2m_\pi}}(P_a^\dg\vec P_b\cdot\vec\nabla)\hat M_{ba}^\dg.\label{eq:72ceq41}
\end{align}
The next-to-leading order terms become
\begin{align}
\frac12\msl_{N.L.O.}^{M_1(\opn{old})}&\to\msl_{N.L.O.}^{M_1}=(-\sigma_1)(\vec P_a^\dg\cdot\vec P_a+P_a^\dg P_a)\nn\\
&\times[(m_u+m_d)-\frac{4}{f_\pi^2}M_q\oneov{2m_\pi}(\hat M_\pi\hat M_\pi^\dg+\hat M_\pi^\dg \hat M_\pi)_{bb}],\label{eq:72ceq42}\\
\frac12\msl_{N.L.O.}^{M_2(\opn{old})}&\to \msl_{N.L.O.}^{M_2}=\frac{\Delta^{(\sigma_1)}}{4}[(-\vec P_a^\dg\cdot\vec P_a+3P_a^\dg P_a)]\nn\\
&\times\left[(m_u+m_d)-\frac{4}{f_\pi^2}M_q\oneov{2m_\pi}(\hat M_\pi\hat M_\pi^\dg+\hat M_\pi^\dg\hat M_\pi)_{bb}\right],\label{eq:72ceq43}
\end{align}
\begin{align}
\msl_{N.L.O.}^{V.R.I_1}&(\{\vec P_a,P_a\}\to e^{\frac{3i\Delta}{4}}\{\vec P_a,P_a\})\nn\\
&\to \msl_{N.L.O.}^{V.R.I_1}=-\oneov{2M_D}\bigg\{
\paren{\frac{3\Delta}{4}}^2(\vec P_a^\dg\cdot\vec P_a+P_a^\dg P_a)-\frac{3\Delta}{2}(\vec P_a^\dg (i\pd^0)\vec P_a\nn\\
&+(m_\pi)\vec P_a^\dg\vec P_a+P_a^\dg(i\pd^0)P_a)+\vec P_a^\dg(i\pd)^2\vec P_a+2\vec P_a^\dg (m_\pi)(i\pd^0)\vec P_a\nn\\
&+\vec P_a^\dg(m_\pi)^2\vec P_a+P_a^\dg(i\pd)^2 P_a\nn\\
&+\frac{i}{f_\pi^2}\frac{3\Delta}{4}(\vec P_a^\dg \vec P_b+P_a^\dg P_b)\oneov{2m_\pi}(im_\pi\hat M_\pi\hat M_\pi^\dg-im_\pi \hat M_\pi^\dg\hat M_\pi\nn\\
&+\hat M_\pi^\dg\overleftrightarrow{\pd_0}\hat M_\pi+\hat M_\pi\overleftrightarrow{\pd_0}\hat M_\pi^\dg)_{ba}\nn\\
&-\frac{i}{f_\pi^2}(-i\pd_\mu\vec P_a^\dg)\vec P_b+\delta_0^\mu(-m_\pi)\vec P_a^\dg\vec P_b+(-i\pd_\mu P_a^\dg)P_b]\nn\\
&\cdot\oneov{2m_\pi}(im_\pi\hat M_\pi\hat M_\pi^\dg\delta_0^\mu-im_\pi\hat M_\pi^\dg \hat M_\pi\delta_0^\mu +\hat M_\pi^\dg\overleftrightarrow{\pd_\mu}\hat M_\pi+\hat M_\pi\overleftrightarrow{\pd_\mu}\hat M_\pi^\dg)_{ba}\bigg\},\label{eq:72ceq44}
\end{align}
in which the $\Delta$s arose from the phase redefinition $\{\vec P_a,P_a\}\to e^{\frac{3i\Delta}{4}}\{\vec P_a,P_a\}$, and
\begin{align}
\frac12\msl_{N.L.O.}^{V.R.I_2}&\to\msl_{N.L.O.}^{V.R.I_2}=\paren{-\frac{g_\pi}{M_D}}\bigg\{\paren{-\oneov{f_\pi}}\oneov{\sqrt{2m_\pi}}(-i\Delta\hat M_\pi+\pd_0\hat M_\pi)_{ba}[(i\vec\nabla\cdot\vec P_a^\dg)P_b\nn\\
&+\vec P_a^\dg\cdot(i\vec\nabla P_b)]+\paren{-\oneov{f_\pi}}\oneov{\sqrt{2m_\pi}}(i\Delta\hat M_\pi^\dg+\pd_0\hat M_\pi^\dg)_{ba}\nn\\
&\cdot[P_a^\dg(i\vec\nabla\cdot\vec P_b)+(i\vec\nabla P_a^\dg)\cdot\vec P_b]\nn\\
&+\oneov{\sqrt{2m_\pi}}\oneov{2m_\pi}\paren{-\frac{i}{f_\pi^3}}[(P_c^\dg\vec P_b)\cdot(i\Delta\hat M_\pi^\dg+\pd_0\hat M_\pi^\dg)_{ba}\nn\\
&+(\vec P_c^\dg P_b)(-i\Delta\hat M_\pi+\pd_0\hat M_\pi)_{ba}]\cdot(\hat M_\pi\overleftrightarrow{\nabla}\hat M_\pi^\dg+\hat M_\pi^\dg\overleftrightarrow{\nabla}\hat M_\pi)_{ac}\nn\\
&-\oneov{\sqrt{2m_\pi}}\oneov{2m_\pi}\paren{-\frac{i}{f_\pi^3}}[(\vec P_c^\dg P_b)\cdot(-i\Delta\hat M_\pi+\pd_0\hat M_\pi)_{ac}\nn\\
&+(P_c^\dg\vec P_b)\cdot(i\Delta\hat M_\pi^\dg+\pd_0\hat M_\pi^\dg)_{ac}]\cdot(\hat M_\pi\overleftrightarrow{\nabla}\hat M_\pi^\dg+\hat M_\pi^\dg\overleftrightarrow{\nabla}\hat M_\pi)_{ba}\nn\\
&+\oneov{\sqrt{2m_\pi}}\oneov{2m_\pi}\paren{-\frac{i}{f_\pi^3}}[(\vec P_c^\dg\cdot P_b)(i\Delta\hat M_\pi^\dg+\pd_0\hat M_\pi^\dg)_{ba}\nn\\
&\cdot(\hat M_\pi\overleftrightarrow{\nabla}\hat M_\pi)_{ac}+(P_c^\dg\vec P_b)(-i\Delta\hat M_\pi+\pd_0\hat M_\pi)_{ba}(\hat M_\pi^\dg\overleftrightarrow{\nabla}\hat M_\pi^\dg)_{ac}]\nn\\
&-\oneov{\sqrt{2m_\pi}}\oneov{2m_\pi}\paren{-\frac{i}{f_\pi^3}}[(\vec P_c^\dg P_b)(i\Delta\hat M_\pi^\dg+\pd_0\hat M_\pi^\dg)_{ac}(\hat M_\pi\overleftrightarrow{\nabla}\hat M_\pi)_{ba}\nn\\
&+(P_c^\dg \vec P_b)(-i\Delta\hat M_\pi+\pd_0\hat M_\pi)_{ac}(\hat M_\pi^\dg\overleftrightarrow{\nabla}\hat M_\pi^\dg)_{ba}]\bigg\},\label{eq:72ceq45}
\end{align}
\begin{align}
\frac12\msl_{N.L.O.}^{\delta_4,\delta_5(\opn{old})}&\to\msl_{N.L.O.}^{\delta_4,\delta_5}=(-1)\paren{\frac{\delta_4+\delta_5}{\Lambda}}(\vec P_a^\dg\cdot\vec P_b+P_a^\dg P_b)\paren{\oneov{f_\pi^2}}\oneov{2m_\pi}\nn\\
&[m_\pi^2(\hat M_\pi\hat M_\pi^\dg+\hat M_\pi^\dg\hat M_\pi)-i\Delta\hat M_\pi\pd_0\hat M_\pi^\dg-i\Delta(\pd_0\hat M_\pi^\dg)\hat M_\pi\nn\\
&+i\Delta\hat M_\pi^\dg\pd_0\hat M_\pi+i\Delta(\pd_0\hat M_\pi)\hat M_\pi^\dg+\pd_0\hat M_\pi\pd_0\hat M_\pi^\dg   
+\pd_0\hat M_\pi^\dg\pd_0\hat M_\pi]_{ba}\nn\\
&+(-1)\frac{\delta_4}{\Lambda}(\vec P_a^\dg\cdot \vec P_b+P_a^\dg P_b)\paren{\oneov{f_\pi^2}}\oneov{2m_\pi}\nn\\
&[\vec\nabla\hat M_\pi\vec\nabla\hat M_\pi^\dg+\vec\nabla \hat M_\pi^\dg\vec\nabla\hat M_\pi]_{ba},\label{eq:72ceq46}
\end{align}

\begin{align}
\frac12\msl_{N.L.O.}^{\delta_6,\delta_7(\opn{old})}&\to\msl_{N.L.O.}^{\delta_6,\delta_7}=\frac{2(\delta_6+\delta_7)}{\Lambda}\paren{\oneov{f_\pi^2}}\oneov{2m_\pi}\nn\\
&\bigg\{\vec P_a^\dg\cdot(\vec\nabla \hat M_\pi)_{ba}\vec P_c\cdot(\vec\nabla \hat M_\pi^\dg)_{cb} 
+\vec P_a^\dg\cdot(\vec\nabla \hat M_\pi^\dg)_{ba}\vec P_c(\vec \nabla\hat M_\pi)_{cb}\nn\\
&-\vec P_a^\dg\cdot(\vec\nabla\hat M_\pi)_{cb}\vec P_c\cdot(\vec\nabla\hat M_\pi^\dg)_{ba}  
-\vec P_a^\dg\cdot(\vec \nabla\hat M_\pi^\dg)_{cb}\vec P_c\cdot(\vec\nabla\hat M_\pi)_{ba}\bigg\},\label{eq:72ceq47}
\end{align}
and
\begin{align}
\frac12\msl_{N.L.O.}^{D\gamma_1}&\to\msl_{N.L.O.}^{D\gamma_1}=\frac{e\beta_1}{2M_D}[2\vec P_{aj}^\dg\vec P_{bk}F_{jk}+i\epsilon_{ijk}(\vec P_a^\dg e^{i\delta t}P_b  
+P_a^\dg e^{-i\delta t}\vec P_b)F_{kl}].\label{eq:72ceq48}
\end{align}

\subsubsection{XEFT Lagrangian with isospin breaking}
\label{sec:II.3.4.1.2}

We match the HHChPT Lagrangian with isospin breaking to the XEFT term by term. First we reduce the spin structure of the HHChPT Lagrangian, as in the last section. The leading-order heavy meson-pion interaction terms are
\begin{align}
\msl_{L.O.}^{D\pi}&=2\vec P_a^\dg[(i\pd^0)\vec P_a-eB^0(Q_{bb}'\vec P_a-Q_{ba}\vec P_b)]\nn\\
&+2P_a^\dg[(i\pd^0)P_a-eB^0(Q_{bb}'P_a-Q_{ba}P_a)]+\frac{2g_\pi}{f_\pi}(\vec P_a^\dg\times\vec P_b)\cdot[\vec\nabla M_{ba}+ie\vec B[Q,M]_{ba}]\nn\\
&+\frac{2g_\pi}{f_\pi}[\vec P_a^\dg\cdot P_b(\vec \nabla M_{ba}+ie\vec B[Q,M]_{ba})]+\frac{2g_\pi}{f_\pi}[P_a^\dg\vec P_b\cdot(\vec \nabla M_{ba}+ie\vec B[Q,M]_{ba})].\label{eq:72ceq49}
\end{align}
The leading-order pion kinetic and mass terms are
\begin{align}
\msl_{L.O.}^\pi&=(\pd_\mu\vec \pi)\pd^\mu\vec \pi+(\pd^\mu\vec \pi)(ieB_\mu[Q,\vec \pi])+ieB^\mu[Q,\vec \pi]\cdot\pd_\mu \vec \pi\nn\\
&-e^2B^\mu B_\mu[Q,\vec \pi]\cdot[Q,\vec\pi]-\oneov{f_\pi}\epsilon_{ijk}\bigg\{\pi^i(\pd_\mu\pi^j)\pd^\mu\pi^k+\pi^i(\pd^\mu\pi^j)(ieB_\mu[Q,\pi^k])\nn\\
&+\pi^i(ieB_\mu[Q,\pi^j])\pd^\mu\pi^k+\pi^i(ieB_\mu[Q,\pi^j])(ieB^\mu[Q,\pi^k])-\pd_\mu\pi^i(\pd^\mu\pi^j)\pi^k\nn\\
&-(ieB_\mu[Q,\mu^i])(\pd^\mu\pi^j)\pd^k-\pd_\mu\pi^i(ieB^\mu[Q,\pi^j])\pi^k-(ieB^\mu[Q,\pi^i])(ieB_\mu[Q,\pi^j])\pi^k\bigg\}\nn\\
&+\frac{f_\pi^2B_0}{2}(m_u+m_d)-B_0(m_u+m_d)\vec\pi^2-iB_0(m_u+m_d)\epsilon^{ij3}\pi^i\pi^j,\label{eq:72ceq50}
\end{align}
where $\pi^i$ is defined in \req{eq:72ceq12}. The remaining leading order terms are
\begin{align}
\msl_{L.O.}^{D^*D}&=\frac{\Delta}{4}(6P_a^\dg P_a-2\vec P_a^\dg\cdot\vec P_a),\label{eq:72ceq51}\\
\msl_{L.O.}^\gamma&=-\frac{1}{4}F^{\mu\nu}F_{\mu\nu}.
\end{align}
The next-to-leading order mass terms are
\begin{align}
\msl_{N.L.O.}^{M_1}&=(-\sigma_1)2(\vec P_a^\dg\cdot\vec P_a+P_a^\dg P_a)\left[(m_u+m_d)-\frac{4}{f_\pi^2}(M_qM_{\pi}^2)_{bb}\right],\label{eq:72ceq53}\\
\msl_{N.L.O.}^{M_2}&=\frac{\Delta^{(\sigma_1)}}{4}(-2\vec P_a^\dg\cdot \vec P_a+6P_a^\dg P_a)\left[(m_u+m_d)-\frac{4}{f_\pi^2}(M_qM_{\pi}^2)_{bb}\right],\label{eq:72ceq54}\\
\msl_{N.L.O.}^{M_3}&=\lambda_1\tr[\bar H_aH_bM_{\xi ba}^{(+)}]\nn\\
&=(-\lambda_1)2(\vec P_a^\dg\cdot \vec P_b+P_a^\dg P_b)\left[(M_q)_{ba}-\frac{4}{f_\pi^2}(M_qM_{\pi}^2)_{ba}\right],\label{eq:72ceq55}\\
\msl_{N.L.O.}^{M_4}&=\frac{\Delta^{(\lambda_1)}}{4}\tr[h_a^\dg\sigma_i h_b\sigma^i M_{\xi ba}^{(+)}]\nn\\
&=\frac{\Delta^{(\lambda_1)}}{4}(-2\vec P_a^\dg\cdot \vec P_b+6P_a^\dg P_b)\left[(M_q)_{ba}-\frac{4}{f_\pi^2}(M_qM_{\pi}^2)_{ba}\right].\label{eq:72ceq56}
\end{align}

The next-to-leading order velocity-reparameterization-invariant terms are
\begin{align}
\msl_{N.L.O.}^{V.R.I_1}&=-\oneov{M_D}\bigg\{\vec P_a^\dg(i\pd)^2\vec P_a+P_a^\dg(i\pd)^2P_a-(-i\pd^\mu\vec P_a^\dg)\paren{\frac{i}{f_\pi^2}M_\pi\overleftrightarrow{\pd_\mu}M_\pi}_{ba}\vec P_b\nn\\
&-(-i\pd_\mu P_a^\dg)\paren{\frac{i}{f_\pi^2}M_\pi\overleftrightarrow{\pd^\mu}M_\pi}_{ab}P_b\nn\\
&+\vec P_a^\dg\paren{e\frac{i}{f_\pi^2}M_\pi\overleftrightarrow{\pd^\mu}M_\pi}_{ab}B_\mu(\vec P_aQ_{cc}'-\vec P_bQ_{cb})-\vec P_a^\dg[ieB_\mu(\pd^\mu\vec P_a Q_{bb}'-\pd^\mu\vec P_bQ_{ba})]\nn\\
&+P_a^\dg\paren{e\frac{i}{f_\pi^2}M_\pi\overleftrightarrow{\pd^\mu}M_\pi}_{ab}B_\mu(P_aQ_{cc}'-P_bQ_{cb})-P_a^\dg[ieB_\mu(\pd^\mu P_aQ_{bb}'-\pd^\mu P_bQ_{ba})]\nn\\
&+(i\pd^\mu\vec P_a^\dg)[ieB_\mu(\vec P_aQ_{bb}'-\vec P_bQ_{ba})]\nn\\
&+(i\pd^\mu P_a^\dg)[ieB_\mu(P_aQ_{bb}'-P_bQ_{ba}]\nn\\
&+\vec P_a^\dg e^2B^\mu B_\mu(Q_{bb}'(\vec P_aQ_{cc}'-\vec P_cQ_{ca})-Q_{ba}(\vec P_bQ_{cc}'-\vec P_cQ_{cb}))\nn\\
&+P_a^\dg e^2B^\mu B_\mu(Q_{bb}'(P_aQ_{cc}'-P_cQ_{ca})-Q_{ba}(P_bQ_{cc}'-P_cQ_{cb}))\nn\\
&+\vec P_a^\dg eB_\mu\paren{Q_{bb}'\left[\paren{\frac{i}{f_\pi^2}M_\pi\overleftrightarrow{\pd^\mu}M_\pi}_{ac}\vec P_c\right]-Q_{ba}\left[\paren{\frac{i}{f_\pi^2}M_\pi\overleftrightarrow{\pd^\mu}M_\pi}_{cb}\vec P_c\right]}\nn\\
&+P_a^\dg eB_\mu\paren{Q_{bb}'\left[\paren{\frac{i}{f_\pi^2}M_\pi\overleftrightarrow{\pd^\mu}M_\pi}_{ac}P_c\right]-Q_{ba}\left[\paren{\frac{i}{f_\pi^2}M_\pi\overleftrightarrow{\pd^\mu}M_\pi}_{cb}P_c\right]}\bigg\},\label{eq:72ceq57}
\end{align}
and
\begin{align}
\msl_{N.L.O.}^{V.R.I_2}&=-\frac{g}{M_D}\bigg\{\frac{i}{f_\pi}(\pd^0M_{ba}\!+\!ieB^0[Q,M]_{ba})\bigl[(i\vec \nabla\!\times\!\vec P_a^\dg-e\vec B\!\times\!(Q_{cc}'\vec P_a^\dg-\vec P_c^\dg Q_{ca})\!\cdot\!\vec P_b\nn\\
&+\vec P_a^\dg\cdot(i\vec \nabla\times\vec P_b-e\vec B\times(Q_{cc}'\vec P_b-\vec P_cQ_{cb}))\bigl]+\frac{(-1)}{f_\pi}(\pd^0M_{ba}+ieB^0[Q,M]_{ba})\nn\\
&[(i\vec \nabla\cdot\vec P_a^\dg-e\vec B\cdot(Q_{cc}'\vec P_a^\dg-\vec P_c^\dg Q_{ca})P_b+P_a^\dg(i\vec\nabla\cdot \vec P_b-e\vec B\cdot(Q_{cc}'\vec P_b-\vec P_c Q_{cb})))]\nn\\
&+\frac{(-1)}{f_\pi}(\pd^0M_{ba}+ieB^0[Q,M]_{ba})\bigg(i\vec\nabla P_a^\dg-e\vec B(Q_{cc}'P_a^\dg-P_c^\dg Q_{ca})\bigg)\cdot\vec P_b\nn\\
&+\vec P_a^\dg\cdot(i\vec\nabla P_b-e\vec B(Q_{cc}'P_b-P_bQ_{cb}))]\nn\\
&+(-i\vec P_c^\dg\!\times\!\vec P_b+\vec P_c^\dg P_b+P_c^\dg\vec P_b)\cdot\!\paren{\frac{i}{f_\pi^2}M_\pi\!\overleftrightarrow{\nabla}\!M_\pi}_{\!ac}\!\!\frac{(-1)}{f_\pi}(\pd^0M_{ba}\!+\!ieB^0[Q,M]_{ba})\nn\\
&-(-i\vec P_c^\dg\!\times\!\vec P_b+\vec P_c^\dg P_b+P_c^\dg\vec P_b)\cdot\!\paren{\frac{i}{f_\pi^2}M_\pi\!\overleftrightarrow{\nabla}\!M_\pi}_{\!ba}\!\!\frac{(-1)}{f_\pi}(\pd^0M_{ac}\!+\!ieB^0[Q,M]_{ac}).\label{eq:72ceq58}
\end{align}

The spin symmetry breaking term is
\begin{equation}\label{eq:72ceq59}
\msl^{\rm spin}_{N.L.O.}=\frac{2g_2}{M_D}\paren{-\oneov{f_\pi}}(-i\vec P_a^\dg\times\vec P_b+\vec P_a^\dg P_b+P_a^\dg \vec P_b)\cdot[\vec\nabla M_{ba}+ie\vec B[Q,M]_{ba}].
\end{equation}
The two Goldstone boson axial operator terms are
\begin{align}
\msl_{N.L.O.}^{\delta_4,\delta_5}&=\frac{\delta_4\!+\!\delta_5}{\Lambda}(-2)(\vec P_a^\dg\!\cdot\!\vec P_b+P_a^\dg P_b)\oneov{f_\pi^2}\bigg[(\pd^0M_\pi)_{bc}(\pd^0M_\pi)_{ca}+ieB^0(\pd^0M_\pi)_{bc}[Q,M_\pi]_{ca}\nn\\
&+ieB^0[Q,M_\pi]_{bc}(\pd^0M_\pi)_{ca}-e^2(B^0)^2[Q,M_\pi]_{bc}[Q,M_\pi]_{ca}\bigg]\nn\\
&+\frac{\delta_4}{\Lambda}(-2)(\vec P_a^\dg\cdot\vec P_b+P_a^\dg P_b)\oneov{f_\pi^2}\bigg[(\vec\nabla M_\pi)_{bc}\cdot(\vec \nabla M_\pi)_{ca}+ie(\vec\nabla M_\pi)_{bc}\cdot\vec B[Q,M_\pi]_{ca}\nn\\
&+ie\vec B[Q,M_\pi]_{bc}\cdot(\vec\nabla M_\pi)_{ca}-e^2(\vec B)^2[Q,M_\pi]_{bc}[Q,M_\pi]_{ca}\bigg],\label{eq:72ceq60}
\end{align}
and
\begin{align}
\msl_{N.L.O.}^{\delta_6,\delta_7}&=\frac{4(\delta_6+\delta_7)}{\Lambda}\bigg\{\vec P_a^\dg\cdot\paren{-\frac{\vec\nabla M_{ba}}{f_\pi}}\vec P_c\cdot\paren{-\frac{\vec\nabla M_{cb}}{f_\pi}}\nn\\
&+\vec P_a^\dg\cdot\paren{-\frac{ie\vec B[Q,M_\pi]_{ba}}{f_\pi}}\vec P_c\cdot\paren{-\frac{-\vec\nabla M_{cb}}{f_\pi}}\nn\\
&+\vec P_a^\dg\cdot\paren{-\frac{\vec\nabla M_{ba}}{f_\pi}}\vec P_c\cdot\paren{-\frac{ie\vec B[Q,M_\pi]_{cb}}{f_\pi}}\nn\\
&+\vec P_a^\dg\cdot\paren{-\frac{ie\vec B[Q,M_\pi]_{ba}}{f_\pi}}\vec P_c\cdot\paren{-\frac{ie\vec B[Q,M_\pi]_{cb}}{f_\pi}}\nn\\
&+(ba\leftrightarrow cb\text{ indices exchanged })\bigg\}\nn\\
&+\frac{4(\delta_6-\delta_7)}{\Lambda}i(\vec P_a^\dg P_c+P_a^\dg \vec P_c)\paren{\oneov{f_\pi^2}}\bigg\{\vec\nabla M_{ba}\times\vec \nabla M_{cb}+ie\vec B[Q,M_\pi]_{ba}\times\vec\nabla M_{cb}\nn\\
&+\vec\nabla M_{ba}\times(ie\vec B[Q,M_\pi]_{cb})-e^2\vec B[Q,M_\pi]_{ba}\times\vec B[Q,M_\pi]_{cb}\bigg\}.\label{eq:72ceq61}
\end{align}
The photon-$D$ meson interaction terms are
\begin{align}
\msl_{N.L.O.}^{D\gamma_1}&=\frac{e\beta_1}{M_D}[2\vec P_{aj}^\dg\vec P_{ak}F_{jk}+i\epsilon_{jkl}(\vec P_{aj}^\dg P_a+P_a^\dg\vec P_{aj})F_{kl}],\label{eq:72ceq62}\\
\msl_{N.L.O.}^{D\gamma_2}&=-\frac{e\beta_2}{M_D}[2\vec P_{aj}^\dg\vec P_{bk}F_{jk}+i\epsilon_{jkl}(\vec P_{aj}^\dg P_b+P_a^\dg\vec P_{bj})F_{kl}]\paren{\oneov{f_\pi^2}}(M_\pi QM_\pi)_{ba}.\label{eq:72ceq63}
\end{align}
The next-to-leading order virtual photon-pion interaction term is
\begin{align}
\msl_{N.L.O.}^{\pi^+\pi^-}&=e^2f_\pi^2\tr[(Q^+\Sigma+\Sigma Q)^2]\nn\\
&=e^2f_\pi^2(\pi^+\pi^-+\pi^-\pi^+).\label{eq:72ceq64}
\end{align}

Now we shift the mass difference between $D^*$ and $D$ as \req{eq:72ceq21} with $\Delta=m_{D^{0*}}-m_{D^0}$. The following operators remain unchanged except for a factor of $1/2$
\begin{align}
\msl_{L.O.}^{D^*D}&\to\frac12\msl_{L.O.}^{D^*D},\,\,\,\, \msl_{N.L.O.}^{M_1}\to\frac12\msl_{N.L.O.}^{M_1},\,\,\, \msl_{N.L.O.}^{M_2}\to\frac12\msl_{N.L.O.}^{M_2}\nn\\
\msl_{N.L.O.}^{M_3}&\to\frac12\msl_{N.L.O.}^{M_3},\,\,\, \msl_{N.L.O.}^{M_4}\to\frac12\msl_{N.L.O.}^{M_4},\,\,\,\, \msl_{N.L.O.}^{V.R.I_2}\to\frac12\msl_{N.L.O.}^{V.R.I_2},\,\,\,\, \msl_{N.L.O.}^{\rm spin}\to\frac12\msl_{N.L.O.}^{\rm spin},\nn\\
\msl_{N.L.O.}^{\delta_4,\delta_5}&\to\frac12\msl_{N.L.O.}^{\delta_4,\delta_5},\,\,\,\, \msl_{N.L.O.}^{\delta_6,\delta_7}\to\msl_{N.L.O.}^{\delta_6,\delta_7},\,\,\,\, \msl_{N.L.O.}^{D\gamma_1}\to\frac12\msl_{N.L.O.}^{D\gamma_1},\,\,\,\, \msl_{N.L.O.}^{D\gamma_2}\to\frac12\msl_{N.L.O.}^{D\gamma_2}.\label{eq:72ceq65}
\end{align}
The photon kinetic term remains unchanged and the rest of the operators change as
\begin{align}
\msl_{L.O.}^{D\pi}&\to\frac12\msl_{L.O.}^{D\pi}+(-\Delta)\vec P_a^\dg \vec P_a,\label{eq:72ceq66}\\
\msl_{N.L.O.}^{V.R.I_1}&\to\frac12\msl_{N.L.O}^{V.R.I_1}-\oneov{2M_D}\bigg\{\paren{\frac{3\Delta}{4}}^2(\vec P_a^\dg\cdot\vec P_a+P_a^\dg P_a)\nn\\
&+\frac{3\Delta}{4}(\vec P_a^\dg \vec P_b+P_a^\dg P_b)(M_\pi\overleftrightarrow{\pd^0}M_\pi)
-\frac{3\Delta}{2}(\vec P_a^\dg(i\pd^0)\vec P_a+P_a^\dg(i\pd^0)P_a)\bigg\}.\label{eq:72ceq67}
\end{align}

Now we reduce the pion to non-relativistic fields  as in Eqs.\,\eqref{eq:72ceq26} to \eqref{eq:72ceq28} and absorb the large phase in pion-$D$ interaction terms into the spin-excited $D$ mesons by redefining $\vec P_a\to e^{-im_\pi t}\vec P_a$. For their leading order Lagrangians, we have
\begin{align}
\msl_{L.O.}^{D\pi}&=\vec P_a^\dg[(i\pd^0)\vec P_a-eB^0(Q_{bb}'\vec P_a-Q_{ba}\vec P_b)]+\vec P_a^\dg (m_\pi)\vec P_a+\paren{\frac{-3\Delta}{4}}(\vec P_a^\dg\vec P_a+P_a^\dg P_a)\nn\\
&+P_a^\dg[(i\pd^0)P_a-eB^0(Q_{bb}'P_a-Q_{ba}P_b)]\nn\\
&+\frac{g_\pi}{f_\pi}(\vec P_a^\dg\cdot P_b)\oneov{\sqrt{2m_\pi}}[(\vec \nabla\hat M_\pi)_{ba}+ie\vec B[Q,\hat M_\pi]_{ba}]\nn\\
&+\frac{g_\pi}{f_\pi}(P_a^\dg \vec P_b)\cdot\oneov{\sqrt{2m_\pi}}[(\vec\nabla \hat M_\pi^\dg)_{ba}+ie\vec B[Q,\hat M_\pi]_{ba}],\label{eq:72ceq68}
\end{align}
\begin{align}
\msl_{L.O.}^{\pi}&=\oneov{2m_\pi}\bigg\{ 2m_\pi^2\vec{\hat \pi}\vec{\hat \pi}^\dg+2\pd_\mu\vec{\hat \pi}\pd^\mu\vec{\hat\pi}^\dg+2\vec{\hat\pi}(-im_\pi)(\pd_0\vec{\hat\pi}^\dg)+2(\pd_0\vec{\hat\pi})(im_\pi\vec{\hat\pi}^\dg)\nn\\
&-\frac{B_0(m_u+m_d)}{m_\pi}\vec{\hat \pi}\vec{\hat \pi}^\dg
+\oneov{2m_\pi}(im_\pi\vec{\hat \pi}^\dg\delta_0^\mu+\pd^\mu\vec{\hat\pi}(ieB_\mu)[Q,\vec{\hat\pi}]\nn\\
&+\oneov{2m_\pi}(ieB^\mu)[Q,\vec{\hat\pi}]\cdot(im_\pi\vec{\hat\pi}^\dg\delta_0^\mu+\pd_\mu\vec{\hat\pi})\nn\\
&+\oneov{2m_\pi}(-im_\pi\vec{\hat \pi}\delta_\mu^0+\pd_\mu\vec{\hat \pi})(ieB^\mu)[Q,\vec{\hat\pi}^\dg]+\oneov{2m_\pi}(ieB^\mu)[Q,\vec{\hat\pi}^\dg](-im_\pi\vec{\hat\pi}\delta_\mu^0+\pd_\mu\vec{\hat\pi})\nn\\
&-e^2B^\mu B_\mu[Q,\vec{\hat\pi}^\dg][Q,\vec{\hat\pi}]-e^2B^\mu B_\mu[Q,\vec{\hat\pi}][Q,\vec{\hat{\pi}}^\dg]\nn\\
&=(\hat\pi^0)^\dg\paren{i\pd_0+\frac{\vnab^2}{2m_\pi}+\delta}\hat\pi^0+(\hat\pi^+)^\dg\paren{i\pd_0+\frac{\vnab^2}{2m_\pi}+\delta}\hat\pi^+\nn\\
&+(\hat\pi^-)^\dg\paren{i\pd_0+\frac{\vnab^2}{2m_\pi}+\delta}\hat\pi^-\nn\\
&+\frac{8ieB_\mu}{2m_\pi}[(\hat\pi^+)^\dg(\overleftarrow{\pd_\mu}-\overrightarrow{\pd_\mu})\hat\pi^+-(\hat\pi^-)^\dg(\overleftarrow{\pd_\mu}-\overrightarrow{\pd_\mu})\hat\pi^-]\nn\\
&+\oneov{2m_\pi}e^2B_\mu B^\mu[(\hat\pi^+)^\dg\hat\pi^++(\hat\pi^-)^\dg\hat\pi^-],\label{eq:72ceq69}
\end{align}
\begin{equation}\label{eq:72ceq70}
\msl_{L.O.}^{D^*D}=\frac{\Delta}{4}(3P_a^\dg P_a-\vec{P}_a^\dg \vec P_a),
\end{equation}
and
\begin{equation}\label{eq:72ceq71}
\msl_{L.O.}^\gamma=-\frac{1}{4}F^{\mu\nu}F_{\mu\nu}.
\end{equation}

The next-to-leading order mass terms are
\begin{align}
\lnlo^{M_1}&=(-\sigma_1)(\vpad \vpa+P_a^\dg P_a)[(m_u+m_d)-\frac{4}{f_\pi^2}M_q\oneov{2m_\pi}(\hmpi\hmpid+\hmpid\hmpi)_{bb}],\label{eq:72ceq72}\\
\lnlo^{M_2}&=\frac{\Delta^{(\sigma_1)}}{4}(-\vpad\vpa+3P_a^\dg P_a)[(m_u+m_d)-\frac{4}{f_\pi^2}M_q\oneov{2m_\pi}(\hmpi\hmpid+\hmpid\hmpi)_{bb}],\label{eq:72ceq73}\\
\lnlo^{M_3}&=(-\lambda_1)(\vpad\vpb+P_a^\dg P_b)[(M_q)_{ba}-\frac{4}{f_\pi^2}\frac{1}{2m_\pi}M_q(\hmpi\hmpid+\hmpid\hmpi)_{ba}],\label{eq:72ceq74}\\
\lnlo^{M_4}&=\paren{\frac{\Delta^{(\lambda_1)}}{4}}(-\vpad\vpb+3P_a^\dg P_b)[(M_q)_{ba}-\frac{4}{f_\pi^2}\oneov{2m_\pi}M_q(\hmpi\hmpid+\hmpid\hmpi)_{ba}].\label{eq:72ceq75}
\end{align}
The next-to-leading order velocity-reparameterization invariant terms are
\begin{align}
\lnlo^{V.R.I_1}&=-\oneov{M_D}\bigg\{\left[\vpad(i\pd)^2\vpa+2\vpad(\Delta)(i\pd)\vpa+\vpad(\Delta^2)\vpa\right]+P_a^\dg(i\pd)^2P_a\nn\\
&+\paren{\frac{3\Delta}{4}}^2(\vpad\cdot\vpa+P_a^\dg P_a)-\frac{3\Delta}{2}(\vpad(i\pd^0)\vpa+(\Delta)\vpad\vpa+P_a^\dg(i\pd^0)P_a)\nn\\
&+\frac{i}{f_\pi^2}\paren{\frac{3\Delta}{4}}(\vpad\vpb+P_a^\dg P_b)\oneov{2m_\pi}(i\Delta\hmpi\hmpid-i\Delta\hmpid\hmpi\nn\\
&+\hmpid \olra{\pd_0}\hmpi+\hmpi\olra{\pd_0}\hmpid)_{ba}\nn\\
&-[(-i\pd^\mu \vpad)\vpb+\delta_0^\mu (-\Delta)\vpad\vpb+(-i\pd^\mu)P_a^\dg P_b]\paren{\frac{i}{f_\pi^2}}\oneov{2m_\pi}\nn\\
&(i\Delta\hmpi\hmpid\delta_\mu^0-i\Delta\hmpid\hmpi\delta_\mu^0+\hmpid\olra{\pd_\mu}\hmpi+\hmpi\olra{\pd_\mu}\hmpid)_{ba}\nn\\
&+\vpad e^2B^\mu B_\mu[Q_{bb}'(\vpa Q_{cc}'-\vec P_cQ_{ca})-Q_{ba}(\vpb Q_{cc}'-\vec P_cQ_{cb})]\nn\\
&+P_a^\dg e^2B^\mu B_\mu[Q_{bb}'(P_aQ_{cc}'-P_c Q_{ca})-Q_{ba}(P_bQ_{cc}'-P_cQ_{cb})]\nn\\
&+(i\pd^\mu\vpad+\delta_0^\mu(-\Delta)\vpad)[ieB_\mu(\vpa Q_{bb}'-\vpb Q_{ba})]\nn\\
&+(i\pd^\mu P_a^\dg)[ieB_\mu(P_aQ_{bb}'-P_bQ_{ba})]-P_a^\dg[ieB_\mu(\pd^\mu P_aQ_{bb}'-\pd^\mu P_bQ_{ba})]\nn\\
&-\vpad[ieB_\mu(\pd^\mu \vpa Q_{bb}'-i\Delta\delta_0^\mu \vpa Q_{bb}'-\pd^\mu\vpb Q_{ba}+i\Delta\delta_0^\mu\vpb Q_{ba})]\nn\\
&+\vpad\paren{e\frac{i}{f_\pi^2}}\oneov{2\Delta}(i\Delta\hmpi\hmpid\delta_\mu^0-i\Delta\hmpid\hmpi\delta_\mu^0+\hmpid\olra{\pd_\mu}\hmpi+\hmpi\olra{\pd_\mu}\hmpid)_{ba}B^\mu\nn\\
&\cdot(\vpb Q_{cc}'-\vec P_c Q_{cb})\nn\\
&+P_a^\dg\paren{e\frac{i}{f_\pi^2}}\oneov{2m_\pi}(i\Delta\hmpi\hmpid\delta_\mu^0-i\Delta\hmpid\hmpi\delta_\mu^0+\hmpid\olra{\pd_\mu}\hmpi+\hmpi\olra{\pd_\mu}\hmpid)_{ba}\nn\\
&\cdot B^\mu (P_bQ_{cc}'-P_cQ_{cb})\nn\\
&+\vpad eB_\mu(Q_{bb}'[\frac{i}{2f_\pi^2m_\pi}(i\Delta\hmpi\hmpid\delta_\mu^0-i\Delta\hmpid\hmpi\delta_\mu^0+\hmpid\olra{\pd_\mu}\hmpi+\hmpi\olra{\pd_\mu}\hmpid)_{ca}\vec P_c])\nn\\
&-Q_{ba}[\frac{i}{2f_\pi^2m_\pi}(i\Delta\hmpi\hmpid\delta_\mu^0-i\Delta\hmpid\hmpi\delta_\mu^0+\hmpid\olra{\pd_\mu}\hmpi+\hmpi\olra{\pd_\mu}\hmpid)_{cb}\vec P_c]\nn\\
&+(\text{ changing }\vpa\to P_a\text{ for the last term}),\label{eq:72ceq76}
\end{align}
and
\begin{align}
\lnlo^{V.R.I_2}&=\frac{-g}{M_D}\bigg\{\paren{-\oneov{f_\pi}}\bigg[(-im_\pi\hmpi+\pd_0\hmpi+ieB_0[Q,\hmpi]_{ba})\cdot(i\vnab\cdot\vpad\nn\\
&-e\vec B(Q_{cc}'\vpad-\vec P_c^\dg Q_{ca})P_b)+(-im_\pi\hmpid+\pd_0\hmpid+ieB_0[Q,\hmpid]_{ba})P_a^\dg\nn\\
&(i\vnab\cdot\vpb-e\vec B\cdot(Q_{cc}'\vpb-\vec P_cQ_{cb}))\bigg]\nn\\
&+\paren{-\oneov{f_\pi}}\bigg[(im_\pi\hmpid+\pd_0\hmpid+ieB_0[Q,\hmpid])_{ba}(i\vnab P_a^\dg+e\vec B(Q_{cc}'P_a^\dg\nn\\
&-P_c^\dg Q_{ca}))\cdot\vpb + (-im_\pi\hmpi+\pd_0\hmpi+ieB_0[Q,\hmpi])_{ba}\vpad\cdot(i\vnab P_b\nn\\
&-e\vec B(Q_{cc}'P_b-P_cQ_{cb}))\bigg]\bigg\}\nn\\
&+\oneov{2m_\pi}\oneov{\sqrt{2m_\pi}}\paren{-\frac{i}{f_\pi^3}}\bigg[(P_c^\dg\vpb)\cdot(im_\pi\hmpid+\pd_0\hmpid+ieB_0[Q,\hmpid])_{ba}\nn\\
&+(\vec P_c^\dg P_b)(-im_\pi\hmpi+\pd_0\hmpi+ieB_0[Q,\hmpi])_{ba}\bigg]\cdot(\hmpi\olra{\nabla}\hmpid+\hmpid\olra{\nabla}\hmpi)_{ac}\nn\\
&-(ba\leftrightarrow ab\text{ for the previous term})\nn\\
&+\oneov{2m_\pi}\oneov{\sqrt{2m_\pi}}\paren{-\frac{i}{f_\pi^3}}\bigg[(\vec P_c^\dg\cdot P_b)\cdot(im_\pi\hmpid+\pd_0\hmpid+ieB_0[Q,\hmpid])_{ba}\nn\\
&\cdot(\hmpi\olra{\nabla}\hmpi)_{ac}\nn\\
&+(P_c^\dg\vpb)(-im_\pi\hmpi+\pd_0\hmpi+eB_0[Q,\hmpi])_{ba}(\hmpid\olra{\nabla}\hmpid)_{ac}\nn\\
&-(ba\leftrightarrow ab\text{ for the previous term}).\label{eq:72ceq77}
\end{align}
The spin symmetry breaking term is
\begin{align}
\lnlo^{\rm spin}&=\frac{g_2}{M_D}\bigg[(\vpad P_b)\paren{-\oneov{f_\pi}}[\vnab \hmpi+ie\vec B[Q,\hmpi]\bigg]_{ba}\oneov{\sqrt{2m_\pi}}\nn\\
&+\frac{g_2}{M_D}\bigg[\vnab\hmpid+ie\vec B[Q,\hmpid]\bigg]_{ba}\oneov{\sqrt{2m_\pi}}.\label{eq:72ceq78}
\end{align}
The two Goldstone-boson-axial operator terms are,
\begin{align}
\lnlo^{\delta_4,\delta_5}&=\frac{\delta_4+\delta_5}{\Lambda}(\vpad\vpb+P_a^\dg P_b)\paren{\oneov{f_\pi^2}}\oneov{2m_\pi}\bigg\{[m_\pi^2(\hmpi\hmpid+\hmpid\hmpi)\nn\\
&-im_\pi\hmpi\pd_0\hmpid-im_\pi(\pd_0\hmpid)\hmpi+im_\pi\hmpid\pd_0\hmpi+im_\pi(\pd_0\hmpi)\hmpid\nn\\
&+\pd_0\hmpi\pd_0\hmpid+(\pd_0\hmpid)(\pd_0\hmpi)]_{ba}+ieB^0(-im_\pi\hmpi+\pd_0\hmpi)_{bc}[Q,\hmpid]_{ba}\nn\\
&+ieB^0(im_\pi\hmpid+\pd_0\hmpid)_{bc}[Q,\hmpi]_{ba}+ieB^0[Q,\hmpid]_{bc}(-im_\pi\hmpi+\pd_0\hmpi)_{ba}\nn\\
&+ieB^0[Q,\hmpi]_{bc}(im_\pi\hmpid+\pd_0\hmpid)_{ba}\nn\\
&-e^2(B^0)^2[Q,\hmpid]_{bc}[Q,\hmpi]_{ca}-e^2(B^0)^2[Q,\hmpi]_{bc}[Q,\hmpid]_{ca}\bigg\}\nn\\
&+\frac{\delta_4}{\Lambda}(\vpad\cdot\vpb+P_a^\dg P_b)\paren{\oneov{f_\pi}}\bigg\{(\vnab\hmpid)_{bc}(\vnab\hmpi)_{ca}+(\vnab\hmpi)_{bc}(\vnab\hmpid)_{ca}\nn\\
&+ie(\vnab\hmpid)_{bc}\cdot\vec B[Q,\hmpi]_{ca}+ie(\vnab\hmpi)_{bc}\cdot\vec B[Q,\hmpid]_{ca}\nn\\
&+ie\vec B[Q,\hmpi]_{bc}\cdot(\vnab\hmpid)_{ca}+ie\vec B[Q,\hmpid]_{bc}\cdot(\vnab\hmpi)_{ca}\nn\\
&-e^2(\vec B)^2[Q,\hmpid]_{bc}[Q,\hmpi]_{ca}-e^2(\vec B)^2[Q,\hmpi]_{bc}[Q,\hmpid]_{ca}\bigg\}\oneov{2m_\pi},\label{eq:72ceq79}
\end{align}
and
\begin{align}
\lnlo^{\delta_6,\delta_7}&=\frac{2(\delta_6+\delta_7)}{\Lambda}\bigg\{\vpad\paren{-\frac{\vnab\hat M_{ba}}{f_\pi}}\vec P_c\paren{-\frac{\vnab\hat M_{cb}^\dg}{f_\pi}}+\vpa\paren{-\frac{\vnab\hat M_{ba}^\dg}{f_\pi}}\vec P_c\paren{-\frac{\vnab\hat M_{ab}}{f_\pi}}\nn\\
&+\vpad\cdot\paren{-\frac{ie\vec B[Q,\hmpi]_{ba}}{f_\pi}}\vec P_c\paren{-\frac{\vnab\hat M_{cb}^\dg}{f_\pi}}\nn\\
&+\vpad\cdot\paren{-\frac{ie\vec B[Q,\hmpid]_{ba}}{f_\pi}}\vec P_c\paren{-\frac{\vnab\hat M_{cb}}{f_\pi}}\nn\\
&+\vpad\paren{-\frac{\vnab\hat M_{ba}}{f_\pi}}\vec P_c\paren{-\frac{ie\vec B[Q,\hmpid]_{cb}}{f_\pi}}\nn\\
&+\vpad\paren{-\frac{\vnab\hat M_{ba}^\dg}{f_\pi}}\vec P_c\cdot\paren{-\frac{ie\vec B[Q,\hmpi]_{cb}}{f_\pi}}\nn\\
&+\vpad\paren{-\frac{ie\vec B[Q,\hmpi]_{ba}}{f_\pi}}\vec P_c\cdot\paren{-\frac{ie\vec B[Q,\hmpid]_{ab}}{f_\pi}}\nn\\
&+\vpad\cdot\paren{-\frac{ie\vec B[Q,\hmpid]_{ba}}{f_\pi}}\vec P_c\cdot\paren{-\frac{ie\vec B[Q,\hmpi]_{cb}}{f_\pi}}\bigg\}\oneov{2m_\pi}.\label{eq:72ceq80}
\end{align}
The photon-D meson interaction terms are
\begin{align}
\lnlo^{D\gamma_1}&=\frac{e\beta_1}{M_D}[\vec P_{aj}^\dg\vec P_{ak}F_{jk}+i\epsilon_{jkl}(\vpad P_ae^{i\delta t}+P_a^\dg\vec P_{ak}e^{-i\delta t})F_{kl}],\label{eq:72ceq81}\\
\lnlo^{D\gamma_2}&=-\frac{e\beta_2}{M_D}[\vec P_{aj}^\dg\vec P_{bk}F_{jk}+i\epsilon_{jkl}(\vpad P_be^{i\delta t}+P_a^\dg \vec P_{bj}e^{-i\delta t})F_{kl}]\nn\\
&\oneov{2m_\pi}\paren{\oneov{f_\pi^2}}[\hmpi Q\hmpid+\hmpid Q\hmpi+e^{-2i\delta t}\hmpi Q\hmpi\nn\\
&+e^{2i\delta t}\hmpid Q\hmpid]_{ba}.\label{eq:72ceq82}
\end{align}
Finally, the next-to-leading order virtual photon-pion interaction term is
\begin{equation}\label{eq:72ceq83}
\lnlo^{\pi^+\pi^-}=\frac{e^2f_\pi^2}{2m_\pi}[(\pi^+)(\pi^+)^\dg + (\pi^-)(\pi^-)^\dg].
\end{equation}

\subsection{Power-counting for \texorpdfstring{$D\pi$}{D-pi} Scattering Amplitude Near \texorpdfstring{$D^*$}{D-star} Threshold}
\label{sec:II.3.5}

We obtained the XEFT Lagrangian up next-to-next-to-next-to-leading order in the previous section and derive the corresponding Feynman rules in Appendix \appx{D.2}.  As we have emphasized several times, an EFT is established to systematically expand the experimental observables into a perturbation series, and using the above Lagrangian we power count the $D\pi$ scattering amplitude near the $D^*$ threshold.  In this section, we first discuss the power-counting scheme for both isospin-conserving and violating cases. Second, we compute the resummed isospin-conserving $D\pi$ scattering amplitude to leading order. After this, we determine the poles of the leading order amplitude in momentum space and discuss the significance of this resummation.

\subsubsection{Isospin-Conserving Terms}

\begin{figure}[th]
\centering
\includegraphics[width=.75\textwidth]{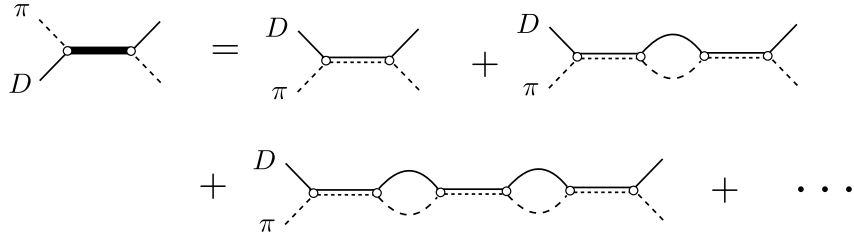}
\caption{Leading order $D\pi$ scattering amplitude in isospin symmetric case}
\label{fig:105fig1}
\end{figure}

We start by treating $D^0$ and $D^+$ as well as $\pi^0, \pi^+$ as identical particles.  The tree-level and one-loop correction to $D\pi$ scattering are represented by the first and second diagrams in the series on the right hand side of \fig{105fig1}.  To obtain the corresponding amplitudes, we need the Feynman rule for the $DD^*\pi$ vertex, which from $\mathcal{L}_{L.O.}^{D\pi}$ in \eq{72ceq41} (or \eq{72ceq68} since the L.O. vertex is the same also with isospin breaking) is 
\begin{equation}
\cL_{D^*D\pi}=ig\frac{i\vec{q}_\pi\cdot\vec\epsilon}{2\sqrt{m_\pi}f_\pi}\,,
\end{equation}
where $g$ is the D-meson axial transition coupling, $\vec\epsilon$ is the polarization vector of $D^{0*}$, and $\vec{q}_\pi$ is the three momentum of the pion. The $D$, $D^*$ and $\pi$ propagators take the form 
\begin{align}
G_D(\vec p)\:&\:=\frac{i}{E-\frac{\vec{p}_{D^0}^2}{2m_{D^0}}+i\epsilon}\,,\nn \\
G_{D^*}(\vec p)&\:=\frac{i}{E_{D^{0*}}}\,, \\
G_{\pi}(\vec p)&\:=\frac{i}{E-\frac{\vec{p}_\pi^2}{2m_{\pi^0}}+\delta+i\epsilon}\,, \nn
\end{align}
where $\delta=\Del-m_\pi$ is the difference between $DD^*$ mass splitting and pion mass and $E_{D^*}=\frac{|\vec p_\pi|^2}{2m_\pi}+\frac{|\vec p_D|^2}{2M_D}-\delta$ is the energy of $D^*$ in the center of mass frame of $D$ and $\pi$. 

The amplitude for tree-level $D\pi$ scattering (represented by the first diagram on the right hand side) is 
\be\label{Dpitree}
\cA_{D\pi\to D\pi}^{(\text{tree})}=\frac{g^2}{f_\pi^2}\frac{\delta}{E_{D^*}}\,,
\ee  
while the amplitude with the one-loop correction to the $D^*$ propagator gives
\be\label{Dpi1loop}
\cA_{D\pi\to D\pi}^{(\text{1-loop})}=\frac{g^2}{f_\pi^2}\frac{\delta}{E_{D^*}}\paren{\frac{g^2(2m_\pi\del)^{3/2}}{4\pi f_\pi^2}}\ov{E_{D^*}}\sim \frac{g^2}{f_\pi^2}\frac{\delta}{E_{D^*}}\frac{1}{E_{D^*}}(g^2\lambda\epsilon^3m_\pi)\,, 
\ee
where the factor inside the bracket is the loop contribution, namely
\be
\mathscr{D}_{D\pi}=\frac{g^2(2m_\pi\delta)^{3/2}}{4\pi f_\pi^2}\sim g^2\epsilon^3\lambda m_\pi\sim 4\pi g^2\alpha\lambda m_\pi
\ee
with a power counting scheme $\epsilon^2\sim e^2$.
As the pion mass is varied between 130\,MeV and 150\,MeV, $\delta$ changes sign from `+' in the physical $D\pi$ scattering case to `-' indicating $D^*$ is a bound state as in the lattice calculations.  Also, $E_{D^*}$ varies around 0.2\,MeV, very close to the $D^*$ threshold, so that $D\pi$ loop corrections to the $D^*$ propagator are leading order and need to be resummed, as shown in \fig{105fig1}. 

\begin{figure}[th]
\centering
\includegraphics[width=.75\textwidth]{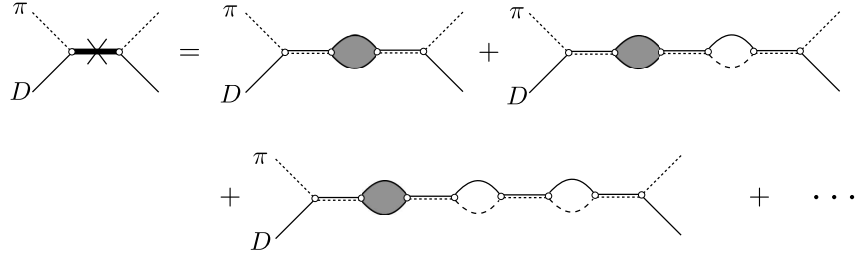}
\caption{Next-to-leading order $D\pi$ scattering near $D^*$ threshold amplitude without isospin breaking}
\label{fig:105fig2}
\end{figure}

\begin{figure}[!h]
\centering
\includegraphics[width=.4\textwidth]{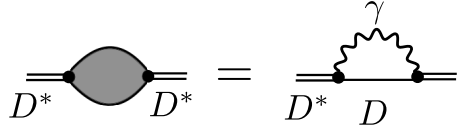}
\caption{One sector next-to-leading order correction to $D^*$ propagator}
\label{fig:105fig3}
\end{figure}

The next-to-leading order $D\pi$ scattering amplitude is shown in \fig{105fig2} where the shaded bubble indicates a next-to-leading order correction to the $D^*$ propagator. In this power counting, the required next-to-leading-order correction is the electromagnetic radiative correction to $D^*$, shown in \fig{105fig3}.  The $D^*\to D\gamma$ Feynman rule is obtained from the $\cL_{N.L.O.}^{D\gamma_1}$ and $\cL_{N.L.O.}^{D\gamma_2}$ terms of the XEFT Lagrangian \eqs{72ceq81}{72ceq82}. This diagram gives
\be
\mathscr{D}_{\rm{D\gamma}}=\ov{9}\frac{e^2}{4\pi}\frac{m_\pi^3}{M_D^2}\paren{\beta_1}^2\simeq \frac{1}{9}\alpha\lambda^2(\beta_1)^2m_\pi\,.
\ee
With the numerical value $(\beta_1)^2\simeq 12$ (\cite{Stewart:1998ke}), $\frac{\lambda(\beta_1)^2}{36\pi g^2}\simeq 0.021$, justifying our power counting. Thus $(\beta_1)^2$ is a natural size constant, and this loop correction is subleading to the $D\pi$ loop,
\be
\frac{\mathscr{D}_{\rm{D\gamma}}}{\mathscr{D}_{D\pi}} \sim\frac{\lambda(\beta_1)^2}{36\pi g^2}, 
\ee
where the contribution of the $D\pi$ loop is isolated in parentheses in \req{Dpi1loop}. Examples of other subleading diagrams can be found in Appendix \appx{D.3}.

\subsubsection{Isospin-Breaking Terms}
Isospin symmetry is broken in the leading order lagrangian  by the electric charge of the $D$ mesons in \eqs{72ceq49}{72ceq50}, but the corresponding amplitudes turn out to be next-to-leading order due to the power counting \eq{72beq81}.  Isospin symmetry is conserved by all leading order $D\pi$ interactions at lagrangian level allowed by charge conservation.  Mass difference-induced isospin breaking operators are next-to-leading order.

\begin{figure}[th]
\centering
\includegraphics[width=.85\textwidth]{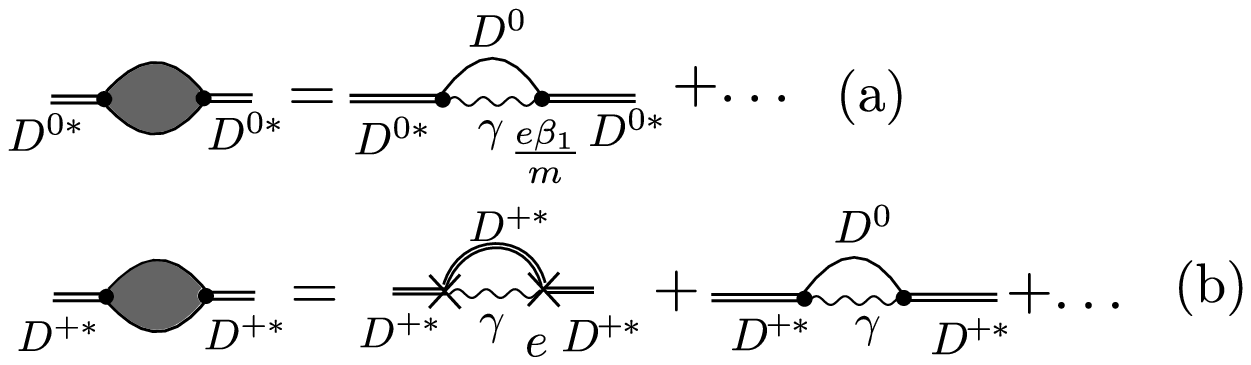}
\caption{Isospin breaking case: next-to-leading order $D^*$ propagator $D\pi$ scattering amplitude}
\label{fig:105fig4}
\end{figure}

To demonstrate the suppression of charge-induced isospin breaking, in this subsection I consider the example from the previous section, the electromagnetic loop correction. Shown in \fig{105fig4}a the neutral $D^*$ states have the same radiative correction as in \fig{105fig3}, and its contribution is next-to-leading order compared to the $D\pi$ loop. In \fig{105fig4}b, I give two examples of the charged $D^*$ states' radiative corrections. On the right-hand side of the equation in \fig{105fig4}b, the second term is the same as $\mathscr{D}_{\rm{D\gamma}}$ and the first term contains the leading order $D^*D\gamma$ interaction induced by the electric charge of the $D^*$. This diagram gives
\be
\mathscr{D}_{D^+\gamma}^{(1)}\sim \alpha\epsilon^2 \lambda m_\pi\,,
\ee
and
\be
\frac{\mathscr{D}_{D^+\gamma}^{(1)}}{\mathscr{D}_{\rm{D\pi}}}\sim\frac{\epsilon^2}{4\pi g^2}.
\ee
Numerically, $\frac{\epsilon^2}{4\pi g^2}\simeq 0.025$ which makes $\mathscr{D}_{D^+\gamma}^{(1)}$ subleading to $\mathscr{D}_{D\pi}$.

More corrections that are higher order compared to \fig{105fig4} are shown in Appendix \appx{D.3}. Therefore, for leading order, we only need to resum the $D\pi$ loop correction to $D^*$ propagator.

\subsection{Pole Hunting for Leading-Order \texorpdfstring{$D\pi$}{D-pi} Scattering Amplitude}
\label{sec:II.3.6}

\begin{figure}[!h]
\centering
\includegraphics[width=.5\textwidth]{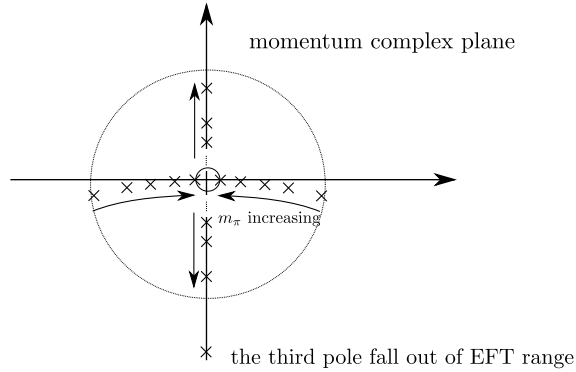}
\caption{Pole structure of $D^*$. The large circle illustrates the range of XEFT. The arrow shows that as $m_\pi$ increases, the two poles within the circle move close to the threshold and then fall away from each other on the imaginary axis. The small circle shows the region through which the lattice extrapolation occurs.}
\label{fig:106fig1}
\end{figure}

As discussed quantitatively in \sec{II.3.2}, the $D\pi$ loops need to be resummed when scattering occurs within 0.2 MeV of the $D^*$ threshold.  Summing the series of $D\pi$ loops, \fig{105fig1}, we obtain
\be\label{eq:106eqDstarpropagator}
ip_{D^*}^{\rm{Full}}=\frac{i}{E_{D^*}+i\frac{g^2(2m_\pi\del+2m_\pi E_{D^*})^{3/2}}{48\pi f_\pi^2}}\,,
\ee
where in the center of mass frame of $D\pi$, $p_\pi=p_D=p$ and
\be\label{eq:106eqDpikinematics}
p^2\left(1+\frac{m_\pi}{M_D}\right)=2m_\pi(\delta+E_{D^*})\,.
\ee
Solving for the poles on the $p$-plane in Appendix \appx{D.4}, we obtain a cubic equation
\be\label{eq:106eqpoleequation}
p^2(1+r)-2m_\pi\del+\frac{ig^2p^3(1+r)^{3/2}m_\pi}{24\pi f_\pi^2}=0\,,
\ee
where $r=\frac{m_\pi}{M_D}$.  Defining
\be 
c:=\frac{24\pi f_\pi^2}{g^2}\,,\qquad  \mu:=\frac{m_\pi M_D}{m_\pi+M_D}\,, \qquad \nu:=\frac{2c^2}{27\mu^3}\,,
\ee
there are two poles close to the threshold within the range $|p|<m_\pi$,
\be\label{eq:106eqDpitwopoles}
p_{1,2}=\pm\sqrt{2\delta\mu}-\frac{i\delta\mu^2}{27c}+\cO\paren{\frac{\del}{\nu}}^{3/2}\,,
\ee
with $\delta/\nu\lesssim 10^{-3}$ for $g\sim 1$.  The third pole is located outside the range $|p|>m_\pi$.  Without the $D\pi$ loop contribution, the pole equation is
\be\label{eq:106eqnaivepoleequation}
\tilde p^2(1+r)-2m_\pi\del=0\,,
\ee
with two poles
\be
\tilde p_{1,2}=\pm\sqrt{2\del\mu}\,,
\ee
which suggests that the imaginary part in \eq{106eqDpitwopoles} is brought in by the $D\pi$ loop and corrects the pole trajectory of $p$. \fig{106fig1} shows the pole trajectories from \eq{106eqpoleequation} as $m_\pi$ is varied.  The pole trajectories without $D\pi$ loop summation (\eq{106eqnaivepoleequation}) follow similar trajectories, but with no imaginary part separating them from the real $p$ axis. Clearly, far away from the $D^*$ threshold, the $D\pi$ loop can be considered a small perturbative correction to the bound state $D^*$ pole. However when in the small neighborhood of the $D^*$ threshold, this correction is no longer small compared to the 0.2 MeV region size required by lattice extrapolation procedure. Therefore, resumming the $D\pi$ loop is necessary.

\section{X(3872) as a \texorpdfstring{$DD^*$}{D-Dstar} Hadronic Molecule}
\label{sec:II.4}

In the previous section, I developed the HHChPT lagrangian systematically to next-to-leading order with isospin breaking and reduced it into the XEFT region.  I used the XEFT Lagrangian to study $D\pi$ scattering.  In this section, I adapt the XEFT to describe the X(3872) as a hadronic molecule by including extra X$DD$ interaction terms.\footnote{This section is based on unpublished work in collaboration with U. van Kolck and S. Fleming.}  I calculate the $DD$ scattering amplitude near the X(3872) threshold and lay out the scheme to fit it to experimental data. With the fitted result, one can use the pole-hunting result and explain the hadronic molecule X(3872) in the picture of scattering theory.

\subsection{X(3872) in Experiments}
\label{sec:II.4.2}
The X(3872) was discovered by the Belle collaboration as a narrow resonance in $e^+e^-$ collisions in the two-step decay $B^\pm\to XK^\pm$ and subsequent $X\to J/\psi \pi^+\pi^-$ (\cite{Choi:2003ue}).  The discovery in the $B^\pm\to KX^\pm$ process was confirmed by the Babar collaboration  (\cite{Aubert:2004ns}), and its existence established in $p\bar p$ collisions by CDF and D$\emptyset$ collaborations (\cite{Acosta:2003zx, Abazov:2004kp}). The average mass obtained from the above measurements is (\cite{Olsen:2004fp})
\begin{equation}\label{eq:74eq1}
m_X=3872.12\pm 0.5\,\text{MeV}\,,
\end{equation}
which is very close to the $\bar D^0\bar D^{*0}$ threshold of $3871.81\pm 0.36$\,MeV (\cite{Cawlfield:2007dw}). An upper bound on the width is $\Gamma_X<2.3$MeV at 90\% confidence level, given by the Belle collaboration (\cite{Choi:2003ue}).

Detections of X(3872) are also made in decays into $J/\psi\gamma$ and $J/\psi\pi^+\pi^-\pi^0$ (\cite{Abe:2005ix}). The branching ratio into three- and two-pion final states is (\cite{Abe:2005ix})
\begin{equation}\label{eq:74eq3}
\frac{Br[X\to J/\psi\pi^+\pi^-\pi^0]}{Br[X\to J/\psi \pi^+\pi^-]}=1.0\pm 0.4\pm 0.3\,,
\end{equation}
which signifies a considerable violation of isospin symmetry since the 2- and 3-pion final state are regarded as having $J/\psi\,\rho$ and $J/\psi\,\omega$ as intermediate states respectively. \cite{Gokhroo:2006bt} report an enhancement near the threshold for $D^0D^0\pi^0$, which is interpreted as the first evidence of $X\to D^0D^0\pi^0$ decay despite peaking at $3875.2\pm0.7_{-1.6}^{+0.3}\pm0.8$\,MeV, i.e. $2\sigma$ above world-average X(3872) mass. The corresponding decay rate $X\to D^0D^0\pi^0$ is observed to be $8.8_{-3.6}^{+3.1}$ times greater than the rate of $X\to J/\psi\pi^+\pi^-$ (\cite{Gokhroo:2006bt}). It was estimated at 90\% C.L. that $Br[X\to J/\psi\pi^+\pi^-]>0.042$ by Babar collaboration (\cite{Mohanty:2005dm, Aubert:2005vi}). For upper limits on the product of $Br[B^\pm\to XK^\pm]$ and other branching fractions of X(3872) including $D^0\bar D^0, D^+D^-$, see \cite{Abe:2003zv}, $\chi_{c_1}\gamma, \chi_{c_2}\gamma, J/\psi\pi^0\pi^0$ see \cite{Abe:2004sd}, and $J/\psi\eta$ see \cite{Aubert:2004fc}. \cite{Yuan:2003yz} and \cite{Dobbs:2004di} have also placed upper limits on partial wave widths for decay $X\to e^+e^-$ and $\gamma\gamma$.

Various experiments have also investigated possible $J^{PC}$ quantum numbers of X(3872). The decay mode $X\to J/\psi\gamma$ implies $C=+$, agreeing with the shape of $\pi^+\pi^-$ invariant mass distributions (\cite{Choi:2003ue, Aubert:2004ns, Abulencia:2005zc}). Angular distributions measured by Belle collaboration prefer $J^{PC}=1^{++}$ (\cite{Abe:2005iya}), while a recent CDF analysis of angular distributions of $J/\psi\pi^+\pi^-$ (\cite{Abulencia:2006ma}) indicate that the only possible quantum numbers are $1^{++}$ or $2^{-+}$.

Hypothesizing that the X(3872) is a $C=+$, S-wave molecular bound state of $D^0\bar D^{*0}+\bar D^0 D^{*0}$ leads to the quantum numbers $J^{PC}=1^{++}$. The shallow molecular state hypothesis is motivated by the proximity of the X(3872) state to the $D^0\bar D^{*0}$ threshold and would justify both the significant isospin violation in pion decays and the large branching ratio for the $D^0\bar D^0\pi^0$ decay mode. It is rare for a conventional charmonium state above $D\bar D$ threshold to have such narrow width and no observation to date of a $X\to \chi_c \gamma$ decay mode. One can deduce the binding energy from \eq{74eq1} and the measurement of the $D^0$ mass in \cite{Cawlfield:2007dw}, giving
\begin{equation}\label{eq:74eq4}
E_X=M_D+m_{D^*}-m_X=0.6\pm0.6\,\text{MeV}\,,
\end{equation}
supporting a bound-state explanation. Nonetheless mass alone does not exclude possibilities of a resonance or `cusp' close to $D^0\bar D^{*0}$ threshold due to its considerable uncertainty (\cite{Bugg:2004rk}). 

In this part the molecular bound state interpretation of X(3872) is adopted, though the method presented also extends to the shallow resonance case.  Based on these measurements of the X(3872) properties, we build an effective field theory description of the nuclear force between $D$ and $D^*$ states.

\subsection{X(3872) in Effective Field Theory Description}
\label{sec:II.4.3}
The experimental evidence strongly suggests that X(3872) is a molecular state of $D$ and $D^*$ with a very small binding energy, as discussed above.  The small binding energy implies that the molecule should possess some universal properties determined by the binding energy.  This makes the $X$ as a $DD^*$ molecule somewhat analogous to the deuteron as a neutron-proton molecule (\cite{Voloshin:2003nt, Voloshin:2005rt}) and can be further utilized via factorization formulae to predict decay rates (\cite{Braaten:2005jj, Braaten:2005ai}). 

\begin{figure}[!h]
\centering
\includegraphics[width=.4\textwidth]{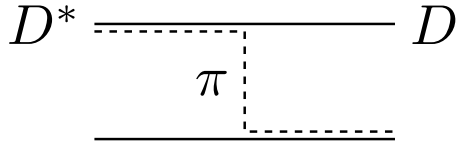}
\caption{$D^{*0}\bar D^0\to D^0\bar D^{*0}$ scattering with one-pion exchange. The solid-plus-dashed line stands for the spin-1 $D^*$ meson, the single solid line for the spin-0 $D$ meson, and the dashed line for $\pi^0$.}
\label{fig:74fig1}
\end{figure}

In this section we study the effect of $\pi^0$ exchange on the properties of X(3872), which is essential to the power counting of XEFT and the analogy between $DD^*$ scattering near the X threshold and proton-neutron scattering near the deuteron threshold. Consider $D^{*0}\bar D^0\to D^0\bar D^{*0}$ scattering as in \fig{74fig1}.  The one-pion exchange amplitude is
\begin{equation}\label{eq:74eq5}
\mathcal{A}_{DD^*\to DD^*}^{(1\pi)}=\frac{g^2}{2f_\pi^2}\frac{\vec \epsilon^*\cdot \vec q\:\vec\epsilon\cdot \vec q}{\vec q^{\,2}-\mu^2}\,,
\end{equation}
where $f_\pi$ is the pion decay constant, $g$ the $D$-meson axial coupling, $\vec \epsilon, \vec \epsilon^*$ are the polarization vectors of incoming and outgoing $D^*$ mesons respectively, $\vec q$ is the transferred momentum and $\mu^2=\Delta^2-m_\pi^2$ with $\Delta$ the $D^*D$ mass splitting and $m_\pi$ the $\pi^0$ mass. Here the hyperfine splitting $\Delta$ enters the pion propagator due to the fact that the exchanged pion has energy $q^0\simeq \Delta$ with momentum $\vec q$. Due to the proximity of $\Delta=142$\,MeV and $m_\pi=135$\,MeV the value of $\mu\simeq 45$\,MeV is unusually small, indicating the pion can have an anomalous long-range effect. This effect should be incorporated as an explicit degree of freedom in depicting the molecule in case the binding energy \eq{74eq4} is not much less than its upper limit.

The key point is that, in contrast to the pion-potential approaches, perturbation theory can be applied to study $\pi^0$ exchange (\cite{Fleming:2007rp}). A simple dimensional-analysis comparison of the two- and one-pion exchange graphs yields
\begin{equation}\label{eq:74eq6}
\frac{\mathcal{A}_{DD^*\to DD^*}^{(2\pi)}}{\mathcal{A}_{DD^*\to DD^*}^{(1\pi)}}
\simeq \frac{g^2M_D\mu}{8\pi f_\pi^2}\lesssim \oneov{10}\,, 
\end{equation}
with $g$ set to $0.5-0.7$ (see \cite{Ahmed:2001xc, Anastassov:2001cw, Fajfer:2006hi}). A similar estimate in the two-nucleon system (\cite{Kaplan:1998tg, Kaplan:1998we}) in contrast gives the ratio
\begin{equation}\label{eq:74eq7}
\frac{\mathcal{A}_{NN\to NN}^{(2\pi)}}{\mathcal{A}_{NN\to NN}^{(1\pi)}}
\simeq\frac{g_A^2M_Nm_\pi}{8\pi f_\pi^2}\approx\oneov{2}\,,
\end{equation}
where $M_N$ is nucleon mass and $g_A=1.25$ is the nucleon axial coupling. The perturbation expansion breaks down for NN scattering in the ${}^3S_1$ channel because of a combination of the not-so-small expansion parameter \eq{74eq7} and large numerical coefficients at NNLO from iteration of the spin-tensor force (\cite{Fleming:1999bs, Fleming:1999ee}). Although \eq{74eq5} also yields a spin-tensor force, the perturbative treatment of pions should suffice due to the smallness of the expansion parameter in \eq{74eq6}, even given large NNLO coefficients similar to those found by \cite{Fleming:1999bs, Fleming:1999ee}.

We have calculated the binding energy of X(3872) in \eq{74eq4} assuming it is a hadronic molecule with positive binding energy composed of a superposition of $D^0\bar D^{*0}$ and $D^{*0}\bar D^0$.  This assumption implies an upper limit on the typical momentum of $D$ and $D^*$ in the bound state:
\begin{equation}
\gamma\equiv (M_DE_X)^{1/2}\le 48\text{MeV}.
\end{equation}
Correspondingly the typical velocity of $D, D^*$ is
\begin{equation}
v_D\simeq \paren{\frac{E_X}{M_D}}^{1/2}\lesssim 0.02\,,
\end{equation}
showing that $D, D^*$ are non-relativistic, allowing use of non-relativistic fields for both.

Furthermore we can treat pion degrees of freedom non-relativistically: in a $X\to D^0D^0\pi^0$ decay the maximum energy of emitted the pion is
\begin{equation}\label{eq:74eq8}
E_\pi=\frac{m_X^2-4M_D^2+m_\pi^2}{2m_X}=142\,\text{MeV}\,,
\end{equation}
a mere 7\,MeV above $m_{\pi^0}=134.98$MeV. This implies the typical velocity of pion is $v_\pi=p_\pi/m_\pi\le 0.34$.  Correspondingly, the maximum pion momentum 44\,MeV is close to both the typical D momentum $p_D\sim \gamma\lesssim 48$MeV and pion-exchange graph typical momentum scale $\mu\sim 45$MeV. This is in contrast to ordinary chiral perturbation theory or NN theory (\cite{Kaplan:1998tg, Kaplan:1998we}).

\subsubsection{XEFT with Transvestite}\label{sec:II.4.4}

Previous work restricted study of the X to a simple bound state composed of a superposition $\ket{D^0\bar D^{*0}}+\ket{D^{*0}\bar D^0}$ (\cite{Fleming:2007rp}).  We have extended that theory to allow the X state to be a virtual bound state or resonance in case further data shows the binding energy is non-positive.\footnote{The name `transvestite' was invented by U. van Kolck to describe the changing (bound state to resonance) character  of the X particle after being dressed by interactions.}  The Lagrangian of \cite{Fleming:2007rp} rewritten with a $1^{++}$ field $\mathbf{X}$ and Galilean invariance (\cite{Braaten:2015tga}) is 
\begin{align}
\cL&=D^\dag\paren{i\pd_0+\frac{\vec\nabla^2}{2M_D}}D+\bar{\mathbf{D}}^\dag \paren{i\pd_0+\frac{\vec\nabla^2}{2(M_D+m_\pi)}}\bar{\mathbf{D}} \nn\\
&+\bar D^\dag\paren{i\pd_0+\frac{\vec\nabla^2}{2M_D}}\bar D+\mathbf{D}\paren{i\pd_0+\frac{\vec\nabla^2}{2m_{D^*}}}\mathbf{D}+\pi^\dg\paren{i\pd_0+\frac{\vec\nabla^2}{2m_\pi}+\delta}\pi\nn\\
&+\mathbf{X}^\dag \sqparen{\sigma\paren{i\pd_0+\frac{\vec\nabla^2}{2M_D}}-b}\mathbf{X}\nn\\
&+\frac{g}{\sqrt{2}f_\pi}\oneov{\sqrt{2m_\pi}}\oneov{M_D\!+\!m_\pi}\!\left[\mathbf{D}^\dag\!\cdot\!(D[M_D \overset{\rightarrow}{\nabla}\!-m_\pi\overset{\leftarrow}{\nabla}] \pi)+\bar{\mathbf{D}}\!\cdot\!(D^\dag[M_D \overset{\rightarrow}{\nabla}\!-m_\pi\!\overset{\leftarrow}{\nabla}] \pi^\dag)+\text{h.c.}\right]\nn\\
&+\frac{y_0}{\sqrt{2}}\left[\mathbf{X}^\dag\cdot(\bar{\mathbf{D}}D+\mathbf{D}\bar D)+\text{h.c.}\right]+\frac{Z_1}{\sqrt{2m_\pi}}\left[\mathbf{X}^\dag\cdot D\bar D\vec\nabla \pi+\text{h.c.}\right]+\ldots\,, \label{eq:78IIeq1}
\end{align} 
where $\delta=\Delta-m_\pi\simeq 7$\,MeV, $\sigma=\pm 1$ and $b$ is the binding energy of the X.  The ellipsis \ldots denotes higher-order interactions.  $g$ is a dimensionless coupling constant, $y_0$ has dimension $-1/2$, and  $Z_1$ has dimension $-5/2$.
We consider the soft scale $Q\sim \mu=\sqrt{\Delta^2-m_\pi^2}\approx \sqrt{2m_\pi\delta}\simeq 45$\,MeV, the break-down scale is $m_\pi$, and we count powers of $Q/m_\pi$.

Expanding $\oneov{M_D+m_\pi}\simeq\oneov{M_D}(1-\lambda)$ where $\lambda=\frac{m_\pi}{M_D}$, we write \eq{78IIeq1} as
\begin{align}
\mcL&=D^\dag \paren{i\pd_0+\frac{\vec\nabla^2}{2M_D}}D+\overline{\vec{D}^\dg}\paren{i\pd_0+\frac{\vec\nabla^2}{2M_D}}\overline{\vec D}+\overline{\vec D^\dg}\paren{-\lambda\frac{\vec\nabla}{2M_D}}\overline{\vec D}\nn\\
&+\overline{D}^\dg\paren{i\pd_0+\frac{\vec\nabla^2}{2M_D}}\bar D+\vec D^\dg\paren{i\pd_0+\frac{\vec\nabla^2}{2M_D}}\vec D+\vec D^\dg\paren{-\lambda\frac{\vec \nabla^2}{2M_D}}\vec D\nn\\
&+\vec X^\dg\paren{\sigma\paren{i\pd_0+\frac{\vec\nabla^2}{2M_D}-b}}\vec X+\pi^\dg\paren{i\pd_0+\frac{\nabla^2}{2m_\pi}+\delta}\pi\nn\\
&+\frac{g}{\sqrt{2}f_\pi}\oneov{\sqrt{2m_\pi}}\left[(D\vec D^\dg\cdot\vec\nabla \pi+D^\dg\overline{\vec D}\cdot\vec\nabla \pi^\dg)+\text{h.c.}\right]\nn\\
&+(-\lambda^2)\frac{g}{\sqrt{2}f_\pi}\oneov{\sqrt{2m_\pi}}\left[(D\vec D^\dg\cdot\vec\nabla \pi+D^\dg \overline{\vec D}\cdot\vec\nabla\pi^\dg)+\text{h.c.}\right] \nn\\
&+(-\lambda)\frac{g}{\sqrt{2}f_\pi}\oneov{\sqrt{2m_\pi}}\left[\vec D^\dag\cdot D(\overset{\leftarrow}{\nabla} \pi)+\overline{\vec D}\cdot(D^\dag\overset{\leftarrow}{\nabla}\pi)+\mathrm{h.c.} \right] \nn\\
&+\frac{y_0}{\sqrt{2}}\left[\vec X^\dg\cdot(\overline{\vec D}D+\vec D D)+\text{h.c.}\right]+\frac{Z_1}{\sqrt{2m_\pi}}\left[\vec X^\dg\cdot D\bar D\vec \nabla \pi+\text{h.c.}\right]+\ldots, \label{eq:3.216}
\end{align}
where terms proportional to $\lambda$ in the $D^*$ kinetic piece are due to the consideration that, at zeroth order in $\epsilon$, $M_{D^*}=M_D+_\pi$.  Also note that the $DD\pi$ interactions terms are consistent with the V.R.I. terms we developed in the previous sections.

Let us check the $D\bar D^*$ loop corrections to the X propagator.  Apart from spin factors, the one-loop $X\to D\bar D^*\to X$ amplitude is $y_0^2$ times the bubble $\sim\alpha Q$, where $\alpha\equiv M_Dy_0^2/4\pi$.  The two-loop bubble in which the $D$ and $\bar D^*$ exchange one pion is $\cO(\alpha Q\beta Q/\mu)$, with $\beta\equiv g^2M_D\mu/8\pi f_\pi^2\ll 1$. We conclude that pions are perturbative at $Q\sim \mu$.

More interesting are the possibilities for the sizes of the parameters $b$ and $\alpha$. The $X$ propagator dressed by bubbles is
\begin{equation}\label{eq:78IIeq2}
G_X(k)\sim \frac{2}{y_0^2}\frac{4\pi}{M_D}\oneov{ik-b/\alpha+\sigma k^2/(\alpha M_D)}\,,
\end{equation}
in the center-of-mass frame where, up to a sign and other dimensionless factors, the energy is $k^2/2M_{DD^*}$ and $M_{DD^*}=M_DM_{D^{*0}}/(M_D+M_{D^{*0}})$ the reduced mass of $D^0$ and $\bar D^{*0}$.  There are two poles in the corresponding amplitude.

First, consider one fine-tuning: $b\sim \mu m_\pi/M_{DD^*}$ and $\alpha\sim m_\pi/M_{DD^*}$. In this case, for $Q\sim \mu$ the range term (the third term in the denominator of \eq{78IIeq2}) is smaller than the other two by $\cO(\mu/m_\pi)$, and is NLO. This is the usual power counting where range corrections are NLO (\cite{Fleming:2007rp}). The LO pole is at $k=ib/\alpha$, representing  a real (virtual) bound state for $b>0$ ($b<0$), which is the situation claimed by \cite{Braaten:2007dw,Braaten:2007ft,Stapleton:2009ey,Hanhart:2007yq}.

Now consider a double fine-tuning: $b\sim \mu^2/M_{DD^*}$ and $\alpha\sim\mu/M_{DD^*}$. In this case, all three terms are $\cO(\mu)$ for $Q\sim \mu$, and there are two low-energy poles at $k=i\sigma \alpha M_{DD^*}(1\pm\sqrt{1-4\sigma b/\alpha^2M_{DD^*}})/2$. For $4\sigma b/\alpha^2M_{DD^*}<1$ the two poles are on the imaginary axis, so they represent two bound states, which is the situation in \cite{Zhang:2009bv}. For $4\sigma b/\alpha^2 M_{DD^*}>1$ and $\sigma=-1$ and the poles are off-axis in the lower complex half-plane and represent one resonance, a situation not so far discussed in the literature.

Finally, consider a stronger double fine-tuning (\cite{Bedaque:2003wa}), $b\sim \mu^2/M_{DD^*}$ and $\alpha\sim\mu^2/m_\pi M_{DD^*}$. In this case, the scattering length and range terms are both $\cO(m_\pi)$ for $Q\sim\mu$, while the bubble is $\cO(\mu)$ and can be treated in perturbation theory. This is a special subcase of the previous case, where at LO there are two poles at $k=\pm\sqrt{\sigma b M_{DD^*}}$, which are a pair of real/virtual bound states if $\sigma b<0$, or a zero-width resonance if $\sigma b>0$.


\subsection{\texorpdfstring{$DD^*$}{D-Dstar} Scattering Power-counting and Amplitude}
\label{sec:II.4.5}

We now discuss $DD^*$ scattering in more detail using the Lagrangian \eq{3.216}.  We will show how the X binding is determined by the resummed series of $DD^*$ scattering loops.  We start by power counting the X interactions assuming it is a bound state as the simplest case, which is the first fine-tuning scenario mentioned in preceding subsection.  This choice does not affect the physics phenomenology outcome, as we will see that next-to-leading order corrections convert the X state to a resonance.  In the subsequent section, we will examine experimental implications in one channel where the X has been detected.

\subsubsection{Power Counting for \texorpdfstring{$DD^*$}{D-Dstar} Scattering Amplitude}
\label{sec:II.4.5.1}

In this section, we first estimate the tree-level leading-order and next-to-leading order $DD^*\to DD^*$ scattering amplitudes, and then estimate the loop corrections. For any loop-corrected amplitudes that are the same order as the tree-level amplitudes, we resum the loop series in the next subsection.

\begin{figure}[!h]
\centering
\includegraphics[width=.6\textwidth]{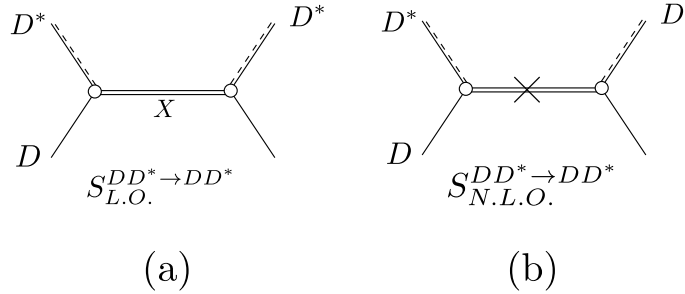}
\caption{Tree-level leading and next-to-leading order $DD^*$ scattering amplitudes}
\label{fig:108fig1}
\end{figure}

\begin{figure}[!h]
\centering
\includegraphics[width=.3\textwidth]{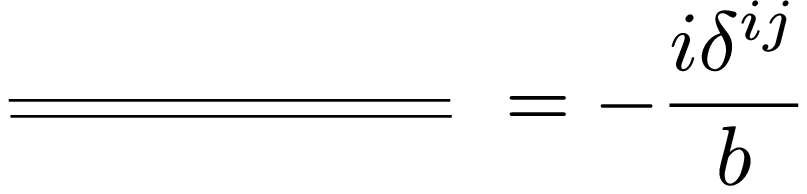}
\caption{Leading order X propagator}
\label{fig:108fig1a}
\end{figure}

\begin{figure}[!h]
\centering
\includegraphics[width=.5\textwidth]{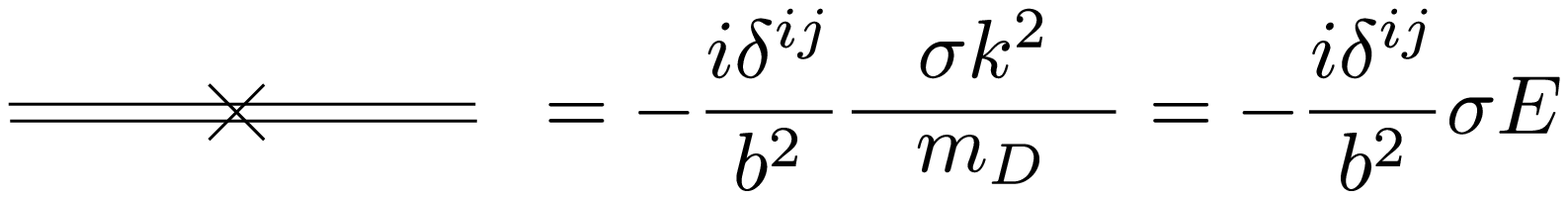}
\caption{Next-to-leading order X propagator}
\label{fig:108fig1b}
\end{figure}

We first discuss the tree-level leading- and next-to-leading order Feynman diagrams for $DD^*$ scattering amplitudes.  The relevant diagrams are shown in \fig{108fig1}, and we will show below that other diagrams are suppressed relative to these.  Considering X(3872) as a bound state of $D$ and $D^*$, we power count the typical momentum $Q$ and binding energy $b$ as being the same order and the relevant soft scale, in which case 
\begin{align}
b\sim \lambda Q,\qquad E_X\sim \frac{2Q^2}{M_D}, \qquad 
k_X\sim Q, \qquad y_0^2\sim \frac{4\pi}{M_D}\,,
\end{align}
where $\lambda=\frac{m_\pi}{M_D}$.
We expand the X propagator in the Lagrangian in \eq{3.216}. The leading order and next-to-leading order X propagators are shown in \fig{108fig1a} and \fig{108fig1b}, respectively.  For $b\gg E=\frac{k_X^2}{M_D}$, the leading order $DD^*$ scattering amplitude is
\be
S_\lo^{\dds\to\dds}=\paren{i\frac{y_0}{\sqrt{2}}}^2\vec\eps_{D^*}^{\:*}\cdot\vec\eps_{D^*}\paren{\frac{i}{-b}} \sim \frac{4\pi}{M_D}\frac{\lambda}{Q}\,,
\ee
where $\vec\eps_{D^*}^{\:*}$ and $\vec \eps_{D^*}$ are the $D^*$ polarization vectors.  The next-to-leading order $DD^*$ scattering amplitude is
\be
S_\nlo^{DD^*\to DD^*}=\paren{i\frac{y_0}{\sqrt{2}}}^2\vec\epsilon_{D^*}^*\cdot\vec \epsilon_{D^*}\paren{\frac{i\sigma E_X}{-b^2}}\sim\paren{\frac{4\pi}{M_D}\frac{\lambda}{Q}}\cdot\frac{2Q}{M_D},
\ee
which is suppressed by $\frac{2Q}{M_D}$ compared to $S_\lo^{DD^*\to DD^*}$.

\begin{figure}[!h]
\centering
\includegraphics[width=0.9\textwidth]{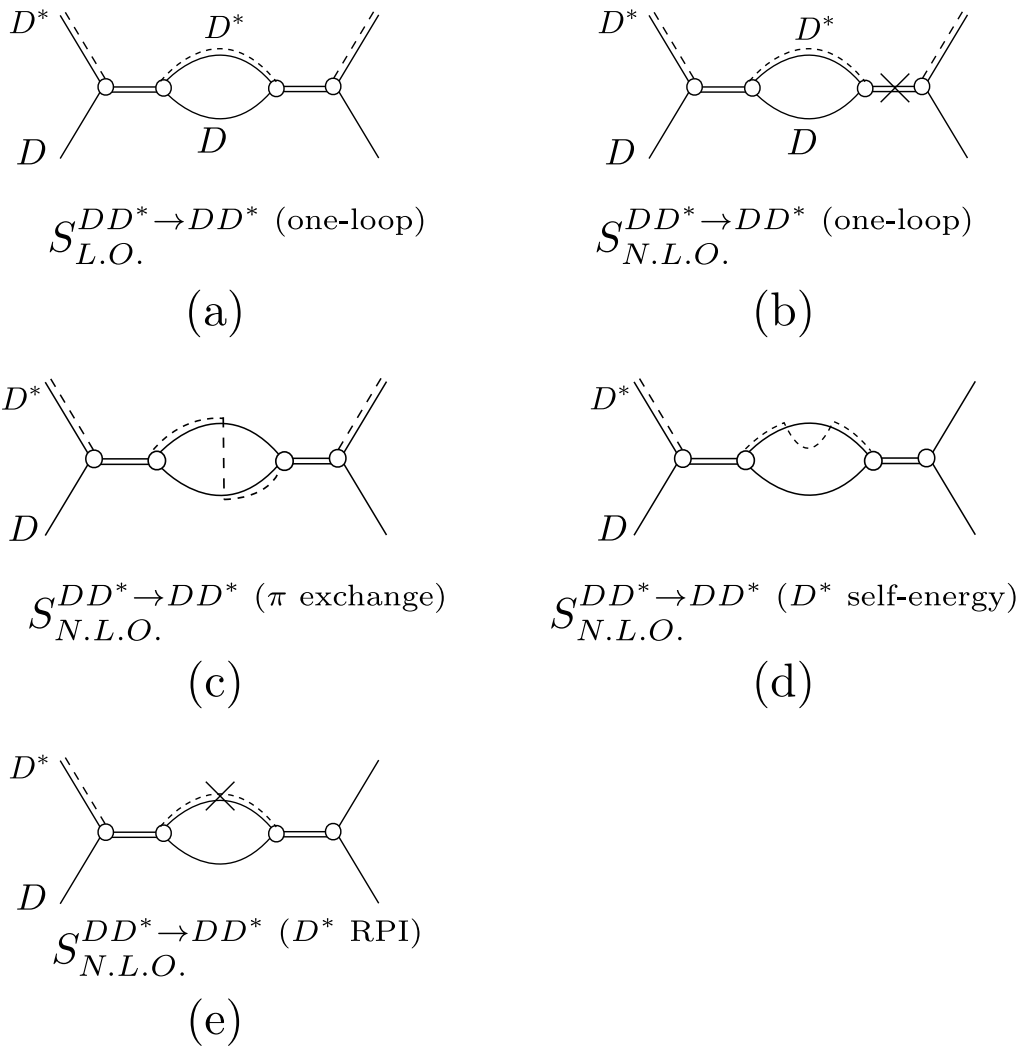}
\caption{Leading and next-to-leading order $D^*D$ loop diagrams to the $DD^*$ scattering amplitude}
\label{fig:108fig2}
\end{figure}

\begin{figure}[!h]
\centering
\includegraphics[width=.95\textwidth]{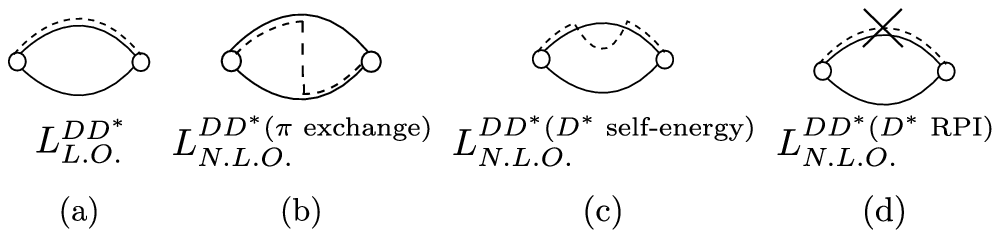}
\caption{Leading and next-to-leading $DD^*$-loop diagrams}
\label{fig:108fig3}
\end{figure}

The leading and next-to-leading loop corrections for $DD^*$ scattering amplitudes are shown in \fig{108fig2}.  Of these, \fig{108fig2}a is the largest contribution, because the $D^* D$ loop inside (shown in \fig{108fig3}a) gives 
\be 
L_\lo^\dds=\paf{iy_0}{\sqrt{2}}^2 i\frac{M_D}{4\pi}ik\,, 
\ee
where $k$ is the typical momentum with the order of $Q$ flowing in the loop.  This implies that the total diagram  \fig{108fig2}a is 
\be
S_\lo^{\dds\to\dds(\rm{one-loop})}=\paren{\frac{iy_0}{\sqrt{2}}}^2\vec\eps_{D^*}^{\:*}\cdot\vec\eps_{D^*}\paren{-\frac{i}{b}}^2\paren{\frac{iy_0}{\sqrt{2}}}^2i\frac{M_D}{4\pi}ik\,,
\ee
which is of order $\sim\frac{2\pi}{M_\dds}\frac{\lambda}{Q}$, the same order as \fig{108fig1}a.  Therefore, we must resum the $DD^*$ loop corrections to the intermediate X propagator.

Containing the next-to-leading order propagator, \fig{108fig2}b is a next-to-leading order correction to $DD^*$ scattering, giving
\begin{align}
S_\nlo^{\dds\to\dds\rm{(one-loops)}}&=\paren{\frac{iy_0}{\sqrt{2}}}^2\vec \eps_{D^*}^*\cdot\vec\eps_{D^*}\paf{i}{-b}\paf{iy_0}{\sqrt{2}}^2i\frac{M_D}{4\pi}ik\paren{-\frac{i\sig k^2}{b^2M_D}} \nn \\
&\sim \frac{4\pi}{M_D}\frac{\lambda}{Q}\paf{2Q}{M_D}\,,
\end{align}
which is the same order as the tree-level N.L.O. amplitude in \fig{108fig1}b. As a result, all the loops in \fig{108fig2}b are required to be resummed to obtain the complete next-to-leading order amplitude.

In diagram \fig{108fig2}c, the one-pion-exchange correction is added to the $DD^*$ loop.  The pion exchange is perturbative as discussed in \sec{II.4.3} and its correction to the $DD^*$ loop (\fig{108fig3}b) is a next-to-leading order correction,
\be
L_\nlo^{\dds\text{ ($\pi$ exchange)}}=\paf{iy_0}{\sqrt{2}}^2\frac{g^2M_D^2}{96\pi^2f_\pi^2}(-ik)^2\sim Q\paren{\frac{M_D}{2\pi}\frac{g^2}{f_\pi^2}Q}\,,
\ee
which is suppressed by a factor of $\left(\frac{M_D}{2\pi f_\pi^2}g^2 Q\right)$ compared to $\msl_\lo^\dds$.
Inserting this loop amplitude between X propagators, \fig{108fig2}c gives
\be
S_\nlo^{\dds\to\dds\text{ ($\pi$ exchange)}}=\paf{iy_0}{\sqrt{2}}^2\frac{g^2M_D^2}{96\pi^2f_\pi^2}((-ik)-\Lam)^2\paf{iy_0}{\sqrt{2}}^2\paf{i}{-b}^2\,,
\ee
where $\Lambda$ is a regulator. This term scales as
\be
S_\nlo^{\dds\to\dds\text{ ($\pi$ exchange)}}\sim
\paren{\frac{g^2M_D Q}{4\pi f_\pi^2}}\frac{4\pi}{M_D}\frac{\lambda}{Q}\,.
\ee
This amplitude is suppressed relative to $S_\lo^{\dsd\to\dsd}$ by a factor $\frac{M_D}{4\pi f_\pi^2}g^2 Q$, but the same order as $S_\nlo^{\dsd\to\dsd\rm{(one-loop)}}$. 

In diagram \fig{108fig2}d, the $D^*$ self-energy correction is added to the $DD^*$ loop. This loop (\fig{108fig3}c) is the same size as $\msl_\nlo^{\dds\text{ ($\pi$ exchange)}}$, so that the estimated contribution to the $DD^*$ scattering amplitude is equal to $S_\nlo^{\dds\to\dds\text{ ($\pi$ exchange)}}$.

Finally in diagram \fig{108fig2}e, the $D^*$ propagator correction in the $DD^*$ loop (\fig{108fig3}d) is brought in by Galilean invariance. The loop \fig{108fig3}d gives
\be
\msl_{\nlo}^{\dds\text{ ($D^*$ V.R.I.)}}=\paren{\frac{iy_0}{\sqrt{2}}}^2\frac{iM_D}{4\pi}ik(-\lambda)\frac{1}{2}\sim \lambda Q\,,
\ee
which is suppressed by $\lambda$ compared to $\msl_\lo^\dds$. Thus the next-to-leading order contribution brought in by this term is
\be
S_\nlo^{\dds\text{ ($D^*$ V.R.I.)}}=\paren{\frac{iy_0}{\sqrt{2}}}^2\vec\epsilon_{D^*}^*\cdot\vec\epsilon_{D^*}\paren{-\frac{i}{b}}^2\paren{\frac{iy_0}{\sqrt{2}}}^2\frac{M_D}{8\pi}\lambda k\sim\paren{\frac{4\pi}{M_D}\frac{\lambda}{Q}}\lambda\,.
\ee
Numerically, $\lambda\simeq 0.07$ while the suppression factor for $S_\nlo^{\dds\text{ ($\pi$ exchange)}}$ is $\frac{M_\dds g^2Q}{2\pi f_\pi^2}\simeq 0.23$. Therefore, it is reasonable to consider \fig{108fig2}e also as a N.L.O. $DD^*\to DD^*$ amplitude.

To summarize so far, the leading order $DD^*$ scattering amplitude is obtained by summing the series of loops \fig{108fig2}a with the tree-level diagram \fig{108fig1}a, while the next-to-order amplitude requires summing the series of loops \fig{108fig2}b with the tree-level diagram \fig{108fig1}b as well as the remaining diagrams \fig{108fig2}c,d, and e.

\begin{figure}[t]
\centering
\includegraphics[width=.2\textwidth]{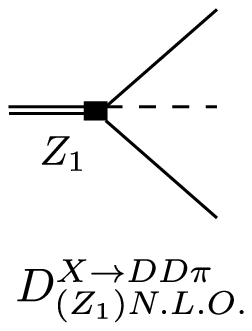}
\caption{N.L.O. vertex for $X\to DD\pi$}
\label{fig:108fig4}
\end{figure}

\begin{figure}[!h]
\centering
\includegraphics[width=0.98\textwidth]{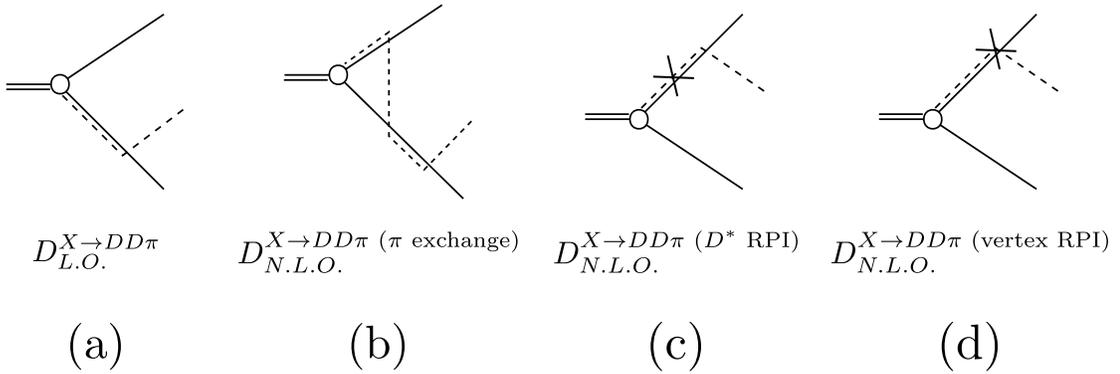}
\caption{Leading and next-to-leading order $X\to DD\pi$. }
\label{fig:108fig6}
\end{figure} 

To ensure we have all the leading and next-to-leading order diagrams, we check that the remaining one-pion exchange loop is next-to-next-to-leading order.  This last loop diagram is generated by the $XDD\pi$ vertex proportional to $Z_1$ in \eq{78IIeq1}.
The vertex is shown separately in \fig{108fig4}, and the corresponding Feynman rule is
\be
D_{(Z_1)\nlo}^{X\to DD\pi}=-\frac{Z_1\vec\eps\cdot\vec p_\pi}{\sqrt{2m_\pi}} \sim 
Z_1Q\ov{\sqrt{m_\pi}}\,.
\ee
The overall magnitude of this diagram is controlled by the size of $Z_1$, and in order to make sure the vertex is next-to-leading order, we estimate the leading and one example of next-to-leading order $X\to DD\pi$ decay diagrams, which are shown in \fig{108fig6}. The leading order \fig{108fig6}a is represented by the two-step decay $X\to D^* D\to DD\pi$ and gives
\be
D_\lo^{X\to DD\pi}=y_0\frac{M_D g(\vec p_\pi\cdot\vec\eps)}{2\sqrt{m_\pi}f_\pi(\vec p_\pi^2-\vec k^2)}
\sim \sqrt{\frac{\pi M_D}{2m_\pi}}\frac{g}{f_\pi}\ov{Q}\,.
\ee
The pion-exchange loop correction shown in \fig{108fig6}b is
\be
D_\nlo^{X\to DD\pi}=\paf{iy_0}{\sqrt{2}}\paren{-i\frac{g}{\sqrt{2}f_\pi}}\frac{\vec\eps\cdot\vec p_\pi}{\sqrt{2m_\pi}}\frac{M_D^2}{-k^2+\vec p_D^2}\frac{g^2}{16\pi f_\pi^2}|\vec p_D|^2\oneov{m_\pi}\,,
\ee
which scales as 
\be
D_\nlo^{X\to DD\pi}\sim \paren{\sqrt{\frac{\pi M_D}{2m_\pi}}\frac{g}{f_\pi}\ov{Q}}\frac{g^2 M_D}{32\pi f_\pi^2}\frac{Q^2}{m_\pi}\,.
\ee
For the decay amplitude $D_{Z_1}^{X\to DD\pi}$ to be the same size as $D_\nlo^{X\to DD\pi}$, $Z_1$ must scale as 
\be
Z_1\sim \frac{g^3 M_D^{3/2}}{32\sqrt{2}\pi f_\pi^3}\ov{Q}\,,
\ee
which is consistent with its dimension.  

\begin{figure}[t]
\centering
\includegraphics[width=.25\textwidth]{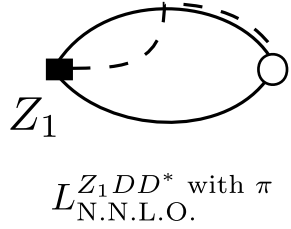}
\caption{One-pion-exchange loop with $Z_1$ vertex}
\label{fig:108fig5}
\end{figure}

Putting this vertex into the $Z_1$ pion-exchange-loop \fig{108fig5}, the loop amplitude becomes
\begin{align}
L_\nnlo^{Z_1\dds\rm{ with }\pi}&=\frac{gZ_1M_D}{48\pi^2\sqrt{2}f_\pi}(\Lam\mu+i\mu^2)(\Lam\gam+\gam^2)\paf{iy_0}{\sqrt{2}}\nn\\
&\sim\paren{Q^2\frac{M_Dg^2}{2\pi f_\pi^2}}\frac{g^2 M_DQ}{768\pi^{3/2}f_\pi^2},
\end{align}
and is next-to-next-to-leading order compared to $L_\lo^\dds$.

All of these power counting results are verified by computing Feynman diagrams explicitly in Appendix \appx{D.5}. 

\subsubsection{Leading Order \texorpdfstring{$\overline{D^*}D\to \overline{D^*}D$}{D-Dstar to D-Dstar} Scattering Amplitude with Renormalization}
\label{sec:II.4.5.1.Renorm}

As we power counted all the leading and next-to-leading order tree-level and loops amplitudes, we can start to calculate them.

\begin{figure}[h]
\centering
\includegraphics[width=.75\textwidth]{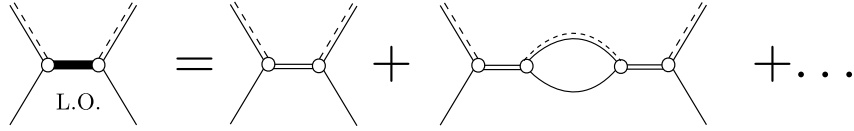}
\caption{Leading order $\overline{D^*}D\to\overline{D^*}D$ scattering.  The double line stands for the $X$ particle, the single solid line for the $D$ meson, the solid-with-dashed line for the $D^*$ meson, and the thick black line for the resummed $X$ propagator in leading order.}
\label{fig:110fig1}
\end{figure}

As we saw in last section, the leading order $DD^*$ scattering amplitude is equal to the sum of the geometric series of $DD^*$ scattering loops shown in \fig{110fig1} with the $DD^*$ loop in \fig{108fig3}a calculated in Appendix \appx{D.5}.  Therefore the complete leading order $D^*D\to D^*D$ scattering amplitude taking into account $X(D^*D)$ as a bound state reads
\begin{align}
S_{L.O.}^{D^*D\to D^*D}&=\frac{4\pi i}{M_D}\oneov{\frac{8\pi b}{y_0^2 M_D}-\sqrt{-\vec k^2-i\epsilon}+\Lambda_{P.D.S.}}\nn\\
&=\frac{4\pi i}{M_D}\oneov{\frac{8\pi b}{y_0^2M_D}+ik+\Lambda_{P.D.S.}}\equiv iP_{L.O.}^{\text{Full}}\,. \label{eq:3.231}
\end{align}
The subtraction in the denominator $\Lambda_{P.D.S}$ is brought in by the power divergence subtraction (P.D.S.) renormalization scheme.  In order to cancel dependence on $\Lambda_{P.D.S}$ in observables, we take
\begin{equation}\label{eq:3.232}
\frac{(y_0^{(0)})^2}{2b^{(0)}}=\frac{4\pi}{M_D}\oneov{\Upsilon-\Lambda_{P.D.S.}}\,,
\end{equation}
where $\Upsilon=-ik|_{\text{pole}}$ is the physical pole position of the full leading-order propagator $P_{L.O.}^{\text{Full}}$, and $y_0^{(0)}$ and $b^{(0)}$ are the leading terms in the expansions of $y_0$ and $b$,
\begin{align}
b&=b^{(0)}+b^{(1)}+\ldots\nn\\
y_0&=y_0^{(0)}+y_0^{(1)}+\ldots\label{eq:ybexpansion}
\end{align}
Plugging \eq{3.232} back into \eq{3.231} shows that $\Lambda_{\rm{P.D.S.}}$ is cancelled, and the renormalized leading order amplitude suggests a bound state picture for X(3872) particle.

\subsubsection{NLO \texorpdfstring{$D^*D\to D^*D$}{Dstar-D to Dstar-D} Scattering Amplitude with Renormalization}
\label{sec:II.4.5.2}

Now we discuss the next-to-leading order $D^*D$ scattering amplitude. 
We first will compute the next-to-leading order amplitudes by resumming all the necessary loops mentioned in \sec{II.4.5.1}. Then we will renoramlize it by absorbing $\Lambda_{\rm{P.D.S}}$ into the coupling constants $b$ and $y_0$ with prescribed physical-pole positions.

\noindent {\bf Next-to-leading Amplitude Calculation}

We organize the contributions into seven sets
\be
S_{DD^*\to DD^*}^{N.L.O.}=S_{N.L.O.}^{(1)}+S_{N.L.O.}^{(2)}+S_{N.L.O.}^{(3)}+S_{N.L.O.}^{(4)}+S_{N.L.O.}^{(5)}+S_{N.L.O.}^{(6)}+S_{N.L.O.}^{(7)}\,.
\ee
First, $S_\nlo^{(1)}$ is described as follows,
\begin{center}
\includegraphics[width=.7\textwidth]{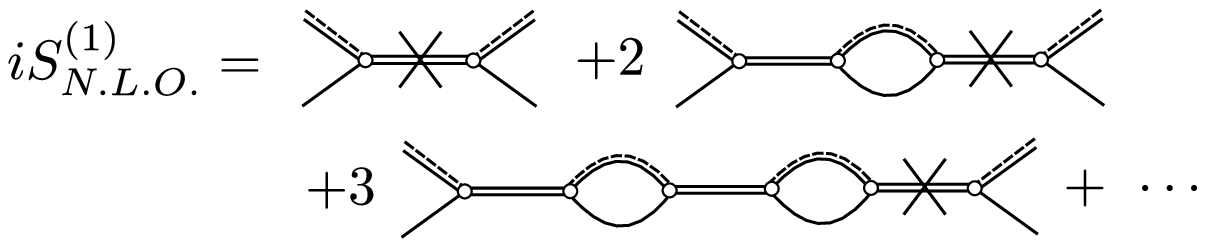}
\end{center}
where the double lines with a cross stand for the next-to-leading order $X$ propagator. Thsi is a geometric series. Computing this series of Feynman diagrams, we have
\begin{align}
iS_\nlo^{(1)}&=\paren{\frac{iy_0}{\sqrt{2}}}^2\Bigg\{\paren{-\frac{i\sig k^2}{M_Db^2}}
\nn\\
&+2\paren{-\frac{i\sig k^2}{M_Db^2}}\paren{-\frac{i}{b}}\paren{\frac{iy_0^2}{\sqrt{2}}}^2\frac{iM_D}{4\pi}(\Lam+ik)\nn\\
& +3\paren{-\frac{i\sig k^2}{M_Db^2}}\paren{-\frac{i}{b}}^2\left[\paren{\frac{iy_0^2}{\sqrt{2}}}^2\frac{iM_D}{4\pi}(\Lam+ik)\right]^2+\ldots\Bigg\}\nn\\
&=\paren{\frac{iy_0}{\sqrt{2}}}^2\paren{-\frac{i\sig k^2}{M_Db^2}}\ov{\paren{1-\paren{-\frac{i}{b}}\paren{\frac{iy_0^2}{\sqrt{2}}}^2\frac{iM_D}{4\pi}(\Lam+ik)}^2}\,,
\end{align}
where we suppress the parenthesis of $\Lambda_{\rm{P.D.S.}}$.
We write  the expression for $P_\lo^\full$ given in the previous section as 
\be
\ov{1+\frac{y_0^2M_D}{8\pi b}(\Lam+ik)}=\frac{2b}{y_0^2}P_\lo^\full\,,
\ee
and use this to rewrite $S_\nlo^{(1)}$ as
\be
S_\nlo^{(1)}=\frac{\sig k^2}{M_Db^2}\frac{y_0^2}{2}\frac{4b^2}{y_0^4}(P_\lo^\full)^2=\frac{2\sig k^2}{M_D y_0^2}(P_\lo^\full)^2\,.
\ee
Second, $S_\nlo^{(2)}$ can be described as
\begin{center}
\includegraphics[width=.85\textwidth]{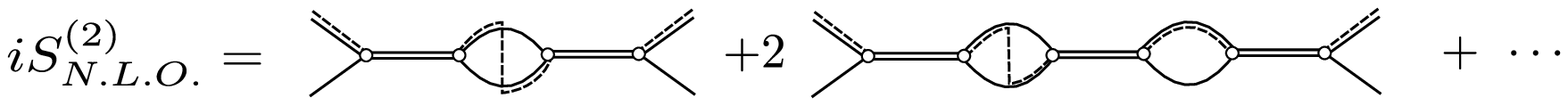}
\end{center}
Computing and resumming this geometric series the same way, we find
\begin{align}
iS_\nlo^{(2)}&=\paren{\frac{iy_0}{\sqrt{2}}}^4\paren{-\frac{i}{b}}^2\paren{-\frac{ig^2M_D^2}{96\pi f_\pi^2}}\left[(\Lam+ik)^2+\mu^2\paren{\log\Lam-\log\paren{-2ik-i\mu}}\right]\nn\\
&\times\Bigg\{1+2\paren{-\frac{i}{b}}\paren{\frac{iy_0}{\sqrt{2}}}^2\frac{iM_D}{4\pi}(\Lam+ik)\nn\\
&+3\paren{-\frac{i}{b}}\sqparen{\paren{\frac{iy_0}{\sqrt{2}}}^2\frac{iM_D}{4\pi}(\Lam+ik)}^2+\ldots\Bigg\}\nn\\
&=\frac{ig^2M_D^2}{96\pi^2 f_\pi^2}\sqparen{(\Lam\!+\!ik)^2+\mu^2\log\frac{\Lam}{-2ik-i\mu}}(P_\lo^\full)^2\,,
\end{align}
and expanding the last line in powers of $(k/\mu)^2$, we have
\be
S_\nlo^{(2)}=(P_\lo^\full)^2\frac{g^2M_D^2}{96\pi^2 f_\pi^2}\sqparen{\Lam^2-\mu^2\log(-i)-\mu^2\log\frac{\mu}{\Lam}+(2i\Lam-2\mu)k+k^2}\,.
\ee
Following the same procedure for the third set,
\begin{center}
\includegraphics[width=.75\textwidth]{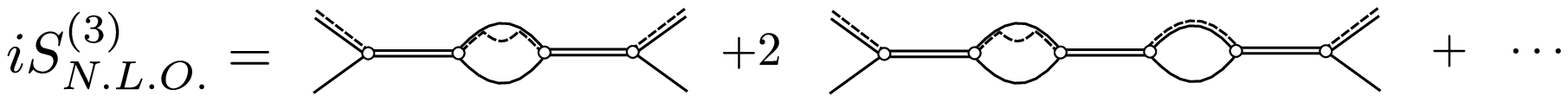}
\end{center}
results in
\be\label{eq:3.215}
S_\nlo^{(3)}=(P_\lo^\full)^2\frac{g^2M_D^2}{192\pi^2f_\pi^2}\frac{\mu^3}{k}
\ee
The fourth, fifth, sixth and seventh classes of diagrams all proceed the same way.  We find
\begin{center}
\includegraphics[width=.65\textwidth]{A001fig4_.eps}
\end{center}
leads to
\be
S_\nlo^{(4)}=\frac{2b^{(1)}}{y_0^2}(P_\lo^\full)^2\,,
\ee
and
\begin{center}
\includegraphics[width=0.27\textwidth]{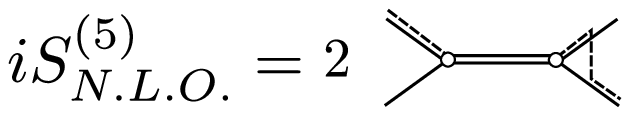}
\end{center}
results in
\begin{align}
S_\nlo^{(5)}&=-\frac{g^2M_D^2}{64\pi^2f_\pi^2}\Bigg[\Upsilon\frac23(\Lam+i\mu)+ik\frac23(\Lam+i\mu)
+\frac{8}{15\mu}\Upsilon k^2 \vec p_{D^*}\cdot\vec\epsilon_{D^*}\Bigg](P_\lo^\full)^2\,,
\end{align}
In the sixth, the $D$ and $D^*$ exchange a pion during the scattering,
\begin{center}
\includegraphics[width=.245\textwidth]{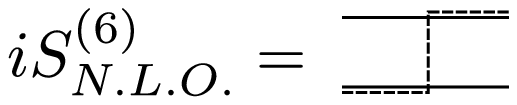}
\end{center} 
This term is
\be
S_\nlo^{(6)}=-\frac{g^2k^2}{3f_\pi^2\mu^2}\frac{M_D^2}{16\pi^2}\Upsilon^2(P_\lo^\full)^2\,.
\ee
Finally,
\begin{center}
\includegraphics[width=0.82\textwidth]{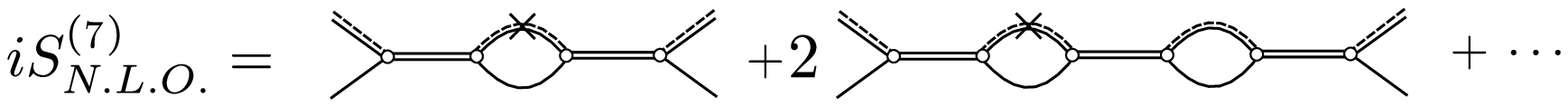}
\end{center}
is
\begin{align}
S_{N.L.O.}^{(7)}&=\paren{\frac{iy_0}{\sqrt{2}}}^2\paren{-\frac{i}{b}}^2\paren{\frac{iM_D}{4\pi}}\paren{\frac{-\lambda}{2}}\paren{\Lambda+ik}\nn\\
&\times\left\{1+2\paren{-\frac{i}{b}}\paren{\frac{iy_0}{\sqrt{2}}}^2\paren{\frac{iM_D}{4\pi}}\paren{\frac{-\lambda}{2}}\paren{\Lambda+ik}+\ldots\right\}\nn\\
&=\paren{\frac{iy_0}{\sqrt{2}}}^2\paren{-\frac{i}{b}}^2\paren{\frac{iM_D}{4\pi}}\paren{\frac{-\lambda}{2}}\paren{\Lambda+ik}\nn\\
&\times\oneov{\paren{1-\paren{-\frac{i}{b}}\paren{\frac{iy_0}{\sqrt{2}}}^2\frac{iM_D}{4\pi}(\Lambda+ik)}^2}\nn\\
&=\frac{M_D}{8\pi}\lambda(\Lambda+ik)(P_{L.O.}^{\rm{Full}})^2
\end{align}

Because we build the XEFT based on the fact that $X$ is a $D^*D$ molecule with small binding energy, we must project the above results to S-wave. We use the Legendre expansion,
\be\label{eq:112eq1}
f_\ell^m=\int_0^{2\pi}d\varphi \int_0^\pi d\theta \sin\theta f(\theta,\varphi)Y_\ell^{m*}(\theta,\varphi)\,,
\ee
setting $\ell=m=0$ and $Y_\ell^{m*}(\theta,\varphi)=\ov{\sqrt{4\pi}}$.  Term-by-term, the above classes of diagrams yield
\begin{align}
S_\nlo^{(1)-S}&=\frac{2\pi}{\sqrt{4\pi}}\frac{2\sig k^2}{M_D y_0^2}(P_\lo^\full)^2(-\cos\theta)\Bigg\vert_0^\pi =\frac{4\sqrt{\pi}\sig k^2}{M_D y_0^2}(P_\lo^\full)^2\,,\nn\\
S_\nlo^{(2)-S}&=\frac{2\sqrt{\pi}g^2M_D^2}{96\pi^2f_\pi^2}\Bigg[\Lam^2+(-i\mu)^2\log\frac{(-i\mu)}{\Lam}+ik(2\Lam+(-i\mu))\nn\\
&+k^2-\ov{2}(-i\mu)^2\paren{\ov{2\eps}+1+\log\paren{e^{\gam_E}\pi}}\Bigg]\,,\nn\\
S_\nlo^{(3)-S}&=\frac{2\sqrt{\pi}g^2M_D^2}{96\pi^2f_\pi^2}\frac{\mu^3}{k}(P_\lo^\full)^2\,,\nn\\
S_\nlo^{(4)-S}&=\frac{4\sqrt{\pi}b}{y_0^2}(P_\lo^\full)^2\,,\nn\\
S_\nlo^{(5)-S}&=-\frac{g^2M_D^2}{64\pi^2f_\pi^2}\Bigg[\Upsilon\frac23(\Lam+i\mu)2\sqrt{\pi}+ik\frac23(\Lam+i\mu)2\sqrt{\pi}\nn\\
&+i\frac{8}{15}\Upsilon k^2\frac{2\sqrt{\pi}}{3}\Bigg](P_\lo^\full)^2\,,\nn\\
S_\nlo^{(6)-S}&=-\frac{2\sqrt{\pi}}{12\pi^2}\frac{g^2k^2}{f_\pi^2\mu^2}M_D^2\Upsilon^2(P_\lo^\full)^2,\nn\\
S_\nlo^{(7)-S}&=\frac{M_D}{4\sqrt{\pi}}\lambda(\Lambda+ik)(P_\lo^\full)^2\,.\label{eq:112eq3}
\end{align}
Now summing them into $S_\nlo^{\dsd\to\dsd}$ and reorganizing in powers of momentum $k$, there are only powers $k^{-1},k^0,k^1$ and $k^2$,
\begin{align}
S_\nlo^{\dsd-S}&=S_\nlo^{(1)-S}+S_\nlo^{(2)-S}+S_\nlo^{(3)-S}+S_\nlo^{(4)-S}+S_\nlo^{(5)-S}+S_\nlo^{(6)-S}+S_\nlo^{(7)-S}\nn\\
&=S_\nlo^{\dsd-S}(k^2)+S_\nlo^{\dsd-S}(k^1)+S_\nlo^{\dsd-S}(k^0)+S_\nlo^{\dsd-S}(k^{-1})\,, \label{eq:112eq4}
\end{align}
where
\begin{align}
S_\nlo^{(k^2)}&=k^2(P_\lo^\full)^2\left(\frac{2\sig}{M_D y_0^2}+\frac{g^2M_D^2}{96\pi^2f_\pi^2}\paren{1+\frac{4}{15\mu}i\Upsilon}-\frac{g^2M_D^2\Upsilon^2}{12\pi^2f_\pi^2\mu^2}\right), \label{eq:112eq5}\\
S_\nlo^{(k^1)}&=k\frac{iM_D}{4\sqrt{\pi}}\lambda(P_\lo^\full)^2,\label{eq:112eq6}\\
S_\nlo^{(k^0)}&=(P_\lo^\full)^2\Bigg(\frac{g^2M_D^2}{96\pi^2f_\pi^2}\Bigg[\Lam^2-\mu^2\log\frac{\mu}{\Lam}-\Upsilon\Lam-\mu^2\log(-i)-i\Upsilon\mu\Bigg]\nn\\
&+\frac{M_D}{4\sqrt{m}}\lambda\Lambda+\frac{2b^{(1)}}{y_0^2}\Bigg),\nn\\
S_\nlo^{(k^{-1})}&=\oneov{k}(P_\lo^\full)^2\paren{\frac{\sqrt{\pi}g^2M_D^2\mu^3}{48\pi^2 f_\pi^2}}\,,\label{eq:112eq7}
\end{align}
where $\Upsilon$ is the physical bound state pole position we introduced in the previous section.

\noindent {\bf Renormalization}

Next we discuss $S_\nlo^{\dds-S}$ term by term to remove $\Lambda_{P.D.S.}$-dependence in the momentum expansion. First recall the S matrix resonance pole is defined as
\be
k\bigl\vert_{\text{resonance pole}}=-i\alpha+\beta\,,
\ee
and the bound state pole as
\be
k\bigl\vert_{\text{bound state pole}}=i\Upsilon\,.
\ee
We claim $|k\bigl\vert_{\text{resonance pole}}|$ is subleading to $|k\bigl\vert_{\text{bound state pole}}|$ according to our power counting. Secondly, recall the expansion of the interaction parameters \eq{ybexpansion}.
The zeroth order pieces in the expansion were constrained in the last section, where at leading order we defined the physical bound state pole to be $\Upsilon$ \eq{3.232}.  This fixed only the relation between $b^{(0)}$ and $y_0^{(0)}$.  We can use the remaining degree of freedom, together with the free parameters $b^(1)$ and $y_0^{(1)}$ and the physical resonance position, to renormalize $S_\nlo^{\dds-S}$. 

To introduce our procedure first recall that in scattering theory, we would parameterize a propagator in terms of the scattering length $a$, scattering range $r$ and  $p$, leading in general to an S matrix with complicated pole structure.  We may expand the propagator under the condition that $-\frac{r^2}{2}k^2+pk^4\ll -\ov{a}-ik$,
\begin{align}
S_0&=-\frac{2\pi}{m}\oneov{-\oneov{a}-ik+\frac{rk^2}{2}-pk^4}\nn\\
&=-\frac{2\pi}{m}\oneov{-\oneov{a}-ik}\paren{1+\oneov{-\oneov{a}-ik}}\paren{-\frac{r^2}{2}k^2+pk^4+\ldots}\,,
\end{align}
which shows $k=i\ov{a}$ is a leading order bound state pole with subleading $-\frac{r^2}{2}k^2+pk^4+\ldots$ as pole position corrections. 

To adapt this general scattering theory to our next-to-leading order result, we add (1) an NLO perturbative correction to the bound state pole position, namely $ik\to i(1-\tilde\lambda)k$, where $\tilde\lambda$ is small compared to 1, and (2) a term $\frac{\eta}{k}$ to the denomenator of $S_0$, where $\eta$ is small compared to $-\frac{1}{a}$. The modified $S_0$ can be expanded as
\begin{align}
\tilde S_0&=-\frac{2\pi}{m}\ov{-\ov{a}-ik-i\tilde\lambda k+\frac{rk^2}{2}+\frac{\eta}{k}}\nn\\
&\simeq \paren{-\frac{2\pi}{m}\ov{-\ov{a}-ik}}+\paren{-\frac{2\pi}{m}\ov{-\ov{a}-ik}}^2\left(-i\tilde\lambda k+\frac{rk^2}{2}+\frac{\eta}{k}+\ldots
\right).
\end{align}
We expand the $D^*D$ scattering amplitude following the same scheme, and the S-wave S matrix element becomes
\begin{align}
S^{\dsd\to\dsd}&=S_\lo^{\dsd\to\dsd}+S_{\nlo}^{\dsd\to\dsd}\nn\\
&=\frac{4\pi}{M_D}\frac{1}{\frac{8\pi b}{y_0^2 M_D}+ik+\Lambda_{P.D.S.}+i\Sigma_{N.L.O.} }  \nn\\
&+(P_\lo^\full)^2\left(\frac{2\sig}{M_D y_0^2}+\frac{g^2M_D^2}{96\pi^2f_\pi^2}\paren{1+\frac{4}{15\mu}i\Upsilon}-\frac{g^2M_D^2\Upsilon^2}{12\pi^2f_\pi^2\mu^2}\right)k^2\nn\\
&+(P_\lo^\full)^2\left(\frac{M_D}{4\sqrt{\pi}}\lambda\right)(ik)\nn\\
&+(P_\lo^\full)^2\Bigg(\frac{g^2M_D^2}{96\pi^2 f_\pi^2}\left[\Lambda^2-\mu^2\log\frac{\mu^2}{\Lambda}-\Upsilon\Lambda-\mu^2\log(-i)-i\Upsilon\Lambda\right]\nn\\
&+\frac{M_D}{4\sqrt{\pi}}\lambda\Lambda+\frac{2b}{y_0^2}\Bigg)+(P_\lo^\full)^2\paren{\frac{\sqrt{\pi}g^2M_D^2}{48\pi^2 f_\pi^2}}\frac{\mu^2}{k}+\ldots
\end{align}
All the terms proportional to $(P_\lo^\full)^2$ are the components of $S_\nlo^{\dds\to\dds}$. Now we discuss these terms one by one.
\begin{description}
\item[(I)] $k^2$ term. At the resonance pole position, we plug $k=-iQ+\beta$ into the above equation and obtain
\begin{align}
0=&-(\Upsilon+\alpha+i\beta)\frac{2\sig}{M_D y_0^2}
+\frac{g^2M_D^2}{96\pi^2f_\pi^2}\paren{\!1+\frac{4}{15\mu}i\Upsilon\!}\nn\\
&+\frac{g^2M_D^2\Upsilon^2}{12\pi^2f_\pi^2\mu^2}\paren{\alpha^2-\beta+2i\alpha\beta}\,.\label{eq:78eq11}
\end{align}
Solving $y_0^{(0)}$ from this equation, we completely determine the leading order parameters $b^{(0)}$ and $y_0^{(0)}$. We can see that $y_0^{(0)}$ is completely $\Lambda$ free, which means only $b^{(0)}$ is $\Lambda$-dependent.
\item[(II)] $k$ term. This term contains a perturbative correction to the bound state pole position, but would not change the bound state into a resonance state.
\item[(III)] $k^0$ term.
We renormalize the $k$-independent term of $S_{N.L.O.}^{DD^*\to DD^*}$ using $b^{(1)}$.  The scheme is to first expand the inverse of the physical binding energy $b$ using
\be
\oneov{b}=\oneov{b^{(0)}}-\oneov{(b^{(0)})^2}b^{(1)}+\ldots\,,
\ee
and corresponding diagrams as NLO corrections to the X propagator,
\begin{center}
\includegraphics[width=.55\textwidth]{78fig2.eps}
\end{center}
The series of diagrams yields
\begin{equation} \label{eq:78eq12}
S^{b^{(1)}}=\paren{\frac{ib^{(1)}}{(y_0^{(0)})^2}}\paren{P_{L.O.}^{\text{Full}}}^2.
\end{equation}
Adding $S^{b^{(1)}}$ into $S_\nlo^{\dsd\to\dsd}$, and taking the only $k^0$ term, we obtain,
\be\label{eq:78eq13}
S_\nlo^{(k^0)}=\left\{\frac{g^2M_D^2}{96\pi^2f_\pi^2}\Bigg[\Lam^2\!-\!\mu^2\log\frac{\mu}{\Lam}-\Upsilon\Lam-\mu^2\log(-i)-i\Upsilon\mu\Bigg]\!+\!\frac{2b^{(1)}}{y_0^2}\right\}P_\lo^\full.
\ee
Setting $S_\nlo^{(k^0)}=0$, we can solve for $b^{(1)}$ to cancel $\Lambda_{P.D.S.}$ dependence in all constant terms brought in by NLO corrections.  This completes the determination of the X binding energy $b$ to NLO.
\item[(IV)] $k^{-1}$ term. This term arises from the $D^*$ self-energy correction and contains no $\Lambda_{\rm{P.D.S.}}$. However at the $X$ threshold, $k=0$ and this term is not finite.  The $D\pi$ loop must then be resummed into the $D^*$ propagator as in \sec{II.3}.
\end{description}

So far we have renormalized the next-to-leading order $D^*D$ scattering amplitude employing only $b^{(0)}, y_0^{(0)}$ and $b^{(1)}$.  The subleading order $D^*DX$ coupling constant $y_0^{(1)}$ remains free. To double check, we write all the $\Lam_{P.D.S.}$-dependent terms in $S_\nlo^{\dsd}$ as
\begin{align}
S_\nlo^{(\Lam)}&=\frac{g^2M_D^2}{96\pi^2 f_\pi^2}\left[\Lam^2-\Upsilon\lambda-\mu^2\log\frac{\mu}{\Lam}\right](P_\lo^\full)^2+\frac{2b^{(1)}}{(y_0^{(0)})^2}(P_\lo^\full)^2\,,
\end{align}
where we can clearly see $b^{(1)}$ absorbed all $\Lambda_{\rm{P.D.S.}}$.

\subsection{XEFT Pole Hunting Preliminary: \texorpdfstring{$B\to DD\pi K$}{B to DDpiK} Line Shape Fitting}
\label{sec:II.4.6}

In this section, we discuss the procedure for $DD^*$ scattering amplitude pole hunting to determine whether X(3872) is a bound state or a resonance of $DD^*$. 
Among X(3872) experiments reviewed in \sec{II.4.3}, the first observable we fit is the line shape obtained in the X particle discovery processes, $B\to D^0\overline{D^*}K$ decays, in which X(3872) is detected as a strong and clear signal, as well as in which the XEFT was first developed.

The parameters obtained by fitting the experimental data should have sizes compatible with our estimates in Sec.\:\ref{sec:II.4.5.1}, and the pole-hunting results should be consistent with our power-counting scheme, namely, the resonance pole positions are obtained by perturbing the bound state pole position. If the results do not fit these expectations, then it would suggest the bound state picture fails to describe the X particle, and we should consider the other fine-tuning power-counting scenarios proposed in \sec{II.4.3}. As in any EFT, the path to incorporating the theoretical calculation into experimental prediction is to separate scales of the physical processes and factorize the observables into single-scale dependent pieces.  Factorization of $B\to D^0\overline{D^*}K$ decays has been previously discussed by \cite{Braaten:2003he,Braaten:2004rw,Braaten:2004fk,Braaten:2005ai,Braaten:2005jj,Braaten:2004jg}.  In this section, we proceed by first discussing how to separate the physical short and long distance scales in the processes of $B\to \overline{D^0}D^{*0}K$. Then we discuss the short and long distance pieces respectively. Last, we discuss the fitting procedure.

\subsubsection{Separation Scales for the Decay $B\to DDK$}
\label{sec:II.4.6.1}

In the hierachy of momentum scales involved in the $B\to DD^*K$ decay, the largest is the mass of the $W$ boson, $M_W\approx 80$ GeV, the intermediate state in the quark decay process $b\to c\bar c s$. The next largest is the scale of the momentum of the $s$ quark that recoils against the $c\bar c$ system, given by $m_b-2m_c\approx 1.5$ GeV. Next is $\Lambda_{\text{QCD}}\approx 200$\,MeV, the momentum scale associated with the wavefunctions of light quarks in the hadrons, and $m_\pi\approx 140$\,MeV associated with the pion-exchange interaction between the $D$ mesons. The smallest momentum scale arises from the scattering length $1/a \simeq 45$\,MeV determining the size of $DD^*$ molecule. This hierarchy of momentum scales is summarized by the inequalities
\begin{equation}
1/a\ll m_\pi \lesssim \Lambda_{\text{QCD}} \ll m_b-2m_c \ll M_W.
\end{equation}

We choose $m_\pi\sim \Lambda_{\text{QCD}}$ as separation scale for this factorization. We refer to processes involving $\vert\vec{q}\vert<\Lambda_{\text{QCD}}\sim m_\pi$ as {\it long-distance} (XEFT) while processes involving momenta $\vert\vec{q}\vert>\lqcd$ are {\it short-distance}

\subsubsection{The Short-distance $B\to DDK$ decay}
\label{sec:II.4.6.2}

\begin{figure}[!h]
\centering
\includegraphics[width=.75\textwidth]{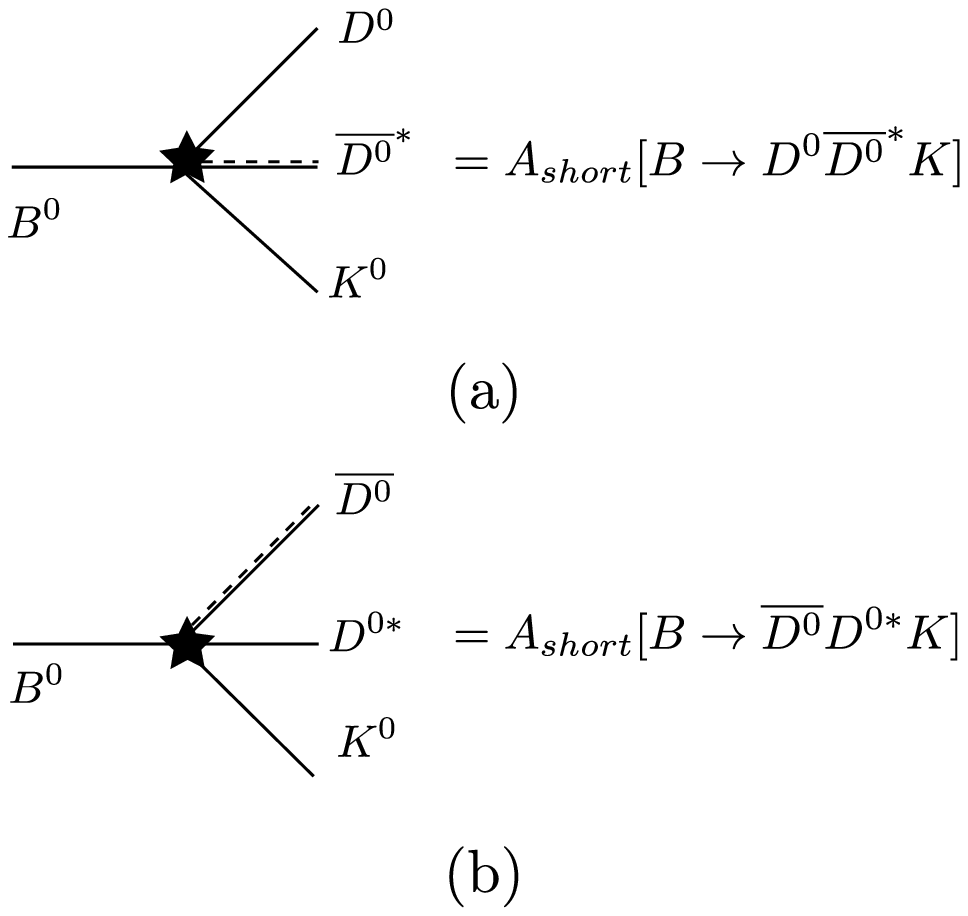}
\caption{Short-distance $B\to DDK$ decay.  The stared vertices represent the short-distance $BDD^*K$ interaction.}
\label{fig:figure003}
\end{figure}

The short-distance vertices we need to introduce are summarized in \fig{figure003}.  Denoting $P$ as the $B^0$ momentum, $\eps^*$ as the $D^*$ polarization vector, we have
\begin{align}\label{BDDKvertex}
A_{short}& \left[B\to D^0\overline{D^{0*}}K\right]=c_1(m_\pi)P\cdot \epsilon^* \,,\nn \\
A_{short}& \left[B\to \overline{D^0} D^{0*}K\right]=c_2(m_\pi)P\cdot\epsilon^* \,,
\end{align}
where $c_1$ and $c_2$ can be extracted from \cite{Aubert:2003jq}.

\subsubsection{Long-Distance \texorpdfstring{$DD^*$}{D-Dstar} Interaction}
\label{sec:II.4.6.3}

The long-distance physics in the $B\to XK\to DD^*K$ process is the $DD^*$ scattering dynamics in which the X particle is created. The nuclear interactions between $DD^*$ and their formation of the X were detailed in \sec{II.4.5}, and we may directly employ the XEFT Lagrangian and the $DD^*$ scattering amplitudes. However, in order to compute $B\to XK\to DD^*K$ amplitude, we also need the decay amplitudes $X\to D^*D$ and $X\to DD\pi$ for fitting experimental data. In this section, we derive these amplitudes and power count the associated Feynman diagrams.

Defining $\mathscr{A}$ as the renormalized X decay amplitude, $Z$ as the counter-term, and $A$ as the unrenormalized $X$ decay amplitude, we expand $A$ and $Z$ order by order,
\begin{align}
\mathscr{A}&=\sqrt{Z}A\nn\\
&=(Z_{L.O.}+Z_{N.L.O.}+\ldots)^{1/2}(A_{L.O.}+A_{N.L.O.}+\ldots)\nn\\
&=(Z_{L.O.})^{1/2}\left(1+\frac{1}{2}\frac{Z_{N.L.O.}}{Z_{L.O.}}+\ldots\right)(A_{L.O.}+A_{N.L.O}+\ldots)\,.
\end{align}
Thus we have leading and next-to-leading order renormalized $X\to DD\pi$ amplitudes,
\begin{align}
\mathscr{A}_{L.O.}&=Z_{L.O.}^{1/2}A_{L.O.},\nn\\
\mathscr{A}_{N.L.O.}&=\frac{1}{2}\frac{Z_{N.L.O.}}{\sqrt{Z_{L.O.}}}A_{L.O.}+Z_{L.O.}^{1/2}A_{N.L.O.}\,,
\end{align}
and the decay (differential) probabilities
\begin{align}
\vert \Gam_{L.O.}\vert^2&=Z_{L.O.}\vert A_{L.O.}\vert^2\nn\\
\vert \Gam_{N.L.O.}\vert^2&=Z_{N.L.O.}\vert A_{L.O.}\vert^2+Z_{L.O.}\left(A_{L.O.}^*A_{N.L.O.}+A_{L.O.}A_{N.L.O.}^*\right)\,,
\end{align}
which need to be integrated over the final state phase space to produce the total decay rate.  The amplitudes and wavefunction renormalization factors are summarized in Appendix \appx{D.7}.

\begin{figure}[!h]
\centering
\includegraphics[width=.8\textwidth]{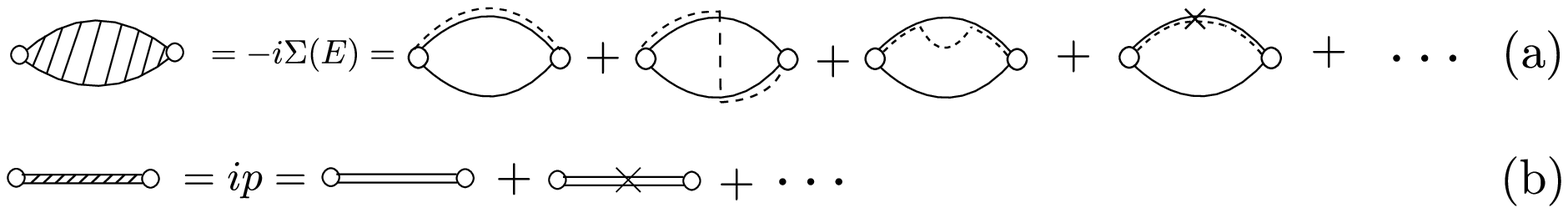}
\caption{$DD^*$ loop and $X$ propagator expansion}
\label{fig:119fig1}
\end{figure}

$Z$ and $A$ are determined by the two-point Green's function,
\be\label{generalGreensfnX}
G(E)=\int d^4x e^{-iEt}\bra{0}X^i(x)X^j(0)\ket{0}=i\del_{ij}\frac{Z(E)}{E+E_X+i\Gam/2}+\ldots
\ee
We define the $DD^*$ loop and $X$ propagator with its expansion by \fig{119fig1}, power counting according to XEFT with X as a bound state at leading order and a resonance at next-to-leading order. $-i\Sig(E)$ is the full $DD^*$ loop: on the right hand side of \fig{119fig1}(a), the first diagram is the leading order $DD^*$ loop, and the second and third are the next-to-leading order diagrams involving one-pion-exchange corrections and $D^*$ propagator correction. $iP$ is the $X$ full propagator, and on the right hand side of \fig{119fig1}(b) the first diagram is the leading order whereas the second is next-to-leading order. In terms of these expansions, the Green's function is
\begin{align}
G(E)&=iP\left(1+\sum_{n=1}^\infty\left(-i\Sigma \cdot iP\right)^n\right)=\frac{iP}{1-P\Sigma}=\frac{iP}{1-P\text{Re}\Sigma-iP\text{Im}\Sigma}\nn\\
&=\frac{iP}{1-P(-E_X)\text{Re}\Sigma(-E_X)-(E+E_X)(P\text{Re}\Sigma)^\prime\vert_{E=-E_X}-iP\text{Im}\Sigma\vert_{E=-E_X}}\,,
\end{align}
where Re$\Sig$ is the real part of $\Sig$, Im$\Sig$ is the imaginary part of $\Sig$, and $(P\rm{Re}\Sig)'\big\vert_{E=-E_X}$ means taking the $E$ derivative of $(P\rm{Re}\Sig)$ and then setting $E=-E_X$. For $X$ considered a bound state at leading order, the expansion of the denominator in the Green's function is done around $E=-E_X$, with the pole mass of $X$ defined from the real part $1-P(-E_X)\rm{Re}\Sig(-E_X)=0$ at $E=-E_X$.  The factor $(P\rm{Re}\Sig)'|_{E=-E_X}$ determines the shape of the $X\to DD\pi$ differential decay rate, as seen by putting the propagator in the form of \req{generalGreensfnX},
\begin{align}
G(E)&=\frac{-iP}{(E+E_X)(P\text{Re}\Sigma)^\prime\vert_{E=-E_X}+iP\text{Im}\Sigma\vert_{E=-E_X}}\nn\\
&=\frac{-i}{E+E_X+\frac{i\text{Im}\Sigma\cdot P}{(P\text{Re}\Sigma)^\prime}\vert_{E=-E_X}}\frac{P}{(P\text{Re}\Sigma)^\prime}\Biggl\vert_{E=-E_X}\,.
\end{align}
We identify the wavefunction renormalization as
\begin{align}
Z(E)&=\frac{P}{(P\text{Re}\Sigma)^\prime}\Biggl\vert_{E=-E_X}\nn\\
&=(P_{L.O.}+P_{N.L.O.})\cdot\left[\left[(P_{L.O.}+P_{N.L.O.})(\text{Re}\Sigma_{L.O.}+\text{Re}\Sigma_{N.L.O.})\right]^\prime \right]^{-1}\nn\\
&=\frac{1}{\text{Re}\Sigma_{L.O.}^\prime}-\frac{\text{Re}\Sigma_{N.L.O.}^\prime}{(\text{Re}\Sigma_{L.O.}^\prime)^2}-\frac{P_{N.L.O.}^\prime\text{Re}\Sigma_{L.O.}}{P_{L.O.}(\text{Re}\Sigma_{L.O.}^\prime)^2}\,,
\end{align}
so that the leading and next-to-leading $Z$ are
\be
Z(E)_{L.O.}=\frac{1}{\text{Re}\Sigma_{L.O.}^\prime}\,,\qquad 
Z(E)_{N.L.O.}=-\frac{\text{Re}\Sigma_{N.L.O.}^\prime}{(\text{Re}\Sigma_{L.O.}^\prime)^2}-\frac{P_{N.L.O.}^\prime\text{Re}\Sigma_{L.O.}}{P_{L.O.}(\text{Re}\Sigma_{L.O.}^\prime)^2}\,,
\ee
and the decay rate is
\be
\Gamma=\frac{P}{(P\text{Re}\Sigma)^\prime}2\text{Im}\Sigma\,.
\ee
We organize all the leading order and next-to-leading order $X\to DD^*K$ decay rates and all Feynman diagram results in \appx{D.7}. With these results, we complete the calculation of the long-distance piece in $B\to DD^*K$ processes.

\subsubsection{Combining Long and Short-Distance Physics: \texorpdfstring{$B\to DD\pi K$}{B to DDpiK} Full Amplitude Power Counting}
\label{sec:II.4.6.4}

\begin{figure}[!h]
\centering
\includegraphics[width=.4\textwidth]{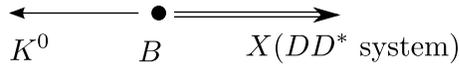}
\caption{$B^0$ rest frame decays into $K^0DD^*$}
\label{fig:figure004}
\end{figure}

To simplify  kinematics, we discuss $B^0\to DD^*\pi K^0$ decay in the $B^0$ rest frame, as shown in \fig{figure004}.  $B^0$ decays such that the $K^0$ and the $X$ ($DD^*$ system) are emitted back to back. The decay vertex is the short-distance operator given in \req{BDDKvertex}, and the $X$ or $DD^*$ system is the long distance dynamics described by XEFT.

\begin{figure}[!h]
\centering
\includegraphics[width=.6\textwidth]{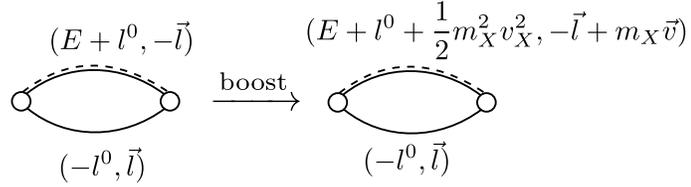}
\caption{Lorentz invariance of leading order $D^*D$ loop}
\label{fig:figure005}
\end{figure}

In $B^0$ rest frame, the momenta are
\be
m_X=M_D+m_{D^*},\quad m_Xv^\mu=p_K^\mu\,,
\ee
where $v^\mu$ is $X$ velocity and $p_K^\mu$ is the momentum of $K$. 
In the previous section, we treated the $X$ in its rest frame. Now we boost it into the $B^0$ rest frame
\be
(E_X,\vec{k}_X)\xrightarrow{\text{boost}}\left(E_X+\frac{1}{2}m_X\vec{v}_X^2,m_X\vec{v}\right)\,,
\ee
where the kinetic energy of $X$ is
\be
E_X=\sqrt{\vec p_X^2+m_X^2}=m_X+\frac{\vec p_X^2}{2m_X}+\ldots
\ee
The correction to the energy component is the order of $(\vec{v}^2/m_X)$ and can be ignored.  This does not make a difference to our results in previous section since constant shifts to the external momenta do not change the values of the loop integrals as is depicted in \fig{figure005}.  The X propagator is also unchanged by the boost from X rest frame to $B^0$ rest frame,
\be
\frac{i\delta^{ij}}{\sigma E-b}\to\frac{i\delta^{ij}}{\sigma\paren{E+\frac{1}{2}m_X\vec{v}_X^2-\frac{\vec{p}_X^2}{2m_X}}-b}=\frac{i\delta^{ij}}{\sigma E-b}\,.
\ee
Thus, being Lorentz-invariant, all the results from XEFT can be used as before.

Now we power count Feynman diagrams for the decay $B\to DD^*\pi K$ in a scheme consistent with the XEFT power counting in the previous section. However, for convenience of utilizing the experimental data, we write the amplitudes for $B\to DD\pi K$ rather than $B\to D^*DK$ from this point, where the $\pi$ is from the $D^*\to D\pi$ decay process.

\begin{figure}[!h]
\centering
\includegraphics[width=.85\textwidth]{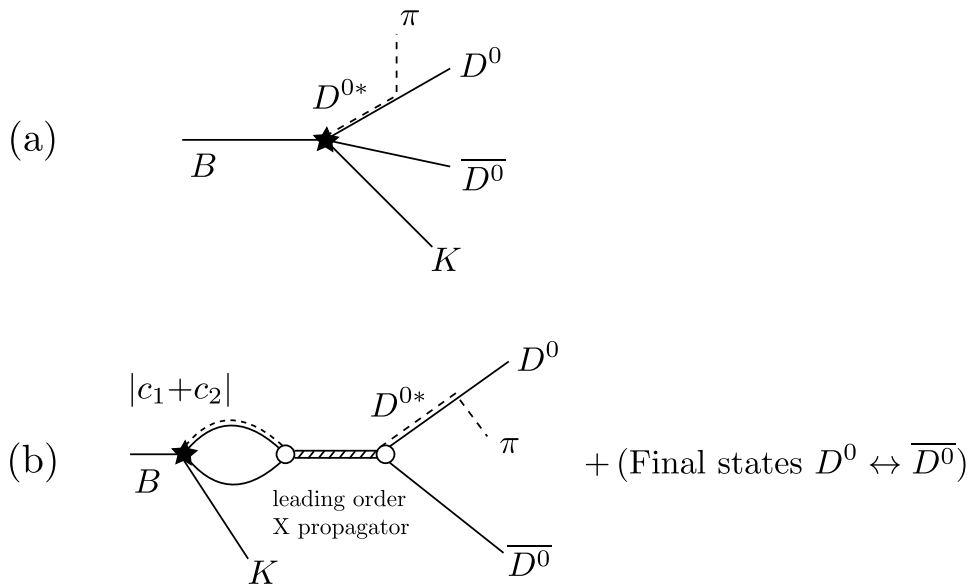}
\caption{Leading order $B\to DD\pi K$ including both short and long-distance physics.}
\label{fig:figure006}
\end{figure}

\fig{figure006} shows the leading order $B\to DD\pi K$ process including both long- and short-distance physics. The starred vertex is the short-distance $B\to DD\pi K$ decay, and the interactions between $DD\pi$ and $X$ are determined by the long-distance XEFT at the lower energy scale. \fig{figure006}a provides the experimental background in the $B\to DD\pi K$ invariant mass spectrum, because there are no interactions between $DD^*$ to form an $X$.  \fig{figure006}b includes an X propagator and provides the experimental signal of the X(3872) as a peak in the $B\to DD\pi K$ invariant mass spectrum. From the power counting in XEFT in Sec.\:\ref{sec:II.4.5}, we learned that the leading order X propagator is composed of a geometric series of $DD^*$ loops.

\begin{figure}[!h]
\centering
\includegraphics[width=.75\textwidth]{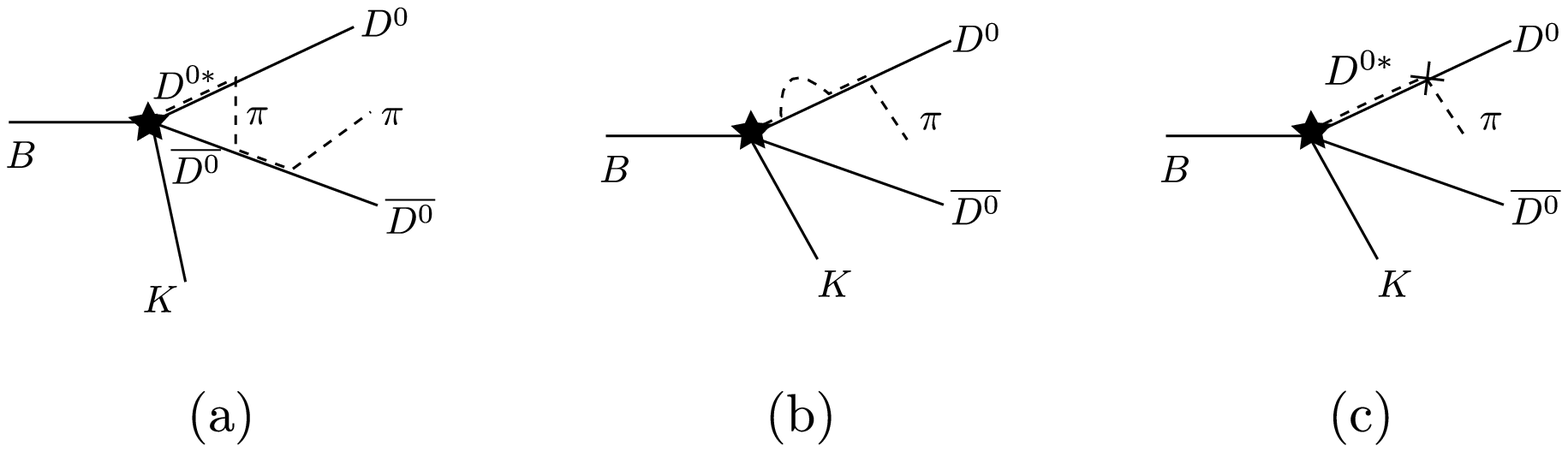}
\caption{Next-to-leading order background corrections to long-distance physics}
\label{fig:figure007a}
\end{figure}

\begin{figure}[!h]
\centering
\includegraphics[width=.75\textwidth]{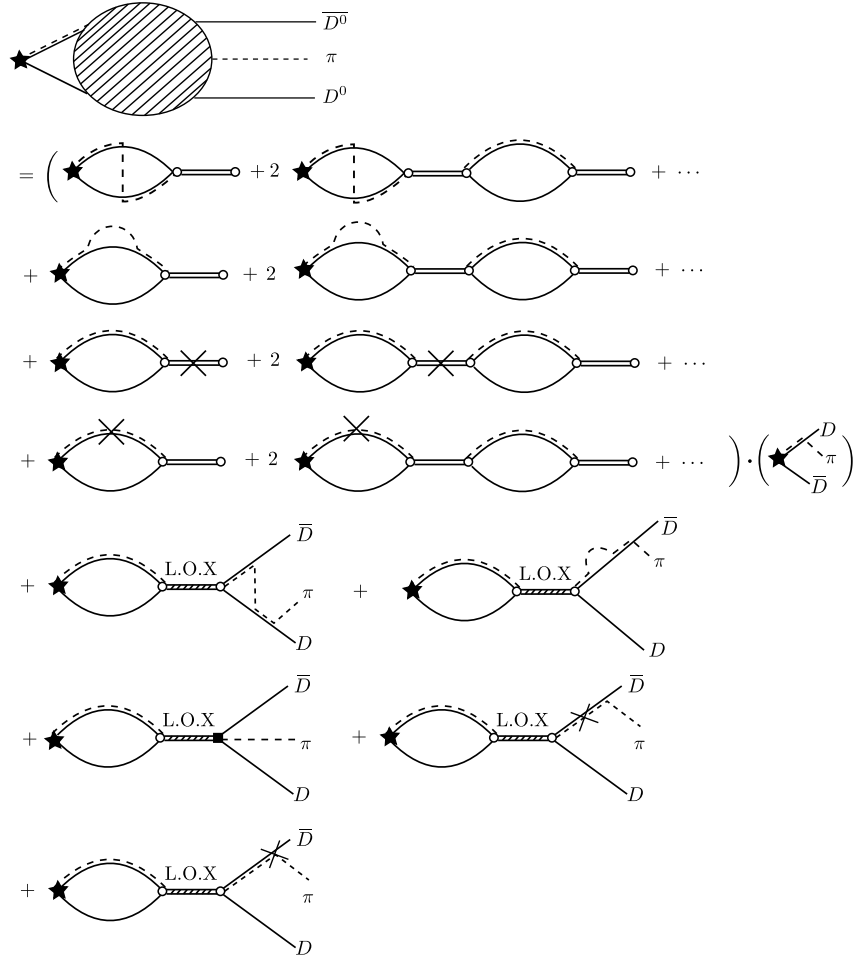}
\caption{Next-to-leading order diagrams correcting the $D\bar D\pi$ final state.}
\label{fig:figure008}
\end{figure}

Next-to-leading long-distance corrections to the background involve pion-exchange and $DD^*\pi$ NLO interaction but no X intermediate, as shown in \fig{figure007a}.  Next-to-leading long-distance corrections to the signal include all the NLO corrections to the X propagator, pion-exchange and $D^*$ propagator corrections to the final state $DD^*$ (with LO X propagator), and the NLO three-body decay $XDD\pi$ vertex in \eq{78IIeq1}.  These diagrams are organized explicitly in \fig{figure008}.

At this point we have power counted and expanded the $B\to DD\pi K$ amplitude by separating the long- and short-distance physics.  We can now organize the analytic results for the total amplitude. The leading order $B\to DD\pi K$ amplitude is
\begin{align}
A_{L.O.}[B\to D^0\overline{D^0}\pi^0 K]&=A_{short}[B\to D^{0*}\overline{D^0}K]A_{L.O.}[D^{*0}\to D^0\pi]\nn\\
&+A_{short}[B\to D^0\overline{D^{*0}}K]A_{L.O.}[\overline{D^{0*}}\to \overline{D^0}\pi]\nn\\
&+A_{short}[B\to D^{0*}\overline{D^0}K]\left(-i\Sigma_{L.O.}^{\overline{D^0}D^{0*}\text{ Loop}}\right)\bigl\vert_{\text{boosted}}\nn\\
&\times A_{L.O.}[\overline{D^0}D^{0*}\to \overline{D^0}D^0\pi]\bigl\vert_{\text{boosted}}\nn\\
&+A_{short}[B\to \overline{D^{0*}}D^0\pi]\left(-i\Sigma_{L.O.}^{\overline{D^{*0}D^0\text{ Loop}}}\right)\bigl\vert_{\text{boosted}}\nn\\
&\times A_{L.O.}[\overline{D^{*0}}D^0\to \overline{D^0}D^0\pi]\bigl\vert_{\text{boosted}}\,,
\end{align}
where the short-distance matrix elements are $A_{short}[B\to D^{*0}\overline{D^0}K],A_{short}[B\to D^0\overline{D^{*0}}K]$ are given in \req{BDDKvertex} and
\begin{align}
A_{L.O.}[D^{*0}\to D^0\pi]&=-i\frac{g(\vec{p}\cdot\vec{\epsilon})}{2\sqrt{m_\pi}f_\pi}\oneov{E_{D^*}-\frac{\vec{p}_{D}^2}{2m_{D}}}\,,\nn\\
A_{L.O.}[\overline{D^{*0}}\to \overline{D^0}\pi]&=-i\frac{g(\vec{p}\cdot\vec{\epsilon})}{2\sqrt{m_\pi}f_\pi}\oneov{E_{\bar D^*}-\frac{\vec{p}_{\bar D}^2}{2m_{\bar D}}}\,,\nn\\
\left(-i\Sigma_{L.O.}^{\overline{D^0}D^{0*}\text{ Loop}}\right)&=\left(-i\Sigma_{L.O.}^{\overline{D^{*0}}D^0\text{ Loop}}\right)=i\frac{M_{D}}{4\pi}\left(-\sqrt{-\vec{k}^2-i\epsilon}+\Lambda_{P.D.S.}\right)\,,\nn\\
A_{L.O.}[\overline{D^0}D^{0*}\to \overline{D^0}D^0\pi]&=S_{L.O.}\bigl\vert_{\text{boosted}}\,.
\end{align}
The $S_\lo\big\vert_{\rm{boosted}}$ can be found in Appendix \appx{D.7}.  The next-to-leading order $B\to DD\pi K$ result is
\begin{align}
A_{N.L.O.}[B\to \overline{D^0}D^0\pi K]&=\left(A_{short}[B\to D^{0*}\overline{D^0}K]+A_{short}[B\to \overline{D^{0*}}D^0K]\right)\nn\\
&\left(i P_{N.L.O.}^{Full}\right)\left(A_{L.O.}[D^{*0}\to D^0\pi]+A_{L.O.}[\overline{D^{*0}}\to \overline{D^0}\pi]\right)\nn\\
&+\left(A_{short}[B\to D^{0*}\overline{D^0}K]+A_{short}[B\to \overline{D^{*0}}D^0K]\right)\nn\\
&\times\left(-i\Sigma_{L.O.}^{\overline{D^0}D^{0*}\text{ Loop}}\right)\Bigg(A_{N.L.O.}[\overline{D^0}D^{*0}\to \overline{D^0}D^0\pi]\nn\\
&+A_{N.L.O.}[D^0\overline{D^{*0}}\to \overline{D^0}D^0\pi]\Bigg)\,,
\end{align}
where
\be
A_{N.L.O.}[\overline{D^0}D^{*0}\to\overline{D^0}D^0\pi]+A_{N.L.O.}[D^0\overline{D^{0*}}\to\overline{D^0}D^0\pi]=S_{N.L.O.}\,.
\ee
All the specific expressions for NLO elements needed above can be found in Appendix \appx{D.7}.  Unsurprisingly, the $\Lambda_{P.D.S.}$ dependence has cancelled in both leading and next-to-leading order decay rates.

\subsection{Experimental Data: Fitting} 
\label{sec:II.4.7}

Having organized the amplitudes for $B\to DD\pi K$, we now present the schematic procedure for fitting $B\to DD\pi K$ data. After obtaining parameters $b^{(0)}, b^{(1)}$, and $y^{(0)}$, we can determine the pole position of X(3872). Although in previous sections we only power counted X as a bound state, we must complete the power counting for the other two scenarios and fit the data in each case to find which hypothesis results in the best fit. The fitting steps are
\begin{description}
\item[Step I.] Extract the invariant mass distribution in the charged channel $B^\pm\to DD\pi K^\pm$, which we use because it exhibits a clearer signal.
\item[Step II.] Fit the short-distance $B\to DD\pi K$ as background, in the range up to the threshold $m_\pi$.  As we discussed before, $m_\pi$ is the scale separating long and short distance physics.
\item[Step III.] Use the long-distance $B\to DD\pi K$ to fit the signal on top of the background, from the lower bound set by the $DD\pi$ threshold up to $m_\pi$.
\item[Setp IV.] The fitting master formula is
\begin{equation*}
N=[\text{bin}]\frac{N^{\text{obs}}}{B}\frac{d\text{Br}(B\to DD\pi K)}{dE},
\end{equation*}
E is the invariant mass of $DD\pi$, the bin width is [bin]$=4.25$\,MeV, $N^{\text{obs}}=17.4\pm 5.2$, $B=(1.07\pm 0.31)\times 10^{-4}$, and
\begin{align*}
\text{Br}&(B^\pm\to DD\pi K^\pm)=\frac{\Gamma(B^\pm\to DD\pi K^\pm)}{\Gamma(B^\pm\to \text{anything})}\,,\\
\Gamma&(B^\pm\to\text{anything})=4.01104\times 10^{-10}\text{MeV}\,.
\end{align*}
To obtain $\Gam(B^\pm\to DD\pi K^\pm)$, we need the leading and next-to-leading order $B\to DD\pi K$ amplitudes integrated over 4-body phase space, which are given in Appendix \appx{D.8}.
\end{description}

\section{Chapter Summary} 
\label{sec:Xsummary}

In this chapter, I have explored hadronic interactions in the low-energy confining region of QCD, through $D\pi$ and $DD^*$ interactions using Heavy Hadron Chiral Perturbation Theory and X Effective Field Theory.  

In the study of $D\pi$ scattering near the $D^*$ threshold, I developed HHChPT with the complete set of isospin symmetry breaking and heavy quark spin-flavor symmetry breaking effects up to next-to-leading order.  This is the first time the full NLO HHChPT Lagrangian has been derived for the $SU(2)$ flavor symmetry case.  I then reduce this theory to the XEFT with non-relativistic, perturbative pions.  Using this version of XEFT, I resummed the $D\pi$ loops near $D^*$ threshold to leading order in order to improve the accuracy of the extrapolation of the D-meson mass spectrum to the physical point from lattice data, which show an unphysical hierarchy in the pion mass being greater than $D$-$D^*$ mass splitting.

In the study of $DD^*$ scattering, I generalized the XEFT Lagrangian to describe the X(3872) as either a real bound state, virtual bound state or resonance and included $XDD$ interaction terms.  I considered the set of physical mass-splittings that suggest the X is, at leading order, a $DD^*$ bound state, power counting accordingly, and renormalized the $DD^*$ scattering amplitude to next-to-leading order by resumming the $DD^*$ loop correction to the X propagator.  After this resummation, the X pole is shifted into the resonance region.  I then used these results in the $B\to DD\pi K$ decay to factorize the short-distance electroweak decay vertex from the long-distance final state $DD\pi$ interactions, described in XEFT.  I provided the formulae to fit the line shape of X in the $B\to DD\pi K$ invariant mass spectrum as a way to determine whether the X is best described as a real or virtual bound state or resonance.

Both $D\pi$ and $DD^*$ scattering involve threshold enhancement effects. In the EFT language, these thresholds contain loop corrections that are the same order as tree-level, and therefore must be resummed into a complete leading order result.  This technique is equivalent to solving the renormalization group equation to resum the leading order large logarithms, which we used extensively in the previous chapter on summing logs in perturbative QCD using SCET.

The efforts in this chapter studying hadronic interactions are directed toward determining the structure of exotic hadrons, whether they are systems of $>3$ quarks, or pairings to two hadrons.  However, the theoretical tools are bottom-up effective theories, which allow for (and usually do not predict)  fine-tunings among the parameters of the theory, and our observables so far are limited to a few scattering parameters: the quantum numbers and the invariant mass spectrum containing a line shape.  Ideally, we would perform experiments to probe the structure of these states directly, just as we use deep inelastic scattering as a microscope to see into the structure of the proton.  To compensate the short lifetime of many of these states, we could imagine a fixed-target experiment with a positron beam.  In the lab frame, the hadronic states of interest are created with relativistic momenta, allowing them to propagate long enough to collide with another electron or nucleus in the target, thus involving a high-energy scattering that would probe the short-distance structure of the state.

%% file: chapter_3.tex
\chapter{Effective Field Theory for Electron Laser System}
\label{sec:III}


In this part, I first introduce the long-existing issue of electron radiation reaction in a strong electromagnetic field, and then review progress made on this issue, particularly the semiclassical approach and its limitations.  Then I present the effective field theory approach to solving this problem, as well as other predictions that this new effective field theory can make to connect to experimental observables.

\section{Current Laser Experiments and Motivation to Develop an EFT for the Electron-Laser System} 
\label{sec:III.2}

I introduce motivations for constructing a new effective field theory for electrons travelling in strong laser fields, driven by both applied and fundamental physics, for facilities around the world to strive to build laser systems with ever higher intensity.  As we will see in the subsequent sections, the intensities of the laser fields in current experiments already enter the domain where quantum processes affect the electron dynamics in strong fields, and if new facilities can create still stronger fields, a systematic quantum field theoretic framework to predict observables will be a necessity.

\subsection{Motivations from Applications and Fundamental Questions}

\begin{figure}[!h]
\centering
\includegraphics[width=0.85\textwidth]{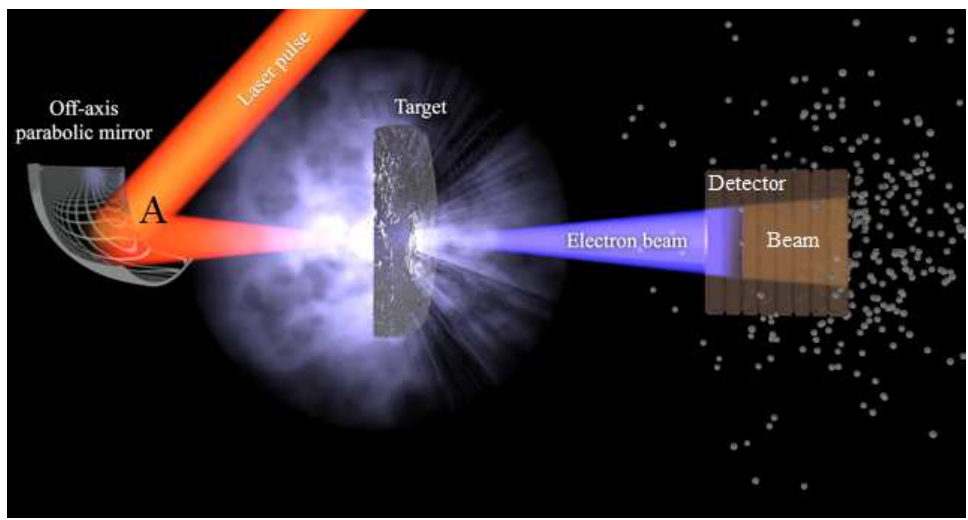}
\caption{Cartoon of a laser-driven high-flux, high-energy particle source (Graphic courtesy of B. M. Hegelich).}
\label{fig:88fig1}
\end{figure}

To date, high-intensity laser experiments in the US mostly aim to develop new particle sources and to study plasma dynamics in strong fields.  A typical experimental setup is shown in Figure \ref{fig:88fig1}.  The laser accelerates electrons out of a solid or gas target.  High momentum $p_e\gtrsim 10-100$\,MeV electrons may then be collided with a second target, as depicted, to produce secondary particles such as neutrons or high-energy photons.  It may also be possible to tailor the structure of the first target to induce the electrons to emit high-energy photons during the acceleration process.  Lower energy electrons remain coupled to the plasma in the target, resulting in nonlinear plasma dynamics that accelerate postively-charged ions in the target to moderately relativistic momenta, from a few percent to several times the ion mass.  Whatever the originating dynamics, the high fluxes of high energy particles produced from laser-matter interactions are anticipated to be useful sources for medical and imaging applications.

The fundamental physics questions originate in some of the earliest studies of classical and quantum field theory, which the advent of high-intensity lasers now provides an experimental tool to test.  The primary questions relevant to our study are: 

\noindent {\bf 1. Particle creation by strong classical fields.}  When the electrostatic potential difference exceeds the mass gap, $e\Delta V> 2m_e$ where $e$ is the elementary charge, the potential spontaneously emits particles (\cite{Klein1929,Dunne:2008kc,Hegelich:2014tda}).  The emission rate depends on the details of the potential but is always nonzero when $e\Delta V>2m_e$.  This spontaneous particle creation was the first example of a nonperturbative effect in quantum theory, because it requires solving the interaction with the potential to all orders, rather than expanding in a power series in the coupling $e$.

Spontaneous particle creation in QED is the prototype to help understand phenomena in other domains.  In quantum chromodynamics, it is thought to help explain the high multiplicity and thermalization of particles created in the early stages of heavy ion collisions (\cite{Andersson:1983ia,Kharzeev:2006zm,Gelis:2014qga,Gelis:2015kya}).  In gravity, the same physics underlies Hawking radiation (\cite{Brout:1995rd,Kim:2007ep}) and particle creation in the expanding universe (\cite{Anderson:2013zia,Anderson:2013ila}), both predictions subject to ongoing investigation and debate.  Magnetic fields of the corresponding strength are believed to exist around neutron stars (\cite{Harding:2006qn}).  Even with such broad impact and long interest, particle production in strong classical fields has yet to be experimentally verified.

\noindent {\bf 2. Electron radiation dynamics in strong fields.} According to Maxwell's equations, electrons radiate when accelerated by strong electromagnetic fields, and conservation of momentum requires that they recoil from that radiation.  However the recoil is not accounted for in the Lorentz force.  This ``radiation reaction'' problem can be perturbatively corrected by subtracting the lost momentum from the Lorentz force (\cite{abraham1905theorie,lorentz1909theory,Dirac:1938nz,landau1975lifshitz}), but even this solution is not complete:  First, the perturbative approach breaks down in strong enough fields, when the momentum in the classical radiation is comparable to the momentum in the radiating electron (\cite{Hadad2010}).  Second, the momentum absorbed by the electron is not removed from the accelerating field. The second problem does not exist in quantum theory, because we use Feynman's Green's function, which is the average of the retarded and advanced functions thus communicating the momentum absorbed by the electron back to the source of the field (\cite{Feynman:1965jda}).  For this reason, radiation reaction is always consistently incorporated in quantum theory, but is not compatible with classical radiation theory, which uses only the retarded Green's function. 

On the other hand, in the same regime where the intensity of classical radiation becomes large, high-energy quantized photon emission also becomes relevant.  The reason is that electrons are accelerated to high momentum $p_e\gg m_e$ in less than one wavelength, corresponding to the coherent absorption of many quanta from the classical field.  Thus, the now-relativistic electron may also re-radiate the momentum into photons with energy much greater than the energy of a single quantum of the classical field.

To describe the high-intensity laser-plasma experiments and push the applications to higher intensity, we need to describe long-wavelength (classical) particle dynamics and radiation in a unified, consistent and systematic framework.

\section{Current theory for electron dynamics in a laser field}

In this part some basics of the dynamics of free electrons in a laser field is reviewed (see also \cite{eberly1969progress}) and connections are made to recent studies. Results in classical and quantum electrodynamics are treated separately. 

\subsection{Laser fields}\label{sec:laserfield}

A laser field is well-approximated by a plane wave, which means it has a lightcone symmetry: The field strength tensor $F^{\mu\nu}$ only depends on a single lightcone coordinate $\bar n\cdot x$ and satisfies $\bar n_\mu F^{\mu\nu}=0$, where $\bar n^\mu=(1,0,0,-1)$ is the lightcone four-vector that was introduced above \eq{lightconep}.  As a consequence, the Lorentz scalar and pseudoscalar constructed from the field tensor,
\begin{align}
\cS(x)&=\oneov{4}F^{\mu\nu}(x)F_{\mu\nu}(x)=-\oneov{2}[E^2(x)-B^2(x)]=0\,,\label{eq:82eq11}\\
\cP(x)&=\oneov{8}F^{\mu\nu}(x)\epsilon_{\mu\nu\alpha\beta}F^{\alpha\beta}(x)=-\vec E\cdot \vec B(x)=0\,,\label{eq:82eq12}
\end{align}
both vanish identically, $\cS=\cP=0$ for a plane wave.  
The vanishing of the pseudoscalar implies there will be no CP violation in the quantum theory.

To describe the corresponding vector potential, we choose to work in the Lorenz and tranverse gauge conditions,
\begin{align}
\partial\cdot A_{cl}&=0, \label{eq:88eq9'}\\
\vec \nabla\cdot\vec A_{cl}&=0. \label{eq:88eq9''}
\end{align}
Decomposing into plane waves with only $\bar n\cdot x$ dependence, we have $A^{\mu}_{cl}(\bar n\cdot x)=\sum_\omega e^{-i\omega \bar n\cdot x}A_{cl,\omega}$, and therefore Eq.\eqref{eq:88eq9'} implies 
\begin{equation}
\bar n\cdot A_{cl}=0. \label{eq:88eq9'''}
\end{equation}
According to Eq.\eqref{eq:88eq9''} we must also have
\begin{equation}
\partial_z A_{cl}^z=i\omega A_{cl}^z=0,\label{eq:88eq9''''}
\end{equation}
for all $\omega$.  Combining \eqs{88eq9'''}{88eq9''''}, we find $-i\omega A_{cl}^0=0$ for all $\omega$, so that $A^{\mu}_{cl}=\vec A_\perp(\bar n\cdot x)$, which means that the laser potential can be written as the field only depending on a perpendicular polarization vector and the single light-cone coordinate $\bar n\cdot x$.

For a laser pulse, we often parametrize the vector potential as
\begin{equation}\label{eq:88eq5}
A_{cl}^\mu(\phi)=\vec A_\perp(\phi)=\vec\epsilon\,|A(n\cdot x)|\,\cos(\omega_{cl}\bar n\cdot x),
\end{equation}
where $\vec \epsilon$ is the transverse polarization vector and $\omega_{cl}$ is the primary frequency, which is typically in the infrared to optical range, $\omega_{cl}\sim 1$\,eV.  $|A(n\cdot x)|$ is a slowly-varying envelope function with the property
\begin{equation}\label{eq:88eq7}
\lim_{|\bar n\cdot x|\to\infty}|A(\bar n\cdot x)|\to 0.
\end{equation}

It is common to discuss the laser field intensity using the parameter
\begin{align}\label{eq:82eq4}
a_0&=\frac{|e|\sqrt{-A_{cl}^2}}{m}=\frac{|e\vec{E}_{cl}|}{m_e\omega_{cl}} \\
&=6.0\sqrt{I_{0}[10^{20}\text{W}/\text{cm}^2]}\lambda[\mu\text{m}]=7.5\sqrt{I_{0}[10^{20}\text{W}/\text{cm}^2]}/\omega_{cl}[eV],
\end{align}
with $I_{0}=|\vec E_{cl}|^2$ the peak intensity.  Note that $a_0$ is gauge and Lorentz invariant (\cite{Heinzl:2008rh}).  Many current laser facilities achieve intensities greater than $10^{22}$ W/cm$^2$, which corresponds to $a_0\simeq 60$.  The Texas Petawatt now achieves $>10^{23}$ W/cm$^2$ at its peak, corresponding to $a_0\simeq 190$.  To describe these experiments, we generally consider the range
\begin{equation}\label{eq:88eq11}
a_0\simeq 10-500 \gg 1.
\end{equation}

\subsection{Electron Classical Dynamics}\label{sec:III.2.2}
Non-relativistic motion of a charged particle in a laser field is, at leading order in the particle velocity, oscillation along the polarization axis of the laser.  Relativistic dynamics additionally involve drift in the laser propagation direction, non-dipole effects such as the famous figure-eight trajectory seen in the drifting frame, and the trajectory sharpening where the velocity component parallel to the laser polarization reverses.  Consequently, relativistic, laser-driven electrons also emit radiation at harmonics of the laser frequency.

\subsubsection{Basic Description without Radiation}
  Using this symmetry, the electron dynamics are easily solved from the action for an electron in a planewave, which we shall need later in any case.  We use $x$ as the electron coordinate, and write the classical action of the electron motion in the laser field as,
\begin{equation}\label{eq:88eq16}
S_{cl}=\int d^4x \left(\frac{m_e}{\gamma}\delta^{(3)}(\vec x-\vec x(\tau))-A_{cl}\cdot j \right),
\end{equation}
where $\gamma=\frac{1}{\sqrt{1-\dot x^2}}$, in natural units $c=1$, $\tau$ is the proper time of the electron and $j^\mu$ is the electron current, which in the classical limit has the form,
\begin{equation}\label{eq:88eq17}
j^\mu=e\dot x^\mu \delta^{(3)}(\vec x-\vec x(\tau)),
\end{equation}
with the electron velocity $\dot x^\mu$ measured in the lab frame.  As in the previous section, we take the four-potential in the Lorenz gauge, in lightcone coordinates $A_{cl}^\mu(\phi)=(0,0,\vec A_\perp(\phi))$.

The classical Lagrangian describing the electron motion in the strong laser field is
\begin{equation}\label{eq:88eq18}
L=m\sqrt{1-\dot x^2}-eA_\perp \dot x_\perp,
\end{equation}
where $\vec A_\perp=(E_{cl}/\omega_{cl})\vec \epsilon_\perp\cos(\phi)$ and
\begin{equation}\label{eq:88eq19}
\dot x^2\equiv (n\cdot \dot x)(\bar n\cdot \dot x)-\dot x_\perp^2.
\end{equation}
From Eq.\eqref{eq:88eq18}, we derive the electron momentum in the light-cone coordinates:
\begin{align}
\frac{\delta L}{\delta \bar n\cdot \dot x}&\equiv n\cdot p=mn\cdot \dot x,\nn\\
\frac{\delta L}{\delta \dot x_\perp}&\equiv p_\perp=\gamma m\cdot x_\perp-e A_\perp, \nn\\
\frac{\delta L}{\delta n\cdot \dot x}&\equiv \bar n\cdot p=m\bar n\cdot \dot x. \label{eq:88eq20}
\end{align}
The Euler-Lagrange equations give the Lorentz force, and for each light-cone direction the resulting equation of motion is
\begin{align}
\frac{d}{d\tau}\frac{\delta L}{\delta \bar n\cdot \dot x}-\frac{\delta L}{\delta \bar n\cdot x}&=\frac{d}{d\tau}n\cdot p-\dot x_\perp\frac{e\partial A_\perp}{\partial \bar n\cdot x}=0,\nn\\
\frac{d}{d\tau}\frac{\delta L}{\delta n\cdot \dot x}-\frac{\delta L}{\delta n\cdot x}&=\frac{d}{d\tau}\bar n\cdot p=0, \nn\\
\frac{d}{d\tau}\frac{\delta L}{\delta \dot x_\perp}-\frac{\delta L}{\delta x_\perp}&=\frac{d}{d\tau}(m\cdot x_\perp-eA_\perp)=0. \label{eq:88eq21}
\end{align}
With the boundary condition that $\tau\to -\infty$, the electron is at rest $P_e^\mu=(m_e,\vec 0)$, the solution to Eq.\eqref{eq:88eq21}, written in lightcone coordinates, is
\begin{equation}
P_e^\mu=\left(\frac{(eA_\perp)^2}{m_e},m_e,eA_\perp\right)\sim m_e(a_0^2,1,a_0),
\end{equation}
which shows that the electron is highly relativistic when $eA_\perp\gg m_e$ since the momentum collinear to the direction the laser wave propagates scales as $a_0^2\gg 1$. Kinematically, the majority of real quantum radiation is smaller than this typical electron momentum in the strong laser field.

In the classical context, $a_0$ can be interpreted as the work done by the laser field on an electron over one wavelength $\lambda_{cl}=2\pi/\omega_{cl}$ in units of the electron mass.  When $a_0$ is of order 1, an electron at rest is accelerated to relativistic speed within one laser wavelength, and the dynamics become nonlinear with respect to the laser field amplitude. For this reason, $a_0$ is also called the classical nonlinearity parameter.  Another way of seeing this is to use \eq{88eq21} to show that an electron gains a drift momentum $\propto a_0^2$ along the propagation direction of the laser field compared to a transverse momentum $\propto a_0$.  In the drifting frame, the classical trajectory forms a figure-eight with transverse extension of order $\lambda_{cl} a_0$ and longitudinal extension of $\lambda_{cl}a_0^2$ indicating that the path departs from one-dimensional oscillation and becomes when $a_0\gtrsim 1$.

The classical action is obtained by substituting the momenta back into \eq{88eq16},
\begin{align}
S_{cl} &= p\cdot x + S_{p}(\bar n\cdot x)\nn \\
S_{p}(\bar n \cdot x)&=\int_{-\infty}^{\bar n\cdot x} \paren{\frac{2p\cdot eA_{cl}(\bar n\cdot y)-e^2A_{cl}^2(\bar n\cdot y)}{2\bar n\cdot p}}d(\bar n\cdot y),\label{eq:82eq7}
\end{align}
where $p\cdot x$ is the free-particle action.  In Appendix \ref{appx:classicalaction}, we obtain this by directly solving the Hamilton-Jacobi equation   (\cite{landau1975lifshitz}), and will play an important role in the semiclassical description below.

More detail than a plane wave is needed to describe realistic laboratory produced laser pulses which have a cross section that varies along the propagation axis. The plane wave is a good approximation when the minimal focusing area (called spot size) is large compared to central wavelength of the pulse. A more precise description is provided by a Gaussian beam in the paraxial approximation with a Gaussian transverse profile.  For experimental purposes, the classical dynamics of electrons are usually calculated by numerically solving the Lorentz equation or using a particle-in-cell simulation code, which solves both the Lorentz force and the Maxwell's equations together on a grid.

\subsubsection{Radiation Reaction Derivation}
\label{sec:III.2.3}
Radiation reaction is the problem in classical dynamics of writing down the equation of motion for a charged particle (e.g. an electron) that accelerates and radiates in an electromagnetic field $F^{\mu\nu}(x)$.  The current best perturbative solution can be viewed as a resummation from the point of quantum theory as we will explain below.

The Lorentz force equation
\be
m\frac{du^\mu}{d\tau}=eF^{\mu\nu}u_\nu
\ee
does not include effect of radiation and loss of energy/momentum by an accelerated electron. Using the Larmor formula for the 4-momentum emitted per unit time by an electron with instantaneous acceleration $du^\mu/d\tau$,
\be\label{eq:larmorrate}
\frac{dP_{rad}^\mu}{d\tau}=\frac{2}{3}e^2 \frac{du^\mu}{d\tau}\frac{du_\mu}{d\tau} u^\mu.
\ee
\cite{lorentz1909theory} (in the nonrelativistic limit) and \cite{abraham1905theorie} reasoned there should be a `damping force' due to the radiation,
\be\label{eq:82eq20}
F_{rad}^\mu=\frac{2}{3}e^2\paren{\frac{d^2u^\mu}{d\tau^2}+\frac{du^\nu}{d\tau}\frac{du_\nu}{d\tau}u^\mu}\,.
\ee
\cite{} provided a mathematical basis to obtain the radiation force $F^\mu_{rad}$ by considering a formal ``self-field'' for the electron, which allowed him to absorb the divergence for a point-like electron into an (infinite) renormalization of the electron mass, see also .  At the end, Dirac justified what is now known as the Lorentz-Abraham-Dirac (LAD) equation:
\be\label{eq:82eq22}
m\frac{du^\mu}{d\tau}=eF^{\mu\nu}u_\nu+\frac{2}{3}e^2\paren{\frac{d^2u^\mu}{d\tau^2}+\frac{du^\nu}{d\tau}\frac{du_\nu}{d\tau}u^\mu}\,.
\ee
The second derivative of the 4-velocity is problematic, and the LAD equation contains unphysical solutions such as the `runaway' solution with exponentially growing electron acceleration (see \cite{hartemann2001high, rohrlich2007classical}).

In the limit of small acceleration, where the radiation correction is small, one can reduce the derivative order of the LAD equation by use of the leading-order equation of motion, i.e. the Lorentz force equation.  Precisely, if the typical wavelength $\lambda$ satisfies
\be\label{eq:82eq23a}
\lambda\gg 1/m_e,
\ee
and the typical strength $F$ of the classical electromagnetic field satisfies
\be\label{eq:82eq23b}
eF \ll  m_e^2,  
\ee
then one may use
\be
m\frac{d^2u^\mu}{d\tau^2}\simeq \frac{d}{d\tau}\left(eF^{\mu\nu}u_\nu\right)
\ee
to substitute into the LAD equation and obtain the Landau-Lifshitz (LL) equation (\cite{landau1975lifshitz}):
\begin{align}
m\frac{du^\mu}{d\tau}&=eF^{\mu\nu}u_\nu+\frac{2}{3}e^2\bigl[\frac{e}{m}(\pd_\alpha F^{\mu\nu})u^\alpha u_\nu\nn\\
&-\frac{e^2}{m^2}F^{\mu\nu}F_{\alpha\nu}u^\alpha+\frac{e^2}{m^2}(F^{\alpha\nu}u_\nu)(F_{\alpha\lambda}u^\lambda)u^\mu\bigl]\,.\label{eq:82eq24}
\end{align}
Now, with vanishing external field, the acceleration of the electron also vanishes, which is a crucial advantage of LL equation over LAD equation.  The conditions \eqs{82eq23a}{82eq23b} must always be satisfied in the electron's instantaneous rest-frame in this classical context so that quantum effects are negligible.  

\begin{figure}[t]
\centering
\includegraphics[width=0.85\textwidth]{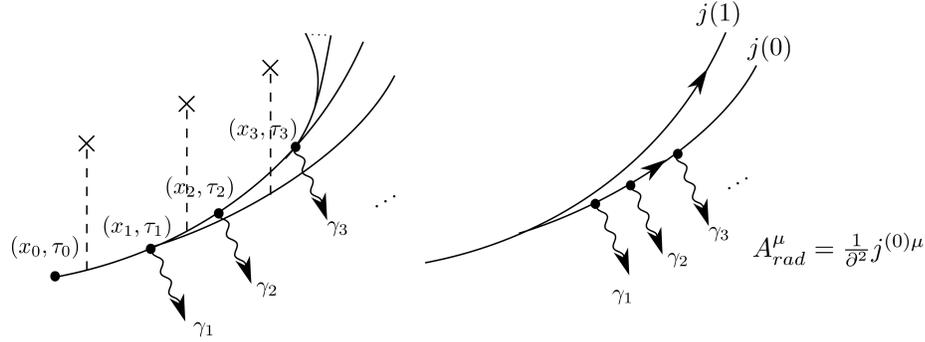}
\caption{Summing perturbative corrections by radiation to the electron trajectory.}
\label{fig:LL}
\end{figure}

In fact, use of the leading-order equation of motion is self-consistent: a perturbative expansion of the particle trajectory is implicit in the Larmor radiation rate \eq{larmorrate} which is derived from Maxwell's equations taking the charged particle trajectory to be known.  From the point of view of quantized radiation, this procedure corresponds to summing emissions along the leading order trajectory and using the total radiated momentum to determine the next-to-leading order trajectory, as depicted in \fig{LL}.

\subsection{Quantum Dynamics}
\label{sec:III.2.2.2}

The current theory for quantum processes by electrons in strong laser fields is based on the study of spontaneous particle creation.  I review the methods in their original context of spontaneous particle creation and critically examine their applicability to laser experiments.  In particular, I highlight their limitations in describing dynamics of real electrons and photons, rather than dynamics of long-wavelength classical fields with no quantized particles present.

\subsubsection{Spontaneous pair creation}

\begin{figure}[t]
\centering
\includegraphics[width=0.75\textwidth]{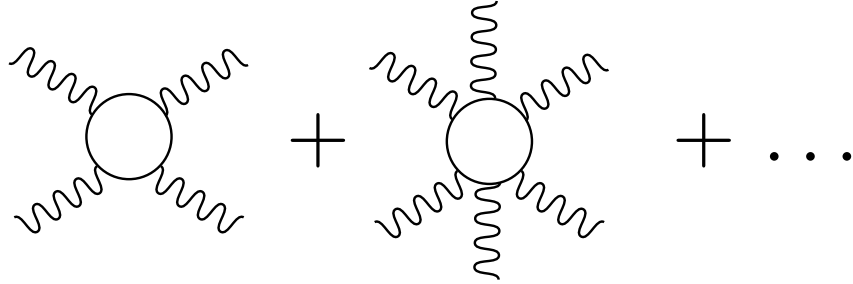}
\caption{Diagrams in the expansion of Heisenberg-Euler effective potential.}
\label{fig:EH}
\end{figure}

Just using dimensional analysis, we can see that the electric field scale $|e\vec E|\sim m_e^2$ should arise in the study of strong electromagnetic fields in QED.  After the calculation by \cite{Euler:1935zz} of effective light-by-light scattering due to an electron loop (first diagram in the series in \fig{EH}), \cite{Heisenberg1936} obtained the first top-down effective field theory studying low-energy $\omega\ll m_e$ light-light interactions.  The Heisenberg-Euler effective potential can be expanded in a power series
\begin{align}
V_{\rm eff}&\simeq \frac{\alpha}{90\pi}\frac{e^2}{m_e^4}\bigg((\vec B^2-\vec E^2)^2+7(\vec E\cdot\vec B)^2\bigg) \nn\\
&+\frac{\alpha}{315\pi^2}\frac{e^4}{m_e^8}\bigg(2(\vec E^2-\vec B^2)^3+13(\vec E\cdot\vec B)^2(\vec E^2-\vec B^2)\bigg)+...,\label{EHVeff}
\end{align}
with each order suppressed by inverse powers of the field scale $m_e^2/e$.  The combinations of electric and magnetic fields that appear are the scalar and pseudoscalar Lorentz-invariants constructed from the field strength tensor,
Since $\cP$ is a pseudoscalar and parity is conserved in QED, the effective potential actually only depends on $\cP^2$.  This power series is represented by diagrams in \fig{EH}, also showing the intermediate electron loop that has been integrated out.

Clearly the power series \req{EHVeff} fails to converge as $|e\vec E|\to m_e^2$.  This breakdown signals the onset of nonperturbative physics (\cite{Dunne:1999uy,Labun2010}).
To see what dynamics it suggests, we consider that in a constant and homogeneous electric field the electrostatic potential difference is $2m_e$ over a length $\Delta z=2m_e/|e\vec E|$.  Using the picture of quantum mechanical barrier scattering, the mass gap $2m_e$ between positive and negative frequency states implies a classically forbidden region with width $\Delta z$ in which the wavefunction of the electron decays exponentially.  The decay length is proportional to the energy of the state and so is  maximized for an electron at rest $\sim 1/m_e$.  Tunneling through the barrier is therefore exponentially suppressed $\sim e^{-m\Delta z}=e^{-\frac{2m_e^2}{|e\vec E|}}$.  
The exponent exhibits what has become known as the critical field
\beq\label{Ecdefn}
E_c=\frac{m_e^2c^3}{e\hbar}=1.32\times 10^{18}~\mathrm{V/m}
\eeq
which sets the scale for pair production to become an order 1 effect and also estimates the breakdown scale as we will see shortly.

\cite{Heisenberg1936} obtained their effective potential by solving the leading order equation of motion, the Dirac equation, in the presence of a classical potential
\beq\label{DiraceqnwithAcl}
(i\slashed{\partial}-e\slashed{A}_{cl}(x)-m)\psi(x)=0.
\eeq
Their procedure was shown to be equivalent to integrating out electrons in the path integral by \cite{schwinger1951gauge}.  This also demonstrates its relation to other examples of integrating out degrees of freedom to obtain an effective theory, such as soft-collinear effective theory and heavy quark effective theory.  
Incorporating the classical potential $A^\mu_{cl}(x)$ into the solution of the Dirac equation assumes that the potential varies slowly relative to the dynamics of the quantized electron. This is the semiclassical condition
\beq\label{semiclassiccond}
\frac{\partial V(x)}{V(x)}\ll \frac{\partial\psi(x)}{\psi(x)} 
\qquad \Leftrightarrow \qquad
k_{cl}^\mu \ll p_e^\mu,
\eeq
which, expanding each field in Fourier modes, is equivalent to the wavelength of the ``classical'' modes being much larger than the wavelength of the quantized modes.  

Equation \eqref{semiclassiccond} is a necessary condition to apply to the semiclassical approximation.  The semiclassical approximation is applied in cases that the potential is leading order.   In contrast, recall that standard perturbation theory assumes the particle is free at leading order and interactions are suppressed by a small coupling.  
The solutions of \req{DiraceqnwithAcl} often display non-polynomial dependence on $eA_{cl}$, though they can be expanded in powers of $eA_{cl}$.  For a plane-wave potential (discussed in the next section), the first terms in the expansions are explicitly verified to agree with a perturbative calculation of the electron scattering from the potential (\cite{Lavelle2013}).  

The approach by \cite{schwinger1951gauge} also allowed the calculation of an imaginary part of the effective potential,
\beq\label{ImVeff}
\mathrm{Im}\,V_{\rm eff} = \frac{\alpha}{2\pi^2}|\vec E|^2\sum_{N=1}^\infty \frac{1}{N^2}e^{-N\pi E_c/|\vec E|}.
\eeq
$V_{\rm eff}$ is complex due to the particle creation dynamics; to create the electric field we must do work to separate charges to $z\to \pm\infty$ and the field decays by converting its energy into electron-positron pairs.  The imaginary part limits the radius of convergence of the power series \req{EHVeff} and ensures that the series is only asymptotic.  

As seen in the power series expansion, the physics of this field scale leads to a breakdown of the low-energy theory.  From the EFT point of view, this breakdown is because low-energy fields produce high-energy radiation of order the cutoff scale $m_e$, i.e. electron-positron pairs.  To show that the theory breaks down at $|\vec E|\sim E_c$ rather than $\alpha^{-1}E_c$ (due to the small coupling), we note the timescale for converting electric field energy to particle rest mass becomes $\sim (\alpha m_e)^{-1}$ at $|\vec E|\sim 5E_c$ (\cite{Labun2009}), meaning that the classical potential is changing as fast the electron can react.  This invalidates the semiclassical approximation solution using to integrate out electron dynamics.

\subsubsection{Semiclassical description of electrons in laser fields}

The semiclassical description of electrons in laser fields is obtained by a similar procedure.  One solves the Dirac equation Eq.\,\eqref{DiraceqnwithAcl} with $A^\mu(x)$ the four potential of a light-like field.   The solution follows from the fact that the classical action \eq{88eq16} for an electron in a plane-wave field can be solved exactly. Let $p^\mu$ and $\sigma_0/2=\pm 1/2$ be electron four-momentum and spin at $\bar n\cdot x\to-\infty$ and with $A^\mu(-\infty)=0$ the positive energy solution is given by (\cite{berestetskii1982quantum, volkov1935class})
\begin{equation}\label{eq:82eq6}
\Psi_{p,\sigma}(x)=\left[1+\frac{\slashed{\bar n} e\slashed{A}_{cl}}{2\bar n\cdot p}\right]e^{-ipx-iS_{p}}\frac{u_{p,\sigma}}{\sqrt{2Vp_0}},
\end{equation}
where $A^\mu_{cl}$ is the classical plane wave potential and $u_{p,\sigma}$ is a positive-energy free spinor ($(\slashed{p}-m)u_p=0$).  The quantization volume $V$, and the zero-component of the four-momentum $p_0$ are standard factors for the normalization.  $S_p(\phi)=S_p(\bar n\cdot x)$ is the plane wave-field-dependent piece of the classical action, given in \eq{82eq7}.  The Volkov state for a positron is given by $p^\mu\to -p^\mu$ and $\sigma\to -\sigma$ in \eq{82eq6} replacing all but the energy in the square root, where the Dirac spinor is replaced with corresponding antiparticle spinor $u_p\to v_p$.   The solutions are orthogonal and complete on $t$-level sets (\cite{Ritus1985, Zakowicz2005, boca2010completeness}).

The completeness of the wavefunctions \eq{82eq6} allows one to construct an S-matrix, with the electron or positron states defined asymptotically $x\to \pm\infty$ where the laser field and perturbative interactions are turned off.  This is known as the Furry picture (\cite{Furry1951}) with $\Psi$ quantized with respect to the classical background plane-wave field, and procedurely involves replacing free states and free propagators in QED amplitude calculations with Volkov states and Volkov propagators (see \cite{Ritus1985}).  

In QED, the parameter $a_0$ can be viewed as the work done by the laser field over the typical QED length scale $\oneov{m}\approx 3.9\times 10^{-11}\text{cm}$ in units of laser photon energy $\omega_{cl}$. This implies that for $a_0\gtrsim 1$ multiphoton interactions become significant and the laser field should be incorporated to all orders in $eA_{cl}$ (\cite{Ritus1985}).  

The probability density $dP/dV dt$ of a quantum process can involve only gauge- and Lorentz-invariant quantities and is thus a function of only the two parameters $a_0$ and 
\beq\label{chidefn}
\chi^2 = \frac{1}{m^4}p_\mu eF^{\mu\nu}eF_{\nu\lambda}p^{\lambda} = p\cdot P_{laser}
\eeq
the Lorentz invariant presenting the center of mass energy in the collision between the electron and the classical field (the momentum density of the classical laser field is multiplied by the Compton volume of the electron $1/m_e^3$ to form a momentum).  The limit $\chi/m_e\to 1$ implies the electric field seen by the electron in its rest frame is equal to the critical field \req{Ecdefn}. 
Generalizing from the plane wave case, the field invariants \eqs{82eq11}{82eq12} can also appear.  Both $\cS,\cP$ are zero for plane waves.  When
\be\label{eq:82eq12a}
|\cS(x)|, |\cP(x)|\ll \min(1,\chi^2(x))F_{cr}^2
\ee
the probability density of a process $dP/dVdt$ is well-approximated by the plane-wave background result.

Defining and calculating a perturbative S-matrix in this way makes at least two key assumptions:
\begin{description}
\item[I.] The classical potential $A_{\rm cl}^\mu(x)$ is known for all $x^\mu$.
\item[II.] Only a few discrete events occur between $t\to-\infty$ and $t\to +\infty$,  such as one or two photon emission or a single pair emission event.  The remainder of the dynamics is elastic.
\end{description}
The first limits semiclassical calculation to low density plasmas, where there is insufficient charge density to significantly modify the classical fields from the (approximately) known input laser field. The second is both technical and subtle.  On the one hand, it can be addressed by brute force calculation of higher and higher order diagrams.  On the other hand, electrons radiate classically in the field throughout their interaction with the laser field, including during any high-energy photon emission.  It is currently an open question how much this low-energy radiation modifies the probability of any quantized process.  My work goes some way to answering this question, as discussed below.

Despite these limitations, one can learn from calculations in semiclassical theory, and I summarize a few relevant results here.  Emission processes have been studied at tree-level, without loop corrections.  The two-loop (order $\alpha^2$) diagrams for self-energy and photon polarization tensor have been studied (\cite{Ritus1985}); however, the vertex correction for photon emission has only been calculated at one-loop order, and its divergent piece was not extracted.

At leading order, the semiclassical approach reproduces properties of QED perturbation theory.  Low-energy radiative corrections factorize and exponentiate, ensuring that rates for hard processes are well-defined in terms of the usual (perturbatively-defined) asymptotic particle states (\cite{Dinu2012,Ilderton:2012qe}).  The low-energy, low-intensity limit of the photon emission probabilities agree with the classical limit (\cite{Ritus1985}).  Additionally, the Ward-Takahashi identity, ensuring gauge invariance of amplitudes involving electron loops constructed from Volkov states, holds (\cite{Meuren2013}), and the optical theorem was verified for the one-loop polarization function for photons (\cite{Meuren2015}).  These results, including the analysis of the propagator (\cite{Lavelle2013}), arise from the fact that the Volkov solution \eq{82eq6} is a Wilson line, as we will show below.

Many articles have predicted phenomenological signatures of high-energy photon emission and pair production.  In a monochromatic plane-wave field, momentum conservation requires that an electron can only radiate photons in integer multiples of the plane-wave wave vector, $k^\mu_{\rm out}=Nk^\mu_{\rm laser}$.  $N=1$ corresponds to perturbative Compton scattering, and $N>1$ is often called non-linear Compton scattering because it requires absorbing $N$ quanta from the classical field and becomes more probable as the laser $a_0\gg 1$.  Many studies focus on the kinematical consequences of short (few cycle) laser pulses, which broadens the momentum distribution of photons for the electron absorb (\cite{Heinzl:2009nd,DiPiazza2010a,Seipt:2010ya,Mackenroth:2010jr,Seipt:2012tn,Harvey:2012ie,Mackenroth:2012rb}).
Pair conversion by photons and pair emission by electrons propagating in the field has been studied under similar conditions
(\cite{Meuren2015,Heinzl2010a,Ilderton2011,Titov:2012rd,King:2013osa,Nousch:2012xe,Meuren:2014kla,Jansen:2015idl}).  Even the effective neutrino-photon coupling has received attention (\cite{Shaisultanov:1997bc,Shaisultanov:2000mg,Gies:2000tc,Gies:2000wc,Meuren:2015iha}).

One phenomenologically important fact is that for  $a_0\gg 1$ pair creation in a plane-wave field is well-approximated by the convolving the local constant crossed-field probability of creation with the classical dynamics of the electrons (\cite{Meuren:2015mra}).  This property is particularly important for constructing a systematic quantum theory; if it were not true, we would have to worry about the length scale over which non-local correlations in the field could impact short-wavelength (quantum) dynamics.

\subsection{Beyond Current Theories}
\label{sec:III.3}

In the semiclassical study of spontaneous pair creation, there are two momentum scales: the frequency of the classical field and the mass of the electron.  The semiclassical condition \req{semiclassiccond} is the statement that these momentum scales are widely separated.  Physically, the electron is a heavy particle, and its point-like fluctuations can be integrated out.

In contrast, processes involving real photons or electrons in the classical field involve at least one additional  momentum scale, the momentum of the particle.  For photons with momentum $k\ll m_e$, electrons are again heavy and can be integrated out.  The Heisenberg-Euler effective potential suffices for this case.  Photons with momentum $k\sim m_e$ and low-momentum electrons however resolve fluctuations around the scale $m_e$ and a full quantum theory with dynamical electron and photon degrees of freedom is necessary. 

The case of greatest interest phenomenologically is photons and electrons with momentum $k,p\gg m_e$, as arise in the final state of a typical acceleration experiment.  In this case, there is a large hierachy of momentum scales
\beq\label{laserhierarchy}
E+p_z \gg |eA_{cl}^\mu| \gg m_e \gg \omega_{cl}.
\eeq
Current high intensity laser systems on which the next experiments will be conducted have
\beq
a_0=\frac{|eA_{cl}^\mu|}{m_e} \sim 10-350
\eeq
at the peak intensity of the laser pulse.

\begin{figure}
\centering
\includegraphics[width=0.3\textwidth]{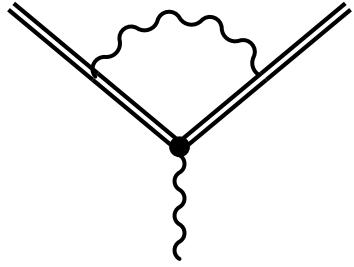}
\caption{One loop correction to the electron-photon vertex.  The double line indicates the semiclassical propagator incorporating the classical potential to all orders.  Details of the calculation will be presented elsewhere.\label{fig:1loopvertex}}
\end{figure}

This hierachical separation of momentum scales is important to account for, because perturbation theory introduces large logarithms of ratios of these scales.  Although the coupling constant is small for QED, a widely-separated hierachy, such as $E+p_z\sim 10^3m_e\sim 10^6\omega_{cl}$ can compensate.  As an example, we consider the radiative corrections to the rate of photon emission.  The largest difference between the leading-order (LO) and next-to-leading order (NLO) rates is a double logarithm involving the scales seen in \req{laserhierarchy}, and we can estimate its size using the vacuum result,
\beq\label{sudakov}
\delta \Gamma(e\to e\gamma) = |\Gamma_{NLO}-\Gamma_{LO}|\underset{-q^2\gg m_e^2}{\simeq} \frac{\alpha}{2\pi}\ln\frac{-q^2}{m^2}\ln\frac{-q^2}{E_d^2},
\eeq
where $-q^2=-(p-p')^2$ is the squared 4-momentum change by the electron.  The coefficient is checked by evaluating the one-loop correction to the electron-photon vertex shown in \fig{1loopvertex}, with the laser field included to all orders by using the semiclassical propagator for the electron.  The momentum scale $E_d$ is the energy resolution of a detector and is an infrared scale distinguishing radiation from the electron (\cite{peskin1995introduction}); in other words it measures the precision to which the electron momentum can be known.  
Here $E_d$ is set by the classical radiation, which having $\omega_{cl.rad.}\ll m_e$ is very low momentum relative to the photon.  We therefore estimate this scale from the cyclotron frequency of the electron in the laser field $\omega_{cl.rad.}\sim \omega_{cyc}=|e\vec B|/p$.  This can be written Lorentz invariantly by going to the instantaneous rest frame of the electron, where the field strength is $|\vec B'|$ and
\beq
E_d\sim\omega_{cyc}=\frac{|e\vec B'|}{m_e} = \chi,
\eeq
with $\chi$ defined above in \req{chidefn}.

For an electron with $\gtrsim 50$ MeV energy co-propagating with the laser, the field in its rest frame is reduced by the Lorentz factor, and $\chi\sim 10^{-6}$.  Then \req{sudakov} implies that the emission probability for a photon with similar energy $k^\mu\sim 50$ MeV is corrected by nearly 50\%.  For a moderate energy photon $k^\mu\sim m_e$, the first logarithm in \req{sudakov} is replaced by a number of order 1, and the correction is 2-5\%, depending on the electron momentum.  Note that this is equal or larger than corrections implied by classical models of radiation reaction-corrected dynamics (\cite{Hadad2010}).  

\begin{figure}[t]
\centering
\includegraphics[width=0.3\textwidth]{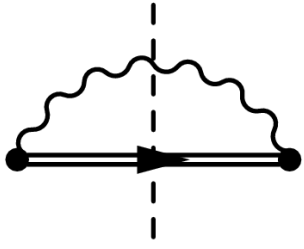}
\caption{Diagram for single-photon emission in a classical laser potential.  The double line indicates the semiclassical propagator incorporating the classical potential to all orders.  The vertical dashed line is the cut, corresponding to taking the imaginary part.  Without the cut, the real part gives the self-energy correction for the electron.  \label{fig:selfenergy}}
\end{figure}

The reason for the magnitude of these corrections is easy to understand from the total radiation rate for an electron propagating in a strong field.  The total rate is the imaginary part of the $\mathcal{O}(\alpha)$ electon self-energy using a semiclassical propagator, the diagram in Figure \ref{fig:selfenergy}.  We work in the quasi-constant field approximation, meaning that our result includes radiation from photon modes $k\gg \omega_{cl}$.  Details of the calculation are given in Appendix \ref{app:fulltheoryselfenergy}.  The result is
\begin{align}
\Gamma_{LO}(e\to e\gamma)\,&=-\frac{1}{\pi}\mathtt{Im}\,\mathrm{tr}\,\Sigma(p-eA_{\rm cl}) \nn\\
&=m_e\left(\frac{m_e}{\chi}\right)^{\!2/3}\frac{4\alpha}{\pi}\int_0^1dx \frac{x(1-\frac{1}{3}x)}{(1-x)^{5/3}}\,\mathrm{Ai}\left(\frac{x m_e^{2/3}}{(1-x)^{2/3}\chi^{2/3}}\right),
\label{ImSigma}
\end{align}
where Ai$(z)$ is the Airy function.
In our calculation, gauge invariance is preserved throughout, and the result depends only on the covariant combination $p-eA$.

\begin{figure}[h]
\centering
\includegraphics[width=0.75\textwidth]{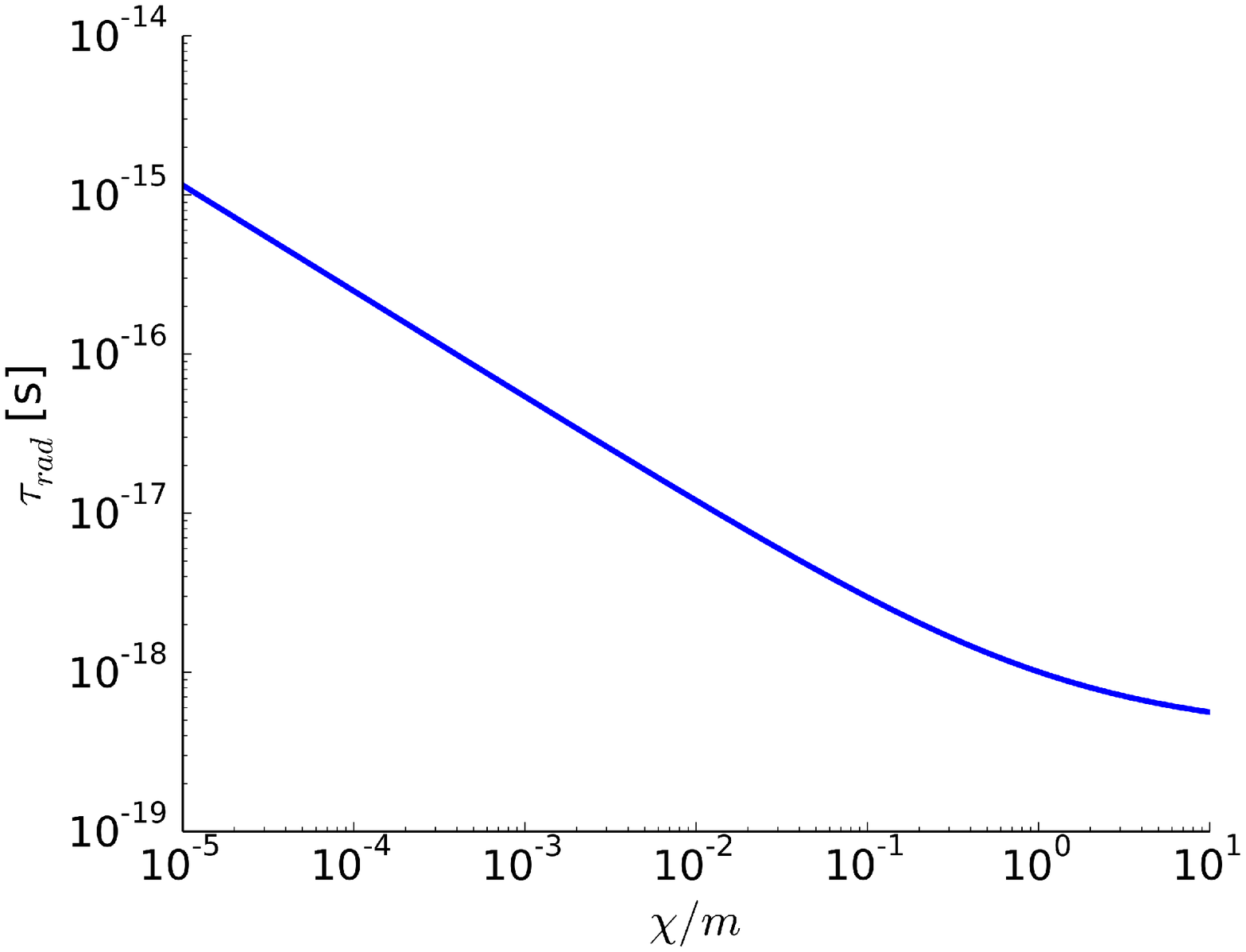}
\caption{Characteristic timescale for a momentum state to decay by radiation of a single photon \req{tauraddefn}.  $\tau_{rad}$ depends on the initial momentum of the electron through the Lorentz invariant $\chi$ \req{chidefn}.  (From \cite{Hegelich:2016}) \label{fig:taurad}}
\end{figure}

As the imaginary part of the self-energy, the radiation rate \req{ImSigma} presents the inverse of the lifetime of a state with momentum $p-eA_{cl}$.  In other words, the number of electrons initially with momentum $p$ decays with time as (\cite{Hegelich:2016})
\beq\label{tauraddefn}
N_e(p,t)\sim e^{-t/\tau_{rad}},\qquad 
\tau_{rad} = \frac{1}{\Gamma\big(e(p)\to e(p')\gamma\big)},
\eeq
with $\tau_{rad}$ plotted as a function of $\chi$ in Fig. \ref{fig:taurad}.
Seeing that the lifetime is smaller than one period of an optical laser pulse, it is consistent to use the quasi-constant field approximation.  We conclude that quantized radiation significantly affects the trajectory of the particle during its interaction.  This is a quantitative demonstration of radiation reaction.  For this reason, a framework is needed to incorporate radiative corrections to the tree-level semiclassical calculations and implement them in simulations.

\section{Processes and Observables for Electrons in Strong Laser Fields}
\label{sec:III.3.2}

We wish describe the interaction between strong laser fields and electrons with effective field theory methods. Before discussing our effective theory, we must take a close look at the experimental conditions and possible observables.  This will provide the basis for an appropriate power counting and beginning construction of relevant observables.

\subsection{Global picture of dynamics in experiment}

\fig{88fig1} shows a typical laser experiment. Point A leads back to the source of the laser pulse. After being focused and amplified, the laser pulse hits the gas or solid target, and it starts to ionize the target before reaching its peak. When the peak arrives, the large ponderomotive potential drives electrons through the ion matrix and out the rear of the target.  Recall that $a_0$ gives the work done by the laser field on an electron in one wavelength in units of electron mass.  Conversely, the electron will typically have a 3-momentum $|\vec p|\sim m_e$ within $\lambda_{cl}/a_0$, where $\lambda_{cl}=\frac{2\pi}{\omega_{\rm cl}}\sim 1\,\mu$m is the wavelength of the laser field.  Thus the electron is relativistic in time and distance much smaller than the length scale of the variation of the classical field. 

By the time laser pulse reaches its peak, the target is fully ionized, since $eA_{cl}$ is much larger than the ionization energy $10$\,eV.  The classical electromagnetic fields and particle densities (at long-wavelength) evolve on the timescale of the plasma frequency of the target, which we estimate using a standard picture.  The spatial derivative of the restoring force inside of the plasma can be written as
\begin{equation}\label{eq:88eq14}
\frac{\partial F}{\partial x}\equiv k=e\vec\nabla\cdot\vec E,
\end{equation}
where $k$ is the restoring coefficient, and $\vec E$ is the electric field arising from the separation of the charges, and satisfies
\begin{equation}\label{eq:88eq15}
\vec \nabla\cdot\vec E=en_e,
\end{equation}
where $n_e$ is the free charge particle density, equal to the electron density. Plugging Eq.\eqref{eq:88eq15} into Eq.\eqref{eq:88eq14}, we have
\begin{equation}\label{eq:plasmarestoringcoefficient}
k=e^2 n_e=m_e\omega_p^2 \quad\Rightarrow\quad \omega_p^2=\frac{e^2n_e}{m_e},
\end{equation}
where $\omega_{pl}$ is the plasma frequency.  From the plasma frequency we define the critical density
\be \label{eq:86eq12}
n_{cr}=\frac{m_e\omega_{\rm{laser}}^2}{e^2},
\ee
which is the threshold density above which the laser wave can propagate freely in the plasma.  When the electrons in the plasma become relativistic, we can replace $m_e\to \gamma m_e$, and the plasma frequency squared and therefore also $n_{cr}$ is reduced by $\gamma$. 
This phenomenon is known as ``relativistic transparency'' and means a portion of the laser pulse can pass through the target only slightly perturbed to continue to accelerate the electrons before they reach to the detector.

The key points of this phenomenology are that: (1) electrons transition from nonrelativistic to relativistic momentum states throughout the laser-plasma interaction, and (2) the laser field can pass through the target, suggesting that the field seen by high-energy electrons approximates a plane wave to the precision required by \eq{82eq12a}.

We want to describe the high-energy radiation from this laser-plasma interaction systematically, which means we must take into account what quantities are actually measured.  Current experiments typically observe the particle spectrum only in a relatively narrow region in phase space: for example, high-energy particle spectrometers have a small angular acceptance (less than a few milliradians from beam direction) despite having large range in one momentum component along the beam direction.  As is well-developed for hadron collisions, the theory should be adapted to the observables.

One experiment so far has provided proof-of-principle that high-energy particle detectors are useful to diagnose laser-particle interactions.  The SLAC E-144 experiment successfully collided the linear accelerator's 40 GeV electron beam with a moderate intensity laser ($I\simeq 10^{19}$ W/cm$^2$).  They detected nonlinear Compton scattering $N\gamma+e\to \gamma+e$ and pair production, achieving agreement with semiclassical predictions in the semi-perturbative regime $N<10$, where the strong classical field is not yet dominant \cite{Bula:1996st,Burke1997,Bamber1999}. 

\section{Constructing Effective Field Theory for Electron-Laser System}
\label{sec:III.4}

In this section, I construct the effective field theory in the standard procedure.\footnote{This section is based on unpublished work with L. Labun.}  First I separate the physical scales in the process, and locate the necessary degrees of freedom and the corresponding symmetries. Then I construct the effective Lagrangian according to the symmetry and match it onto the underlying theory, namely QED with a classical potential. After this, I prove the non-perturbative part in final cross-section is universal and compute its evolution with increasing energy. Finally, I give the prediction of wide photon emission in this process, and some prospectives of this effective field theory in future.

 We focus on experimental conditions such that a (focused) laser pulse interacts with a low-density electron gas.  Here low density means vanishing chemical potential; both plasma dynamics and classical radiation may modify the laser field and are incorporated into the nonperturbative classical radiation function, as we will show. 

\subsection{Separation of Scales}
\label{sec:III.4.2}

First we estimate the electron final momentum achieved in the strong laser acceleration experiments. As we saw above, solving the classical motion in a plane-wave field shows that an electron starting from rest at $t\to-\infty$ has 4-momentum
\be\label{PeA}
P_e^\mu=\left(\frac{(eA_\perp)^2}{m_e},m_e,eA_\perp\right)\,.
\ee
We shall see below this momentum appears as a large phase in the wave functions of the electrons dressed by the potential.  It also corresponds to the phenomenological observation of highly relativistic electrons exiting approximately collinear to the laser pulse, that is $P^{\rm (final)}\sim P_e$ of \req{PeA}.  Initially the electrons are approximately at rest in the lab frame, with 4-momentum $\sim m_e$.  We will discuss the applicability of our theory to other kinematics after demonstrating on this example.

For the typical acceleration kinematics therefore, the experiments are summarized by a hierarchy of scales
\be
p_+^{\rm (final)}\gg |eA_{\rm cl}^\mu|\sim p_\perp^{\rm (final)} \gg m_e\sim p^{\rm (initial)} \gg a_0\omega_{\rm laser}\gg \omega_{\rm laser}.
\ee
If we choose our expansion parameter to be $\lambda\sim 1/a_0$, then the three highest momentum scales are each separated by a power of $\lambda$:
\be
m_e\sim \lambda|eA_{\rm cl}|\sim \lambda^2p_+^{\rm (final)}.
\ee
As described above, the laser frequency is typically about 1\,eV$\sim 10^{-6}m_e$, which for $\lambda\sim 1/a_0\sim 1/100$ implies $\omega_{\rm laser}\sim \lambda^3m_e$ in this power counting.  However, differences between the momentum scales below $m_e$ will not enter our discussion, and for concreteness we will consider all these scales to be of order $\lambda^4p_+$.  This power counting is conservative but sufficient to show how ultrasoft corrections do not influence the quantum dynamics we calculate corrections to.

\subsection{Electron-Laser Effective Lagrangian}
After separating the scales in the dynamics of the electron travelling and radiating in a strong plane-wave field, we can fix the corresponding degrees of freedom at each of the scale. We have light-like electrons, which are accelerated to be ultra-relativistic and called `collinear' electrons, as they travel collinear to the laser beam direction.  Collinear electron momentum scales as $P_c\sim(1,\lam^2,\lam)Q$, where $Q=n\cdot p=\frac{(eA_\perp)^2}{m_e}$ is the largest scale in our theory, and $\lam Q\simeq (eA_\perp)$ is the second largest scale. We remove both scales from this theory by integrating them out from the full theory Lagrangian.

We have `soft' electrons, referring to electrons with low momentum before being accelerated to become light-like particles. The momentum of soft electrons scales as $P_s\sim(\lam^2,\lam^2,\lam^2)Q$, and their Lagrangian can be obtained by integrating out the background dressing of the classical field, which is expected to be pure QED Lagrangian.

We separate photons into soft and collinear modes, corresponding to the respective electrons, and having the same momentum scalings as soft and collinear electrons. Both types of photons enter the electron Lagrangians via the interaction terms preserving the gauge symmetry. We derive the photon kinetic term in the same way we derive the kinetic term for gauge bosons in the SCET Lagrangian.

Lastly we verify another essential symmetry besides gauge symmetry which is required of this EFT -- the reparametrization invariance.

\subsubsection{Collinear Electron Lagrangian}
\label{sec:III.4.3.1}
In this part we derive the effective Lagrangian for collinear electrons.  The full theory Lagrangian is,
\be \label{eq:89eq0}
\cL=\bar\psi(i\Dslash-e\Aslash_{cl}-m)\psi,
\ee
where ${D}^\mu$ is the covariant derivative, ${A}_{cl}$ is the classical potential field, and $m$ is the electron mass.

We treat the classical field ${A}_{cl}^\mu$ as a non-dynamical field, in which we include both the driving laser and the coherent ultrasoft radiation with energy of the same order as the laser frequency, $\omega_{\rm cl}$, emitted by the accelerating electron. Based on this simplification, we solve the equation of motion to this Lagrangian by introducing a Wilson line
\be
W_P^{cl}\equiv e^{-iS_{P}+\frac{\slashed{n}}{2}\frac{e\slashed{A}_{cl}}{i\bn\cdot P}},
\ee
where $S_{P}(\bar n\cdot x)$ is field-dependent part of the classical action of an electron in a classical potential traveling from $-\infty$ to $x$,
\be
S_{cl}(\bar n\cdot x)=\int_{-\infty}^x d\bn\cdot x'\frac{2P\cdot eA_{cl}{(x')}-(eA_{cl})^2(x')}{2\bn\cdot P},
\ee
and $P$ is the electron momentum which we identify at $-\infty $ when the laser is shut off.  $W_P^{cl}$ satisfies the Wilson line (operator) equation of motion:
\be\label{WpclEoM}
(i\slashed{\partial}-e\slashed{A}_{cl})W_P^{cl} =0\, .
\ee
The solution to the equation of motion in the full theory is given by attaching this Wilson line to the ordinary planewave solutions of QED,
\be \label{eq:89eq1}
\psi(x)=\sum_P\,e^{-iS_{cl}+\frac{\bar\nslash}{2}\frac{e\Aslash_{cl}}{\bar n\cdot P}}e^{-iP x}u_P=\sum_P W_P^{cl}(x)e^{-iPx}u_P,
\ee
where $u_P$ is a Dirac spinor, satisfying
\be\label{planewavesoln}
(\slashed{P}-m)u_P=0,
\ee
and the plane-wave phase $e^{-iPx}$ is brought in by Fourier transformation of $u_P$ in position space.  Combining \req{WpclEoM} with \req{planewavesoln}, one sees we are using the Volkov solutions \eq{82eq6} to incorporate the leading order dynamics of the potential, and we will improve on the semiclassical approach by implementing our power counting to obtain a systematic expansion and develop perturbation theory.

For collinear electrons, the momentum along the laser beam direction $n\cdot P$ and the transverse momentum $P_\perp$ are at the same scale as the classical-field dressing,
\be \label{eq:89eq2}
P_\perp\sim eA_{cl}\sim \lambda,\quad n\cdot P\sim\frac{2P_\perp\cdot eA_{cl}-(eA_{cl})^2}{\bar n\cdot P}\sim \lambda^0,
\ee
which must be integrated out in the EFT.
Because $A_{cl}{(x)}$ is a slowly-varying background field with frequency $\omega_{cl}$, we expand $A_{cl}{(x)}$ as
\be
A_{cl}{(x)}=A_{cl}(0)+\left.x\cdot\frac{d}{dx}A_{cl}\right\vert_{x=0}+\ldots,
\ee
where $\left.\frac{dA_{cl}}{dx}\right\vert_{x=0}$ is at order of $\omega_{\rm laser}\cdot A_{cl}\sim \lam^4 A_{cl}$. Replacing $P$ with the covariant derivative $D$ in the classical action $S_{P}(x)$, we define the leading order part of $S_{cl}(x)$ as $\hat K$,
\begin{align}\label{eq:89eq5}
S_{P}^{L.O.}(x)=&\left(\oneov{i\bar n\cdot D}\left\{eA^\mu_{cl},iD^\mu\right\}-\frac{(eA_{cl})^2}{i\bar n\cdot D}\right)\frac{\bar n \cdot x}{2} \nn \\
=&\left(\left\{eA_{cl}^\mu, iD_\perp^\mu\right\}\oneov{i\bar n\cdot D}-\frac{(eA_{cl})^2}{i\bar n\cdot D}\right)\frac{\bar n \cdot x}{2}
\equiv \hat K\frac{\bar n\cdot x}{2}.
\end{align}
We can rewrite $S_{cl}(x)$ as
\be
S_{P}(x)=\frac{1}{2}\hat K \bar n\cdot x+\ln Y_p(x),
\ee
where $Y_p(x)$ is the Wilson line containing the ultra-soft contribution from the background and is defined as
\begin{align}
\ln Y_p(x)&=-i\int_{-\infty}^x\! \frac{d(\bn\cdot x)}{2}\frac{2(x'\cdot\pd)eA_{cl}^\mu D_\mu-(x'\cdot\pd)|A_{cl}|^2}{\bn\cdot D}  
+\cO(\pd^2 A_{cl}).
\end{align}
Decomposing the plane wave phase $e^{-iPx}$ into
\be
e^{-iPx}=e^{-iP_L x}e^{-iP_rx},
\ee
where $P_L$ scales as $(1,0,\lam)Q$ and $P_r$ is the residual momentum and scales as $(\lam^2,\lam^2,\lam^2)Q$. Plugging the classical solution $\psi$ and $\bar\psi$ into the full theory Lagrangian with the reduced $S_{cl}(x)$ we have
\begin{align}
\cL&=\sum_{P,P'}\bar u_{P'}e^{iP'_L\cdot x}e^{iP'_rx}Y_{P'}^\dg e^{\frac{i}{2}\hat K^\dag \bar n\cdot x}\Bigg(i\Dslash -\frac{\slashed{\bn}}{2}\left\{\frac{e\Aslash_{cl}}{i\bar n\cdot D},i\Dslash_\perp\right\}\nn\\
&+\frac{\slashed{\bn}}{2}\frac{(eA_{cl})^2}{i\bar n\cdot D}-m\del_{PP'}\Bigg)e^{-\frac{i}{2}\hat K\bar n\cdot x}Y_Pe^{-iP_L\cdot x}e^{-iP_rx}u_P.\label{eq:89eq6}
\end{align}
The $\dg$ on $\hat K$ is a convention to indicate the derivative operators inside act to the left.
We define $P_L$ as the `label momentum', corresponding to discretizing momentum space into bins each labeled by a distinct $P_L$, and $P_r$ as the residual momentum a continuous variable correcting the value of the label momentum inside of each bin. Thus Fourier transformation for a QED spinor becomes
\begin{align}
\psi_{QED}&=\int \frac{d^4P}{(2\pi)^4} e^{-iPx} u_P\nn\\
&=\sum_{P_L}e^{-iP_Lx}\paren{\int \frac{d^4P_r}{(2\pi)^4} e^{-iP_rx}u_{P_r}},
\end{align}
which means in order to sum all over all momenta, we must first sum the label momentum discretely over all bins, and then integrate over the continuous residual momentum inside each bin. As a result, the sum over momenta in the Lagrangian becomes
\be
\sum_{P'P}\to \sum_{P_L'P_L}\int \frac{d^4P'_r}{(2\pi)^4}\frac{d^4P_r}{(2\pi)^4}.
\ee

Next we need to separate the collinear and soft photons in the covariant derivative. We rewrite the covariant derivative as
\be
iD^\mu=i\pd^\mu -eA_c^\mu -eA_s^\mu,
\ee
where $A_s^\mu$ stands for soft photons and $A_c^\mu$ stands for collinear photons. We introduce the Wilson lines for collinear photons
\begin{align}\label{eq:Wndefn}
W_n(x)&=\exp\paren{-i\int_{-\infty}^x eA_c\cdot n ds},\nn\\
W_\bn(x)&=\exp\paren{-i\int_{-\infty}^x eA_c\cdot \bn ds},\nn\\
W_\perp(x)&=\exp\paren{-i\int_{-\infty}^x eA_c\cdot a ds},
\end{align}
where $a^\mu$ is the unit polarization vector for the background,
\be\label{eq:Aclpolarization}
a^\mu=\frac{eA_{cl}^\mu}{|eA_{cl}^\mu|}.
\ee
The equations of motion for the collinear Wilson lines are
\begin{align}
i\bn\cdot D_c W_\bn&=(i\bn\cdot\pd-e\bn\cdot A_c)W_\bn=0,\nn\\
a\cdot iD_{c\perp}W_\perp&=(a\cdot i\pd_\perp-ea\cdot A_{c\perp})W_\perp=0,
\end{align}
which induce,
\begin{align}
W_\bn\ov{i\bn\cdot\pd}W_\bn^\dg&=\ov{i\bn\cdot D_c},\qquad [i\bar n\cdot D_c,W_{\bn}]=0, \nn\\
W_\perp\ov{ia\cdot\pd_\perp}W_\perp^\dag&=\ov{ia\cdot D_{c\perp}}, \qquad [iD_{c\perp},W_\perp]=0.
\end{align}
We also introduce Wilson lines for soft photons,
\begin{align}\label{eq:Sbndefn}
S_\bn&=\exp\paren{-i\int_{-\infty}^x e\bn\cdot A_s ds}, \\ \label{eq:Sperpdefn}
S_\perp&=\exp\paren{-i\int_{-\infty}^x ea\cdot A_s\,ds},
\end{align}
with
\be
i\bn\cdot D_s S_\bn=(i\bn\cdot\pd-e\bn\cdot A_s)S_\bn=0,
\ee
which gives
\be
S_\bn\ov{i\bn\cdot\pd}S_\bn^\dg=\ov{i\bn\cdot D_s},
\ee
and
\be
S_\bn^\dg i\slashed{D}S_\bn=S_\bn^\dg(i\slashed{D}_c-e\slashed{A}_s)S_\bn=i\slashed{D}_c.
\ee
As we are in QED with an Abelian gauge group, different types of Wilson lines commute with each other and the collinear derviatives are not acting on the soft phase that the only non-trivial commutators are
\begin{align}\label{eq:Wcommutator1}
[W_\bn^\dg, iD_\perp^\mu]&=\paren{-\frac{\pd_\perp^\mu}{\bn\cdot\pd}e\bn\cdot A_c}W_\bn^\dg ,
\\ \label{eq:Wcommutator2}
[iD_\perp^\mu, W_\bn]&=W_\bn\paren{\frac{\pd_\perp^\mu}{\bn\cdot\pd}e\bn\cdot A_c} ,
\end{align}
and
\be \label{eq:Wcommutator3}
[W_\bn^\dg, iD_\perp^\mu]^\dg=[W_\bn^\dg,iD_\perp^\mu].
\ee
Following the above rules, we are able to remove the covariant derivatives from the classical Wilson line
\be\label{eq:89eq9}
\hat K=S_\bn W_\bn W_\perp \hat K_c W_\perp^\dg W_\bn^\dg S_\bn^\dg,
\ee
with
\begin{align}\label{hatKcdefn}
\hat K_c&=-\frac{(eA_{cl})^2}{i\bn\cdot\pd}+\oneov{i\bn\cdot\pd}2 e A_{cl}\cdot i\pd-eA_{cl}^2 W_\perp^\dg\left[\oneov{i\bn\cdot\pd},W_\perp\right]+\frac{2 e A_{cl}^\mu}{i\bn\cdot\pd}[W_\bn^\dg,i\pd_\perp^\mu]W_\bn\nn\\
&+W_\perp^\dg\left[\oneov{i\bn\cdot\pd},W_\perp\right]2 e A_{cl}\cdot i\pd+W_\perp^\dg\left[\oneov{i\bn\cdot\pd},W_\perp\right]2 eA_{cl}^\mu[W_\bn^\dg,i\pd_\perp^\mu]W_\bn+...,
\end{align}
where the $...$ stands for terms that are higher order in $\lambda$ because they involve  soft photon fields.  
Now we have the lagrangian in the form
\begin{align}
\cL&=\sum_{P_L P'_L}\int dP' dP\: \bar u_{P'}e^{iP_r'x}e^{iP_L'x}Y_{P'}^\dg S_\bn W_\bn W_\perp e^{\frac{i}{2}\hat K_c^\dg \bn\cdot x} W_\perp^\dg W_\bn^\dg S_\bn^\dg 
\nn\\
&\times \bigl[ i\slashed{D}-\frac{\nslash}{2}\left\{\frac{e\Aslash_{cl}}{i\bar n\cdot D},i\Dslash_\perp\right\}+\frac{\nslash}{2}\frac{(eA_{cl})^2}{i\bar n\cdot D}-m\bigl] 
\nn\\
& \times S_\bn W_\bn W_\perp e^{-\frac{i}{2}\hat K_c\bn\cdot x} W_\perp^\dg W_\bn^\dg S_\bn^\dg Y_{P}e^{-iP_Lx}e^{-iP_rx}u_P.\label{eq:collinearLagstep1}
\end{align}
The coupling to the ultra-soft and soft photons can be removed from the collinear electron lagrangian at this point with a BPS phase redefinition of collinear photon field
\be\label{eq:Acredef}
A_c=S_\bn Y_{P'} A_c^{(0)} Y_{P}^\dg S_\bn^\dg,
\ee
and electron fields
\be\label{eq:uPredef}
u_P = S_\bn u_P^{(0)}, \qquad \bar u_{P'} = \bar u_{P'}^{(0)}S_\bn^\dg,
\ee
using also the fact that
\be
Y_P(x)Y_P^\dg(x)=Y_{P'}(x)^\dg Y_{P'}(x)=S_\bn^\dg(x)S_\bn(x)=1.
\ee
(The original substitution of the $W_P^{cl}$ Wilson lines already accomplished the analogous separation of the $A_{cl}$ potential from the spinors.)

Using the commutator identities Eqs.\eqref{eq:Wcommutator1},\,\eqref{eq:Wcommutator2} and \eqref{eq:Wcommutator3}, interactions with $\bn$- and $\perp$-direction collinear photons are moved into commutators of Wilson lines and derivatives.  This step is just a book-keeping to track of all the interaction terms.  It is now easier to show that passing the large phases $\hat K_c,P_L$ through the derivatives cancels all leading-order in $\lambda$ terms involving $A_{cl}$ combined with covariant derivatives.  Thus we obtain
\begin{align}
\cL&=\sum_{P_L P'_L}\int dP' dP \bar u_{P'}^{(0)}e^{iP_r'x}W_\bn W_\perp e^{i(P_L'-P_L)x+\frac{i}{2}(K^\dg-K)\bn\cdot x}\bigg(i\bn\cdot\pd\frac{\nslash}{2}+in\cdot D\frac{\nslash}{2}
\nn\\
&+i\slashed{D}_{\perp}'+W_\perp^\dg[i\bn\cdot\pd,W_\perp]\frac{\nslash}{2}+[W_\perp^\pd,in\cdot D]W_\perp\frac{\nslash}{2}\nn\\
&+W_\bn^\dg[in\cdot D,W_\bn]\frac{\nslash}{2}+W_\bn^\dg[i\slashed{D}_{\perp}',W_\bn]-m\bigg)W_\perp^\dg W_\bn^\dg e^{-iP_rx}u_P^{(0)},\label{eq:collinearLagstep2}
\end{align}
where $i\slashed{D}_\perp'$ is defined as
\be
i\slashed{D}'_{\perp}=a\cdot \gam a\cdot i\pd+b\cdot\gam ib\cdot D_{c\perp},
\ee
because the $a^\mu$-direction pieces of the collinear photons have been moved into $W_\perp$ lines.  Recall that $a^{\mu}$ is the polarization vector of $A_{cl}^\mu$ \eq{Aclpolarization}, and $b^\mu$ is chosen to complete the basis is the transverse space
\be
g_\perp^{\mu\nu}=a^\mu a^\nu +b^\mu b^\nu\quad a\cdot b=0.
\ee 
We can integrate over the short-distance scale $\bn\cdot x \sim 1/Q$ to yield a $\delta$-function conserving the large momentum components.  We will suppress the large momentum phases and use `label' momentum conservation as in SCET.   

We restore the $\perp$ and $\bn$ collinear photons into the usual lagrangian interactions by pushing the $W_\perp$ and $W_\bn$ back through the derivatives, reversing the previous step.  We then have the leading order EFT Lagrangian as,
\begin{align}
\cL&=\sum_{P_L P'_L}\int dP' dP\, \bar u_{P'}^{(0)}e^{iP'_rx}W_\bn W_\perp\bigl[i\bn\cdot D_c\frac{\nslash}{2}+in\cdot D\frac{\nslash}{2}+i\Dslash_\perp'+W_\perp^\dg[i\bn\cdot\pd,W_\perp]\frac{\nslash}{2}\nn\\
&+[W_\perp^\dg,in\cdot D]W_\perp\frac{\nslash}{2}+W_\bn^\dg[in\cdot D,W_\bn]\frac{\nslash}{2}+W_\bn^\dg[i\Dslash_\perp',W_\bn]-m\bigl]W_\perp^\dg W_\bn^\dg e^{-iP_rx}u_P^{(0)}.\label{eq:89eq22}
\end{align}
Here the subscript `$c$' on the $\bn$-direction covariant derivative indicates that only the collinear photon remains,
\be\label{eq:Dcdefn}
D_c^\mu=\partial^\mu+ieA_c^\mu\,,
\ee
as the $\bn$-direction photon was removed by the BPS phase redefinition Eqs.\,\eqref{eq:Acredef} and \eqref{eq:uPredef}. We will see shortly that this means that soft photons have been decoupled completely from the collinear lagrangian at leading power in $\lambda$.

Next, as in SCET, we decompose $u_P^{(0)}$ as two-spinors
\begin{align}
\xi_\bn&=\frac{\slashed{\bn}\slashed{n}}{4}u_P^{(0)}, \qquad
\chi_n=\frac{\slashed{n}\slashed{\bn}}{4}u_P^{(0)},
\end{align}
which is complete since $\frac{\slashed{\bn}\slashed{n}}{4}$ and $\frac{\slashed{n}\slashed{\bn}}{4}$ are complementary projectors,
\be
\frac{\slashed{\bn}\slashed{n}}{4}+\frac{\slashed{n}\slashed{\bn}}{4}=1,\qquad u_P^{(0)}=\chi_n+\xi_\bn.
\ee
Following the same procedure of obtaining the SCET leading order Lagrangian and integrating out the small spinor components $\chi_n$, we have
\begin{align}
\cL_n^{(0)}&=\bar\xi_\bn \left(i\bn\cdot D_c \frac{\nslash}{2}  
+(i\Dslash_{c\perp})\oneov{in\cdot D_c}i\Dslash_{c\perp} \frac{\nslash}{2}\right)  \xi_\bn.\label{eq:collinearLagstep4}
\end{align}
The $n$-direction Wilson lines \eq{Wndefn} are introduced to remove the $n$-direction collinear photons from the denominator, using the identity
\begin{align}
\ov{i n\cdot D_c}= W_n\ov{in\cdot\pd}W_n^\dg  .
\end{align}
Thus we obtain
\begin{align}
\cL_n^{(0)}&=\bar\xi_\bn \bigl(i\bn\cdot D_c 
+i\Dslash_{c\perp}W_n\oneov{in\cdot \partial}W_n^\dg i\Dslash_{c\perp} \bigl)\frac{\nslash}{2}  \xi_\bn.\label{eq:89eq26}
\end{align}
Soft photons do not contribute to the $D_\perp$ and $n\cdot D$ terms at leading power in $\lambda$.  At this point we can see, that by redefining the fields as well as introducing Wilson lines, we obtain a gauge invariant collinear electron Lagrangian with soft photons decoupled.


\subsubsection{Soft Electron Lagrangian}
\label{sec:III.4.3.2}

In this part we derive soft electron Lagrangian. Soft electron momentum $P_s$ scales as $(\lam^2,\lam^2,\lam^2)Q$, which are the electrons not accelerated to the high energy mode. Starting with the full theory,
\be\label{eq:90eq0}
\cL=\bar\psi(i\slashed{D}-e\slashed{A}_{cl}-m)\psi,
\ee
with the classical solution,
\be\label{eq:90eq1}
\psi=\sum_P -e^{iS_{cl}+\frac{\slashed{n}}{2}\frac{e\slashed{A}_{cl}}{i\bn\cdot P_s}}e^{-iP_sx}u_{P_s}\equiv W_{P_s}^{cl}e^{-iP_sx}u_{P_s}.
\ee
The covariant derivative picks out the gauge covariant residual momentum of the field operator, which gives the equation of motion,
\be\label{eq:90eq2}
(i\slashed{D}_s-e\slashed{A}_{cl})W_{P_s}^{cl}e^{-iP_sx}u_{P_s}=W_{P_s}^{cl}i\slashed{D}_se^{-iP_sx}u_{P_s}.
\ee
We want to expand the quantum fluctuations around this semiclassical solution, which is at the soft scale in our EFT. The kinematic momentum, also determined by the current with semiclassical states $\langle j^\mu\rangle=\langle \bar u_PW_P^\dg \gamma^\mu W_P u_P\rangle$ is
\be\label{eq:90eq3}
\Pi^\mu=P_s^\mu-eA_{cl}^\mu+\frac{\bn^\mu}{2}\frac{P_{s\perp}\cdot eA_{cl}-(eA_{cl})^2}{\bn\cdot P_s}.
\ee
For soft electrons the momentum is subleading compared to the classical dressing, as $P_{s\perp}^\mu\ll eA_{cl}$, $n\cdot P_s\ll n\cdot \Pi\simeq \frac{P_\perp \cdot eA_{cl}-(eA_{cl})^2}{\bn\cdot P_s}$. Here $P_s^\mu$ is a residual parameter compared to
\be
\hat K_s=-\frac{(eA_{cl})^2}{i\bn\cdot D_s}+\oneov{i\bn\cdot D_s}\left\{iD_{s\perp},eA_{cl}\right\},
\ee
where the first term has the scale of $\lam^0$, and depends on the soft derivative non-perturbatively. In order to separate the soft modes from the large scale in $\hat K_s$,  we use $S_\bn,S_\perp$ Wilson lines \eq{Sbndefn} and \eq{Sperpdefn} obeying all the properties in the previous section.
Thus we have
\be\label{eq:89eq9soft}
\hat K=S_\bn S_\perp \hat K_s S_\perp^\dg S_\bn^\dg,
\ee
with
\begin{align}
\hat K_s&=-\frac{(eA_{cl})^2}{i\bn\cdot\pd}+\oneov{i\bn\cdot\pd}2 e A_{cl}\cdot i\pd-eA_{cl}^2 S_\perp^\dg\left[\oneov{i\bn\cdot\pd},S_\perp\right]+\frac{2 e A_{cl}^\mu}{i\bn\cdot\pd}[S_\bn^\dg,i\pd_\perp^\mu]S_\bn\nn\\
&+S_\perp^\dg\left[\oneov{i\bn\cdot\pd},S_\perp\right]2 e A_{cl}\cdot i\pd+S_\perp^\dg\left[\oneov{i\bn\cdot\pd},S_\perp\right]2 eA_{cl}^\mu[S_\bn^\dg,i\pd_\perp^\mu]S_\bn.
\end{align}
Inserting the semiclassical solution and this separation of Wilson lines from $\hat K$ into the lagrangian \eq{90eq0}, we obtain
\begin{align}
\cL&=\sum_{P,P'}\bar u_{P'}e^{iP'_rx}Y_{P'}^\dg S_\bn S_\perp e^{\frac{i}{2}\hat K_s^\dag \bar n\cdot x}S_\perp^\dg S_\bn^\dg\Bigg(i\Dslash -\frac{\slashed{\bn}}{2}\left\{\frac{e\Aslash_{cl}}{i\bar n\cdot D},i\Dslash_\perp\right\}\nn\\
&+\frac{\slashed{\bn}}{2}\frac{(eA_{cl})^2}{i\bar n\cdot D}-m\del_{PP'}\Bigg)S_\bn S_\perp e^{-\frac{i}{2}\hat K_s\bar n\cdot x}S_\perp^\dg S_\bn^\dg Y_Pe^{-iP_rx}u_P,\label{eq:89eq6soft}
\end{align}
analogous to the intermediate step in the collinear lagrangian \eq{89eq22}, but without the large momentum $P_L,P_L'$ to be summed over discretely.

We now decouple the ultrasoft Wilson lines by a BPS phase redefinition
\be\label{eq:softfieldredef}
u_P = Y_P^\dg u_P^{(0)}, \qquad A_s = Y_{P'}A^{(0)}_s Y_P^\dg,
\ee
using as well $Y_P(x)Y_P^\dg(x)=1$.

Following the same procedure in the previous section, we can simplify the Lagrangian by first moving soft photon interaction terms out of the covariant derivatives and into $S_\perp, S_\bn$, then commuting the large phase $K_s$ through the derivatives to cancel the interaction terms involving $A_{cl}$ and eventually obtain
\begin{align}
\cL&=\sum_{P',P}\bar u_{P'}^{(0)} S_\bn S_\perp e^{\frac{i}{2}(K_s^\dg-K_s)\bn\cdot x}\bigg(in\cdot D \frac{\slashed{n}}{2}+i\bn\cdot\pd \frac{\slashed{n}}{2}+i\slashed{D}_\perp'+S_\perp^\dg[i\bn\cdot\pd,S_\perp] \frac{\slashed{n}}{2}\nn\\
&+[S_\perp^\dg,in\cdot D]S_\perp\frac{\slashed{n}}{2}+S_\bn^\dg[in\cdot D,S_\bn]\frac{\slashed{n}}{2}+S_\bn^\dg[i\slashed{D}_\perp',S_\bn]  
-m\bigg)S_\perp^\dg S_\bn^\dg u_P^{(0)}\,.
\end{align}
We integrate over $\bn\cdot x$ at short distance scale to enforce the large momentum conservation $\delta(K_s-K_s^\dag)$, and suppress these large momenta from now.
Restoring $\bn\cdot A_s, a\cdot A_s$ to covariant derivative rather than Wilson lines by inserting $1=S_\perp^\dg S_\perp$ and $1=S_\bn^\dg S_\bn$ the Lagrangian reads,
\be
\cL=\sum_{P'P}\delta_{K_s',K_s}\bar u_{P'}\paren{i\slashed{D}-m}u_P\,,
\ee
which is just the usual QED Lagrangian.  The information of the external potential is contained in the conservation of the large momentum $K_s$.

\subsubsection{Photon Lagrangian}
\label{sec:III.4.3.3}

In this part we derive photon Lagrangian. The soft photon Lagrangian is the same as QED, so we only focus on the collinear photon one. Field strength tensor is given in terms of commutator of covariant derivatives
\be
F^{\mu\nu}=\frac{i}{g}[D^\mu,D^\nu].
\ee
We expand the covariant derivative as,
\begin{align}
iD^\mu&=\frac{\bn^\mu}{2}\paren{\hat P_n gn\cdot A_c}+\hat P_\perp^\mu- gA_{c\perp}^\mu\nn\\
&+\frac{n^\mu}{2}\paren{i\bn\cdot\pd-g\bn\cdot A_c-g\bn\cdot A_s}+\frac{\bn^\mu}{2}\paren{in\cdot \pd-gn\cdot A_s}\nn\\
&-gA_{s\perp}^\mu,
\end{align}
and $A_c^\mu\sim (1,\lam^2,\lam)Q$ is collinear photon field, $A_s^\mu\sim (\lam^2,\lam^2,\lam^2)Q$ as soft field, $\hat P^\mu$ is the operator projecting the large label momentum out of the collinear photons whose scale is $(1,0,\lam)Q$. The first term scales as $\lam^0$, the second term scales as $\lam^1$, and rest of the terms scales as $\lam^2$.
With the commutator properties,
\begin{align}
\left[\hat P_n-gn\cdot A_c,\hat P_n-gn\cdot A_c\right]&=-g\hat P_nn\cdot A_c-(-g)\hat P_nn\cdot A_c=0\nn\\
\left[-\hat P_n-gn\cdot A_c,\hat P_\perp^\nu-gA_{c\perp}^\nu\right]&=-g\hat P_nA_{c\perp}^\nu+g\hat P_\perp^\nu n\cdot A_c\nn\\
\Bigg[\hat P_n-gn\cdot A_c,i\bn\cdot\pd-g\bn\cdot A_n&-g\bn\cdot A_s\Bigg]=-g\hat P_n\bn\cdot A_c+gi\bn\cdot\pd n\cdot A_c\nn\\
\left[\hat P_n-gn\cdot A_c,in\cdot\pd-gn\cdot A_s\right]&=+gin\cdot \pd n\cdot A_c\nn\\
\left[\hat P_\perp^\mu-gA_{c\perp}^\mu,\hat P_\perp^\nu-gA_{c\perp}^\nu\right]&=-g\hat P_\perp^\mu A_{c\perp}^\nu+g\hat P_\perp^\nu A_{c\perp}^\mu\nn\\
\Bigg[i\bn\cdot\pd-g\bn\cdot A_c-g\bn\cdot A_s,i\bn\cdot\pd&-g\bn\cdot A_c-g\bn\cdot A_s\Bigg]=0\nn\\
\Bigg[i\bn\cdot\pd-g\bn\cdot A_c-g\bn\cdot A_s,in\cdot \pd&-gn\cdot A_s\Bigg]=-gi\bn\cdot \pd n\cdot A_s+gin\cdot \pd \bn\cdot A_c\nn\\
&+gin\cdot \pd \bn\cdot A_s,
\end{align}
we keep the leading order Lagrangian which is $\lam^4 Q$ order, 
\begin{align}
\cL_c^\gam&=-\ov{4}F_c^{\mu\nu}F_c^{\mu\nu}=\paren{i\bn\cdot\pd n\cdot A_c-\hat P_n \bn\cdot A_c}^2+\ov{4}\paren{\hat P_\perp^\mu A_{c\perp}^\nu-\hat P_\perp^\nu A_{c\perp}^\mu}^2\nn\\
&+\ov{2}\paren{\hat P_n A_{c\perp}^\mu}\paren{i\bn\cdot \pd A_{c\perp}^\mu}-\ov{2}\paren{P_\perp^\mu n\cdot A_c}\paren{in\cdot \pd A_{c\perp}^\mu}\nn\\
&+\ov{2}\paren{\hat P_n A_{c\perp}^\mu}\paren{i\bn\cdot \pd A_{c\perp}^\mu}-\ov{2}\paren{P_\perp^\mu n\cdot A_c}\paren{i\bn\cdot \pd A_{c\perp}^\mu}\nn\\
&+\ov{2}\paren{\hat P_\perp^\mu n\cdot A_c}\paren{\hat P_\perp^\mu\bn\cdot A_c}-\ov{2}\paren{\hat P_n A_{c\perp}^\mu}\paren{\hat P_\perp^\mu \bn\cdot A_c}.
\end{align}



\subsubsection{Reparametrization Invariance in Leading Order Lagrangian}
\label{sec:III.4.3.4}

Recall that in reparameterization invariance arises in SCET due to the artificial separation of collinear momentum into `label' and `residual' parts.  Because the large `label' momentum is light-like, just as in SCET, RPIs in electro-laser effective theory consequently take the same three forms, written in terms of the change in the collinear direction vector $n^\mu$,
\begin{align}
\rm{Type~I}:\quad & n^\mu\to n^\mu+\Delta_\perp^\mu ,\quad \bn^\mu~\rm{ unchanged,}\label{eq:eLETRPI1}\\
\rm{Type~II}:\quad & \bn^\mu \to\bn^\mu +\eps_\perp^\mu, \quad n^\mu ~\rm{ unchanged,}\label{eq:eLETRPI2}\\
\rm{Type~III}:\quad & n^\mu\to e^\alpha n^\mu, \qquad \bn^\mu \to e^{-\alpha}\bn^\mu,\label{eq:eLETRPI3}
\end{align}
where $\Delta_\perp\sim \lam$ and $\alpha,\eps_\perp\sim\lam^0$.  The properties of these transformations were given above in Section \ref{sec:SCETRPI}.  In particular, the parameters $\Delta^\mu,\eps^\mu$ are perpendicular vectors, and 4-vectors are unchanged under all three types of transformations.

The leading-order collinear and soft electron lagrangians satisfy all the restrictions imposed by invariance under RPI.  For instance, Type III (\eq{eLETRPI3}) requires $n^\mu$ in an operator is always accompanied either by an $\bn^\mu$ or an $n^\mu$ in the denominator.  The fields transform as
\begin{align}
n\cdot D&\to n\cdot D+\Delta_\perp\cdot D_\perp\,, \\
D_\perp^\mu& \to D_\perp^\mu-\frac{\Delta_\perp^\mu}{2}\bn\cdot D-\frac{\bn^\mu}{2}\Delta_\perp\cdot D_\perp \,,\\
\xi_\bn&\to \paren{1+\frac{\slashed{\Delta}_\perp}{2}\oneov{in\cdot D}i\slashed{D}_\perp}\xi_\bn\,,  \\
W_n&\to \paren{1-\frac{1}{in\cdot D}i\Delta_\perp\cdot D_\perp}W_n \,,
\end{align}
and $\bn\cdot D$ is invariant under Type I (\eq{eLETRPI1}). Under Type II (\eq{eLETRPI2}), the fields transform as
\begin{align}
\bn\cdot D&\to \bn\cdot D+\eps_\perp\cdot D_\perp\,,\label{eq:91eq7.43eLET}\\
D_\perp^\mu&\to D_\perp^\mu-\frac{\eps_\perp^\mu}{2}n\cdot D-\frac{n^\mu}{2}\eps_\perp\cdot D_\perp\,, \\
\xi_\bn&\to \paren{1+\frac{\slashed{\eps}_\perp\slashed{n}}{4}}\xi_\bn\,,
\end{align}
and $n\cdot D$ is invariant.

\subsubsection{Gauge Invariance in Leading Order Lagrangian}
\label{sec:III.4.3.5}
As in SCET, the gauge invariance of QED is decomposed into invariances for collinear and soft degrees of freedom.  The effective Lagrangian for each sector is invariant only under residual gauge transformations because the effective field operators only describe modes with specific momentum scaling. The residual gauge symmetries satisfy either the soft scaling $(\bar n\cdot\partial,n\cdot\partial,\partial^\perp)U_s(x)\sim Q(\lambda^2,\lambda^2,\lambda^2)U_s(x)$ or the collinear scaling $(\bar n\cdot\partial,n\cdot\partial,\partial^\perp)U_n(x)\sim Q(1,\lambda^2,\lambda)U_n(x)$.  The rules of transformation for all the fields involved are obtained from the full gauge symmetries of all the operators and given in Table \ref{tab:eLETgaugesym}

To define a collinear gauge invariant operators, we use collinear Wilson line (\cite{bauer2001invariant})
\begin{equation}
W_n(x)=P\exp\left[-ig\int_{-\infty}^0ds\bar n\cdot A_n(s\bar n+x)\right],
\end{equation}
with the transformation law $W_n\to U_nW_n$ under collinear gauge symmetry. Since $\xi_n$ obeys the same transformation law, the combination $W_n^\dagger\psi_n$ is gauge invariant.  Owing to the photon field redefinitions \eq{Acredef} and \eq{softfieldredef}, we define a collinear gauge-invariant photon field 
\begin{equation}
B_n^\mu=e^{-1}W_n^\dagger iD_n^\mu W_n\,,
\end{equation}
though the possible coupling between collinear and soft photons does not yet play a role in the electron-laser theory, since there are no lagrangian interactions between soft and collinear photons  at leading power.
One uses these gauge-invariant combinations to build collinear operators in the electron laser effective theory.

Due to the additional background field $A_{cl}$, we have an additional class of gauge transformations.  As $A_{cl}$ is classical and a background to all the quantized fields in our theory, the transformations are simple.  First, note that $A_{cl}$ transforms under a restricted class of gauge transformations
\be\label{eq:Acltransform}
A_{cl}\to U_{cl}^\dg(A_{cl}+\frac{1}{e}\partial) U_{cl}, 
\ee
with $U_{cl}$ a function only of $\bn\cdot x$ to preserve the symmetries of the background; the complete set of transformations can be restored at higher order.  This implies $\partial^2 U_{cl}=(\partial U_{cl})^2=0$, and it is a simple calculation to determine that the classical Wilson line transforms as
\be
W_P^{cl}\to W_P^{cl}U_{cl}.
\ee
However, we also decomposed the Wilson line into a large phase piece and residual ultrasoft Wilson line.  We may similarly decompose $U_{cl}$, writing
\be
\frac{i}{e}\ln U_{cl}(\bn\cdot x)=\ln U_{cl}(0)+\frac{\bn\cdot x}{2}n\cdot\partial \ln U_{cl}(0)+\frac{(\bn\cdot x)^2}{8}(n\cdot\partial)^2\ln U_{cl}(0).
\ee
The first term in the expansion cancels in all transformations.  Note that in order to not change the frequency composition of $A_{cl}$ (therefore its power counting), $n\cdot\partial \ln U_{cl}(0)$ is a constant at the length scales of the effective theory, just as is $A_{cl}$.  Therefore, by comparison to 
\be
W_P^{cl}=e^{-\frac{i}{2}\hat K\bn\cdot x}Y_P(\bn\cdot x)\,,
\ee
the second term is just a constant shift to both $\hat K_{c,s}$ and $\hat K^\dg_{c,s}$, which consequently does not change physical kinematics.  The higher derivatives as the transformation of $Y_P$.  Thus, invariance under the classical gauge transformations is ensured by fermion fields always appearing together with the ultrasoft line $Y_P$.

\begin{table}
\centering
\begin{tabular}{lcccc}
\hline
& Field & Scaling & $U_c$& $U_{s}$ \\
 & $\xi_n$ & $\lambda$ & $U_c\xi_n$ & $U_{s}\xi_n$ \\
collinear & $A_c^\mu$ & $(\lambda^2,1,\lambda)$ & $U_cA_c^\mu U_c^\dag+\frac{i}{g}U_c[iD_c^\mu,U_c^\dag]$ & $U_{s}A_c^\mu U_{s}^\dag$ \\
 & $W_n$ & 1 & $U_cW_n$ & $U_{s}W_nU_{s}^\dag$ \\
\hline
& $q$ & $\lambda^3$ & $q$ & $U_{s}q$ \\
soft & $A_{s}^\mu$ & $\lam^2$ & $A_{s}^\mu$ & $U_{s}(A_{s}^\mu+\frac{i}{g}\pd^\mu)U_{s}^\dag$ \\
& $S$ & 1 & $S$ & $U_{s}S$\\
\hline
\end{tabular}
\caption{Rules for collinear and ultrasoft gauge transformations and scaling dimensions of EFT fields}
\label{tab:eLETgaugesym}
\end{table}


\subsection{Matching Soft-to-Collinear Current from EFT to Full Theory}
\label{sec:III.4.4}

\begin{figure}
\centering
\includegraphics[width=.8\textwidth]{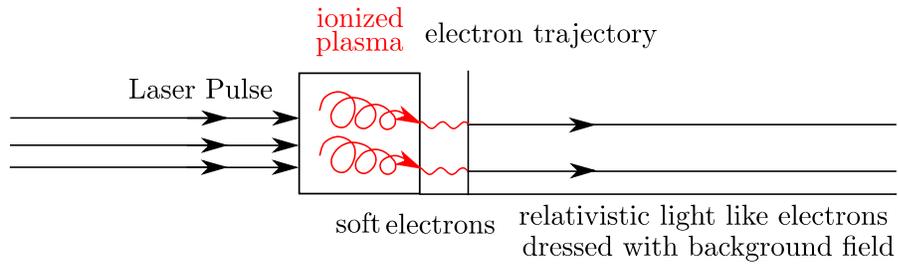}
\caption{Schematic electron trajectory in laser acceleration experiments.}
\label{fig:200fig1}
\end{figure}

In this section, we construct a soft electron to collinear electron current, and match this current onto the full theory. We simplify the electrons' trajectory in \fig{200fig1}, which shows that electrons are first traveling in a convoluted motion after being ionized in the target but before achieving high momentum.  At some point, the electron absorbs momentum from the background and travels as a light-like particle in an approximately straight line varying with the laser wave length $\lam_{\text{laser}}\sim\ov{\omega_{\rm laser}}$ which is Lorentz dilated in lab frame, representing very long distance physics.

\begin{figure}
\centering
\includegraphics[width=.8\textwidth]{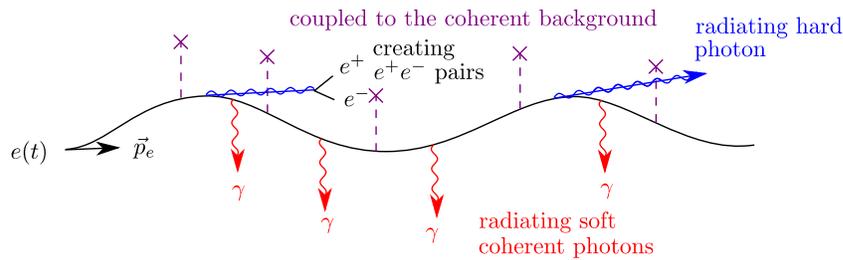}
\caption{Quasi-Feynman diagram for the electron dynamics with radiation reaction and pair production. }
\label{fig:200fig2}
\end{figure}

\begin{figure}
\centering
\includegraphics[width=.9\textwidth]{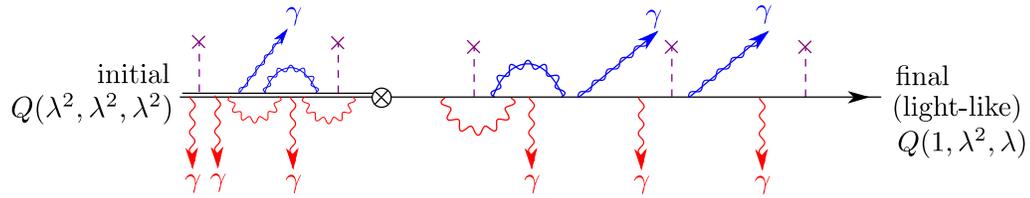}
\caption{Soft-to-collinear current without field decoupling}
\label{fig:200fig3}
\end{figure}

We can demonstrate this picture in a Feynman diagram fashion, shown in \fig{200fig2}, which depicts the electron travelling along a slowly-varying classical trajectory.  Dashed lines with crosses stands for coupling with the coherent classical background, wavy lines represents soft radiation photons, and wavy lines with a straight line through them represent collinear photons, capable of creating $e^+e^-$ pairs. In our EFT, we can first drop the electron classical trajectory since it only varies with long distance physics.  We decompose the electron into soft and collinear modes representing the not-yet-accelerated electrons and fully dressed electrons respectively (see \fig{200fig3}).  The circle connecting the two types of electrons is a vertex with Lorentz structure.

\begin{figure}
\centering
\includegraphics[width=.6\textwidth]{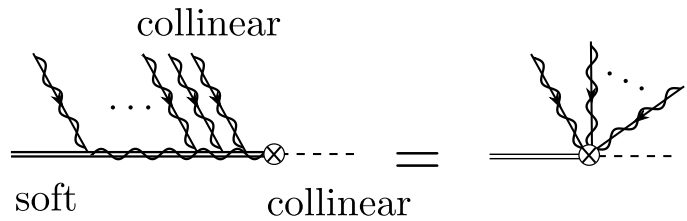}
\caption{Decoupling collinear Wilson lines from the soft electron}
\label{fig:200fig4}
\end{figure}

\begin{figure}
\centering
\includegraphics[width=.6\textwidth]{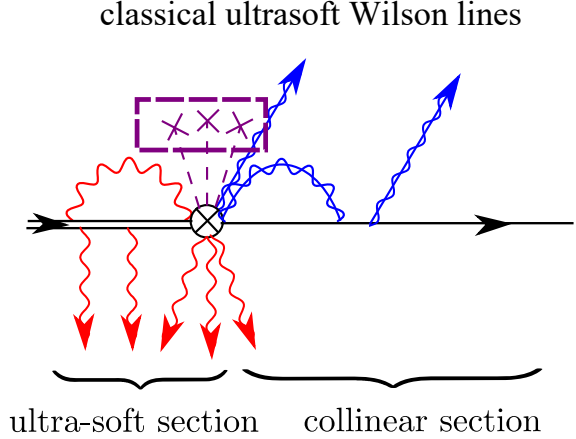}
\caption{Decoupled soft-to-collinear current}
\label{fig:200fig5}
\end{figure}

Similar to construction of EFT in the previous section, we first integrate out the large classical background dressing for both soft and collinear electrons and then decouple the rest of ultra soft background. Then we integrate out heavy initial states that are dressed with collinear photons, which leaves a sum of such photons attached to the vertex, as is shown in \fig{200fig4}. In this fashion we sum all the non-perturbative effects from initial state electrons into the vertex and leave them as pure QED electrons.  This represents the fact that the absorption of momentum from the background occurs coherently and cannot be identified with any specific point $x$ at shorter distance scale than $\lambda_{\text{laser}}$.  Then we redefine the collinear electron field in the final states by decoupling all of the soft photons and summing them into soft photon Wilson lines attached to the vertex. At this point we have decoupled the soft and collinear as well as ultrasoft background contribution from each other as is shown in \fig{200fig5}.

The above physics is organized into a gauge invariant current operator which in momentum space is written as
\be
J_{\rm EFT}^\mu=\paren{\bar\xi_c W_n S_\bn Y_{P'_c}^\dg\gam^\mu Y_{P_s}\psi_s}C(P_s,P'_c;\mu),
\ee
where $\psi_s$ is the soft field and $\bar\xi_c$ is the collinear field, and $C(P_s,P'_c;\mu)$ is the Wilson coefficient that we can obtain from matching.  A more precise derivation of this current will be given below during factorization, see \eq{EFTcurrent}; here we emphasize the operator composition.  The combination $\xi_c W_n$ is collinear gauge invariant, $S_\bn Y_{P'}^\dag Y_P\psi_s$ is soft gauge invariant, and $\xi_c Y_{P'}^\dg$ and $Y_{P}\psi_s$ are classical gauge invariant. We define
\be
\Gamma^\mu=Y_{P'_c}^\dg\gam^\mu Y_{P_s}.
\ee

We rewrite the current in position space as,
\begin{align}
J_{\rm EFT}^\mu(x)&=\del(P_c-q)\int dP'_r dP_s e^{-i(P_s-P'_r)x} 
[\bar \xi_{c\bn}W_nS_\bn](x)\Gam^\mu(x)\psi_s(x) C(P_c;P_s;\mu),
\end{align}
where $q$ is the transferred momentum entering through the vertex.

\begin{figure}
\centering
\includegraphics[width=.5\textwidth]{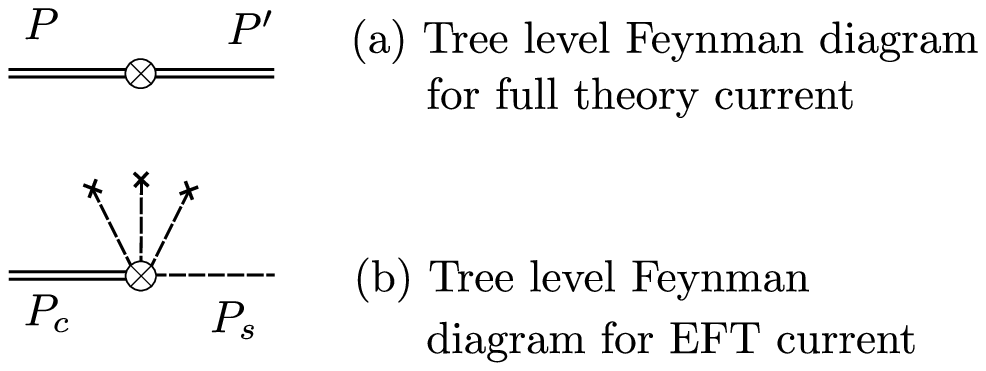}
\caption{Full-theory current at tree level.}
\label{fig:200fig6}
\end{figure}

Next we match this current onto the full theory. Full theory current reads,
\be
J_{\rm{full}}^\mu(x)=\int dP'dP \:e^{-iqx}\bar u_{P'}\paren{W_{P'}^{cl}}^\dg \gam^\mu W_P^{cl}u_P(x).
\ee
The tree level matching is shown in \fig{200fig6} and gives $C=1$.

\begin{figure}
\centering
\includegraphics[width=.75\textwidth]{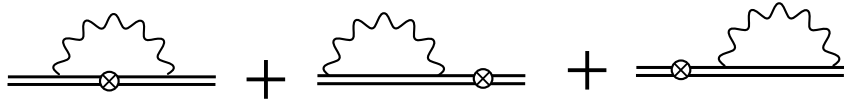}
\caption{One-loop corrections to full theory current.}
\label{fig:200fig7}
\end{figure}

The one-loop diagrams for full theory current are shown in \fig{200fig7}. The results are absent from previous literature, but as expected the short-distance behavior of the semiclassical results reproduces no-background-field QED.  We give a detailed calculation of the diagrams in Appendix \ref{app:fulltheoryselfenergy}.  Combining the results,
\begin{align}
J_{\text{full}}^{\text{1-Loop}}&=-\frac{\alpha}{4\pi}\Bigg\{\ln\frac{-q^2}{m_e^2}\ln\frac{-q^2}{m_\gamma^2}+2\ln\frac{-q^2}{m_e^2}-\frac{1}{2}\ln\frac{-q^2}{\mu^2}-\frac{1}{2}\ln\frac{m_e^2}{\mu^2}\Bigg\}
+H_{\rm{Full}}(k^2,m^2),
\end{align}
where $-q^2=-(p-p')^2$ is the squared momentum transfer and $m_e$ is the electron mass, which we distinguish now with the explicit subscript.  The infrared scale $m_\gamma^2$ is of order $\omega_{\rm laser}$ or $a_0\omega_{\rm laser}$ arising from the precision to which the electron momentum can be known in the presence of the background field.  $H_{\rm{Full}}(k^2,m^2)$ is a finite, analytic function containing the background information and is subleading to the first term. 

\begin{figure}
\centering
\includegraphics[width=.75\textwidth]{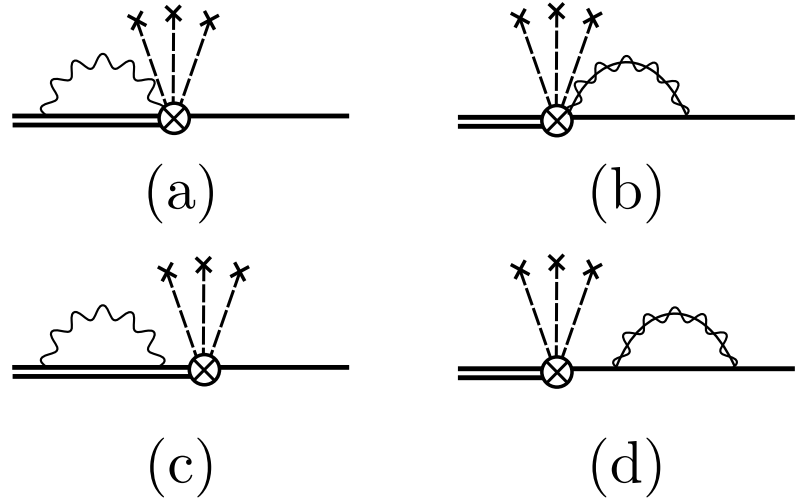}
\caption{One-loop corrections to effective theory current.}
\label{fig:200fig8}
\end{figure}

The one-loop diagrams of the EFT current are shown in \fig{200fig8}. \fig{200fig8}(a) is the soft one-loop correction and \fig{200fig8}(c) is the wave function renormalization diagram to the soft electrons. These two diagrams give, 
\begin{align}
J_{\rm EFT}^a&=\frac{\alpha}{2\pi}\left\{-\ov{\eps}-\ln\frac{\mu^2}{m_e^2}-\frac{1}{2}\ln\frac{\mu^2}{m_e^2}\ln\frac{\mu^2}{m_\gamma^2}\right\},\nn\\
J_{\rm EFT}^c&=\frac{\alpha}{4\pi}\paren{-\ov{\eps}+\ln\frac{m^2}{\mu^2}},
\end{align}
\fig{200fig8}(b) is the collinear one-loop correction and \fig{200fig8}(d) is the wave function renormalization diagram to the collinear electrons.
With the zero-bin subtraction, which is required here as in SCET, these two diagrams give,
\begin{align}
J_{\rm EFT}^b&=\frac{\alpha}{2\pi}\paren{\frac{2}{\eps^2}+\frac{2}{\eps}\ln\frac{\mu^2}{-q^2}-\frac{1}{2}\ln^2\frac{\mu^2}{-q^2}},\nn\\
J_{\rm EFT}^d&=-\frac{\alpha}{4\pi}\paren{\ov{\eps}+\ln\frac{-q^2}{\mu^2}}.
\end{align}
As a result, the EFT current one-loop correction is
\begin{align}
J_{\rm EFT}^{\text{1-loop}}&=J_{\rm EFT}^a+J_{\rm EFT}^b-\frac{1}{2}J_{\rm EFT}^c-\frac{1}{2}J_{\rm EFT}^d\nn\\
&=-\frac{\alpha}{4\pi}\Bigg(\frac{4}{\eps^2}+\ov{\eps}+\frac{4}{\eps}\ln\frac{\mu^2}{-q^2}+2\ln^2\frac{-q^2}{\mu^2}+\ln\frac{\mu^2}{m_e^2}\ln\frac{\mu^2}{m_\gamma^2} 
\nn \\
&~+2\ln\frac{\mu^2}{m_e^2}-\ln\frac{-q^2}{m_e^2}\Bigg)\,,
\end{align}
in which the $m_\gamma^2$ term agrees exactly with the full theory result, showing that we have repeated the infrared physics of the full theory.  Therefore the matching coefficient at one-loop order is
\be
C^{(1)}(\mu)=\frac{\alpha}{4\pi}\left(2\ln\frac{-q^2}{\mu^2}+\ln\frac{-q^2}{\mu^2}\ln\frac{-q^2}{m_e^2}\right).
\ee
The counter-term for the EFT current is defined as
\be
Z_u=1+\frac{\alpha}{\pi}\paren{\ov{\eps^2}+\ov{\eps}+\ov{\eps}\ln\frac{\mu^2}{-q^2}},
\ee
from which we extract the anomalous dimension 
\be
\gam_u=\mu\frac{d}{d\mu}Z_u=-\frac{2\alpha}{\pi}\left(\ln\frac{\mu^2}{-q^2}+1\right),
\ee
using
\be
\mu\frac{d\alpha}{d\mu}=-2\eps\alpha+\cO(\alpha^2).
\ee
We resum the large logarithm of $\ln\frac{\mu^2}{-q^2}$ to the arbitrary scale $\mu$
by solving the RG equation,
\be
\mu\frac{d}{d\mu}C(\mu)=\gam_uC(\mu),
\ee
and obtain
\be
C(\mu,k)=\exp\left(-\frac{2\alpha}{\pi}\big(\ln^2\frac{\mu}{\sqrt{-q^2}}+\ln\frac{\mu}{\sqrt{-q^2}}\big)\right)C(k).
\ee
where $\mu$ is the matching scale and the matching boundary condition determines $C(-q^2)=1+\cO(\alpha)$.

We will use this current in the next section to compute a real process.

\section{Factorization Theorem for Photon Emission}
\label{sec:III.4.5}

In this part we show how the effective theory allows factorization of $e\to e\gamma$.  This is an important process to consider first for both theoretical and experimental reasons.  Current experiments are equipped to detect high energy $2\,\text{MeV}\lesssim E_{\gamma}\lesssim 200\,\text{MeV}$ photons and electrons across a broad range of energies.  For example, the on-going experiment at the Texas Petawatt deploys electron and photon spectrometers on the beam axis and at small to moderate angles $\theta\lesssim \pi/4$ from the beam axis.  Thus it is important to understand the reliability of theoretical predictions by studying next-to-leading order corrections, especially as it is a relatively simple process that has been studied extensively in the semiclassical method.

 We start with the kinematics, evaluated in the lab frame.  Initial state electrons are at rest up to soft corrections, corresponding to momentum
\be\label{eq:94eq0a}
P\sim(m,m,0).
\ee
The final state electron has been accelerated to high momentum collinear to the laser pulse,
\be\label{eq:94eq0b}
P'\simeq \paren{n\cdot P',\frac{(P'_\perp)^2}{n\cdot P},P'_\perp}\sim Q(1,\lambda^2,\lambda),
\ee
where the large momentum scale is
\be
Q=-\frac{(eA_{cl})^2}{m}\,,
\ee
being the typical momentum imparted by the classical field.  For the Texas Petawatt photon emissions are most likely to occur near the peak of the pulse, where $a_0\sim 1/\lambda\gtrsim 50$.  Thus photons emitted at angles $\theta\sim\lambda \gtrsim 1/50$ are considered separable from the final electron.  Momentum conservation in the full theory requires allows arbitrary $n$-direction momentum to be absorbed from the field, but $\bn$-direction and $\perp$ momenta are conserved, up to ultrasoft corrections that are not accounted for in the semiclassical calculation.  Therefore we set $q_\perp=-P_\perp'\gg \lambda^2$ and consider the photon momentum to have scaling
\be
q=\paren{n\cdot q,\frac{q_\perp^2}{n\cdot q},q_\perp}\sim Q(1,\lambda^2,\lambda),
\ee
such that the final electron direction is distinguishable from the photon direction.

\begin{figure}[!h]
\centering
\includegraphics[width=.45\textwidth]{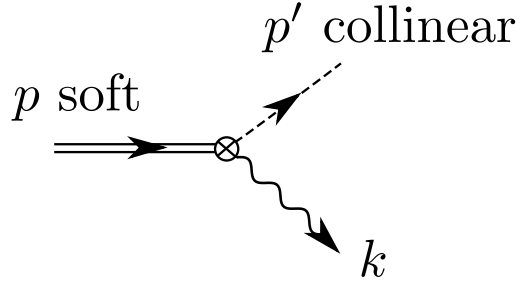}
\caption{Kinematics in laboratory frame}
\label{fig:94fig1}
\end{figure}

The probability of photon emission, averaged over the initial electron spin, is
\be\label{eq:94eq1}
d\mathcal{P}[e\to e\gamma]=\oneov{2}\sum_{\rm{pol, spins}}\frac{\dbar P}{2P_0}\frac{\dbar P'}{2P_0'}\frac{\dbar q}{2|\vec q|}|i\cM(e\to e\gamma)|^2 ,
\ee
where $\dbar P=d^3P/(2\pi)^3$ (see for example Eq.(A13) of \cite{Ilderton:2012qe}).  The $e\to e\gamma$ matrix element involves the current we studied in the preceding section:
\be\label{eq:94eq2}
i\cM(e\to e\gamma)=-ie\int d^4x\bra{e(P')\gamma(q)}J_{PP'}^\mu(x) \epsilon_\mu e^{iqx}\ket{e(P)}.
\ee
Summing over photon polarizations $\sum_{\rm{pol.}}\epsilon^\nu\epsilon^{*\mu}=-g^{\mu\nu}$,
\begin{align}
\frac{d\mathcal{P}}{\dbar q}&=-\frac{e^2}{2}\sum_{\rm{spins}}\frac{\dbar P}{2P_0}\frac{\dbar P'}{2P_0'}\int \! d^4x d^4y\, e^{-iq(y-x)}  
\bra{e(P')}J_{PP'}^\mu(x)\ket{e(P)}\bra{e(P)}J_{PP'}^{\dag\mu}(y)\ket{e(P')}. \label{eq:94eq3}
\end{align}
In order to define an inclusive observable, we generalize the outgoing electron to allow for arbitrary final states with total momentum $P'$ and spin $\sigma'$
\be
\sum_{\sigma'}\frac{\dbar P'}{2 P_0'}\to\sum_X\text{ for X collinear}.
\ee
We can then sum over the final states
\begin{align}
\frac{d\mathcal{P}}{\dbar q}&=-\frac{e^2}{2}\sum_\sigma\sum_X\frac{\dbar P}{2P_0}\int d^4xd^4y e^{iq(x-y)}\bra{e(P)}J_{PP'}^{\mu\dag}(y)\ket{X}\bra{X}J_{PP'}^\mu(x)\ket{e(P)}\nn\\
&=-\frac{e^2}{2}\sum_\sigma\frac{\dbar P}{2P_0}\int d^4xd^4y \,e^{iq(x-y)}\bra{e(P,\sigma)}J_{PP'}^{\mu\dag}(y)J_{PP'}^\mu(X)\ket{e(P,\sigma)}.\label{eq:94eq4}
\end{align}

Next we go through the detailed steps of matching the current at leading order.  The operator that creates a collinear field with total $n$-collinear momentum $n\cdot P'+K_{P'}$ is
\be\label{eq:94eq5}
\bar\chi_{\bn,\omega}S_\bn Y_{P'}^\dg e^{(i/2)\omega\bn\cdot x+iP_\perp\cdot x}\quad \bar\chi_{\bn,\omega}=[\bar\chi_\bn W_n\delta(\omega-n\cdot \hat P^\dg)].
\ee
The $\omega$ here is a variable that will summed over and fixed during matching (not to be confused with $\omega_{\rm laser}$). The operator $\hat P$ is a `label' operator projecting out the large component of the momentum, and the $\dagger$ indicates that it acts to the left inside the bracket.  The corresponding operator in the soft sector,
\be\label{eq:94eq6}
e^{-(i/2)\omega\bn\cdot x}Y_P\psi_{s,\omega},\quad \psi_s=[\delta(\omega- \hat P)\psi_s],
\ee
creates a soft field that comes with the large momentum component $K_s=-\frac{(eA_{cl})^2}{\bn\cdot P}$ due to the classical potential. The effective current is then
\begin{align}
J_{PP'}^\mu(x)&=[\bar\psi_{P'}\gamma^\mu\psi_P](x) \to
\bar \chi_{\bn,\omega_2}S_\bn Y_{P'}^\dg \gamma^\mu Y_P\psi_{\omega_1}e^{(i/2)\omega_2\bn\cdot x+iP_\perp\cdot x}e^{-(i/2)\omega_1\bn\cdot x},
\end{align}
and the variables $\omega_1,\omega_2$ must be summed over.  This introduces the convolution that allows factorization and running later.  Explicitly, with the sum, we have
\begin{align}\label{eq:EFTcurrent}
J_{\rm{eff}}^\mu(x)&=\sum_{\omega_1,\omega_2}C(\omega_1,\omega_2;\mu)\Bigg(\bar\chi_{\bn,\omega_2}S_\bn Y_{P'}^\dg\paren{\frac{\slashed{n}}{2}\bn^\mu+\gamma_\perp^\mu}Y_P
\psi_{\omega_1}e^{\frac{i}{2}(\omega_2-\omega_1)\bn\cdot x+iP_\perp x}+\rm{h.c.}\Bigg).
\end{align}
Inserting this current in \eq{94eq4}
\begin{align}
\frac{d\mathcal{P}}{\dbar q}&=-\frac{e^2}{2}\sum_\sigma\frac{\dbar P}{2P_0}\int d^4x d^4y\,e^{iq(x-y)}
\sum_{\omega_1,\omega_2,\omega_1',\omega_2'}C(\omega_1,\omega_2;\mu)C^*(\omega_1',\omega_2';\mu)\nn\\
&\times \bra{e(P,\sigma)}\left[\bar \psi_{\omega_1'}Y_P^\dg\paren{\frac{\slashed{n}}{2}\bn^\mu+\gamma_\perp^\mu}Y_{P'}S_\bn\chi_{\bn,\omega_2'}(y)\right]\, e^{\frac{i}{2}(\omega_1'-\omega_2')\bn\cdot y-iP_\perp'y}\nn\\
&\times\left[\bar\chi_{\bn,\omega_2}S_\bn^\dg Y_{P'}^\dg\paren{\frac{\slashed{n}}{2}\bn^\mu+\gamma_\perp^\mu}Y_P\psi_{\omega_1}(x)\right]\,e^{\frac{i}{2}(\omega_2-\omega_1)\bn\cdot x+iP_\perp'x}\ket{e(P,\sigma)}.\label{eq:94eq8}
\end{align}
In the contraction of Lorentz indices, only the $\gamma_\perp^\mu\gamma_\perp^\mu$ component survives
\begin{align}
\frac{d\mathcal{P}}{\dbar q}&=-e^2\sum_{\omega_1,\omega_2,\omega_1',\omega_2'}C(\omega_1,\omega_2;\mu)C^*(\omega_1',\omega_2';\mu) 
\nn\\
&\times\int d^4x d^4y e^{iq(x-y)}e^{-\frac{i}{2}(\omega_2'-\omega_1')\bn\cdot y-iP_\perp'y}e^{\frac{i}{2}(\omega_2-\omega_1)\bn\cdot x+iP_\perp'\cdot x}
\nn\\ 
&\times\oneov{2}\sum_{P,\sigma}\bra{e(P,\sigma)}\overline T\left[\bar\psi_{\omega_1'}Y_P^\dg \gamma_\perp^\mu Y_{P'}S_\bn \chi_{\bn,W_2'}(y)\right]\, T\left[\bar\chi_{\bn,\omega_2}S_\bn^\dg Y_{P'}^\dg\gamma_\perp^\mu Y_P\psi_{\omega_1}(x)\right]\ket{e(P,\sigma)}.\label{eq:94eq9}
\end{align}
Here the $T$ is the time-ordering operator and $\overline T$ is the anti-time ordering operator.  The $x,y$ integrals at scale $\lambda^0$ enforce label momentum conservation, giving $\delta$ functions, $\delta(k_++\omega_2-\omega_1)\delta(k_++\omega_2'-\omega_1')$.
Integrating over $x$ at the scale $\lambda^{-1}$ allows separating the emitted photon from the final state electron (plus radiation).  Thus $P_\perp'$ is fixed by the outgoing photon $q_\perp$ though we did not introduce an extra convolution variable for it.  Now,
\begin{align}
\frac{d\mathcal{P}}{\dbar q}&=-e^2\sum_{\omega_1,\omega_1'}C(\omega_1,\omega_1-q_+;\mu)C^*(\omega_1',\omega_1'-q_+;\mu)
\int d^4xd^4y\,e^{\frac{i}{2}q_-n\cdot(x-y)}\oneov{2}\sum_{P,\sigma} \nn\\
&\times\bra{e(P,\sigma)}\overline{T}\big[\bar \psi_{\omega_1'}Y_P^\dg \gamma_\perp^\mu Y_{P'}S_\bn^\dg\chi_{\bn,\omega_1'-q_+}(y)\big] T\big[\bar \chi_{\bn,\omega_1-q_+}S_\bn Y_{P'}^\dg \gamma_\perp^\mu Y_P\psi_{\omega_1}(x)\big]\ket{e(P,\sigma)}.  \label{eq:94eq10}
\end{align}
Using a Fierz identity standard in SCET, we move the collinear fields into one Dirac structure and soft into the other
\begin{align}
\frac{d\mathcal{P}}{\dbar q}&=-\frac{e^2}{2}\sum_{\omega_1,\omega_1'}C(\omega_1,\omega_1-q_+;\mu)C^*(\omega_1',\omega_1'-q_+;\mu)\int d^4xd^4y \,e^{\frac{i}{2}q_-n\cdot(x-y)}\nn\\
&\times\oneov{2}\sum_{P,\sigma}\bra{e(P,\sigma)}\bar\psi_{\omega_1}(x)S_\bn^\dg\frac{\slashed{\bn}}{2}S_\bn \psi_{\omega_1}(x)\ket{e(P,\sigma)}\nn\\
&\times \bra{0}\frac{\slashed{n}}{2}\chi_{\bn,\omega_1'-q_+}(y)\bar\chi_{\bn,\omega_1-q_+}(x)\ket{0}
\bra{0} \overline{T}[Y_P^\dg Y_{P'}(y)]\, T[Y_{P'}^\dg Y_P(x)]\ket{0}.\label{eq:94eq11}
\end{align}
The ultrasoft Wilson lines depend only on one light cone coordinate $\bn\cdot x$, and we define
\be\label{eq:94eq12}
\bra{0}\bar T[Y_P^\dg Y_{P'}(y)]\,T[Y_{P'}^\dg Y_P(x)]\ket{0}=\int d\ell \,e^{(i/2)\ell\bn\cdot(x-y)}\mathcal{Y}_u(\ell;\mu).
\ee
The final state collinear function is also written in momentum space
\be\label{eq:94eq13}
\bra{0}\frac{\slashed{n}}{2}\chi_{\bn,W_1'-k_+}(y)\bar \chi_{\bn,W_1-k_+}(x)\ket{0}=\int\frac{d^4r}{(2\pi)^4}e^{ir\cdot(x-y)}\mathcal{J}(r^\nu;\mu).
\ee
However, we will only measure the $z$-momentum along the beam-axis ($n$-direction) so that in effect $\mathcal{J}(r^\nu;\mu)\to \mathcal{J}(n\cdot r;\mu)$ a function of $r_+$ only. Then
\begin{align}
\bra{0}\frac{\slashed{n}}{2}\chi_{\bn,\omega_1'-k_+}(y)\bar\chi_{\bn,\omega_1-k_+}(x)\ket{0}&=\delta_\perp^2(x-y)\delta(n\cdot(x-y))\nn\\
&\times\int \frac{dr_+}{2(2\pi)}e^{\frac{i}{2}r_+\bn\cdot(x-y)}J(r_+;\mu),\label{eq:94eq14}
\end{align}
where label conservation is used to set $\omega_1'-q_+=\omega_1-q_+$. For the remaining soft matrix element, label conservation implies $\omega_1=\omega_1'$.  We insert a $\delta$-function to make this explicit
\begin{align}
\bra{e(P,\sigma)}\bar\psi_{\omega_1'}&S_\bn(y)\frac{\slashed{n}}{2}S_\bn^\dg\psi_{\omega_1}(x)\ket{e(P,\sigma)}\nn\\
&=\delta_{\omega_1,\omega_1'}\bra{e(P,\sigma)}\bar\psi S_\bn(y)\frac{\slashed{\bn}}{2}\delta_{\hat P_+,2\omega}S_\bn^\dg \psi(x)\ket{e(P,\sigma)},\label{eq:94eq15}
\end{align}
with $\hat P_+=n\cdot \hat P+n\cdot \hat P^\dg$ the sum of label operators. Inserting \eqss{94eq12}{94eq14}{94eq15} into \eq{94eq11} we get
\begin{align}
\frac{d\mathcal{P}}{\dbar q}&=
-\frac{e^2}{2}\sum_{\omega_1}|C(\omega_1,\omega_1-q_+;\mu)|^2\int d^4xd^4y \,e^{\frac{i}{2}q_-(x-y)}
\nn\\ &\times 
\oneov{2}\sum_{P,\sigma}\bra{e(P,\sigma)}\bar\psi S_\bn(y)\frac{\slashed{\bn}}{2}\delta_{\hat P_+,2\omega}S_\bn^\dg \psi(x)\ket{e(P,\sigma)} 
  \int d\ell \,e^{\frac{i}{2}\ell\bn\cdot(x-y)}\mathcal{Y}_u(\ell;\mu) 
\nn\\\label{eq:94eq16} &\times
\delta_\perp^2(x-y)\delta(n\cdot(x-y))\int\frac{dr_+}{4\pi}e^{\frac{i}{2}r_+\bn\cdot(x-y)}\mathcal{J}(r_+;\mu) \\
&=-\frac{e^2}{2}\sum_\omega|C(\omega,\omega-q_+;\mu)|^2
\int\frac{d\bn\cdot x}{2}\frac{d\bn\cdot y}{2}
\int\frac{dr_+}{4\pi}\int d\ell e^{\frac{i}{2}(r_++\ell)\bn\cdot (x-y)}  
\nn\\&\times
\mathcal{J}(r_+;\mu)\mathcal{Y}_u(\ell;\mu) 
\oneov{2}\sum_{P,\sigma}\bra{e(P,\sigma)}\bar\psi S_\bn(\tilde y)\frac{\slashed{\bn}}{2}\delta_{\hat P,2W}S_\bn^\dg \psi(\tilde x)\ket{e(P,\sigma)},
\label{eq:94eq17}
\end{align}
where $\tilde x^\mu=(0,\bn\cdot x,0)$, $\tilde y^\mu=(0,\bn\cdot y,0)$.

To proceed we must consider the final state collinear momentum known to order $\lambda^0$, to fix $\omega-q_+$.  Recall in contrast, that in DIS or Drell-Yan processes considered in SCET, the initial state is known to this order.  We set $\omega-q_+\equiv P_+'$, which reduces the sum over $\omega$.
Switching to relative and average coordinates
\be\label{eq:94eq18}
X=\frac{x+y}{2}\quad z=x-y\quad d(\bn\cdot x) d(\bn\cdot y)=d(\bn \cdot X) d(\bn\cdot z),
\ee
we now have
\begin{align}
\frac{d\mathcal{P}}{\dbar q}&=-\frac{e^2}{2}H(P_+',q_+;\mu)\int \frac{d\bn\cdot X d\bn\cdot z}{4}\int \frac{dr_+}{4\pi}\int d\ell \nn\\
&\times e^{(\frac{i}{2}(r_++\ell)\bn\cdot z}\mathcal{J}(r^+;\mu)\mathcal{Y}_u(\ell;\mu) (\ell;\mu)\nn\\
&\times\oneov{2}\sum_{P,\sigma}\bra{e(P,\sigma)}\bar\psi S_\bn\big(\bn\cdot\big(X-\textstyle{\frac{z}{2}})\big)\frac{\slashed{\bn}}{2}\delta_{\hat P,2W}S_\bn^\dg\psi\big(\bn\cdot\big(X+\textstyle{\frac{z}{2}})\big)\ket{e(P,\sigma)},\label{eq:94eq19}
\end{align}
where $H(P_+',q_+;\mu)=|C(P_+'+q_+',P_+';\mu)|^2$ is the hard function.
Working on the soft matrix element
\begin{align}
\int d\bn\cdot z &\, e^{\frac{i}{2}p\bn\cdot r}\bra{e(P,\sigma)}\bar\psi S_\bn\paren{\bn\cdot X-\textstyle{\frac{\bn\cdot z}{2}}}\frac{\slashed{\bn}}{2}\delta_{\hat P,2W}S_\bn^\dg\psi\paren{\bn\cdot X+\textstyle{\frac{\bn\cdot z}{2}}}\ket{e(P,\sigma)}    \nn \\
&=\bra{e(P,\sigma)}\bar\psi S_\bn(\bn\cdot X)\frac{\slashed{\bn}}{2}\delta_{\hat P,2W}\delta(p-in\cdot \pd)S_\bn^\dg\psi(\bn\cdot X)\ket{e(P,\sigma)}\label{eq:94eq21}\\
&\equiv f_{P,\sigma}(\bn\cdot X;\mu),
\end{align}
which is the electron distribution function in the initial state. Thus we arrive at the final factorized probability
\begin{align}
\frac{d\mathcal{P}}{\dbar q}&=-\frac{e^2}{2}H(P_+',q_+;\mu)\int\frac{dr}{4\pi}\int\frac{d\ell}{2}
 \mathcal{J}(r;\mu)\mathcal{Y}_u(\ell;\mu)
 \oneov{2}\sum_{\bn\cdot X}\oneov{2}\sum_{P,\sigma}f_{P,\sigma}(\bn\cdot X,q+\ell;\mu)
\nn\\&
=-\alpha H(P_+',k_+;\mu)\int\frac{dr}{2}\int\frac{d\ell}{2}\mathcal{J}(r;\mu)\mathcal{Y}_u(\ell;\mu)
\oneov{2}\sum_{\bn\cdot X}\oneov{2}\sum_{P,\sigma}f_{P,\sigma}(\bn\cdot X,r+\ell;\mu).
\end{align}

This factorization realizes the separation of quantum and classical physics latent in the semiclassical calculations.  The hard function $H$ characterizes the large coherent momentum transfer from the background field and the transition to a high-energy state.  This is clearly a nonperturbative process and can only depend on the local value of the potential -- crudely speaking the number of low-energy photons available to the electron at its position.  The initial state composed of relatively low-energy $P\sim m_e$ electrons is described by the electron distribution function $f_{P,\sigma}$.  Because electrons of this momentum scale remain more strongly-coupled to the plasma, this function encodes the nonperturbative dynamics of the plasma by its dependence on position the average position $X$.  The electron distribution function is universal in that we anticipate this same operator appears in all processes for which the initial state is a low-energy electron.  On the other hand, it is natural that the highly-relativistic, accelerated electron decouples from the plasma.  Thus, the final state collinear function describes the high-energy electron after the momentum transfer at the vertex.  In general, such a high-momentum final state need not be exactly on mass-shell after a large transfer of (light-like) momentum from the potential and emission of a high-energy photon.  The final-state electron would then need to emit one or more photons before being detected, and these fragmentation dynamics are incorporated in $\mathcal{J}$.  At longest wavelength, the electron may absorb more or less ultra-low energy (classical) radiation than the observed momentum transfer implies.  As a result, there are generally classical radiation corrections to the process, which are included in the ultra-soft function $\mathcal{Y}_u$.
 
We may now compute corrections to each of these functions separately, and use the renormalization group to sum the logarithms arising from the separation of physics scales.

\subsection{One-Loop Renormalization to the Soft and Collinear Functions}
\begin{figure}
\centering
\includegraphics[width=.6\textwidth]{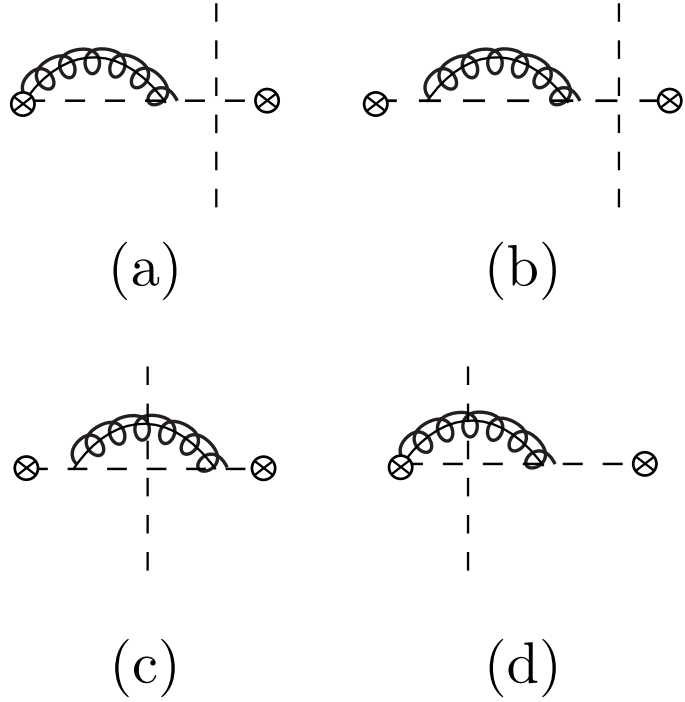}
\caption{One-loop correction to collinear function}
\label{fig:200fig9}
\end{figure}

The one loop corrections to the collinear function in the factorized cross section are shown in \fig{200fig9}.  The first two, \fig{200fig9}(a) and (b), were calculated during the matching procedure. \fig{200fig9}(c) and \fig{200fig9}(d) are the same as the SCET real jet functions.

Combining the results of (a), (b), (c), (d) and their mirror diagrams, we have the total collinear function to one loop,
\begin{align}
\mathcal{J}^{\text{one-loop}}&=\frac{\alpha}{2\pi}\Bigg\{\del(r)\Bigg[\frac{2}{\eps^2}+\ov{\eps}\paren{\frac{3}{2}+2\ln\frac{\mu^2}{-q^2}}+\frac{3}{2}\ln\frac{\mu^2}{-q^2}+\ln^2\frac{\mu^2}{-q^2}+\frac{7}{2}-\frac{\pi^2}{2}\Bigg]\nn\\
&-\paren{\ov{r}}_+ \paren{\frac{2}{\eps}+2\ln\frac{\mu^2}{-q^2}+\frac{3}{2}}+2\paren{\frac{\ln r}{r}}_+\Bigg\}.
\end{align}
The counter term is obtained from the $\epsilon$ poles in the one-loop result,
\begin{align}
Z_{\mathcal{J}}&=1-\mathcal{J}_{\eps\text{ dependent terms}}^{\text{one-loop}}\nn\\
&=1-\frac{\alpha}{2\pi}\Bigg\{\del(r)\Bigg[\frac{2}{\eps^2}+\ov{\eps}\paren{\frac{3}{2}+2\ln\frac{\mu^2}{-q^2}}\Bigg]-\frac{2}{\eps}\paren{\ov{r}}_+\Bigg\},
\end{align}
and the anomalous dimension its differential with respect to the arbitrary scale $\mu$,
\be
\gamma_\mathcal{J}=\frac{dZ_{\cal J}}{d\ln\mu}=\del(r)\paren{-\frac{\alpha}{2\pi}}\paren{-2\ln\frac{\mu^2}{-q^2}}\,.
\ee
To sum the large logarithms arising from the ratio of the matching scale to the `natural scale' of the collinear function, which is of order the momentum transfer $-q^2$, we solve the collinear RG equation,
\be
\frac{d\mathcal{J}(\mu)}{d\ln\mu}=\gamma_{\cal J} \mathcal{J}(\mu),
\ee
and obtain
\be
\mathcal{J}(\mu,-q^2)=\exp\left(\frac{\alpha}{\pi}\ln^2\frac{\mu^2}{-q^2}\right)\mathcal{J}(-q^2),
\ee
with $\mathcal{J}(-q^2)=1+\cO(\alpha)$.

\begin{figure}
\centering
\includegraphics[width=.6\textwidth]{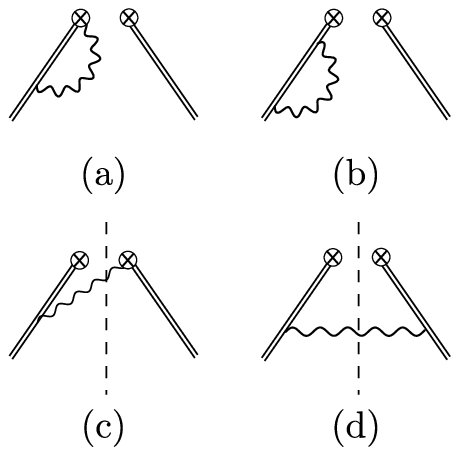}
\label{fig:200fig10}
\end{figure}

One-loop corrections to the electron distribution function are shown in \fig{200fig10}.  \fig{200fig10}(a) and (b) have been calculated during the matching procedure. \fig{200fig10}(c) and (d) give
\begin{align}
f_{\rm{Real}}^{\text{1-loop,c}}&=\paren{-\frac{\alpha}{2\pi}}\ov{p_+}\Bigg\{\del\!\paren{\frac{\ell_+}{p_+}}\left[\ov{\eps^2}-\ov{\eps}\ln\frac{m_e^2}{\mu^2}-\frac{1}{\eps}-\frac{1}{2}\ln\frac{\mu^2}{m_e^2}\ln\frac{m_\gamma^2}{m_e^2}\right]
\nn \\ &
+\paren{\frac{p_+}{\ell_+}}_+\left[-\ov{\eps}-1-\ln\frac{\mu^2}{m_e^2} \right] +\left(\frac{\ln(\ell_+/p_+)}{\ell_+/p_+}\right)_+\Bigg\} \\
f_{\rm{Real}}^{\text{1-loop,d}}&=\paren{-\frac{\alpha}{4\pi}}\ov{p_+}\Bigg\{\ov{\eps}\left[1-\del\paren{\frac{\ell_+^2}{p_+^2}}\right]+\paren{\frac{p_+^2}{\ell_+^2}}_+-1+\ln\frac{\mu^2}{m^2}\Bigg\},
\end{align}
recalling that the momentum on the soft line is $p\to P_s\sim\lambda^2$.
Combining (a), (b), (c) and (d) with their mirror images, we have the soft function one-loop correction
\be
f^{\text{one-loop}}=2J_{EFT}^{(a)}-\ov{2}J_{EFT}^{(c)}+2f_{\rm{Real}}^{\text{one-loop c}}+f_{\rm{Real}}^{\text{one-loop d}},
\ee
and we define the counter term of the soft function from the divergent pieces,
\begin{align}
Z_{\rm{edf}}&=1-\frac{\alpha}{2\pi}\Bigg\{-\frac{2}{\eps^2}+\frac{3}{4\eps}+\frac{2}{\eps}\ln\frac{m_e^2}{\mu^2}+\frac{2}{\eps}\paren{\frac{\ell_+}{p_+}}_+\Bigg\}.
\end{align}
The anomalous dimension is
\be
\gam_{\rm{edf}}=-\frac{\alpha}{\pi}\Bigg[1-\ln\frac{\mu^2}{m^2}-\ov{2}\del\paren{\frac{\ell_+^2}{p_+^2}}\Bigg].
\ee
We solve the RG equation for the soft function,
\be
\mu\frac{df(\mu)}{d\mu}=\gam_{\rm soft} f(\mu),
\ee
and obtain
\be
f(\mu,\ell)=\exp\left(\frac{\alpha}{\pi}\bigg(\ln^2\frac{\mu^2}{m_\gamma^2}-\frac{1}{4}\ln\frac{\mu^2}{m_\gamma^2}\bigg)\right)f(m_\gamma),
\ee
where $\mathcal{Y}_u(m_\gamma)=1+\cO(\alpha)$.

\section{Chapter Summary}

In this chapter, I have reviewed the state of the art in predicting quantum emission processes in high-intensity laser experiments and given a brief introduction to the current experimental efforts.  We saw that a more systematic theoretical framework is required, improving on the preceding semiclassical approaches.  As the laser-plasma system involves a wide separation of scales, up to 9 orders of magnitude between the high momentum of accelerated electrons and the laser frequency, effective field theories are natural tools.  Taking into account the experimental conditions and potential observables, I have developed an effective field theory for high-energy and low-energy electrons in the presence of a high-intensity laser potential, and shown that it can be used to sum large logarithmic corrections to photon emission by an accelerated electron.

So far I have only studied one process, considered one of the most important for the first experiments.  There remain many other pieces to study: pair emission by an accelerated electron, the kinematic end-point of photon and pair emission, the effect of restricted phase space on observables, electron-ion scattering in the background.  These processes might be studied already with the framework developed here.  An important frontier beyond this theory is its adaptation to improve numerical plasma simulations.  Current simulations use a simple model of QED emissions based on semiclassical results, and some simulations suggest that experiments can be arranged such that the QED processes have a large effect on the plasma dynamics.  Simulating these conditions requires having reliable and systematically-improvable local rates for QED processes that can be implemented in real-time.  Achieving this requires at least matching this theory, at the simulation resolution length-scale, onto an effective theory that represents the long-wavelength physics as implemented in the plasma simulation codes.

Beyond the work to be done within the high-intensity laser-plasma domain, I anticipate this theory may aid the development of other effective theories for other quantum processes in classical potentials.  Semiclassical methods have been used to predict nonperturbative effects in strong classical fields, and the results considered phenomenologically very important both within QED, in the search for spontaneous pair production, and outside QED, in gravitational potentials and even heavy-ion collisions.  However, quantum corrections to these processes have not been systematically calculated or summed before now, because they are typically considered ``small'', usually due to the size of the coupling.  This theory may provide inspiration as to how such calculations may be accomplished. 

%% file: chapter_4.tex
\chapter{Summary and Future Prospectives on Effective Field Theory \label{partIV}}

\section{Overview} \label{sect:IV.1}
In this chapter, I first review the effective field theories developed in this dissertation and the highlights of their applications.  Then I discuss the limits of effective field theory methods with specific examples demonstrating circumstances under which they become less useful.  Finally, I provide a few opinions on the future development and applications of effective field theory methods.

\section{Summary of this work}

In this thesis, I first introduced the general concepts and principles of effective field theory methods.  Typically, as long as the physical scales are well-separated in a process, we can establish an effective field theory to organize the observables into a controlled perturbative series, systematically expanding in powers of small parameters determined by hierarchies of the physical scales.  In case we know the full theory, we can take the top-down approach, and I gave a toy example reducing $\phi^4$ theory to effective theories in the non-relativistic and ultra-relativistic momentum limits.  This example explicitly showed how the effective theory inherits the infrared physics of the underlying theory through matching coefficients, reorganizes the perturbative series of amplitudes  by resumming contributions that are leading order in the power counting, and relates the new perturbative expansions to experimental observables.  In case we do not know the full theory, we can take the bottom-up approach, in which one systematically accounts for all interactions allowed by the symmetries and organizes them into a perturbative expansion based on a power-counting scheme that relates the experimental energy scale to the heavy (breakdown) scale in the system.

I applied these principles and procedures to three EFTs, among which two (Soft Collinear Effective Theory in Chapter \ref{partI} and Electron Laser Effective Theory in Chapter \ref{sec:III}) are top-down type EFTs for ultra-relativistic particles, whose operators and coefficients are obtained by matching from the QCD and QED lagrangians respectively.  The third effective theory, X-EFT in Chapter \ref{sec:II} is a bottom-up type EFT for non-relativistic particles, because it describes hadron scattering in the low-energy region of QCD, which is highly nonperturbative.  For this EFT, we must write down the most general set of operators according to the low-energy symmetries of QCD and strive to obtain the matching coefficients from lattice calculations or experimental data.  This procedure is more akin to building an effective theory for beyond Standard Model physics.

I briefly review the results in each of these theories.

\subsection{Soft Collinear Effective Theory}
Collider and jet physics in vacuum and in heavy-ion collisions depend on the hierarchically-separated scales: $Q$ the hard collision energy, $p_T$ the typical transverse momentum of collinear modes, the soft radiation with momentum $k_s$, and the hadronization scale $\Lambda_{\rm QCD}$. SCET facilitates the factorization of physical observables depending on these many scales into single-scale functions, the resummation of large logarithms of ratios of the scales via renormalization group evolution of these functions, and the proof of universality of non-perturbative effects on these observables in many cases.  I applied SCET in the following two projects.

\subsubsection{Rapidity divergences in Semi-inclusive DIS and DY processes in the endpoint region} 
By developing the effective theory description of the semi-inclusive or elastic limits in hard scattering processes, I provided a new tool to systematically study the transition from short-distance to long-distance interactions in QCD.  This provides a novel angle to understand how quarks and gluons are confined in nucleons.  

Perturbative calculations receive logarithmic enhancements near the elastic limits, which have long remained unexplained.  Using SCET, I identified the source of the logarithmic enhancement as the rapidity scale arising from approaching the elastic limit showing that the non-perturbative effects are associated with the rapidity scale and advanced the resummation of logarithmic corrections at next to leading order

Taking into account the non-perturbative effects associated with the endpoint rapidity scale,  I redefined the Parton Distribution Function (PDF) for DIS and extended it to DY which allowed identification of the origin of the partonic threshold singularity with the single parton endpoint singularity, and discovered that in the elastic limit the two colliding protons interfere with each other via the soft radiation. 

\subsubsection{Universality of soft radiation in electron-positron, electron-proton and proton-proton collisions}

Separating QCD backgrounds in new particle searches requires high-accuracy DIS and DY differential cross sections.  These cross sections are expressed in terms of infrared-safe QCD observables, such as event shape variables that characterize QCD events and contain only a single scale and single large logarithm.
I have improved the DIS and DY cross section accuracy to next-to-next-to-next-to-leading log (N$^3$LL) and  order, same order as $e^+e^-$, by proving to order $\alpha_s^2$ and $\alpha^3$ the equality of matrix elements describing the low energy, wide angle radiation associated with the collinear particles.  This soft radiation is determined by correlations of momentum flow with each other and with Wilson lines that describe the coupling of soft, wide-angle radiation to jets and neutralize the colors of jets. By establishing the universality of this soft radiation, I provided the last ingredient needed for resummation of DIS and DY event shapes.  Resumming this single log then leads to more accurate predictions and more precise extractions of the strong coupling $\alpha_s$ and leading nonperturbative moment on event shapes.  

\subsection{Heavy hadron chiral perturbation theory and $X$ effective theory}

For a new window onto the nuclear force, which is traditionally studied via nucleon-nucleon scattering, I investigated the scattering of heavy mesons such as $D$ and its excited state $D^*$.  By including the $X(3872)$ as a molecular state of $D$ and $D^*$, the X effective theory (XEFT) for $DD^*$ scattering has a similar structure to $NN$ scattering with the deuteron included.  
In the XEFT, pions are non-relativistic pions and the fine-tuning of the $D$-$D^*$ mass splitting being nearly equal to the pion mass is systematically accounted for.  This makes it an excellent tool to study another pion exchange process: $D-\pi$ scattering near $D^*$ threshold, which is analogous to nucleon-pion scattering, and since pion exchange provides the long range component of the nuclear force, is a good theoretical laboratory for understanding the non-perturbative structure in hadron interactions. 

I advanced the XEFT by expanding the HH$\chi$PT Lagrangian to complete next-to-leading order including isospin breaking effects and resumming the $D-\pi$ interaction near $D^*$ threshold, which is also of great use for improving the extrapolation of the $D$-meson mass spectrum on lattice.  This improved XEFT can be used study other heavy meson molecules, known as exotic, $X,Y$ and $Z$ states, and strengthen the connection to lattice calculations in particular by varying the pion mass to improve extrapolations of heavy meson properties to the physical point. 

\subsection{Electron Laser Effective Theory}

I developed the electron laser effective theory to predict the high-energy photon emission and electron-positron pair production rates from highly-relativistic electrons traveling in a strong background laser field.  Defining a small expansion parameter $\lambda$, the hierarchy of relevant physical quantities is (from highest to lowest): the large electron momentum component collinear to the laser direction $Q$, the classical background potential $|eA_{\rm cl}^\mu|$ which is the same order the electron transverse momentum $|eA_{\rm cl}^\mu|\sim p_\perp\sim\lambda Q$, the electron mass $m_e\sim \lambda^2 Q$ also the scale to consider the photon emission a quantized process and the typical momentum of the un-accelerated electrons, and the laser frequency scale $\omega_{\rm laser}\sim \lambda^5 Q$, which is the infrared physics scale.  After projecting and integrating out the largest scale $Q$ in this theory, I restore the gauge and Lorentz symmetry in the manner similar to SCET.  

My first result from this effective theory is to factorize physical observables into different functions that each describe physics at one scale, e.g. the electron distribution function encoding the initial state classical plasma dynamics before the high-energy emission event.  

Before the introduction of this effective theory, there were no effective field theory methods applied to strong field QED processes.  Previous efforts focussed on solving the electron wave function in the semiclassical approximation, which separates the electron mass scale from the laser frequency scale and sums the interaction with the laser field into a leading order effect.   Calculating processes with the semiclassical wave functions can be considered as a fixed-scale fixed-order theoretical calculation and should be equivalent to summing the large logarithms in the separation of scales between $m_e$ and $\omega_{\rm laser}$.  The relativistic electron momentum scale was ignored in earlier work and therefore no high-energy photon $k\gtrsim m_e$ or electron-positron pair observables can be calculated in terms of experimental observables.  Moreover, the quantum fluctuations of the non-perturbative functions, which are crucial to make experimental predictions for these processes, were not taken into account.

These limitations of the previous methods are addressed by the effective theory constructed in Chapter \ref{sec:III}. After constructing and proving the factorization theorem for these physical processes, I first separate the universal non-perturbative effects: the classical background function and electron distribution function.  The classical background can be considered as a constant factor not changing with energy, because it contains the large-amplitude classical radiation modes, whose quantum fluctuations are suppressed.  The electron distribution function incorporates the quantum fluctuations of soft photon modes and evolves with the energy scale of the high-energy emission process that probes the electron.  With the help of these separate universal non-perturbative functions, I predict the distribution of wide-angle photon emission.

\subsection{Summary of the summary}
In summary, this thesis has explored two effective field theories for QCD, SCET and X-EFT.  Both offer insight to the structure of nucleons and their interactions.  I used $\SCETa$ to improve predictions of QCD observables in hadron colliders to N$^3$LL accuracy, which is essential for new physics searches, and I used $\SCETb$ to study semi-inclusive nucleon scattering processes by separating collinear and soft radiation scales and demonstrating the resummation of logarithmic enhancements in the transition to the elastic scattering limit.  I used X-EFT to study low-energy $p\ll \LQCD$ nuclear interactions between the $D$ and $D^*$ hadronic states, pions and the X(3872) interpreted as a hadronic molecule by summing the enhancements in $D\pi$ scattering near the $D^*$ threshold and $DD^*$ scattering near the X(3872) threshold.  The third EFT I study is a newly-developed electron laser effective theory, which is inspired by SCET and describes relativistic electrons traveling in a strong laser field and radiating high-energy photons.  I used this EFT to make measurable predictions for high-intensity laser experiments.

\section{Boundaries and Frontiers in Effective Field Theory Methods}

Reflecting on the three EFTs in this thesis, I will now discuss some of the current limits on the application of effective theory methods.  These limits do not appear to be fundamental to the EFT approach that was outlined in the introduction, but rather simply are not yet thoroughly addressed in an EFT framework.  

{\bf I.}  The number of operators and associated Wilson/matching coefficients increases drastically when the EFT expansion is taken to one higher order.  This arises together with the symmetries that EFTs must preserve.  For example, in order to preserve Lorentz symmetry at next-to-leading order in highly-relativistic or non-relativistic particle EFTs, we must include all repararmeterization invariant operators with their coefficients to the same order.  The number of such operators generically increases factorially.

{\bf II.}  Observables must be infrared safe.  This issue arises with the separation of scales by EFTs.  When the hierarchy of scales is large enough, infrared scales (such as the light quark masses in QCD) approach zero, bringing in infrared divergences.  Most EFT power-counting schemes are dedicated to resolving this issue.  For example, resumming the enhancement in $DD^*$ scattering near the X(3872) threshold or introducing a new scale in the endpoint region of proton scattering to resum large logarithms.  However, in many cases, such as LHC processes involving many physical scales and large strong-interaction coupling constant, infrared safety of the observables is not guaranteed to all orders in the power counting.

{\bf III.}  Parameters in EFTs may be fine-tuned.  This is a long-standing question in the development of effective theory methods.  Fine-tuning can arise from many sources, and here I give an example how it enters via unknown physics beyond the underlying theory.  There is a well-known EFT argument why the sky is blue: photons scattering from neutral atoms in their ground state have much lower energy than excitation energy of the atoms $\Delta E_{\rm atom}$, the momentum corresponding to the inverse size of the atom $a_{\rm Bohr}^{-1}$ and of course the mass of the atom $M_{\rm atom}$.  This hierarchy is summarized
\begin{equation}
E_{\gamma}\ll \Delta E_{\rm atom}\ll a_{\rm Bohr}^{-1}\ll M_{\rm atom}
\end{equation}
Defining the neutral atom field as $\phi$, its 4-velocity $v^\mu$, and the $F^{\mu\nu}$ the photon field strenght tensor, we write the gauge-invariant Hermition effective operators for the photon-atom interaction as
\begin{equation}
\cL = c_1\phi^\dag\phi F_{\mu\nu}F^{\mu\nu}+c_2\phi^\dag \phi v^\alpha F_{\alpha\mu} v_\beta F^{\beta\mu}
\end{equation}
where $c_1,c_2$ have mass dimension $-3$ and can be interpreted as the volumes of the atoms, suggesting $c_1,c_2\sim a_0^{-3}$.  Since the cross section for photon-atom scattering must have mass dimesion -2 and involve the square of the operators above, we infer that
\begin{equation}
\sigma \propto |c_i E_\gamma^2|^2\sim E_\gamma^4 a_0^6
\end{equation}
This estimate yields the correct power for the cross section energy dependence, as is known from Rayleigh scattering calculations.  This scaling only shows that higher energy photons are more likely to be scattered, but does not explain why the sky is blue rather than purple.  The only way to determine the color of the sky is to also know the spectrum of photons emitted by the source, which is information even beyond the theory (QED) underlying this EFT.  Thus, the sky being blue appears as a fine-tuning problem.

{\bf IV.}  Scales may not be well-separated.  Some dynamics preclude the separation of scales, as in turbulent flows, where modal instabilities allow for dynamical couplings between long-wavelength and short-wavelength modes.

\section{Future of Effective Field Theories}

Despite these challenges, Effective Field Theory methods provide state of the art tools to explore new physics in all fields.  In particle physics, we consider the Standard Model as the leading order lagrangian in EFT expansion and construct the mass-dimension-5 and -6 operators as subleading terms.  These operators take into account the evidence (such as neutrino masses and baryon number asymmetry) that physics violating the symmetries of the Standard Model must be present, possibly lying in the energy region between the electroweak scale $\sim 1\,$TeV and the Planck scale $\sim 10^{15}\,$TeV.  By writing this physics into low-energy (Standard Model energy scale) operators, we can study its impact on experimental observables.

In this program, continuing effort on developing $\SCETa$ as an EFT for QCD in accelerators is essential for the corresponding improvement in accuracy of QCD predictions of backgrounds to heavy particle production.  In nuclear physics, there are also many mysteries, such as how to describe hadronization ab initio in QCD and how quarks are bound inside exotic hadrons, as systems of $>3$ quarks or longer-distance molecules of hadrons?  To answer these questions $\SCETb$ and X-EFT, as EFTs of semi-inclusive and exclusive processes, are of great help, especially summing threshold logarithmic enhancements near infrared singularities.

Development of the electron laser effective theory provides the first rigorous mathematical basis for the quantum-classical separation implicit in numerical simulation of high-intensity laser-plasma experiments, in which the collective plasma dynamics are treated classically and QED processes are implemented via a local, stochastic Monte Carlo radiation algorithm.  It can be developed similarly to SCET to define and compute new observables in high-intensity laser plasma experiments that can therefore be used as more reliable tests of the underlying theory.  For example, it allows investigate integrated observables such as the amount of energy in low-energy (classical) radiation entering a specified cone.  Definition of such observables and their measurement in high-intensity laser experiments feeds back into the understanding of those analogous quantities in QCD.  Furthermore, the electron-laser theory establishes the foundation to investigate more complicated systems, involving both electrons and ions, to describe the ionization processes and high-Z atomic spectra in strong classical fields.

As the first example of an effective theory incorporating nonperturbative effects of a strong classical potential, the electron-laser effective theory may also open new avenues to study quantum effects in other classical backgrounds.  Important issues here include the dynamics of Hawking radiation, particle production in the expanding universe and degrees of freedom arising from the breaking of conformal symmetry by vacuum polarization in the classical field.

%% file: appendix_B.tex
\chapter{Useful Results for Soft Radiation in \texorpdfstring{$e^+e^-$, $ep$, $pp$}{e+e-, ep, pp} Scattering Processes}
\label{appx:softresults}


\section{Hemisphere Soft Function from General Soft Function}\label{app:hemispheresoft}

The soft function depends on the momenta $k_{B,J}$ projected onto the $n_{a,b}$ directions in the regions $\cH_{B,J}$, respectively. The shape of these regions in turn depends on the vectors $\qBJ =\omega_{a,b}n_{a,b}/2$ in the definition of the 0-jettiness:
\begin{align}
\tau_0&=\frac{2}{Q^2}\sum_i\opn{min}\left\{q_a\cdot p_i,q_b\cdot p_i\right\} \nn \\
q_{ab}&=\oneov{2}\chi_{ab}E_{cm}n_{a,b}
\end{align}
Indicating this dependence explicitly, we express the soft function as
\begin{align} \label{softdef}
S(k_1,k_2,\qJ, \qB,\mu) &= \frac{1}{N_C}\opn{tr} \sum_{X_s}
  \abs{\bra{X_s}[Y_{\nJ}^\dag Y_{\nB}](0)\ket{0}}^2 
\nn \\ &\times
  \: \delta\Bigl(k_1 - \sum_{i\in X_s} 
    \theta(\qB\mcdot p_i - \qJ\mcdot p_i)\nJ\mcdot p_i\Bigr) 
 \nn \\
&\times \delta\Bigl(k_2 - \sum_{i\in X_s}
  \theta(\qJ\mcdot p_i - \qB\mcdot p_i)\nB\mcdot p_i\Bigr) 
 \,.
\end{align}
The soft function for DIS involves the square of one incoming and one outgoing Wilson line, and hence differs from that for $e^+e^-\to {\rm dijets}$, which has two outgoing lines, and for $pp\to L+0$-jets which has two incoming lines.  We can relate \req{softdef} to the usual hemisphere soft function for DIS by generalizing an argument given in \cite{Kang:2013nha}. Note that the Wilson lines $Y_n$ are invariant under rescaling of $n$ (boost invariance):
\begin{equation}
\label{Yrescaling}
Y_{\beta n_a} = P\exp\left[ ig\int_{-\infty}^0 ds\,\beta n_a\mcdot A_s(\beta n_a s)\right] = P\exp\left[ ig\int_{-\infty}^0 ds\, n_a\mcdot A_s( n_a s)\right] = Y_{n_a}\,,
\end{equation}
and similarly for the lines extending from $0$ to $+\infty$, $Y_{\beta n_b}=Y_{n_b}$.  Define
\begin{equation}
\label{RJRBdefs}
 R_b = \sqrt{\frac{\wB \nB\mcdot \nJ}{2\wJ}}
 \,,\quad\qquad 
 R_a = \sqrt{\frac{\wJ \nJ\mcdot \nB}{2\wB}} 
 \,,
\end{equation}
so that for the rescaled four-vectors $\nJ' = \nJ/R_b$ and $\nB' = \nB/R_a$, we have $(q_a-q_b)\cdot p_i = \frac12 \omega_a R_a\, (n_a'-n_b')\cdot p_i$ since $\omega_b R_b=\omega_a R_a$.  This implies that the same partitioning defined in \req{softdef} can be expressed with $\theta(n_a'\cdot p_i - n_b'\cdot p_i)$ and $\theta(n_b'\cdot p_i- n_a'\cdot p_i)$. Furthermore $\nB'\mcdot \nJ' = 2$. Thus expressing \req{softdef} in terms of the rescaled vectors, $\nJ'$ and $\nB'$, we obtain
\begin{align} \label{softrescaled}
S(k_1,k_2,\qJ, \qB,\mu) &= \frac{1}{N_C R_b R_a}\opn{tr} \sum_{X_s}
 \abs{\bra{X_s}[Y_{\nJ'}^\dag Y_{\nB'}](0)\ket{0}}^2 
\nn \\&\times 
 \delta\Bigl(\frac{k_1}{R_b} - \sum_{i\in X_s} 
 \theta(\nB'\mcdot p_i - \nJ'\mcdot p_i)\nJ'\mcdot p_i\Bigr)
  \nn \\
& \times \delta\Bigl(\frac{k_2}{R_a} - \sum_{i\in X_s}
  \theta(\nJ'\mcdot p_i - \nB'\mcdot p_i)\nB'\mcdot p_i\Bigr)  
 \nn \\
 &= \frac{1}{R_b R_a}\: S_2\Bigl(\frac{k_1}{R_b},\frac{k_2}{R_a},\mu\Bigr)
\,.
\end{align}
The regions $\cH_{a,b}$ are hemispheres because they are separated by $n_{a,b}'$ whose spatial components are perpendicular to the same plane.

\section{General Properties of Hemisphere Soft Functions}
\label{appx:softgeneralproperties}

Some properties of these soft functions can be deduced even before performing explicit computations. Some of these are more easily expressed in position space:
\begin{align}\label{eq:softeqA.1}
\wt S_2(x_1,x_2,\mu) = \int d\ell_1 \,d\ell_2 \,S_2(\ell_1,\ell_2,\mu) \,e^{-i\ell_1 x_1} e^{-i\ell_2 x_2}\,,
\end{align}
where the momentum-space soft function $S_2$ and its Fourier transform $\wt S_2$ in this equation may stand for the $e^+e^-$, DIS, or $pp$ soft functions. It follows from the definitions \eqss{softeq7}{softeq18}{softeq25} that $S_2,\wt S_2$ are symmetric in their arguments:
\begin{align}\label{eq:softeqA.2}
S_2(\ell_1,\ell_2) = S_2(\ell_2,\ell_1)\,,\qquad \wt S_2(x_1,x_2) = \wt S_2(x_2,x_1)\,.
\end{align}
The renormalization conditions for the soft function satisfy some nontrivial properties. From the factorization theorems for doubly-differential distributions such as in the invariant masses $m_1^2,m_2^2$ in the two separate hemispheres of $e^+e^-$ collisions, one can derive from consistency of the hard, jet, and soft anomalous dimensions that the soft function renormalization itself takes a factorized form (\cite{Hoang:2007vb,Hornig:2011iu}):
\be \label{eq:softeqA.3}
\wt S_2(x_1,x_2,\mu) = \wt Z_S^{-1}(x_1,\mu)\wt Z_S^{-1}(x_2,\mu) \wt S_2^{\text{bare}}(x_1,x_2)\,.
\ee
The renormalized soft function satisfies the renormalization group evolution equation (RGE),
\be
\label{eq:softeqA.4}
\mu\frac{d}{d\mu} \ln \wt S_2(x_1,x_2,\mu) = \gamma_S(x_1,\mu) + \gamma_S(x_2,\mu) \,,
\ee
where each piece of the total anomalous dimension on the right-hand side is given by
\be
\label{eq:softeqA.5}
\gamma_S(x,\mu) = -\frac{d\ln\wt Z(x,\mu)}{d\ln \mu} = -\Gamma_\cusp[\as] \ln(ie^{\gamma_E} x\mu) + \gamma_S[\as]\,,
\ee
where $\Gamma_\cusp[\as]$ is the cusp anomalous dimension, known to three loops (\cite{korchemsky1987renormalization,Korchemskaya:1992je}), and $\gamma_S[\as]$ is the non-cusp anomalous dimension. 

The solution to the RGE \eq{softeqA.4} is 
\be
\label{eq:softeqA.6}
\wt S_2(x_1,x_2,\mu) = U_S(x_1,\mu,\mu_0)U_S(x_2,\mu_2,\mu_0) \wt S_2(x_1,x_2,\mu_0)\,,
\ee
giving $\wt S_2$ at one scale $\mu$ in terms of $\wt S_2$ at another scale $\mu_0$. The evolution kernels $U_S$ are
\be
\label{eq:softeqA.7}
U_S(x,\mu,\mu_0) = e^{K(\Gamma_\cusp,\gamma_S,\mu,\mu_0)}(ie^{\gamma_E}x\mu_0)^{\eta(\Gamma_\cusp,\mu,\mu_0)}\,,
\ee
where the functions $K,\eta$ are
\begin{subequations}
\label{eq:softeqA.8}
\begin{align}
K(\Gamma,\gamma,\mu,\mu_0) &= \int_{\mu_0}^\mu \frac{d\mu'}{\mu'}\Bigl( -2\Gamma_\cusp[\as(\mu')] \ln\frac{\mu'}{\mu_0} + \gamma_S[\as(\mu')]\Bigr)\,, \\
\eta(\Gamma,\mu,\mu_0) &= -2 \int_{\mu_0}^\mu \frac{d\mu'}{\mu'} \Gamma_\cusp[\as(\mu')]\,.
\end{align}
\end{subequations}
Using the solution \eq{softeqA.6} one can predict all the $\mu$-dependent terms that appear in $\wt S(x_1,x_2,\mu)$ to all orders in $\as$, up to the logarithmic accuracy to which the anomalous dimensions $\Gamma_\cusp$ and $\gamma_S$ are known.

\cite{Hoang:2008fs} showed that the scale dependence in \eq{softeqA.6} could be slightly reorganized to yield the form
\be \label{eq:softeqA.9}
\wt S_2(x_1,x_2,\mu) = U_S(x_1,\mu,\mu_{x_1})U_S(x_2,\mu,\mu_{x_2}) e^{\wt T(x_1,x_2)}\,,
\ee
where $\mu_{x_i} = (ie^{\gamma_E} x_i)^{-1}$.  The exponent $\wt T(x_1,x_2)$ is independent of $\mu$ and exponentiates in this way based on non-Abelian exponentiation (\cite{Gatheral:1983cz,Frenkel:1984pz}). Using the definitions \eqs{softeqA.7}{softeqA.8}, we obtain the very simple exponentiated result,
\be
\label{eq:softeqA.10}
\wt S_2(x_1,x_2,\mu) = e^{K(x_1,\mu) + K(x_2,\mu)} e^{\wt T(x_1,x_2)}\,,
\ee
where $K(x_i,\mu)\equiv K(\Gamma_\cusp,\gamma_S,\mu,\mu_{x_i})$. 

Using symmetry, \eq{softeqA.2}, and that $\wt T$ must be dimensionless, \cite{Hoang:2008fs} argued that, to $\cO(\as^2)$, we can deduce
\be
\label{eq:softeqA.11}
\wt T(x_1,x_2) = \frac{\as(\mu_{x_1})}{4\pi} t_1 + \frac{\as(\mu_{x_2})}{4\pi} t_1 + 2\Bigl(\frac{\as}{4\pi}\Bigr)^2 t_2(x_1/x_2)\,,
\ee
where $t_1$ is a pure constant, and $t_2(b)$ is a dimensionless function of the dimensionless ratio $b=x_1/x_2$, satisfying $t_2(b) = t_2(1/b)$.

We know that the anomalous dimensions of the hemisphere soft functions for $S_2^{ee,ep,pp}$ are all the same, and thus the $\mu$-dependent terms predicted by \eq{softeqA.10} will all be the same. The change in directions of Wilson lines in the different cases could, in principle, lead to different results for the $\mu$-independent function $\wt T(x_1,x_2)$.  Having proven equality of the soft functions to $\cO(\as^3)$ in the text, I have determined that $\wt T(x_1,x_2)$ is the same at least up to $\cO(\as^3)$.

\section{One-loop Soft Function}
\label{appx:soft1loop}

The one-loop result for the soft function $S_2$ can be computed from the diagrams illustrated in \fig{softfig1}.  The sum over squared amplitudes contained in \eq{softeq29} then gives the result, here written for $e^+e^-$, 
\begin{align}
\label{eq:softeqA.12}
S_2^{(1)}(\ell_1,\ell_2) &= \frac{2g^2\mu^{2\epsilon}C_F}{(2\pi)^{D-1}} \int\! d^Dk\, \delta(k^2)\theta(k^0) \,\cM(\ell_1,\ell_2;k)
\frac{1}{\bn \cdot k - i\epsilon} \frac{1}{n\cdot k +i\epsilon} + \text{c.c.}
\end{align}
However, the delta function and measurement function in \eq{softeqA.12} renders all $k$ integrals real and the signs of the $i\epsilon$'s irrelevant. Once these are used, the integration regions do not cross the poles in $k^\pm\pm i\epsilon$, so the $i\epsilon$'s can be dropped. 
 So the result for the one-loop diagrams in \fig{softfig1} for all three cases $ee,ep,pp$ gives the well-known result \eq{softeq37} for the bare soft function at $\cO(\as)$,
which also implies that the constant $t_1$ in \eq{softeqA.11} for the renormalized soft function is
\be
t_1 = -C_F \frac{\pi^2}{2}.
\ee 

\section{Two-loop Soft Function}
\label{appx:soft2loop}

Here we give known results for the pieces of the $\cO(\as^2)$ hemisphere soft function in \eq{softeq39}.
The first set of terms $R_c$ can be deduced from the known soft anomalous dimension,
\be
\Gamma_{\text{cusp}}[\as] = \sum_{k=0}^\infty \Bigl(\frac{\as}{4\pi}\Bigr)^{k+1} \Gamma^k \,,\quad \gamma_S[\as] = \sum_{k=0}\Bigl(\frac{\as}{4\pi}\Bigr)^{k+1} \gamma_S^k\,,
\ee
where to $\cO(\as^2)$
\be
\Gamma^0 = 4C_F\,,\quad \Gamma^1 = 4C_F C_A\Bigl( \frac{67}{9} - \frac{\pi^2}{3}\Bigr) - C_F T_R n_f \frac{80}{9}\,,
\ee
and
\be
\gamma_S^0 = 0 \,,\ \gamma_S^1 = C_F C_A \Bigl(-\frac{808}{27} + \frac{11\pi^2}{9} + 28\zeta_3\Bigr) + C_F T_R n_f\Bigl(\frac{224}{27} - \frac{4\pi^2}{9}\Bigr)\,,
\ee
and the beta function,
\be
\beta[\as] = -2\as \sum_{k=0}^\infty \Bigl(\frac{\as}{4\pi}\Bigr)^{k+1} \beta_k \,,
\ee
where to $\cO(\as^2)$\,,
\be
\beta_0 = \frac{11 C_A - 4 T_R n_f}{3}\,,\quad \beta_1 = \frac{34 C_A^2}{3} - \frac{20}{3}C_A T_R n_f - 4 C_F T_R n_f\,.
\ee

\subsection{Momentum space}
\label{appx:softmomentumspace}
The values of the anomalous dimensions above imply that at two-loop order (\cite{Hoang:2008fs}),
\be
\label{eq:softeqA.19}
\begin{split}
R_c^{(2)}(\ell_1,\ell_2,\mu) &= 8 C_F^2 (L_1^4 + L_2^4) + 16 C_F^2 L_1^2 L_2^2 \\
&+ \Bigl( \frac{88}{9}C_F C_A - \frac{32}{9} C_F T_R n_f\Bigr) (L_1^3 + L_2^3) \\
&+ \Bigl[ - \frac{20\pi^2}{3} C_F^2 + C_F C_A \Bigl( \frac{4\pi^2}{3} - \frac{268}{9}\Bigr) + \frac{80}{9} C_F T_R n_f\Bigr] (L_1^2 + L_2^2) \\
&+ \Bigl[ 64\zeta_3 C_F^2 + C_FC_A \Bigl( \frac{808}{27} - \frac{22\pi^2}{9} - 28\zeta_3\Bigr) \\
&~~ - C_F T_R n_f \Bigl( \frac{224}{27} - \frac{8\pi^2}{9}\Bigr)\Bigr] (L_1 + L_2)\,,
\end{split}
\ee
where $L_{1,2} = \ln (k_{1,2}/\mu)$. 

The non-global terms $S_{\text{NG}}$ receive contributions from all the real gluon diagrams, but \cite{Hornig:2011iu} showed that they could be deduced just from the two-real-gluon graphs where the two gluons go into opposite hemispheres. The nontrivial dependence on both $k_{1,2}$ arises from these configurations. IR divergences in the opposite-hemisphere two-gluon diagrams are cancelled by the real gluon graphs in which both gluons go into one hemisphere or where there is a single real gluon.  These same diagrams contribute to the constant term $c_S^{(2)}$. The result from \cite{Hornig:2011iu} for the non-constant terms can be expressed
\begin{align}
\label{eq:softeqA.20}
S_{\text{NG}}^{(2)} (\ell_1,\ell_2)&= - \frac{\pi^2}{3} C_F C_A \ln^2\frac{\ell_1}{\ell_2} \\
&+ \Bigl( C_F C_A \frac{11\pi^2 \minus 3 \minus 18\zeta_3}{9} + C_F T_R n_f \frac{6-4\pi^2}{9} \Bigr) \ln\frac{\ell_1/\ell_2 \plus \ell_2/\ell_1}{2} \nn \\
&+ C_F C_A \Bigl[ f_N\Bigl(\frac{\ell_1}{\ell_2}\Bigr) + f_N\Bigl(\frac{\ell_2}{\ell_1}\Bigr) - 2f_N(1)\Bigr] \nn \\
&+ C_F T_R n_f \Bigl[ f_Q\Bigl(\frac{\ell_1}{\ell_2}\Bigr) + f_Q\Bigl(\frac{\ell_2}{\ell_1}\Bigr) - 2f_Q(1)\Bigr]\,, \nn
\end{align}
where the functions $f_{N,Q}$ are given by
\begin{align}
\label{eq:softeqA.21}
f_Q(a) &= \Bigl[ \frac{2\pi^2}{9} - \frac{2}{3(a+1)}\Bigr] \ln a - \frac{4}{3}\ln a \Li_2(-a) + 4\Li_3(-a) \nn \\
&+ \frac{2\pi^2-3}{9} \ln\Bigl(a+\frac{1}{a}\Bigr)\,, \nn \\
f_N(a) &= -4\Li_4\Bigl(\frac{1}{a+1}\Bigr) - 11\Li_3(-a) + 2\Li_3\Bigl(\frac{1}{a+1}\Bigr) \ln\frac{a}{(a+1)^2} \nn \\
&\ + \Li_2\Bigl(\frac{1}{a+1}\Bigr) \Bigl[ \pi^2 - \ln^2(a+1) - \frac{1}{2}\ln a \ln\frac{a}{(a+1)^2} + \frac{11}{3}\ln a\Bigr] \nn \\
&\ + \Bigl[ \frac{11}{12}\ln\frac{a}{(a+1)^2} - \frac{1}{4} \ln\frac{a+1}{a} \ln(a+1) + \frac{\pi^2}{24}\Bigr] \ln^2 a \nn \\
&\ - \frac{1}{6}\frac{a-1}{a+1} \ln a + \frac{5\pi^2}{12}\ln\frac{a+1}{a} \ln (a+1) - \frac{11\pi^4}{180} \nn \\
&\ - \frac{11\pi^2 -3 -18\zeta_3}{18} \ln\Bigl( a + \frac{1}{a}\Bigr)\,.
\end{align}
These functions are bounded and vanish as $a\to0,\infty$. Their values at $a=1$ are
\be
\begin{split}
2f_Q(1) &= -6\zeta_3 + \frac{2}{9}(2\pi^2 -3)\ln 2 \\
2f_N(1) &= -8\Li_4\frac{1}{2} + \zeta_3\Bigl(\frac{33}{2}-5\ln2\Bigr) + \frac{\ln 2 - \ln^4 2}{3} \\
&\quad + \frac{2\pi^4}{45} + \frac{\pi^2}{3} \Bigl(\ln^2 2 - \frac{11}{3}\ln2\Bigr)\,,
\end{split}
\ee
and are subtracted out of the last two lines of \eq{softeqA.20} so that $S_{NG}$ vanishes at $\ell_1=\ell_2$. The functions $f_{Q,N}$ satisfy the rather remarkable approximation
\be\label{eq:softeqA.23}
f_{Q,N}(a) + f_{Q,N}(1/a) \approx 2f_{Q,N}(1) \frac{4a}{(1+a)^2}\,,
\ee
which can serve as a fairly good replacement rule for \eq{softeqA.21} (see Fig.~4 in \cite{Hornig:2011iu}). 

The momentum-space result \eqs{softeqA.20}{softeqA.21} are consistent with those given by \cite{Kelley:2011ng}. \cite{Hornig:2011iu} also computed the analogous results in position space, which we give below.

The constant term $c_S^{(2)}$ in \eq{softeq39} requires a  full computation of all the one- and two-real-gluon-emission diagrams. This was performed by \cite{Kelley:2011ng,Monni:2011gb}, with the result
\be
\label{eq:softeqA.24}
\begin{split}
c_S^{(2)} &= C_F^2 \frac{\pi^4}{8} + C_F C_A \biggl[ - \frac{508}{81} - \frac{871}{216}\pi^2 + \frac{4}{9}\pi^4 + \frac{22}{9} \zeta_3 \\
&\quad  - 7\zeta_3 \ln2 + \frac{\pi^2}{3} \ln^2 2 - \frac{1}{3}\ln^4 2 - 8\Li_4\Bigl(\frac{1}{2}\Bigr) \biggr] \\
&\quad + C_F T_R n_f \biggl( -\frac{34}{81} + \frac{77}{54}\pi^2 - \frac{8}{9}\zeta_3 \biggr) \,.
\end{split}
\ee
Thus the final result for the $\cO(\as^2)$ hemisphere soft function in $e^+e^-$ is given by \eq{softeq39} with the three individual pieces given by \eqss{softeqA.19}{softeqA.20}{softeqA.24}.

In this section, we have reviewed the previously known results for the $e^+e^-$, DIS, and $pp$ hemisphere soft functions at $\cO(\as)$ and the $e^+e^-$ soft function at $\cO(\as^2)$, which we have shown in the paper is also equal to the DIS and $pp$ results at $\cO(\as^2)$.

\subsection{Position-space soft function}
\label{appx:softpositionspace}

The position-space soft function is defined by the Fourier transform from momentum space \eq{softeqA.1}.
All the non-constant terms at $\cO(\as^2)$ were computed in \cite{Hornig:2011iu}. The constants at $\cO(\as^2)$ can be obtained analytically from the momentum-space results of \cite{Kelley:2011ng}. 

To $\cO(\as^2)$ the renormalized soft function \eq{softeqA.1} takes the form
\begin{align}
\label{eq:softeqA.25}
\wt S(x_1,x_2,\mu) = 1 &+ \frac{\as(\mu)C_F}{\pi}\Bigl( \tilde L_1^2 + \tilde L_2^2 + \frac{\pi^2}4\bigr) \\
&+ \frac{\as(\mu)^2}{4\pi^2} \Bigl[ \wt R(x_1,x_2,\mu) + \frac{1}{2} t_2\Bigl(\frac{x_1}{x_2} \Bigr) + \tilde c_S^{(2)}\Bigr]\,. \nn
\end{align}
The $\mu$-dependent logs are in $\wt R$,
\begin{align}
\wt R(x_1,x_2,\mu) &= \frac{\as(\mu)^2}{16\pi^2} \biggl\{ C_F^2 ( 8 \tL_1^4 + 8\tL_2^4 + 16\tL_1^2\tL_2^2) \\
&+ \Bigl( -\frac{88}{9} C_F C_A + \frac{32}{9} C_F T_R n_f\Bigr) (\tL_1^3 + \tL_2^3) \nn \\
& + \Bigl[ 4\pi^2 C_F^2 - 4C_F C_A \Bigl( \frac{67}{9} - \frac{\pi^2}{3}\Bigr) + \frac{80}{9} C_F T_R n_f\Bigr] (\tL_1^2 + \tL_2^2) \nn \\
&+ \Bigl[ C_F C_A\Bigl( 28\zeta_3 - \frac{808}{27} - \frac{22}{9}\pi^2\Bigr) + C_F T_R n_f \Bigl( \frac{224}{27} + \frac{8}{9}\pi^2\Bigr)\Bigr] (\tL_1 + \tL_2)\biggr\}\,. \nn
\end{align}
The non-global terms are in $t_2$,
\begin{align}
t_2(b) &= - C_F C_A\frac{2\pi^2}{3} \ln^2 b \\
&+ 2 \Bigl( C_F C_A \frac{11\pi^2 \minus 3 \minus 18\zeta_3}{9} + C_F T_R n_f \frac{6-4\pi^2}{9}\Bigr)\ln\frac{b + 1/b}{2} \nn \\
&+ 2 C_F T_R n_F \Bigl[ F_Q(b) + F_Q(1/b) - 2F_Q(1)\Bigr] \nn \\
&+ 2C_F C_A \Bigl[ F_N(b) + F_N(1/b) - 2F_N(1)\Bigr] \,, \nn
\end{align}
where the functions $F_{Q,N}$ are given by
\begin{align}
\label{eq:softeqA.28}
F_Q(b) &= \frac{2\ln b}{3(b-1)} - \frac{b\ln^2 b}{3(b-1)^2} - \frac{3-2\pi^2}{9} \ln\Bigl( b + \frac{1}{b}\Bigr) \nn \\
&\quad + \frac{2}{3}\ln^2 b \ln(1-b) + \frac{8}{3} \ln b \Li_2(b) - 4\Li_3(b) \nn \\
F_N(b) &= -\frac{\pi^4}{36} - \frac{\ln b}{3(b-1)} + \frac{b\ln^2 b}{6(b-1)^2} + \frac{\ln^4 b}{24} \nn \\
&\quad + \frac{3\minus11\pi^2\plus18\zeta_3}{18} \ln\Bigl( b+\frac{1}{b}\Bigr)  - \frac{11}{6}\ln^2 b \ln(1-b)  \nn \\
&\quad - \frac{\pi^2}{3} \Li_2(1-b) + [\Li_2(1-b)]^2 - \frac{22}{3}\ln b \Li_2(b) \nn \\
&\quad + 2\ln b \Li_3(1-b) + 11\Li_3(b)\,.
\end{align}
The values of these functions at $b=1$ are subtracted in \eq{softeqA.28} so that $t_2\to 0$ at $b=1$. They are given by
\be
\begin{split}
2F_Q(1) &= \frac{2}{3} + \Bigl( \frac{4\pi^2}{9} - \frac{2}{3}\Bigr) \ln2 - 2\zeta_3 \\
2F_N(1) &= - \frac{1}{3} - \frac{\pi^4}{18} + \Bigl( \frac{1}{3} - \frac{11\pi^2}{9} + 2\zeta_3\Bigr) \ln 2 + 22\zeta_3\,.
\end{split}
\ee
Then the constant in \eq{softeqA.25} is given by
\begin{align}
\tilde c_S^{(2)} &= C_F^2 \frac{\pi^4}{8} + C_F C_A \Bigl(-\frac{535}{81} - \frac{871}{216}\pi^2 + \frac{7}{30}\pi^4 + \frac{143}{18}\zeta_3\Bigr) \nn \\ 
&\quad + C_F T_R n_f \Bigl( \frac{20}{81} + \frac{77}{54}\pi^2 - \frac{26}{9}\zeta_3\Bigr)\,.
\end{align}


\section{Three-gluon Vertex Diagram for \texorpdfstring{$ep$, $pp$}{ep, pp}}
\label{appx:soft3gluon}

Here we provide an explicit computation of the amplitude in \fig{softfig7}(C), the result of which is given by \eqs{softeq48}{softeq65}, for the $ee,ep,pp$ soft functions. This provides an alternate derivation of the $ee$ result given in \cite{Kelley:2011ng}, and in addition shows explicitly how the imaginary $i\pi$ term in \eq{softeq65} arises. 

The integral $\cI_C^\cT(k)$ appearing in the amplitude \eq{softeq48} is given for the three different soft functions by (we will drop the ${}_\cC$ and ${}^\cT$ sub/superscripts in this appendix):
\be
\label{eq:softeqB.1}
\begin{split}
\cI_{ee}(k) \equiv \int \!\! \frac{d^D q}{(2\pi)^D} \frac{1}{q^2 \plus i\e}\frac{1}{(k \minus q)^2 +i\e} \frac{1 }{ (k \minus q)^++i\e} \frac{1}{q^- +i\e}\,, \\
\cI_{ep}(k) \equiv \int \!\! \frac{d^D q}{(2\pi)^D} \frac{1}{q^2 \plus i\e}\frac{1}{(k \minus q)^2 +i\e} \frac{1 }{(k \minus q)^++i\e} \frac{1}{q^- -i\e}\,, \\
\cI_{pp}(k) \equiv \int \!\! \frac{d^D q}{(2\pi)^D} \frac{1}{q^2 \plus i\e}\frac{1}{(k \minus q)^2 +i\e} \frac{1 }{(k \minus q)^+-i\e} \frac{1}{q^- -i\e}\,,
\end{split}
\ee
with the sign of the $i\e$ in one of the eikonal propagators switching in each line, as one Wilson line flips each time from outgoing to incoming.

\begin{figure}
\centering
\includegraphics[width=0.48\textwidth]{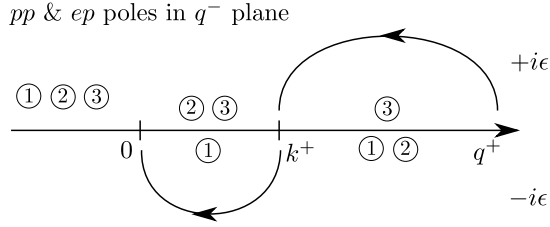}
\caption{Pole Positions and Contour Choice}
\label{fig:softfigB.9}
\end{figure}

The $\cI_{ee}$ and $\cI_{ep}$ integrands have the same pole structure in $q^+$, while $\cI_{ep}$ and $\cI_{pp}$ have the same pole structure in $q^-$. Namely, $\cI_{ee,ep}$ have poles in $q^+$ at:
\be
\label{eq:softeqB.2}
q^+ = k^+ + i\e\,, \frac{\vect{q}_\perp^2 -i\e}{q^-} \,, \frac{\vect{q}_\perp^2 - 2\vect{q}_\perp\mcdot\vect{k}_\perp + q^- k^+ - i\e}{q^- - k^-}\,,
\ee
while $\cI_{ep,pp}$ have poles in $q^-$ at:
\be
\label{eq:qminuspoles}
q^- =  i\e\,, \frac{\vect{q}_\perp^2 -i\e}{q^+} \,, \frac{\vect{q}_\perp^2 - 2\vect{q}_\perp\mcdot\vect{k}_\perp + q^+ k^- - i\e}{q^+ - k^+}\,.
\ee
The locations of these poles as a function of $q^-$ (or $q^+$) are shown in \fig{softfigB.9}.

First we will perform the $\cI_{pp/ep}$ integrals by contour integration in $q^-$, and then compute $\cI_{ee/ep}$ by contour integration in $q^+$, and obtain the result in \eq{softeq65}.

\subsection{Integrals for $ep,pp$}
\label{appx:softintegralseppp}

In the $\cI_{ep,pp}$ integrals in \eq{softeqB.1}, we perform the $q^-$ integration by contours. For $q^+<0$, all three poles are in the upper half complex plane, and so closing the contour in the lower half plane, we obtain zero. The nonzero contributions come from the other two regions in $q^+$:
\be
\label{eq:softeqB.4}
\cI_{ep,pp}(k) = -\frac{i}{4\pi} \int\frac{d^dq_\perp}{(2\pi)^d} \frac{1}{\vect{q}_\perp^2} \int_0^\infty dq^+ \frac{ F(q^+,\vect{q}_\perp,k)}{q^+ - k^+ \mp i\e}\,,
\ee
where $d = 2-2\e$. The upper (lower) sign in $\mp i\e$ in the $q^+$ eikonal propagator is for $ep$ ($pp$), and the function $F$ is given by
\be
\label{eq:softeqB.5}
F(q^+,\vect{q}_\perp,k) \equiv \frac{q^+}{k^+} \frac{\theta(k^+ - q^+)}{\bigl( \vect{q}_\perp - \frac{q^+}{k^+}\vect{k}_\perp\bigr)^2} \\
+ \frac{\theta(q^+ - k^+) }{\bigl(\vect{q}_\perp - \vect{k}_\perp)^2 + \vect{k}_\perp^2 (\frac{q^+}{k^+} - 1)}\,.
\ee
We used the on-shell condition $k^- = \vect{k}_\perp^2/k^+$ to eliminate $k^-$ from this expression. 
The $i\e$s in the $\vect{q}_\perp$-dependent propagators in \eqs{softeqB.4}{softeqB.5} can be dropped since the denominators are $\geq 0$, and the integral over $\vect{q}_\perp$ does not cross any singularities. The $q^+$ integral in \eq{softeqB.4}, however, encounters the singularity at $q^+ = k^+$, and we use the prescription
\be
\label{eq:softeqB.6}
\frac{1}{q^+ - k^+ \mp i\e} = \PV\frac{1}{q^+-k^+} \pm i\pi \delta(q^+ - k^+)
\ee
to perform the integral. The function $F$ is finite and continuous at $q^+ = k^+$:
\be\label{eq:softeqB.7}
F(k^+,\vect{q}_\perp,k) = \frac{1}{(\vect{q}_\perp - \vect{k}_\perp)^2}\,.
\ee
The result of using this prescription in \eq{softeqB.4} can be expressed
\be
\label{eq:softeqB.8}
\cI_{ep,pp}(k) = -\frac{i}{4\pi} A \pm \frac{1}{4} B\,,
\ee
where
\begin{subequations}
\begin{align}
A &\equiv  \int \frac{d^d q_\perp}{(2\pi)^d} \frac{1}{\vect{q}_\perp^2} \int_0^\infty dq^+ F(q^+,\vect{q}_\perp,k)  \PV \frac{1}{q^+-k^+}  \\
B &\equiv \int \frac{d^d q_\perp}{(2\pi)^d} \frac{1}{\vect{q}_\perp^2 (\vect{q}_\perp - \vect{k}_\perp)^2}\,.
\end{align}
\end{subequations}
$B$ is easily evaluated. Combining denominators using a Feynman parameter and then completing the integrations, we obtain 
\be
B = \frac{1}{(4\pi)^{1-\e}} \frac{\Gamma(1+\e) B(-\e,-\e)}{(\vect{k}_\perp^2)^{1+\e}}\,,
\ee
where $B(a,b)$ is the beta function. 

To evaluate $A$ we must regulate the singularity at $q^+ = k^+$ consistently with the principal value prescription. This can be done with equidistant cutoffs around $q^+ = k^+$, or, conveniently, we can insert a factor (similar to, but not directly associated with, the rapidity regulator of \cite{Chiu:2011qc,Chiu:2012ir}):
\be
\label{eq:softeqB.11}\!\!
A = \lim_{\eta\to 0} \int\frac{d^dq_\perp}{(2\pi)^d} \frac{1}{\vect{q}_\perp^2} \int_0^\infty\!\! dq^+ \biggl(\frac{\nu}{\abs{q^+-k^+}}\biggr)^\eta \frac{F(q^+,\vect{q}_\perp,k)}{q^+-k^+}\,.
\ee
Using the changes of variables
\be
q^+ \to q'= \abs{k^+ - q^+}\,, \quad q' = k^+ u\,,
\ee
and combining the $\vect{q}_\perp$ denominators in \eqs{softeqB.11}{softeqB.5} using a Feynman parameter, we obtain from the $\vect{q}_\perp$ integral
\be
\begin{split}
A &= \frac{1}{(4\pi)^{1-\e}} \frac{\Gamma(1+\e)}{ (\vect{k}_\perp^2)^{1+\e}} \biggl(\frac{\nu}{k^+}\biggr)^\eta \int_0^1 \frac{dx}{x^{1+\e}} \\
&\qquad\times \biggl\{ -\int_0^1 \frac{du}{u^{1+\eta}}\, \frac{1}{(1-x)^{1+\e} (1-u)^{1+2\e}} 
 + \int_0^\infty \frac{du}{u^{1+\eta}}\, \frac{1}{x^{1+\e}  (1-x+u)^{1+\e}}  \biggr\}\,.
\end{split}
\ee
In the final line we use the change of variables $u = u'(1-x)$ and the $x$ and $u(')$ integrals both immediately give beta functions,
\be
\begin{split}
A &= \frac{1}{(4\pi)^{1-\e}} \frac{\Gamma(1+\e)}{ (\vect{k}_\perp^2)^{1+\e}}\biggl(\frac{\nu}{k^+}\biggr)^\eta 
 \Bigl[ - B(-\eta,-2\e) B(-\e,-\e)  \\
&\hspace{2.2cm}+ B(-\e,-\e-\eta) B(1+\e+\eta,-\eta)\Bigr]\,.
\end{split}
\ee
The $1/\eta$ poles of the two terms in brackets in cancel, and we can take the $\eta\to 0$ limit to obtain
\be
A = \frac{1}{(4\pi)^{1-\e}} \frac{\Gamma(1+\e)}{ (\vect{k}_\perp^2)^{1+\e}}B(-\e,-\e) \frac{\pi}{\tan(\pi\e)}\,,
\ee 
Thus the result for the sum of $A,B$ terms in the integral \eq{softeqB.8} is
\be
\label{eq:softeqB.16}
\begin{split}
\cI_{ep,pp} &= -\frac{i}{16\pi^2} (4\pi)^\e\frac{\Gamma(1+\e)}{ (\vect{k}_\perp^2)^{1+\e}}B(-\e,-\e)  \frac{\pi e^{\pm i\pi\e}}{\sin(\pi\e)}\,.
\end{split}
\ee
Plugging this integral back into the amplitude \eq{softeq48}, multiplying by the sum of conjugates of the 1-gluon tree-level amplitudes from \eq{softeq34}, and summing over final-state polarizations and integrating over the final-state gluon momentum $k$ in \eq{softeq29}, we obtain for this contribution to the soft function,
\begin{align}
\cS_{ep,pp} &= \frac{1}{N_C}\Tr \int\frac{d^D k}{(2\pi)^D} 2\pi\delta(k^2) \theta(k^0) \cM_{\ell_1\ell_2}(k) 
\cA_{ep,pp}^\cT(k) [\cA_{1n}^\dag(k) + \cA_{1\bn}^\dag(k)] \nn \\
&=\frac{\as^2\, C_A C_F }{16\pi^2}\mu^{4\e} \left[ \frac{\delta(\ell_2)}{\ell_1^{1+4\e} } +\frac{\delta(\ell_1)}{\ell_2^{1+4\e}}  \right] \nn\\
&~~ \times \frac{1}{\e}\biggl\{ - \frac{2}{\e^2} + \pi^2 + \frac{16\zeta_3}{3} \e - \frac{\pi^4}{60}\e^2 
\pm i\pi \biggl(-\frac{2}{\e} + \frac{\pi^2}{3}\e + \frac{16\zeta_3}{3}\e^2\biggr)\biggr\}  \,.
\end{align}
Upon adding the complex conjugate diagrams, $\cS_{ep,pp} + \cS_{ep,pp}^*$, the imaginary parts cancel, and the real parts combine to reproduce the result for these diagrams in the $ee$ soft function in \cite{Kelley:2011ng}.

\subsection{Integrals for $ee,ep$}
\label{appx:softintegralseeep}

The integrals $\cI_{ee,ep}$ in \eq{softeqB.1} share the same poles in $q^+$ given in \eq{softeqB.2}, and according to \eqs{softeq54}{softeq56} should give exactly the same result. For $q^- < 0$ all three poles in $q^+$ lie in the upper half complex plane, and we can close the $q^+$ contour in the lower half plane to obtain zero. The nonzero contributions come from the regions $q^->0$:
\be
\label{eq:softeqB.18}
\begin{split}
\cI_{ee} = -\frac{i}{4\pi} \int \frac{d^d q_\perp}{(2\pi)^d}\int_0^\infty dq^- \frac{G(q^-,\vect{q}_\perp,k)}{\vect{q}_\perp^2 - \frac{q^-}{k^-}\vect{k}_\perp^2  - i\e}\,,
\end{split}
\ee
where
\be
\label{eq:softeqB.19}
G(q^-,\vect{q}_\perp,k) \equiv  \frac{ \theta(k^- - q^-)/k^-}{( \vect{q}_\perp - \frac{q^-}{k^-}\vect{k}_\perp)^2 -i\e} 
+  \frac{\theta(q^- - k^-)/q^-}{(\vect{q}_\perp -\vect{k}_\perp)^2 - i\e}\,.
\ee
We could evaluate \eq{softeqB.18} by applying the prescription \eq{softeqB.6} to the $q^-$-dependent propagator in \eq{softeqB.18}, but since the singularity at $q^- = k^- (\vect{q}_\perp^2/\vect{k}_\perp^2)$ is not necessarily at the point $q^- = k^-$ where the step in the theta functions in \eq{softeqB.19} occurs, we find it more convenient to evaluate \eq{softeqB.18} directly as a complex integral, keeping the $i\e$s where they appear.

Making the changes of variables,
\be
\begin{split}
u &= \frac{q^-}{k^-} \,, \qquad  
\vect{q}_\perp' = 
\begin{cases}
\vect{q}_\perp - u \vect{k}_\perp & \mbox{for } q^- < k^- \\
\vect{q}_\perp - \vect{k}_\perp & \mbox{for }  q^- > k^-
\end{cases}\,,
\end{split}
\ee
then combining $\vect{q}_\perp'$ denominators using a Feynman parameter and performing the $\vect{q}_\perp'$ integral, we obtain the intermediate result
\begin{align}
\cI_{ee}  &= \frac{i}{16\pi^2} (4\pi)^\e \frac{\Gamma(1+\e)}{ (\vect{k}_\perp^2)^{1+\e}}e^{i\pi\e} \int_0^1 dx\,(1-x)^{-1-\e} \\
&\quad\times \biggl\{ \int_0^1 du\,  [u (1-ux)]^{-1-\e} + \int_1^\infty \frac{du}{u} (u-x)^{-1-\e}\biggr\}. \nn
\end{align}
In the second integral over $u$, we make the further change of variables $u\to 1/u$, and the two integrals combine into one between $0<u<1$,
\be
\begin{split}
\cI_{ee}  &= \frac{i}{16\pi^2} (4\pi)^\e\frac{\Gamma(1+\e)}{ (\vect{k}_\perp^2)^{1+\e}}  e^{i\pi\e} \int_0^1 dx\,(1-x)^{-1-\e} \\
&\quad\times \biggl\{ \int_0^1 du \, (1-ux)^{-1-\e} (u^{-1-\e} + u^\e ) \biggr\}.
\end{split}
\ee
Performing the $u$ integral first, we obtain
\begin{align}
\cI_{ee}  &= \frac{i}{16\pi^2} (4\pi)^\e\frac{\Gamma(1+\e)}{ (\vect{k}_\perp^2)^{1+\e}}  e^{i\pi\e} \int_0^1 dx\,(1-x)^{-1-\e} \\
&\quad\times \biggl\{ x^\e B(x,-\e,-\e) + \frac{1}{1+\e} \F (1+\e,1+\e,2+\e,x)\biggr\}\,, \nn
\end{align}
where $B(x,a,b)$ is the incomplete beta function and $\F$ is a hypergeometric function. The $x$ integral can also be performed in terms of generalized hypergeometric functions,
\begin{align}
\cI_{ee}  = \frac{i}{16\pi^2} (4\pi)^\e \frac{\Gamma(1+\e)}{ (\vect{k}_\perp^2)^{1+\e}} e^{i\pi\e} 
 &\biggl[  \frac{1}{\e^2} \Fiii \bigl(1 , -\e,1+\e; 1-\e,1-\e ; 1\bigr) \nn\\
& - \frac{1}{\e(1+\e)} \Fiii \bigl( 1 , 1+\e,1+\e; 1-\e,2+\e; 1\bigr)\biggr ]\,, \nn
\end{align}

The hypergeometric functions can be expanded in powers of $\e$ with the help of the Mathematica package \texttt{HypExp} (\cite{Huber:2005yg,Huber:2007dx}), with the result
\be
\label{eq:softeqB.25}
\cI_{ee}  = \frac{i}{16\pi^2} (4\pi)^\e \frac{\Gamma(1+\e)}{ (\vect{k}_\perp^2)^{1+\e}} e^{i\pi\e} \biggl( \frac{2}{\e^2} - 4\zeta_3 \e - \frac{\pi^4}{15}\e^2\biggr)\,.
\ee
The result \eq{softeqB.25} is now in a similar form to \eq{softeqB.16}. 
It is equal to the result in \eq{softeqB.16} for $\cI_{ep}$, since we have the expansion 
\be
\label{eq:softeqB.26}
B(-\e,-\e) \frac{\pi }{\sin(\pi\e)}
=-\biggl( \frac{2}{\e^2} - 4\zeta_3 \e - \frac{\pi^4}{15}\e^2\biggr)\,.
\ee
This confirms the result \eq{softeq56} of the proof that the amplitudes in \fig{softfig7}(C) are equal for the $ee$ and $ep$ soft functions. 

The equality of the series expansions in \eqs{softeqB.25}{softeqB.26} suggests the functional identity
\begin{align}
B(-\e,-\e) \frac{\pi }{\sin(\pi\e)}  =& -   \frac{1}{\e^2} \Fiii \bigl( 1 , -\e,1+\e; 1-\e,1-\e; 1\bigr) \nn \\
& + \frac{1}{\e(1+\e)} \Fiii \bigl( 1 , 1+\e,1+\e; 1-\e,2+\e; 1\bigr)\end{align}
which we find to be numerically exact for any $\e<0$.  (The form on the right-hand side is defined only for $\e<0$.)


%% file: appendix_C.tex
\chapter{Rapidity Integrals, Anomalous Dimensions and Renormalization Group Equations}
\label{appx:B}

\section{Rapidity Anomalous Dimensions and RG Equations at 2-loops in TMDPDF}
\label{appx:B.1}

The functions $K_{\Gamma}(\mu_0, \mu)$, $\eta_{\Gamma}(\mu_0, \mu)$, $K_{\gamma} (\mu_0, \mu)$ in the RGE solutions 
\eqss{34eq4.14}{34eq4.20}{34eq4.27}  are defined as
\begin{align} \label{eq:34eqA.1}
K_{\Gamma}(\mu_0, \mu)
&\equiv \int_{\mu_0}^{\mu} \df(\ln \mu') \Gamma_\cusp\big[\as(\mu')\big] \, \ln \frac{\mu'}{\mu_0}
= \int_{\as(\mu_0)}^{\as(\mu)}\!\frac{\df\as}{\beta(\as)}\,
\Gamma_\cusp(\as) \int_{\as(\mu_0)}^{\as} \frac{\df \as'}{\beta(\as')}
\,,\nn\\
\eta_{\Gamma}(\mu_0, \mu)
&\equiv \int_{\mu_0}^{\mu} \df(\ln \mu') \Gamma_\cusp \big[\as(\mu')\big] 
= \int_{\as(\mu_0)}^{\as(\mu)}\!\frac{\df\as}{\beta(\as)}\, \Gamma_\cusp(\as)
\,, \nn\\
K_{\gamma}(\mu_0, \mu)
&\equiv \int_{\mu_0}^{\mu} \df(\ln \mu') \gamma \big[\as(\mu')\big] 
= \int_{\as(\mu_0)}^{\as(\mu)}\!\frac{\df\as}{\beta(\as)}\, \gamma(\as)
\,.\end{align}
Expanding the beta function and anomalous dimensions in powers of $\as$,
\begin{gather}
\label{eq:34eqA.2}
\beta(\as) =
- 2 \as \sum_{n=0}^\infty \beta_n\Bigl(\frac{\as}{4\pi}\Bigr)^{n+1}
\,, \\
\Gamma_\cusp(\as) = \sum_{n=0}^\infty \Gamma_n \Bigl(\frac{\as}{4\pi}\Bigr)^{n+1}
\,, \ \gamma(\as) = \sum_{n=0}^\infty \gamma_n \Bigl(\frac{\as}{4\pi}\Bigr)^{n+1}
\,, \nn
\end{gather}
their explicit expressions to NNLL accuracy are
\begin{subequations}
\begin{align} \label{eq:34eqA.3}
K_\Gamma(\mu_0, \mu) &= -\frac{\Gamma_0}{4\beta_0^2}\,
\biggl\{ \frac{4\pi}{\as(\mu_0)}\, \Bigl(1 - \frac{1}{r} - \ln r\Bigr)
   + \biggl(\frac{\Gamma_1 }{\Gamma_0 } - \frac{\beta_1}{\beta_0}\biggr) (1-r+\ln r)
   + \frac{\beta_1}{2\beta_0} \ln^2 r
\nn\\ & 
+ \frac{\as(\mu_0)}{4\pi} 
 \bigg[ \biggl(\frac{\beta_1^2}{\beta_0^2} - \frac{\beta_2}{\beta_0} \biggr) \Bigl(\frac{1 - r^2}{2} + \ln r\Bigr)
  + \biggl(\frac{\beta_1\Gamma_1 }{\beta_0 \Gamma_0 } - \frac{\beta_1^2}{\beta_0^2} \biggr) (1- r+ r\ln r)
\nn\\&  - \biggl(\frac{\Gamma_2 }{\Gamma_0} - \frac{\beta_1\Gamma_1}{\beta_0\Gamma_0} \biggr) \frac{(1\!-\! r)^2}{2}
     \bigg] \biggr\}
\,, \\
\eta_\Gamma(\mu_0, \mu) &=
 - \frac{\Gamma_0}{2\beta_0}\, \biggl[ \ln r
+ \frac{\as(\mu_0)}{4\pi}\biggl(\frac{\Gamma_1 }{\Gamma_0 }
 - \frac{\beta_1}{\beta_0}\biggr)(r\!-\!1)
 \nn\\ & 
+ \left(\frac{\as(\mu_0)}{4\pi}\right)^2\! \bigg(
 \frac{\Gamma_2 }{\Gamma_0 } - \frac{\beta_1\Gamma_1 }{\beta_0 \Gamma_0 }
      + \frac{\beta_1^2}{\beta_0^2} -\frac{\beta_2}{\beta_0} \bigg) \frac{r^2\!-\!1}{2}
    \biggr]
\,, 
\end{align}
and
\begin{align}
K_\gamma(\mu_0, \mu) &=
 - \frac{\gamma_0}{2\beta_0}\, \biggl[ 
\ln r + \frac{\as(\mu_0)}{4\pi}\, \biggl(\frac{\gamma_1 }{\gamma_0 } - \frac{\beta_1}{\beta_0}\biggr)(r-1)
 \biggr]
\,.
\end{align}
\end{subequations}
Here, $r = \as(\mu)/\as(\mu_0)$, and the running coupling is given to 3-loop order by
\begin{align} \label{eq:34eqA.4}
\frac{1}{\as(\mu)} &= \frac{X}{\as(\mu_0)}
  +\frac{\beta_1}{4\pi\beta_0}  \ln X
  + \frac{\as(\mu_0)}{16\pi^2} \biggr[
  \frac{\beta_2}{\beta_0} \Bigl(1-\frac{1}{X}\Bigr)
  + \frac{\beta_1^2}{\beta_0^2} \Bigl( \frac{\ln X}{X} +\frac{1}{X} -1\Bigr) \biggl]
,\end{align}
where $X\equiv 1+\as(\mu_0)\beta_0 \ln(\mu/\mu_0)/(2\pi)$.
In our numerical analysis we use the full NNLL expressions for $K_{\Gamma,\gamma},\eta_\Gamma$ in \eq{34eqA.3}, but to be consistent with the value of $\as(\mu)$ used in the NLO PDFs we only use the two-loop truncation of \eq{34eqA.4}, dropping the $\beta_2$ and $\beta_1^2$ terms, to obtain numerical values for $\as(\mu)$. 
Up to three loops, the coefficients of the beta function (\cite{Tarasov:1980au, Larin:1993tp}) and cusp anomalous dimension (\cite{Korchemsky:1987wg, Moch:2004pa}) in $\overline{\mathrm{MS}}$ are
\begin{align} \label{eq:34eqA.5}
\beta_0 &= \frac{11}{3}\,C_A -\frac{4}{3}\,T_F\,n_f
\,,\nn\\
\beta_1 &= \frac{34}{3}\,C_A^2  - \Bigl(\frac{20}{3}\,C_A\, + 4 C_F\Bigr)\, T_F\,n_f
\,, \nn\\
\beta_2 &=
\frac{2857}{54}\,C_A^3 + \Bigl(C_F^2 - \frac{205}{18}\,C_F C_A
 - \frac{1415}{54}\,C_A^2 \Bigr)\, 2T_F\,n_f
 + \Bigl(\frac{11}{9}\, C_F + \frac{79}{54}\, C_A \Bigr)\, 4T_F^2\,n_f^2
\,,\nn\\[2ex]
\Gamma_0 &= 4C_F
\,,\nn\\
\Gamma_1 &= 4C_F \Bigl[\Bigl( \frac{67}{9} -\frac{\pi^2}{3} \Bigr)\,C_A  -
   \frac{20}{9}\,T_F\, n_f \Bigr]
\,,\nn\\
\Gamma_2 &= 4C_F \Bigl[
\Bigl(\frac{245}{6} -\frac{134 \pi^2}{27} + \frac{11 \pi ^4}{45}
  + \frac{22 \zeta_3}{3}\Bigr)C_A^2 
  + \Bigl(- \frac{418}{27} + \frac{40 \pi^2}{27}  - \frac{56 \zeta_3}{3} \Bigr)C_A\, T_F\,n_f
\nn\\* & \hspace{8ex}
  + \Bigl(- \frac{55}{3} + 16 \zeta_3 \Bigr) C_F\, T_F\,n_f
  - \frac{16}{27}\,T_F^2\, n_f^2 \Bigr]
\,.\end{align}

The $\overline{\mathrm{MS}}$ non-cusp anomalous dimension $\gamma_H = 2\gamma_C$ for the hard function $H$ can be obtained (\cite{Idilbi:2006dg, Becher:2006mr}) from the IR divergences of the on-shell massless quark form factor $C(q^2,\mu)$ which are known to three loops (\cite{Moch:2005id}).
Here we write results up to 2 loops
\begin{align} \label{eq:34eqA.6}
\gamma_{H\,0} &=2 \gamma_{C\,0} = -12 C_F
\,,\nn\\
\gamma_{H\,1} &=2\gamma_{C\,1}
= -2 C_F 
\Bigl[
  \Bigl(\frac{82}{9} - 52 \zeta_3\Bigr) C_A
+ (3 - 4 \pi^2 + 48 \zeta_3) C_F
+ \Bigl(\frac{65}{9} + \pi^2 \Bigr) \beta_0 \Bigr]
\,.\end{align}

The non-cusp $\mu$- and $\nu$-anomalous dimensions for the TMD PDF and gluon soft function at 1 loop were calculated by \cite{Chiu:2012ir}.  By replacing $C_A$ with $C_F$, we obtain anomalous dimensions for the quark functions.  These results agree with the result from explicit calculation of the quark TMD PDF in \sec{I.5.2.6}.  However, as shown in \eq{34eq4.30} knowing $\gamma_H=-\gamma_S-2 \gamma_f$ is enough information, though their one loop results are known separately  $\gamma_{f\,0} = 6 C_F$ and $\gamma_{S\,0} =0$.
The $\nu$-anomalous dimensions up to 2 loops are given by
\begin{align}\label{eq:34eqA.7}
\gamma_{R\, S\,0} & =-2 \gamma_{R\,f\,0} = 0
\,,\nn\\
 \gamma_{R\, S\,1} & =-2 \gamma_{R\,f\,2} =-2C_F \Big[  \Big(  \frac{64}{9}-28 \zeta_3   \Big) C_A +\frac{56}{9}  \beta_0 \Big] 
\,.\end{align}
Note that the 2-loop result $\gamma_{R\, S\,1}$ is the same as that for the thrust soft function (\cite{Kang:2013nha,Stewart:2010pd}) except that \eq{34eqA.7} does not contain a $\pi^2\beta_0$ term.

\subsection{Rapidity anomalous dimension at 2-loop}
\label{appx:B.1.1}
In order to achieve NNLL accuracy the value of the 2-loop anomalous dimension $\gamma_{R\, S}$ is necessary.
It can be determined by comparing the factorized cross section 
\be\label{appeq:factorizedTMDsigma}
\sigma(\mu,\nu) =H(Q^2,\mu) \tilde S(\mu,\nu) \tilde f^\perp (\mu,\nu) \tilde f^\perp (\mu,\nu)
\ee
to the singular limit of the 2-loop QCD cross section (\cite{Ellis:1981hk,Arnold:1988dp,Gonsalves:1989ar}).
Alternatively, it can extracted from the singular behavior of the differential soft function $\tS(b^+,b^-,b_\perp)$ in \cite{Li:2011zp}.
\subsubsection{Fixed order cross section}
\label{appx:B.1.1.1}
To find $\gamma_{R\,S}$, we consider the expression for  the fixed order cross section is simplest at the scale $\mu_b=2/(b e^{\gamma_E})$, which kills all terms involving the logarithm $\ln(\mu b e^{\gamma_E}/2)$ in the soft function and PDF generated by $\mu$ evolution. At this scale, the 2-loop expressions for H, $\tS$, $\tft$ are
\begin{align}
H(\mu_b,Q)=& 1+ \frac{\as(\mu_b)}{4\pi}\bigg[  -\frac{\Gamma_0}{2}L_Q^2+\frac{\gamma_{H0}}{2} L_Q +c_{H 1}\Big]
\nn \\&
+\left(\frac{\as(\mu_b)}{4\pi}\right)^{\!2} \Big[ 
	\frac{\Ga_0^2}{8}L_Q^4
	-\frac{\Ga_0}{12}(2\beta_0+3\gamma_{H0})L_Q^3
\nn \\ &	
	+\Bigg(-\frac{\Ga_1+c_{H1} \Ga_0}{2}+\frac{\gamma_{H0} }{4}\Big(\beta_0+\frac{\gamma_{H0}}{2} \Big)
		\Bigg) L_Q^2
\nn \\&
	+\bigg(\gamma_{H1}  +c_{H1}\Big(\beta_0+\frac{\gamma_{H0}}{2} \Big)\bigg)L_Q +c_{H2} \bigg]
\,,\label{eq:34eqB.1}
\\ 
\tS(\mu_b,\nu)
 =&1 +\frac{\as(\mu_b)}{4\pi} c_{S1} +\left(\frac{\as(\mu_b)}{4\pi}\right)^{\!2} \Big[ \frac{\gamma_{R\,S1}}{2} L_\nu + c_{S2} \Big]
\,,\label{eq:34eqB.2}
\\ 
\tft_q(\mu_b,\nu)
=&
\Big[ 1  +\left(\frac{\as(\mu_b)}{4\pi}\right)^2  \frac{\gamma_{R\,f1}}{2} (L_\nu+L_Q) \Big] f_{j/P}(\mu_b)
\nn \\
&+\Big[ \frac{\as(\mu_b)}{4\pi}  c^j_{f1}   +\left(\frac{\as(\mu_b)}{4\pi}\right)^{\!2} \ c^j_{f2}  \Big] \otimes_z f_{j/P}(\mu_b)
\,, \label{eq:34eqB.3}
\end{align}
where $L_Q=\ln(\mu_b^2/Q^2)$ and $L_\nu=\ln (\nu^2/\mu_b^2)$.  In $\tft_q(\mu_b,\nu)$, $\otimes_z$ denotes the convolution with respect to $z$. The constants $c_{S1}$ and $c_{f1}$ are given in the one-loop results for the soft and collinear functions, while $c_{S2}$ and $c_{f2}$ are not known. 
We already used $\gamma_{R\,S0}=-2\gamma_{R\,f0}=0$.

Inserting \eqss{34eqB.1}{34eqB.2}{34eqB.3} into \eq{34eq4.1}, we obtain the fixed-order cross section at 2 loops. The $\alpha_s^2 L_Q$ term contains $\gamma_{R\, S1}$ 
\begin{align}\label{eq:34eqB.4}
\sigma^\text{fix}(\mu_b,Q) \ni
\left(\frac{\as(\mu_b)}{4\pi} \right)^{\!2}\frac{L_Q}{2}\Bigg[&
\Big[\gamma_{H1} -\gamma_{R\, S1} +\gamma_{H0} c_{S1}+ c_{H1} (2\beta_0+\gamma_{H 0})
	\Big]  \, f_i f_j
\nn \\&	
+\gamma_{H0} \Big[ (c^i_{f1}\otimes_{z_1}f_i ) f_j + i\leftrightarrow j \Big]
\Bigg]
\,,\end{align}
where we omit arguments of the PDF $f_i(z_1)$ and $f_j(z_2)$ and the constants $c^i_{f1}(z_1)$ and $c^j_{f1}(z_2)$ for simplicity.

\subsubsection{Differential soft function}
\label{appx:B.1.1.2}
An alternate way to determine the anomalous dimension $\gamma_{R\, S}$ is to explore the singular behavior of the differential soft function $\tS(b^+,b^-,b_\perp)$ in the limit $b^+b^-\to 0$, which corresponds to the collinear limit $p^+ p^- \to \infty$ where the rapidity divergence arises.  In this limit, the product $b^+b^-$ regulates the rapidity and  variation of the soft function respect to $b^+b^-$ corresponds to the rapidity anomalous dimension in a scheme different from the $\eta$ regulator.  However different schemes should reproduce the same $\gamma_{R\, S}$ term in \eq{34eqB.4}, and this implies that the anomalous dimension is uniquely determined by QCD  and independent of the choice of regulator.  We therefore write RRG equations for $\tS(b^+,b^-,b_\perp)$ determining $\gamma_{R\,S}$ as
\begin{align}\label{eq:34eqB.5}
\gamma_{R\, S}=\frac{D(b^+b^-)}{\tS_\text{sing} (b^+,b^-,b_\perp)}\frac{\df \tS_\text{sing} (b^+,b^-,b_\perp)}{\df \ln  b^+b^-} 
\,,\end{align}
where $\tS_\text{sing}=\lim_{b^+ b^- \to 0}\tilde S$ is the singular part of the differential soft function and $D(b^+b^-)=-2$ is dimensional factor taking into account difference between differentiation respect to $\nu$ and to $b^+b^-$.

The 2-loop soft function in $b$ space $\tS(b^+,b^-,b_\perp)$ has been calculated in \cite{Li:2011zp}.
In the singular limit, the 1- and 2-loop results are
\begin{align}\label{eq:34eqB.6}
\tS_\text{sing}& =1+ \frac{\as C_F}{\pi}\tS_{1}+ \frac{\as^2 C_F}{\pi^2}\Bigg[C_F\, \tS_{ C_F2}+n_f \,\tS_{CF NF}+C_A \,\tS_{CF CA}\Bigg]
\,,\\\label{eq:34eqB.7}
&\tS_{1}=L_b \ln\frac{b^+b^-}{\vec{b}_\perp^2}+\frac{L_b^2}{2} -\frac{\pi^2}{12}
\,,\qquad \tS_{ C_F2}=\frac{\tS_{1}^2}{2}
\,,\\ \label{eq:34eqB.8}
&\tS_{CF NF} =\Bigg( -\frac{1}{12}L_b^2-\frac{5}{18} L_b -\frac{7}{27}  \Bigg) \ln\frac{b^+b^-}{\vec{b}_\perp^2} 
\,,\\ \label{eq:34eqB.9}
&\tS_{CF CA} =\Bigg( \frac{11}{24}L_b^2+\Big( \frac{67}{36}-\frac{\pi^2}{12} \Big) L_b +\frac{101}{54} -\frac{7}{4}\zeta_3   \Bigg) \ln\frac{b^+b^-}{\vec{b}_\perp^2} 
\,,\end{align}
where $L_b=\ln \mu^2/\mu_b^2$ and $\mu_b=2/( |\vec{b}_\perp| e^{\gamma_E} )$.
Note that keeping a constant term in $\ln (b^+b^-/b_\perp^2)$ is necessary to ensure cancellation of $C_F^2$ terms in \eq{34eqB.5}.
By inserting \eq{34eqB.6} into \eq{34eqB.5}, the cusp and non-cusp part at 2-loop
\begin{align}\label{eq:34eqB.10}
\gamma_{R\, S}(\mu)\vert_{\cusp}
 &= \frac{\as(\mu)C_F}{\pi} \Bigg\{-2L_b
 +\frac{\as(\mu)}{\pi} \Bigg[ \Big(\frac{\pi^2}{6}-\frac{2}{3} \Big) L_b   C_A -\Big(\frac{L_b^2}{4}+\frac56 L_b\Big) \beta_0 \Bigg]
 \Bigg\}
\,,\\ \label{eq:34eqB.11}
\gamma_{R\, S}(\as(\mu))
&=-2 C_F\Big( \frac{\as(\mu)}{4\pi}\Big)^2 
\Bigl[\Bigl(\frac{64}{9} - 28 \zeta_3\Bigr) C_A + \frac{56}{9}  \beta_0 \Bigr]
\,.\end{align}
\eq{34eqB.10} agrees with the fixed-order expansion of the cusp part  $-4\eta_\Ga (\mu_b,\mu)$ in \eq{34eq4.25}.  \eq{34eqB.11} is consistent with zero at $\cO(\as)$ given in \sec{I.5.2.6} and \sec{I.5.2.6.3}, and agrees with Eq.~(51) given in \cite{Becher:2010tm}.  With \eq{34eqB.11} resummation at NNLL accuracy is achieved.

\section{Initial and Final State Soft Wilson Lines in Soft Functions}
\label{appx:B.2}

In this appendix, we prove \req{eq:44eq57}, based on the appendix of \cite{bauer2004enhanced}.  We start with a general event-shape function,
\begin{align}\label{eq:44eqA1}
S(k) &= \frac{1}{N_c}\int\frac{du}{(2\pi)}e^{iku}\langle 0| \bar T[(Y_{\bar n}^{\dagger})_d^{e}(Y_n)_e^a](un/2)T[(Y_{n}^{\dagger})^c_a(Y_{\bar n})_c^d](0)|0\rangle\,.
\end{align}
The Wilson lines in this expression can be divided into $N$ infinitesimal segments of length $ds$ with a subscript denoting their space-time position along the integration path,
\begin{align}
\label{eq:44eqA3} (Y_{n})_e^a& = \overline{P}\exp\bigg(-ig\int_0^\infty ds n\cdot A_s\bigg) = (e^{-igA_1ds})_e^{b_1}\ldots(e^{-igA_Nds})_{b_{N-1}}^a \,, \\
\label{eq:44eqA3dag} (Y_{n}^{\dagger})_a^c &= {P}\exp\bigg(ig\int_0^\infty ds n\cdot A_s\bigg) = (e^{igA_Nds})_a^{b_{N-1}}\ldots(e^{igA_1ds})_{b_1}^c \,, \\
(Y_{\bar n})_c^d &= {P}\exp\bigg(ig\int_{-\infty}^0 ds \bar n\cdot A_s\bigg) = (e^{-igA_1(\bar n)ds})_a^{b_1}\ldots(e^{-igA_N(\bar n)ds})_{b_{N-1}}^c \,, \\
(Y_{\bar n}^\dagger)_d^{e} &= \overline{P}\exp\bigg(ig\int_{-\infty}^0 ds \bar n\cdot A_s\bigg) = (e^{igA_N(\bar n)ds})_d^{b_{N-1}} \ldots (e^{igA_1(\bar n)ds})_{b_1}^e \,.
\end{align}
Among these Wilson lines, \req{eq:44eqA3} and \req{eq:44eqA3dag} are sums of outgoing gluons, which represent final state gluons.  Applying time-ordering and anti-time-ordering operators, we obtain
\begin{align}
T(Y_{n}^{\dagger})_a^c&=(Y_{n}^{\dagger})_a^c \,,
\end{align}
and
\begin{align}
\bar T(Y_{n})_e^a&=(Y_{n})_e^a \,.
\end{align}
For the other two we find
\begin{align}
T(Y_{\bar n})_c^d &= (e^{-igA_N(\bar n)ds})_{b_{N-1}}^d\ldots(e^{-igA_1(\bar n)ds})_c^{b_1} \nn\\
&=(e^{-igA_N^T(\bar n)ds})_d^{b_{N-1}}\ldots(e^{-igA_1^T(\bar n)ds})_{b_1}^c \nn\\
&=(e^{ig\bar n\cdot \bar A_N ds})_d^{b_{N-1}}\ldots(e^{ig\bar n\cdot A_1ds})_{b_1}^c = (\bar{Y}_{\bar n}^{\dagger})_d^c\,, \\
\bar T(Y_{\bar n}^\dagger)_d^{e} &= (e^{igA_1(\bar n)ds})_{b_1}^e\ldots(e^{igA_N(\bar n)ds})_d^{b_{N-1}} \nn\\
&=(e^{igA_1^Tds})_e^{b_1}\ldots(e^{igA_N^T(\bar n)ds})_{b_{N-1}}^d \nn \\
&=(e^{-ig\bar A_i\bar nds})_e^{b_1}\ldots(e^{-ig\bar n\cdot \bar A_Nds})_{b_{N-1}}^d=\bar Y_{\bar ne}^d \,.
\end{align}
Applying the above identities to the expression in Eq.~(\ref{eq:44eqA1}) gives
\begin{align}
S(k)&=\frac{1}{N_c}\int\frac{du}{(2\pi)}e^{iku}\langle 0| (\overline{Y}_{\bar n}^{\dagger})_e^d(Y_{n})_e^{a'}(un/2)\delta^a_{a'} (Y_{n}^\dagger)_a^{c}(\overline{Y}_{\bar n})_d^c(0)|0\rangle \,.
\end{align}
Now consider two infinite Wilson lines
\begin{align}
(Y_\infty)_{a'}^f&=P\exp\bigg(ig\int_{-\infty}^\infty \!ds\,n\cdot A_s\big({\textstyle \frac{un}{2}}\big)\bigg)_{a'}^f=P\exp\bigg(ig\int_{-\infty}^\infty \!ds\,n\cdot A_s(0)\bigg)_{a'}^f \nn \\
&=\{(e^{igA_N\cdot nds})_{a'}^{c_{N-1}}\ldots (e^{igA_1\cdot nds})_{c_1}^{c_0}\}\{(e^{igA_{-1}\cdot nds})_{c_0}^{c_1}\ldots (e^{igA_{-N}\cdot nds})_{c_{N+1}}^f\}\,, \\
(Y_\infty^\dagger)_f^a &= \overline{P}\exp\left(-ig\int_{-\infty}^{\infty}\!ds\, n\cdot A_s(un/2)\right)^{a'}_f=\overline{P}\exp\bigg(-ig\int_{-\infty}^{\infty}\!ds\, n\cdot A_s(0)\bigg)^{a'}_f \nn \\
&= \{(e^{-igA_{-N}\cdot nds})_f^{c_{N+1}}\cdot (e^{-igA_{-1}\ldots nds})_{c_{-1}}^{c_0}\}\{(e^{-igA_1\cdot nds})_{c_0}^{c_1}\ldots (e^{-igA_N\cdot nds})_{c_{N-1}}^a\} \,,
\end{align}
which have the property that 
\begin{align}
(Y_\infty)_{a'}^f(Y_\infty^\dagger)_f^a &=\delta_{a'}^a \,.
\end{align}
We can use this property to replace the identity $\delta_{a'}^a$ in $S(k)$ with the pair of infinite Wilson lines above
\begin{align}
S(k)&=\frac{1}{N_c}\int\frac{du}{(2\pi)}e^{iku}  \langle 0|(\bar Y_{\bar n})_e^d (Y_{n})_e^{a'}(un/2)\delta_{a'}^a (Y_{n}^{\dagger})_a^c(\bar Y_{\bar n}^{\dagger})_d^c(0)|0\rangle \nn\\
&=\frac{1}{N_c}\int\frac{du}{(2\pi)}e^{iku}   \notag \\
&\times\langle 0|\{(e^{-ig\bar A_N\cdot\bar nds})_e^{b_1}\ldots(e^{-ig\bar n\cdot A_1 ds})_{b_{N-1}}^d\}\{(e^{-igA_1\cdot nds})_e^{b_1}\ldots(e^{-igA_N\cdot nds})_{b_{N-1}}^{a'}\}({\textstyle \frac{un}{2}}) \notag \\
&\times\{(e^{igA_N\cdot nds})_{a'}^{c_{N-1}}\ldots(e^{igA_1\cdot nds})_{c_1}^{c_0}\}\{(e^{igA_{-1}\cdot nds})_{c_0}^{c_{-1}}\ldots (e^{igA_{-N}\cdot nds})_{c_{N+1}}^f\}({\textstyle \frac{un}{2}})  \notag \\
&\times\{(e^{-igA_{-N}\cdot nds})_f^{c_{-N+1}}\ldots(e^{-igA_{-1}\cdot nds})_{c_{-1}}^{c_0}\}\{(e^{-igA_1\cdot nds})_{c_0}^{c_1}\ldots(e^{-igA_N\cdot nds})_{c_{N-1}}^a\}(0)   \notag \\
&\times\{(e^{igA_N\cdot n ds})_a^{b_{N-1}}\ldots (e^{igA_1\cdot nds})_{b_1}^c\}\{(e^{ig\bar n\cdot A_N ds})_d^{b_{N-1}}\ldots(e^{ig\bar n\cdot A_1ds})_{b_1}^c\}(0)  |0\rangle   \notag \\
&=\frac{1}{N_c}\int\frac{du}{(2\pi)}e^{iku} \notag \\
&\times\!\langle 0| \{(e^{-ig\bar A_N\cdot\bar n ds})_e^{b_1}\ldots(e^{-ig\bar n\cdot A_1ds})_{b_{N-1}}^d\}\{(e^{igA_{-1}\cdot nds})_e^{c_{-1}}\ldots(e^{igA_{-N}\cdot nds})_{c_{-N+1}}^f\}({\textstyle \frac{un}{2}})  \notag \\
&\times\{(e^{-igA_{-N}\cdot nds})_f^{c_{-N+1}}\ldots(e^{-igA_{-1}\cdot nds})_{c_{-1}}^c\}\{(e^{ig\bar n\cdot A_Nds})_d^{b_{N-1}}\ldots(e^{ig\bar A_1\cdot\bar nds})_{b_1}^c\}(0) |0\rangle \notag \\
\label{eq:44eqA21}
&= \frac{1}{N_c}\int\frac{du}{(2\pi)}e^{iku}  \langle 0| (\overline{Y}_{\bar n})_e^d(Y_n)_e^f(un/2)(Y^\dagger_n)^f_c(Y^\dagger_{\bar n})_d^c(0) |0\rangle\,,
\end{align}
in which
\begin{align}
(Y_n)^f_e\big({\textstyle \frac{un}{2}}\big) &= \left( e^{igA_{-1}\cdot nds}\right)^{c_{-1}}_e\ldots \left(e^{igA_{-N}\cdot nds}\right)^f_{c_{-N+1}} \\ & \notag
= P \exp\!\left(ig\int_{-\infty}^0\!\!ds\, n\!\cdot\! A_s\right)  \\
(Y^\dagger_n)_c^f(0) &=   \left( e^{-igA_{-N}\cdot nds}\right)^{c_{-N+1}}_f\ldots \left(e^{-igA_{-1}\cdot nds}\right)^c_{c_{-1}} \\ \notag  &
= \overline{P} \exp\!\left(\!-ig\int_{-\infty}^0\!\!ds \,n\!\cdot\! A_s\right) 
\end{align}
are incoming gluon lines.  Thus, from \req{eq:44eqA1} to \req{eq:44eqA21}, we show that by inserting the identity operator for infinite Wilson lines, we change the final state Wilson lines in the soft function into initial state Wilson lines.


\section{Endpoint DIS and DY Rapidity Divergence-Related Results}
\label{appx:B.3}

\subsection{DIS Final Jet Function to \texorpdfstring{$\cO(\alpha_s)$}{O(alpha)} with Delta Regulator}
\label{appx:B.3.1}

\begin{figure}[h!]
\centering
\includegraphics[width=0.7\textwidth]{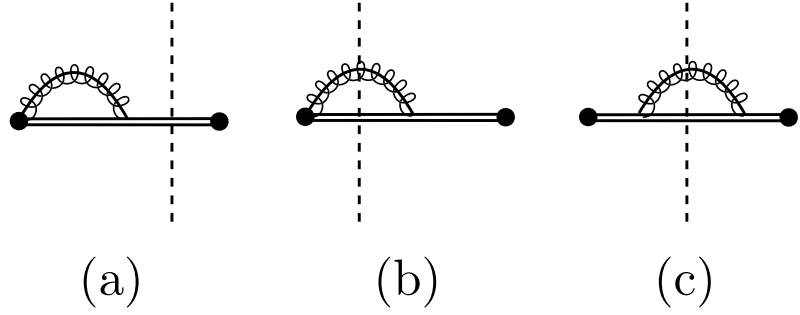}
\caption{$\mathcal{O}(\alpha_s)$ Feynman diagrams for the $\bar n$ jet function}
\label{fig:35fig4}
\end{figure}

In this section, we calculate the DIS jet function with the Delta regulator. The final jet function is defined in Eq.~\eqref{eq:35eq7} and has been previously calculated to $\mathcal{O}(\alpha_s)$ by \cite{Manohar:2003vb,Chay:2013zya,Hornig:2009kv,Bauer:2010vu,Becher:2006qw} with different regulators. Here we use the Delta-regulator prescription introduced by \cite{Chiu:2009yx} with $m_g^2$ in the gluon propagator and two Delta regulators for two Wilson lines. The Delta regulators are added to the collinear and soft Wilson lines the same way as in \sec{I.5.3.1.5a}. The $\mathcal{O}(\alpha_s)$ Feynman diagrams for the DIS jet function are shown in \fig{35fig4} where we omit the mirror images of \fig{35fig4}(a) and (b).

The naive amplitude for virtual gluon emission in \fig{35fig4}(a) is
\begin{align}\label{eq:35eqA1}
\hat M_a^{\text{jet}}=&
\:2ig^2C_F\mu^{2\epsilon}\delta(r)\int\!\frac{d^Dq}{(2\pi)^D}
\frac{n\!\cdot\!(p-q)}{q^2\!-\!m_g^2\!+\!i\epsilon}\oneov{(p\!-\!q)^2\!-\Delta_1\!+\!i\epsilon}\oneov{n\!\cdot\! q+\delta_2\!+\!i\epsilon}\,,
\end{align}
where $p_X$ is the DIS final jet momentum. We let $p_X^\mu=(p_X^+,p_X^-,p_{X\perp})=(Q,r,0)$ and \req{eq:35eqA1} becomes
\begin{align}
\hat M_a^{\text{jet}}&=\paren{-\frac{\alpha_sC_F}{2\pi}}\delta(r)\bigg\{-\oneov{\epsilon}\paren{\ln\frac{\delta_2}{p^+}+1}-\ln\frac{\mu^2}{m_g^2}\paren{\ln\frac{\delta_2}{p^+}+1}\nonumber\\
&-\left[\ln\paren{1-\frac{\Delta_1}{m_g^2}}\ln\frac{\Delta_1}{m_g^2}+1-\frac{\Delta_1/m_g^2}{\frac{\Delta_1}{m_g^2}-1}\ln\frac{\Delta_1}{m_g^2}+Li_2\paren{\frac{\Delta_1}{m_g^2}}-\frac{\pi^2}{6}\right]\bigg\}\,,\label{eq:35eqA2}
\end{align}
which has the same form as the naive amplitude of the DIS $n$-collinear function \fig{35fig2}(a). The zero-bin for \fig{35fig4}(a) is
\begin{align}
\hat M_{a\phi}^{\text{jet}}&=
(2ig^2C_F)\mu^{2\epsilon}\delta(r)\int\!\frac{d^Dq}{(2\pi)^D}\oneov{q^+}\oneov{\paren{q^--\frac{q_\perp^2+m_g^2-i\epsilon}{q^+}}}\oneov{q^-+\delta_1-i\epsilon}\oneov{-q^+-\delta_2-i\epsilon}\nonumber \\
&=-\frac{\alpha_s C_F}{2\pi}\delta(r)\bigg\{ \oneov{\epsilon^2}+\oneov{\epsilon}\ln\frac{\mu^2}{\delta_1\delta_2}+\ln\!\paren{\frac{\mu^2}{m_g^2}}\ln\!\paren{\frac{\mu^2}{\delta_1\delta_2}}-\oneov{2}\ln^2\paren{\frac{\mu^2}{m_g^2}}
\nonumber \\&
-Li_2\paren{1-\frac{\delta_1\delta_2}{m_g^2}}+\frac{\pi^2}{12}\bigg\}\,,\label{eq:35eqA3}
\end{align}
which, as expected, has the same form as the zero-bin amplitude of DIS $n$-collinear function \fig{35fig2}(a).
Including the mirror image diagram, the amplitude for the virtual gluon loop correction to the final jet function is
\begin{align}
M_a^{\text{jet}}&=2(\hat M_a^{\text{jet}}-\hat M_{a\phi}^{\text{jet}})
\nonumber \\
&=-\frac{\alpha_sC_F}{2\pi}2\delta(r)\bigg\{-\oneov{\epsilon^2}-\bigg(\oneov{\epsilon}+\ln\frac{\mu^2}{m_g^2}\bigg)\!\bigg(\!\ln\frac{\mu^2}{\Delta_1}+1\bigg)-\ln\!\bigg(1-\frac{\Delta_1}{m_g^2}\bigg)\ln\frac{\Delta_1}{m_g^2}
\nonumber\\&
+\oneov{2}\ln^2\!\paren{\frac{\mu^2}{m_g^2}}+\frac{\Delta_1/m_g^2}{1-\frac{\Delta_1}{m_g^2}}\ln\frac{\Delta_1}{m_g^2}
+Li_2\!\paren{\frac{\Delta_1}{m_g^2}}-Li_2\!\paren{1-\frac{\delta_1\delta_2}{m_g^2}}+1-\frac{\pi^2}{12}\bigg\}\,.\label{eq:35eqA4}
\end{align}

The naive amplitude for the real gluon emission in \fig{35fig4}(b) is
\begin{align}
\hat M_b^{\text{jet}}&=
(4\pi g^2C_F)\frac{\mu^{2\epsilon}\,n\cdot p_X}{p_X^2-\Delta_1+i\epsilon} \nn\\
&\times\int\! \frac{d^Dq}{(2\pi)^D}\frac{n\cdot(p-q)}{n\cdot q+\delta_2}
\delta(q^2\!-\!m_g^2)\delta[(p-q)^2-\Delta_1]\theta(p^+\!-q^+)\theta(p^-\!-q^-)\,,\label{eq:35eqA5}
\end{align}
where we use $\Delta_1$ to regulate $\bar n$-direction final jets.  Integrating yields
\begin{equation}\label{eq:35eqA6}
\hat M_b^{\text{jet}}=\frac{\alpha_sC_F}{2\pi Q}\left\{\delta(z)\ln\!\bigg(\!-\frac{\delta_1}{Q}\bigg)\ln\frac{\delta_2}{p^+}+\delta(z)\ln\!\bigg(\!-\frac{\delta_1}{Q}\bigg)-\paren{\oneov{z}}_+\bigg(\ln\frac{\delta_2}{p^+}+1\bigg)\right\}\,.
\end{equation}
The zero bin for this amplitude is
\begin{align}
\hat M_{b\phi}^{\text{jet}}&=(-4\pi g^2C_F)\mu^{2\epsilon}\frac{n\cdot p_X}{p_X^2-\Delta_1}\int\frac{d^Dq}{(2\pi)^D}\frac{\delta(q^2-m_g^2)}{n\cdot q+\delta_2}
\theta(p^--q^-)\delta\paren{p^--q^--\frac{\Delta_1}{p^+}} 
\nonumber\\&
=-\frac{\alpha_sC_F}{2\pi Q}\bigg\{\oneov{\epsilon}\left[-\delta(z)\ln\!\bigg(-\frac{\delta_1}{Q}\bigg)+\paren{\oneov{z}}_+\right]
\nonumber \\ &
-\ln\paren{\frac{\delta_2 Q}{\mu^2}}\left[\delta(z)\ln\!\bigg(\!-\frac{\delta_1}{Q}\bigg)-\paren{\oneov{z}}_{\!+}\,\right]
+\delta(z)\,Li_2\!\bigg(\frac{Q}{\delta_1}\bigg)+\paren{\frac{\ln z}{z}}_+\bigg\}\,.\label{eq:35eqA7}
\end{align}
Including the mirror image diagram, the amplitude of final jet function for real gluon emission in \fig{35fig4}(b) is
\begin{align}
M_b^{\text{jet}}&=2(\hat M_b^{\text{jet}}-\hat M_{b\phi}^{\text{jet}})
\nonumber \\ &
=\frac{\alpha_sC_F}{2\pi Q}2\bigg\{
\oneov{\epsilon}\left[-\delta(z)\ln(-\frac{\delta_1}{Q})+\paren{\oneov{z}}_+\right] \nn\\
&+\delta(z)\bigg[-\oneov{2}\ln^2\frac{\delta_1}{Q}+i\pi\ln\frac{\delta_1}{Q} +\ln\!\bigg(\!\!-\frac{\delta_1}{Q}\bigg)\Big(1+\ln \frac{\mu^2}{n\cdot p Q}\Big)\bigg] \nn \\
&+\delta(z)\frac{\pi^2}{3}-\bigg(\oneov{z}\bigg)_{\!+}\Big(1+\ln \frac{\mu^2}{n\cdot p Q}\Big) +\paren{\frac{\ln z}{z}}_+ \bigg\}\,.\label{eq:35eqA8}
\end{align}
The naive amplitude for real gluon emission in \fig{35fig4}(c) is
\begin{align}
\hat M_c^{\text{jet}}&=\paren{\frac{-g^2C_F}{2\pi}}\frac{(in\cdot p_X)^2}{(p_X^2-\Delta_1)^2}(D-2)\mu^{2\epsilon}\int\frac{d^Dq}{(2\pi)^D}(i)(-2\pi i)\delta(q^2-m_g^2)\nonumber\\
&\times(-2\pi i)\delta[(p-q)^2-\Delta_1]\frac{in\cdot(p-q)\,q_\perp^2}{[n\cdot(p-q)]^2}\theta(p^+-q^+)\theta(p^--q^-) \notag \\
&=\paren{\frac{\alpha_sC_F}{2\pi Q}}\oneov{2}\left[-\delta(z)\ln\frac{\delta_1}{Q}+\paren{\oneov{z}}_+\right]\,.\label{eq:35eqA9}
\end{align}
The zero bin for this diagram is
\begin{align}
\hat M_{c\phi}^{\text{jet}}&=\paren{\frac{-g^2C_F}{2\pi}}\frac{(in\cdot p_X)^2}{(p_X-\Delta_1)^2}(D-2)\mu^{2\epsilon}\int\frac{d^Dq}{(2\pi)^D}i(-2\pi i)\delta(q^2-m_g^2)\nonumber \\
&\times\frac{q_\perp^2}{(n\cdot p)^2}(in\cdot p)(-2\pi i)\delta(p^-(p^+-q^+)-\Delta_1)\theta(p^--q^-)\nonumber \\
&=0 \,.\label{eq:35eqA10}
\end{align}
There is no mirror image for \fig{35fig4}(c), so
\begin{equation}\label{eq:35eqA12}
M_c^{\text{jet}}=\hat M_c^{\text{jet}}\,.
\end{equation}
The wavefunction contribution to the final jet function is
\begin{equation}\label{eq:35eqA13}
M_{\text{wave}}^{\text{jet}}=\frac{\alpha_sC_F}{2\pi Q}\paren{\oneov{\epsilon}-1-\ln\frac{m_g^2}{\mu^2}}\delta(z)\,.
\end{equation}
Combining all the results above, the $\mathcal{O}(\alpha_s)$ expression for the final jet function is
\begin{align}
M^{\text{jet}}&=M_a^{\text{jet}}+M_b^{\text{jet}}+M_c^{\text{jet}}-\oneov{2}M_{\text{wave}}^{\text{jet}}\nonumber \\
&=\paren{\frac{\alpha_sC_F}{\pi Q}}\bigg\{\oneov{\epsilon^2}+\oneov{\epsilon}\left[\frac{3}{4}\delta(z)+\ln\frac{\mu^2}{\Delta_1}\delta(z)+\ln(-\frac{\delta_1}{Q})-\paren{\oneov{z}}_+\right]\nonumber \\
&+\delta(z)\Bigg[\frac{3}{4}\ln\frac{\mu^2}{m_g^2}+\frac{3}{4}\ln(-\frac{\delta_1}{Q})
+\ln\!\paren{1-\frac{\Delta_1}{m_g^2}}\ln\frac{\Delta_1}{m_g^2}+\frac{7}{8}+\frac{\Delta_1/m_g^2}{1-\frac{\Delta_1}{m_g^2}}\ln\frac{\Delta_1}{m_g^2}+\frac{\pi^2}{4}\Bigg] \nn \\
&+\frac34\paren{\oneov{z}}_++\paren{\frac{\ln z}{z}}_++\ln\paren{\frac{n\cdot p Q}{\mu^2}}\paren{\oneov{z}}_+
\bigg\}\,.
\end{align}
This result is independent of $\delta_2$ for the same reason the $n$-collinear function in Eq.~\eqref{eq:35eq130} is independent of $\delta_1$.
The counter-term for the final jet function is
\begin{equation}\label{eq:35eqA14}
\mathcal{Z}_{\bar n}^{\text{jet}}=\delta(z)+\frac{\alpha_sC_F}{2\pi}\left(\delta(z)\left(\frac34+\ln\frac{\mu^2}{\Delta_1}+\ln\left(-\frac{\Delta_1}{n\cdot p}\right)\right)-\left(\frac{1}{z}\right)_+\right)\,.
\end{equation}
With the choice $-\delta_1 Q=m_g^2$, we have the renormalized jet function,
\begin{align}
M_{\text{jet}}^R=\frac{\alpha_sC_F}{\pi}\frac{1}{Q}\bigg\{&
\delta(z)\left[\frac34\ln\!\left(\!-\frac{\mu^2}{Q^2}\right)+\frac78+\frac{\pi^2}{4}\right]
+\left(\frac{\ln z}{z}\right)_{\!+}\! +\bigg[\ln\frac{Q^2}{\mu^2}+\frac{3}{4}\bigg]\left(\frac{1}{z}\right)_{\!+}
 \bigg\}\,. \label{eq:35eqA15}
\end{align}


\subsection{Kinematic Constraints of the Zero-bin Subtraction with the Rapidity Regulator}
\label{appx:B.3.2}

The gauge-invariant rapidity regulator automatically ensures the zero-bins of the following forms of integrals are scaleless:
\begin{enumerate}
\item integrals in virtual diagrams, and
\item integrals in real diagrams with measurement functions only involving $\vec k_\perp$\,.
\end{enumerate}
However, we encounter integrals for both DIS and DY real soft functions that are not included in the above cases. As a result, we must examine the zero-bin subtraction prescriptions for each of these soft functions carefully to determine whether or not any momenta run into the collinear region.

\begin{figure}[h!]
\includegraphics[width=.65\textwidth]{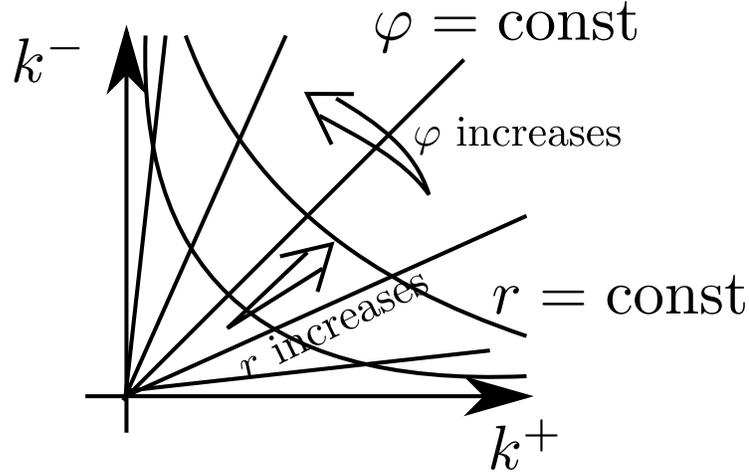}
\centering
\caption{The integration area of $k^+,k^-$ and $r,\varphi$}
\label{fig:35fig5}
\end{figure}

After integrating over the perpendicular momentum, the real soft functions for DIS and DY have the following forms respectively
\begin{align}
I^{\text{DY}}&=\int_0^\infty\!\! dk^+\int_0^\infty \!\!dk^- \frac{|k^+k^-\!-m_g^2|^{-\epsilon}}{k^+k^-}\theta(k^+k^-\!-\!m_g^2)|k^+\!-\!k^-|^{-\eta}\delta(2l-k^+-k^-)\,, \label{eq:35eqB1}\\
I^{\text{DIS}}&=\int_0^\infty\!\! dk^+\int_0^\infty\!\! dk^- \frac{|k^+k^--m_g^2|^{-\epsilon}}{k^+k^-}\theta(k^+k^--m_g^2)|k^+-k^-|^{-\eta}\delta(l-k^-)\,.\label{eq:35eqB2}
\end{align}
In order to illustrate the origins of the rapidity divergences and the zero bins, we choose a different set of the variables,
\begin{equation}\label{eq:35eqB3}
k^+=re^\varphi, \quad k^-=re^{-\varphi} 
\end{equation}
so that 
\begin{align}
I^{\text{DY}}&=\int_{-\infty}^\infty d\varphi \int_{m_g}^\infty dr\frac{2^{1-\eta}}{r^{1+\eta}}\frac{|r^2-m_g^2|^{-\epsilon}}{|\sinh\varphi|^\eta}\frac{1}{2\cosh\varphi}\delta\left(r-\frac{l}{\cosh\varphi}\right)\label{eq:35eqB4}\\
I^{\text{DIS}}&=\int_{-\infty}^\infty d\varphi\int_{m_g}^\infty dr\frac{2^{1-\eta}}{r^{1+\eta}}\frac{|r^2-m_g^2|^{-\epsilon}}{|\sinh\varphi|^\eta}\delta\left(r-le^\varphi\right)e^\varphi\,. \label{eq:35eqB5}
\end{align}
As we can see in Eq.\,\eqref{eq:35eqB1} and Eq.\,\eqref{eq:35eqB2}, $|k^+-k^-|\to\infty$ can bring in both a rapidity divergence and an ultraviolet divergence. We separate these two types of divergences by working with the $r$ and $\varphi$ variables, because the rapidity divergence is only brought in by $|\sinh\varphi|\to\infty$, and the infrared regulator $m_g^2$ distinguishes an infrared divergence from a rapidity divergence. We illustrate the relations of these two sets of variables in \fig{35fig5}. The hyperbolas show the on-shell
condition $k^+k^-=k_\perp^2+m_g^2$, and the zero bins are the rapidity regions $k^+\gg k^-$, $k^-\gg k^+$, or $\varphi\gg 0$, which is also known as the collinear contribution to the soft function.

\begin{figure}[h!]
\includegraphics[width=.65\textwidth]{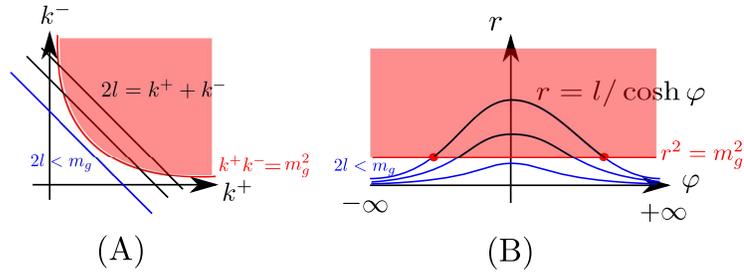}
\centering
\caption{The kinematic constraints for DY real soft function. (A) is kinematic constraint in $(k^+,k^-)$ space; (B) is kinematic constraint in $(r,\varphi)$ space.}
\label{fig:35fig6}
\end{figure}

\begin{figure}[h!]
\centering
\includegraphics[width=.65\textwidth]{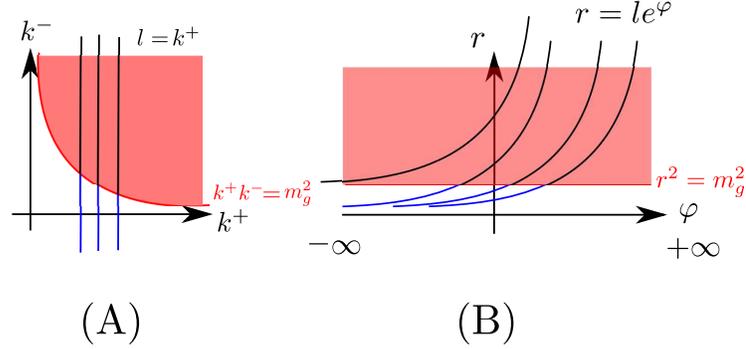}
\caption{The kinematic constraints for DIS real soft function. (A) is kinematic constraint in $(k^+,k^-)$ space; (B) is kinematic constraint in $(r,\varphi)$ space.}
\label{fig:35fig7}
\end{figure}

The kinematic constraints are shown in Fig.~\ref{fig:35fig6} and Fig.~\ref{fig:35fig7}.
In Fig.\,\ref{fig:35fig6}, the (red) shaded part is the integration area, which is constrained by the infrared regulator $m_g^2$. The black lines are the constraints brought in by the measurement function. In Fig.\,\ref{fig:35fig6}(A), while $l$ becomes large, it is difficult to tell whether the zero bins $k^+\gg k^-$ or $k^-\gg k^+$ contribute to the naive soft function integral. However, in Fig.\,\ref{fig:35fig6}(B), it is very clear that when integrating over the black curve $r\cosh\varphi=l$, $r^2=m_g^2$ cuts off all the collinear contributions from $\varphi\to +\infty$ or $\varphi\to-\infty$.

Therefore, we can conclude that there is no rapidity divergence in the DY real soft function. Interestingly because of the constraint from the measurement function, $r$ is always bounded by $l$, which suggests we do not have the ultraviolet divergence for this function either.

We analyze the DIS real function in a similar manner in Fig.~\ref{fig:35fig7}.  Because the infrared regulator  does not exclude the region $\varphi\to\infty$, collinear momenta contribute to the integral, which brings in the rapidity divergence and requires the zero-bin subtraction.

Carrying out the integrals for the DY  real soft function
\begin{align}
I^{\text{DY}}&=\int_{-\arccosh(l/m_g)}^{\arccosh(l/m_g)}\frac{1}{2^\eta l^{1+\eta}}\frac{\cosh^\eta\varphi}{|\sinh\varphi|^\eta}\left|\frac{l^2}{\cosh^2\varphi}-m_g^2\right|^{-\epsilon}d\varphi
\nonumber \\
&=\frac{\Gamma(\epsilon)\Gamma(1-\epsilon)}{2^\eta (m_g^2)^\epsilon}\frac{1}{(l^2-m_g^2)^{\frac{1+\eta}{2}}}-\frac{\Gamma(1-\epsilon)\Gamma(\frac{1-\eta}{2})}{\epsilon 2^\eta\Gamma\left(\frac{1-2\epsilon-\eta}{2}\right)}\frac{1}{(l^2-m_g^2)^{\frac{1+2\epsilon+\eta}{2}}}+..., 
\label{eq:35eqB6}
\end{align}
and DIS real soft function
\begin{align}
I^{\text{DIS}}&=\int_{\ln(m_g/l)}^\infty \frac{2^{1-\eta}}{l^{1+\eta}}\frac{e^\varphi}{e^{\varphi(1+\eta)}}\frac{|l^2e^{2\varphi}-m_g^2|^\epsilon}{|\sinh\varphi|^\eta}d\varphi 
\nonumber \\&
=\frac{2^{-\eta}}{l^{1+\eta}(m_g^2)^{\epsilon}}\Gamma(\epsilon)\,.\label{eq:35eqB7}
\end{align}
The $...$ in the DY integral are terms $\mathcal{O}\left(\frac{l^2}{m_g^2}-1\right)^{-3/2}$ and higher, which do not contribute to the singularities.  For DY, Eq.\,\eqref{eq:35eqB6} shows that the $\epsilon$ ultraviolet poles cancel between the two terms, and $\eta$ and $\epsilon$ do not regulate $l$ in the factors $(l^2-m_g^2)^{-(1+\eta)/2}$ and $(l^2-m_g^2)^{-(1+2\epsilon+\eta)/2}$. However for DIS, we can extract both rapidity and ultraviolet poles in Eq.\,\eqref{eq:35eqB7}. This analysis clearly shows that the zero bin subtraction is required only in the presence of the rapidity divergences.

The kinematic constraints seen in Fig.~\ref{fig:35fig7} actually produce two distinct zero-bin subtractions in DIS: the first is the ``intuitive'' collinear area in which $k^-\gg k^+$ with $l$ fixed.  This case corresponds to $\varphi\to-\infty$ with $r$ fixed.  The second collinear area occurs when $k^+\gg k^-$, because $l$ is large and the measurement function $\delta(l-k^+)$ fixes $k^+=l$.   In DY, we cannot separate the limits $k^+\gg k^-$ and $k^-\gg k^+$ in the integrand of \eq{35eqB6} because this requires letting $l$ become large, which opens up phase space at \emph{both} $\varphi$ large and positive and $\varphi$ large and negative, see Fig \ref{fig:35fig6}(B). Therefore DY does not have distinct $k^+\gg k^-$ and $k^-\gg k^+$ areas, which is equivalent to the statement that there is no zero bin subtraction.  


\subsection{Alternative Method for One-loop DIS Real Soft Function Integral}
\label{appx:B.3.3}


We start from the integral of the form\footnote{This section is based on discussions and correspondence with S. Fleming, A. Hoang, P. Pietrulewicz and D. Samitz.}
\begin{align}
S_r&=32\pi^2\alpha_s C_F\tilde \mu^{2\epsilon}\nu^\eta\int \frac{d^dk}{(2\pi)^d}\delta(k^2-m_g^2)\delta(\ell-k^+)\frac{|k^+\!-\!k^-|^{-\eta}}{k^+k^-}\nn\\
&=\frac{\as C_F}{\pi}\frac{\paren{\mu^2 e^{\gamma_E}}^\epsilon \nu^\eta}{\Gamma(1-\epsilon)}\int \!\!dk^+ dk^-dk_\perp^2\,\Theta(k_\perp^2)\delta\paren{k_\perp^2\!-\!(k^+k^-\!-\!m_g^2)}\delta(\ell\!-\!k^+)\frac{|k^+\!-\!k^-|^{-\eta}}{(k_\perp^2)^{\epsilon}\,k^+k^-}\nn\\
&=\frac{\as C_F}{\pi}\frac{\paren{\mu^2 e^{\gamma_E}}^\epsilon \nu^\eta}{\Gamma(1-\epsilon)}  \oneov{\ell}\int \!dk^-\, \Theta(\ell k^--m_g^2)\frac{(\ell k^--m_g^2)^{-\epsilon}|\ell-k^-|^{-\eta}}{k^-}\nn\\
&=\frac{\as C_F}{\pi}\frac{\paren{\mu^2 e^{\gamma_E}}^\epsilon \nu^\eta}{\Gamma(1-\epsilon)}\oneov{\ell^{1+\epsilon}}
\nn\\&\times
\int\! dk^-\,\Theta\!\paren{\!k^-\!-\!\frac{m_g^2}{\ell}}\frac{\!\Theta(\ell\!-\!k^-)(\ell\!-\!k^-)^{-\eta}+\Theta(k^-\!-\!\ell)(k^-\!-\!\ell)^{-\eta}}{\paren{k^--\frac{m_g^2}{\ell}}^{\epsilon} k^-},\label{eq:appB.3.3.1eq1}
\end{align}
which corresponds to Eq.~(76) in \cite{Hoang:2015iva} and Eq.~(30) in \cite{Fleming:2016nhs}. The theta-functions can be rewritten as
\begin{align}
\Theta&\paren{k^--\frac{m_g^2}{\ell}}\left[\Theta(\ell-k^-)(\ell-k^-)^{-\eta}+\Theta(k^--\ell)(k^--\ell)^{-\eta}\right]\nn\\
&=\Theta(\ell-m_g)\left[\Theta\paren{k^--\frac{m_g^2}{\ell}}\Theta(\ell-k^-)(\ell-k^-)^{-\eta}+\Theta(k^--\ell)(k^--\ell)^{-\eta}\right]\nn\\
&+\Theta(m_g-\ell)\left[\Theta\paren{k^--\frac{m_g^2}{\ell}}(k^--\ell)^{-\eta}\right]\,,
\label{eq:appB.3.3.1eq2}
\end{align}
which leads to the following integrals:
\begin{subequations}
\begin{align}
I_{A+B}=&\:
\frac{\Theta(\ell-m_g)\nu^\eta}{\ell^{1+\epsilon}}\Bigg[\int_{\frac{m_g^2}{\ell}}^\ell dk^-\frac{\paren{k^--\frac{m_g^2}{\ell}}^{-\epsilon}(\ell-k^-)^{-\eta}}{k^-}
\nn\\
&+\int_\ell^\infty dk^-\frac{\paren{k^--\frac{m_g^2}{\ell}}^{-\epsilon}(k^--\ell)^{-\eta}}{k^-}\Bigg] 
\end{align}
and
\begin{align}
I_C=&\:\frac{\Theta(m_g-\ell)\nu^\eta}{\ell^{1+\epsilon}}\int_{\frac{m_g^2}{\ell}}^\infty dk^-\frac{\paren{k^--\frac{m_g^2}{\ell}}^{-\epsilon}(k^--\ell)^{-\eta}}{k^-}\,.
\end{align}
\end{subequations}
The integrals in $I_{A+B}$ do not lead to any rapidity divergences, so the regulator $\eta$ can be dropped and the two integrals can be merged into one:
\begin{align}
I_{A+B}&=\frac{\Theta(\ell-m_g)}{\ell^{1+\epsilon}}\int_{\frac{m_g^2}{\ell}}^\infty dk^-\frac{\paren{k^--\frac{m_g^2}{\ell}}^{-\epsilon}}{k^-}\nn\\
&=\Theta(\ell-m_g)\Gamma(\epsilon)\Gamma(1-\epsilon)(m_g^2)^{-\epsilon}\oneov{\ell}\nn\\
&=\Theta(\ell-m_g)\Gamma(\epsilon)\Gamma(1-\epsilon)(m_g^2)^{-\epsilon}\oneov{\nu}\left[\frac{\Theta(\tilde\ell)}{\tilde\ell}\right]_+\,,\label{eq:appB.3.3.1eq4}
\end{align}
with $\tilde\ell=\ell/\nu$. In the last step the term $1/\ell$ was rewritten as a plus-distribution, which can be done because the step function in front ensures that $\ell>m_g$ and so the pole at $\ell=0$ does not contribute. In this form it will be easier to combine this with the result of the remaining integral $I_C$.

For $I_C$ we get
\begin{align}
I_C&=\Theta(m_g-\ell)\Gamma(1-\epsilon)(m_g^2)^{-\epsilon-\eta}\frac{\nu^\eta}{\ell^{1-\eta}}\frac{\Gamma(\epsilon+\eta)}{\Gamma(1+\eta)}{}_2F_1\paren{\eta,\epsilon+\eta,1+\eta,\frac{\ell^2}{m_g^2}}\nn\\
&=\Theta(m_g-\ell)\Gamma(1-\epsilon)(m_g^2)^{-\epsilon}\oneov{\nu}\paren{\frac{\nu^2}{m_g^2}}^{\eta}\oneov{\tilde\ell^{1-\eta}}\frac{\Gamma(\epsilon+\eta)}{\Gamma(1+\eta)}{}_2F_1\paren{\eta,\epsilon+\eta,1+\eta,\frac{\tilde\ell^2\nu^2}{m_g^2}}\,.\label{eq:appB.3.3.1eq5}
\end{align}
Expanding in $\eta$ yields,
\begin{equation}\label{eq:appB.3.3.1eq6}
I_C=\Theta(m_g-\ell)\Gamma(\epsilon)\Gamma(1-\epsilon)(m_g^2)^{-\epsilon}\oneov{\nu}\left\{\delta(\tilde\ell)\paren{\oneov{\eta}+\ln\frac{\nu^2}{m_g^2}+H_{\epsilon-1}}\!+\!\left[\frac{\Theta(\tilde\ell)}{\tilde\ell}\right]_+\!+\cO(\eta)\right\},
\end{equation}
where $H_\alpha=\psi(1+\alpha)+\gamma_E$ is the Harmonic number. For the delta-distribution the step function $\Theta(m_g-\ell)$ can be dropped, and the term with the plus-distribution matches exactly the result of $I_{A+B}$ such that $\Theta(m_g-\ell)+\Theta(\ell-m_g)=1$.

The final result for the real radiation soft function diagram with a massive gluon using the $\eta$ rapidity regulator is:
\begin{align}
S_r&=\frac{\as C_F}{\pi}\frac{(\mu^2e^{\gamma_E})^{\epsilon}}{\Gamma(1-\epsilon)}\paren{I_{A+B}+I_C}\nn\\
&=\frac{\as C_F}{\pi}\Gamma(\epsilon)\paren{\frac{\mu^2e^{\gamma_E}}{m_g^2}}^\epsilon\oneov{\nu}\left\{\delta(\tilde\ell)\paren{\oneov{\eta}+\ln\frac{\nu^2}{m_g^2}+H_{\epsilon-1}}+\left[\frac{\Theta(\tilde\ell)}{\tilde\ell}\right]_++\cO(\eta)\right\}\,.\label{eq:appB.3.3.1eq7}
\end{align}



\begin{figure}
\centering
\includegraphics[width=.55\textwidth]{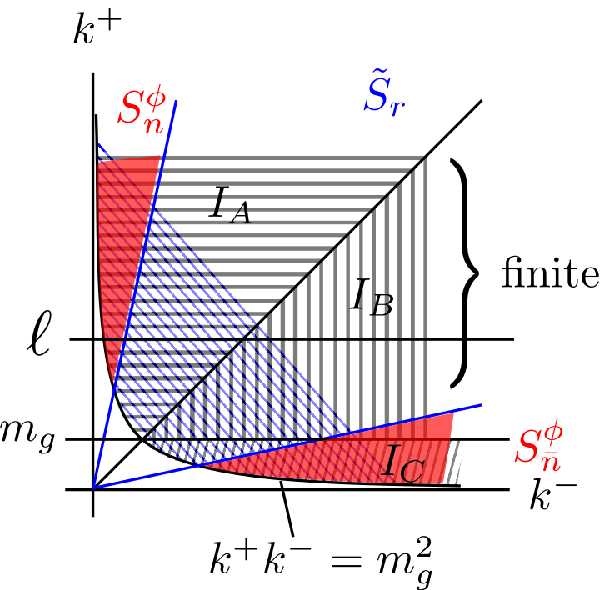}
\caption{ Separation of integration regions in \eq{appB.3.3.1eq8}.  The total integral $I_A+I_B+I_C$ is equivalent to $\tilde S_r-S_n^\phi-S_{\bar n}^\phi$. }
\label{fig:appB.3.3.1fig1}
\end{figure}

In the above, the integral is decomposed by partitioning the integration region into three parts, $S=I_A+I_B+I_C$ as shown above by three different shaded areas, while in \cite{Fleming:2016nhs} we have 
\begin{equation}\label{eq:appB.3.3.1eq8}
S=-S_{\bar n}^\phi+\mathcal{O}\paren{\frac{m_g}{l}}
\end{equation}
where we subtract the large rapidity integrand which approximately amounts to cutting off the regions shown above in red, beyond the two blue lines. $S_n^\phi$ differs from $S_r$ only by $\mathcal{O}(m_g/l)$.  Therefore our integral is $-S_{\bar n}^\phi$ which accounts for part of $I_B$ and $I_C$, and as we can see from the diagram above, the difference is arbitrarily small when $m_g$ is sufficiently small.


%% file: appendix_ChPT.tex
\chapter{Chiral Perturbation Theory and Heavy Quark Effective Theory}

\section{Chiral Perturbation Theory}
\label{appx:ChPT}

In this appendix, we briefly introduce the motivation and setup of chiral perturbation theory.

In studying the hadron spectrum, the relevant separation is the large gap between the pion mass and the masses of the vector mesons such as $\rho(770)$ and $\omega(782)$. Thus one takes the pion mass as the soft scale $Q\sim m_\pi$ and the $\rho$ mass as the hard scale $m_\rho\sim 700\,\text{MeV}\sim\Lambda_\chi $, the chiral symmetry-breaking scale. It is then natural to consider an expansion in terms of the ratio $Q/\Lambda_\chi$.  For the ground state and low-lying excitations in the spectrum of nuclei and conventional nuclear forces, the relevant degrees of freedom are nucleons and mesons (and possibly also low energy resonances) instead of quarks and gluons. 

To establish a link to QCD and ensure it is not just another phenomenology, the EFT incorporates all relevant symmetries of the underlying theory, as described in the introduction for bottom-up type effective theories.  Seeing that the scale of nonperturbative strong interaction physics is large compared to the `light' quark $u,d$ masses, we can study the approximation of QCD with $m_u,m_d$ set to zero and investigate perturbation theory in terms of $m_q$.  The relevant isospin symmetry of the D-meson system I study involves only the $u,d$ quarks, and I will consider only the two-flavor chiral symmetry.  The strange quark is sometimes also considered light, as $m_s\sim 100$ MeV\,$<\Lambda_\chi\sim 600-1000$ MeV.  Study of heavy hadron chiral perturbation theory in the three-flavor case can be found in \cite{Stewart:1999um}.

In the $m_q\to 0$ limit the light quark Lagrangian for the isospin doublet $q^T=(u~d)$ 
\begin{equation}\label{eq:69eq1.93}
\cL_{\text{light quarks}}=\bar qi\slashed{D} q=\bar q_Li\slashed{D} q_L+\bar q_Ri\slashed{D} q_R\,,
\end{equation}
exhibits chiral symmetry $SU(2)_L\times SU(2)_R$ where the two factors transform left- and right-handed quark fields $q$ differently:
\begin{equation}\label{eq:69eq1.94}
q_L\to L q_L, \qquad q_R\to R q_R\,.
\end{equation}
For this reason, $m_q\to 0$ is also known as the chiral limit.
The Lagrangian \eq{69eq1.93} also exhibits an $U(1)$ baryon number symmetry with phase transformation common to both $L$ and $R$ quarks, and an axial $U(1)$ symmetry with phase transformation opposite for $L$ and $R$. The axial $U(1)$ however alters the measure of the path integral (\cite{Fujikawa:2004cx}), and therefore at quantum level is no longer a symmetry of QCD, a phenomenon known as the axial anomaly.

The chiral symmetry $SU(2)_L\times SU(2)_R$ of massless QCD with 3 flavors of quarks undergoes spontaneous breaking by non-perturbative strong interaction dynamics, which give rise to the vacuum expectation value (VEV)
\begin{equation}\label{eq:69eq1.95}
\langle \bar q_R^a q_L^b\rangle =v\delta^{ab}\,,
\end{equation}
where $a,b$ are flavor indices with values 1,2 standing for $u,d$ respectively, color indices are suppressed, and $v\sim \cO(\lqcd^3)$ is a constant. Under an $SU(2)_L\times SU(2)_R$ transformation represented by $q\to q'$, the VEV transforms as
\begin{equation}\label{eq:69eq1.96}
\langle \bar q_R^{'a}q_L^{'b}\rangle =v(LR^\dag)^{ab}\,.
\end{equation}
The VEV is invariant under the transformation only with $L=R$, and thus chiral symmetry is spontaneously broken to its diagonal subgroup $SU(2)_V$. The composite field $\bar q_R^a q_L^b$ is transformed along symmetry directions by the three generators of the $SU(2)_V$, implying the potential energy is unchanged by $SU(2)_V$ transformations.  Thus excitations along these directions give rise to 3 Goldstone bosons. These fields are denoted by the $2\times 2$ special unitary matrix $\Sigma(x)$ representing the possible low-energy long-wavelength excitations of $\bar q_R q_L$.   $\Sigma$ transforms as
\begin{equation}\label{eq:69eq1.97}
\Sigma\to L\Sigma R^\dag\,,
\end{equation}
and $v\Sigma_{ab}(x)\sim \bar q_R^a(x)q_L^b(x)$ provides local orientation of quark condensate in $SU(2)_V$ space, so that the expectation value of $\Sigma$ is $\langle\Sigma\rangle=\mathbb{I}$.

At leading power in the expansion in small quark masses, the low energy interactions concerning these Goldstone bosons are given by the most general effective Lagrangian for $\Sigma$, invariant under \eq{69eq1.97}:
\begin{equation}\label{eq:69eq1.98}
\cL_{eff}=\frac{f_\pi^2}{8}\tr\partial^\mu\Sigma\partial_\mu\Sigma^\dagger+\text{higher derivative terms,}
\end{equation}
with $f_\pi$ a constant having dimension of mass. As $\tr\,\Sigma\Sigma^\dagger=\tr\,\mathbb{I}=2$, a constant, only terms with derivatives yield dynamics. Higher derivative terms are suppressed by $p_{typ}^2/\Lambda_{\chi}^2$, provided the typical momentum $p_{typ}$  remains smaller than the chiral symmetry breaking scale.

Rather than the field $\Sigma_{ab}(x)$, it is more convenient to work with 
\begin{equation}\label{eq:72beq4}
\xi_{ab}=\exp\paren{iM_{ab}/f_\pi}=\sqrt{\Sigma_{ab}}
\end{equation}
with $M_{ab}$ traceless being an element of the Lie algebra associated to the $SU(2)$ group.  The coefficient $2/f_\pi$ is included to give kinetic energy terms in Lagrangian \eq{69eq1.98} the usual normalization.  $\xi$ transforms as
\begin{equation}\label{eq:72beq6}
\xi \to L\xi U^\dagger=U\xi R^\dagger
\end{equation}
which defines $U\in SU(2)_L\times SU(2)_R$. Under the reduced symmetry, $SU(2)_V$, we set $L=R=U=V\in SU(2)_V$ and obtain the simpler transformation
\begin{equation}\label{xitransform}
\xi\to V\xi V^\dagger\,.
\end{equation}
Therefore, $\Sigma\to V\Sigma V^\dagger$ and $M\to VMV^\dagger$ as well, showing that $M_{ab}$ lives in the adjoint representation. $M_{ab}$ can also be expressed in terms of the three Goldstone fields
\begin{equation}\label{eq:69eq1.100}
M_{ab}=\pi^i_{ab}\sigma^i=\begin{pmatrix}
\pi^0/\sqrt{2} & \pi^+ \\
\pi^- & -\pi^0/\sqrt{2}
\end{pmatrix}\,,
\end{equation}
showing that the $ab$ subscript is the light-flavor index.  

To see how non-zero quark mass effects are included, recall the mass term for the light quarks in the QCD Lagrangian is
\begin{equation}\label{eq:69eq1.101}
\cL_{mass}=\bar q_L m_q q_R+h.c.\,,
\end{equation}
which transforms under $SU(2)_L\times SU(2)_R$ in the representations $(\bar 2_L,2_R)+(2_L,\bar 2_R)$.  The light quark mass matrix is
\begin{equation}\label{eq:72beq18}
m_q=\begin{pmatrix}
m_u & \\ & m_d\end{pmatrix}\,.
\end{equation}
By adding terms linear in quark mass to \eq{69eq1.98}, we incorporate effect of quark mass to first order in the interaction of pseudo-Goldstone bosons.  Alternatively one can assume quark matrix itself obeys the transformation rule $m_q\to Lm_qR^\dag$ under $SU(2)_L\times SU(2)_R$, making the Lagrangian \eq{69eq1.101} invariant under chiral transformations.  Applying this rule, we can add to \eq{69eq1.98} terms linear in $m_q, m_q^\dag$ that are invariant under $SU(2)_R\times SU(2)_L$ giving
\begin{equation}\label{eq:69eq1.102}
\cL_{eff}=\frac{f_\pi^2}{8}\,\tr\left[\partial^\mu\Sigma\partial_\mu\Sigma^\dag\right]+v\,\tr\left[m_q^\dag\Sigma+m_q\Sigma^\dag\right]+\ldots
\end{equation}
where terms in \ldots have more derivatives or more $m_q$s.  The mass matrix $m_q$ is real except in the presence of CP violation.  We will not consider CP violation effects explicitly, but continue to write $m_q,m_q^\dag$ separately to manifest the symmetry in later analysis of allowed operators.  The linear term containing $m_q$ above provides masses to Goldstone bosons, given by
\begin{align}
m_{\pi^\pm}^2&=\frac{4v}{f_\pi^2}(m_u+m_d)\,.\label{eq:69eq1.103} 
\end{align}
In this case, the pions are called pseudo-Goldstone bosons, since the symmetry was only approximate before the spontaneous symmetry breaking.  Phenomenologically, we define $m_\pi^2\sim B_0(m_u+m_d)$, which suggests $B_0\sim 3$\,GeV.

There are two low energy constants in the chiral Lagrangian \eq{69eq1.102}: $v$ with dimension of $(\text{mass})^3$ and $f_\pi$ with dimension of mass. In the effective Lagrangian in \eq{69eq1.102}, quark masses are always accompanied by the parameter $v$ and as a result the effective theory of low-energy interactions of the pseudo-Goldstone bosons is only capable of providing ratios of quark masses.

\begin{figure}[!h]
\centering
\includegraphics[width=.25\textwidth]{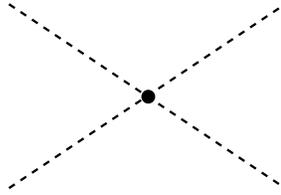}
\caption{$\pi\pi$ scattering, tree-level contribution}
\label{fig:69fig1.6}
\end{figure}

One can use the above effective theory to compute $\pi\pi$ scattering processes and similar processes involving the pseudo-Goldstone bosons. An expansion of $\Sigma$ in terms of the meson fields the factor $\tr\partial_\mu\Sigma\partial^\mu\Sigma^\dagger$ gives the four-meson interaction
\begin{equation}\label{eq:69eq1.107}
\cL_{int}=\oneov{6f_\pi^2}\tr\left([M,\partial_\mu M][M,\partial^\mu M]\right)+...\,,
\end{equation}
the tree-level matrix element of which (see \fig{69fig1.6}) gives a contribution to the $\pi\pi$ scattering amplitude of the form
\begin{equation}\label{eq:69eq1.108}
\cM\sim \frac{p_{typ}^2}{f_\pi^2}\,,
\end{equation}
where the power of $p_{typ}$ is due to the fact that the $\pi^4$ vertex contains two derivatives. The mass term also contributes a factor of similar size for $p_{typ}^2\sim m_\pi^2$.  Here we see that the cutoff scale $\Lambda_\chi\sim 4\pi f_\pi$.

\begin{figure}[!h]
\centering
\includegraphics[width=.4\textwidth]{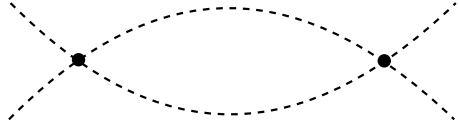}
\caption{$\pi\pi$ scattering, one-loop level contribution}
\label{fig:69fig1.7}
\end{figure}

Considering loop diagrams, there are one-loop diagrams with two $\pi^4$ vertices such as \fig{69fig1.7}. The two propagators contribute a factor of $p^{-4}$ while each vertex provides a factor of $p^2/f_\pi^2$ and a factor of $p^4$ arises from the loop integration. Therefore the amplitude in the $\MSbar$ scheme has the form
\begin{equation}\label{eq:69eq1.109}
\cM\sim\frac{p_{typ}^4}{16\pi^2f_\pi^4}\ln\frac{p_{typ}^2}{\mu^2}\,,
\end{equation}
where the numerator $p_{typ}^4$ is necessary by dimensional analysis to compensate the factor of $f_\pi^4$ in the denominator.  The subtraction point $\mu$ (also with dimension of mass) only appears inside logarithms and the factor $16\pi^2$ is typical of one-loop diagrams.  In this momentum expansion the 1-loop result contribution is of the same order as operators in the chiral Lagrangian with four derivatives or two insertions of quark mass matrix. At order $p^4$ the total amplitude is a sum of 1-loop graphs with $\cO(p^2)$ vertices and tree diagrams associated to $p^4$ term in the Lagrangian. Note also that the $\mu$ dependence in \eq{69eq1.109} is canceled by the $\mu$ dependence in the coefficient of the $p^4$ term resulting in a $\mu$-independent total $p^4$ amplitude.

The above analysis applies more generally: diagrams with more loops contribute at the same order as higher derivative terms in the Lagrangian. It is easy to show that a diagram will produce an amplitude of order
\begin{equation}\label{eq:69eq1.110}
p^D=p^{2+2L+\sum_k (k-2)n_k}\,,
\end{equation}
where $L$ is number of loops, $n_k$ is number of vertices with order $p^k$ and each vertex insertion contributes a factor  of $p^{k-2}$. Each term in the exponent of \eq{69eq1.110} is positive since Lagrangian starts from $p^2$ order. If the mass scale $\Lambda_{\chi}$ responsible for suppression of higher order derivative terms is $\Lambda_\chi\simeq 4\pi f_\pi$, loop corrections will be of similar importance to higher derivative operators. 

This computation of pseudo-Goldstone boson interactions using the effective Lagrangian under a momentum expansion is called chiral perturbation theory ($\chi$PT).

\section{Symmetries in Heavy Hadron Chiral Perturbation Theory}

In this appendix, we verify several charge, parity and time-reversal symmetries of the Heavy Hadron Chiral Perturbation Theory lagrangian we derive in Chapter \ref{sec:II.3.3}.

\subsection{Charge Conjugation and Heavy Anti-mesons}
\label{appx:antiheavymeson}

With the standard charge conjugation phase conventions, the anti-meson states should be related to the meson states by
\begin{align}
P_a^{(anti)*\mu}&=-\mathcal{C} P_a^{*\mu} \mathcal{C}^{-1}\,,\nn\\
P_a^{(anti)}&=\mathcal{C}P_a\mathcal{C}^{-1}\,,\label{eq:72beq14}
\end{align}
and the heavy anti-meson field related to the heavy meson field by
\begin{equation}\label{eq:72beq15}
H^{(anti)}_a=\mathcal{C}H_a\mathcal{C}^{-1}\,.
\end{equation}
$H^{(anti)}_a$ is a $4\times 4$ matrix in Dirac spinor space and transforms as the form above with $\mathcal{C}\to C=i\gamma^2\gamma^0$, the charge conjugation matrix for Dirac spinors.  The heavy field velocity changes like a momentum $Cv^\mu C^{-1}=v^\mu\,.$  Applying charge conjugation to Eq.(\ref{eq:72beq11}), we verify that
\begin{align}
\mathcal{C}H_a\mathcal{C}^{-1}&=[\mathcal{C}P_{a^{*\mu}}\mathcal{C}^{-1}(C\gamma_\mu^T C^{-1})-\mathcal{C}P_a\mathcal{C}^{-1}(C\gamma_5^T C^{-1})]C\paren{\frac{1+\slashed{v}}{2}}^\dg C^{-1}\nn\\
&=[P_a^{(anti)*\mu}\gamma_\mu-P^{(anti)}_a\gamma_5]\frac{1-\slashed{v}}{2}\,.\label{eq:72beq16}
\end{align}
consistent with \req{antiHdefn}.

In the rest frame, the heavy anti-field simplifies to
\begin{align}
H^{(anti)}_a&=-\begin{pmatrix} 0 & \overrightarrow{P}_a^{(anti)*}\cdot \vec\sigma+P^{(anti)}_a \\ 0 & 0 \end{pmatrix}\equiv -\begin{pmatrix} 0 & \overline{h_a}^{(anti)} \\ 0 & 0 \end{pmatrix}\,,\nn\\
\overline{H}_a^{(anti)}&=\begin{pmatrix} 0 & 0 \\ \overrightarrow{\mathcal{P}}_a^{*\dg}\cdot\vec\sigma+\mathcal{P}_a^\dg & 0 \end{pmatrix}\equiv \begin{pmatrix} 0 & 0 \\ \overline{h_a}^\dg & 0 \end{pmatrix}\,.\label{eq:72beq17e}
\end{align}
$H^{(anti)}_a$ obeys the same symmetry rules as $H_a$, and the Lagrangian from Eq.(\ref{eq:72beq9p})
remains unchanged for $H^{(anti)}_a$, except for the definition of the anti-matter fields. In the rest frame of the anti-meson, $H^{(anti)}_a$ and $\overline{H}_a^{(anti)}$ are reduced into the same form of Eq.(\ref{eq:72beq11}). Defining $h_a^{(anti)}$ and $h_a^{(anti)\dg}$ as the anti-particle fields of $h_a$ and $h_a^\dg$, we have the leading order Lagrangian for anti-mesons,
\begin{align}
\mathcal{L}_{L.O.}^{\bar D-\pi}&=-i\tr[\overline{H}_a^{(anti)} v_\mu D_{ab}^\mu H_b^{(anti)}] 
+g_\pi\tr[\overline{H}_a^{(anti)} H^{(anti)}_b\gamma_\mu\gamma_5\mba_{ba}^\mu]\nn\\
&=\tr[h_a^{(anti)\dg} iD_{ba}^0 h^{(anti)}_b]-g_\mu\tr[h_a^{(anti)\dg} h^{(anti)}_b\vec\sigma\cdot \vec\mba_{ba}]\,.\label{eq:72beq17}
\end{align}

\subsection{C,P, and T invariance constraints on the HHChPT Lagrangian}\label{app:CPTonHHChPT}

Because we work in the region of $M_D$, which is much smaller than the CP violation scale $\sim M_{W,Z}$, our effective field theory must preserve C, P and T invariance separately. Under parity transformations $D^*$, $D$ and pions are all parity odd,
\begin{align}
\mathcal{P}^{-1}P_a^{*\mu}\mathcal{P}&=(-1)^\mu P_a^{*\mu} \,,\nn \\
\mathcal{P}^{-1}P_a\mathcal{P}&=(-1)P_a \,, \nn\\
\mathcal{P}^{-1}M_\xi\mathcal{P}&=-M_\xi\,, \label{eq:72beq37}
\end{align}
and this leads to
\begin{align}
\mathcal{P}^{-1}H_a\mathcal{P}&=H_a\,, \nn \\
\mathcal{P}^{-1}\bar{H}_a\mathcal{P}&=\bar{H}_a\,, \nn \\
\mathcal{P}^{-1}\xi\mathcal{P}&=\xi^\dagger\,, \nn\\
\mathcal{P}^{-1}\mba^\mu\mathcal{P}&=-(-1)^\mu\mba^\mu\,.  \label{eq:72beq38}
\end{align}
Choosing the parity transformation Dirac Matrix representation as $P=\gamma^0$, combining with the Hermitian condition for all effective operators we find that mass terms of the form
\begin{equation}\label{appeq:72beq39}
\msl\sim\tr\left[\bar H_aH_a(\xi M_q\xi-\xi^\dg M_q\xi^\dg)\right]
\end{equation}
are forbidden.  

It is a nontrivial check of Eqs.\,\eqref{eq:72beq61} and \eqref{eq:72beq62} that the velocity reparameterization invariance-induced operators preserve C, P and T transformation separately as well as being Hermitian conjugate of themselves. Defining $\mathcal{T}$ as the time-reversal operator, we have
\begin{align}
\mathcal{T}P_a^{*\mu}\mathcal{T}^{-1}&=(-1)^\mu P_a^{*\mu}\,,\nn\\
\mathcal{T}P_a\mathcal{T}^{-1}&=(-1)P_a\,,\nn\\
\mathcal{T}M_\xi\mathcal{T}^{-1}&=-M_\xi\,.\label{eq:72beq64}
\end{align}
This leads to
\begin{align}
\mathcal{T}H_a\mathcal{T}^{-1}&=\frac{1-\slashed{v}}{2}[P_a^{*\mu}\gamma_\mu+P_a\gamma_5]=\bar H_a\,,\nn\\
\mathcal{T}\bar H_a\mathcal{T}^{-1}&=H_a\,,\nn\\
\mathcal{T}\xi\mathcal{T}^{-1}&=\xi^*\,,\label{eq:72beq65}
\end{align}
choosing the time reversal operator to be represented by the Dirac matrix $T=\gamma^1\gamma^3$, and
\begin{align}
\mathcal{T}\mba^\mu\mathcal{T}^{-1}&=(-1)^\mu\mba^\mu\,,\nn\\
\mathcal{T}V^\mu\mathcal{T}^{-1}&=-(-1)^\mu V^\mu\,,\nn\\
\mathcal{T}D^\mu\mathcal{T}^{-1}&=-(-1)^\mu D^\mu\,, \label{eq:72beq66}
\end{align}
which make it easier to see the two operators in Eqs.\,\eqref{eq:72beq61} and \eqref{eq:72beq62} are even under time reversal:
\begin{align}
\mathcal{T}\mathcal{L}_{N.L.O.}^{VRI_2}\mathcal{T}^{-1}=&\frac{g}{M_D}[\tr[\mathcal{T}\bar H_c\mathcal{T}^{-1}(-1)(-1)^\mu i\overleftarrow{D}_{ac}^\mu\nn\\
&\times(-1)(-1)(-1)^\mu(-1)^\mu v\cdot\mba_{ba}\gamma_\mu\gamma_5(-1)^\mu\mathcal{T}H_b\mathcal{T}^{-1}]+\text{h.c.}\nn\\
=&\mathcal{L}_{N.L.O.}^{VRI_2}\,.\label{eq:72beq67'}
\end{align}
Similarly, using the parity properties listed in \req{eq:72beq36}, combined with
\begin{equation}\label{eq:72beq68}
\mathcal{P}\mba^\mu\mathcal{P}^{-1}=(-1)(-1)^\mu\mba^\mu\,,
\end{equation}
we can safely conclude $\mathcal{L}_{NLO}^{VRI_1}$ and $\mathcal{L}_{NLO}^{VRI_2}$ are parity even as well. From the charge conjugation properties in \req{eq:72beq16}, we have
\begin{equation}\label{eq:72beq69}
\mathcal{C}\xi \mathcal{C}^{-1}=\xi^\dagger\,,
\end{equation}
and correspondingly
\begin{align}\label{eq:72beq70}
\mathcal{C}\mba^\mu\mathcal{C}^{-1}&=(-1)\mba^\mu\,, \nn\\
\mathcal{C}D^\mu\mathcal{C}^{-1}&=D^\mu\,.
\end{align}
With the CPT transformation properties verified, we have established \req{eq:72beq62} is the effective Lagrangian for anti-heavy mesons to this order with the replacement of the meson fields $H$ by the anti-meson forms $\cal H$

Now we prove the first term of $\mathcal{L}_{N.L.O.}^{VRI_2}$ in \req{eq:72beq62} is invariant under Hermitian conjugation:
\begin{align}
\left(\mathcal{L}_{N.L.O.}^{VRI_2}\right)^\dagger&=\frac{g}{M_D}\tr[\gamma_5^\dg\gamma_\mu^\dg H_b^\dg(v\cdot\mba_{ba})^\dg(iD_{ac}^\mu)^\dg(\gamma_0H_c^\dg \gamma_0)^\dg]\nn\\
&=\frac{g}{M_D}\tr[\gamma_5\gamma_0\gamma_\mu\gamma_0H_b^\dg(v\cdot\mba_{ba})(-iD_{ac}^\mu)\gamma_0H_c\gamma_0]\nn\\
&=\frac{g}{M_D}\tr[\bar H_b(v\cdot\mba_{ba})(-iD_{ac}^\mu)H_c\gamma_0\gamma_5\gamma_0\gamma_\mu]\nn\\
&=\frac{g}{M_D}\tr[\bar H_b(v\cdot \mba_{ba})(iD_{ac}^\mu)H_c\gamma_5\gamma_\mu]\nn\\
&=\mathcal{L}_{N.L.O.}^{VRI_2}\,.\label{eq:72beq71}
\end{align}
For heavy mesons, in rest frame, $\mathcal{L}_{NLO}^{VRI_1}$ and $\mathcal{L}_{NLO}^{VRI_2}$ reduce to
\begin{align}
\mathcal{L}_{N.L.O.}^{VRI}&=-\frac{1}{2M_D}\tr[h_a^\dg (iD)_{ba}^2h_b]   
-\frac{g}{M_D}\tr[h_c^\dg(\overleftarrow{i\vec{D}\cdot\vec{\sigma}})_{ac}\mba_{ba}^0- \mba_{ac}^0(i\vec{D}\cdot\vec{\sigma})_{ba} h_b]\,.   
\end{align}

To show that the electromagnetic interactions in HHChPT satisfy the C, P,  and T invariances, we need the charge, parity and time reversal transformation properties of the photon fields:
\begin{align}
\mathcal{C}^{-1}B^\mu \mathcal{C}&=-B^\mu\,, \nn\\
\mathcal{P}^{-1}B^\mu \mathcal{P}&=(-1)^\mu B^\mu\,, \label{eq:72beq89}\\
\mathcal{T}^{-1}B^\mu \mathcal{T}&=(-1)^\mu B^\mu\,, \nn
\end{align}
Consequently, the photon field strength transforms as
\begin{align}
\mathcal{C}^{-1}F^{\mu\nu}\mathcal{C}&=(-1)F^{\mu \nu}\,,\nn\\
\mathcal{P}^{-1}F^{\mu\nu}\mathcal{P}&=F^{\mu\nu}\,, \label{eq:72beq90}\\
\mathcal{T}^{-1}F^{\mu\nu}\mathcal{T}&=-F^{\mu\nu}\,,\nn
\end{align}
showing that $\mathcal{L}^{D-\gamma_2}_{N.L.O.}$ \req{eq:72beq87} preserves C, P and T transformations separately.

\section{Heavy Quark Effective Theory: Integrating Out Heavy Modes in the Form of Path-Integral}
\label{appx:D.1}

To derive the leading-order heavy quark Lagrangian, we start from the full relativistic Dirac lagrangian for the heavy quark field $Q$,
\begin{align}\label{DiracLagQ}
\cL&=\bar Q(i\slashed{D}-m_Q)Q.
\end{align}
We define two component spinor fields by projecting out particle and anti-particle pieces using the 4-velocity $v^\mu$,
\begin{align}
\phi&=\frac{1+\slashed{v}}{2}Q,\qquad 
\chi&=\frac{1-\slashed{v}}{2}Q, 
\end{align}
recalling from \req{HQparticleprojector} that in the rest frame $v^\mu\to (1,\vec 0)$, $\phi$ contains the particle components and $\chi$ the anti-particle components.  Since $m_Q$ is a heavy scale and the largest contribution to the energy, we separate that piece of the phase in each field by defining
\begin{align}
\phi&=e^{-im_Q(v\cdot x)}h_v,\qquad 
\chi&=e^{-im_Q(v\cdot x)}H_v
\end{align}
Plugging these fields into the lagrangian \req{DiracLagQ},
\begin{align}\label{LagQstep1}
\cL=\bar h_vi(v\cdot D)h_v -\bar H_v \big(i(v\cdot D)+2m_Q\big)H_v 
     +\bar h_v i\slashed{D}^\perp H_v +\bar H_v i\slashed{D}^\perp h_v,
\end{align}
where $v^\mu$ has led to a natural decomposition of the covariant derivative
\begin{align}
\slashed{D}=\slashed{v}(v\cdot D)+\slashed{D}^\perp,\\
\slashed{D}^\perp=\gamma^\mu(g_{\mu\nu}-v_\mu v_\nu)D^\perp
\qquad \{\slashed{D}^\perp,\slashed{v}\}&=0.
\end{align}
Now we can decouple the anti-particle components by redefining the field
\begin{equation}\label{Hvshift}
H\to H-\frac{i\slashed{D}^\perp}{2m_Q+i(v\cdot D)}h,\qquad
\bar H\to \bar H-\bar h\frac{i\slashed{D}^\perp}{2m_Q+i(v\cdot D)}
\end{equation}
which is just a shift of the Grassman variable in the path integral measure.  Inserting this into \req{LagQstep1}, 
\begin{align}
\cL&= \paren{\bar H_v - \bar h_v\frac{i\slashed{D}^\perp}{2m_Q+i(v\cdot D)}}\big(i(v\cdot D)+2m_Q\big)\paren{H_v-\frac{i\slashed{D}^\perp}{2m+i(v\cdot D)}h_v}\nn\\
&+\bar h_vi(v\cdot D)h_v + \bar h_ i\slashed{D}^\perp\paren{H_v-\frac{i\slashed{D}^\perp}{2m_Q+(v\cdot D)}h_v}\nn\\
&+\left(\bar H_v-\bar h_v\frac{i\slashed{D}^\perp}{2m_Q+i(v\cdot D)}\right)i\slashed{D}^\perp h_v\nn\\
\label{LagQstep2}
&=\bar H_v\big(iv\cdot D+2m_Q\big)H_v +\bar h_v iv\cdot Dh_v  
-\bar h_v i\slashed{D}^\perp \frac{1}{2m_Q+iv\cdot D}i\slashed{D}^\perp h_v.
\end{align}
The covariant derivatives only pick up soft momenta $p^\mu\ll m_Q$.  Therefore, the $iv\cdot D$ term is much smaller than the $2m_Q$ term between the $\bar H_v$ and $H_v$ fields.  At leading order in $p/m_Q$, the $H_v$ field is non-dynamical and can be trivially integrated out using the standard integration rules for Grassmann numbers.  Thus, we obtain the lagrangian for the $h_v$ fields at leading power in $p/m_Q$,
\begin{align}\label{hvLeff}
\cL=\bar h_v iv\cdot Dh_v +\mathcal{O}(iv\cdot D/m_Q)
\end{align}  
Note that a contribution of the heavy $H_v$ field to the $h_v$ field dynamics has been obtained from the field redefinition \req{Hvshift} and is suppressed by $p/m_Q$ relative to the kinetic term for the $h_v$ fields, as seen in the third term of \req{LagQstep2}.  However, in general other next-to-leading order operators are allowed by the symmetries, and this procedure does not necessarily yield the complete NLO lagrangian.  A complete set of allowed NLO operators must be written and their coefficients in the effective theory obtained by matching an observable computed to the same order in the full theory.

\section{Velocity Reparameterization Invariance}\label{appx:VRI}

Velocity reparameterization invariance restores the Lorentz invariance of the full theory order by order in the expansion in $v$.  To see how, we first study the properties of scalar particles under the non-relativistic expansion, then extend it to heavy fermions.  

A Lorentz invariant scalar theory can be written as 
\begin{equation}\label{eq:72beq44}
\mathcal{L}=D^\mu \phi^* D_\mu\phi-M^2\phi^*\phi\,.
\end{equation}
$\phi$ is the (composite) scalar field, $D_\mu$ is the covariant derivative and $M$ is the $\phi$ field mass. Setting $M$ as the break down scale of an effective field theory, we write the momentum of the heavy $\phi$ field as $p=Mv+k$, with the four velocity $v$ satisfying $v^2=1$ and the residual momentum $k\ll M$ the soft scale in the effective theory.  The velocity dependent effective field is
\begin{equation}\label{eq:72beq45}
\phi_v(x)=\sqrt{2M}e^{iMv\cdot x}\phi(x)\,.
\end{equation}
Similar to above, we derive the effective Lagrangian for the low-energy dynamics
\begin{equation}\label{eq:72beq45b}
\mathcal{L}_{eff}=\sum_v\phi_v^*(iv\cdot D)\phi_v+\mathcal{O}\left(\frac{k}{M}\right)\,.
\end{equation}
We now consider change of frame, equivalent to an infinitesimal shift in the velocity of the heavy field, parameterized by a momentum $q\sim k$,
\begin{equation}\label{eq:72beq42}
v\to v+\frac{q}{M}\,,
\end{equation}
such that the invariant square of the velocity still satisfies
\begin{equation}\label{eq:72beq43}
v^2=(v+q/M)^2=1
\end{equation}
to order $1/M$.  Under this shift, the field transforms as
\begin{equation}\label{eq:72beq46}
\phi_w(x)=e^{iq\cdot x}\phi_v(x)\,,\qquad w=v+\frac{q}{M}\,.
\end{equation}
The fact that the effective Lagrangian Eq.\,\eqref{eq:72beq45b} is invariant under Eq\.\,\eqref{eq:72beq46}, while the full Lagrangian \eqref{eq:72beq44} is Lorentz invariant, suggests the heavy field momentum is shifted
\begin{equation}\label{eq:72beq47}
(v,k)\to (v+q/M,k-q)\,,
\end{equation}
exhibiting the arbitrariness of the separation between $Mv$ and $k$. This invariance is Velocity Reparameterization Invariance (VRI).

Applying the velocity transformation Eq.\,\eqref{eq:72beq46} to the leading-order effective Lagrangian Eq.\,\eqref{eq:72beq44}, we have
\begin{equation}\label{eq:72beq48}
\mathcal{L}_{eff}=\sum_w\phi_w^*(w-q/M)(iD+q)\phi_w\,,
\end{equation}
which involves an $\mathcal{O}(1/M)$ term.  Therefore invariance under Eq.\,\eqref{eq:72beq46} constrains operators of different orders in the $1/M$ expansion.  The most general $\mathcal{O}(1/M)$ term is
\begin{equation}\label{eq:72beq49}
\mathcal{L}_{\mathcal{O}(1/M)}=\sum_v -\frac{A}{2M}\phi_v^*D^2\phi_v
\end{equation}
with $A$ a constant to be determined.  This term is unique, since the other possible structure, $\phi_v^*(v\cdot D)^2\phi_v$, is eliminated by the leading order equation of motion.  Under Eq.\,\eqref{eq:72beq46}, $\mathcal{L}_{\mathcal{O}(1/M)}$ becomes
\begin{equation}\label{eq:72beq50}
\mathcal{L}_{\mathcal{O}(1/M)}=\sum_w -\frac{A}{2M}\phi_w^*(D-iq)^2\phi_w\,.
\end{equation}
Therefore, to the first order in $q$, we have
\begin{align}
\delta\mathcal{L}&=\sum_w\phi_w^*\left\{(w-q/M)\cdot(iD+q) \right\}\phi_w-\frac{A}{2M}\phi_w^2(D-iq)^2\phi_w\nn \\
&-\sum_v\phi_v^*(iv\cdot D)\phi_v+\frac{A}{2M}\phi_v^*D^2\phi_v\nn \\
&=(A-1)\phi_v^*\frac{q\cdot D}{M}\phi_v+\mathcal{O}\left(q^2,\frac{1}{M^2}\right)\,,\label{eq:72beq51}
\end{align}
up to $\mathcal{O}(1/M)$, and we conclude that the effective Lagrangian
\begin{equation}\label{eq:72beq52}
\mathcal{L}_{eff}=\sum_v\phi_v^*(iv\cdot D)\phi_v-\frac{A}{2M}\phi_v^*(D^2)\phi_v
\end{equation}
is Lorentz symmetry reparameterization invariant if and only if $A=1$. The VRI combination of $v_\mu$ and $D_\mu$ up to $\mathcal{O}(1/M)$ is fixed to
\begin{equation}\label{eq:72beq53}
V_\mu=v_\mu+\frac{iD_\mu}{M}\,.
\end{equation}

Reparameterization invariance for fermions must involve the spinor structure, and the velocity reparameterization takes the general form
\begin{equation}\label{eq:72beq54}
\psi_w(x)=e^{iq\cdot x}R(w,v)\psi_v(x)\,,
\end{equation}
where $R(w,v)$  is a Lorentz transformation that we can work out explicitly by boosting first to the rest frame of the particle and then to the new frame (at velocity $w$)
\begin{equation}\label{eq:72beq55}
R(w,v)=\tilde\Lambda(w,p/M)\tilde\Lambda^{-1}(v,p/M)\,,
\end{equation}
with $\tilde\Lambda(w,v)$ the Lorentz boost in the spinor representation
\begin{equation}\label{eq:72beq56}
\tilde\Lambda(w,v)=\frac{1+\slashed{w}\slashed{v}}{\sqrt{2(1+w\cdot v)}}\,.
\end{equation}
To construct the most general Lagrangian invariant under Eq.\,\eqref{eq:72beq54}, it is convenient to define the reparameterization covariant spinor field
\begin{equation}\label{eq:72beq57}
\Psi_v(x)=\tilde \Lambda(p/M,v)\psi_v(x)\,,
\end{equation}
which transforms as
\begin{equation}\label{eq:72beq58}
\Psi_w(x)=e^{iq\cdot x}\Psi_v(x)\,.
\end{equation}
In this form, the result of Eq.\,\eqref{eq:72beq51} can be immediately applied, and we infer that  at $\mathcal{O}(1/M)$ we can write the field $\Psi$ as
\begin{equation}\label{eq:72beq59}
\Psi_v(x)=\left(1+\frac{i\slashed{D}}{2M}\right)\psi_v(x)\,.
\end{equation}

To order $1/M$, the standard fermion bilinear VRI forms are
\begin{align}
\bar\Psi_v\Psi_v&=\psi_v\psi_v\,,\nn\\
\bar\Psi_v\gamma_5\Psi_v&=0\,,\nn\\
\bar\Psi_v\gamma^\mu\Psi_v&=\bar\psi_v\left(v^\mu+\frac{iD^\mu}{M}\right)\psi_v+\mathcal{O}(1/M^2)\,,\nn\\
\bar\Psi_v\gamma^\mu\gamma_5\Psi_v&=\bar\psi_v\left(\gamma^\mu\gamma_5-v^\mu\frac{i\slashed{D}}{M}\gamma_5\right)\psi_v+\mathcal{O}(1/M^2)\,,\nn\\
\bar\Psi_v\sigma^{\alpha\beta}\Psi_v&=\epsilon^{\alpha\beta\lambda\sigma}\bar\Psi_v\gamma_\sigma\gamma_5V_\lambda\Psi_v\,.\label{eq:72beq60}
\end{align}

%% file: appendix_D.tex
\chapter{Useful Formulae of X-Effective Field Theory}
\label{appx:D}

\section{Feynman Rules and Power Counting for XEFT}
\label{appx:D.2}

In this section we power count amplitudes for scattering processes in the XEFT and its fine-tuning region.

\subsection{Isospin-conserving operators}
\label{appx:D.2.1.1}

First we reorganize the reduced XEFT Lagrangian we derived in Sec.\:\ref{sec:II.3.4} from the leading order to the next-to-leading order based on the XEFT power counting. Recall the small parameters are the HHChPT expansion parameter, $\lambda=q/M_D$ where $q$ is the typical momentum, and the maximum kinetic energy of the $D^*$ near its threshold in $D\pi$ scattering, namely $\epsilon=|\delta|/m_\pi$ where $\delta=m_{D^{0*}}-m_{D^0}-m_\pi$ from \req{deltaDstardefn}.  Additionally, we define  $\lambda_{ph}=\Delta/M_D$ with $\Delta=m_{D^{0*}}-m_{D^0}$.  Throughout, the $\sim$ symbol means `scales as' in our power counting.

\vspace{2pc}
\begin{longtable}{| c | p{.4\textwidth} | p{.4\textwidth} |}
\hline 
HHChPT order & Operator & Contribution at Tree Level \\
\hline 
L.O. & 
$\arraycolsep=1.4pt\def\arraystretch{2}
\begin{array}{c} 
\hat \pi^{0\dg}(i\pd_0)\hat \pi^0 \\
(\hat\pi^+)^\dg(i\pd_0)\hat\pi^+ \\
(\hat\pi^-)^\dg(i\pd_0)\hat\pi^-
\end{array}$ & 
$\dfrac{|\vec p_\pi|^2}{2m_\pi}\sim\epsilon^2 m_\pi$ \\
\hline 
L.O. & 
$\arraycolsep=1.4pt\def\arraystretch{2}
\begin{array}{c}
(\hat \pi^0)^\dg\paren{\dfrac{\vec\nabla^2}{2m_\pi}}\hat \pi^0\\
(\hat \pi^-)^\dg\paren{\dfrac{\vec\nabla^2}{2m_\pi}}\hat\pi^-
\end{array}$ & 
$\dfrac{|\vec p_\pi|^2}{2m_\pi}\sim\epsilon^2m_\pi$ \\
\hline 
L.O. & 
$\arraycolsep=1.4pt\def\arraystretch{2}
\begin{array}{c}
\vec P_a^\dg(i\pd_0)\vec P_a\\
P_a^\dg (i\pd^0)P_a
\end{array}$  & 
$\dfrac{|\vec p_D|^2}{2M_D}\sim\epsilon^2\lambda m_\pi$ \\
\hline 
L.O. & $\hat\pi^+\delta\hat\pi$ &
$\Delta-m_\pi=\delta$ \Tstrut\Bstrut \\
\hline 
L.O. & 
$\arraycolsep=1.4pt\def\arraystretch{2}
\begin{array}{c}
\dfrac{g_\pi}{f_\pi}\oneov{\sqrt{2m_\pi}}(\vec P_a^\dg P_b)\cdot\vec\nabla \hat M_{\pi ba} \\
\dfrac{g_\pi}{f_\pi}\oneov{\sqrt{2m_\pi}}(P_a^\dg \vec P_b)\cdot\vec\nabla \hat M_{\pi ba}^\dg 
\end{array}
$ & 
\begin{tabular}{l}
\includegraphics[width=.25\linewidth]{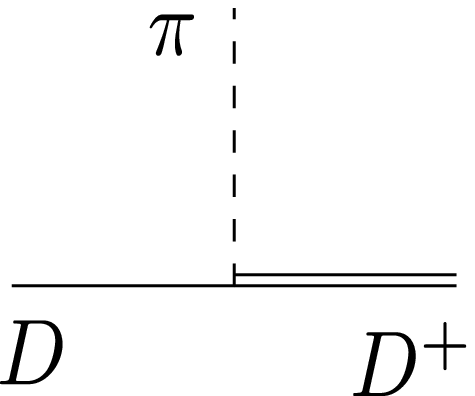}
\\
$\dfrac{g_\pi}{f_\pi}\oneov{\sqrt{2m_\pi}}|\vec p_\pi|\sim\epsilon\oneov{\sqrt{2m_\pi}}$
\end{tabular}
 \\
\hline 
\begin{tabular}{l}
N.L.O.\\
$(M_1)$
\end{tabular} &
$\arraycolsep=1.4pt\def\arraystretch{2}
\begin{array}{c}
(-\sigma_1)(m_u+m_d)\vec P_a^\dg\cdot \vec P_a\\
(-\sigma_1)(m_u+m_d)P_a^\dg P_a
\end{array}
$ &
$\arraycolsep=1.4pt\def\arraystretch{2}
\begin{array}{c}
(-\sigma_1)(m_u+m_d)\\
\sim(-\sigma_1)\lambda m_\pi
\end{array}
$ \\
\hline 
\begin{tabular}{l}
N.L.O.\\
$(M_2)$
\end{tabular} &
$-\dfrac{\Delta^{(\sigma_1)}}{4}(m_u+m_d)\vec P_a^\dg\cdot\vec P_a$ & 
$-\dfrac{\Delta^{(\sigma_1)}}{4}\lambda m_\pi$ \\
\hline 
\begin{tabular}{l}
N.L.O.\\
$(M_2)$
\end{tabular} &
$\dfrac{3\Delta^{(\sigma_1)}}{4}(m_u+m_d)P_a^\dg P_a$ &
$\dfrac{3\Delta^{(\sigma_1)}}{4}\lambda m_\pi$ \\
\hline 
\begin{tabular}{l}
N.L.O.\\
$(M_1)$
\end{tabular} &
$\arraycolsep=1.4pt\def\arraystretch{2}
\begin{array}{l}
(\sigma_1)\paren{\dfrac{4}{f_\pi^2}\dfrac{1}{2m_\pi}}(\vec P_a^\dg\cdot\vec P_a)\\
\times M_q(\hat M_\pi\hat M_\pi^\dg+\hat M_\pi^\dg\hat M_\pi)_{bb}\\
(\sigma_1)\paren{\dfrac{4}{f_\pi^2}\dfrac{1}{2m_\pi}}(P_a^\dg P_a)\\
\times M_q(\hat M_\pi\hat M_\pi^\dg+\hat M_\pi^\dg \hat M_\pi)_{bb}
\end{array}
$ &
\begin{tabular}{l}
\\
\includegraphics[width=.25\linewidth]{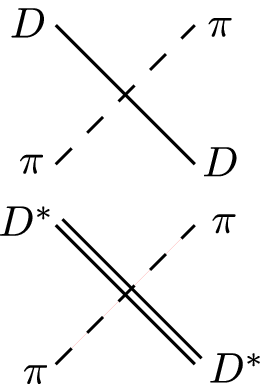}\\
$\dfrac{2(\sigma_1)}{f_\pi^2m_\pi}(m_u+m_d) \sim\dfrac{2\sigma_1}{f_\pi^2}\lambda$
\end{tabular} \\
\hline 
\begin{tabular}{l}
N.L.O.\\
$(M_2)$
\end{tabular} & 
$\arraycolsep=1.4pt\def\arraystretch{2}
\begin{array}{l}
\paren{\dfrac{\Delta^{(\sigma_1)}}{4}}\paren{\dfrac{4}{2f_\pi^2m_\pi}}\vpad\cdot\vpa\\
\times M_q(\hmpi\hmpid+\hmpid\hmpi)_{bb}
\end{array}
$ &
\begin{tabular}{ll}
\\
\includegraphics[width=.25\linewidth]{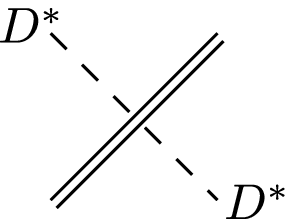} & $\dfrac{\Delta^{(\sigma_1)}}{2f_\pi^2m_\pi}(m_u+m_d)$ \\
 & $\sim\dfrac{\Delta^{(\sigma_1)}}{2f_\pi^2}\lambda$
\end{tabular} \\
\hline 
\begin{tabular}{l}
N.L.O.\\
$(M_2)$
\end{tabular} & 
$\arraycolsep=1.4pt\def\arraystretch{2}
\begin{array}{l}
\paren{-\dfrac{\Delta^{(\sigma_1)}}{4}\dfrac{12}{2f_\pi^2m_\pi}P_a^\dg P_a}\\
\cdot M_q(\hmpi\hmpid+\hmpid\hmpi)_{bb}
\end{array}
$ &
\begin{tabular}{ll}
\\
\includegraphics[width=.25\linewidth]{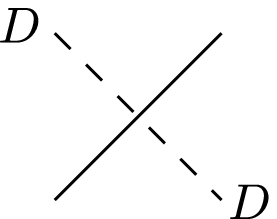} & $-\dfrac{3\Delta^{(\sigma_1)}}{2f_\pi^2m_\pi}(m_u+m_d)$ \\
& $\sim\dfrac{3\Delta^{(\sigma)}}{2f_\pi^2}\lambda$
\end{tabular} \\
\hline 
\begin{tabular}{l}
N.L.O.\\
$(V.R.I_1)$
\end{tabular} & 
$\arraycolsep=1.4pt\def\arraystretch{2}
\begin{array}{l}
\paren{-\dfrac{1}{2M}}\paren{\dfrac{3\Delta}{4}}^2(\vec P_a^\dg\cdot\vec P_a) \\
\paren{-\dfrac{1}{2M}}\paren{\dfrac{3\Delta}{4}}^2(P_a^\dg P_a)
\end{array}$ & 
$-\dfrac{9\Delta^2}{32M}\sim-\dfrac{9}{32}\lambda_{ph}\Delta$ \\
\hline 
\begin{tabular}{l}
N.L.O.\\
$(V.R.I_1)$
\end{tabular} & 
$\arraycolsep=1.4pt\def\arraystretch{2}
\begin{array}{l}
\paren{\dfrac{1}{2M}}\paren{\dfrac{3\Delta}{2}}\vec P_a^\dg (i\pd^0)\vec P_a \\
\paren{\dfrac{1}{2M}}\paren{\dfrac{3\Delta}{2}}P_a^\dg(i\pd^0)P_a
\end{array}$ &
$\arraycolsep=1.4pt\def\arraystretch{2}
\begin{array}{l}
\dfrac{3\Delta}{4M}\dfrac{|\vec p_D|^2}{M_D}\\
\sim\dfrac{3}{4}\epsilon^2\lambda\lambda_{ph}m_\pi
\end{array}$ \\
\hline 
\begin{tabular}{l}
N.L.O.\\
$(V.R.I_1)$
\end{tabular} & 
$\paren{\dfrac{1}{2M}}(\Delta)\vec P_a^\dg\cdot\vec P_a\paren{\dfrac{3\Delta}{2}}$ & 
$\arraycolsep=1.4pt\def\arraystretch{2}
\begin{array}{l}
-\dfrac{\Delta}{2M_D}\dfrac{3\Delta}{2}\\
\sim\paren{-\dfrac{3}{4}}\lambda_{ph}\Delta
\end{array}$ \\
\hline 
\begin{tabular}{l}
N.L.O.\\
$(V.R.I_1)$
\end{tabular} &
$\arraycolsep=1.4pt\def\arraystretch{2}
\begin{array}{l}
\paren{-\dfrac{1}{2M}}\vpad(i\vnab)^2\vpa\\
\paren{-\dfrac{1}{2M}}P_a^\dg (i\vnab)^2P_a
\end{array}$ & 
$\arraycolsep=1.4pt\def\arraystretch{2}
\begin{array}{l}
\paren{-\dfrac{1}{2M}}|\vec P_D|^2\\
\sim-\epsilon^2\lambda m_\pi
\end{array}$ \\
\hline 
\begin{tabular}{l}
N.L.O.\\
$(V.R.I_1)$
\end{tabular} & 
$\arraycolsep=1.4pt\def\arraystretch{2}
\begin{array}{l}
\paren{-\dfrac{1}{2M}}\vpad(i\pd^0)^2\vpa\\
\paren{-\dfrac{1}{2M}}\vpad(i\pd^0)^2\vpa
\end{array}$ & 
$\arraycolsep=1.4pt\def\arraystretch{2}
\begin{array}{l}
\paren{-\dfrac{1}{2M}}\paren{\dfrac{|\vec p_D|^2}{2M_D}}^2\\
\sim-\epsilon^4\lambda^3m_\pi
\end{array}$ \\
\hline 
\begin{tabular}{l}
N.L.O.\\
$(V.R.I_1)$
\end{tabular} & 
$\paren{\dfrac{1}{M}}(\Delta)\vpad (i\pd_0)\vpa$ & 
$\arraycolsep=1.4pt\def\arraystretch{2}
\begin{array}{l}
\paren{-\dfrac{1}{M}}\Delta\dfrac{|\vec p|^2}{M_D}\\
\sim-\epsilon^2\Delta\lambda_{ph}^2
\end{array}
$ \\
\hline 
\begin{tabular}{l}
N.L.O.\\
$(V.R.I_1)$
\end{tabular} & 
$\paren{-\dfrac{1}{2M}}(m_\pi)^2\vpad\vpa$ & 
$\paren{-\dfrac{1}{2M}}(m_\pi)^2\sim -\lambda m_\pi/2$ \\
\hline 
\begin{tabular}{l}
N.L.O.\\
$(V.R.I_1)$
\end{tabular} & 
$\arraycolsep=1.4pt\def\arraystretch{2}
\begin{array}{l}
\dfrac{i}{f_\pi^2}\dfrac{3\Delta}{8m_\pi}(\vpad\vpb)(im_\pi)\\
\times(\hmpi\hmpid-\hmpid\hmpi)_{ba}\paren{-\dfrac{1}{2M}}\\
\dfrac{i}{f_\pi^2}\dfrac{3\Delta}{8m_\pi}(im_\pi)(P_a^\dg P_b)\\
\times(\hmpi\hmpid-\hmpid\hmpi)_{ba}\paren{-\dfrac{1}{2M}}
\end{array}$ & 
\begin{tabular}{l}
\includegraphics[width=.25\textwidth]{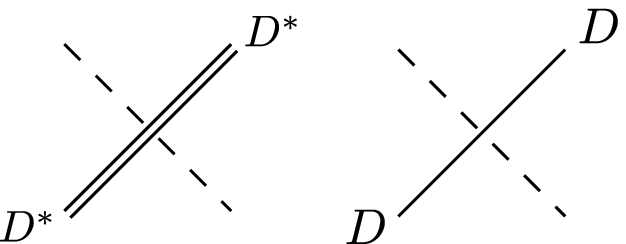}\\
$\dfrac{1}{16M}\dfrac{3\Delta}{f_\pi^2}\sim\dfrac{3}{16}\lambda_{ph}\dfrac{1}{f_\pi^2}$
\end{tabular} \\
\hline 
\begin{tabular}{l}
N.L.O.\\
$(V.R.I_1)$
\end{tabular} & 
$\arraycolsep=1.4pt\def\arraystretch{2}
\begin{array}{l}
\paren{-\dfrac{1}{2M}}\dfrac{i}{f_\pi^2}\dfrac{3\Delta}{4}\dfrac{1}{2m_\pi}(\vpad\vpb) \\
\times(\hmpi\olra{\pd_0}\hmpid+\hmpid\olra{\pd_0}\hmpi)_{ba} \\
\paren{-\dfrac{1}{2M}}\dfrac{i}{f_\pi^2}\dfrac{3\Delta}{4}\dfrac{1}{2m_\pi}(P_a^\dg P_b)\\
\times(\hmpi\olra{\pd_0}\hmpid+\hmpid\olra{\pd_0}\hmpi)_{ba}
\end{array}$ & 
\begin{tabular}{l}
\includegraphics[width=.25\textwidth]{72cfigf5.eps}\\
$\arraycolsep=1.4pt\def\arraystretch{2}
\begin{array}{l}
-\dfrac{i3\Delta}{16Mm_\pi f_\pi^2}\dfrac{|p_\pi|^2}{2m_\pi}\\
\sim-\dfrac{i3}{32}\lambda_{ph}\epsilon^2\dfrac{1}{f_\pi^2}
\end{array}
$
\end{tabular} \\
\hline 
\begin{tabular}{l}
N.L.O.\\
$(V.R.I_1)$
\end{tabular} &
$\arraycolsep=1.4pt\def\arraystretch{2}
\begin{array}{l}
\paren{-\dfrac{1}{2M}}\dfrac{im_\pi}{f_\pi^2}\vpad\vpb\\
\times(\hmpi\hmpid-\hmpid\hmpi)_{ba}
\end{array}$ & 
\begin{tabular}{l}
\includegraphics[width=.125\textwidth]{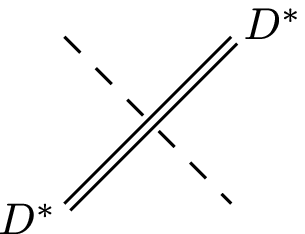} \\ 
$\sim-\dfrac{im_\pi}{2M_D f_\pi^2}\sim(-i)\lambda\dfrac{1}{f_\pi^2}$
\end{tabular} \\
\hline 
\begin{tabular}{l}
N.L.O.\\
$(V.R.I_1)$
\end{tabular} & 
$\arraycolsep=1.4pt\def\arraystretch{2}
\begin{array}{l}
\paren{-\dfrac{1}{2M}}\dfrac{im_\pi}{f_\pi^2}\dfrac{1}{2m_\pi}\vpad\vpb \\
\times(\hmpi\olra{\pd_0}\hmpid-\hmpid\olra{\pd_0}\hmpi)
\end{array}$ & 
\begin{tabular}{ll}
\includegraphics[width=.125\textwidth]{72cfigf6.eps} & $\sim\dfrac{-i}{4f_\pi^2M}\dfrac{|\vec p_\pi|^2}{2m_\pi}$\\
& $\sim\dfrac{-i}{4}\epsilon^2\lambda\dfrac{1}{f_\pi^2}$
\end{tabular} \\
\hline 
\begin{tabular}{l}
N.L.O.\\
$(V.R.I_1)$
\end{tabular} & 
$\arraycolsep=1.4pt\def\arraystretch{2}
\begin{array}{l}
\paren{\dfrac{1}{2M_D}}\dfrac{1}{f_\pi^2}(\vnab\vpad)\vpb\dfrac{1}{2m_\pi}\\
\times(\hmpid\olra{\nabla}\hmpi+\hmpi\olra{\nabla}\hmpid)_{ba}\\
\paren{\dfrac{1}{2M_D}}\dfrac{1}{f_\pi^2}\dfrac{2m_\pi}(\vnab P_a^\dg)P_b\\
\times(\hmpid\olra{\nabla}\hmpi+\hmpi\olra{\nabla}\hmpid)_{ba}
\end{array}$ &
\begin{tabular}{l}
\includegraphics[width=.25\textwidth]{72cfigf5.eps}\\
$\arraycolsep=1.4pt\def\arraystretch{2}
\begin{array}{l}
\dfrac{1}{4M_D}\dfrac{1}{M_\pi}\dfrac{1}{f_\pi^2}|\vec p_D||\vec p_\pi|\\
\sim\dfrac{1}{4}\epsilon^2\lambda\dfrac{1}{f_\pi^2}
\end{array}$
\end{tabular} \\
\hline 
\begin{tabular}{l}
N.L.O.\\
$(V.R.I_1)$
\end{tabular} & 
$\arraycolsep=1.4pt\def\arraystretch{2}
\begin{array}{l}
\paren{\dfrac{1}{2M_D}}\dfrac{f_\pi^2}(\pd_0\vpad)\vpb\dfrac{1}{2M_\pi}\\
\cdot(\hmpid\olra{\pd_0}\hmpi+\hmpi\olra{\pd_0}\hmpid)_{ba}\\
\paren{\dfrac{1}{2M}}\dfrac{1}{f_\pi^2}(\pd_0P_a^\dg)P_b\dfrac{1}{2m_\pi}\\
\cdot(\hmpid\olra{\pd_0}\hmpi+\hmpi\olra{\pd_0}\hmpid)_{ba}
\end{array}$ & 
\begin{tabular}{l}
\includegraphics[width=.25\textwidth]{72cfigf5.eps}\\
$\arraycolsep=1.4pt\def\arraystretch{2}
\begin{array}{l}
\dfrac{1}{4M_D}\dfrac{1}{m_\pi}\dfrac{1}{f_\pi^2}\dfrac{|\vec p_D|^2}{2M_D}\dfrac{|\vec p_\pi|^2}{2m_\pi}\\
\sim\dfrac{1}{4}\epsilon^4\lambda^2\dfrac{1}{f_\pi^2}
\end{array}$
\end{tabular} \\
\hline 
\begin{tabular}{l}
N.L.O.\\
$(V.R.I_2)$
\end{tabular} &
$\arraycolsep=1.4pt\def\arraystretch{2}
\begin{array}{l}
\paren{\dfrac{g_\pi}{M}} \dfrac{1}{f_\pi} \paren{\dfrac{i\Delta}{\sqrt{2m_\pi}}}\\
\cdot(-\hmpi+\hmpid)_{ba}[(i\vnab\cdot\vpad)P_b\\
+\vpad\cdot(i\vnab P_b)]
\end{array}$ &
\begin{tabular}{ll}
\includegraphics[width=.125\textwidth]{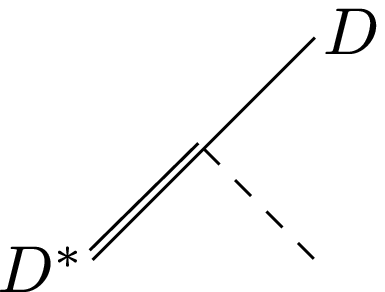} & 
$\sim\dfrac{g_\pi}{M}\dfrac{1}{f_\pi}\dfrac{i\Delta}{\sqrt{2m_\pi}}|\vec p_D|$\\
& $\sim\epsilon\dfrac{1}{\sqrt{2m_\pi}}\lambda_{ph}$
\end{tabular} \\
\hline 
\begin{tabular}{l}
N.L.O.\\
$(V.R.I_2)$
\end{tabular} &
$\arraycolsep=1.4pt\def\arraystretch{2}
\begin{array}{l}
\paren{\dfrac{g_\pi}{M}} \dfrac{1}{f_\pi} \paren{\dfrac{1}{\sqrt{2m_\pi}}}\\
\cdot[\pd_0\hmpi+\pd_0\hmpid]_{ba}\\
\cdot[(i\vnab\cdot\vpad)P_b+\vpad\cdot(i\vnab P_b)]
\end{array}$ & 
\begin{tabular}{l}
\includegraphics[width=.125\textwidth]{72cfigf7.eps} \\
$\sim\dfrac{g_\pi}{M}\dfrac{1}{f_\pi}\dfrac{1}{\sqrt{2M_\pi}}\dfrac{|\vec p_\pi|^2}{2m_\pi}|\vec p_D|\sim\epsilon^3\lambda\dfrac{1}{\sqrt{2m_\pi}}$
\end{tabular} \\
\hline 
\begin{tabular}{l}
N.L.O.\\
$(V.R.I_2)$
\end{tabular} &
$\arraycolsep=1.4pt\def\arraystretch{2}
\begin{array}{l}
\paren{\dfrac{g_\pi}{M}}\dfrac{1}{\sqrt{2m_\pi}}\dfrac{1}{2m_\pi}\paren{\dfrac{i}{f_\pi^3}}(i\Delta)\\
\cdot[P_c^\dg \vpb(\hmpid)_{ba}-\vec P_c^\dg P_b(\hmpi)_{ba}]\\
\cdot[\hmpid\olra{\nabla}\hmpid+\hmpid\olra{\nabla}\hmpi]_{ac}\\
\paren{\dfrac{g_\pi}{M}}\dfrac{1}{\sqrt{2m_\pi}}\dfrac{1}{2m_\pi}\paren{\dfrac{i}{f_\pi^3}}(i\Delta)\\
\cdot[P_c^\dg \vpb(\hmpid)_{ba}-\vec P_c^\dg P_b(\hmpi)_{ba}]\\
\cdot[\hmpid\olra{\nabla}\hmpid+\hmpi\olra{\nabla}\hmpi]_{ac}\\
\text{and indices exchanging terms}
\end{array}$ & 
\begin{tabular}{l}
\includegraphics[width=.15\textwidth]{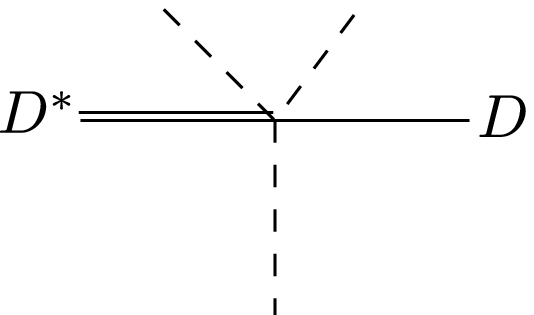}\\
$\arraycolsep=1.4pt\def\arraystretch{2}
\begin{array}{l}
\sim\paren{-\dfrac{g_\pi}{M}\dfrac{1}{2\sqrt{2m_\pi}}}\dfrac{1}{f_\pi^3}|\vec p_\pi|\dfrac{\Delta}{m_\pi}\\
\sim-\dfrac{1}{2\sqrt{2m_\pi}}\epsilon\lambda_{ph}\dfrac{g}{f_\pi^3}
\end{array}
$
\end{tabular} \\
\hline 
\begin{tabular}{l}
N.L.O.\\
$(V.R.I_2)$
\end{tabular} &
$\arraycolsep=1.4pt\def\arraystretch{2}
\begin{array}{l}
\paren{\dfrac{g_\pi}{M}}\dfrac{1}{\sqrt{2m_\pi}}\dfrac{1}{2m_\pi}\paren{\dfrac{i}{f_\pi^3}}\\
\,[P_c^\dg\vec P_b(\pd_0\hmpid)_{ba}\\
+\vec P_c^\dg P_b(\pd_0\hmpi)_{ba}]\cdot[\hmpi\olra{\nabla}\hmpid\\
+\hmpid\olra{\nabla}\hmpi]_{ac}\\
\paren{\dfrac{g_\pi}{M}}\dfrac{1}{\sqrt{2m_\pi}}\dfrac{1}{2m_\pi}\paren{\dfrac{i}{f_\pi^3}}\\
\,[P_c^\dg\vec P_b(\pd_0\hmpid)_{ba}\\
+\vec P_c^\dg P_b(\pd_0\hmpi)_{ba}]\cdot[\hmpid\olra{\nabla}\hmpid\\
+\hmpi\olra{\nabla}\hmpi]_{ac}\\
\text{and indices exchaning term}
\end{array}$ & 
\begin{tabular}{l}
\includegraphics[width=.15\textwidth]{72cfigf8.eps} \\
$\arraycolsep=1.4pt\def\arraystretch{2}
\begin{array}{l}
\sim\paren{\dfrac{g_\pi}{M}}\dfrac{1}{2m_\pi}\dfrac{1}{\sqrt{2m_\pi}}\dfrac{i}{f_\pi^3}\dfrac{|\vec p_\pi|^2}{2m_\pi}\cdot|\vec p_\pi|\\
\sim\epsilon^3\dfrac{g_\pi}{2\sqrt{2m_\pi}}\dfrac{i}{f_\pi^3}\lambda
\end{array}$
\end{tabular} \\
\hline 
\begin{tabular}{l}
N.L.O.\\
$(\delta_4,\delta_5)$
\end{tabular} & 
$\arraycolsep=1.4pt\def\arraystretch{2}
\begin{array}{l}
\paren{-\dfrac{\delta_4+\delta_5}{\Lambda}}\dfrac{1}{f_\pi^2}\dfrac{1}{2m_\pi}\Delta^2\\
(\vpad\cdot\vpb+P_a^\dg P_b)\cdot(\hmpi\hmpid\\
+\hmpid\hmpi)_{ba}
\end{array}$ &
\begin{tabular}{l}
\includegraphics[width=.25\textwidth]{72cfigf5.eps}\\
$\arraycolsep=1.4pt\def\arraystretch{2}
\begin{array}{l}
\paren{-\dfrac{\delta_4+\delta_5}{\Lambda}}\dfrac{1}{f_\pi^2}\dfrac{\Delta^2}{2m_\pi}\\
\sim 2(-\delta_4-\delta_5)\lambda\dfrac{1}{f_\pi^2}\lambda_{ph}\dfrac{\Delta}{m_\pi}
\end{array}$
\end{tabular} \\
\hline 
\begin{tabular}{l}
N.L.O.\\
$(\delta_4,\delta_5)$
\end{tabular} & 
$\arraycolsep=1.4pt\def\arraystretch{2}
\begin{array}{l}
\paren{-\dfrac{\delta_4+\delta_5}{\Lambda}}\dfrac{1}{f_\pi^2}\dfrac{2}{2m_\pi}(\vpad\cdot\vpb\\
+P_a^\dg P_b)\cdot[-i\Delta\hmpi\pd_0\hmpid\\
+i\Delta\hmpid\pd_0\hmpi]_{ba}
\end{array}$ &
\begin{tabular}{l}
\includegraphics[width=.25\textwidth]{72cfigf5.eps}\\
$\arraycolsep=1.4pt\def\arraystretch{2}
\begin{array}{l}
\paren{-\dfrac{\delta_4+\delta_5}{\Lambda}}\dfrac{1}{f_\pi^2}(i)\dfrac{|\vec p_\pi|^2}{2m_\pi^2}\Delta\\
\sim (-\delta_4-\delta_5)(i)\epsilon^2\dfrac{1}{f_\pi^2}\lambda_{ph}
\end{array}$
\end{tabular} \\
\hline 
\begin{tabular}{l}
N.L.O.\\
$(\delta_4,\delta_5)$
\end{tabular} & 
$\arraycolsep=1.4pt\def\arraystretch{2}
\begin{array}{l}
\paren{-\dfrac{\delta_4+\delta_5}{\Lambda}}\dfrac{1}{f_\pi^2}\dfrac{1}{2m_\pi}(\vpad\cdot\vpb\\
+P_a^\dg P_b)\\
\cdot[\pd_0\hmpi\pd_0\hmpid+\pd_0\hmpid\pd_0\hmpi]_{ba}
\end{array}$ &
\begin{tabular}{l}
\includegraphics[width=.25\textwidth]{72cfigf5.eps}\\
$\arraycolsep=1.4pt\def\arraystretch{2}
\begin{array}{l}
\paren{-\dfrac{\delta_4+\delta_5}{\Lambda}}\dfrac{1}{f_\pi^2}\dfrac{1}{2m_\pi}\dfrac{|\vec p_\pi|^4}{4m_\pi^2}\\
\sim (-\delta_4-\delta_5)\dfrac{1}{8}\epsilon^4\lambda\dfrac{1}{f_\pi^2}
\end{array}$
\end{tabular} \\
\hline 
\begin{tabular}{l}
N.L.O.\\
$(\delta_4,\delta_5)$
\end{tabular} & 
$\arraycolsep=1.4pt\def\arraystretch{2}
\begin{array}{l}
\paren{-\dfrac{\delta_4}{\Lambda}}\dfrac{1}{f_\pi^2}\dfrac{1}{2m_\pi}(\vpad\cdot\vpb\\
+P_a^\dg P_b)\cdot[\vnab\hmpi\vnab\hmpid\\
+\vnab\hmpid\vnab\hmpi]_{ba}
\end{array}$ &
\begin{tabular}{l}
\includegraphics[width=.25\textwidth]{72cfigf5.eps}\\
$\arraycolsep=1.4pt\def\arraystretch{2}
\begin{array}{l}
\paren{-\dfrac{\delta_4}{\Lambda}}\dfrac{1}{f_\pi^2}\dfrac{1}{2m_\pi}|\vec p_\pi|^2\\
\sim (-\delta_4)\dfrac{1}{2f_\pi^2}\epsilon^2\lambda
\end{array}$
\end{tabular} \\
\hline 
\begin{tabular}{l}
N.L.O.\\
$(\delta_6,\delta_7)$
\end{tabular} & 
$\arraycolsep=1.4pt\def\arraystretch{2}
\begin{array}{l}
\paren{-\dfrac{\delta_6+\delta_7}{\Lambda}}\dfrac{1}{f_\pi^2}\dfrac{1}{M_\pi}\\
(\vpad\cdot(\vnab\hmpi)_{bc}\vec P_c\cdot(\vnab\hmpid)_{cb}\\
+\vpad\cdot(\vnab\hmpid)_{bc}\vec P_c\cdot(\vnab\hmpi)_{cb})\\
\text{and indices exchaning terms}
\end{array}$ &
\begin{tabular}{l}
\includegraphics[width=.25\textwidth]{72cfigf5.eps}\\
$\arraycolsep=1.4pt\def\arraystretch{2}
\begin{array}{l}
\paren{-\dfrac{\delta_6+\delta_7}{\Lambda}}\dfrac{1}{f_\pi^2}\dfrac{1}{m_\pi}|\vec p_\pi|^2\\
\sim (\delta_6+\delta_7)\epsilon^2\lambda\dfrac{1}{2f_\pi^2}
\end{array}$
\end{tabular} \\
\hline 
\end{longtable}

\subsubsection{Isospin breaking case}
\label{appx:D.2.1.2}

Here we list the extra terms arising from isospin breaking effects. It is convenient to introduce the dimensionless constant $\epsilon_{ph}=\dfrac{p_\pi}{f_\pi}$.
\begin{longtable}{|c| p{.4\textwidth} | p{.4\textwidth} |}
\hline 
HH$\chi$PT order & Operator & Contribution at Tree Level \\
\hline 
L.O. & 
$\arraycolsep=1.4pt\def\arraystretch{2}
\begin{array}{l}
\dfrac{-e}{\sqrt{2m_\pi}}\vec P_a^\dg[B^0(Q_{bb}'\vec P_a-Q_{ba}\vec P_b)]\\
\dfrac{-e}{\sqrt{2m_\pi}}P_a^\dg[B^0(Q_{bb}'P_a-Q_{ba}\vec P_b)]
\end{array}
$ &
\begin{tabular}{ll}
\includegraphics[width=.125\textwidth]{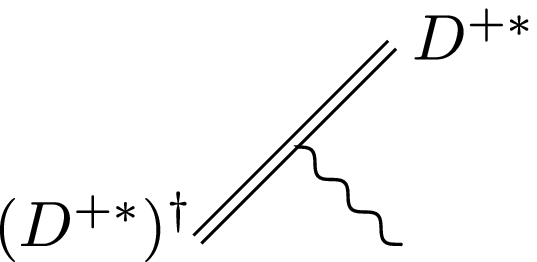} &
\includegraphics[width=.125\textwidth]{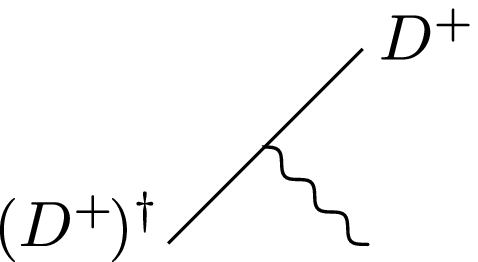} \\
\multicolumn{2}{l}{
$
\dfrac{(-e)}{\sqrt{2m_\pi}}\sim\paren{-\dfrac{\sqrt{\lambda}}{\sqrt{2m_\pi}}}\dfrac{p_\pi}{f_\pi}
$} \\
\multicolumn{2}{l}{
$\sim-\dfrac{\sqrt{\lambda}\epsilon_{ph}}{\sqrt{2m_\pi}}$
}
\end{tabular}
 \\
\hline  
L.O. &
$\arraycolsep=1.4pt\def\arraystretch{2}
\begin{array}{l}
\paren{\dfrac{g_\pi}{f_\pi}}\dfrac{1}{\sqrt{2m_\pi}}(ie)(\vec P_a^\dg\cdot P_b)\cdot\vec B\\
\times[Q,\hat M_\pi]_{ba}\\
\paren{\dfrac{g_\pi}{f_\pi}}\dfrac{1}{\sqrt{2m_\pi}}(ie)(P_a^\dg \vec P_b)\cdot\vec B\\
\times[Q,\hmpid]_{ba}
\end{array}
$ &
\begin{tabular}{ll}
\includegraphics[width=.125\textwidth]{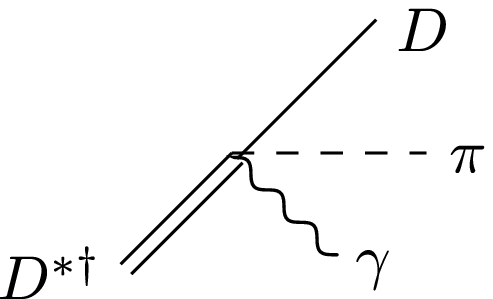} &
\includegraphics[width=.125\textwidth]{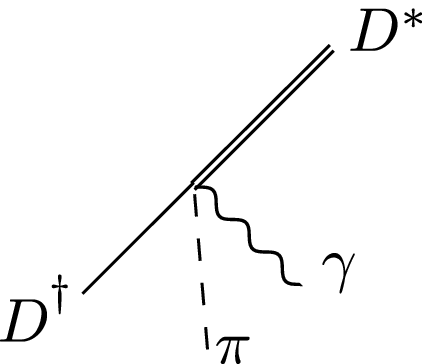} \\
\multicolumn{2}{c}{
$\arraycolsep=1.4pt\def\arraystretch{2}
\begin{array}{l}
\paren{\dfrac{g_\pi}{f_\pi}}\dfrac{1}{\sqrt{2m_\pi}}(ie)\\
\sim-i\dfrac{g_\pi}{f_\pi}\dfrac{1}{\sqrt{2m_\pi}}\sqrt{\lambda}\dfrac{p_\pi}{f_\pi}\\
\sim(-i)\sqrt{\lambda}\epsilon\dfrac{1}{\sqrt{2m_\pi}}\dfrac{1}{f_\pi}
\end{array}
$
}
\end{tabular} \\
\hline 
L.O. & 
$\arraycolsep=1.4pt\def\arraystretch{2}
\begin{array}{l}
\dfrac{8ieB_\mu}{m_\pi}[(\hat \pi^+)^\dg\olra{\pd_\mu}\hat \pi^+
-(\hat\pi^-)^\dg\olra{\pd_\mu}\hat\pi^-]
\end{array}$ & 
\begin{tabular}{ll}
$\dfrac{8ie}{2m_\pi}|\vec p_\pi|$ &
\includegraphics[width=.1\textwidth]{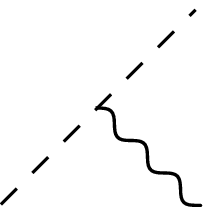}\\
\multicolumn{2}{l}{
$\sim 4i\sqrt{\lambda}\dfrac{|\vec p_\pi|}{f_\pi}\epsilon\sim 4i\sqrt{\lambda}\epsilon\epsilon_{ph}$
}
\end{tabular} \\
\hline 
L.O. &
$\arraycolsep=1.4pt\def\arraystretch{2}
\begin{array}{l}
\dfrac{e^2B_\mu B^\mu}{2m_\pi}[(\hat\pi^+)^\dg\hat \pi^+\\
+(\hat \pi^-)^\dg\hat \pi^-]
\end{array}$ & 
\begin{tabular}{l}
$\arraycolsep=1.4pt\def\arraystretch{2}
\begin{array}{l}
\dfrac{e^2}{2m_\pi}\sim\lambda\dfrac{\vec p_\pi^2}{f_\pi^2}\dfrac{1}{2m_\pi}\\
\sim\lambda\dfrac{\epsilon_{ph}^2}{2m_\pi}
\end{array}$\\
\includegraphics[width=.1\textwidth]{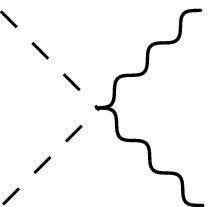}
\end{tabular} \\
\hline 
\begin{tabular}{l}
N.L.O.\\
$(M_3)$
\end{tabular} & 
$\arraycolsep=1.4pt\def\arraystretch{2}
\begin{array}{l}
(-\lambda_1)(m_u)(\vec D_0^\dg\vec D_0)\\
(-\lambda_1)(m_u)(D_0^\dg D_0)
\end{array}$ & 
$\arraycolsep=1.4pt\def\arraystretch{2}
\begin{array}{l}
(-\lambda_1)m_u\\
\sim(-\lambda_1)\lambda m_u
\end{array}$ \\
\hline 
\begin{tabular}{l}
N.L.O.\\
$(M_3)$
\end{tabular} & 
$\arraycolsep=1.4pt\def\arraystretch{2}
\begin{array}{l}
(-\lambda_1)(m_d)(\vec D_+^\dg \vec D_+)\\
(-\lambda_1)(m_d)(D_+^\dg D_+)
\end{array}$ & 
$\arraycolsep=1.4pt\def\arraystretch{2}
\begin{array}{l}
(-\lambda_1)m_d\\
\sim(-\lambda_1)\lambda m_\pi
\end{array}$ \\
\hline 
\begin{tabular}{l}
N.L.O.\\
$(M_3)$
\end{tabular} & 
$\arraycolsep=1.4pt\def\arraystretch{2}
\begin{array}{l}
\dfrac{4\lambda_1}{f_\pi^2 2m_\pi}(m_u)(\vec D_0^\dg \vec D_0)\\
\cdot(\hat\pi^0\hat \pi^{0\dg}+\hat \pi^+(\hat\pi^+)^\dg+(\hat\pi^-)(\hat\pi^-)^\dg)\\
\dfrac{4\lambda_1}{f_\pi^2 2m_\pi}(m_u)(D_0^\dg D_0)\\
\cdot(\hat \pi^0\hat\pi^{0\dg}+\hat\pi^+(\hat\pi^+)^\dg+(\hat\pi^-)(\hat\pi^-)^\dg)
\end{array}$ &
\begin{tabular}{l}
\includegraphics[width=.25\textwidth]{72cfigf5.eps}\\
$\sim\dfrac{4\lambda_1}{2m_\pi f_\pi^2}m_u$\\
$\sim\dfrac{4\lambda_1}{2f_\pi^2}\lambda$
\end{tabular} \\
\hline 
\begin{tabular}{l}
N.L.O.\\
$(M_3)$
\end{tabular} &
$\arraycolsep=1.4pt\def\arraystretch{2}
\begin{array}{l}
\dfrac{4\lambda_1}{f_\pi^2 2m_\pi}(m_d)(\vec D_+^\dg \vec D_+)\\
\cdot(\hat\pi^0\hat \pi^{0\dg}+\hat \pi^+(\hat\pi^+)^\dg+(\hat\pi^-)(\hat\pi^-)^\dg)\\
\dfrac{4\lambda_1}{f_\pi^2 2m_\pi}(m_d)(D_+^\dg D_+)\\
\cdot(\hat \pi^0\hat\pi^{0\dg}+\hat\pi^+(\hat\pi^+)^\dg+(\hat\pi^-)(\hat\pi^-)^\dg)
\end{array}$ &
\begin{tabular}{l}
\includegraphics[width=.25\textwidth]{72cfigf5.eps}\\
$\sim\dfrac{4\lambda_1}{2f_\pi^2}\lambda$
\end{tabular} \\
\hline 
\multirow{2}{*}{
\begin{tabular}{l}
N.L.O.\\
$(M_4)$
\end{tabular}
} & 
$\phantom{\Bigg\Vert}\paren{-\dfrac{\Delta^{(\lambda_1)}}{4}}(m_u)(\vec D_0^\dg\vec D_0)$ & $\phantom{\Bigg\Vert}-\dfrac{\Delta^{(\lambda_1)}}{4}\lambda m_\pi$ \\
\cline{2-3}
& $\phantom{\Bigg\Vert}\paren{\dfrac{3\Delta^{(\lambda_1)}}{4}}(m_u)( D_0^\dg D_0)$ & $\phantom{\Bigg\Vert}\dfrac{3\Delta^{(\lambda_1)}}{4}\lambda m_\pi$ \\
\hline 
\multirow{2}{*}{
\begin{tabular}{l}
N.L.O.\\
$(M_4)$
\end{tabular}
} & 
$\phantom{\Bigg\Vert}\paren{-\dfrac{\Delta^{(\lambda_1)}}{4}}(m_d)(\vec D_0^\dg\vec D_0)$ & $\phantom{\Bigg\Vert}-\dfrac{\Delta^{(\lambda_1)}}{4}\lambda m_\pi$ \\
\cline{2-3}
& $\phantom{\Bigg\Vert}\paren{\dfrac{3\Delta^{(\lambda_1)}}{4}}(m_d)( D_0^\dg D_0)$ & $\phantom{\Bigg\Vert}\dfrac{3\Delta^{(\lambda_1)}}{4}\lambda m_\pi$ \\
\hline 

\multirow{2}{*}{
\begin{tabular}{l}
N.L.O.\\
$(M_4)$
\end{tabular}
} & 
$\arraycolsep=1.4pt\def\arraystretch{2}
\begin{array}{l}
\dfrac{\Delta^{(\lambda_1)}}{f_\pi^2 2m_\pi}(m_u)(\vec D_0^\dg \vec D_0)\\
\cdot(\hat\pi^0\hat\pi^{0\dg}+\hat\pi^+(\hat\pi^+)^\dg+(\hat \pi^-)(\hat \pi^-)^\dg)
\end{array}
$ & 
\begin{tabular}{ll}
\includegraphics[width=.125\textwidth]{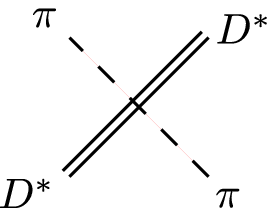} & 
$\sim\dfrac{\Delta^{(\lambda_1)}}{2f_\pi^2}\lambda$
\end{tabular} \\ \cline{2-3}
& 
$\arraycolsep=1.4pt\def\arraystretch{2}
\begin{array}{l}
-\dfrac{3\Delta^{(\lambda_1)}}{2m_\pi f_\pi^2}(m_u)(D_0^\dg D_0)\\
\cdot(\hat\pi^0\hat\pi^{0\dg}+\hat\pi^+(\hat\pi^+)^\dg+(\hat \pi^-)(\hat \pi^-)^\dg)
\end{array}$ & 
\begin{tabular}{ll}
\includegraphics[width=.125\textwidth]{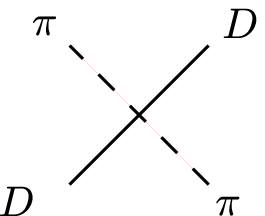} &
$\sim-\dfrac{3\Delta^{(\lambda_1)}}{2f_\pi^2}\lambda$
\end{tabular} \\
\hline 
\multirow{2}{*}{
\begin{tabular}{l}
N.L.O.\\
$(M_4)$
\end{tabular}
} & 
$\arraycolsep=1.4pt\def\arraystretch{2}
\begin{array}{l}
\dfrac{\Delta^{(\lambda_1)}}{f_\pi^2 2m_\pi}(m_d)(\vec D_+^\dg \vec D_+)\\
\cdot(\hat\pi^0\hat\pi^{0\dg}+\hat\pi^+(\hat\pi^+)^\dg+(\hat \pi^-)(\hat \pi^-)^\dg)
\end{array}
$ & 
\begin{tabular}{ll}
\includegraphics[width=.125\textwidth]{72cfigf15.eps} & 
$\sim\dfrac{\Delta^{(\lambda_1)}}{2f_\pi^2}\lambda$
\end{tabular} \\ \cline{2-3}
& 
$\arraycolsep=1.4pt\def\arraystretch{2}
\begin{array}{l}
-\dfrac{3\Delta^{(\lambda_1)}}{2m_\pi f_\pi^2}(m_d)(D_+^\dg D_+)\\
\cdot(\hat\pi^0\hat\pi^{0\dg}+\hat\pi^+(\hat\pi^+)^\dg+(\hat \pi^-)(\hat \pi^-)^\dg)
\end{array}$ & 
\begin{tabular}{ll}
\includegraphics[width=.125\textwidth]{72cfigf16.eps} &
$\sim-\dfrac{3\Delta^{(\lambda_1)}}{2f_\pi^2}\lambda$
\end{tabular} \\
\hline 
\begin{tabular}{l}
N.L.O.\\
$(V.R.I_1)$
\end{tabular} &
$\arraycolsep=1.4pt\def\arraystretch{2}
\begin{array}{l}
\vec D_+^\dg e^2 B^\mu B_\mu \vec D_+\paren{-\dfrac{1}{M}}\\
D_+^\dg e^2 B^\mu B_\mu D_+\paren{-\dfrac{1}{M}}
\end{array}$ &
\begin{tabular}{ll}
\includegraphics[width=.125\textwidth]{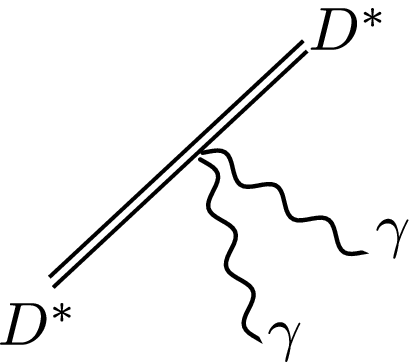} &
\includegraphics[width=.125\textwidth]{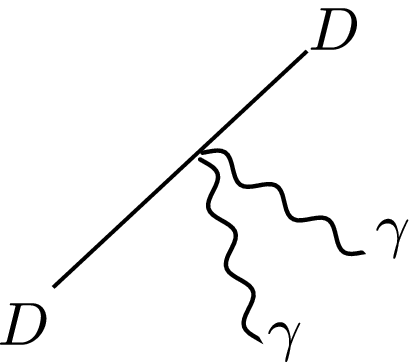}\\
\multicolumn{2}{l}{
$\sim\dfrac{e^2}{M}\sim\dfrac{\lambda}{M}\dfrac{p_\pi}{f_\pi}\sim\dfrac{\lambda}{M}\epsilon_{ph}$
}
\end{tabular} \\
\hline 
\begin{tabular}{l}
N.L.O.\\
$(V.R.I_1)$
\end{tabular} &
$\arraycolsep=1.4pt\def\arraystretch{2}
\begin{array}{l}
\pd^\mu\vec D_+^\dg (-e) B_\mu \vec D_+\paren{-\dfrac{1}{M}}\\
\pd^\mu D_+^\dg (-e) B_\mu D_+\paren{-\dfrac{1}{M}}
\end{array}$ &
\begin{tabular}{l}
\includegraphics[width=.25\textwidth]{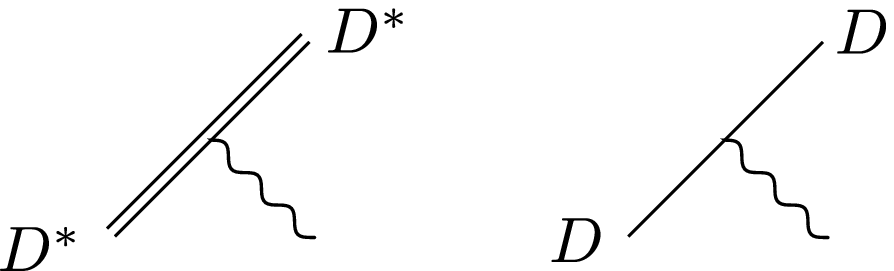}\\
$\sim-\dfrac{e|\vec p_\pi|}{M}\sim-\sqrt{\lambda}\epsilon\epsilon_{ph}\lambda$ 
\end{tabular} \\
\hline 
\begin{tabular}{l}
N.L.O.\\
$(V.R.I_1)$
\end{tabular} &
$\dfrac{m_\pi}{M}\vec D_+^\dg(ie)B^0(\vec D_+)
$ &
\begin{tabular}{l}
\includegraphics[width=.125\textwidth]{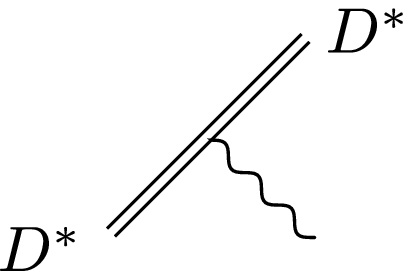}\\
$\sim(ie)\dfrac{m_\pi}{M}\sim i\sqrt{\lambda}\lambda \epsilon_{ph}$ 
\end{tabular} \\
\hline 
\begin{tabular}{l}
N.L.O.\\
$(V.R.I_1)$
\end{tabular} &
$\arraycolsep=1.4pt\def\arraystretch{2}
\begin{array}{l}
\paren{\dfrac{1}{M}}D_+^\dg (ie)B_\mu \pd^\mu D_+\\
\paren{\dfrac{1}{M}}\vec D_+^\dg (ie)B_\mu \pd^\mu \vec D_+
\end{array}$ &
\begin{tabular}{l}
\includegraphics[width=.25\textwidth]{72cfigf19.eps}\\
$\sim(ie)\dfrac{|p_\pi|}{M}\sim i\sqrt{\lambda}\dfrac{m_\pi}{m_\pi}\dfrac{p_\pi^2}{f_\pi}\dfrac{1}{M}$\\
$\sim i\sqrt{\lambda}\epsilon\epsilon_{ph}\lambda$
\end{tabular} \\
\hline 
\begin{tabular}{l}
N.L.O.\\
$(V.R.I_1)$
\end{tabular} &
$\paren{-\dfrac{1}{M}}(im_\pi)(ie)\vec D_+^\dg B^0\vec D_+
$ &
\begin{tabular}{l}
\includegraphics[width=.125\textwidth]{72cfigf20.eps}\\
$\sim\dfrac{m_\pi}{M}e\sim\lambda\sqrt{\lambda}\epsilon_{ph}$
\end{tabular} \\
\hline 
\begin{tabular}{l}
N.L.O.\\
$(V.R.I_1)$
\end{tabular} &
$\arraycolsep=1.4pt\def\arraystretch{2}
\begin{array}{l}
\paren{-\dfrac{1}{M}}\vec D_+^\dg\paren{\dfrac{ie}{f_\pi^2}}\dfrac{im_\pi}{2m_\pi}B^0\\
\cdot[\hat\pi^-(\hat \pi^-)^\dg-(\hat \pi^+)(\hat\pi^+)^\dg]\vec D_+\\
\paren{-\dfrac{1}{M}}D_+^\dg\paren{\dfrac{ie}{f_\pi^2}}\dfrac{im_\pi}{2m_\pi}\\
B^0[\hat\pi^-(\hat\pi^-)^\dg\\
-\hat\pi^+(\hat\pi^+)^\dg]D_+
\end{array}$ &
\begin{tabular}{l}
\includegraphics[width=.25\textwidth]{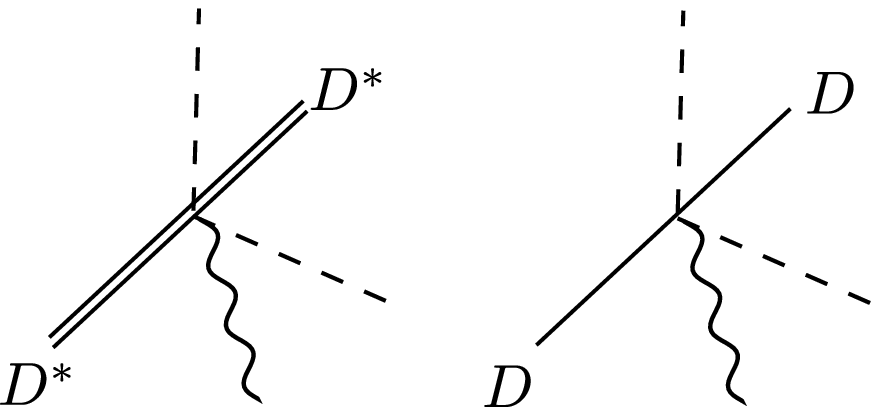}\\
$\arraycolsep=1.4pt\def\arraystretch{2}
\begin{array}{l}
\sim\dfrac{e}{2f_\pi^2}\dfrac{1}{M}\sim\paren{-\dfrac{1}{2}}\sqrt{\lambda}\dfrac{p_\pi}{m_\pi}\dfrac{m_\pi}{M}\dfrac{1}{f_\pi^3}\\
\sim\paren{-\dfrac{1}{2}}\sqrt{\lambda}\epsilon\lambda\dfrac{1}{f_\pi^3}
\end{array}$
\end{tabular} \\
\hline 
\begin{tabular}{l}
N.L.O.\\
$(V.R.I_1)$
\end{tabular} &
$\arraycolsep=1.4pt\def\arraystretch{2}
\begin{array}{l}
\paren{-\dfrac{1}{M}}\vec D_+^\dg\paren{\dfrac{ie}{f_\pi^2}}\paren{\dfrac{1}{2m_\pi}}\\
\,[(\hat\pi^-)\olra{\pd^\mu}(\hat\pi^-)^\dg+\dfrac{1}{2}(\hat \pi^0)\olra{\pd^\mu}(\hat\pi^0)^\dg\\
+(\hat\pi^+)\olra{\pd^\mu}(\hat\pi^+)^\dg]B_\mu\vec D_+\\
\paren{-\dfrac{1}{M}}D_+^\dg\paren{\dfrac{ie}{f_\pi^2}}\paren{\dfrac{1}{2m_\pi}}\\
\,[(\hat\pi^-)\olra{\pd^\mu}(\hat\pi^-)^\dg\\
+\dfrac{1}{2}(\hat\pi^0)\olra{\pd^\mu}(\hat\pi^0)^\dg+\dfrac{1}{2}(\hat\pi^+)\olra{\pd^\mu}\\
(\hat\pi^+)^\dg]B_\mu D_+
\end{array}$ 
& 
\begin{tabular}{l}
\includegraphics[width=.25\textwidth]{72cfigf21.eps}\\
$\arraycolsep=1.4pt\def\arraystretch{2}
\begin{array}{l}
\sim\dfrac{e}{2f_\pi^2}\dfrac{p_\pi}{M}\dfrac{1}{m_\pi}\\
\sim\paren{-\dfrac{1}{2}}\sqrt{\lambda}\dfrac{p_\pi^2}{f_\pi^3}\dfrac{1}{M}\dfrac{m_\pi}{m_\pi^2}\\
\sim\paren{-\dfrac{1}{2}}\sqrt{\lambda}\lambda\epsilon^2\dfrac{1}{f_\pi^3}
\end{array}
$
\end{tabular} \\
\hline 
\begin{tabular}{l}
N.L.O.\\
$(V.R.I_1)$
\end{tabular} & 
$\arraycolsep=1.4pt\def\arraystretch{2}
\begin{array}{l}
(eB_0)\dfrac{m_\pi}{\sqrt{2f_\pi^2m_\pi}}\paren{\dfrac{1}{M}}(\vec D_+^\dg\vec D^0)\\
\cdot[(\hat \pi^-)^\dg (\hat \pi^0)-(\hat\pi^0)^\dg(\hat\pi^+)]\\
(eB_0)\dfrac{m_\pi}{\sqrt{2}f_\pi^2}\dfrac{1}{m_\pi}\dfrac{1}{M}(\vec D_+^\dg\vec D_+)\\
\cdot[(\hat\pi^-)(\hat \pi^-)^\dg-(\hat\pi^+)(\hat\pi^+)^\dg]\\
\text{and $\vec D D$ exchange terms}
\end{array}$ & 
\begin{tabular}{l}
\includegraphics[width=.125\textwidth]{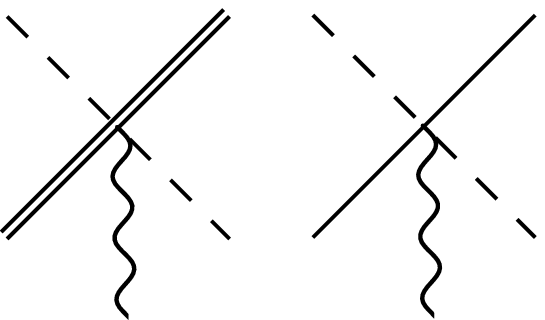}\\
$\arraycolsep=1.4pt\def\arraystretch{2}
\begin{array}{l}
\dfrac{e}{\sqrt{2}f_\pi^2}\dfrac{1}{M}\sim\dfrac{\sqrt{\lambda}}{\sqrt{2}}\dfrac{p_\pi}{f_\pi}\dfrac{1}{M}\dfrac{1}{f_\pi^2}\\
\sim\dfrac{\sqrt{\lambda}}{\sqrt{2}}\lambda\epsilon\dfrac{1}{f_\pi^3}
\end{array}$
\end{tabular} \\
\hline 
\begin{tabular}{l}
N.L.O.\\
$(V.R.I_2)$
\end{tabular} &
$\arraycolsep=1.4pt\def\arraystretch{2}
\begin{array}{l}
\dfrac{1}{\sqrt{2m_\pi}}\paren{\dfrac{g_\pi}{M}} \dfrac{1}{f_\pi} (ie)(\hat\pi^+)\\
(i\vec\nabla\cdot\vec D_+^\dg)D_0B_0\\
\dfrac{1}{\sqrt{2m_\pi}}\paren{\dfrac{g_\pi}{M}} \dfrac{1}{f_\pi} (ie)B_0(\hat\pi^-)\\
(i\vnab\cdot\vec D_0^\dg)D_+\\
\text{and h.c. terms + $\vec D,D$ exchange}
\end{array}$ & 
\begin{tabular}{l}
\includegraphics[width=.125\textwidth]{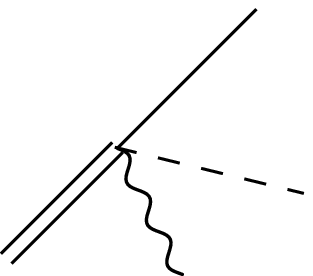}\\
$\arraycolsep=1.4pt\def\arraystretch{2}
\begin{array}{l}
-\dfrac{1}{\sqrt{2m_\pi}}\paren{\dfrac{g_\pi}{f_\pi}}\dfrac{1}{M}e|\vec p_0|\\
\sim\paren{-\dfrac{g_\pi}{f_\pi}}\sqrt{\lambda}\dfrac{|\vec p_\pi|^2}{f_\pi}\dfrac{1}{M}\dfrac{1}{\sqrt{2m_\pi}}\\
\sim(-\sqrt{\lambda})\epsilon^2\lambda\dfrac{1}{f_\pi}\dfrac{1}{\sqrt{2m_\pi}}
\end{array}$
\end{tabular} \\
\hline 
\begin{tabular}{l}
N.L.O.\\
$(V.R.I_2)$
\end{tabular} &
$\arraycolsep=1.4pt\def\arraystretch{2}
\begin{array}{l}
\dfrac{1}{\sqrt{2m_\pi}}\paren{\dfrac{g_\pi}{M}} \dfrac{1}{f_\pi} \\
(-im_\pi)(-e)\hat\pi^+ D^0\vec D_+^\dg\cdot\vec B\\
\dfrac{1}{\sqrt{2m_\pi}}\paren{\dfrac{g_\pi}{M}} \dfrac{1}{f_\pi} \\
(-im_\pi)(-e)\hat\pi^0 D_+\vec D_+^\dg\cdot\vec B\\
\text{plus h.c. and $\vec D,D$ exchange terms}
\end{array}$ &
\begin{tabular}{l}
\includegraphics[width=.125\textwidth]{72cfigf23.eps}\\
$\arraycolsep=1.4pt\def\arraystretch{2}
\begin{array}{l}
\dfrac{i}{\sqrt{2m_\pi}}\dfrac{g_\pi}{f_\pi}m_\pi\sqrt{\lambda}\dfrac{|\vec p_\pi|}{f_\pi}\dfrac{1}{M}\\
\sim\dfrac{i}{\sqrt{2m_\pi}}\epsilon\lambda\sqrt{\lambda}\dfrac{1}{f_\pi}\dfrac{1}{\sqrt{2m_\pi}}
\end{array}$ 
\end{tabular} \\
\hline 
\begin{tabular}{l}
N.L.O.\\
$(V.R.I_2)$
\end{tabular} &
$\arraycolsep=1.4pt\def\arraystretch{2}
\begin{array}{l}
\dfrac{1}{\sqrt{2m_\pi}}\paren{\dfrac{g_\pi}{M}} \dfrac{1}{f_\pi} (-e)\\
\pd_0\hat\pi^+ D^0\vec D_+^\dg\cdot\vec B\\
\dfrac{1}{\sqrt{2m_\pi}}\paren{\dfrac{g_\pi}{M}} \dfrac{1}{f_\pi} (-e)\\
\pd_0\hat\pi^0 D_+\vec D_+^\dg\cdot\vec B\\
\text{plus h.c. and $\vec D,D$ exchange terms}
\end{array}$ &
\begin{tabular}{l}
\includegraphics[width=.125\textwidth]{72cfigf23.eps}\\
$\arraycolsep=1.4pt\def\arraystretch{2}
\begin{array}{l}
\dfrac{-1}{\sqrt{2m_\pi}}\dfrac{g_\pi}{f_\pi}\dfrac{1}{M}\sqrt{\lambda}\dfrac{|\vec p_\pi|}{f_\pi}\dfrac{|\vec p_\pi|^2}{2m_\pi}\\
\sim(-\sqrt{\lambda})\epsilon^3\lambda\dfrac{1}{f_\pi}\dfrac{1}{\sqrt{2m_\pi}}
\end{array}$
\end{tabular} \\
\hline 
\begin{tabular}{l}
N.L.O.\\
$(V.R.I_2)$
\end{tabular} &
$\arraycolsep=1.4pt\def\arraystretch{2}
\begin{array}{l}
\dfrac{1}{\sqrt{2m_\pi}}\dfrac{g_\pi}{M}\dfrac{1}{f_\pi}(ie)B_0 D_+\\
(\hat \pi^-)(-e)\vec B\cdot\vec D_0^\dg\\
\dfrac{1}{\sqrt{2m_\pi}}\dfrac{g_\pi}{f_\pi}\dfrac{1}{M}(ie)B_0D_0\\
(\hat\pi^+)(-e)\vec B\cdot\vec D_+^\dg\\
\text{and h.c. terms }\\
\text{plus $\vec D,D$ exchange terms}
\end{array}$ &
\begin{tabular}{l}
\includegraphics[width=.125\textwidth]{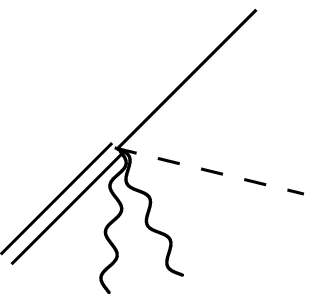}\\
$\arraycolsep=1.4pt\def\arraystretch{2}
\begin{array}{l}
(-ie)^2\dfrac{g_\pi}{M}\dfrac{1}{f_\pi}\dfrac{1}{\sqrt{2m_\pi}}\\
\sim(-i)\lambda\dfrac{|\vec p_\pi|^2}{f_\pi^2}\dfrac{g_\pi}{M}\dfrac{1}{f_\pi}\dfrac{1}{\sqrt{2m_\pi}}\\
\sim(-i)\lambda^2\epsilon^2\dfrac{1}{f_\pi}\dfrac{1}{\sqrt{2m_\pi}}
\end{array}$
\end{tabular} \\
\hline 
\begin{tabular}{l}
N.L.O.\\
$(\delta_4,\delta_5)$
\end{tabular} &
$\arraycolsep=1.4pt\def\arraystretch{2}
\begin{array}{l}
\paren{\dfrac{\delta_4+\delta_5}{\Lambda}}\dfrac{1}{f_\pi^2}\oneov{2m_\pi}(ie)(B^0)\\
\bigg(\vec D^{0\dg}\vec D^0(\pi^+)^\dg\pd_0(\pi^+)\\
+\vec D^{0\dg}\vec D^0(\pi^+)\pd_0(\pi^+)^\dg\\
+\vec D^{0\dg}\vec D^0(\pi^-)^\dg\pd_0(\pi^-)\\
+\vec D^{0\dg}\vec D^0(\pi^-)\pd_0(\pi^-)^\dg\\
+\vec D^{+\dg}\vec D^+(\pi^+)^\dg\pd_0(\pi^+)\\
+\vec D^{+\dg}\vec D^+(\pi^+)\pd_0(\pi^+)^\dg\\
+\vec D^{+\dg}\vec D^+(\pi^-)^\dg\pd_0(\pi^-)\\
+\vec D^{+\dg}\vec D^+(\pi^-)\pd_0(\pi^-)^\dg\bigg)\\
\text{ and $\vec D,D$ exhange terms}
\end{array}$ &
\begin{tabular}{l}
\includegraphics[width=.25\textwidth]{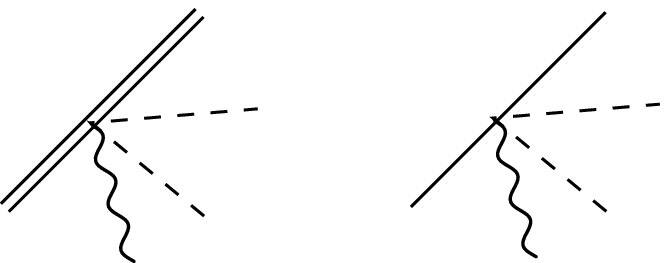}\\
$\arraycolsep=1.4pt\def\arraystretch{2}
\begin{array}{l}
\paren{\dfrac{\delta_4+\delta_5}{\Lambda}}\dfrac{1}{f_\pi^2}\oneov{2m_\pi}(ie)\dfrac{p_\pi^2}{2m_\pi}\\
\sim \paren{\dfrac{\delta_4+\delta_5}{\Lambda}}\dfrac{i}{f_\pi^2}\sqrt{\lambda}\dfrac{p_\pi}{f_\pi}\dfrac14\epsilon^2\\
\sim\dfrac{(\delta_4+\delta_5)}{f_\pi^3}i\sqrt{\lambda}\dfrac{\lambda}{4}\epsilon^3
\end{array}$
\end{tabular} \\
\hline 
\begin{tabular}{l}
N.L.O.\\
$(\delta_4,\delta_5)$
\end{tabular} &
$\arraycolsep=1.4pt\def\arraystretch{2}
\begin{array}{l}
\paren{\dfrac{\delta_4+\delta_5}{\Lambda}}\dfrac{1}{f_\pi^2}\oneov{2m_\pi}(-e^2)(B^0)^2\\
\bigg[\vec D^{0\dg}\vec D^0(\pi^+(\pi^+)^\dg+\pi^-(\pi^-)^\dg)\\
+\vec D^{+\dg}\vec D^+(\pi^+(\pi^+)^\dg+\pi^-(\pi^-)^\dg)\bigg]\\
\text{and $\vec D,D$ exchange terms}
\end{array}$ &
\begin{tabular}{l}
\includegraphics[width=.25\textwidth]{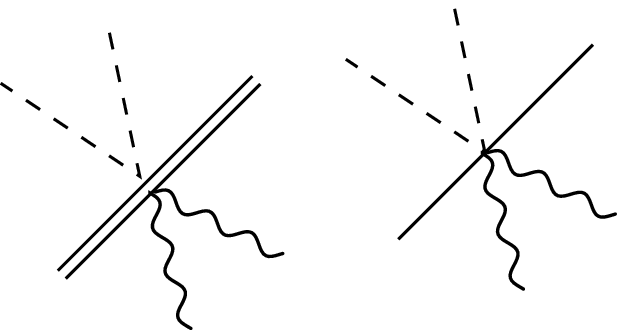}\\
$\arraycolsep=1.4pt\def\arraystretch{2}
\begin{array}{l}
\paren{\dfrac{\delta_4+\delta_5}{\Lambda}}\dfrac{1}{f_\pi^2}\oneov{2m_\pi}(-e^2)\\
\sim \paren{\dfrac{\delta_4+\delta_5}{\Lambda}}\dfrac{1}{f_\pi^2}\oneov{2m_\pi}(-\lambda)\dfrac{p_\pi^2}{f_\pi^2}\\
\sim(-\delta_4-\delta_5)\lambda^2\epsilon^2\oneov{f_\pi^4}
\end{array}$
\end{tabular} \\
\hline 
\begin{tabular}{l}
N.L.O.\\
$(\delta_4,\delta_5)$
\end{tabular} &
$\arraycolsep=1.4pt\def\arraystretch{2}
\begin{array}{l}
\paren{\dfrac{\delta_4}{\Lambda}}\dfrac{1}{f_\pi^2}\oneov{2m_\pi}(ie)\vec B\cdot\\
\,[\vec D^{0\dg}\vec D^0(\pi^+)^\dg\vnab\pi^+\\
+\vec D^{0\dg}\vec D^0(\pi^+)(\vnab\pi^+)^\dg\\
+\vec D^{-\dg}\vec D^0(\pi^-)^\dg\vnab\pi^-\\
+\vec D^{0\dg}\vec D^0(\pi^-)(\vnab\pi^-)^\dg]\\
\text{and $D^0,D^+$ exchange terms}\\
\text{and $\vec D,D$ exchange terms}
\end{array}$ &
\begin{tabular}{l}
\includegraphics[width=.25\textwidth]{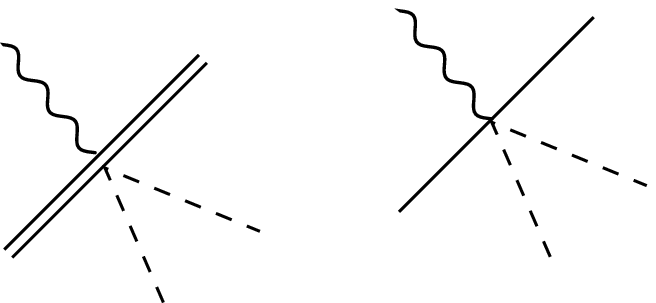}\\
$\arraycolsep=1.4pt\def\arraystretch{2}
\begin{array}{l}
\dfrac{\delta_4}{\Lambda}\oneov{f_\pi^2}\oneov{2m_\pi}(ie)|\vec p_\pi|\\
\sim\dfrac{\delta_4}{\Lambda}\oneov{f_\pi^2}\dfrac{i}{2m_\pi}\sqrt{\lambda}\dfrac{|\vec p_\pi^2|}{f_\pi}\\
\sim(i\delta_4)\sqrt{\lambda}\lambda\epsilon^2\oneov{f_\pi^3}
\end{array}$
\end{tabular} \\
\hline 
\begin{tabular}{l}
N.L.O.\\
$(\delta_4,\delta_5)$
\end{tabular} &
$\arraycolsep=1.4pt\def\arraystretch{2}
\begin{array}{l}
\paren{\dfrac{\delta_4}{\Lambda}}\dfrac{1}{f_\pi^2}\oneov{2m_\pi}(-e^2)\vec B^2\\
\cdot[\vec D^{0\dg}\vec D^0(\pi^+(\pi^+)^\dg+\pi^-(\pi^-)^\dg)]\\
\text{and $D^0D^+$ exchange terms}\\
\text{and $\vec D,D$ exchange terms}
\end{array}$ & 
\begin{tabular}{l}
\includegraphics[width=.25\textwidth]{72cfigf26.eps}\\
$\arraycolsep=1.4pt\def\arraystretch{2}
\begin{array}{l}
\dfrac{\delta_4}{\Lambda}\oneov{f_\pi^2}\oneov{2m_\pi}(ie)|\vec p_\pi|\\
\sim\dfrac{\delta_4}{\Lambda}\oneov{f_\pi^2}\dfrac{1}{2m_\pi}(-e^2)\\
\sim(-\delta_4)\lambda^2\epsilon^2\oneov{f_\pi^4}
\end{array}$
\end{tabular} \\
\hline 
\begin{tabular}{l}
N.L.O.\\
$(\delta_6,\delta_7)$
\end{tabular} &
$\arraycolsep=1.4pt\def\arraystretch{2}
\begin{array}{l}
2\dfrac{2(\delta_6+\delta_7)}{\Lambda}(ie)\vec B\\
\cdot[\vec D_+^\dg\pi^+\vec D_+\vnab(\pi^+)^\dg\\
+\vec D_+^\dg(\vnab\pi^+)\vec D_+(\pi^+)^\dg\\
+\vec D_0^\dg(\pi^+)\vec D_0(\vnab\pi^+)^\dg\\
+\vec D_0^\dg(\vnab\pi^+)\vec D_0(\pi^+)^\dg]\oneov{f_\pi^2}\oneov{2m_\pi}\\
\text{and $\pi^+\pi^-$ exchange terms}
\end{array}$ &
\begin{tabular}{l}
\includegraphics[width=.125\textwidth]{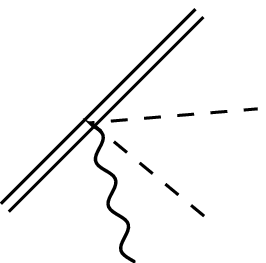}\\
$\arraycolsep=1.4pt\def\arraystretch{2}
\begin{array}{l}
\dfrac{4(\delta_6+\delta_7)}{\Lambda}ie\oneov{f_\pi^2}\oneov{2m_\pi}\\
\sim 2(\delta_6+\delta_7)i\sqrt{\lambda}\dfrac{p_\pi}{f_\pi^3}\oneov{\Lambda}\oneov{m_\pi}\\
\sim 2(\delta_6+\delta_7)i\sqrt{\lambda}\epsilon\lambda\oneov{f_\pi^3}
\end{array}$
\end{tabular} \\
\hline 
\begin{tabular}{l}
N.L.O.\\
$(\delta_6,\delta_7)$
\end{tabular} &
$\arraycolsep=1.4pt\def\arraystretch{2}
\begin{array}{l}
\dfrac{2(\delta_6+\delta_7)}{\Lambda}(-e^2)(\vec B)^2\oneov{f_\pi^2}\oneov{2m_\pi}\\
\,[\vec D^{+\dg}\pi^+\vec D^+(\pi^+)^\dg\\
+\vec D^{0\dg}\pi^-(\pi^-)^\dg\vec D^0]\\
\text{and $\pi^+\pi^-$ exchange terms}
\end{array}$ & 
\begin{tabular}{l}
\includegraphics[width=.125\textwidth]{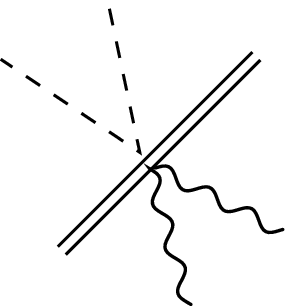}\\
$\arraycolsep=1.4pt\def\arraystretch{2}
\begin{array}{l}
\sim\dfrac{(\delta_6+\delta_7)}{\Lambda}\oneov{m_\pi}(-e^2)\oneov{f_\pi^2}\\
\sim-(\delta_6+\delta_7)\oneov{\Lambda}\lambda\dfrac{p_\pi^2}{f_\pi^2}\oneov{f-\pi^2}\oneov{2m_\pi}\\
\sim(-\delta_6-\delta_7)\lambda^2\epsilon^2\oneov{f_\pi^4}
\end{array}$
\end{tabular}  \\
\hline 
\begin{tabular}{l}
N.L.O.\\
$(\pi^+\pi^-)$
\end{tabular} &
$\dfrac{e^2F_\pi^2}{2m_\pi}[(\pi^+)(\pi^+)^\dg+(\pi^-)(\pi^-)^\dg]$ &
$\arraycolsep=1.4pt\def\arraystretch{2}
\begin{array}{l}
\dfrac{e^2f_\pi^2}{2m_\pi}\sim\lambda\dfrac{p_\pi^2}{f_\pi^2}\dfrac{f_\pi^2}{2m_\pi}\\
\sim\dfrac{\lambda}{2}\epsilon^2 m_\pi
\end{array}$ \\
\hline 
\end{longtable}


\section{Higher Order \texorpdfstring{$D^*$}{D star} Propagator Corrections}
\label{appx:D.3}



In this part, we give a few examples of beyond next-to-leading order $D\pi$ scattering Feynman amplitudes. We organize these diagrams by classes and give one example per class.

\begin{description}
\item[(I)] Contact vertex for $D\pi$ scattering.  The diagram and its size are
\begin{center}
\includegraphics[width=.5\textwidth]{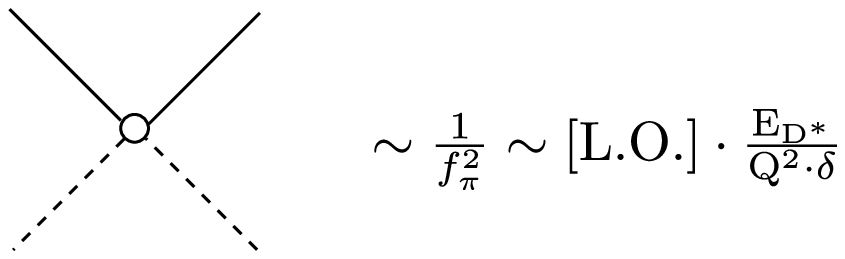}
\end{center}
\item[(II)] Two-loop diagram with pion exchange.  For example,
\begin{center}
\includegraphics[width=.5\textwidth]{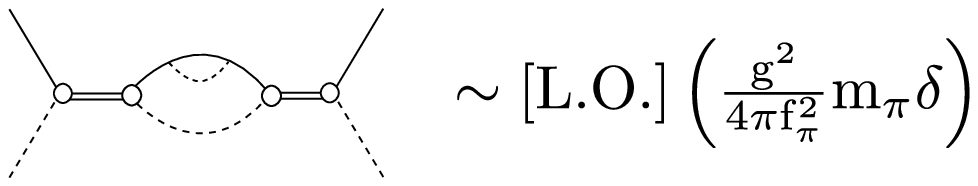}
\end{center}
\item[(III)] Two-loop diagrams with photon-pion interaction.  For example,
\begin{center}
\includegraphics[width=.5\textwidth]{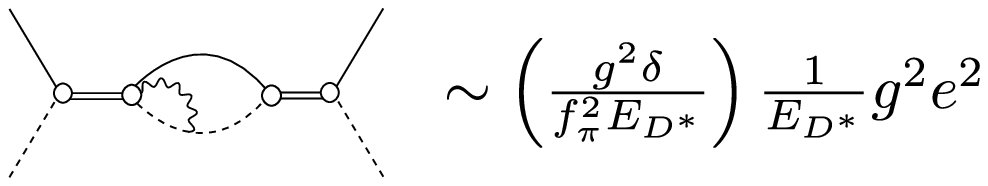}
\end{center}
\end{description}

\section{Mathematica Solutions to Leading Order \texorpdfstring{$D^*$}{D star} Pole}
\label{appx:D.4}

We solve the algebraic equation $p^2(1+\lambda)-2m_\pi\delta+ip^3b=0$ for the pole location with
\begin{align*}
M_D=1800\, \mev, \quad\Delta=142\,\mev,\quad f_\pi=93\,\mev, \quad g=0.6, \quad
\delta=\Delta-m_\pi,
\end{align*}
and we define dimensionless constants for notational simplicity
\begin{align*}
b=\frac{m_\pi}{c},\quad c=\frac{24\pi f_\pi^2}{g^2},\quad \lambda=\frac{m_\pi}{M_D},\quad
\mu=\frac{m_\pi M_D}{m_\pi+M_D},\quad \nu=\frac{2c^2}{27\mu^3}
\end{align*}
where $\mu$ is reduced mass of the $D\pi$ system.  For $m_\pi\le\Delta$ the solutions are
\begin{align*}
p_1&=i\left(c^{1/3}\sqrt[3]{\left(\delta+\frac{\nu}{2} \right )+\sqrt{\delta(\delta+\nu)}}+\sqrt[3]{\left(\delta+\frac{\nu}{2} \right )-\sqrt{\delta(\delta+\nu)}}+\frac{c}{3\mu}\right)\\
p_2&=i\left(c^{1/3}\omega\sqrt[3]{\left(\delta+\frac{\nu}{2} \right )+\sqrt{\delta(\delta+\nu)}}+\omega^2\sqrt[3]{\left(\delta+\frac{\nu}{2} \right )-\sqrt{\delta(\delta+\nu)}}+\frac{c}{3\mu}\right)\\
p_3&=i\left(c^{1/3}\omega^2\sqrt[3]{\left(\delta+\frac{\nu}{2} \right )+\sqrt{\delta(\delta+\nu)}}+\omega\sqrt[3]{\left(\delta+\frac{\nu}{2} \right )-\sqrt{\delta(\delta+\nu)}}+\frac{c}{3\mu}\right).
\end{align*}
in which $\omega=e^{2\pi i/3}=-\oneov{2}+\frac{\sqrt{3}}{2}i$ is the complex cube-root of 1.  On the other hand, for $m_\pi>\Delta$ the solutions are
\begin{align*}
p_1&=i\left(c^{1/3}2\text{Re}\left[\sqrt[3]{\left(\delta+\frac{\nu}{2} \right )+\sqrt{\delta(\delta+\nu)}}\right]
+\frac{c}{3\mu}\right)\\
p_2&=i\left(c^{1/3}2\text{Re}\left[\omega\sqrt[3]{\left(\delta+\frac{\nu}{2} \right )+\sqrt{\delta(\delta+\nu)}}\right]
+\frac{c}{3\mu}\right)\\
p_3&=i\left(c^{1/3}2\text{Re}\left[\omega^2\sqrt[3]{\left(\delta+\frac{\nu}{2} \right )+\sqrt{\delta(\delta+\nu)}}\right]
+\frac{c}{3\mu}\right).\\
\end{align*}
These roots have been verified with Mathematica, but the labeling is by no means stable across $m_\pi=\Delta$ as we do not control which cube root Mathematica takes. To obtain analytically the roots of a cubic equation, $x^3+px+q=0$, we insert the Ans\"{a}tze: (1) $x=u+v$, so that $x^3=(u+v)^3=u^3+v^3+3uv(u+v)$ and $px=p(u+v)$, and (2) $p=-3uv$ and $q=u^3+v^3$.  This yields a quadratic equation in $u^3, v^3$. So To ensure we add something with real coefficient (when $p,q\in \mbr$), we need to make sure we add corresponding conjugate pairs.  

The behavior of roots in the range $m_\pi\in[130,250]$ splits into two cases.  When $m_\pi\le \Delta$, there are two complex roots symmetric with respect to imaginary axis with small negative real part (i.e. $i$ times these are complex conjugate pairs), and a pure imaginary root.  As $m_\pi$ approaches $\Delta$ from below the two complex roots come together toward origin.  When $m_\pi>\Delta$, there are three imaginary roots, two of which are nearly the negative of each other.

\section{XEFT \texorpdfstring{$DD^*$}{D-Dstar} Scattering Diagrams without Isospin Breaking}
\label{appx:D.5}

In this part, I present the leading and next-to-leading orders of $DD^*$ loop results with detailed calculations.

\subsection{Leading Order \texorpdfstring{$DD^*$}{D-Dstar} Loop Integral}
\label{appx:D.5.1}

\begin{figure}[!h]
\centering
\includegraphics[width=.25\textwidth]{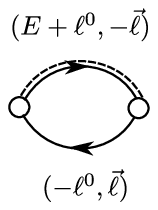}
\caption{Leading order $DD^*$ loop}
\label{fig:109fig1}
\end{figure}

\fig{109fig1} shows the leading order $D^*D$ loop diagram.  The loop amplitude is
\begin{align}
iL_\dds&=\paf{\Lam}{2}^{4-D}\paf{iy_0}{\sqrt{2}}^2\int\frac{d^D\ell}{(2\pi)^D}\frac{i}{\paren{E+\ell^0-\frac{\vec\ell^2}{2m_{D}}}+i\eps}\frac{i}{\paren{-\ell^0-\frac{\vec\ell^2}{2M_D}}+i\eps}\nn\\
&=\paf{y_0^2}{2}(-i)\paf{\Lam}{2}^{4-D}\int\frac{d^{D-1}\ell}{(2\pi)^{D-1}}\ov{E-\frac{\vec\ell^2}{4}+i\eps}\nn\\
&=\paf{y_0^2}{2}(-i)\paf{\Lam}{2}^{4-D}M_D\int\frac{d^{D-1}\ell}{(2\pi)^{D-1}}\ov{\vec\ell^2-\vec k^2-i\eps}\nn\\
&=-\frac{y_0^2}{2}iM_D\frac{(-\vec k^2-i\eps)^{(d-3)/2}\Gam\paren{\frac{3-d}{2}}}{(4\pi)^{(d-1)/2}}\paf{\Lam}{2}^{4-d}.
\end{align}
In the Power Divergence Subtraction (P.D.S.) scheme, the $d=4$ and $d=3$ pole subtractions are
\begin{align}
d=3:& \quad -\frac{y_0^2}{2}(i)M_D\paf{\Lam}{2}\ov{4\pi}\paf{2}{3-d},\nn\\
d=4:& \quad -\frac{y_0^2}{2}(i)M_D(-\vec k^2-i\eps)^{1/2}\Gam\paren{-\ov{2}}\ov{(4\pi)^{3/2}}=\frac{y_0^2}{2}i(M_D)\sqrt{-\vec k^2-i\eps}\ov{4\pi}.
\end{align}
After the subtractions, the self-energy becomes 
\begin{align}
-i\Sig_\lo
&=-\paf{iy_0}{\sqrt{2}}^2i\frac{M_D}{4\pi}\paren{\sqrt{-\vec k^2-i\eps}-\Lam}
\end{align}
and, evaluating its derivative using that $\vec k^2=M_DE$, we find
\begin{align}
\Sig_\lo'&=\frac{d\Sig_\lo}{dE}=\frac{d}{dE}\paren{\frac{M_D}{4\pi}\paren{\sqrt{-\vec k^2-i\eps}-\Lam}\paf{iy_0}{\sqrt{2}}^2}\nn\\
&=\paf{iy_0}{\sqrt{2}}^2\frac{M_D^2}{8\pi}\frac{-1}{\sqrt{-M_D E-i\eps}}.
\end{align}
Then since the bound state energy is $E=-E_X$ with $E_X>0$, 
\be
\sqrt{-M_D E-i\eps}=\sqrt{M_D E_X-i\eps}=\sqrt{M_D E_X}.
\ee

To check the sign of $\sqrt{-\vec k^2-i\eps}$ gives, we have two equivalent methods:
\begin{description}
\item[(I)] Carrying out the loop integral setting $D-1=3$ from the beginning yields
\begin{align}
\int\frac{d^3\vec \ell}{(2\pi)^3}\oneov{\vec\ell^2-\vec k^2-i\epsilon}&=\int_0^\infty \frac{4\pi d\ell}{(2\pi)^3}\frac{\ell^2-k^2+k^2}{\ell^2-k^2-i\epsilon}\nn\\
&=\int_0^\infty 4\pi d\ell\oneov{(2\pi)^3}+\int_0^\infty \frac{4\pi d\ell}{(2\pi)^3}\frac{k^2}{\ell^2-k^2-i\epsilon}\nn\\
&\to\int_0^\infty \frac{4\pi d\ell}{(2\pi)^3}\frac{k^2}{(\ell-k-i\epsilon)(\ell+k+i\epsilon)}\nn\\
&=\oneov{2}\frac{4\pi}{(2\pi)^3}(2\pi i)\frac{(k+i\epsilon)^2}{(2k+2i\epsilon)}=\frac{i}{4\pi}k
\end{align}
\item{(II)} Carrying out the loop integral in $D-1$ dimensions, we find
\begin{align}
\int\frac{d^{D-1}\ell}{(2\pi)^D}\oneov{\ell^2-k^2-i\epsilon}&=\frac{\paren{-\vec k^2-i\epsilon}^{\frac{d-3}{2}}\Gamma\paren{\frac{3-d}{2}}}{(4\pi)^{\frac{(d-1)}{2}}}\nn\\
&=-\frac{\sqrt{-\vec k^2-i\epsilon}}{4\pi}\to\frac{i}{4\pi}k,
\end{align}
in which we set $d=4$ from the first to the second line and require that $\sqrt{-\vec k^2-i\epsilon}\to -ik$ in transitioning to the final result.
\end{description}
Therefore, the two methods are consistent when $\sqrt{-\vec k^2-i\epsilon}= -ik$.

\subsection{One-Pion Exchange 2-Loop Diagram}
\label{appx:D.5.2}

\begin{figure}[!h]
\centering
\includegraphics[width=.5\textwidth]{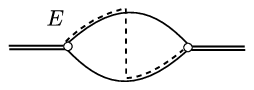}
\caption{Next-to-leading order pion exchange diagram}
\label{fig:80fig4}
\end{figure}

Figure \ref{fig:80fig4} is one of the two one-pion exchange $D^*D$ loop diagrams, which we denote $i\cM_{1PE}\paf{iy_0}{\sqrt{2}}^2$. 
The amplitude for $i\cM_{1PE}$ is
\begin{align}
i\cM_{1PE}&=\paren{\frac{\Lam}{2}}^{2(4-d)}\int \frac{d^dk}{(2\pi)^d}\int\frac{d^dq}{(2\pi)^d}\paren{\frac{ig}{\sqrt{2}f_\pi}\oneov{\sqrt{2m_\pi}}}^2\frac{i(q-k)^i(q-k)^j}{-q^0-\frac{\vec q^2}{2m_{D}}+i\eps}\nn\\
&\times\frac{i\eps^i\eps^j}{E+q^0-\frac{\vec q^2}{2m_{D}}+i\epsilon}\frac{i}{E+k^0-\frac{\vec k^2}{2m_{D}}+i\eps}\frac{i}{-k^0-\frac{\vec k^2}{2M_D}+i\eps}\nn\\
&\times\frac{i}{(q^0-k^0)-\frac{(\vec q-\vec k)^2}{2m_\pi}+\del+i\eps}\nn\\
&=-\frac{ig^2}{2f_\pi^2}\frac{M_{D}^2}{3}\paren{I_1+\mu^2 I_2},
\end{align}
where
\be
I_1=\paren{\frac{\Lam}{2}}^{2(4-d)}\left[\int\frac{d^{d-1}k}{(2\pi)^{d-1}}\oneov{\vec k^2-M_{D}E-i\eps}\right]^2.
\ee
In P.D.S. scheme, the two integrals are
\begin{align}
I_1&=\paren{\ov{2\pi}\Lam_\pds+\frac{ik}{4\pi}}^2,\nn\\
I_2&=\oneov{16\pi^2}\left\{\oneov{4\eps}+\oneov{2}-\log(-i)-\log\frac{2k+\mu}{\Lam}+\oneov{2}\log\pi-\frac{\gam_E}{2}\right\}.
\end{align}
In the $\MSbar$ scheme, all but the fourth term in the parentheses are absorbed into the constant $\Lam_{P.D.S.}$.

\begin{figure}[!h]
\centering
\includegraphics[width=.2\textwidth]{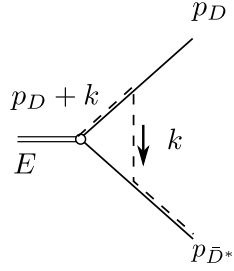}
\caption{One-pion-exchange correction to $X\to DD^*$ vertex}
\label{fig:80fig5}
\end{figure}

Before continuing the second pion-loop correction to the $DD^*$ loop, we consider the the one pion exchange correction to the $X\to DD^*$ decay, \fig{80fig5}, which is closely related to the loop we just computed. We denote \fig{80fig5} as $i\cV_{1PE}$.  The corresponding amplitude is
\begin{align}
i\cV_{1PE}&=\paren{\frac{\Lam}{2}}^{4-d}\int \frac{d^dk}{(2\pi)^d}\int\frac{d^dq}{(2\pi)^d}\paren{\frac{ig}{\sqrt{2}f_\pi}}^2\oneov{2m_\pi}\frac{-ik^ik^j}{k^0+E_D-\frac{(\vec k+\vec p_D)^2}{2m_{D}}+i\eps}\nn\\
&\times\frac{i}{-\frac{\vec k^2}{2m_\pi}+\del+i\eps+k^0}\frac{i\eps^i\eps^j}{E-(E_D+k^0)-\frac{(\vec p_D+\vec k)^2}{2M_D}+i\eps}\paren{\frac{-iy_0}{\sqrt{2}}}\nn\\
&=\paren{\frac{-iy_0}{\sqrt{2}}}\frac{g^2}{2f_\pi^2m_\pi}I_3,
\end{align}
where
\begin{align}
I_3&=M_D\paren{\frac{\Lam}{2}}^{4-d}\int\frac{d^{d-1}k}{(2\pi)^{d-1}}\frac{k^ik^j}{\vec k^2-\mu^2-i\eps}\frac{\eps^i\eps^j}{(\vec p_D+\vec k)^2-\tilde k^2-i\eps}\nn\\
&=M_D\paren{\frac{\Lam}{2}}^{\!4-d}\!\int_0^1 dx\Bigg\{\frac{d-1}{2(4\pi)^{\frac{d-1}{2}}}\Gam\!\paren{\frac{3-d}{2}}\oneov{3}
\oneov{[\vec p_D^2x (1\!-\!x)-\tilde k^2 x-\mu^2(1\!-\!x)-i\eps]^{\frac{3-d}{2}}}\nn\\
&+\frac{(\vec p_D\cdot\vec\eps)^2}{(4\pi)^{\frac{d-1}{2}}}\Gam\paren{\frac{5-d}{2}}\oneov{[\vec p_D^2x (1-x)-\tilde k^2 x-\mu^2(1-x)-i\eps]^{(5-d)/2}}\Bigg\},
\end{align}
with $\tilde k^2=M_D E$. The $d=3$ pole is
\begin{align}
\left.I_3\right|_{d\to 3}&=\paf{\Lam}{2}\int_0^1 dx\oneov{4\pi}\ov{2}\frac{2}{3-d}\ov{3}(d-1)\nn\\
&\to\paf{\Lam}{2}\ov{4\pi}\frac23=\paf{\Lam}{8\pi}\frac23 M_{D}.
\end{align}
At $d=4$, the integral to be evaluated is
\begin{align}
\left.I_3\right|_{d\to 4}&=\frac{M_D}{8\pi}\int_0^1dx\Bigg(-\sqrt{\vec p_D^2x (1-x)-\tilde k^2 x-\mu^2(1-x)-i\eps}\nn\\
&+\frac{x^2(\vec p_D\cdot\vec\epsilon)^2}{\sqrt{\vec p_D^2x (1-x)-\tilde k^2 x-\mu^2(1-x)-i\eps}}\Bigg),
\end{align}
and we define
\begin{align}
h_1(p_D)&=I_{3a}=\int_0^1 dx\sqrt{ax^2+bx+c}\nn\\
&=\frac{(\mu^2-k^2-\vec p_D^2)(ik)}{4\vec p_D^2}-\frac{(\mu^2+\vec p_D^2-k^2)(i\mu)}{4\vec p_D^2}\nn\\
&+i\frac{(\vec p_D^2-\vec k^2+\mu^2)^2-4\vec p_D^2\mu^2}{8|\vec p_D|^3}\log\paf{|\vec p_D|+|\vec k|+\mu}{|\vec k|-|\vec p_D|+\mu} \\
h_2(p_D)&=I_{3b}=\int dx\frac{x^2}{\sqrt{ax^2+bx+c}},
\end{align}
with the kinematic variables 
\be
a=-\vec p_D^2-i\eps, \quad b=\vec p_D^2-\tilde k^2+\mu^2,\quad c=-\mu^2-i\eps.
\ee
Then the second integral becomes
\begin{align}
I_{3b}&=\frac{(2a-3b)\sqrt{a+b+c}}{4a^2}+\frac{3b\sqrt{c}}{4a^2}\nn\\
&+\frac{4ac-3b^2}{8a^{5/2}}\log\paf{b+2\sqrt{ac}}{2a+b+2\sqrt{a}\sqrt{a+b+c}}.
\end{align}
Following the above result, we set
\be
\sqrt{-\tilde k^2-i\eps}=-i\tilde k,\quad \sqrt{-\vec p_D^2-i\eps}=-i\vec p_D,\quad \sqrt{-\mu^2-i\eps}=-i\mu,
\ee
and summing the kinematic variables in $I_{3b}$, we find
\begin{align}
I_{3b}&=(ik)\frac{5\vec p_D^{\:2}-3\vec k^2+3\mu^2}{4\vec p_D^4}+\frac34(i\mu)\frac{\vec k^{\:2}-\vec p_D^{\:2}-\mu^2}{\vec p_D^4}\nn\\
&+i\frac{3(\vec p_D^{\:2}-\tilde k^{\:2}+\mu^2)^2-4\vec p_D^{\:2}\mu^2}{8|\vec p_D|^5}\log\paf{|\vec p_D|+|\vec k|+\mu}{|\vec k|-|\vec p_D|+\mu}
\end{align}
Putting the pieces back together, the total integral is
\be
I_3=\frac{M_D}{8\pi}\left[\frac23\Lam-h_1(p_D)+(\vec p_D\cdot\vec\eps)^2h_2(p_D)\right].
\ee

\begin{figure}[!h]
\centering
\includegraphics[width=.2\textwidth]{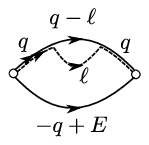}
\caption{$D^*$ self-energy loop}
\label{fig:80fig1}
\end{figure}

Figure \ref{fig:80fig1} displays the other pion correction to the $DD^*$-loop diagram, which we denote as $2i\cM_{SE}\paf{iy_0}{\sqrt{2}}^2$.  The amplitude is
\begin{align}
i\cM_{SE}&=\paren{\frac{\Lambda}{2}}^{2(4-d)}\int \frac{d^d\ell}{(2\pi)^d}\int\frac{d^d q}{(2\pi)^q}\frac{i}{E-q^0 -\frac{\vec q^2}{2M_D}+i\epsilon}\left[\frac{i}{q^0-\frac{\vec q^2}{2m_{D}}+i\epsilon}\right]^2\nn\\
&\times \frac{i}{q^0-\ell^0-\frac{(\vec q-\vec \ell)^2}{2M_D}+i\epsilon}\frac{i}{-\frac{\vec\ell^2}{2m_\pi}+\delta+i\epsilon}\ell^i\ell^j\paren{\frac{ig}{\sqrt{2}f_\pi}\oneov{\sqrt{2m_\pi}}}^2.
\end{align}

\begin{figure}[t]
\centering
\includegraphics[width=.18\textwidth]{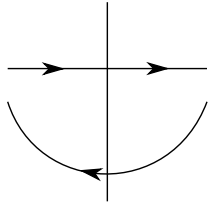}
\caption{Contour integral where $\ell^0=\rho e^{i\phi}$ and $d\ell^0=\rho e^{i\phi}id\phi$. At infinity the integral $\rho\to \infty$ keep limits only $\phi$ change from $2\pi\to \pi$.}
\label{fig:80fig2}
\end{figure}

We perform the $\ell$ integral first,
\begin{align}
I_4&=\int \frac{d^d\ell}{(2\pi)^d}\frac{\ell^i\ell^j\epsilon^i\epsilon^j}{q^0-\ell^0-\frac{(\vec q-\vec \ell)^2}{2M_D}+i\eps}\oneov{-\frac{\vec\ell^2}{2m_\pi}+\del+i\eps}\nn\\
&=2m_\pi\int\frac{d^{d-1}\ell}{(2\pi)^{d-1}}\int_{-\infty}^{+\infty}\frac{d\ell^0}{2\pi}\frac{\ell^i\ell^j\eps^i\eps^j}{\ell^0-q^0+\frac{(\vec q-\vec \ell)^2}{2M_D}-i\epsilon}\oneov{\vec\ell^2-\mu^2-i\eps}.
\end{align}
The $\ell_0$ integral contains only one pole, which we calculate by contour integration, closing the contour in the lower half plane, as shown in Fig. \ref{fig:80fig2}.  The result is
\be\label{eq:ellzerosinglepole}
\int_{-\infty}^\infty\frac{d\ell^0}{(2\pi)}\oneov{\ell^0-q^0+\frac{(\vec q-\vec \ell)^2}{2M_D}-i\eps}=\frac{i}{2}
\ee
If we keep the pion energy, then there are two poles in the $\ell_0$ integral and the result would be two times \req{eq:ellzerosinglepole}.

For the integral of $\vec \ell$, we obtain
\begin{align}
I_4&=im_\pi\int \frac{d^{d-1}\ell}{(2\pi)^{d-1}}\oneov{d-1}\frac{\ell^2-\mu^2+\mu^2}{\vec\ell^2-\mu^2-i\eps}\nn\\
&=im_\pi\frac{1}{d-1}\left[\int\frac{d^{d-1}\ell}{(2\pi)^{d-1}}+\int\frac{d^{d-1}\ell}{(2\pi)^{d-1}}\frac{\mu^2}{\ell^2-\mu^2-i\epsilon}\right]\nn\\
&=m_\pi\oneov{d-1}\mu^2\oneov{(4\pi)^{\frac{d-1}{2}}}\Gamma\paren{\frac{3-d}{2}}\paren{\mu^2}^{\frac{d-3}{2}}.
\end{align}
To restore the contribution of the pion $\ell^0$, this result is multiplied by two. The $d=3$ subtraction is identical to the $d=3$ pole in the self-energy counter term. For $d=4$, the integral yields
\be
I_4=m_\pi\oneov{3}\frac{\mu^2}{(4\pi)^{3/2}}\Gamma\paren{-\oneov{2}}(\mu^2)^{1/2}=\frac{\mu^3}{12\pi}m_\pi.
\ee
Thus so far we have,
\begin{align}
i\cM_{SE}&=\paren{\frac{\Lambda}{2}}^{\!2(4-d)}\!\!\int\!\frac{d^dq}{(2\pi)^d}\oneov{E-q^0-\frac{\vec q^2}{2M_D}+i\eps}\paren{\oneov{q^0-\frac{\vec q^2}{2m_{D}}}+i\eps}^2 
\frac{-ig^2}{4f_\pi^2 m_\pi} I_4.
\end{align}

The $q$ integral is computed similarly, starting with the performing the $q^0$ integral by contours,
\begin{align}
\int\!\frac{d^dq}{(2\pi)^d}\oneov{E-q^0-\frac{\vec q^2}{2M_D}+i\eps}\paren{\oneov{q^0-\frac{\vec q^2}{2M_{D}}}+i\eps}^2
&=i\!\int\!\frac{d^{d-1}q}{(2\pi)^{d-1}}\left[\oneov{E-\frac{\vec q^2}{M_{D}}+i\eps}\right]^2\nn\\
&=iM_{D}^2\!\int\!\frac{d^{d-1}q}{(2\pi)^{d-1}}\left[\oneov{\vec q^2+\gam^2-i\eps}\right]^2\nn\\
&=\frac{-iM_{D}^2}{(4\pi)^{\frac{d-1}{2}}}\Gamma\paren{\frac{5-d}{2}}\paren{\gamma^2-i\eps}^{\frac{d-5}{2}}.
\end{align}
The integral picks up three simple poles, none of which are in Minkowski space.  As a consequence, the kinematic variables retain their complex offset, $(+\gamma^2-i\eps)$ and $(-\mu^2-i\eps)$, and there are no poles in $d=3$. Therefore
\begin{align}
i\cM_{SE}&=\paren{\frac{\Lam}{2}}^{2(4-d)}\frac{-M_{D}^2}{(4\pi)^{(d-1)/2}}\Gam\paren{\frac{5-d}{2}}\paren{\gam^2-i\eps}^{(d-5)/2}\nn\\
&\times \frac{g^2}{4f_\pi^2 m_\pi} \frac{m_\pi\mu^2}{(d-1)(4\pi)^{(d-1)/2}}\Gam\paren{\frac{3-d}{2}}\mu^{d-3},
\end{align}
with
\be
\gam^2=-M_DE=-k^2, \quad (\gam^2-i\eps)^{1/2}=-ik, \quad (\gam^2+i\eps)^{1/2}=ik.
\ee
The P.D.S. pole subtractions amount to
\be
\paren{\frac{\Lam}{2}}(-\gam^{-2})(M_{D})^2\frac{1}{3(4\pi)^2}\frac{g^2}{4f_\pi^2}\frac{2\mu^2}{3-d},
\ee
and the final result for the amplitude is
\begin{align}
i\cM_{SE}&=-\frac{g^2}{4f_\pi^2}\frac{\mu^2M_{D}^2}{16\pi^2}\left\{\paren{\frac{\Lam}{2}}^2\oneov{\gam^2}+\frac{\mu}{6\gam}\right\}.
\end{align}
Multiplying this amplitude by two and subtracting the pole leads to
\be
-i\Sigma_{NLO}^{(3)}=2i\cM_{SE}=-\frac{g^2\mu^3 M_D^2}{192\pi^2f_\pi^2\gam}.
\ee

\begin{figure}[!h]
\centering
\includegraphics[width=.2\textwidth]{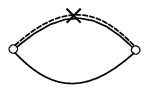}
\caption{Next-to-leading-order $D^*$ propagator correction diagram}
\label{fig:80fig2point5}
\end{figure}

Figure \ref{fig:80fig2point5} shows the $DD^*$ loop diagram with the next-to-leading order $D^*$ propagator correction. We denote the corresponding amplitude $i\mathcal{M}_{D^*}\paren{\frac{i y_0}{\sqrt{2}}}^2$, and evaluate it as
\begin{align}
i\mathcal{M}_{D^*}\paren{\frac{i y_0}{\sqrt{2}}}^2&=\paf{\Lambda}{2}^{4-D}\int \frac{d^D\ell}{(2\pi)^D}\frac{i}{E+\ell^0-\frac{\vec\ell^2}{2M_D}+i\eps}\paren{\frac{-i\lambda\vec\ell^2}{2M_D}}\nn\\
&\times\frac{i}{E+\ell^0-\frac{\vec\ell^2}{2M_D}+i\eps}\frac{i}{\paren{-\ell^0-\frac{\vec\ell^2}{2M_D}}+i\eps}\paren{\frac{iy_0}{\sqrt{2}}}^2\nn\\
&=\frac{y_0^2}{2}\paf{-i\lambda}{2M_D}\paf{\Lambda}{2}^{4-D}\frac{-M_D^2}{(4\pi)^{\frac{D-1}{2}}}\frac{D-1}{2}\nn\\
&\times\frac{\Gam\paren{1-\frac{D-1}{2}}}{\Gam(2)}\paren{\ov{-\vec k^2-i\eps}}^{1-\frac{D-1}{2}}\nn\\
&=\frac{y_0^2}{2}\frac{i\lambda M_D}{2}\paf{\Lambda}{2}^{4-D}\frac{D-1}{2(4\pi)^{\frac{D-1}{2}}}\Gam\!\paf{3-D}{2}\paf{1}{-\vec k^2-i\eps}^{\frac{3-D}{2}}.
\end{align}
In P.D.S., the $d=3$ and $d=4$ pole subtractions are
\begin{align}
d=3: & \qquad \frac{i y_0^2}{2}\frac{M_D\lambda}{2}\paf{\Lambda}{2}\paf{1}{4\pi}\paf{2}{3-D}\nn\\
d=4: & \qquad \frac{i y_0^2}{2}\frac{M_D\lambda}{2}\frac32\oneov{(4\pi)^{3/2}}\Gam\paren{-\ov{2}}(-\vec k^2-i\eps)^{1/2}
 =-i\frac{y_0^2}{4}\frac{M_D\lambda}{4\pi}\sqrt{-\vec k^2-i\eps}.
\end{align}
After making the subtractions, the result for the next-to-leading order self-energy is
\begin{align}
i\mathcal{M}_{D^*}\paf{i y_0^2}{2}^2
&=\frac{i y_0^2}{2}^2\frac{iM_D\lambda}{8\pi}\paren{\sqrt{-\vec k^2-i\eps}-\Lambda}\equiv -i\Sigma_{\nlo}^{(D^*)},
\end{align}
and its derivative is
\begin{align}
\frac{d\Sigma_\nlo^{D^*}}{dE}
&=\frac{d}{dE}\paren{-\paf{-i y_0}{\sqrt{2}}^2\frac{M_D\lambda}{8\pi}\sqrt{-M_DE-i\eps}-\Lambda }\nn\\
&=\paf{i y_0}{\sqrt{2}}^2\frac{M_D^2 \lambda}{16\pi}\ov{\sqrt{-M_D E-i\eps}},
\end{align}
in which $\sqrt{-\vec k^2-i\eps}=-ik$ and $\sqrt{-M_D E-i\eps}=\sqrt{M_D E_X}$.


\subsection{An Example of Next-To-Next-To-Next-To-Leading-Order Pion Exchange Diagram with \texorpdfstring{$Z_1$}{Z1} Vertex}
\label{appx:D.5.4}

\begin{figure}[!h]
\centering
\includegraphics[width=.3\textwidth]{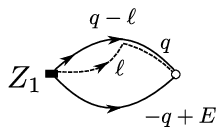}
\caption{Next-to-next-to-next-to-leading-order pion exchange diagram}
\label{fig:80fig3}
\end{figure}

Figure \ref{fig:80fig3} is the $Z_1$ interaction two-loop $DD^*$ diagram, which we denote $i\cM_{Z_1}\paf{iy_0}{\sqrt{2}}$.  The amplitude is
\begin{align}
i\cM_{Z_1}&=\paren{\frac{\Lam}{2}}^{2(4-d)}\int \frac{d^d\ell}{(2\pi)^d}\int\frac{d^dq}{(2\pi)^d}\frac{i}{E-q^0 -\frac{\vec q^2}{2M_D}+i\epsilon}\frac{i}{q^0-\frac{\vec q^2}{2M_{D}}+i\epsilon}\nn\\
&\times\paren{-\frac{Z_1\vec\eps_X\cdot\vec\ell}{\sqrt{2m_\pi}}}\frac{ig}{\sqrt{2}f_\pi}\frac{1}{\sqrt{2m_\pi}}\frac{i}{q^0-\ell^0-\frac{(\vec q-\vec \ell)^2}{2M_D}+i\epsilon}\frac{(-i\vec\ell\cdot\vec\eps_{D^*})i}{\ell^0-\frac{\vec\ell^2}{2m_\pi}+\del+i\eps}
\end{align}
Evaluating the $\ell_0$ and $q_0$ integrals by contours and picking up the poles at $\ell^0=\vec\ell^2/(2m_\pi)-\del-i\eps$ and $q^0=-\vec q^2/(2M_{D})-i\eps$, we find
\begin{align}
i\cM_{Z_1}&=\frac{gZ_1\eps_X^i\eps_{D}^j}{2\sqrt{2}m_\pi f_\pi}\paren{\frac{\Lam}{2}}^{2(4-d)}\int\frac{d^{d-1}\ell}{(2\pi)^{d-1}}\int\frac{d^{d-1}q}{(2\pi)^{d-1}}\nn\\
&\times\frac{\ell^i\ell^j}{E-\frac{\vec q^2}{2M_{D}}-\frac{\vec q^2}{2M_D}+i\eps}\frac{1}{\frac{\vec q^2}{2m_{D}}-\frac{\vec \ell^2}{2m_\pi}+\del-\frac{(\vec q-\vec \ell)^2}{2M_D}+i\eps}.
\end{align}
To the order of $(m_\pi/M_D)$, 
\begin{align}
i\cM_{Z_1}&=\frac{(gZ_1\eps_X^i\eps_{D^*}^j)}{\sqrt{2}m_\pi f_\pi}\paren{\frac{\Lam}{2}}^{2(4-d)}\int\frac{d^{d-1}\ell}{(2\pi)^{d-1}}\int\frac{d^{d-1}q}{(2\pi)^{d-1}}
\frac{\ell^i\ell^j(M_{D})}{(\vec q^2+\gam^2-i\eps)(\vec \ell^2-\mu^2-i\eps)}\nn\\
&=\frac{gZ_1(M_{D})}{\sqrt{2}f_\pi}I_5I_6,
\end{align}
where
\begin{align}
I_5&=\paren{\frac{\Lam}{2}}^{4-d}\int\frac{d^{d-1}\ell}{(2\pi)^{d-1}}\frac{\ell^i\ell^j}{\vec\ell^2-\mu^2-i\eps},\\
I_6&=\paren{\frac{\Lam}{2}}^{4-d}\int\frac{d^{d-1}q}{(2\pi)^{d-1}}\oneov{\vec q^2+\Upsilon^2-i\eps}.
\end{align}
The $\ell$  integral gives
\be
I_5=\paren{\frac{\Lam}{2}}^{4-d}\int\frac{d^{d-1}\ell}{(2\pi)^{d-1}}\frac{\vec\ell^2-\mu^2+\mu^2}{\vec\ell^2-\mu^2-i\eps}\oneov{D-1}=\oneov{12\pi}\paren{\Lam\mu+i\mu^2},
\ee
while the $q$  integral gives
\be
I_6=\paren{\frac{\Lam}{2}}^{4-d}\int\frac{d^{d-1}q}{(2\pi)^{d-1}}\oneov{q^2+\Upsilon^2-i\eps}=\frac{\Lam}{4\pi}\Upsilon+\oneov{4\pi}\Upsilon^2.
\ee
Putting these integrals together, we find
\begin{align}
i\cM_Z&=\frac{gZ_1(M_{D}/2)}{24\pi^2\sqrt{2}f_\pi}\paren{\Lam\mu+i\mu^2}\paren{\Lam\Upsilon+\Upsilon^2},\nn\\
-i\Sigma_{N.L.O.}^{(Z_1)}&=(i\cM_{Z_1})\paren{\frac{iy_0}{\sqrt{2}}},\nn\\
\rm{Re}\paren{\Sigma_{N.L.O.}^{(Z_1)}}
&=\frac{-gZ_1M_D}{48\pi^2\sqrt{2}f_\pi}(i\mu^2+\Lam\mu)(\Lam\Upsilon+\Upsilon^2)\paren{\frac{iy_0}{\sqrt{2}}}.
\end{align}
Taking its derivative with respect to $E$ and setting $E=-E_X$ gives
\be
\mathrm{Re}\paren{\frac{d\Sigma_{N.L.O.}^{(Z_1)}}{dE}\Bigg\vert_{E=-E_X}}=\frac{g Z_1 M_D^2}{96\pi^2\sqrt{2}f_\pi}(i\mu^2+\Lam\mu)\paren{\frac{i\, y_0}{\sqrt{2}}}\paren{\frac{\Lam}{\Upsilon}+2}
\ee


\section{Summary of \texorpdfstring{$X\to DD\pi$}{X to DDpi} Results in XEFT}
\label{appx:D.7}
Here we summarize and collate the results of the above calculations.
\subsection{Feynman diagrams and amplitudes}
\label{appx:D.7.1}

\begin{center}
\includegraphics[width=.75\textwidth]{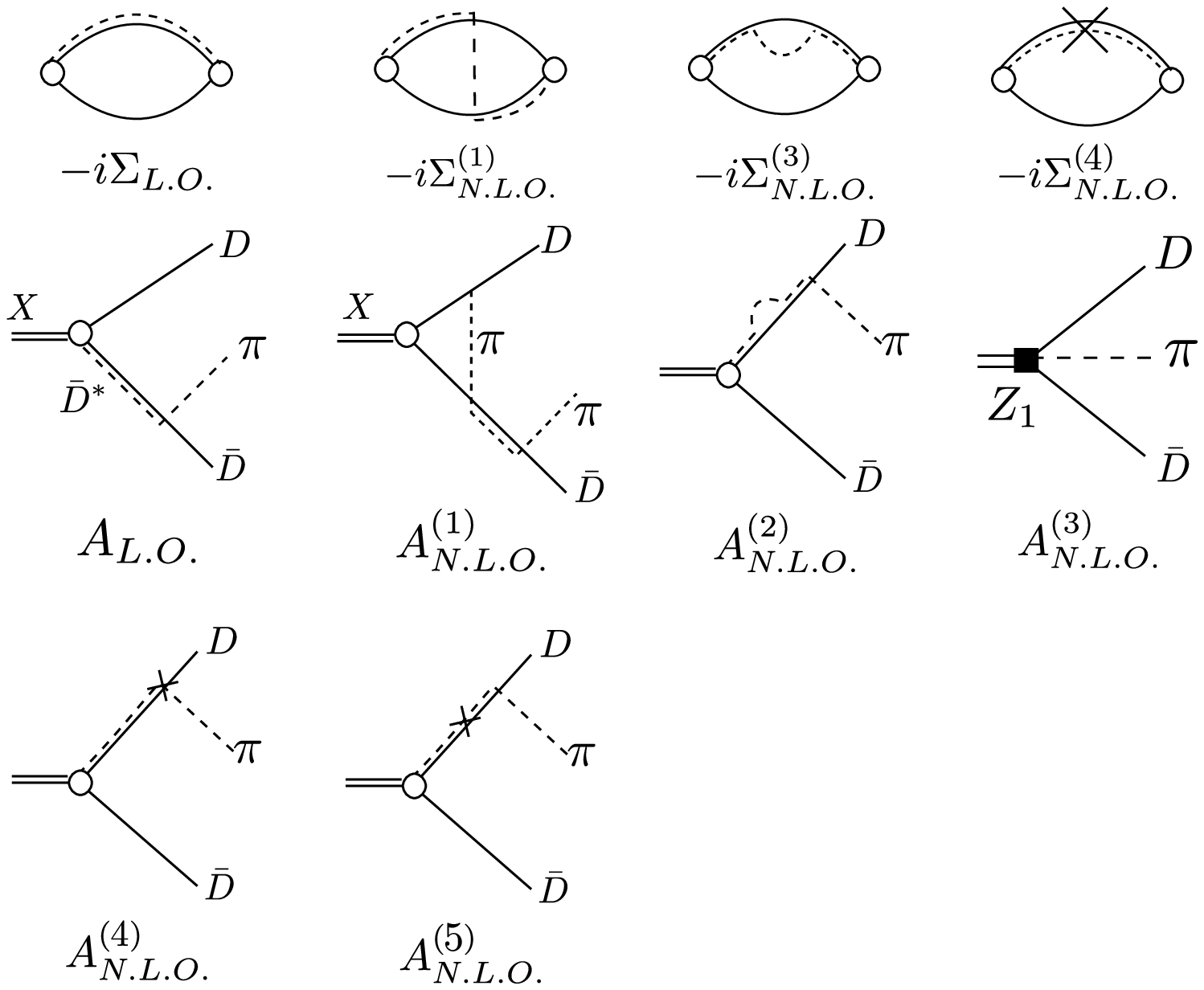}
\end{center}

Leading order $DD^*$ Loop Integral:
\begin{align*}
-i\Sigma_{L.O.}&=\left(\frac{iy_0}{\sqrt{2}}\right)^2i\frac{M_{D}}{4\pi}(-\sqrtke+\Lambda)\\
\frac{d\Sigma_{L.O.}}{dE}&=\left(\frac{iy_0}{\sqrt{2}}\right)^2\left(-\frac{(M_D/2)^2}{2\pi}\right)\frac{1}{\sqrt{-2\mdd E-i\epsilon}}
\end{align*}

Leading order X decay amplitude:
\begin{equation*}
A_{L.O.}=\frac{y_0M_{D}g(\vec{p}_\pi\cdot\vec{\epsilon})}{2\sqrt{2m_\pi}f_\pi}\oneovponeov
\end{equation*}

1-pion exchange correction to the $DD^*$ loop:
\begin{align*}
-i\Sigma_{N.L.O.}^{(1)}&=\left(\frac{iy_0}{\sqrt{2}}\right)^2\left(-i\frac{g^2M_D^2}{96\pi^2f_\pi^2}\right)\nn\\
&\times\left\{
(\lpds-\sqrtke)^2+\mu^2\left[
\log\lpds-\log\left(2\sqrtke-i\mu\right)\right]\right\}\nn\\
\text{Re}\!\left(\frac{d\Sigma_{N.L.O.}^{(1)}}{dE}\Biggl\vert_{E=-E_x}\right)&=\left(\frac{iy_0}{\sqrt{2}}\right)^2\left(\frac{g^2M_D^2}{96\pi^2f_\pi^2}\right)M_D\nn\\
&\times\left\{(\lpds-\sqrt{M_D E_X})\frac{1}{\sqrt{M_D E_X}}+\frac{2\mu^2}{4M_D E_X+\mu^2}\right\}
\end{align*}

Self-energy correction to the $DD^*$ loop:
\begin{align*}
-i\Sigma_{N.L.O.}^{(3)}&=-\frac{g^2\mu^3M_D^2}{192\pi^2f_\pi^2}\frac{1}{\sqrtke}\ynot\\
\frac{d\Sigma_{N.L.O.}^{(3)}}{dE}&=\left(\frac{iy_0}{\sqrt{2}}\right)^2\left(-\frac{ig^2\mu^3M_D^3}{384\pi^2f_\pi^2}\right)\frac{1}{(-\vec{k}^2-i\epsilon)^{3/2}}
\end{align*}

$D^*$ propagator correction to $DD^*$ loop
\begin{align}
-i\Sigma_\nlo^{(4)}&=\paf{i y_0}{\sqrt{2}}^2\paf{iM_D\lambda}{8\pi}\paren{\sqrt{-\vec k^2-i\eps}-\Lambda}\nn\\
\frac{d\Sigma_\nlo^{(4)}}{dE}&=\paf{i y_0}{\sqrt{2}}^2\frac{M_D^2\lambda}{16\pi}\ov{\sqrt{-\vec k^2-i\eps}}
\end{align}

N.L.O. decay amplitude: pion exchange correction
\begin{align*}
A_{N.L.O.}^{(1)}&=\left(\frac{iy_0}{\sqrt{2}}\right)\Biggl[\frac{-ig}{\sqrt{2}f_\pi}\frac{\vec{\epsilon}_X\cdot\vec{p}_\pi}{\sqrt{2m_\pi}}\frac{M_D^2}{-k^2+\vec{p}_D^2}\left(\frac{g^2}{16\pi f_\pi^2}\right)\left(\frac{2}{3}\Lambda-h_1(p_D)\right)\\
&+\left(\frac{-ig}{\sqrt{2}f_\pi}\frac{\vec{\epsilon}_X\cdot\vec{p}_D\vec{p}_D\cdot\vec{p}_\pi}{\sqrt{2m_\pi}}\right)\frac{M_D^2}{-\vec{k}^2+\vec{p}_D^2}\frac{g^2}{16\pi f_\pi^2}h_2(p_D) 
+(\vec{p}_D\leftrightarrow \vec{p}_{\bar D})\Biggl]
\end{align*}

N.L.O. decay amplitude: $D^*$ self-energy correction
\begin{equation*}
A_{N.L.O.}^{(2)}=\left(\frac{iy_0}{\sqrt{2}}\right)\frac{ig^3}{8f_\pi^3}\frac{1}{\sqrt{m_\pi}}\frac{i\vec{p}_\pi\cdot\vec{\epsilon}}{12\pi}\frac{M_D^2\mu^3}{(\vec{p}_{\bar D}^2+\gamma^2)^2}+(\vec{p}_D\leftrightarrow\vec{p}_{\bar D})
\end{equation*}

N.L.O. decay amplitude: $Z_1$ correction
\begin{equation}
A_{N.L.O.}^{(3)}=i\frac{Z_1}{\sqrt{2m_\pi}}i\vec{\epsilon}\cdot\vec{p}_\pi
\end{equation}

N.L.O. decay amplitude: $DD\pi$ vertex correction
\be
A_\nlo^{(4)}=(-2\lambda)\frac{y_0M_D g(\vec p_\pi\cdot\vec\eps)}{2\sqrt{2m_\pi}f_\pi}\paren{\ov{\vec k^2-\vec p_D^2}+\ov{\vec k^2-\vec p_{\bar D}^2}}
\ee

N.L.O. decay amplitude:
\be
A_\nlo^{(5)}=(-\lambda)\frac{y_0 g(\vec p_\pi\cdot\vec \eps)M_D}{2\sqrt{2m_\pi}f_\pi}\paren{\frac{|\vec p_D|^2}{(\vec k^2-\vec p_D^{\:2})^2}+\frac{|\vec p_{\bar D}|^2}{(\vec k^2-\vec p_{\bar D}^{\:2})^2}}
\ee

\subsection{Leading Order \texorpdfstring{$X\to DD\pi$}{X to D D pi} Differential Decay Rate} 
The leading order differential decay rate is
\begin{align*}
\frac{d\Gamma_{L.O.}}{dp_D^2 dp_{\bar{D}}^2}&=(2m_\pi)\frac{1}{3\pi^3}Z_{L.O.}\vert A_{L.O.}\vert^2,
\end{align*}
for which we need the wavefunction normalization and the squared amplitude,
\begin{align*}
Z_{L.O.}&=
\text{Re}\left(\frac{d\Sigma_{N.L.O.}^{(1)}}{dE}\Biggl\vert_{E=-E_x}\right)^{-1}
=(-1)\frac{8\pi\sqrt{M_DE_X}}{M_D^2}\left(\frac{iy_0}{\sqrt{2}}\right)^{-2}\\
\vert A_{L.O.}\vert^2&=\frac{y_0^2M_D^2g^2(\vec{p}_\pi\cdot \vec{\epsilon})^2}{8m_\pi f_\pi^2}\left(\frac{1}{-k^2+\vec{p}_D^2}+\frac{1}{-k^2+\vec{p}_{\bar D}^2}\right)^2.
\end{align*}
Inserting these in the decay rate yields
\begin{align*}
\frac{d\Gamma_{L.O.}}{dp_D^2 dp_{\bar D}^2}&=\frac{\sqrt{M_DE_X}}{8\pi^2f_\pi^2}g^2(\vec{p}_\pi\cdot\vec{\epsilon})^2\left(\frac{1}{-k^2+\vec{p}_D^2}+\frac{1}{-k^2+\vec{p}_{\bar D}^2}\right)^2.
\end{align*}

\subsection{Next-to-leading order $X\to DD\pi$ differential decay rate}

For the next-to-leading order decay rate, we need the follow combinations of the renormalization factor and squared amplitude:
\begin{align*}
Z_{N.L.O.}|A_{L.O.}|^2
&=\Big[\frac{g^4 E_XM_D^2}{3f_\pi^4 2m_\pi}\bigg(\frac{\lpds}{\sqrt{M_D E_X}}-\frac{4M_D E_X-\mu^2}{4M_D E_X+\mu^2}\bigg)+\frac{2\pi g^2}{m_\pi f_\pi^2}\lambda\sqrt{M_DE_X}\\
&+\sigma\frac{4\pi E_X g^2}{m_\pi f_\pi^2b}(\sqrt{M_D E_X}-\lpds) \Big]\ppie^2\oneovponeov^2\\
2Z_{L.O.}\re A_{L.O.}A_{N.L.O.}&=\Bigg\{
\frac{g^4M_D}{8\pi f_\pi^4 m_\pi}\Big(
  \frac{h_1(p_D)}{k^2-\vec{p}_D^2}+\frac{h_1(p_{\bar D})}{k^2-\vec{p}_{\bar D}^2}\\
  &-\frac{\vec{\epsilon}\cdot\vec{p}_D\vec{p}_D\cdot\vec{p}_\pi}{k^2-\vec{p}_D^2}-\frac{\vec{\epsilon}\cdot\vec{p}_{\bar D}\vec{p}_{\bar D}\cdot\vec{p}_\pi}{k^2-\vec{p}_{\bar D}^2}
  -\frac{2}{3}\lpds\oneovponeov \Big) \\
&+\frac{2Z_1g}{m_\pi f_\pi y_0 M_D}\Bigg\}\oneovponeov \ppie^2 2\pi\sqrt{M_D E_X}.
\end{align*}
Then the decay rate to next-to-leading order is
\begin{align*}
\frac{d \Gamma _{N.L.O.}}{d p_D^2d p_{\bar{D}}^2}&=\frac{1}{32\pi^3}\Bigg\{\Big[\frac{g^4E_XM_D^2\ppie^2}{3f_\pi^4}\frac{\lpds}{\sqrt{M_D E_X}}\\
&-\frac{g^4E_XM_D^2\ppie^2}{3f_\pi^4}\frac{4M_D E_X-\mu^2}{4M_D E_X+\mu^2}\\
&+\sigma\frac{8\pi E_X g^2\ppie^2}{bf_\pi^2}(\sqrt{M_D E_X}-\lpds)\Big]\oneovponeov^2\\
&-\frac{g^4\sqrt{M_D E_X}M_D \ppie^2}{3f_\pi^4}\lpds\oneovponeov^2\\
&+\frac{-g^4\sqrt{M_D E_X}}{3f_\pi^4}\oneovponeov M_D\ppie \\
&\times\left[\frac{1}{k^2-\vec{p}_D^2}(\ppie h_1(p_D)-\vec{\epsilon}\cdot\vec{p}_D\vec{p}_\pi\cdot\vec{p}_D)+\ddbarex \right]\\
&+\frac{8\pi Z_1 g\ppie^2\sqrt{M_D E_X}}{f_\pi y_0(M_D/2)}\oneovponeov\Bigg\},
\end{align*}
which is actually independent of $\Lambda_{P.D.S.}$,
\begin{align*}
\frac{d \Gamma _{N.L.O.}}{d p_D^2d p_{\bar{D}}^2}&=
\frac{1}{32\pi^3}\Bigg\{  \frac{-2g^4\sqrt{M_D E_X}M_D\ppie}{6f_\pi^4}\\
&\times\Big[\frac{1}{k^2-\vec{p}_D^2}(\vec{p}_\pi\cdot\vec{\epsilon}h_1(p_D)-\vec{\epsilon}\cdot\vec{p}_D\vec{p}_\pi\cdot\vec{p}_D)+\ddbarex\Big]\\
&+\frac{16\pi Z_1 g\sqrt{M_D E_X}\ppie^2}{f_\pi^2 y_0M_D}\Bigg\}\oneovponeov\\
&+\frac{g^2M_D\sqrt{M_D E_X}}{12\pi f_\pi^2}\frac{\mu^2-4M_D E_X}{\mu^2+4M_D E_X}\frac{d\Gamma_{L.O.}}{dp_D^2dp_{\bar D}^2}\\
&+\sigma\frac{2E_X}{b}\left(1-\frac{\lpds}{\sqrt{E_XM_D}}\right)\frac{d\Gamma_{L.O.}}{dp_D^2 dp_{\bar D}^2},
\end{align*}
where we drop $A_{N.L.O.}^{(4)}$ and $A_{N.L.O.}^{(5)}$ corrections because they contribute at next-to-next-to-leading order to the decay rate.

\section{Four-body decay integral}
\label{appx:D.8}

The phase space integral for a 4-body decay $i \to 1+2+3+4$ in the decay-particle rest frame is
\begin{equation*}
\int d\Phi_4(P;p_1,p_2,p_3,p_4)=\prod_1^4 \int \frac{d^3\vec{p}_i}{(2\pi)^32E_i}(2\pi)^4\delta^4(P-p_1-p_2-p_3-p_4),
\end{equation*}
where $P=(\sqrt{s},\vec{0})$. We decompose this integral into a sequence of two-body decay phase space integrals with the order:
\begin{equation*}
P\to p_1+q_{234} \to p_1+(p_2+q_{34}) \to p_1+(p_2+(p_3+p_4)),
\end{equation*} 
where $q_{234}, q_{34}$ are intermediate variables.
Using the identities
\begin{align*}
1&=\int\frac{d^4q}{(2\pi)^4}(2\pi)^4\delta^4(q-p)\theta(q^0),\\
1&=\int\frac{ds}{2\pi}(2\pi)\delta(s-q^2),
\end{align*}
and by multiplying and integrating out $q^0$ we have
\begin{align*}
1&=1\cdot 1=\int \frac{ds}{2\pi}\int\frac{d^4q}{(2\pi)^4}(2\pi)^5\delta^4(q-p)\theta(q^0)\delta(s-q^2)\\
&=\int\frac{ds}{2\pi}\int\frac{dq^0}{2\pi}\int\frac{d^3\vec{q}}{(2\pi)^32E}(2\pi)^4\delta^4(q-p)
\end{align*}
where $E=\sqrt{\vec{q}^2+s}$. 
We insert similar factors into $\int d\Phi_4$, exchanging orders of integration and integrating out redundant variables and using the well-known 2- and 3-body decay formulae
\begin{align*}
\int d\Phi_2(P;p_1,p_2)&=\frac{\bar\beta\paren{\frac{m_1^2}{s},\frac{m_2^2}{s}}}{8\pi}\int\frac{d\cos\theta}{2}\int\frac{d\phi}{2\pi}\\
\int d\Phi_3(P;p_1,p_2,p_3)&=\int\frac{ds_{23}}{2\pi}d\Phi_2(p_1,q_{23})d\Phi_2(p_2,p_3),
\end{align*}
where $\theta,\phi$ are the usual spherical coordinate angles of 3-momentum of either particle 1 or sum of particle 2 and 3 in the rest frame, and 
\begin{align*}
\bar\beta(x,y)&\equiv\sqrt{1-2(x+y)+(x-y)^2},\\
\bar\beta_1&=\bar\beta\paren{\frac{m_1^2}{s},\frac{s_{234}}{s}},\\
\bar\beta_2&=\bar\beta\paren{\frac{m_2^2}{s_{234}},\frac{s_{34}}{s}},\\
\bar\beta_3&=\bar\beta\paren{\frac{m_3^2}{s_{34}},\frac{m_4^2}{s_{34}}}.
\end{align*}
The four-body decay phase integral can now be written as:
\begin{align*}
\int\!\! d\Phi_4(P;p_1,p_2,p_3,p_4)
&=\prod_1^4\int\!\frac{d^3\vec{p}_i}{(2\pi)^32E_i}\int\!\frac{ds_{234}}{2\pi}\int\!\frac{d^3\vec{q_{234}}}{(2\pi)^32E_{234}}(2\pi)^4\delta^4(P-p_1-q_{234})\\
&\times (2\pi)^4\delta^4(q_{234}-p_2-p_3-p_4)\\
&=\int\!\frac{ds_{234}}{2\pi}\int d\Phi_2(P;p_1,q_{234})\int d\Phi_3(q_{234};p_2,p_3,p_4)\\
&=\int\!\frac{ds_{234}}{2\pi}\frac{\bar\beta_1}{8\pi}\int\!\frac{d\cos\hat{\theta}_1}{2}\int\!\frac{d\hat\phi_1}{2\pi}\frac{ds_{34}}{2\pi}\int\! d\Phi_2(q_{234};p_2,q_{34})\int\! d\Phi_2(q_{34};p_3,p_4)\\
&=\int\!\frac{ds_{234}}{2\pi}\frac{\bar\beta_1}{8\pi}\int\!\frac{ds_{34}}{2\pi}\frac{\bar\beta_2}{8\pi}\int\!\frac{d\cos\hat{\theta}_2}{2}\int\!\frac{d\hat{\phi}_2}{2\pi}\frac{\bar\beta_3}{8\pi}\int\!\frac{d\cos\hat{\theta}_3}{2}\int\frac{d\hat{\phi}_3}{2\pi}\\
&=\int\!\frac{ds_{234}}{2\pi}\frac{\bar\beta_1}{8\pi}\int\frac{ds_{34}}{2\pi}\frac{\bar\beta_2}{8\pi}\int\frac{d\cos\hat{\theta}_2}{2}\frac{\bar\beta_3}{8\pi}\int\frac{d\cos\hat{\theta}_3}{2}\int\frac{d\hat{\phi}_3}{2\pi},
\end{align*}
where $\hat{\theta}_2, \hat{\phi}_2; \hat{\theta}_3, \hat{\phi}_3$ are spherical angles of 3-momentum particle 2 and 3 in rest frame of $q_{234}$ and $q_{34}$ respectively, and in the last few steps in the above derivation we have assumed isotropy in the first decay step: $i\to 1+(234)$, i.e. that $|\mathcal{M}|^2$ is independent of the angles $\hat{\theta}_1,\hat{\phi}_1$ and $\hat{\phi}_2$. In summary, we have under those conditions,
\begin{equation*}
\frac{d\Gamma}{ds_{234}}=\frac{\bar\beta_1}{16\pi^2}\int\frac{ds_{34}}{2\pi}\frac{\bar\beta_2}{8\pi}\int\frac{d\cos\hat{\theta}_2}{2}\frac{\bar\beta_3}{8\pi}\int\frac{d\cos\hat{\theta}_3}{2}\int\frac{d\hat{\phi}_3}{2\pi}|\mathcal{M}|^2,
\end{equation*}
where without loss of generality in the rest frame of $q_{234}$ we assume 3-momentum of particle 2 parallel to +z axis and 3-momentum $q_{34}$ lies in x-z plane.

We also need formulae to convert between the remaining three angles and Lorentz-invariant inner products of final state 4-momenta $p_i\cdot p_j$. We use the following notation:
\begin{align*}
\eta_1^{\epsilon_1\epsilon_2}&=1+\epsilon_1\frac{m_1^2}{s}+\epsilon_2\frac{s_{234}}{s},\\
\eta_2^{\epsilon_2\epsilon_2}&=1+\epsilon_1\frac{m_2^2}{s_{234}}+\epsilon_2\frac{s_{34}}{s_{234}},\\
\eta_3^{\epsilon_2\epsilon_2}&=1+\epsilon_1\frac{m_3^2}{s_{34}}+\epsilon_2\frac{m_4^2}{s_{34}},
\end{align*}
where $\epsilon_i=\pm$, for instance
\begin{equation*}
\eta_1^{+-}=1+\frac{m_1^2}{s}-\frac{s_{234}}{s}.
\end{equation*}
In the rest frame of decay particle, we have
\begin{align*}
p_1&=\frac{\sqrt{s}}{2}(\eta_1^{+-},0,0,-\bar\beta_1),\\
q_{234}&=\frac{\sqrt{s}}{2}(\eta_1^{-+},0,0,\bar\beta_1).
\end{align*}
The boost from the rest frame of the decaying particle to the rest frame of $q_{234}$ is, in matrix form,
\begin{equation*}
\Lambda_1=\frac{\sqrt{s}}{2\sqrt{s_{234}}}\begin{pmatrix}
\eta_1^{-+} & 0 & 0 & -\bar\beta_1\\
0 & 1 & 0 & 0 \\
0 & 0 & 1 & 0 \\
-\bar\beta_1 & 0 & 0 & \eta_1^{-+}
\end{pmatrix},
\end{equation*}
and after this boost the momenta of particles 1 and 2 are
\begin{align*}
p_1^\prime&=\Lambda_1 p_1=\frac{s}{2\sqrt{s_{234}}}(\eta_1^{--},0,0,-\bar\beta_1),\\
p_2^\prime&=\Lambda_1 p_2=\frac{\sqrt{s_{234}}}{2}(\eta_2^{+-},\bar\beta_2\sin\hat{\theta}_2,0,\bar\beta_2\cos\hat{\theta}_2).
\end{align*}
In terms of these variables then,
\begin{equation*}
p_1\cdot p_2=p_1^\prime \cdot p_2^\prime=\frac{s}{4}(\eta_2^{+-}\eta_1^{--}+\bar\beta_1\bar\beta_2\cos\hat{\theta}_2).
\end{equation*}
Without loss of generality, before boosting to rest frame of $q_{34}$ we rotate coordinates so that $p_2^\prime$ points to +z axis.  The required rotation matrix is
\begin{equation*}
R=\begin{pmatrix}
1 & 0 & 0 & 0 \\
0 & \cos\hat\theta_2 & 0 & -\sin\hat\theta_2 \\
0 & 0 & 1 & 0 \\
0 & \sin\hat\theta_2 & 0 & \cos\hat\theta_2
\end{pmatrix},
\end{equation*}
and the boost to the rest frame of $q_{34}$ is now in simpler form:
\begin{equation*}
\Lambda_2=\frac{\sqrt{s_{234}}}{2\sqrt{s_{34}}}\begin{pmatrix}
\eta_2^{-+} & 0 & 0 & \bar\beta_2\\
0 & 1 & 0 & 0 \\
0 & 0 & 1 & 0 \\
\bar\beta_2 & 0 & 0 & \eta_2^{-+}
\end{pmatrix}.
\end{equation*}
In the rest frame of $q_{34}$, the particle momenta are 
\begin{align*}
p_1^{\prime\prime}&=\Lambda_2 R p_1^\prime=\frac{s}{4\sqrt{s_{34}}}\paren{\eta_2^{-+}\eta_1^{--}-\bar\beta_1\bar\beta_2\cos\hat\theta_2,\bar\beta_1\sin\hat\theta_2,0,\bar\beta_2\eta_1^{--}-\eta_2^{-+}\bar\beta_1\cos\hat\theta_2},\\
p_2^{\prime\prime}&=\Lambda_2 R p_2^\prime=\frac{s_{234}}{2\sqrt{s_{34}}}(\eta_2^{--},0,0,\bar\beta_2),\\
p_3^{\prime\prime}&=\frac{\sqrt{s_{34}}}{2}(\eta_3^{+-},\bar\beta\cos\hat\phi_3\sin\hat\theta_3,\bar\beta_3\sin\hat\phi_3\sin\hat\theta_3,\bar\beta_3\cos\hat\theta_3),
\end{align*}
and the Lorentz-invariant inner products are
\begin{align*}
p_2\cdot p_3=p_2^{\prime\prime}\cdot p_3^{\prime\prime}&=\frac{s_{234}}{4}(\eta_3^{+-}\eta_2^{--}-\bar\beta_2\bar\beta_3\cos\hat\theta_3)\\
p_1\cdot p_3=p_1^{\prime\prime}\cdot p_3^{\prime\prime}&=\frac{s}{8}\bigl(\eta_3^{+-}(\eta_2^{-+}\eta_1^{--}-\bar\beta_1\bar\beta_2\cos\hat\theta_2)-\bar\beta_1\bar\beta_3\sin\hat\theta_2\sin\hat\theta_3\cos\hat\phi_3\\
&-\eta_1^{--}\bar\beta_2\bar\beta_3\cos\hat\theta_3+\eta_2^{-+}\bar\beta_1\bar\beta_3\cos\hat\theta_2\bigl).
\end{align*}
Then the Jacobian for the variable change $d\cos\hat\theta_2 d\cos\hat\theta_3 d\hat\phi_3 \to d(p_1\cdot p_2)d(p_2\cdot p_3)d(p_1\cdot p_3)$ is
\begin{equation*}
d\cos\hat\theta_2 d\cos\hat\theta_3d\hat\phi_3=\frac{64}{s^2s_{234}\bar\beta_1^2\bar\beta_2^2\bar\beta_3^2\sin\hat\theta_2\sin\hat\theta_3\sin\hat\phi_3}d(p_1\cdot p_2)d(p_1\cdot p_3)d(p_2\cdot p_3),
\end{equation*}
where $\cos\hat\theta_{2,3}$ and $\sin\hat\phi_3$ are obtained from the above conversion formulae.

\subsection{Range of four-body decay integral}
\label{appx:D.8.1}

Similar to three-body decay, the range of the integral is set by fixing several invariant masses and the ranges of the trigonometric functions. First of all, the ranges of $s_{234}$ and $s_{34}$ are
\begin{align*}
(m_2+m_3+m_4)^2&\le s_{234} \le (\sqrt{s}-m_1)^2,\\
(m_3+m_4)^2&\le s_{34} \le (\sqrt{s_{234}}-m_4)^2.
\end{align*}
With $s_{234}, s_{34}$ fixed, the range of $p_1\cdot p_2$ and $p_2\cdot p_3$ are separately determined:
\begin{align*}
\frac{s}{4}\paren{\eta_2^{+-}\eta_1^{--}-\bar\beta_1\bar\beta_2}&\le p_1\cdot p_2\le \frac{s}{4}\paren{\eta_2^{+-}\eta_1^{--}+\bar\beta_1\bar\beta_2},\\
\frac{s_{234}}{4}\paren{\eta_3^{+-}\eta_2^{--}-\bar\beta_2\bar\beta_3}&\le p_2\cdot p_3\le \frac{s_{234}}{4}\paren{\eta_3^{+-}\eta_2^{--}+\bar\beta_2\bar\beta_3}.
\end{align*}
With $p_1\cdot p_2$, and $p_2\cdot p_3$ fixed, using the above expression for $p_1\cdot p_3$, and expressing $\cos\hat\theta_{2,3}$ and $\sin\hat\theta_{2,3}$ in terms of $p_1\cdot p_2$ and $p_2\cdot p_3$, we find
\begin{align*}
(p_1\cdot p_3)_{\text{max}}&=\frac{s}{8}\Bigl(\eta_3^{+-}\paren{\eta_2^{-+}\eta_1^{--}-\bar\beta_1\bar\beta_2\cos\hat\theta_2(p_1\cdot p_2)}-\eta_1^{--}\bar\beta_2\bar\beta_3\cos\hat\theta_3(p_2\cdot p_3)\\
&+\eta_2^{-+}\bar\beta_1\bar\beta_3\cos\hat\theta_2(p_1\cdot p_2)\Bigl)+\frac{s}{8}\bar\beta_2\bar\beta_3|\sin\hat\theta_2(p_1\cdot p_2)\sin\hat\theta_3(p_2\cdot p_3)|,\\
(p_1\cdot p_3)_{\text{min}}&=\frac{s}{8}\Bigl(\eta_3^{+-}\paren{\eta_2^{-+}\eta_1^{--}-\bar\beta_1\bar\beta_2\cos\hat\theta_2(p_1\cdot p_2)}-\eta_1^{--}\bar\beta_2\bar\beta_3\cos\hat\theta_3(p_2\cdot p_3)\\
&+\eta_2^{-+}\bar\beta_1\bar\beta_3\cos\hat\theta_2(p_1\cdot p_2)\Bigl)-\frac{s}{8}\bar\beta_2\bar\beta_3|\sin\hat\theta_2(p_1\cdot p_2)\sin\hat\theta_3(p_2\cdot p_3)|.\\
\end{align*}

\subsection{An example of Dalitz plot in four-body decay}
\label{appx:D.8.2}

As an example of four-body decay, I plot the phase space of $B^0\to K^0+D^0+D^0+\pi^0$ in the 3 dimensional space ($p_1\cdot p_2, p_2\cdot p_3, p_1\cdot p_3$), where $p_1, p_2, p_3$ are four momenta of $K^0$ and the two $D^0$'s respectively, with $s_{234}, s_{34}$, the invariant mass of $D^0D^0\pi^0$ and $D^0\pi^0$ fixed near the masses of $X(3872)$ and $\overline{D^{*0}}$ respectively:
\begin{align*}
m_1=&497.614\,\mev, \quad m_2=m_3=1864.86\,\mev, \quad m_4=134.9766\,\mev, \\
s=&(5279.58\,\mev)^2, \quad s_{234}=(3871.68\,\mev)^2, \quad s_{34}=(2006\,\mev)^2.
\end{align*}
The allowed phase space is the pillow-shaped region displayed here:
\begin{center}
\includegraphics[width=.55\textwidth]{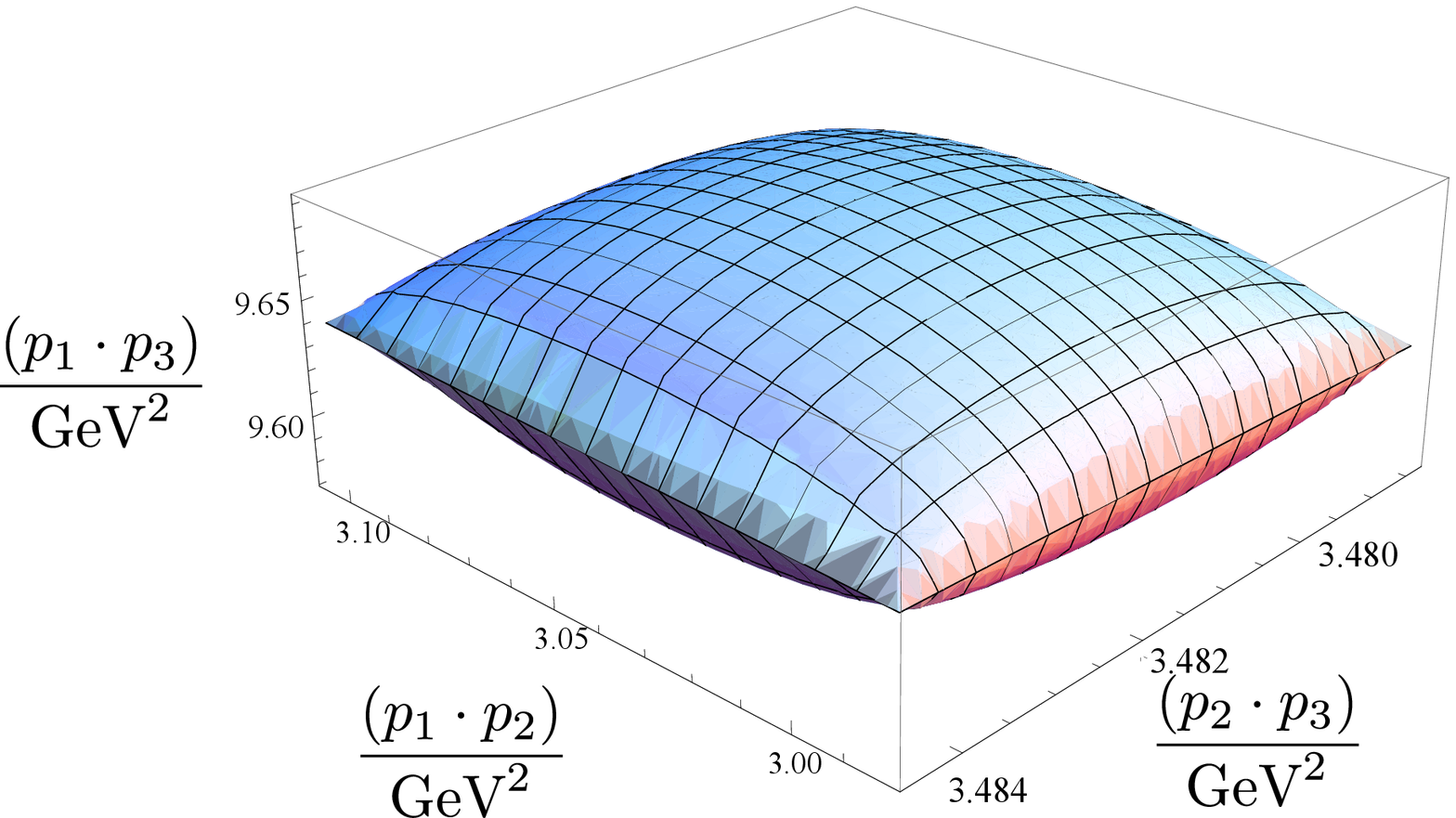}
\end{center}

%% file: appendix_E.tex
\chapter{Useful Formula in Laser EFT}
\label{appx:LaserEFTformulae}

\section{Classical action of a charged scalar particle in a plane wave potential}\label{appx:classicalaction}

We consider a vector potential of the plane wave form, with the Lorenz and tranverse gauge conditions, as discussed in Sec.\:\ref{sec:laserfield},
\be
A^\mu_{\rm cl} = A^\mu_{\rm cl,\perp}(\bar n \cdot x),
\ee
a function of $\bar n\cdot x$ only.  The classical action for a scalar particle with charge $\pm e$ and mass $m$ satisfies the Hamilton-Jacobi equation,
\be\label{HJeqn}
(\partial^\mu S_{cl}\pm eA_{\rm cl}^\mu)(\partial^\nu S_{cl}\pm eA^\nu_{\rm cl})g_{\mu\nu}-m^2 =0,
\ee
where the metric is the usual Minkowski metric $g_{\mu\nu}=$diag($1,-1,-1,-1$), as used throughout.  We set the Ansatz 
\be
S_{cl}=p\cdot x+S_p(\bar n\cdot x), \qquad p^2=m^2
\ee
which is the sum of a free-particle piece and a function of $p^\mu$ that has the same coordinate dependence as the potential.  Inserting this Ansatz in \req{HJeqn}, we have
\begin{align}
0&=(p^\mu+\partial^\mu S_p(\bar n\cdot x)\pm eA^\mu_{\rm cl})(p^\nu+\partial^\nu S_p(\bar n\cdot x)\pm eA^\nu_{\rm cl})g_{\mu\nu}-m^2 \nn\\
&=2p\cdot \bar n S_p'\pm 2p\cdot eA_{\rm cl}+(S_p')^2\bar n\cdot\bar n+(eA_{\rm cl})^2\pm 2e\bar n\cdot A_{\rm cl} S_p' \nn\\
&=2p\cdot \bar n S_p'\pm 2p\cdot eA_{\rm cl}+(eA_{\rm cl})^2,
\end{align}
using that $\bar n^2=0$ and $\bar n\cdot A_{\rm cl}=0$.  Thus, we have a first-order differential equation for $S_p(\bar n\cdot x)$,
\be
\frac{\partial S_p}{\partial \bar n\cdot x}=\frac{\mp 2p\cdot eA_{\rm cl}-(eA_{\rm cl})^2}{2\bar n \cdot p}.
\ee
Note that $p_\perp,\bar n\cdot p$ are constants of motion and therefore have no dependence on $\bar n\cdot x$, for which the conjugate momentum is $n\cdot p$.  The integration can be performed using either future $\bar n\cdot x\to +\infty$ or past $\bar n\cdot x\to -\infty$ boundary conditions,
\begin{align}
S_p(\bar n\cdot x)&=\int_{-\infty}^{\bar n\cdot x}\frac{\mp 2p\cdot eA_{\rm cl}(\bar n\cdot y)-(eA_{\rm cl})^2(\bar n\cdot y)}{2\bar n \cdot p}d(\bar n\cdot y) \\
&=-\int^{\infty}_{\bar n\cdot x}\frac{\mp 2p\cdot eA_{\rm cl}(\bar n\cdot y)-(eA_{\rm cl})^2(\bar n\cdot y)}{2\bar n \cdot p}d(\bar n\cdot y).
\end{align}
For electrons, we take the upper sign, corresponding to charge $-e<0$, while for positrons, we take the lower sign, corresponding to charge $e>0$.

\section{Proper time representation of propagator in plane wave fields}

Here we briefly reproduce the derivation of the gauge-invariant propagator in a plane-wave potential $A^\mu(\bar n\cdot x)$ depending only on one lightcone coordinate $\bar n\cdot x$, following the method of \cite{schwinger1951gauge}.  In $x$ space, we write the propagator as an integral over a ``proper time'' $s$ of a kernel $K(x',x;s)$,
\be\label{eq:A004eq0}
G(x',x)=i\int_0^\infty K(x',x;s)ds.
\ee
The Green's function is defined by the operator equation
\be\label{eq:A004eq1}
(i\slashed{D}_{x'}-m)G(x',x)=\del(x-x'),
\ee
so that the proper time kernel satisfies
\be\label{eq:A004eq2}
(i\slashed{D}_{x'}-m)K(x',x;s)=i\pd_s K(x',x;s),
\ee
with boundary condition
\be\label{eq:A004eq3}
K(x',x;s=0)=\del(x'-x).
\ee
To obtain a representation of $K(x',x;s)$ in terms of field operators, consider $G(x',x)$ as a matrix element
\be\label{eq:A004eq4}
G(x',x)=\bra{x'}\ov{i\slashed{D}_{x'}-m}\ket{x},
\ee
where $\braket{x'}{x}=\delta(x'-x)$, and introduce an integral representation of the operator inverse
\be\label{eq:A004eq5}
\ov{i\slashed{D}_{x'}-m}=(i\slashed{D}_{x'}+m)(-i)\int_0^\infty\exp\left[i\paren{(iD)^2-\frac{e\sig F}{2}-m^2+i\eps}s\right]ds.
\ee
This shows that the kernel $K$ is the matrix element of an exponentiated operator,
\be\label{eq:A004eq6}
K(x',x;s)=(-i\slashed{D}_{x'}+m)e^{-im^2 s}\bra{x'}e^{-i\cH s}\ket{x},
\ee
and the matrix element can be evaluated by solving the Hamiltonian dynamics in the ``time'' $s$:
\begin{align}
\cH &=-(i\slashed{D})^2=-(iD)^2+\frac{e\sig F}{2}=-\Pi^2+\frac{e\sig F}{2},\nn\\
i\pd_s x^\mu(s)&=[x^\mu,\cH]=2i\Pi^\mu(s),\nn\\
i\pd_s \Pi^\mu(s)&=[\Pi^\mu,\cH] 
=i2eF^{\mu\nu}\Pi^\nu+i^2\pd^\nu eF^{\mu\nu}+\frac{i}{2}e\sig_{\alpha\beta}\pd^\mu F^{\alpha\beta}.
\end{align}
The kernel is then obtained by noting that
\be\label{eq:A004eq9}
i\pd_s K(x',x;s)=i\pd_s\bra{x'(s)}e^{-i\cH s}\ket{x(0)}=\bra{x'(s)}\cH\ket{x(0)}.
\ee
The boundary condition becomes $\left\langle x'(s=0) \vert x(0)\right\rangle=\del(x'-x)$. Additionally, this matrix element in $s$ must satisfy
\begin{align}
i\slashed{D}_{x'}\left\langle x'(s)\big\vert x(0)\right\rangle&=\bra{x'(s)}\slashed{\Pi}(s)\ket{x(0)},\nn\\
-iD_x^\mu\left\langle x'(s)\big\vert x(0)\right\rangle&=\bra{x'(s)}\Pi^\mu(0)\ket{x(0)}\paren{=\bra{x'(s)}\ket{x(0)}i\overleftarrow{D}_x}. \label{eq:Wilsonlinkdiffeq}
\end{align}

Now specializing to plane wave fields, the field tensor can be written only depends on one lightcone coordinate,
\be\label{eq:A004eq11}
F^{\mu\nu}=f^{\mu\nu}F(\varphi),\quad \varphi=\bn\cdot x,\quad \pd^\nu\varphi=\bn^\nu,
\ee
in which we separate the tensor structure from a scalar function $F(\varphi)$ containing the coordinate dependence.  The tensor structure vanishes when contracted with the same lightcone vector, $\bn_\mu f^{\mu\nu}=0 \bn_\nu f^{\mu\nu}=0$.  Using the lightcone decomposition of the metric $g^{\mu\nu}=\frac{n^\mu\bn^\nu}{2}+\frac{\bn^\mu n^\nu}{2}+g_\perp^{\mu\nu}$, one can show that
\begin{align}\label{eq:A004eq12}
f^{\mu\lam}f^{\kappa\nu}g_{\lam\kappa}
=f^{\mu\lam}g_{\lam\kappa}^\perp f^{\kappa\nu}=-\bn^\mu \bn^\nu,
\end{align}
to satisfy $\bn\cdot f \cdot f=0$ (using matrix notation) and $f^2=0$.  When contracted with the dual tensor $f^{*\mu\nu}=\epsilon^{\mu\nu\alpha\beta}f_{\alpha\beta}/2$, we find
\begin{align}\label{eq:A004eq13b}
f^{\mu\kappa}f^{*\lam\nu}g_{\kappa\lam}
&=f^{\mu\kappa}\paren{\frac{n_\kappa \bn_\lam}{2}+g_{\kappa\lam}^\perp}\frac{1}{2}\eps^{\lam\nu\alpha\beta}f_{\alpha\beta}
=f^{\mu\kappa}g_{\kappa\lam}^\perp f^{*\lam\nu}=0,
\end{align}
to satisfy $\bn\cdot f\cdot f^*=0$ and $f\cdot f^*=0$.
Now the proper time equations of motion in the case of plane wave become
\begin{align}
\pd_s x^\mu&=2\Pi^\mu,\nn\\
\pd_s\Pi^\mu
&=2ef^{\mu\nu}\Pi^\nu F(\varphi)+\frac{e\sig_{\alpha\beta}}{2}f^{\alpha\beta}F'(\varphi)\bn^\mu.
\end{align}
As anticipated from the symmetry, $\bn\cdot \Pi$ is conserved:
\begin{align}
\pd_s(\bn\cdot \Pi)&=2e\bn\cdot f\cdot \Pi F(\varphi)+\frac{e\sig_{\alpha\beta}f^{\alpha\beta}}{2}F'(\varphi)\bn^2=0 ,
\end{align}
and therefore $\bn\cdot\Pi$ is a constant and we can immediately integrate $\pd_s(\bn\cdot x)=\pd_s\varphi=2\bn\cdot\Pi$ to give
\begin{align}
\bn\cdot(x(s)-x(0))=2\bn\cdot\Pi(s)\cdot s=2\bn\cdot\Pi(0)s.
\end{align}
Another conserved quantity is $f_{\mu\nu}^*\Pi^\nu$, as we can see by
\begin{align}
f_{\mu\nu}^*\pd_s\Pi_\nu&=f_{\mu\nu}^*2ef_{\nu\lam}\Pi_\lam F(\varphi)+\frac{e\sig \cdot f}{2}F'(\varphi)f_{\mu\nu}^* \bn_\nu=0.
\end{align}
We can simplify the $\Pi^\mu$ equation of motion by identifying another constant of motion by first contracting with the unit tensor containing the field structure,
\begin{align}
f_{\mu\nu}\pd_s\Pi^\nu&=f_{\mu\nu}2ef^{\nu\lam}\Pi^\lam F(\varphi)+\frac{e\sig\cdot f}{2}F'(\varphi)f_{\mu\nu}\bn^\nu=-2\bn_\mu (\bn\cdot\Pi) eF(\varphi).
\end{align}
We define a scalar function
\begin{align}
\pd_s A(\varphi)=A'(\varphi)=F(\varphi),
\end{align}
which, together with $2\bn\cdot\Pi=\pd_s\varphi$,  allows us to write
\begin{align}
-\bn_\mu (2\bn\cdot\Pi) eF(\varphi)=-\bn_\mu(\pd_s\varphi)eF(\varphi)=-\bn_\mu e\pd_sA(\varphi),
\end{align}
and implies that the combination is conserved,
\begin{align}
\pd_s\paren{f_{\mu\nu}\Pi^\nu+\bn_\mu eA(\varphi)}=0.
\end{align}
Note that the operator for the lightcone coordinate commutes along the path,
\begin{align}
[\pd_s\varphi,\varphi]=0=[\varphi(s),\varphi(0)].
\end{align}
We can therefore use $d\varphi=2\bn\cdot\Pi ds$ as a valid change of variables.  
Define the constant vector
\be
C_\mu=f_{\mu\nu}\Pi^\nu+\bn_\nu eA(\varphi),
\ee
which satisfies $\bn\cdot C=0$, $f_{\mu\nu}^*C_\nu=0$, and 
\begin{align}
f_{\mu\nu}C_\nu&=-\bn_\mu \bn\cdot\Pi,\nn\\
C_\mu C^\mu&=\Pi^\nu\paren{-f_{\nu\mu}f^{\mu\lam}}\Pi_\lam=(\bn\cdot\Pi)^2.
\end{align}
This constant vector allows rewriting the right-hand side of the $\Pi^\mu$ equation,
\begin{align}
\pd_s\Pi^\mu&=2eF(\varphi)f^{\mu\nu}\Pi^\nu+\frac{e\sig \cdot F}{2}F'(\varphi) \bn^\mu\nn\\
&=(2\bn\cdot\Pi)^{-1}\pd_s\left[2e AC^\mu-e^2 A^2\bn^\mu+\frac{e\sig f}{2}F\bn^\mu\right],
\end{align}
which can clearly be integrated.  We obtain
\begin{align}
\ov{2}\frac{dx^\mu}{ds}&=\Pi^\mu=\ov{2\bn\cdot\Pi}\paren{2eAC^\mu-e^2A^2\bn^\mu+\frac{e\sig f}{2}F\bn^\mu}+C_0^\mu,
\end{align}
in which the constant of integration can be isolated by $f_{\mu\nu}^*\Pi^\nu=f_{\mu\nu}^* C_0^\mu$, and is of course independent of $s$.  Integrating again,
\begin{align}
x^\mu(s)-x^\mu(0)&=\int_0^s ds\left[\frac{2eAC^\mu-(e^2A^2)\bn^\mu+\frac{e\sig f}{2}F\bn^\mu}{\bn\cdot\Pi}+2C_0^\mu\right]\nn\\
&=\int_{\varphi(0)}^{\varphi(s)}\frac{d\varphi}{2(\bn\cdot\Pi)^2}\left[2eAC^\mu-(e^2A^2)\bn^\mu+\frac{e\sig\cdot f}{2}F\bn^\mu +2C_0^\mu s\right].
\end{align}
This determines the constant of integration, $C_0^\mu$. Now to evaluate the matrix element, \eq{A004eq9} put $\cH$ in terms of the $x^\mu(s)$ and $x^\mu(0)$ with all operators evaluated at $s$ standing to the left of operators evaluated at $s=0$.  To this end, we note the following identities
\begin{align}
C_\mu&=\frac{f_{\mu\nu}(x_s^\nu-x_0^\nu)}{2s}+\frac{\bn_\mu}{\varphi_s-\varphi_0}\int_{\varphi_0}^{\varphi_s}eA(\varphi)d\varphi,\nn\\
f_{\mu\nu}\Pi^\nu(s)&=C_\mu-\bn_\mu eA(\varphi_s)=f_{\mu\nu}\frac{(x_s^\nu\!-\!x_0^\nu)}{2s}+\ov{s}\int_{\varphi_0}^{\varphi_s}\frac{eA(\varphi)}{2\bn\cdot\Pi}\bn^\mu d\varphi-eA(\varphi_s)\bn^\mu,\nn\\
f_{\mu\nu}(x_s^\nu-x_0^\nu)&=2s\paren{f_{\mu\nu}\Pi^\nu(s)+eA(\varphi_s)\bn_\mu-\int_{\varphi_0}^{\varphi_s}eA(\varphi)\frac{d\varphi}{2\bn\cdot\Pi}\frac{\bn^\mu}{s}}.
\end{align}
After many lines of algebra, we obtain
\begin{align}
\cH&=-\frac{(x_s-x_0)^2}{4s^2}+\ov{(\bn\cdot(x_s-x_0))^2}\left[\int_{\varphi_0}^{\varphi_s}eAd\varphi\right]^2
 \nn\\ &+\ov{\bn\cdot(x_s-x_0)}\int_{\varphi_0}^{\varphi_s}\frac{e\sig f}{2}F-(eA)^2d\varphi.
\end{align}
To commute the operators, we use $[\Pi^\mu(s),x^{\mu}(s)]=i$ to show that
\begin{align}
[x^\mu(0),x^\mu(s)]&=-8is,
\end{align}
and finally have
\begin{align}
\cH&=-\ov{4s^2}\paren{x_s^2-2x_sx_0+x_0^2}-\frac{2i}{s}+\ov{(\varphi_s-\varphi_0)^2}\left[\int_{\varphi_0}^{\varphi_s}eAd\varphi\right]^2 \nn\\
&+\ov{\varphi_s-\varphi_0}\int_{\varphi_0}^{\varphi_s}\frac{\sig f}{2}eF-(eA)^2d\varphi.
\end{align} 
Inserting this into \eq{A004eq9} gives a differential equation in $s$ for the matrix element (equivalently the kernel $K$),
\begin{align}
i\pd_s\left\langle x'\big\vert x\right\rangle
&=\Bigg\{-\ov{4s^2}(x'-x)^2-\frac{2i}{s}+\ov{(\bn\cdot(x'-x))^2}\left[\int_{\bn\cdot x}^{\bn\cdot x'}eAd\varphi\right]^2\nn\\ &+\ov{\bn\cdot(x'-x)}\int_{\bn\cdot x}^{\bn\cdot x'}\frac{\sig f}{2}eF-(eA)^2d\varphi\Bigg\}\left\langle x'\big\vert x\right\rangle.
\end{align}
The solution is
\begin{align}
\left\langle x'\bigg\vert x\right\rangle &=C(x',x)s^{-2}\exp\Bigg(-i\frac{(x'-x)^2}{4s}-is\ov{(\bn\cdot(x'-x))^2}\left[\int_{\bn\cdot x}^{\bn\cdot x'}eAd\varphi\right]^2\nn\\
&-\frac{is}{\bn\cdot(x'-x)}\int_{\bn\cdot x}^{\bn\cdot x'}\frac{\sig f}{2}eF-(eA)^2d\varphi\Bigg).\label{eq:A004kernelalmost}
\end{align}
where $C(x',x)$ is independent of $s$.  To determine the function $C(x',x)$, we use \eq{Wilsonlinkdiffeq}, which gives a differential equation,
\begin{align}
\bigg[i\pd_{x'}^\mu-eA^\mu(x')&-f^{\mu\nu}(x'-x)_\nu\frac{eA(x')}{\bn\cdot(x'-x)} \nn\\
&+f^{\mu\nu}(x'-x)_\nu\ov{(\bn\cdot(x'-x))^2}\int_{\bn\cdot x}^{\bn\cdot x'}eAd\varphi\bigg]C(x',x)=0,
\end{align}
which is solved by a Wilson line connecting $x,x'$ due to antisymmetry of $f^{\mu\nu}$:
\be
C(x',x)=C_0\exp\paren{-ie\int_x^{x'}A^\mu dy_\mu}.
\ee
Lastly, $C_0$ is a normalization coefficient that is determined by the boundary condition in $s$, \eq{A004eq3}.
Setting $s=0$ in \eq{A004kernelalmost}, we find a delta-function if $C_0=-\ov{4\pi^2}$.

All together, we have
\begin{align}
G(x',x)&=(i\slashed{D}_{x'}+m)\frac{-1}{4\pi^2}\exp\paren{-ie\int_x^{x'}A^\mu dy_\mu}\int_0^\infty\frac{ds}{s^2}e^{-im^2s}\nn\\
&\times \exp\Bigg(-i\frac{(x'-x)^2}{4s}-is\left[\ov{\bn\cdot(x'-x)}\int_{\bn\cdot x}^{\bn\cdot x'}eAd\varphi\right]^2 \nn\\
&-\frac{is}{\bn\cdot(x'-x)}\int_{\bn\cdot x}^{\bn\cdot x'}\frac{\sig f}{2}eF-(eA)^2d\varphi\Bigg).
\label{eq:Apropagatorfinal}
\end{align}
Note that all gauge dependence is in the Wilson line $e^{-ie\int A\cdot dx}$. The scalar $A(\bn\cdot x)$ in the exponent is defined as the integral of the field tensor $F^{\mu\nu}=f^{\mu\nu}F(\bn\cdot x)$, $A=\int F(\bn\cdot x)d\bn\cdot x$, and the corresponding ambiguity in $A$ (by constant shift) does not change the kernel $K(x',x;s)$.

For small separations, we can expand the integrals in the exponent, and the kernel of the integral reduces to that of a constant field
\begin{align}
\ov{\bn\cdot (x'-x)}&\int_{\bn\cdot x}^{\bn\cdot x'}(eA)^2d\varphi-\left[\ov{\bn\cdot (x'-x)}\int_{\bn\cdot x}^{\bn\cdot x'}eAd\varphi\right]^2 \nn\\
&=-\ov{12}(x'-x)_\mu eF^{\mu\lam}eF_{\lam\nu}(x'-x)^\nu+\ldots\label{eq:smallsepexpansion}
\end{align}
in which, for plane wave fields, all matrix contractions involving more than two $eF^{\mu\nu}$ vanish, due to \eq{A004eq12}.

\section{Self-energy in semiclassical QED: planewave background}\label{app:fulltheoryselfenergy}

In this section, we calculate the 1-loop self-energy diagram for an electron in plane-wave classical electromagnetic field.  The diagram and its amplitude are
\be\label{eq:FTSEU.0}
\includegraphics[width=.2\textwidth]{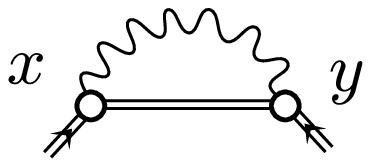} = i\sum_{p',p}\overline{W_{p'}}(y)(-i e\gamma^\mu)G_F(y,x)(-i e\gamma^\nu)D_{F\mu\nu}(y-x)W_p(x),
\ee
in which the electron propagator is given by \eq{Apropagatorfinal}.  The incoming Wilson line is
\be\label{eq:FTSEU.2}
W_p(x)=e^{\frac{\slashed{n}e\slashed{A}}{2\bar n\cdot p}}\exp\paren{-ipx-i\int_{-\infty}^{\bar n\cdot x}2p\cdot e A-(e A)^2\frac{d\varphi}{2\bar n\cdot p}},
\ee
and the outgoing Wilson line is
\be\label{eq:FTSEU.3}
\overline{ W_{p'}}(y)=e^{\frac{e\slashed{A}\slashed{\bar n}}{2\bar n\cdot p'}}\exp\paren{+ip'\cdot y-i\int_{\bar n\cdot y}^\infty 2p'\cdot e A-(e A)^2\frac{d\varphi}{2\bar n\cdot p'}}.
\ee
The photon propagator is also put in the form of a proper time integral,
\be\label{eq:FTSEU.4}
\int\frac{d^4k}{(2\pi)^4}\frac{-ig_{\mu\nu}}{k^2+i\eps}=-\int_0^\infty dt\int\frac{d^4k}{(2\pi)^4}e^{-ik(y-x)}e^{i(k^2+i\eps)t}g_{\mu\nu}.
\ee
Inserting these into the amplitude, we have
\begin{align}
i\Sig(x,y)&=i(-i e)^2\sum_{p';p}e^{ip'y-i\int_{\bar n\cdot y}^\infty 2p'\cdot e A-(e A)^2\frac{d\varphi}{\bar n\cdot p'}}\paren{1+\frac{e\slashed{A}\slashed{\bar n}}{2\bar n\cdot p'}}\gam^\mu\frac{1}{16\pi^2}e^{-i e\int_x^y A\cdot dy}\nn\\
&\times\!\int_0^\infty\!\frac{ds}{s^2}(i\slashed{\partial}_y\!+\!m)\!\left[e^{-i\frac{(y-x)^2}{4s}-is\left[\ov{\bar n\cdot (y-x)}\int_{\bar n\cdot x}^{\bar n\cdot y}e Ad\varphi\right]^2-\frac{is}{\bar n\cdot(y-x)}\int_{\bar n\cdot x}^{\bar n\cdot y}\frac{e\sigma F}{2}-(e A)^2d\varphi}\right]e^{-i(m^2-i\eps)s}\nn\\
&\times\gam_\mu\paren{1+\frac{\slashed{n}}{2}\frac{e\slashed{A}}{\bar n\cdot p}}e^{-ipx-i\int_{-\infty}^{\bar n\cdot x}2p\cdot e A-(e A)^2\frac{d\varphi}{\bar n\cdot p}}\int_0^\infty \!\!dt\int\!\frac{d^4k}{(2\pi)^4}e^{-ik(y-x)+i(k^2+i\eps)t}\,,\label{eq:FTSEU.5}
\end{align}
in which the derivative acts only within the square brackets. 

We change coordinates in order to look at small separations,
\begin{align}
r&=y-x\qquad y=X+r/2\nn\\
X&=\frac{y+x}{2}\qquad x=X-r/2,\label{eq:FTSEU.6}
\end{align}
with the result that
\begin{align}
i\Sig(X,r)&=-i\frac{(-i e)^2)}{16\pi^2}\sum_{p',p,k}e^{ip'(X+r/2)-i\int_{\bar n\cdot (X+r/2)}^\infty 2p'\cdot e A-(e A)^2\frac{d\varphi}{2\bar n\cdot p'}}
\nn \\
&\times \paren{1+\frac{e\slashed{A}(X+r/2)}{\bar n\cdot p'}\frac{\slashed{n}}{2}}\gam^\mu e^{-i e\int_x^y A\cdot dy}\nn\\
&\times\int_0^\infty \frac{ds}{s^2}(i\slashed{\partial}_y+m)\left[e^{-\frac{ir^2}{4s}-is\left[\ov{\bar n\cdot r}\int_{\bar n\cdot(X-r/2)}^{\bar n\cdot (X+r/2)}e Ad\varphi\right]^2-\frac{is}{\bar n\cdot r}\int_{\bar n\cdot(X-r/2)}^{\bar n\cdot(X+r/2)}\frac{e\sigma F}{2}-(e A)^2d\varphi}\right]\nn\\
&\times e^{-i(m^2-i\eps)s}\gam_\mu\paren{1+\frac{\slashed{n}}{2}\frac{e\slashed{A}(X-r/2)}{\bar n\cdot p}}
\nn \\
&\times e^{-ip(X-r/2)-i\int_{-\infty}^{\bar n\cdot(X-r/2)}2p\cdot e A-(e A)^2\frac{d\varphi}{2\bar n\cdot p}\int_0^\infty dt e^{-ik(y-x)}e^{i(k^2+i\eps)t}}.\label{eq:FTSEU.7}
\end{align}
Note there is only dependence on $\bar n\cdot X$, as there is translation invariance in the $n\cdot X, X_\perp$ dierctions. In the limit of small separations, $r\ll \lambda_{cl}$ the plane-wave wavelength, we expand electron Green's function using \eq{smallsepexpansion} and
\begin{align}
\ov{\bar n\cdot(y-x)}&\int_{\bar n\cdot x}^{\bar n\cdot y}e F(\varphi)d\varphi\simeq e F+\ldots\label{eq:FTSEU.8b}
\end{align}
with the field evaluated at the midpoint $\bar n\cdot X=\frac{\bar n\cdot(y+x)}{2}$. We also expand the Wilson line for the incoming and outgoing $e^-$ by writing
\be\label{eq:FTSEU.9}
-i\!\int_{-\infty}^{\bar n\cdot (X-r/2)}\!\!2p\cdot e A-(e A)^2d\varphi=-i\!\int_{-\infty}^{\bar n\cdot X}\!\!2p\cdot e A-(e A)^2d\varphi+i\!\int_{\bar n\cdot(X-r/2)}^{\bar n\cdot X}\!\!2p\cdot e A-(e A)^2d\varphi,
\ee
and expanding the second integral,
\begin{align}\label{eq:FTSEU.10}
\int_{\bar n\cdot (X-r/2)}^{\bar n\cdot X}&2p\cdot e A-(e A)^2d\varphi \\  \nn
&\simeq(2p\cdot e A-(e A)^2)\frac{\bar n\cdot r}{2}-(p-e A)\cdot e A'\frac{(\bar n\cdot r)^2}{4}-\ov{3}(e A')^2\frac{(\bar n\cdot r)^3}{8},
\end{align}
dropping higher than first derivatives in $A$.  Similarly, for the integral in the outgoing Wilson line we have
\begin{align}\label{eq:FTSEU.11}
\int_{\bar n\cdot X}^{\bar n\cdot (X+r/2)}&2p'\cdot e A-(e A)^2d\varphi \\ \nn 
&\simeq (2p'\cdot e A-(e A)^2)\frac{\bar n\cdot r}{2}+(p'-e A)\cdot e A'\frac{(\bar n\cdot r)^2}{4}+\ov{3}(e A')^2\frac{(\bar n\cdot r)^3}{8}.
\end{align}
Substituting these Eqs.\:\eqref{eq:FTSEU.8b},\:\eqref{eq:FTSEU.10}, and \eqref{eq:FTSEU.11} into \eq{FTSEU.7}, we find
\begin{align}
i\Sig(X,r)&\simeq i\frac{(-i e)^2}{16\pi^2}\sum_{p,p',k}e^{ip'X-i\int_{\bar n\cdot X}^\infty 2p'\cdot e A-(e A)^2\frac{d\varphi}{2\bar n\cdot p'}}e^{ip'r/2+\frac{i}{2}(2p'\cdot e A-(e A)^2)\frac{\bar n\cdot r}{2\bar n\cdot p'}}\nn\\
&\times \exp\paren{\frac{i}{2}(p'-e A)\cdot e A'\frac{(\bar n\cdot r)^2}{4\bar n\cdot p'}+\frac{i}{2}(e A')^2\frac{(\bar n\cdot r)^3}{24\bar n\cdot p'}}\paren{1+\frac{e\slashed{A}+e\slashed{A}\frac{\bar n\cdot r}{2}}{\bar n\cdot p'}\frac{\slashed{\bar n}}{2}}\nn\\
&\times \gam^\mu e^{-i e A\cdot r}\int_0^\infty \frac{ds}{s^2}\paren{i\slashed{\partial}_y+m}\left[e^{-\frac{ir^2}{4s}-\frac{is}{12}r(e F)^2r-i\frac{e\sigma F}{2}s}\right]e^{-i(m^2-i\eps)s}\gam_\mu\nn\\
&\times\paren{1+\frac{\slashed{\bar n}}{2}\frac{e\slashed{A}-e\slashed{A}'\frac{\bar n\cdot r}{2}}{\bar n\cdot p}}\exp\bigg(\frac{i}{2}(2p\cdot e A-(e A)^2)\frac{\bar n\cdot r}{2\bar n\cdot p}\bigg) \nn\\
&\times \exp\bigg(-\frac{i}{2}(p-e A)\cdot e A'\frac{(\bar n\cdot r)^2}{4\bar n\cdot p}-\frac{i}{2}(e A')^2\frac{(\bar n\cdot r)^3}{24\bar n\cdot p}\bigg)\nn\\
&\times e^{ipr/2-i p X}\int_0^\infty dt e^{-i k r+i(k^2+i\eps)t}e^{-i\int_{-\infty}^{\bar n\cdot X}2p\cdot e A-(e A)^2\frac{d\varphi}{\bar n\cdot p}}.\label{eq:FTSEU.12}
\end{align}
Note that $(e A')^2(\bar n\cdot r)^2=-r(e F)^2r$.  Focussing on the $r$-dependence, we can go to momentum space. Define
\begin{align}
i\Sig(X)&=\int d^4r i\Sig(X,r)\nn\\
&=\int d^4r \int\frac{d^4p d^4p'}{(2\pi)^8}e^{i p' X-i\int_{\bar n\cdot X}^\infty 2 p'\cdot e A-(e A)^2\frac{d\varphi}{2\bar n\cdot p'}}i\Sig(p,p',r)\nn\\
&\times e^{-i p X-i\int_{-\infty}^{\bar n\cdot X}2p\cdot e A-(e A)^2\frac{d\varphi}{2\bar n\cdot p}}\label{eq:FTSEU14}\\
&=\int\frac{d^4p d^4p'}{(2\pi)^8}e^{ip\cdot X}\bar W_{p'}(\bar n\cdot X)\int\frac{d^4k}{(2\pi)^4}i\Sig(p',p,k)e^{-i p X}W_p(\bar n\cdot X), \label{eq:FTSEU.15}
\end{align}
where
\begin{align}
i\Sig(p',p,k)&=i\frac{(-i e)^2}{16\pi^2}\int d^4r e^{ip'r/2+\frac{i}{2}(2p'\cdot e A-(e A)^2)\frac{\bar n\cdot r}{2\bar n\cdot p'}+\frac{i}{2}(p'-e A)\cdot e A'\frac{(\bar n\cdot r)^2}{4\bar n\cdot p}+\frac{i}{2}(e A')^2\frac{(\bar n\cdot r)^3}{24\bar n\cdot p}}\nn\\
&\times\paren{1+e\slashed{A}'\frac{\bar n\cdot r}{2\bar n\cdot p'}\frac{\slashed{\bar n}}{2}}\gam^\mu e^{-i e A\cdot r}\int_0^\infty dt e^{-i k r+i(k^2+i\eps)t}\nn\\
&\times\int_0^\infty \frac{ds}{s^2}(i\slashed{\partial}_r+m)\left[e^{-ir^2/4s-\frac{is}{12}r(e F)^2r}\right]e^{-i\frac{e\sigma F}{2}s-i(m^2-i\eps)s}\gam_\mu\nn\\
&\times\paren{1+\frac{\slashed{\bar n}}{2}e\slashed{A}'\frac{\bar n\cdot r}{2\bar n\cdot p}}e^{i p r/2+\frac{i}{2}(2p\cdot e A-(e A)^2)\frac{\bar n\cdot r}{2\bar n\cdot p}-\frac{i}{2}(p-e A)\cdot e A'\frac{(\bar n\cdot r)^2}{4\bar n\cdot p}-\frac{i}{2}(e A')^2\frac{(\bar n\cdot r)^3}{24\bar n\cdot p}}.\label{eq:FTSEU.16}
\end{align}
To do the $r^\mu$ integral, we first go to momentum space in the proper time representation of the propagator,
\be\label{eq:FTSEU.17}
e^{-i\frac{r^2}{4s}-i s r(e F)^2r}\equiv e^{-ir\frac{M}{4s} r}=\int \frac{d^4q}{(2\pi)^4} e^{-i q r} g(q),
\ee
in which the matrix is defined
\begin{align}
M^{\mu\nu}=g^{\mu\nu}+\frac{s^2}{3}e F^{\mu\lambda} eF_\lambda^{\phantom{\lambda}\nu}.
\end{align}
Inverting the Fourier transform is straightforward, and gives
\begin{align}
g(p)&=\int d^4r e^{i q r} e^{-\frac{i}{4s} r\cdot M \cdot r}\qquad \nn\\
&=e^{i s q\cdot M^{-1}\cdot q}\paren{\pi\frac{4s}{i}}^2\paren{\det M}^{-1/2}=-16s^2\pi^2 e^{i s q\cdot M^{-1}\cdot q}.\label{eq:FTSEU.18}
\end{align}
The determinant and inverse of $M$ are easily evaluated using the special properties of plane-wave fields that the scalar and pseudoscalar invariants both vanish, $F^2=FF^*=0$. Now
\begin{align}
i\Sig(p',p,k)&=-i(-i e)^2\int d^4r \frac{d^4q}{(2\pi)^4}e^{i\paren{\frac{p+p'}{2}-e A}r-i k r \int_0^\infty dt e^{i(k^2 +i\eps) t}}\nn\\
&\times \exp\!\left[\frac{i}{2}(2p'\!\cdot\! e A-(e A)^2)\frac{\bar n\!\cdot\! r}{2\bar n\!\cdot\! p'}+\frac{i}{2}(p'\!-\!e A)\!\cdot\! e A'\frac{(\bar n\cdot r)^2}{4\bar n\cdot p}+\frac{i}{2}(e A')^2\frac{(\bar n\!\cdot\! r)^3}{24\bar n\cdot p}\right]\nn\\
&\times \exp\!\left[\frac{i}{2}(2p\!\cdot\! e A-(e A)^2)\frac{\bar n\!\cdot\! r}{2\bar n\!\cdot\! p}-\frac{i}{2}(p\!-\!e A)\!\cdot\! e A'\frac{(\bar n\!\cdot\! r)^2}{4\bar n\!\cdot\! p}-\frac{i}{2}(e A')^2\frac{(\bar n\!\cdot\! r)^3}{24\bar n\!\cdot\! p}\right]\nn\\
&\times\paren{1+e\slashed{A}'\frac{\bar n\cdot r}{2\bar n\cdot p'}\frac{\slashed{\bar n}}{2}}\gam^\mu\int_0^\infty ds(i\slashed{\partial}_r+m)\left[e^{-i q r+i p \cdot M^{-1}\cdot p s}\right]e^{-i\frac{e \sigma F}{2}s-i(m^2-i\eps)s}\nn\\
&\times\gam_\mu\paren{1+\frac{\slashed{\bar n}}{2}e\slashed{A}'\frac{\bar n\cdot r}{2\bar n\cdot p}}.\label{eq:FTSEU.19}
\end{align}
As we are interested in the divergent pieces and the imaginary part, we take the limit $p'\to p$,
\begin{align}
i\Sig(p)&=-i(-i e)^2\int\frac{d^4 k}{(2\pi)^4}\int d^4r \frac{d^4q}{(2\pi)^4}e^{i(p-e A)r-i(k+q)r}\int_0^\infty dt e^{i (k^2+i\eps) t}\nn\\
&\times \exp\paren{i(2 p\cdot e A-(e A)^2)\frac{\bar n\cdot r}{2\bar n\cdot p}}\paren{1+e\slashed{A}'\frac{\bar n\cdot r}{2\bar n\cdot p}\frac{\slashed{\bar n}}{2}}\gam^\mu\nn\\
&\times\int_0^\infty ds (\slashed{q}+m) e^{i q \cdot M^{-1}\cdot q s-i\frac{e\sigma F}{2}s-i(m^2-i\eps)s}\gam_\mu\paren{1+\frac{\slashed{\bar n}}{2}e \slashed{A}'\frac{\bar n\cdot r}{2\bar n\cdot p}}.\label{eq:FTSEU.20}
\end{align}

To do the $r$-integral, first note we can write $r^\mu=e^{i k\cdot r}i\frac{\pd}{\pd k_\mu}\left[e^{-i k\cdot r}\right]$ and so replace any $\bar n\cdot r$ not appearing in an exponent with a derivative with respect to $k_+$,
\begin{align}
i\Sig(p)&=-i (-i e)^2\int\frac{d^4k}{(2\pi)^4}\int_0^\infty ds\int_0^\infty dt e^{i(k^2+i\eps)t}\int\frac{d^4q}{(2\pi)^4}d^4r e^{\frac{i}{2}(-\Pi^2+m^2)\frac{\bar n\cdot r}{\bar n\cdot\Pi}}\nn\\
&\times\paren{1+\frac{i e\sigma F}{2}\ov{2\bar n\cdot p}\frac{\pd}{\pd k_+}}\gam^\mu(\slashed{q}+m)e^{i q\cdot M^{-1}\cdot q s-i\frac{e\sigma F}{2}s-i(m^2-i\eps)s}\gam_\mu\nn\\
&\times \paren{1-\frac{i e \sigma F}{2}\ov{\bar n\cdot p}\frac{\pd}{\pd k_+}}e^{i(\Pi-k-q)\cdot r}.\label{eq:FTSEU.22}
\end{align}
Now the $r$ integral yields a delta function
\begin{align}
\int &d^4 r\exp\paren{i\frac{2 p\cdot e A-(e A)^2}{2\bar n\cdot \Pi}\bar n\cdot r+i(\Pi-k-q)^\mu r_\mu}\nn\\
&=(2\pi)^4\delta\paren{\frac{2 p\cdot e A-(e A)^2}{\bar n\cdot\Pi}+n\cdot(\Pi-k-q)}\delta(\bar n\cdot(\Pi-k-q))\delta_\perp^2(\Pi-k-q),\label{eq:FTSEU.23}
\end{align}
which can be used to perform the $q$ integral, setting
\begin{align}
q_+&=n\cdot q=n\cdot \Pi-n\cdot k+\frac{2 p\cdot e A-(e A)^2}{\bar n\cdot\Pi},\nn\\
q_-&=\bar n\cdot q=\bar n\cdot(\Pi-k),\qquad q_\perp=(\Pi-k)_\perp\,.\label{eq:FTSEU.24}
\end{align}
Note at this point, the expression only depends on the kinetic momentum on the external electron leg.

Keeping track of the position of the derivatives in $k_+$ as implied by the substitution above,
\begin{align}
i\Sig(p)&=-i(-i e)^2\int\frac{d^4k}{(2\pi)^4}\int_0^\infty ds\int_0^\infty dt e^{i(k^2+i\eps)t}e^{-i(m^2-i\eps)s}\nn\\
&\times\bigg\{\paren{1+\frac{i e \sigma F}{2}\ov{\bar n\cdot p}\frac{\pd}{\pd k_+}}\gam^\mu(\gam^\alpha)e^{-i\frac{e \sigma F}{2}s}\gam_\mu\paren{1-\frac{i e \sigma F}{2}\ov{\bar n\cdot p}\frac{\pd}{\pd k_+}}q_\alpha e^{i q\cdot M^{-1}\cdot q s}\nn\\
&+\paren{1+\frac{i e \sigma F}{2}\ov{\bar n\cdot p}\frac{\pd}{\pd k_+}}\gam^\mu m e^{-i\frac{e \sigma F}{2}s}\gam_\mu\paren{1-\frac{i e \sigma F}{2}\ov{\bar n\cdot p}\frac{\pd}{\pd k_+}}e^{i q\cdot M^{-1}\cdot q s}\bigg\}\,,\label{eq:FTSEU.25}
\end{align}
we now simplify the Dirac structure.  Using $\gam^\mu\sig^{\alpha\beta}\gam_\mu=0$, the second line in brackets simplifies,
\begin{align}
\paren{1+\frac{i e \sigma F}{2}\ov{\bar n\cdot p}\frac{\pd}{\pd k_+}}&\gam^\mu e^{-\frac{i e \sigma F}{2}s}\gam_\mu\paren{1-\frac{i e \sigma F}{2}\ov{\bar n\cdot p}\frac{\pd}{\pd k_+}}\nn\\
&=\paren{1+\frac{i e \sigma F}{2}\ov{\bar n\cdot p}\frac{\pd}{\pd k_+}}\gam^\mu \gam_\mu\paren{1-\frac{i e \sigma F}{2}\ov{\bar n\cdot p}\frac{\pd}{\pd k_+}}=4\,,\label{eq:FTSEU.26}
\end{align}
since $(\sigma F)^2=F^{\mu\nu}F_{\mu\nu}+\frac{i}{2}F^{\mu\nu}\eps_{\mu\nu\kappa\lambda}F^{\kappa\lambda}\gam_s\to 0$ for plane waves. The first line in brackets is
\begin{align}
\paren{1+\frac{i e \sigma F}{2}\ov{\bar n\cdot p}\frac{\pd}{\pd k_+}}& \gam^\mu \gam^\alpha e^{-i \frac{e \sigma F}{2}s}\gam_\mu\paren{1-\frac{i e\sigma F}{2}\ov{\bar n\cdot p}\frac{\pd}{\pd k_+}}\nn\\
&=-2\Bigg(e^{i\frac{e\sigma F}{2}s}\gamma^\alpha+\ov{\bar n\cdot p}\frac{\pd}{\pd k_+}\frac{i e}{2}4 i\gam_\beta F^{\beta\alpha} \nn\\
&+\frac{e^2}{4}\paren{\ov{\bar n\cdot p}\frac{\pd}{\pd k_+}+s}\ov{\bar n\cdot p}\frac{\pd}{\pd k_+}8\gam_\beta F^{\beta\kappa} F_\kappa^{\phantom{\kappa}\alpha}\Bigg)\label{eq:FTSEU.27}
\end{align}
See Dirac matrix identities above. All other combinations of field tensors vanish due to plane wave properties.  Then we have,
\begin{align}
i\Sig(p)&=-i (-i e)^2\int \frac{d^4k}{(2\pi)^4}\int_0^\infty ds\int_0^\infty dt e^{i(k^2+i\eps)t}e^{-i(m^2-i\eps)s}\nn\\
&\times\bigg\{4 m e^{i q\cdot M^{-1}\cdot q s}-2\bigg[ e^{i\frac{e\sigma F}{2}s}\gam^\alpha+\ov{\bar n\cdot p}\frac{\pd}{\pd k_+}(-2 e\gam\cdot F)^\alpha 
\nn\\
&+2e^2(\gam\cdot F\cdot F)^\alpha\paren{\ov{\bar n\cdot p}\frac{\pd}{\pd k_+}+s}\ov{\bar n\cdot p}\frac{\pd}{\pd k_+}\bigg] q_k e^{i q\cdot M^{-1}\cdot q s}\bigg\}
\end{align}
with $q^\mu$ given by \eq{FTSEU.24}.  Now to evaluate the $k_+$ derivatives, we use that the inverse of $M$ is
\begin{align*}
M_{\alpha\beta}^{-1}&=g_{\alpha\beta}-\frac{s^2}{3}e F_{\alpha\kappa}e F^{\kappa}_{\phantom{\kappa}\beta}\nn\\
&=\frac{\bar n^\alpha n^\beta}{2}+\frac{n^\alpha \bar n^\beta}{2}+g_{\alpha\beta}^\perp-\frac{s^2}{3}\bar n^\alpha \bar n^\beta (e F)^2,
\end{align*}
and this yields
\begin{align}
\fpp{}{k_+}\left[q_\alpha e^{i q\cdot M^{-1}\cdot q s}\right]&=\frac{\bar n^\alpha}{2}e^{i q\cdot M^{-1}\cdot q s}+(q^\alpha \bar n\cdot q i s)e^{i q \cdot M^{-1}\cdot q s}\label{eq:FTSEU.33}\\
\fpp{{}^2}{k_+^2}\left[q_\alpha e^{i q\cdot M^{-1}\cdot q s}\right]
&=i s \bar n\cdot q(\bar n^\alpha+q^\alpha \bar n\cdot q i s)e^{i q \cdot M^{-1}\cdot q s}.\label{eq:FTSEU.34}
\end{align}
Using the basic property of plane wave fields that $\bar n_\mu F^{\mu\nu}=0$, we simplify to
\begin{align}
i\Sig(p)&=-i(-i e)^2\int_0^\infty ds\int_0^\infty dt \int\frac{d^4 k}{(2\pi)^4}e^{i(k^2+i\eps)t-i(m^2-i\eps)s+i(k-\Pi)\cdot M^{-1}(k-\Pi)s}\nn\\
&\times\bigg\{ 4 m-2\bigg[e^{i\frac{e\sigma F}{2}s}\paren{\slashed{\Pi}-\slashed{k}}-2\gam\cdot eF\cdot (\Pi-k)\frac{\bar n\cdot(\Pi-k)}{\bar n\cdot\Pi}i s\nn\\
&+2 i s^2 \gam\cdot e F\cdot e F\cdot(\Pi-k)\frac{\bar n\cdot(\Pi-k)}{\bar n\cdot \Pi}\paren{i\frac{\bar n\cdot(\Pi-k)}{\bar n\cdot\Pi}+1}\bigg]\bigg\}\,,\label{eq:FTSEU.36}
\end{align}
and change variables of integration, $u=s+t, v=s-t\in[-u,u]$ with $ds dt=\frac{du dv}{2}$, to obtain
\begin{align}
\Sig(p)&=+e^2\int_0^\infty \frac{du}{2}\int_{-u}^u dv\int\frac{d^4k}{(2\pi)^4}e^{i(k^2+i\eps)\frac{u-v}{2}-i(m^2-i\eps)\frac{u+v}{2}+i(\Pi-k)\cdot M^{-1}\cdot(\Pi-k)\frac{u+v}{2}}\nn\\
&\times\bigg\{4 m-2\bigg[e^{i\frac{e\sigma F}{2}\frac{u+v}{2}}(\slashed{\Pi}-\slashed{k})-2i\frac{u+v}{2}\gam\cdot e F\cdot(\Pi-k)\frac{\bar n\cdot(\Pi-k)}{\bar n\cdot\Pi}\nn\\
&+2i s^2\gam \cdot e F\cdot e F\cdot(\Pi-k)\frac{\bar n\cdot(\Pi-k)}{\bar n\cdot \Pi}\paren{i\frac{\bar n\cdot(\Pi-k)}{\bar n\cdot\Pi}+1}\bigg]\bigg\}\,.\label{eq:FTSEU.37}
\end{align}
Changing variables again $v=u(1-2x)$, $dv=-2u dx$ $[-u,u]\mapsto [1,0]$ and $k-\Pi=-k'$ with $d^4k=d^4k'$, we have
\begin{align}
\Sig(p)&=e^2\int_0^\infty du u\int_0^1 dx\int\frac{d^4 k'}{(2\pi)^4}e^{i\paren{(-k'+\Pi)^2+i\eps}u x-i(m^2-i\eps) u(1-x)+ik'\frac{\text{tanh} e F u(1-x)}{e F u (1-x)}k' u (1-x)}\nn\\
&\times\bigg\{ 4m-2\bigg[ e^{i\frac{e\sigma F}{2} u(1-x)}(+\slashed{k}')-2\gam\cdot e F\cdot (+k') \frac{\bar n\cdot (+k')}{\bar n\cdot\Pi}i u(1-x)\nn\\
&+2 i (u(1-x))^2 \gam\cdot e F\cdot e F\cdot(+k')\frac{\bar n\cdot(+k')}{\bar n\cdot\Pi}\paren{i\frac{\bar n\cdot(+k')}{\bar n\cdot\Pi}+1}\bigg]\bigg\}\,.\label{eq:FTSEU.38}
\end{align}
After completing the square in $k'$ in the exponent,
\begin{align}
\Sig(p)&=e^2\int_0^\infty du u\int_0^1 dx\int\frac{d^4 k''}{(2\pi)^4}e^{i(k'' Bk''+i\eps)u x+i(\Pi^2+i\eps)u x-i(\Pi B^{-1}\Pi)u x}\nn\\
&e^{-i(m^2-i\eps) u (1-x)}\bigg\{4 m-2\bigg[ e^{i\frac{e\sigma F}{2}u (1-x)} (k''+\Pi\cdot B^{-1})\cdot\gam \nn\\
&-2 i u(1-x) \gam\cdot e F\cdot(k''+B^{-1}\!\cdot\Pi)\frac{\bar n\cdot(k''+B^{-1}\cdot\Pi)}{\bar n\cdot\Pi}\nn\\
&+2i u^2 (1-x)^2\gam\!\cdot\! (e F)^2\!\cdot(k''+B^{-1}\!\cdot\Pi)\frac{\bar n\cdot(k''+B^{-1}\Pi)}{\bar n\cdot\Pi}\paren{i\frac{\bar n\cdot(k''+ B^{-1}\!\cdot\Pi)}{\bar n\cdot\Pi}+1}\bigg]\bigg\}\,,\label{eq:FTSEU.39}
\end{align}
with $k''=k'-\Pi\cdot B^{-1}$ and $B=1+M^{-1}$.  Note that for simplicity of notation we have redefined
\be
n\cdot \Pi=\Pi_+\to \Pi_+ +\frac{2p\cdot e A-(e A)^2}{\bar n \cdot \Pi}
\ee
when inserting $q^\mu$ from \eq{FTSEU.24}. 
Dropping the primes on $k$, and terms odd in $k$, which vanish under integration, we simplify the remaining terms using that
\begin{align}
B^{-1}&=\ov{1+\frac{\text{tanh e F u (1-x)}}{e F u x}}=x\paren{1+\frac{(e F)^2}{3}u^2 (1-x)^2}\label{eq:FTSEU.41a}\\
1-B^{-1}&=(1-x)-x\frac{(e F)^2}{3}u^2 (1-x)^3\label{eq:FTSEU.41b}\\
\det(-i u x B)^{-1/2}&
=(-i u)^{-d/2}\label{eq:FTSEU.41c}
\end{align}
in matrix notation for $B$.  The expression for $B^{-1}$ implies that
\begin{align*}
e F\cdot B^{-1}\cdot \Pi&=e F\cdot\paren{1+\frac{(e F)^2}{3}u^2(1-x)^3}x\cdot\Pi=e F\cdot\Pi x\\
\frac{\bar n\cdot B^{-1}\cdot\Pi}{\bar n\cdot\Pi}&=\frac{\bar n\cdot\Pi x}{\bar n\cdot\Pi}=x
\end{align*}
since $e F\cdot (e F)^2=0$.  Then we obtain,
\begin{align}
\Sig(p)&=e^2\int_0^\infty du u \int_0^1 dx \int \frac{d^4k}{(2\pi)^4} e^{i (k\cdot B\cdot k+i\eps)u x+i(\Pi(1-B^{-1})\Pi u x-i (m^2-i\eps)u (1-x)}\nn\\
&\times \bigg\{ 4m-2\bigg[ e^{i\frac{e\sigma F}{2}u(1-x)}\Pi\cdot\paren{1+\frac{(e F)^2}{3}u^2(1-x)^3}\cdot \gam x\nn\\
&-2 i u (1-x)\paren{\gam \cdot e F\cdot n\frac{(\bar n\cdot k)^2}{2\bar n\cdot\Pi}+\gam\cdot e F\cdot\Pi x^2}\nn\\
&+2 i u^2(1-x)^2\bigg( \gam\cdot (e F)^2\cdot n\frac{(\bar n\cdot k)^2}{2\bar n\cdot \Pi}(i x+1)+\gam\cdot(e F)^2\cdot n\frac{\bar n\cdot k}{2}x i \frac{\bar n\cdot k}{\bar n\cdot \Pi}\nn\\
&+\gam\cdot (e F)^2\cdot \Pi i\paren{\frac{\bar n\cdot k}{\bar n\cdot\Pi}}^2+\gam\cdot (e F)^2\cdot \Pi x(i x+1)\bigg)\bigg]\bigg\} \label{eq:FTSEU.42}
\end{align}

Now we use that $\int d^4k e^{i k^2 u x}k^\mu k^\nu=\int\frac{d^4k}{(u x)^2}e^{i k^2}\frac{k^2}{u x}\frac{g^{\mu\nu}}{d}$, and 
since $\bar n\cdot B^{-1}=\bar n$ and $\bar n \cdot e F=0$, several terms vanish, leaving
\begin{align}
\Sig(p)
&=\frac{e^2}{(4\pi)^2}\int_0^\infty \frac{du}{u}\int_0^1 dx e^{i\Pi\paren{1-\frac{(e F u)^2}{3}x (1-x)^2}\Pi u x (1-x)-i(m^2-i\eps)u(1-x)}\nn\\
&\times\bigg\{4 m-2 x e^{i\frac{e \sigma F}{2}u(1-x)}\slashed{\Pi}+4 i u(1-x) x^2\gam\cdot e F\cdot \Pi
\nn\\
&-2 u^2(1-x)^2x\paren{\frac{1-x}{3}e^{i\frac{e\sigma F}{2}u(1-x)}+2i(i x+1)}\gam\cdot(e F)^2\cdot\Pi\bigg\}.\label{eq:FTSEU.45}
\end{align}
This is our final result for the self-energy in the presence of classical plane-wave.

We first check the limit $e F\to 0$,
\be\label{eq:FTSEU.46}
\Sig(p)\Bigl\vert_{e F=0}=\frac{\alpha}{4\pi}\int_0^\infty\frac{du}{u}\int_0^1 dx e^{i p^2 u x (1-x)-i(m^2-i\eps)u (1-x)}(-2\slashed{p}x+4 m),
\ee
verifying agreement with zero-field QED result. For the wave function renormalization, we need
\begin{align}
\frac{d\Sigma}{d\slashed{\Pi}}&=\frac{\alpha}{4\pi}\int_0^\infty \frac{du}{u}\int_0^1 dx\bigg\{-\!2 x e^{i\frac{e\sigma F}{2}u(1-x)} \nn \\
&+\paren{2i\slashed{\Pi}u x (1\!-\!x)-\frac{2i}{3}\gam\cdot \!(e F)^2\!\cdot \Pi u^3(1\!-\!x)^2 x^2}\times\nn\\
&\times \bigg[ 4m-2x e^{i\frac{e\sigma F}{2}u(1-x)}\slashed{\Pi} \nn\\
&-2 u^2(1-x)^2x\paren{\frac{1\!-\!x}{3}e^{i\frac{e\sigma F}{2}u(1-x)}+(i x+1)2 i}\gam\cdot(e F)^2\cdot \Pi\nn\\
&+4i u(1-x)x^2\gam\cdot e F\cdot\Pi\bigg]\bigg\} e^{i\Pi \paren{1-\frac{(e F u)^2}{3}x(1-x)^2}\Pi u x (1-x)-i(m^2-i\eps)u(1-x)},\label{eq:FTSEU.47}
\end{align}
in which we have used
\begin{align}
\frac{d}{d\slashed{\Pi}}\gam\cdot e F\cdot\Pi=0=\frac{d}{d\slashed{\Pi}}\gam\cdot (e F)^2\cdot \Pi,
\nn\\
 \frac{d}{d\slashed{\Pi}}\Pi^2=2\slashed{\Pi},\qquad \frac{d}{d\slashed{\Pi}}\Pi\cdot (e F)^2\cdot\Pi=2\gam \cdot (e F)^2\cdot\Pi. \label{eq:FTSEU.48}
\end{align}
For soft electrons, set $\slashed{\Pi}=m$,
\begin{align}
\frac{d\Sig}{d\slashed{\Pi}}\Bigl\vert_{\slashed{\Pi}=m}&=\frac{\alpha}{4\pi}\int_0^\infty \frac{du}{u}\int_0^1 dx\bigg\{ -2x e^{i\frac{e\sigma F}{2}u(1-x)} \nn\\
&+2i u x (1-x)\paren{m-\ov{3}\gam\cdot (e F)^2\cdot\Pi u^2(1-x)^2 x}\bigg[4m-2 m x e^{i\frac{e\sigma F}{2}u(1-x)}\nn\\
&-2u^2(1-x)^2 x\paren{\frac{1\!-\!x}{3}e^{i\frac{e\sigma F}{2}u(1-x)}+(i x+1)2 i}\gam\cdot\! (eF)^2\!\cdot\Pi\nn\\ 
&+4 i u(1-x) x^2\gam\cdot e F\cdot \Pi\bigg]\bigg\} e^{i (m^2+i\eps)(u x(1-x)-u(1-x))-i\frac{u^3}{3}(1-x)^3 x^2 \chi^2 m^4}.\label{eq:FTSEU.49}
\end{align}
The log divergence is isolated by integrating by parts in $x$ on the first term in braces.  The second term in braces is finite. We scale out an arbitrary momentum scale from the integration variable, setting $u\to u/\mu^2$.  Additionally changing variables trivially in the $x$ integral $x\mapsto 1-x$, we focus on the first term and obtain
\begin{align}
\frac{d\Sig}{d\slashed{\Pi}}\Bigl\vert_{\slashed{\Pi}=m}
&=\frac{\alpha}{4\pi}\int_0^\infty \frac{du}{u}(-1)e^{-i\frac{m^2-i\eps}{\mu^2}u}\nn\\
&+\frac{\alpha}{4\pi}\int_0^\infty \!i du\int_0^1 \!dx(x^2-2 x)\bigg[\frac{m^2}{\mu^2}2 x+\frac{\chi^2 m^4}{\mu^6}\frac{u^2}{3}(1-x)x^2(3-5x)\bigg] \nn \\
&\times e^{-i\frac{m^2}{\mu^2} u x^2-i\frac{u^3}{3}(1-x)^2 x^3\frac{\chi^2 m^4}{\mu^6}}+\textrm{additional~finite~terms}.\label{eq:FTSEU.51}
\end{align}
The term on the first line is a representation of a logarithm, in exact agreement with vacuum result
\be\label{eq:FTSEU.52}
\frac{d\Sig}{d\slashed{\Pi}}\bigg\vert_{\slashed{\Pi}=m}=\frac{\alpha}{4\pi}\ln\frac{m^2-i\eps}{\mu^2}+(\text{finite}),
\ee
while all the remaining terms are finite.

For high momentum electrons, the hierarchy of momentum scales
\be\label{eq:FTSEU.54'}
p^2\gg m^2\gg (p\cdot e F)^{2/3}=m^2\paren{\frac{\chi}{m}}^{2/3}
\ee
shows that dropping $m^2$ terms requires also dropping $\sim p\cdot e F$ terms.  With this simplification, we have
\begin{align}
\frac{d\Sig}{d\slashed{\Pi}}\bigg\vert_{m^2,p\cdot e F\to 0}&=\frac{\alpha}{4\pi}\int_0^\infty \frac{du}{u}\int_0^1 dx\bigg\{-2 x e^{i\frac{e\sigma F}{2}u(1-x)} \nn\\
&+2i\slashed{\Pi}u x (1-x)\paren{-2 x e^{i\frac{e\sigma F}{2}u(1-x)}\slashed{\Pi}}\bigg\} e^{i\Pi^2 u x(1-x)}\nn\\
&=\frac{\alpha}{4\pi}\paren{\ln\frac{-\Pi^2+i\eps}{\mu^2}+\int_0^1 \!dx\, 2 x \ln[x(1-x)]+2+4\slashed{\Pi}\frac{e\sigma F}{2}\slashed{\Pi}\frac{-1}{(-\Pi^2)^2}}\label{eq:FTSEU.54}
\end{align}
again in agreement with the zero-field QED result.

Finally, to determine the total rate of photon emission, we calculate the imaginary part, corresponding to the cut self-energy diagram,
\be\label{eq:FTSEU.55}
\Gam(e\to e \gam)=\includegraphics[width=.25\textwidth]{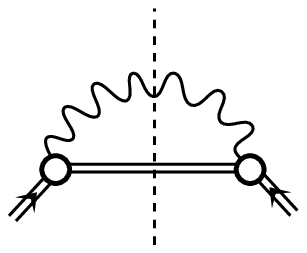}=-\ov{\pi}\text{Im}\,\text{tr}\,\Sig(p).
\ee
First observe that since the Dirac trace of any odd number of $\gamma$ matrices vanishes, the trace of $\Sigma(p)$ simplifies,
\begin{align}
\tr\Sig(p)
&=\frac{4\alpha}{\pi}m\int_0^\infty\frac{du}{u}\int_0^1 dx e^{i\Pi^2 u x (1-x)-i(m^2-i\eps)u(1-x)-i\chi^2m^4\frac{u^3}{3}x^2(1-x)^3},\label{eq:FTSEU.56}
\end{align}
using that $\tr\mathbb{I}=4$. Setting $\Pi^2=m^2$ for on-shell electrons and changing variables $x\to 1-x$, we integrate by parts in $x$ to remove the log-divergent piece, which does not contribute to the imaginary part.  Then,
\begin{align}
\ov{m}\tr\Sig(p)&=\frac{4\alpha}{\pi}\int_0^\infty\frac{du}{u}\int_0^1 dx e^{-i u x^2 m^2-i\frac{u^3}{3}x^3(1-x)^2\chi^2m^4}\nn\\
&=-\frac{4\alpha}{\pi}\ln\frac{m^2}{\mu^2} +\frac{4\alpha}{\pi}\int_0^\infty \!i du\int_0^1 \!\!dx\, x^2\paren{2m^2+\frac{u^3}{3}x(1-x)(3-5x)\chi^2m^4}\nn\\
&\times e^{-i u x^2m^2-i\frac{ u^3}{3}x^3(1-x)^2\chi^2}.
\end{align}
To obtain the imaginary part, we use that the pieces even in $u$ are real while the pieces odd in $u$ contributes the imaginary part. Hence we extend the integration over the whole real line, $u\in (-\infty,\infty)$ and divide by $2i$ (\cite{dittrich2000probing}),
\begin{align}
\ov{m}\text{Im}\tr\Sig(p)&=\frac{4\alpha}{\pi}\ov{2i}\int_{-\infty}^\infty i du\int_0^1 dx x^2\paren{2m^2+\frac{u^2}{3}x(1-x)(3-5x)\chi^2m^4} \nn\\
&\times e^{-iu x^2m^2-i\frac{u^3}{3}x^3(1-x)^2\chi^2m^4}\nn\\
&=\frac{2\alpha}{\pi}\ov{i}\int_{-\infty}^\infty i du\int_0^1 dx x^2\paren{2+\frac{u^2}{3}x(1-x)(3-5x)\frac{\chi^2}{m^2}} \nn\\
&\times e^{-i u x^2-\frac{i u^3}{3}x^3(1-x)^2\paren{\frac{\chi}{m}}^2},\label{eq:FTSEU.58}
\end{align}
which corresponds to the imaginary part being defined by $m^2-i\epsilon$ and closing the $u$-contour in the lower half plane.  In transiting the second line, we scaled $u$ by $m^2u\to u$.  Now from \cite{abramowitz1964handbook}, we recognize an integral definiton of the Airy function,
\be\label{eq:FTSEU.59}
(3a)^{-1/2}\pi\mathrm{Ai}\paren{\pm(3 a)^{-1/2}z}=\int_0^\infty \cos(a t^3\pm z t)dt=\ov{2}\int_{-\infty}^\infty e^{\mp i z t-i a t^3}dt.
\ee
We will rewrite the expression for $\mathrm{Im}\tr\Sigma$ using the right-most integral, as well as the related expression,
\begin{align}
\int_{-\infty}^\infty t^2 e^{-i z t-i a t^3/3}dt
&=-\frac{2\pi}{a}\mathrm{Ai}''\paren{a^{1/3}z}=-\frac{2\pi}{a}\paren{a^{-1/3}z}\mathrm{Ai}(a^{-1/3}z)\,,\label{eq:FTSEU.61}
\end{align}
where in the second equality what used the fundamental differential equation for the Airy function
\begin{align}
\mathrm{Ai}''(z)-z\,\mathrm{Ai}(z)=0.
\end{align}
Therefore, we have 
\begin{align}
\int_{-\infty}^\infty du e^{-i u x^2-\frac{i u^3}{3}x^3 (1-x)^2\frac{\chi^2}{m^2}}
&=\frac{2\pi}{x(1-x)^{2/3}}\paren{\frac{\chi}{m}}^{-2/3}\mathrm{Ai}\paren{\frac{x m^{2/3}}{(1-x)^{2/3}\chi^{2/3}}},\label{eq:FTSEU.60}
\\
\int_{-\infty}^\infty u^2 e^{-i u x^2-i\frac{u^3}{3}x^3(1-x)^2\frac{\chi^2}{m^2}}du&=-\frac{2\pi}{x^3(1-x)^2}\paren{\frac{m}{\chi}}^2\mathrm{Ai}''\paren{\frac{x m^{2/3}}{(1-x)^{2/3}\chi^{2/3}}}. \label{eq:FTSEU.62}
\end{align}
Using \eqs{FTSEU.60}{FTSEU.62} in \eq{FTSEU.58}, we obtain
\begin{align}
\frac{1}{\pi}\text{Im}\ov{m}\tr\Sig(p)
&=\frac{4\alpha}{\pi}\paf{\chi}{m}^{-2/3}\int_0^1 d x \frac{x(1-\frac{1}{3}x)}{(1-x)^{5/3}}\mathrm{Ai}\paren{\frac{x}{(1-x)^{2/3}}\paf{\chi}{m}^{-2/3}},\label{eq:FTSEU.63}
\end{align}
completing the derivation.  

\section{Wilson lines in the EFT}

In this section, we separate nonperturbative couplings to photons into Wilson lines. Before integrating out the large momentum $\hat K$ in the electron lagrangian, we have,
\begin{align}
\cL&=\sum_{P,P'}\bar u_{P'}e^{iP'_L\cdot x}e^{iP'_rx}Y_{P'}^\dg e^{\frac{i}{2}\hat K^\dag \bar n\cdot x}\Bigg(i\Dslash -\frac{\slashed{\bn}}{2}\left\{\frac{e\Aslash_{cl}}{i\bar n\cdot D},i\Dslash_\perp\right\}\nn\\
&+\frac{\slashed{\bn}}{2}\frac{(eA_{cl})^2}{i\bar n\cdot D}-m\del_{PP'}\Bigg)e^{-\frac{i}{2}\hat K\bar n\cdot x}Y_Pe^{-iP_L\cdot x}e^{-iP_rx}u_P.\label{app:89eq6}
\end{align}
where
\be\label{eq:B003eq1}
\hat K=\left\{eA_{cl}^\mu, iD_\perp^\mu\right\}\oneov{i\bar n\cdot D}-\frac{(eA_{cl})^2}{i\bar n\cdot D}.
\ee
Recall that $i\bn\cdot\pd \hat K=0$ by symmetry. Define a Wilson line for the $\bn$ direction collinear photons, satisfying
\be
\bn\cdot DW_\bn=0, \qquad W_\bn^\dg(i\bn\cdot D)W_\bn=i\bn\cdot \pd ,
\ee
which implies
\be
W_\bn\ov{i\bn\cdot\pd}W_\bn^\dg=\ov{i\bn\cdot D}.
\ee
An analogous Wilson line is defined for the $\bn$ direction soft photons, $S_\bn$, with an identical form.  Using this definition to rewrite $\hat K$, we find
\begin{align}
\hat K&=\left\{W_\bn\frac{eA_{cl}^\mu}{i\bn\cdot\pd}W_\bn^\dg,iD_\perp^\mu\right\}-W_\bn\frac{(eA_{cl)^2}}{i\bn\cdot\pd}W_\bn^\dg\label{eq:B003eq3}\\
&=W_\bn\frac{eA_{cl}}{i\bn\cdot\pd}W_\bn^\dg i\slashed{D}_\perp^\mu+iD_\perp W_\bn\frac{eA_{cl}}{i\bn\cdot\pd}W_\bn^\dg-W_\bn\frac{(eA_{cl})^2}{i\bn\cdot\pd}W_\bn^\dg\nn\\
&=W_\bn\paren{\frac{eA_{cl}^\mu}{i\bn\cdot \pd}iD_\perp^\mu+iD_\perp^\mu\frac{eA_{cl}^\mu}{i\bn\cdot\pd}}W_\bn^\dg+W_\bn\frac{eA^\mu}{i\bn\cdot\pd}[W_\bn^\dg ,iD_\perp^\mu]\nn\\
&+[iD_\perp^\mu,W_\bn]\frac{eA_{cl}}{i\bn\cdot\pd}W_\bn^\dg-W_\bn\frac{(eA_{cl})^2}{i\bn\cdot\pd}W_\bn^\dg.
\end{align}
We need several commutators,
\begin{align}
([W_\bn^\dg,iD_\perp^\mu])^\dg&=(W_\bn^\dg iD_\perp-iD_\perp W_\bn^\dg)^\dg=iD_\perp^\dg W_\bn-W_\bn iD_\perp^\dg,\label{eq:B003eq4}\\
&=[iD_\perp^\dg,W_\bn]=-[iD_\perp,W_\bn],\nn\\
[W_\bn^\dg,iD_\perp^\mu]&=(-iD_\perp^\mu W_\bn^\dg)=\paren{-iD_\perp^\mu\frac{-1}{i\bn\cdot\pd}e\bn\cdot A_\bn}W_\bn^\dg,\label{eq:B003eq5}\\
&=\paren{i\pd_\perp^\mu\ov{i\bn\cdot\pd}e\bn\cdot A_\bn}W_\bn^\dg=\paren{-\frac{\pd_\perp^\mu}{\bn\cdot\pd}e\bn\cdot A_\bn}W_\bn^\dg,\nn\\
[iD_\perp^\mu,W_\bn]&=(iD_\perp W_\bn)=W_\bn\paren{i\pd_\perp^\mu\ov{i\bn\cdot\pd}e\bn\cdot A_\bn}.\label{eq:B003eq6}
\end{align}
Using these commutators, we push the Wilson lines out to the left and right, obtaining
\begin{align}
\hat K&=W_\bn\paren{\left\{\frac{eA_{cl}^\mu}{i\bn\cdot\pd},iD_\perp^\mu\right\}-\frac{(eA_{cl})^2}{i\bn\cdot\pd}}W_\bn^\dg
\nn\\&
+W_\bn\frac{eA_{cl}^\mu}{i\bn\cdot\pd}\paren{\frac{\pd_\perp^\mu}{i\bn\cdot\pd}e\bn\cdot A_\bn}W_\bn^\dg+W_\bn\frac{\pd_\perp^\mu}{\bn\cdot\pd}e\bn\cdot A_\bn\frac{eA_{cl}^\mu}{i\bn\cdot\pd}W_\bn^\dg,\label{eq:B003eq7}
\end{align}
in which the terms on the second line are seen as corrections to $\hat K$.  We define $\hat K\equiv W_\bn \tilde{\hat K}W_\bn^\dg$ and have
\begin{align}\label{eq:B003eq8}
\exp\paren{-\frac{i}{2}\hat K\bn\cdot x}&=\exp\paren{-\frac{i}{2}W_\bn\tilde{\hat K} W_\bn^\dg \bn\cdot x}=W_\bn\exp\paren{-\frac{i}{2}\tilde\hat{K}}W_\bn^\dg.
\end{align}
When we perform the same manipulations for $S_\bn$, note that the commutator of $S_\bn$ with $iD_\perp^\mu$ is next-to-leading order in $\lambda$.  Thus, they separate similarly, $\exp\paren{-\frac{i}{2}\hat K\bn\cdot x}\to S_\bn W_\bn \exp\paren{-\frac{i}{2}\tilde{\hat K}\bn\cdot x}W_\bn^\dg S_\bn^\dg$. 

Next, define $\perp$ Wilson lines according to
\begin{align}
\eps_{cl}^\mu iD_\perp^\mu W_\perp&=\eps_{cl}^\mu\paren{i\pd^\mu-eA_{\bn\perp}^\mu}W_\perp=0\label{eq:B003eq9}
\end{align}
where $\eps_{cl}$ is the polarization vector of the classical laser field.  Using this Wilson to separate the photons from the $D_\perp$ covariant derivatives, we have
\begin{align}
\tilde {\hat K}
&=W_\perp\paren{\left\{\frac{eA_{cl}^\mu}{i\bn\cdot\pd},i\pd_\perp^\mu\right\}}W_\perp^\dg+\left[\frac{eA_{cl}^\mu}{i\bn\cdot\pd},W_\perp\right]i\pd_\perp^\mu W_\perp^\dg+W_\perp i\pd_\perp^\mu\left[W_\perp^\dg,\frac{eA_{cl}^\mu}{i\bn\cdot\pd}\right]\nn\\
&-W_\perp\frac{(eA_{cl})^2}{i\bn\cdot\pd}W_\perp^\dg+W_\perp\left[\frac{(eA_{cl})^2}{i\bn\cdot\pd},W_\perp^\dg\right].\end{align}
We need the additional commutators,
\begin{align}
\left[\ov{i\bn\cdot\pd},W_\perp\right]&=\ov{i\bn\cdot\pd}W_\perp=W_\perp\ov{i\bn\cdot\pd}\ov{\eps\cdot\hat P_\perp}e\eps\cdot A_\bn,\label{eq:B003eq11}\\
\left[W_\perp^\dg,\ov{i\bn\cdot\pd}\right]&=\paren{-\ov{i\bn\cdot\pd}W_\perp^\dg}
=\paren{\ov{i\bn\cdot\pd}\ov{\eps\cdot\hat P_\perp}e\eps\cdot A_\bn}W_\perp^\dg,
\end{align}
after which we obtain
\begin{align}
\tilde{\hat K}&=W_\perp\paren{\left\{\frac{eA_{cl}^\mu}{i\bn\cdot\pd},i\pd_\perp^\mu\right\}-\frac{(eA_{cl}^2)}{i\bn\cdot\pd}}W_\perp^\dg+W_\perp\paren{\ov{i\bn\cdot\pd}\ov{\hat P_\perp}e\eps\cdot A_\bn}eA_{cl}^\mu i\pd_\perp^\mu W_\perp^\dg \nn\\
&+W_\perp eA_{cl}^\mu \hat P_\perp^\mu\paren{\ov{i\bn\cdot\pd}\ov{\hat P_\perp}e\eps\cdot A_\bn}W_\perp^\dg+W_\perp (eA_{cl})^2\paren{\frac{-1}{i\bn\cdot\pd}\ov{\hat P_\perp}e\eps\cdot A_\bn}W_\perp^\dg\nn\\
&+\ldots
\end{align}
As above, we may now write,
\begin{align}
\tilde{\hat K}&=W_\perp \hat K_c W_\perp^\dg\nn\\
\exp\paren{-\frac{i}{2}\tilde{\hat K}\bn\cdot x}&=\exp\paren{-\frac{i}{2}W_\perp \hat K_c W_\perp^\dg \bn\cdot x}=W_\perp\exp\paren{-\frac{i}{2}\hat K_c\bn\cdot x}W_\perp^\dg,
\end{align}
in which $\hat K_c$ is given by \req{hatKcdefn}.  This same separation works on the denominator of the effective lagrangian.

\section{Identities Used in Factorization}
\subsection{Fierz Identity for Light-like Currents}\label{app:Fierzid}

Next we present a side calculation used factorization by applying Fierz identity to separate spin components. 
All $4\times 4$ matrices can be written as linear combination of the matrices
\be\label{Gammabasis}
\Gamma=\left\{1,\gamma^\mu,\frac{\sigma^{\mu\nu}}{\sqrt{2}},i\gamma^\mu\gamma^5,\gamma^5\right\}.
\ee
Using the general Fierz decomposition, we have
\begin{align}
[\bar\psi\gamma_\perp^\mu\chi][\bar\chi\gamma_\perp^\mu\psi]&=-\oneov{4}\Bigg([\bar\psi\psi][\bar\chi\chi]4-2[\bar\psi\gamma^\mu\psi][\bar\chi\gamma^\mu\chi]+0+2[\bar\psi i\gamma^\mu\gamma^5\psi][\bar\chi i\gamma^\mu\gamma^5\chi]-4[\bar\psi\gamma^5 \psi][\bar\chi\gamma^5 \chi]\Bigg)
\end{align}
with terms following the order of the list \req{Gammabasis}.  When $\chi$ is a collinear spinor, most of these terms vanish:
\begin{align}
\bar\chi\chi&=\bar\chi\frac{\slashed{\bn}\slashed{n}}{4}\chi=0\nn \\
\bar\chi\gamma^5\chi&=\bar\chi\frac{\slashed{n}\slashed{\bn}}{4}\gamma^5\chi=0\nn\\
\bar\chi i\gamma^\mu\gamma^5\chi&=\bar\chi i\frac{\slashed{n}\slashed{\bn}}{4}\gamma^\mu\gamma^5\chi
=\bar\chi i\paren{\frac{\slashed{\bn}\slashed{n}}{4}\frac{\slashed{n}}{2}\bn^\mu+\gamma_\perp^\mu\frac{\slashed{n}\slashed{\bn}}{4}}\gamma_5\chi\nn\\
&=-i\bn^\mu\bar\chi\frac{\slashed{n}\slashed{\bn}}{4}\gamma_5\frac{\slashed{n}}{2}\chi=0.
\end{align}
The last is seen to vanish on writing out $\gamma^5=i\epsilon^{\mu\nu\rho\sigma}\gamma_\mu\gamma_\nu\gamma_\rho\gamma_\sigma/4!$.  Therefore, we obtain
\begin{align}
[\bar\psi\gamma_\perp^\mu\chi][\bar\chi\gamma_\perp^\mu\psi]=\oneov{2}\left[\bar\psi\frac{\slashed{\bar n}}{2}n^\mu\psi\right]\left[\bar\chi\frac{\slashed{n}}{2}\bn^\mu\chi\right],
\end{align}
as used in the text.